\documentclass{report}
\usepackage{myjournal}
\usepackage{amsmath}
\usepackage{epsfig}
\usepackage[usenames]{color}
\usepackage{setspace}
\usepackage[printonlyused,withpage]{acronym}
\usepackage[top=2.5cm, bottom=2.5cm, left=3.8cm, right=3.8cm]{geometry}
\usepackage{url}

\title{Search for the Standard Model Higgs boson produced in association with a $W$ Boson in the isolated-track charged-lepton channel using the  Collider Detector at Fermilab}
\author{Adrian Buzatu\\
  Department of Physics,\\
  McGill University, Montreal\\
  \ \\
  \ \\
	August 2011
\ \\
  \ \\
  \ \\
A thesis submitted to McGill University in partial fulfilment\\ of the requirements of the degree of Doctor of Philosophy\\
\ \\
\ \\\copyright\  Adrian Buzatu, 2011}

\begin{document}
\pagenumbering{roman}

\date{} %this does not put automatically the data on the draftttt

\maketitle

\ \\
\clearpage

\tableofcontents
\listoftables
\listoffigures
\clearpage{\pagestyle{empty}\cleardoublepage}

\addcontentsline{toc}{chapter}{ABSTRACT}
\chapter*{Abstract\label{chapter:Abstract-En}}

\ \\The Higgs boson is the only elementary particle predicted by the Standard Model (SM) that has not yet been observed experimentally. If it exists, it explains the spontaneous electroweak symmetry breaking and the origin of mass for gauge bosons and fermions. We test the validity of the SM by performing a search for the associated production of a Higgs boson and a $W$ boson in the channel where the Higgs boson decays to a bottom-antibottom quark pair and the $W$ boson decays to a charged lepton and a neutrino (the $WH$ channel). We study a dataset of proton-antiproton collisions at a centre-of-mass energy $\sqrt{s}=1.96\ \tev$ provided by the Tevatron accelerator, corresponding to an integrated luminosity of 5.7 fb$^{-1}$, and recorded using the Collider Detector at Fermilab (CDF). We select events consistent with the signature of exactly one charged lepton (electron or muon), missing transverse energy due to the undetected neutrino (MET) and two collimated streams of particles (jets), at least one of which is required to be identified as originating from a bottom quark. We improve the discrimination of Higgs signal from backgrounds through the use of an artificial neural network. Using a Bayesian statistical inference approach, we set for each hypothetical Higgs boson mass in the range 100 - 150 $\gevcc$ with 5 $\gevcc$ increments a $95\%$ credibility level (CL) upper limit on the ratio between the Higgs production cross section times branching fraction and the SM prediction.

\ \\Our main original contributions are the addition of a novel charged lepton reconstruction algorithm with looser requirements (ISOTRK) with respect the electron or muon tight criteria (TIGHT), as well as the introduction of a novel trigger-combination method that allows to maximize the event yield while avoiding trigger correlations and that is used for the ISOTRK category. 

\ \\The ISOTRK candidate is a high-transverse-momentum good-quality track isolated from other activity in the tracking system and not required to match a calorimeter cluster, as for a tight electron candidate, or an energy deposit in the muon detector, as for a tight muon candidate. The ISOTRK category recovers real charged leptons that otherwise would be lost in the non-instrumented regions of the detector. This allows the reconstruction of more $W$ boson candidates, which in turn increases the number of reconstructed $WH$ signal candidate events, and therefore improves the sensitivity of the $WH$ search. 

\ \\For the TIGHT charged lepton categories, we employ charged-lepton-dedicated triggers to improve the rate of WH signal acceptance during data taking. Since there is no ISOTRK-dedicated trigger at CDF, for the ISOTRK charged lepton category we employ three MET-plus-jets-based triggers. For each trigger we first identify the jet selection where the trigger efficiency is flat with respect to jet information (transverse energy and direction of motion in the transverse plane for the two jets in the event) and then we parametrize the trigger efficiency as a function of trigger MET. On an event-by-event basis, for each trigger we compute a trigger efficiency as a function of trigger parametrization, trigger MET, jet information, trigger prescale and information about whether the trigger is defined or not. For the ISOTRK category we combine the three triggers using a novel method, which allows the combination of any number of triggers in order to maximize the event yield while avoiding trigger correlations. On an event-by-event basis, only the trigger with the largest efficiency is used. By avoiding a logical ``OR'' between triggers, the loss in the yield of events accepted by the trigger combination is compensated by a smaller and easier-to-compute corresponding systematic uncertainty. 

\ \\The addition of the ISOTRK charged lepton category to the TIGHT category produces an increase of 33\% in the $WH$ signal yield and a decrease of 15.5\% to 19.0\% in the median expected 95\% CL cross-section upper limits across the entire studied Higgs mass interval. The improvement in analysis sensitivity is smaller than the improvement in signal yield because the ISOTRK category has a smaller signal over background ratio than the TIGHT category, due to the looser ISOTRK reconstruction criteria. The observed (median expected) 95\% CL SM Higgs upper limits on cross section times branching ratio vary between 2.39 $\times$~SM (2.73 $\times$~SM) for a Higgs mass of 100 $\gevcc$ to 31.1 $\times$~SM (31.2 $\times$~SM) for a Higgs mass of 150 $\gevcc$, while the value for a 115 $\gevcc$ Higgs boson is that of 5.08 $\times$~SM (3.79 $\times$~SM).

\ \\The novel trigger combination method is already in use by several CDF analyses. It is applicable to any analysis that uses triggers based on MET and jets, such as supersymmetry searches at the ATLAS and CMS experiments at the Large Hadron Collider. In its most general form, the method can be used by any analysis that combines any number of different triggers. 

\clearpage{\pagestyle{empty}\cleardoublepage}

\addcontentsline{toc}{chapter}{R\'ESUM\'E}
\chapter*{Abr\'eg\'e\label{chapter:Abstract-Fr}}

\ \\Le boson de Higgs est la seule particule \'el\'ementaire pr\'edite par le Mod\`ele Standard qui n'a jamais \'et\'e observ\'ee exp\'erimentalement. S'il existe, il explique la brisure spontan\'ee de la sym\'etrie \'electrofaible, ainsi que la masse des bosons $W$ et $Z$ et de tous les fermions. On v\'erifie la validit\'e du Mod\`ele standard en effectuant une recherche sur la production associ\'ee d'un boson de Higgs et d'un boson $W$ dans le cas o\`u le boson de Higgs se d\'esint\`egre en une paire de quarks bottom-antibottom et le boson $W$ se d\'esint\`egre en un lepton charg\'e et un neutrino (le mode $WH$). Nos donn\'ees furent accumul\'ees en \'etudiant  des collisions proton-antiproton \`a une \'energie au centre de masse de $\sqrt{s}=1.96\ \tev$ produites par l'acc\'el\'erateur Tevatron, \`a une luminosit\'e integr\'ee de 5.7 $\invfb$ et collect\'ees par le d\'etecteur Collider Detector at Fermilab (CDF). On s\'electionne des \'ev\'enements avec une signature correspondante  \`a exactement un lepton charg\'e (\'electron ou muon), de l'\'energie manquante transversale \`a cause du neutrino qui s'\'echappe du d\'etecteur (MET) et deux jets de particules, dont au moins un doit provenir d'un quark bottom. On am\'eliore la discrimination entre le signal Higgs et le bruit de fond \`a l'aide d'un r\'eseau de neurones artificiels. En utilisant une inf\'erence statistique bayesienne, on calcule pour chaque masse hypoth\'etique du boson de Higgs dans l'intervalle 100-150 $\gevcc$, avec des increments de 5 $\gevcc$, une limite sup\'erieure de 95\% d'intervalle de credibilit\'e (CL), sur le rapport entre la section efficace multipli\'ee par le rapport d'embranchement et celle pr\'edite par le Mod\`ele Standard. 

\ \\Notre contribution principale est l'introduction d'un nouvel algorithme de reconstruction d'un lepton charg\'e avec des crit\`eres plus l\^aches (ISOTRK) par rapport aux crit\`eres stricts de reconstruction des candidats d'\'electrons et de muons (TIGHT). La deuxi\`eme contribution majeure consiste en l'introduction d'une nouvelle m\'ethode pour combiner des d\'eclencheurs diff\'erents permetant de maximiser le nombre d'\'ev\'enements s\'electionn\'es et en m\^eme temps que d'\'eviter les corr\'elations entre les d\'eclencheurs.

\ \\Un candidat de ISOTRK est une trajectoire qui correspond aux crit\`eres de qualit\'e, qui a une large quantit\'e de mouvement transverse, qui est isol\'ee d'autres activit\'es dans le d\'etecteur de trajectoires et qui ne doit pas se prolonger dans une zone active du calorim\`etre (d\'etecteur de muons), comme pour un candidat d'\'electron (muon). Les candidats de ISOTRK r\'ecup\`erent de vrais leptons charg\'es qui seraient autrement perdus dans les zones non instrument\'ees du d\'etecteur. L'ajout de la cat\'egorie ISOTRK \`a la cat\'egorie TIGHT permet de reconstruire plusieurs bosons $W$ r\'eels et par la suite de r\'ecup\'erer plusieurs \'ev\'enements $WH$, ce qui am\'eliore la sensitivit\'e de la recherche de $WH$.

\ \\Pour la cat\'egorie TIGHT des leptons charg\'es, on utilise des d\'eclencheurs d\'edi\'es aux leptons charg\'es. Comme \`a CDF il n'y a pas de d\'eclencheurs d\'edi\'es au leptons charg\'es ISOTRK, on utilise pour la cat\'egorie ISOTRK trois d\'eclencheurs bas\'es sur MET et jets. Pour chaque d\'eclencheur, on identifie la s\'election des jets pour laquelle l'efficacit\'e du d\'eclencheur est constante par rapport \`a l'information des jets (l'\'energie transverse et la direction de deplacement dans le plan transverse pour les deux jets de l'\'ev\'enement). Ensuite, on param\'etrise l'efficacit\'e du d\'eclencheur en fonction du MET au niveau du d\'eclencheur. Pour chaque \'ev\'enement, on calcule l'efficacit\'e du d\'eclencheur en fonction de la parametrisation du d\'eclencheur, le MET au niveau du d\'eclencheur, l'information des jets, le facteur de r\'eduction des d\'eclencheurs et l'information si le d\'eclencheur est d\'efini ou pas. Pour la cat\'egorie ISOTRK, on combine les trois d\'eclencheurs avec une nouvelle m\'ethode qui permet de maximiser le nombre d'\'ev\'enements accumul\'es tout en \'evitant les correlations des d\'eclencheurs. Pour chaque \'ev\'enement, ce n'est que le d\'eclencheur avec la plus grande efficacit\'e qui est utilis\'e. Le nombre d'\'ev\'enements ramass\'es est l\'eg\`erement inferieur \`a celui d'un ``OR'' logique entre les d\'eclencheurs, mais cela est compens\'e par une erreur syst\'ematique qui est \`a la fois moins importante et plus facile \`a \'evaluer.

\ \\L'utilisation de la cat\'egorie ISOTRK en plus de la cat\'egorie TIGHT augmente de 33\% le nombre d'\'ev\'enements $WH$ s\'electionn\'es et d\'ecroit de 15.5\% \`a 19.0\% la limite sup\'erieure m\'ediane attendue, exprim\'ee \`a un niveau de cr\'edibilit\'e de 95\%, calcul\'ee sur tout l'intervalle \'etudi\'e des masses du boson de Higgs. L'am\'elioration dans la sensitivit\'e de l'analyse est moins importante que l'am\'elioration du nombre d'\'ev\'enements collect\'es parce que la cat\'egorie ISOTRK a un plus faible rapport du signal sur le bruit de fond que la cat\'egorie TIGHT, \`a cause des crit\`eres de s\'election plus l\^aches pour la cat\'egorie ISOTRK. La limite sup\'erieure observ\'ee (attendue) du rapport entre la section efficace multipli\'ee par le rapport d'embranchement et celle pr\'edite par le Mod\`ele Standard, exprim\'ee en un niveau de cr\'edibilit\'e de 95\%, varie entre 2.39 $\times$~SM (2.73 $\times$~SM) pour un boson de Higgs de 100 $\gevcc$ jusqu'\`a 31.1 $\times$~SM (31.2 $\times$~SM) pour un boson de Higgs de 150 $\gevcc$. En m\^eme temps, pour un boson de Higgs de 115 $\gevcc$, la valeur est 5.08 $\times$~SM (3.79 $\times$~SM).  

\ \\La nouvelle m\'ethode de combiner des d\'eclencheurs diff\'erents est d\'ej\`a utilis\'ee par plusieurs analyses effectu\'ees \`a CDF. Elle peut \^etre utilis\'ee par toute analyse qui combine plusieurs d\'eclencheurs bas\'es sur le MET et les jets, comme par exemple la recherche de la supersym\'etrie \`a l'aide des d\'etecteurs ATLAS et CMS de l'acc\'el\'erateur Large Hadron Collider. Dans sa forme la plus g\'en\'erale, la m\'ethode peut \^etre utilis\'ee par toute analyse qui utilise un nombre variable de d\'eclencheurs diff\'erents.  

\clearpage{\pagestyle{empty}\cleardoublepage}
%\end{document}

\addcontentsline{toc}{chapter}{DEDICATION}
\chapter*{Dedication\label{chapter:Dedication}}

\ \\I dedicate this PhD thesis to the people that influenced me the most while I grew up in Romania: my parents, Dumitru Stefan Buzatu and Mariana Buzatu; my teachers from ``Mircea Eliade'' school, Craiova, especially my elementary school teacher, Maria Popescu, and later my maths teacher, Florin Benea; my teachers from ``Fratii Buzesti'' high school, Craiova, especially my physics teacher, Stefan Brinzan. 

%\ \\I dedicate this PhD thesis to the high school that formed me, ``Colegiul National Fratii Buzesti'', Craiova, Romania. I am especially thankful for the opportunity it offered me to meet my physics professor, Stefan Brinzan, who changed my life on more than one level. The physics and way of thinking he taught me allowed me to study in France at undergraduate level, which lead me to McGill University for my graduate studies. The sport I made in high school at his advice allowed me to solve a heart problem I used to have, thus offering me the chance to have a real life.

\clearpage{\pagestyle{empty}\cleardoublepage}

\addcontentsline{toc}{chapter}{ACKNOWLEDGEMENTS}
\chapter*{Acknowledgements\label{chapter:Acknowledgement}}

\ \\This work could not have been possible without the support and help I received from many people, including, but not limited to, my supervisor, my colleagues from McGill University and the Collider Detector at Fermilab experiment, my professors from university and high school, family and friends. 

%Andreas
\ \\I am most grateful for the support I received during my MSc and PhD graduate studies at McGill University from my research supervisor, Andreas Warburton. He encouraged me to choose the topics of research that I am most enthusiastic about from the analysis topics available to me as a member of the Collider Detector at Fermilab collaboration. Therefore, I chose the low mass Standard Model Higgs boson search, the topic of this PhD thesis. His office door was always open for me to present updates on my research progress. His feedback was always prompt, encouraging and constructive. He always answered to my emails in the shortest delays, even late at night or on weekends. The grant applications that Andreas and his faculty colleagues from CDF Canada applied for were always successful with NSERC, the national agency that funds particle physics research in Canada. As a result, I enjoyed the opportunity to travel frequently to Fermilab, the US particle physics laboratory where the research presented in this thesis was performed. For the same reason, I was also able to present my research results almost every year to the Canadian Association of Physicists Congress and attend several particle physics symposiums and schools. The graduate level particle physics class I took at McGill University was also taught very well by Andreas. I enjoyed the class very much and learnt the basic theoretical and experimental physics that I used afterwards in my research. I was also inspired by Andreas' contributions to science outreach and to the physics community in Canada. Following his example, I had a three-year term as graduate student representative in the Council of the Canadian Association of Physicists. In that period I had the opportunity to lead the local organization of a physics graduate student conference with students from across North America. Two years later I acted as the Canadian student contact for the Mexico edition the same conference. I am also very grateful for the many comments Andreas offered me in preparation for conference or regular research presentations. This allowed me to win the best oral presentation prize in the particle physics division at the Canadian Association of Physicists Congress in 2009. I also appreciate the thorough comments Andreas offered to my CDF internal documentation, to my MSc and PhD theses. Andreas' insights about career paths possible in particle physics were very helpful to me, too. Andreas is an excellent researcher, teacher and supervisor. I learnt a lot through interaction with him. Not least, he was also a good friend to me, supporting and understanding me in difficult personal times, such as a serious car accident I had as a passenger at the beginning of my MSc studies at McGill.  

%Nils 
\ \\This work could not have been possible without the help and guidance of Nils Krumnack while he was postdoctoral researcher at CDF, affiliated with Baylor University, USA. Nils taught me sound programming skills and guided my steps in implementing in code the parametrization of trigger efficiency turnon curves for three missing-transverse-energy-plus-jets triggers that I used in my analysis. It was also Nils who presented me the idea of the method to combine an unlimited number of triggers without having a logical ``OR'' between them in order to increase the signal acceptance while avoiding hard-to-estimate correlations between triggers. He also supported me while I coded the software package ABCDF for CDF that implements this idea. Nils also created the code that measures the data to simulated events efficiency scale factors for isolated track events. I maintained this code after Nils left CDF and used it to measure the isolated track scale factor for our analysis. These are key to my original contribution to the CDF analysis that searches for the associated production of a $W$ boson and a Higgs boson, namely the introduction of a new charged lepton category with looser reconstruction criteria (isolated tracks), in order to improve the sensitivity of the analysis. I am very grateful to Nils Krumnack for helping me understand triggers and for teaching me a lot of excellent coding skills. 

%Yoshikazu, Martin, Timo, Fede, Homer
\ \\I am also in debt to my fellow student colleagues from CDF: Yoshikazu Nagai (Tsukuba University, Japan), Martin Frank (Baylor University, USA), Federico Sforza (Pisa University, Italy) and Timo Aaltonen (Helsinki University, Finland). We worked together as part of the WH subgroup at CDF. I learnt a lot from interaction with them, either live at Fermilab or through video conferences and email exchanges. We used to offer each other advice on C++/ROOT coding and shell scripting, as well as on data analysis techniques. I interacted a lot with Yoshikazu during my effort to reproduce his $WH$ analysis in a new data analysis framework called WHAM. Yoshikazu has answered all my questions very patiently during all this time. He also offered me his scripts, which I personalized and included in WHAM. I interacted a lot with Martin in the process of building in WHAM the data ntuple tree used in this analysis. I collaborated a lot with Federico while validating the background estimation code that he introduced in WHAM based on the latest CDF background estimation methodology. I communicated a lot with Timo about the physics interpretation of the control plots for this analysis. I am also very grateful to Homer Wolfe, postdoctoral researcher at CDF (Ohio State University, USA) who created the foundations and leads the effort of development for WHAM.  

%Wei-Ming, Craig, Eric, Ben, and staff
\ \\I owe a lot to the leaders of the WH subgroup at CDF (Wei-Ming Yao, BNL, USA and Craig Group, Virginia University, USA) and the conveners of the Higgs group at CDF (Eric James, Fermilab, USA, and Ben Kilminster, Fermilab, USA) who patiently followed my numerous research talks at CDF and offered me encouragement and constructive feedback and advice. I also want to thank the numerous staff personnel at Fermilab who created a great atmosphere for research.  

%Paul Mercure
\ \\In addition, I am very thankful to the staff of the physics department at McGill University for all their support during my studies. Their help was essential to running things smoothly and allowing me to spend more energy on research. I would mention especially Paula Domingues, our secretary, who helped me most with the paperwork, Sonia Vieira, who helped me with my travel reimbursement claims, and Paul Mercure, who solved promptly any computing problem we encountered. 

%Teo 
\ \\I am grateful to Teodora Dan for her help and advice in coding, especially my first classes in C++, which became the skeleton of the ABCDF package.

%Takao Inagaki, Jean-Yves Hostachy
\ \\I am also grateful to Takao Inagaki, KEK, Japan, spokesperson of the E391a experiment, for accepting me in July 2004 for a one-month-funded internship at their experiment, where I reviewed the literature on a scintillating crystal they were planning to use in their particle detector. I am also in debt to Jean-Yves Hostachy, Laboratoaire de Physique Subatomique et Cosmologie, Grenoble, France, who coordinated my ``these de licence'' at Joseph Fourier University, Grenoble, France, where I described the ATLAS detector at LH, CERN. 

\ \\I would certainly not have succeeded in studying abroad if it were not for the primary and secondary education I received in my native Romania. I owe an enormous debt to my teachers from ''Mircea Eliade'' school and ``Fratii Buzesti'' high school in Craiova, Romania. My life-long passion for studying was introduced in my heart by Maria Popescu, my teacher from elementary school. My mathematical skills were trained by Florin Benea, my maths teacher. My personal education reached its apogee thanks to the ``Fratii Buzesti'' high school, that allowed me to have contact with many great teachers. I am especially grateful to have met my physics teacher, Stefan Brinzan. Not only did he offer me a sound education in the theory and history of physics, but he also saved my life by encouraging his students to do sport daily, especially jogging. At the moment, I was forbidden by doctors to do any sport due to a heart problem that plagued my life until then. However, I followed my teacher's advice in a progressive manner and within half a year I was jogging 5 km daily and within one year and a half my heart problem was confirmed by doctors to have disappeared. In this process, I was able to avoid a scheduled heart surgery at the end of high school and start a real life for myself. As a side effect, my mind was fresh to grasp efficiently the excellent knowledge that the teachers at ``Fratii Buzesti'' high school were offered me. Especially I was able to win prizes to national competitions and olympiads at Physics and French, which won me a partial scholarship to do my university studies in France. I owe my health and the school successes to Stefan Brinzan.

\ \\Finally, I could not stand here without the love and support of my parents, Dumitru Stefan Buzatu and Mariana Buzatu, as well as of my sister, Simona Buzatu. The moral and life values I received from them made me who I am. Without them, I would certainly not be writing this PhD thesis today.  

\clearpage{\pagestyle{empty}\cleardoublepage}

\addcontentsline{toc}{chapter}{OVERVIEW}
\chapter*{Thesis Overview\label{chapter:Overview}}

\ \\Chapter~\ref{chapter:Theory} presents the theory of elementary particles and their interactions (the Standard Model) in general and the spontaneous symmetry breaking through the Higgs mechanism in particular. Possible mechanisms to achieve the spontaneous symmetry breaking in theories beyond the Standard Model are also briefly discussed.

\ \\Chapter~\ref{chapter:Searches} reviews the direct Standard Model Higgs boson searches at the LEP, Tevatron and LHC particle accelerators and the indirect constraints on the Higgs boson mass from fits of precision electroweak data to the Standard Model theory predictions. The direct Higgs boson search presented in this thesis, as well as its motivation, are also introduced.

\ \\Chapter~\ref{chapter:Experiment} presents the experimental infrastructure used for this analysis, especially the Fermilab particle accelerator complex, including the Tevatron accelerator, and the Collider Detector at Fermilab, especially its tracking, calorimeter and muon subdetector systems.

\ \\Chapter~\ref{chapter:Object} presents the high level object reconstruction, such as tracks, primary vertices, calorimeter clusters, charged lepton candidates (electron, muon, isolated track), missing transverse energy, jets. The algorithms used in this analysis that identify jets originating in bottom quarks are also described.  

\ \\Chapter~\ref{chapter:Simulation} introduces the signal and background processes used by this analysis, describes how an event is simulated using a Monte Carlo generator and enumerates the generators used to simulate every relevant physics process. 

\ \\Chapter~\ref{chapter:Selection} details the online (trigger) and offline (analysis) event selection. The analysis uses electron or muon triggers for the tight charged lepton categories and the missing transverse energy plus jets for the isolated track category. The baseline event selection, the several $b$-tagging categories and the non-$W$ (QCD) background veto are also described. 

\ \\Chapter~\ref{chapter:Signal} presents the computation of the predicted number of $WH$ and $ZH$ signal events, and enumerates various sources of systematic uncertainties for the event yield calculation.

\ \\Chapter~\ref{chapter:Background} presents the complex methodology used to compute the estimated event yield for each background process, as well as the tables for background, signal and data event yields. 

\ \\Chapter~\ref{chapter:Discriminant} describes the artificial neural network (ANN) used as a final discriminant in this analysis. After an introduction to ANNs, the variables used in the ANN training are enumerated, an overtraining check is presented and the ANN output shapes are overlaid for different signal and background processes. The second part of the chapter describes a second ANN used in this analysis to correct the jet energies to their true parton-level energies.

\ \\Chapter~\ref{chapter:Limit} presents the final result of this analysis, namely the ratio of the 95\% credibility level upper limits on the Standard Model Higgs boson cross section times branching ratio to the SM predicted values. 

\ \\Chapter~\ref{chapter:Conclusion} concludes this dissertation with a review of the analysis, its methodology, our original contributions and the results achieved. Future plans and possible improvements are also discussed. 

\ \\Appendix~\ref{chapter:METTriggers} describes in detail one of our major contributions to the analysis, namely the parameterization of each of the three missing-transverse-energy-plus-jets triggers and the measurement of the trigger prescale. 

\ \\Appendix~\ref{chapter:CombineTriggers} presents the details of another major contribution of ours to the analysis, namely the novel method to combine any number of triggers in order to maximize the event yield while avoiding the correlations between triggers. 

\ \\Appendix~\ref{chapter:WHAM} describes the general structure, features and advantages of the data analysis software package used for this analysis, for which I was one of the three main developers. 

\ \\Appendix~\ref{chapter:ControlPlots} enumerates the control plots relevant to proving that the analysis uses variables for which the background modelling in Monte Carlo simulated events agrees very well with the data measurements. 

\ \\The results presented in this dissertation have undergone multiple stages of intensive review within the Collider Detector at Fermilab collaboration. As such, the findings of this study may be disseminated publicly. They have already been presented in a talk at the New Perspectives conference on May 31 2011 and in a poster at the Fermilab Users Meeting on June 01 2011, both at Fermilab, USA.  

\clearpage{\pagestyle{empty}\cleardoublepage}

\addcontentsline{toc}{chapter}{ORIGINAL CONTRIBUTION}
\chapter*{Original Contributions\label{chapter:OriginalContributions}}

\ \\This dissertation presents an experimental search for the Standard Model Higgs boson produced in association with a $W$ boson in proton-antiproton collisions at the Tevatron accelerator at Fermilab and recorded with the Collider Detector at Fermilab. The goal is to observe or exclude the Standard Model Higgs boson and thus confirm or refute the Higgs Mechanism. Research in experimental particle physics is a collaborative effort and this $WH$ search makes no exception. This work could not have been possible without the contribution of the approximately 600 members of the Collider Detector at Fermilab (CDF) collaboration in general and without that of the approximately two dozen members of the $WH$ CDF group in particular. In the sections below, I will emphasize my own original contributions to the $WH$ search at CDF and present the sections in the dissertations that explain these in more detail, as well as the CDF published results that employ my work. Also, the images employed in the thesis must be assumed to be my own contribution, unless mentioned differently in the caption by a reference or credit statement. 

\section*{A New Loose Charged Lepton Channel: ``Isolated Track''}

\ \\I improved the sensitivity of the $WH$ search by making possible the introduction of a novel looser criterion to reconstruct charged lepton candidates. High-transverse-momentum good quality tracks isolated from other activity in the tracking chamber that are not required to match a calorimeter cluster or a muon chamber deposit are called ``isolated tracks'' or ``ISOTRK'' candidates. These loose charged lepton candidates form an orthogonal sample to the tight charged lepton candidates (which we call ``TIGHT'' candidates) and reconstruct also real charged leptons that arrive in the non-instrumented regions of the detector and would otherwise have been lost (Chapter \ref{chapter:Object} and Chapter \ref{chapter:Selection}). Thus, the number of $W$ boson candidate events is increased, which in turn increases the expected number of $WH$ signal candidate events in our analysis (Chapter \ref{chapter:Signal}). Although the signal over background ratio is worse for the ISOTRK category than for the TIGHT lepton category (Chapter \ref{chapter:Background}), the expected upper limit on the cross section times branching ratio for the rare $WH$ signal is improved for all Higgs boson mass points when both categories are used in the limit calculation (Chapter \ref{chapter:Limit}). The final result of this dissertation is the Higgs boson upper limit with TIGHT and ISOTRK alone and with TIGHT and ISOTRK combined. 

\ \\The $WH$ analysis presented in this thesis is not the first one at CDF. I continuously improved the trigger parametrization and therefore the signal yield for the ISOTRK category, applying it to the $WH$ searches that used 2.7 $\invfb$, 4.3 $\invfb$ and 5.7 $\invfb$ of integrated luminosity in different years. The $WH$ search presented in this dissertation introduces for the first time a novel method to combine different triggers. This method is subsequently applied to two previously used triggers and one new trigger to arrive at an updated set of results in a sample of 5.7 $\invfb$ of integrated luminosity. 

\section*{Three ``MET + Jets'' Trigger Parameterization}

\ \\Since the CDF detector does not have a dedicated-``isolated track'' trigger, we use triggers based on the event information orthogonal to the charged lepton information, i.e. missing transverse energy (MET) and jets. We have three such MET-based triggers at CDF. In order for these to be usable for analyses, one needs to parametrize for each trigger the efficiency turnon curves for each trigger level, compute the trigger prescale and identify the jet selection needed so that the trigger efficiency is flat with respect to jet information. These measurements are done in data samples. This complex parametrization is used to compute for each Monte-Carlo-simulated event a trigger efficiency using event information. The simulated events are weighted by the trigger efficiency. One of my important contributions to the $WH$ search was determining the MET-based trigger parametrization in new datasets and continuously improving the parametrization methodology. 

\ \\My parametrization of a first MET-based trigger was used for the ISOTRK charged lepton channel of the $WH$ analysis with 2.7 $\invfb$ that is described in detail in the PhD dissertation of Jason Slaunwhite from the summer of 2008 \cite{JasonSlaunwhiteThesis}. A Physical Review D paper draft based on this analysis is currently under the second and last review of the CDF collaboration. Our analysis is a $WH$ analysis using an artificial neural network. There is also another $WH$ analysis using matrix element computations as inputs to a multivariate technique involving a boosted decision tree approach. The two analyses using the same event selection and an integrated luminosity of 2.7 $\invfb$ were combined in the summer of 2008. The combination increased the $WH$ search sensitivity by 10\% over the best of the two analyses. In this combination, both analyses used the ISOTRK charged lepton channel and the trigger parametrization I had measured. The result was presented at the summer conferences of 2008 and published in a Physical Review Letters paper \cite{WHCombination2.7fb-1}. 

\ \\I improved the methodology for trigger parametrization and two MET-based triggers were used for the ISOTRK charged lepton category for the $WH$ analysis with 4.3 $\invfb$ that is described in detail in the PhD dissertation of Yoshikazu Nagai from the summer of 2009 \cite{YoshikazuNagaiThesis}. The combination of the two triggers avoided correlations between triggers since the events were divided in two orthogonal kinematic regions based on jet information. Each trigger was used for all the events in a given kinematic region and ignored for the events in the other kinematic region. The result was one of the inputs for the limit calculation for the CDF combination of August 2009 presented in Figure \ref{figure:HiggsCombinationAllMassAug2009CDF} and for the Tevatron combination of November 2009 presented in the bottom part of Figure~\ref{figure:HiggsCombinationAllMassTevatron}. The results were presented at the summer conferences of 2009.

\ \\I used the same methodology to parametrize the trigger efficiency turnon curves using data collected with a total integrated luminosity of 5.7 $\invfb$. These were used to update the $WH$ analysis, whose result was again one of the inputs for the limit calculation for the CDF combination of July 2010 presented in Figure \ref{figure:HiggsCombinationAllMassJuly2010CDF} and for the Tevatron combination of July 2010 presented in the top part of Figure~\ref{figure:HiggsCombinationAllMassTevatron}. The results were presented at the summer conferences of 2010. A Physical Review D paper draft based on this analysis is currently under development. 

\ \\I later parametrized a third additional MET-based trigger that did not exist from the beginning of Run II of the Tevatron, but was introduced after about 2.7 $\invfb$ had already been collected, thanks to the effort of the CDF Higgs Trigger Task Force. For each of the three MET-based triggers, I measured the prescale and the necessary jet variable selection so that the parametrization is done as a function of missing energy only. A detailed description of the MET-based trigger parametrization is given in Appendix \ref{chapter:METTriggers}.

\ \\For three triggers, it is more complex to divide the jet kinematic phase space in orthogonal regions and study which trigger is on average more efficient for each region. It is in this context that I introduced a novel method to combine any number of triggers in order to maximize the event yield and yet not have an ``OR'' between the triggers in order to avoid trigger correlations and thus systematic uncertainty estimation difficulties.

\section*{Novel Method to Combine Triggers}

\ \\The novel method I introduced generalizes the previous method by considering each individual event as an infinitesimally small kinematic region. Just as before, only one trigger is assigned to all the events in this kinematic region. A study is performed to check which trigger is more efficient in this region. However, since the region is formed of only one event, the study consists simply of comparing the trigger efficiencies of the three triggers for this event and assigning to the event only the trigger with the largest efficiency. For data events, it is checked if the event has fired the chosen trigger. If it does, then the event is kept. If it does not, the event is rejected, without checking if the event has fired other triggers, which makes sure correlations between triggers are avoided. For a Monte-Carlo simulated event, the trigger is always assumed to fire and the event is assigned a weight equal to the efficiency of the chosen trigger.

\ \\Although the event yield is slightly smaller than in the case when we take a simple ``OR'' between triggers, a reduction in the systematic uncertainty of the event yield is expected due to avoiding correlations between the triggers. It also becomes easier to evaluate the systematic uncertainty. To achieve this, I also parametrized the trigger efficiency turnon curves in several bins of several variables. A detailed description of the novel method to combine triggers is given in Appendix \ref{chapter:CombineTriggers}.

\ \\I also developed a software package called ABCDF, which is easily portable in any CDF analysis (and in fact in any high-energy-physics data analysis framework). ABCDF will take as input all the relevant information from a given event and return the trigger efficiency (event weight due to the trigger parametrization). The method works with an unlimited number of triggers, so other triggers can in principle be added in order to increase the signal acceptance.

\ \\The analysis presented in this thesis uses the same integrated luminosity as of the summer of 2010, but introduces for the first time the third MET-based trigger and the novel methodology to combine triggers, which improves the signal acceptance and the analysis sensitivity even more. As soon as this PhD dissertation is submitted, I will continue to work on the $WH$ analysis as a postdoctoral URA Visiting Scholar Fellow in order to make use of the latest available integrated luminosity and add further improvements to be part of the Summer 2011 CDF and Tevatron combinations and to be shown at the Summer 2011 conferences. 

\section*{Main Author of a New Data Analysis Framework}

\ \\ As part of the subgroup that searches for the Higgs boson produced in association with a W boson and decaying into a bottom quark pair (WH group), I am one of the main authors that developed a new data analysis software framework, called WHAM, that performs several analyses that use the signature of exactly one charged lepton plus missing transverse energy plus a number of tight jets, such as $WH$, $WZ$ and Technicolor searches and single-top measurements. WHAM allows all these analyses to share common tools in order to produce a larger number of scientific results with less manpower. WHAM will play a crucial role from now until the end of the CDF analysis effort, when the CDF collaboration manpower decreases. Since I integrated my ABCDF software package fully into WHAM, most of these ongoing analyses are benefiting from the ISOTRK channel or from other loose muon channels that make use of the three MET-based triggers combined with the novel method. For these reasons, I will be a main author for the subsequent publications of these analyses in the $WH$ group at CDF. WHAM is described in detail in Appendix \ref{chapter:WHAM}.

\ \\The work on WHAM has delayed my thesis submission by about a year. Yet, it was time well spent and a huge investment in the analysis power in the lepton plus jet signature at CDF, as well as in my coding and data analysis skills.  

\section*{``Isolated Track'' Scale Factor Measurement}

\ \\As for all the charged lepton categories, also ISOTRK candidates have a different reconstruction efficiency in Monte Carlo simulated events and in data events. A postdoctoral researcher (Nils Krumnack) built a code to measure these efficiencies for each jet multiplicity and compute their ratio, which is also known as the ``scale factor'' between data and Monte Carlo (Subsubsection \ref{subsubsection:ISOTRKScaleFactor}). For the past two years I have maintained this code and updated the scale factor value and its control plots for new data periods being processed for the collaboration. 

%The result is used in this analysis, but is also being used in other analyses that now use ISOTRK charged leptons at CDF. 

\clearpage{\pagestyle{empty}\cleardoublepage}

\pagenumbering{arabic}

\chapter{The Standard Model and the Higgs Mechanism\label{chapter:Theory}}

\section{Elementary Particle Physics}

\ \\Humans have always asked themselves how the world works. They looked around them and observed a large diversity of plants, animals and minerals. They wanted to know what all these are made of and how the constituent elements stayed together. In other words, they wanted to know what are the fundamental ingredients of matter and what is the recipe used by Nature to produce out of the fundamental ingredients all the diversity of things, both alive and inert, that exist.

\ \\Elementary particle physics\footnote{Elementary particle physics is also called particle physics, subatomic physics or high energy physics.} is a domain of physics that uses the scientific method to describe the fundamental building blocks of matter and the elementary interactions between them. The current elementary particle physics theory is called the Standard Model (SM). Before we examine the Standard Model in more detail, let us have a short incursion into the history of the human attempts to answer these fundamental questions about the Universe. Let us examine briefly the advance of science that lead to the advent of elementary particle physics.

\subsection{Short History of Quest for Ingredients of Matter}

\ \\At first, humans attributed supernatural causes to everyday phenomena. Their explanations touched on superstition and religion. 

\ \\However, about 2,500 years ago, people living in Ancient Greece and impassioned about understanding the Universe, who called themselves philosophers, started to search for natural causes for natural phenomena. Therefore, they created science and created the first schools where science was studied and promoted. Two schools of thought in Ancient Greece proposed two different answers to what the fundamental ingredients of the Universe were. One proposed that everything is made of various combinations of four fundamental elements: earth, water, air and fire\footnote{This line of thought was first mentioned in the work of the Greek philosopher Empedocles around 450 BC.}. Another one proposed that everything is made up of many types of indivisible spheres called atoms\footnote{This school of thought was introduced by the Ancient Greece philosophers Leucippus, Democritus and Epicurus.}. Both were closer to philosophy than to scientific theories and both had no idea of what the underlying laws of the Universe could be.

\ \\The next big step happened around 500 years ago in Western Europe when ancient science evolved into modern science. Modern science used both experiments and mathematical formalism to advance knowledge of natural phenomena. Modern science made the jump from simple observation of the world to performing reproducible experiments. The advancement of mathematics allowed for a mathematical formulation of scientific ideas. Scientific theories used mathematics to make numerical predictions about observable quantities in nature. Scientific experiments measured these quantities. If there was an agreement, the theory was potentially confirmed. Otherwise, the theory was contradicted. It is experiment that is the supreme judge of truth in the scientific method\footnote{However, it is possible for the outcomes of scientific experiments to agree with theories that are incorrect.}.

\ \\Modern science was indeed successful. First\footnote{The generation of chemists lead by the French chemist Antoine-Laurent de Lavoisier (1743 - 1794) founded modern chemistry by discovering the chemical elements.}, it showed that the fundamental four elements of Ancient Greece were not fundamental after all, but rather made of other elements. Tens of such elements were discovered and studied by a scientific field called chemistry. Therefore, fundamental elements of the Universe were thought to be these chemical elements, but no fundamental interaction was known between these. Next\footnote{The British chemist John Dalton (1766 - 1844) discovered the atoms.}, chemists showed that each chemical element is made of a certain type of atom, an idea also originating in Ancient Greece, but whose experimental demonstration came only in the 19$^{th}$ century. 

\ \\Then, another field of science called physics showed that even atoms are not elementary, but rather they have a structure. Inside atoms there are elementary particles called electrons that have a negative electric charge. It is the electric interaction between electrons that keeps atoms together in molecules and molecules together in inorganic matter or cells of living organisms. Therefore, for the first time, a recipe of the Universe was proposed: the electric force. Physics pushed more and discovered inside an atom also a nucleus positively charged. The same electric force now explained the structure and stability of atoms. 

\ \\Later\footnote{The British theoretical physicist James Clerk Maxwell (1831 - 1879) built in 1865 the theory of electromagnetism that described the electric force, the magnetic force and light.} it was shown that the electric force and the magnetic force are two different aspects of one fundamental force: the electromagnetic force. Next\footnote{The New Zealander experimental physicist Ernest Rutherford (1871 - 1937) discovered the nucleus in 1910 and the proton in 1919. His student, British experimental physicist James Chadwick (1891 - 1974), discovered the neutron in 1932.}, physics showed that even nuclei have a structure, as they are made of protons and neutrons. A new sub domain of physics was formed: nuclear physics. In order to explain how protons and neutrons are kept together inside a nucleus, a new interaction was introduced, namely the strong nuclear interactions. In order to explain why certain types of atoms are unstable and decay into other type of atoms plus some radiation, a third interaction was introduced, namely the weak nuclear interaction. It was the second time that fundamental recipes of the Universe were introduced. At that stage the fundamental ingredients of the Universe were the proton, the neutron and the electron and the fundamental interactions of the Universe the electromagnetic force, the gravitational force, the strong nuclear force and the weak nuclear force. 

\ \\ But physicists did not stop there. They showed that not even protons and neutrons are elementary particles. Instead, they are each formed by three main quarks and a multitude of other quarks and antiquarks, all kept together by the exchange of new elementary particles, gluons. A new sub domain of physics is thus formed: particle physics. This is the domain of science discussed in this thesis.

\subsection{Current Paradigm in Particle Physics}

\ \\The current paradigm in particle physics is expressed in the theory of the Standard Model of particle physics. The SM describes that matter is formed by six types of quarks and six types of leptons. These elementary particles interact with each other through exchanges of other elementary particles called gauge bosons. There are four fundamental interactions in nature, but the gravitational force is not described by the SM because of the very small strength of the gravitational force at the energies currently available in our particle accelerators. The SM describes the strong force mediated by gluons, the weak force mediated by the $Z^0$, $W^+$ and $W^-$ bosons and the electromagnetic force mediated by the photon ($\gamma$). 

\ \\However, cosmological experimental data in the last decade has shown that the ordinary matter described by the SM represents in fact only about 4$\%$ of the matter-energy content of the Universe, while a currently unknown type of matter forms about 22$\%$ (dark matter) and an unknown type of energy forms about 74$\%$ (dark energy). 

\subsection{Accelerator and Cosmic-Ray Particle Physics}

\ \\Particle physics also has two sub domains: accelerator-based particle physics and cosmic-ray particle physics. In accelerator-based particle physics, subatomic particles such as protons, antiprotons, electrons and/or positrons are produced and accelerated to very high energies, up to velocities very close to the speed of light in vacuum. These accelerated beams of particles are collided either head-on with other beams of particles (collider particle physics) or with a fixed target material (fixed-target particle physics). In these collisions, the large kinetic energy of incoming particles is transformed into mass of new elementary particles, which decay into other elementary particles that are recorded by large particle detectors that surround the collision region.

\ \\Cosmic-ray particle physics studies collisions of very energetic cosmic subatomic particles (neutrons, light nuclei) with protons and neutrons from the atoms from Earth's atmosphere. These collisions produce a shower of particles that propagate in the atmosphere and are detected by ground-based particle detectors. The advantage of cosmic-ray collisions is that they are sometimes a lot more energetic than the current particle accelerators on Earth can offer. The disadvantage is that the experimental conditions are not reproducible and the data rates are low.

\ \\Some of the first subatomic particles were discovered in cosmic-ray collisions: the positron, the muon, the $\pi$ meson. Then accelerator-based particle physics started to dominate the advancement of particle physics. In this thesis we will be discussing collider particle physics. 

\subsection{Elementary Particles and Interactions}

\ \\Elementary particles\footnote{Elementary particles are also called fundamental particles.} come in two types: fermions and bosons~\cite{PDG}. Fermions have semi-integer spins and obey the statistics of Fermi-Dirac. Bosons have integer spins and obey the statistics of Bose-Einstein. Matter elementary particles are fermions and they also come in two categories: leptons and quarks. Leptons and quarks interact with each other through the exchange of force carrier particles, which are bosons. 

\ \\There are six types of leptons grouped in three weak isospin doublets. Each pair contains a negatively charged lepton and a neutrally charged lepton generically called neutrino. The first doublet is formed by an electron ($e^-$) and an electron neutrino ($\nu_e$). The second doublet is formed by a muon ($\mu^-$) and a muon neutrino ($\nu_\mu$). The third doublet is formed by a tau lepton ($\tau^-$) and a tau neutrino ($\nu_\tau$). Leptons interact only through the electromagnetic and the weak forces.

\ \\There are also six types of quarks grouped in three pairs as well. Unlike the leptons, quarks have fractional electric charge. However, the electric charge difference between the members of the pair is also a unit of one. Each pair contains a quark with electric charge equal to $+\frac{2}{3}|e|$ and another quark with electric charge equal to $-\frac{1}{3}|e|$, where $e$ is the electric charge of the electron. The first pair is formed by the quarks up ($u$) and down ($d$). The second pair is formed by the quarks charm ($c$) and strange ($s$). The third pair is formed by the quarks top ($t$) and bottom ($b$). Quarks interact through all the three elementary interactions, including the strong force which is specific only to them. Each quark possesses a quantum number called ``color'' that can have the values of red, green or blue\footnote{These names have nothing to do with the colours from everyday life. They originate in an analogy with light from everyday life, where the colours red, green and blue produce the colour white when combined. The colour white is considered neutral. If for the electric force there is a need of two different charges to produce a neutral object, the strong force needs three. This is why they bear the names of the three colours that need to be combined to produce a neutral colour (white).}.

\ \\The three families of leptons and quarks respectively are also called generations. On average, the particles of the second generations are more massive than the ones from the first generation and less massive than the ones in the third generation. It is currently believed that when the Universe was still relatively small and hot, all these elementary particles were produced. However, as the Universe expanded and cooled, the particles from the third generation decayed into particles of inferior generations and as the Universe expanded and cooled even more, the particles of the second generation decayed into particles of the first generation. Therefore, at current energies in the Universe, all matter is made up of particles from the first generation: the up and down quarks form protons and neutrons, electrons are part of the atoms and electronic neutrinos are emitted in nuclear radioactive decay. 

\ \\The electromagnetic interaction is mediated by a photon ($\gamma$). The electroweak interaction is mediated by the $Z^0$, $W^+$ and $W^-$ bosons. The strong interaction is mediated by eight type of gluons. Each gluon has a combination of a color and an anticolor. 

\ \\In Table~\ref{table:SM} we present a summary of the elementary particles of the Standard Model and their properties.

\begin{center}
\begin{table}[h] % [t] puts at top of page
\begin{center}
\begin{tabular}{|c|c|c|c|c|c|c|c|}
\hline \hline
Fermions& $1^{st}$ Gen. & $2^{nd}$ Gen. & $3^{rd}$ Gen. & Interaction(s) & Q & Spin\\
\hline \hline
Leptons & electron ($e^{-}$) & muon ($\mu^{-}$) & tau ($\tau^{-}$)& EM, Weak & $-1$ & 1/2\\
& e-neutrino ($\nu_{e}$)& $\mu$-neutrino ($\nu_{\mu}$)& $\tau$-neutrino ($\nu_{\tau}$)& Weak & $0$ & 1/2\\
Quarks & up ($u$) & charm ($c$) & top ($t$) & EM, Weak, Strong & $+2/3$ & 1/2\\
& down ($d$) & strange ($s$) & bottom ($b$) & EM, Weak, Strong & $-1/3$ & 1/2\\
\hline \hline
& Name & Force & Coupling & Mass ($\gevcc$) & Q & Spin\\
\hline \hline
Gauge & photon ($\gamma$) & EM & $10^{-2}$ & 0 & 0 & 1\\
Bosons & W boson & Weak & $10^{-13}$ & 80.4 & $\pm 1$ & 1\\
& Z boson & Weak & $10^{-13}$ & 91.2 & 0 & 1\\
& gluon (g) & Strong & 1 & 0 & 0 & 1\\
\hline\hline
\end{tabular}
\caption[Table of Standard Model elementary particles and their properties]{Table of Standard Model elementary particles and their properties. Q means the electric charge.}
\label{table:SM}
\end{center}
\end{table}
\end{center}

\subsection{Antiparticles}

\ \\For almost every elementary particle there is an antiparticle with the same mass, spin, lifetime and decay width, but with opposite electric charge and other quantum numbers. The particles that are their own antiparticles are the photon and the $Z^0$ boson. Experiments have not yet shown if the neutrinos are the same particle as the antineutrinos, or if they are different particles, as the Standard Model assumes.

\ \\Unless otherwise explicitly mentioned, throughout this thesis all statements referring to particles are also valid for their corresponding antiparticles.

\section{Particle Physics Theories}

\subsection{Global Gauge Theories}

\ \\The Standard Model is a local quantum field theory with a local gauge symmetry. The equations of motion are deduced using a least action principle. 

\ \\Historically speaking, a theory with a global gauge symmetry was first proposed by Paul Adrien Maurice Dirac who described the equations of motion of an electron with electric charge and spin that did not interact with other particles (free particles). The Dirac free Lagrangian describes in fact all free fermions fields: 

\begin{equation}\label{LagrangianDirac}
{L}(x)=\bar{\psi}(x)(i\gamma^{\mu}\partial_{\mu}- m)\psi(x)\ \rm{.}
\end{equation}

\ \\In equation ~\ref{LagrangianDirac}, $\psi$ represents a Dirac field of mass $m$ and $\gamma^\mu$ are Dirac's matrices. This equation is invariant under a global U(1) transformation described by

\begin{equation}
\psi(x)\to e^{iQ\alpha}\psi(x)\ \rm{,}
\end{equation}

\ \\where $Q$ is the electric charge, $x$ is a space-time 4-vector and $\alpha$ is a parameter that does not vary with spatial coordinates. There is a theorem called Noether's theorem~\cite{Weinberg} that states that for every symmetry there is a conserved quantity. The U(1) symmetry results in a conserved 4-vector current $j^{\mu} = -Q\bar{\psi}\gamma^{\mu}\psi$:

\begin{equation}
\partial_{\mu}j^{\mu} = 0\ \rm{.}
\end{equation}

\ \\It means that the charge Q is conserved in time, i.e. the integral over the space coordinates for $j^\mu$.

\subsection{Local Gauge Theories}

\ \\As real elementary particles interact with one another, interactions need to be introduced in the free Lagrangian from equation \ref{LagrangianDirac}. This is obtained by using a parameter $\alpha$ that is space dependent. The U(1) transformation becomes a local U(1) transformation. 

\begin{equation}
\psi(x)\to e^{iQ\alpha(x)}\psi(x)\rm{.}
\end{equation}

\ \\Adding simply this local gauge transformation in the free Lagrangian does not keep the Lagrangian invariant. In order to do so, the partial derivative is also transformed into a covariant derivative $D_{\mu}$ that is defined by the following equation:

\begin{equation}
\partial_{\mu}\to D_{\mu}=\partial_{\mu}+iQA_{\mu}\ \rm{,} 
\end{equation}

\ \\where $A_{\mu}$ is a new 4-vector field that is transformed like this:

\begin{equation}
A_{\mu}\to A_{\mu}-\partial_{\mu}\alpha(x)\ \mathrm{.}
\end{equation}

\ \\It results that 
\begin{equation}
D_{\mu}\psi(x)\to e^{iQ\alpha(x)}D_{\mu}\psi(x)\ \mathrm{.}
\end{equation}

\ \\ By replacing all these in the Dirac Lagrangian formula in equation \ref{LagrangianDirac}, we obtain the Lagrangian for the interacting fermions. This theory is called Quantum ElectroDynamics (QED):

\begin{equation}\label{Lagrangian_QED}
L_{QED}=\bar{\psi}(x)(i\gamma^{\mu}D_{\mu}- m)\psi(x) - \frac{1}{4}F_{\mu\nu}F^{\mu\nu}\ \rm{,}
\end{equation}

\ \\where $F_{\mu\nu}\equiv\partial_{\mu}A_{\nu}-\partial_{\nu}A_{\mu}$ is the 
covariant kinetic term of $A_{\mu}$.

\ \\The QED theory describes the electromagnetic interaction. If we apply the Euler-Lagrange least action principle\cite{Weinberg}  on \ref{Lagrangian_QED} then we obtain the equation of motion for a field $\psi$ undergoing electromagnetic interaction through the exchange of a massless vector field $A_\mu$: 

\begin{equation}
(i\gamma^{\mu}\partial_{\mu}- m)\psi(x) = Q\gamma^{\mu}A_{\mu}\psi(x)\ \rm{.}
\end{equation}

\ \\Indeed, this Lagrangian does not have any mass term of the type  $\frac{1}{2}m^{2}A_{\mu}A^{\mu}$, as this would break the Lorentz invariance. Therefore the vector field $A_\mu$ is massless. This is in complete agreement with the fact that the gauge boson mediator for the electromagnetic interaction, the photon, is massless. This is due to the fact that the electromagnetic force's range of action is infinite. 

\subsection{Standard Model Theory}

\ \\Local gauge theories are also used to describe the other fundamental interactions of elementary particles. The weak interaction has already been unified theoretically with the electromagnetic interaction described above. The new fundamental interaction called electroweak is described by the Standard Model. 

\ \\The groups $SU(2)_L \times U(1)_L$ describe the electroweak interaction which is spontaneously broken into the weak interaction described by the V-A theory and the electromagnetic interaction described by Quantum Electrodynamics (QED). SU(2) is a non-Abelian group of spin algebra (weak isospin group) and it has three gauge vector fields. 

\ \\A simplified model with two half-integer-spin massless fermions $f$ and $f'$ can explain the electroweak interaction such as the $Q_{f}=Q_{f'}+1$, where $Q$ is the electric charge. The V-A currents explain the weak interactions of leptons using a left-handed doublet field and two right-handed singlet fields.

\begin{equation}\label{ElectroweakLeftRight}
\psi_{1} \equiv \left( \begin{array}{c} f_{L}(x) \\ f'_{L}(x) \end{array}\right), \qquad \psi_{2} \equiv f_{R}(x) 
\qquad \psi_{3} \equiv f'_{R}(x)\ \rm{,}
\end{equation}
with:
\begin{equation}
f_{L,R}(x)=\frac{1}{2}(1\pm\gamma_{5})f(x), \qquad 
\bar{f}_{L,R}(x)=\frac{1}{2}\bar{f}(x)(1\pm\gamma_{5})\ \rm{,}
\end{equation}
\begin{equation}
f'_{L,R}(x)=\frac{1}{2}(1 \pm \gamma_{5})f'(x), \qquad 
\bar{f'}_{L,R}(x)=\frac{1}{2}\bar{f'}(x)(1 \pm \gamma_{5})\ \rm{.}
\end{equation}

\ \\It is helpful to define $T_3$ as the third component of the weak isospin from the $SU(2)$ group and Y the hypercharge from the $U(1)$ group. Then the electric charge Q, $T_3$ and $Y$ are connected by the Gell-Mann-Nishijima equation:

\begin{equation}
Q = T_{3}+\frac{Y}{2}\ \rm{.}
\end{equation}

\ \\We can see that the left-handed doublet with $T_{3}=\pm 1/2$ and $Y = 1$ describes a charged lepton $f$ and a neutrino $f'$. However, the right-handed singlet with $T_{3}=0$, $Y = -2$ contains only the charged lepton. This means that there is no right-handed component of the neutrino.

\ \\In order to express the Lagrangian for the full electroweak interaction with three generations of massless lepton pairs, we start with the free field Lagrangian:

\begin{equation}\label{LagrangianElectroweakFree}
L(x)=\sum_{j=1}^{3}i\bar{\psi}_{j}(x)\gamma^{\mu}\partial_{\mu}\psi_{j}(x)\ \rm{.}
\end{equation}

\ \\Introducing now in the free Lagrangian a $SU(2)\otimes U(1)$ gauge transformation given by

\begin{equation}\psi_{j}(x)\to\psi'_{j}(x)= e^{i\frac{\tau}{2}\cdot\vec{\alpha}(x) + iY_{j}\beta(x)}\psi_{j}(x)\ \rm{,}
\end{equation}

\ \\and a covariant derivative given by

\begin{equation}\label{ElectroweakBosons}
D_{\mu}^{j}= \partial_{\mu}-ig\frac{\tau}{2}\cdot\vec{W}_{\mu}(x)-ig'Y_{j}B_{\mu}(x)\ \rm{,}
\end{equation}

\ \\we obtain the following interaction Lagrangian:

\begin{equation}\label{LagrangianElectroweakInteraction}
L_{I}(x)=\sum_{j=1}^{3}i\bar{\psi}_{j}(x)D_{\mu}^{j}\partial_{\mu}\psi(x)_{j}\ \rm{.}
\end{equation} 

\ \\In equation \ref{ElectroweakBosons} there are three vector bosons ($\vec{W_{\mu}}$) from the $SU(2)$ generators, one ($B_{\mu}$) from the $U(1)$ generator, and four coupling constants ($g$ and $g'Y_{j}$, where $j=1,2,3$). 

\ \\After algebraic calculations, the electroweak interaction Lagrangian~\ref{LagrangianElectroweakInteraction} can be decomposed in a charge current term ($L_{CC}$) and a neutral current term ($L_{NC}$):

\begin{equation}
L_{I}(x)= L_{CC}(x) + L_{NC}(x)\ \rm{,}
\end{equation} 

\ \\The charge current term contains only terms from the left doublet field

\begin{equation}\label{ElectroweakLagrangianInteractionChargedCurrent}
L_{CC}(x)=\frac{g}{2\sqrt{2}}\Big\{\bar{f}(x)\gamma^{\mu}(1-\gamma_{5})f'(x)\frac{1}{\sqrt{2}}W_{\mu}^{+}(x) + \rm{Hermitian\ conjugate}\Big\}\ \mathrm{.}
\end{equation}

\ \\We recognize in this expression that $W_{\mu}^{+}(x)$ is a linear combination of $W_{\mu}^{1}(x)$ and $W_{\mu}^{2}(x)$. This Lagrangian describes the fermion interactions mediated by the $W^+$ and $W^-$ bosons.

\ \\Since the neutral current term of the electroweak interaction Lagrangian contains a linear combination of the neutral vector fields $B_{\mu}(x)$
and $W_{\mu}^{0}(x)$, it can be also written as the sum of two terms:

\begin{equation}
L_{NC}(x)= L_{NC}^{A}(x) + L_{NC}^{Z}(x)\ \rm{.}
\end{equation}

\ \\The first term can be written as
\begin{equation}
L_{NC}^{A}(x)=\sum^{3}_{j=1}\bar{\psi}_{j}(x)\gamma^{\mu}\big[g\frac{\tau_{3}}{2}\sin \theta_{W}
+ g'Y_{j}\cos \theta_{W}\big]\psi_{j}(x)A_{\mu}(x)\ \rm{,}
\label{EquationLNCA}
\end{equation}

\ \\and the second term can be written as
\begin{equation}
L_{NC}^{Z}(x)=\sum^{3}_{j=1}\bar{\psi}_{j}(x)\gamma^{\mu}\big[g\frac{\tau_{3}}{2}\cos \theta_{W}
+ g'Y_{j}\sin \theta_{W}\big]\psi_{j}(x)Z_{\mu}(x)\ \rm{.}
\label{EquationLNCZ}
\end{equation}

\ \\In the equations~\ref{EquationLNCA} and \ref{EquationLNCZ} $\theta_{W}$ is the Weinberg angle. We then write the following four equations:

\begin{equation}
g\sin \theta_{W} = e\mathrm{,}
\end{equation}

\begin{equation}
g'\cos \theta_{W}Y_{1}=e(Q_{f}-1/2)\ \mathrm{,}
\end{equation}

\begin{equation}
g'\cos \theta_{W}Y_{2}=eQ_{f}\ \mathrm{,}
\end{equation}

\begin{equation}
g'\cos \theta_{W}Y_{3}=eQ_{f'}\ \mathrm{.}
\end{equation}

\ \\These equations represent the fundamentals of the Standard Model, to which we have to add the equations that explain the gluon exchange interactions between the quarks. The equations above explain the fermion interaction, as the coupling constants appear in these equations. However, in these equations all fermions and bosons are massless. We know from experiment that is not the case. In section \ref{Section:SpontaneousSymmetryBreaking} we will discuss how a spontaneous symmetry breaking and the Higgs mechanism generate the mass for leptons and gauge bosons without breaking the gauge invariance of the electroweak interaction Lagrangian. 

\begin{figure}[!ht]
\begin{center}
\includegraphics[width=0.9\textwidth]{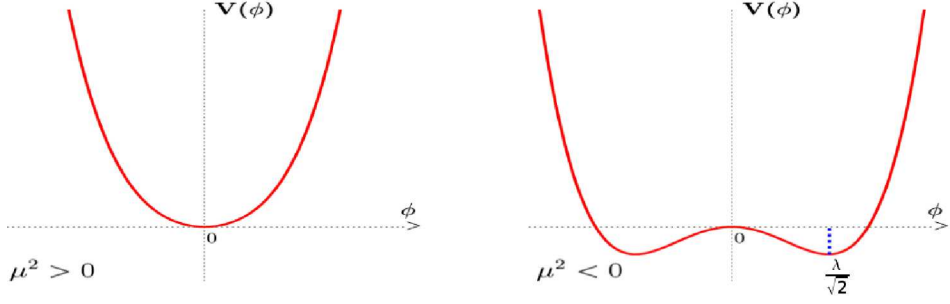}
\end{center}
\caption[Symmetry breaking depending on $\mu^{2}$ parameter]{Symmetry breaking depending on $\mu^{2}$ parameter: $\mu^{2}>0$ on 
the left, $\mu^{2}<0$ on the right \cite{FedericoSforzaMScThesis}.}\label{figure:HiggsMechanism}
\end{figure}

\section{Spontaneous Symmetry Breaking\label{Section:SpontaneousSymmetryBreaking}}

\ \\Systems with infinite degrees of freedom and a Lagrangian invariant under a group $G$ of transformations can have non symmetric states through spontaneous symmetry breaking, or a Higgs mechanism.

\ \\Starting with the electrodynamics Lagrangian we can imagine a toy Higgs mechanism~\cite{HiggsPaper}~\cite{EnglertBroutPaper}~\cite{GuralnikHagenKibblePaper}:

\begin{eqnarray}\label{LagrangianToyHiggsMechanism}
L(x)&=&-\frac{1}{4}F_{\mu\nu}(x)F^{\mu\nu}(x)+\\
&&\big[(\partial_{\mu}+ieA_{\mu}(x))\phi^{\dagger}(x)\big]\cdot\big[(\partial^{\mu}-ieA^{\mu}(x))\phi(x)\big]\nonumber\\
&&-\mu^{2}\phi^{\dagger}(x)\phi(x)-h\big[\phi^{\dagger}(x)\phi(x)\big]^{2}\nonumber\ \mathrm{,}
\end{eqnarray}

\ \\where $\phi(x)$ is the scalar field of electromagnetic interaction via the Abelian gauge field $A^{\mu}(x)$ and where  $h>0$, $\mu^{2}<0$. It can be checked easily that equation~\ref{LagrangianToyHiggsMechanism} is invariant for the following gauge transformations:

\begin{equation}
\phi(x)\to\phi'(x)=e^{i\alpha(x)}\phi(x)\mathrm{,}\qquad A_{\mu}(x)\to A'_{\mu}(x)=A_{\mu}(x)-\frac{1}{e}\ \partial_{\mu}\alpha(x)\ \mathrm{.}
\end{equation}

\ \\The equations of motion produced using this Lagrangian have solutions that correspond to the minimal energy, thus to the vacuum expectation values in the lowest order perturbation theory. Since $\mu^{2}<0$, besides the trivial solution $\phi(x)=0$ there  exists a set of degenerate solutions with $|\phi^{2}|=\frac{-\mu^{2}}{h}=\frac{\lambda^{2}}{2}$ due to the underlying gauge symmetry $\phi(x)= \frac{\lambda}{\sqrt{2}}e^{i\alpha(x)}$(see Fig.~\ref{figure:HiggsMechanism}). We are allowed to choose  such that $\phi'(x)$ is real. It implies that the lowest energy state is $\phi(x)=\frac{\lambda}{\sqrt{2}}$. 

\ \\Then at first order we can write 

\begin{equation}\label{LagrangianToyHiggsMechanismGaugeFix}
\phi'(x)=\frac{1}{\sqrt{2}}[\lambda+\phi_{1}(x)],\quad\phi_{2}(x)=0,\quad A'_{\mu}(x)=B_{\mu}(x)\ \mathrm{.}
\end{equation}

\ \\ Now we can substitute equation \ref{LagrangianToyHiggsMechanismGaugeFix} into equation \ref{LagrangianToyHiggsMechanism} and arrange the Lagrangian in powers of $\phi_{1}(x)$:

\begin{eqnarray}\label{LagrangianToyHiggsMechanism2}
L(x)&=&-\frac{1}{4}B_{\mu\nu}(x)B^{\mu\nu}(x)+\frac{1}{2}e^{2}\lambda^{2}B_{\mu}(x)B^{\mu}(x)\\
&&+e^{2}\lambda B_{\mu}(x)B^{\mu}\phi_{1}(x)+\frac{1}{2}e^{2}\lambda B_{\mu}(x)B^{\mu}\phi_{1}^{2}(x)\nonumber\\
&&+\frac{1}{2}\big[\partial^{\mu}\phi_{1}(x)\partial_{\mu}\phi_{1}(x) + 2\mu^{2}\phi_{1}^{2}(x)\big]\nonumber\\
&&+\frac{\mu^{2}}{\lambda}\phi_{1}^{3}(x)+\frac{\mu^{2}}{4\lambda^{2}}\phi_{1}^{4}(x)-\frac{1}{4}\lambda^{2}\mu^{2}\mathrm{.}\nonumber
\end{eqnarray} 

\ \\We can now find physical meaning in each of these lines. The first line describes a massive vector field with mass $|e\lambda|$ (the previously massless gauge boson acquired mass). The second line describes the interaction term between the massive vector field and the neutral scalar field with coupling strengths $e^{2}\lambda$ and $\frac{1}{2}e^{2}$. The third line corresponds to a free scalar particle called the Higgs particle with a mass  $M_{H}=\sqrt{-2\mu^{2}}$. The fourth line describes the self interaction of the scalar field.

\ \\The effect of the spontaneous symmetry breaking is that the initial four degrees of freedom in equation~\ref{LagrangianToyHiggsMechanism} (two in the initial complex scalar field and two in the massless vector field) transformed into four other degrees of freedom (a real neutral scalar particle and a massive gauge boson). 

\section{The Higgs Mechanism and the Higgs Boson}

\ \\In a similar way, the Higgs mechanism can be applied to equation~\ref{ElectroweakLeftRight} to allow the $W^{\pm}$ and $Z^{0}$ bosons to acquire mass. Two complex scalar fields introduced to break spontaneously the symmetry of the gauge groups $SU(2)\otimes U(1)$ form an isodoublet with respect to the $SU(2)$ group:

\begin{equation}\label{PhiDoublet}
\phi(x)\equiv\left(\begin{array}{c} \phi^{+}(x) \\\phi^{0}(x) \end{array}\right)\mathrm{,}
\end{equation}

\ \\where the charged (neutral) component of the doublet is $\phi^{+}(x)$ ($\phi^{0}(x)$), respectively. This creates a Higgs field potential where $h>0$ and $\mu^{2}<0$:

\begin{equation}
 V_{H}(x)\equiv -\mu^{2}\phi^{\dagger}(x)\phi(x)-h\big[\phi^{\dagger}(x)\phi(x)\big] ^{2}\ \mathrm{.}
\end{equation}

\ \\From equation~\ref{LagrangianToyHiggsMechanismGaugeFix} we know that the neutral scalar field  $\phi^{0}(x)$ has a vacuum expectation value of $\frac{\lambda}{\sqrt{2}}$. Therefore, at first order we can rewrite equation~\ref{PhiDoublet} to

\begin{equation}\label{PhiDoubletFirstOrder}
\phi(x)=e^{\frac{i}{\lambda}\vec{\tau}\cdot\vec{\theta(x)}}\left(\begin{array}{c} 0 \\\frac{1}{\sqrt{2}}(\lambda + \chi(x)) \end{array}\right)\ \mathrm{.}
\end{equation}

\ \\In this equation there is an explicit $SU(2)$ gauge freedom, as three of the four components of the field $\phi(x)$ are gone and only one real scalar field $\phi^{0}(x)$ remains, with $\phi^{0}(x)= \frac{1}{\sqrt{2}}(\lambda + \chi(x))$. 

\ \\By putting together all the previous equations we obtain the SM part of the Lagrangian that produces the mass of the $W^{\pm}$ and $Z^{0}$ bosons: 

\begin{eqnarray}\label{LagrangianMassWZ}
L(x)&=& \frac{1}{4}g^{2}\lambda^{2}W_{\mu}^{\dagger}(x)W^{\mu}(x) + \frac{1}{1}(g^{2}+g'^{2})\lambda^{2}Z_{\mu}(x)Z^{\mu}\\
&&+\frac{1}{2}g^{2}\lambda W^{\dagger}_{\mu}(x)W^{\mu}(x)\chi(x)+\frac{1}{4}g^{2}W^{\dagger}_{\mu}W^{\mu}\chi^{2}(x)\nonumber\\
&&+\frac{1}{4}(g^{2}+g'^{2})\lambda Z_{\mu}(x)Z^{\mu}(x)\chi(x)+\frac{1}{8}g^{2}Z_{\mu}(x)Z^{\mu}(x)\chi^{2}(x)\nonumber\\
&&+\frac{1}{2}\big[\partial^{\mu}\chi(x)\partial_{\mu}\chi(x) + 2\mu^{2}\chi^{2}(x)\big]\nonumber\\
&&+\frac{\mu^{2}}{\lambda}\chi^{3}(x)+\frac{\mu^{2}}{4\lambda^{2}}\chi^{4}(x)-\frac{1}{4}\lambda^{2}\mu^{2}\mathrm{.}\nonumber
\end{eqnarray}

\ \\Equation~\ref{LagrangianMassWZ} tells us that the $W^\pm$ bosons have acquired mass:

\begin{equation}\label{doublet}
m_{W^+}=m_{W^-}=\frac{1}{2}\lambda g\ \mathrm{,}
\end{equation}

\ \\and that also the $Z^0$ bosons have acquired mass:

\begin{equation}\label{doublet}
m_{Z}=\frac{1}{2}\lambda \sqrt{g^{x}+g'^{x}}=\frac{1}{2}\frac{\lambda g}{\cos \theta_{w}}\ \mathrm{.}
\end{equation}

\ \\We can see that some Standard Model parameters are now constrained by theory, such as:

\begin{equation}
m_{Z}=\frac{m_{W}}{\cos \theta_W} \ge m_{W}\ \mathrm{,}
\end{equation}

\begin{equation}
\frac{G_{F}}{\sqrt{2}}=\frac{g^{2}}{8m_{W}^{2}}\ \mathrm{.}
\end{equation}

\ \\However, the mass of the Higgs boson ($m_{\chi}=\sqrt{-2\mu^{2}}$, sometimes written as $m_{H}$ as well) is not constrained by theory and has to be measured by experiments once the Higgs boson is observed experimentally. 

\ \\If a Yukawa coupling is added such as

\begin{eqnarray}
\L_{f}(x)&=&c_{f'}\Bigg[(\bar{f}(x),\bar{f}'(x))_{L} \left(\begin{array}{c} \phi^{+}(x) \\\phi^{0}(x) \end{array}\right)\Bigg]f_{R}'(x)\\
&&+ c_{f}\Bigg[(\bar{f}(x),\bar{f}'(x))_{L} \left(\begin{array}{c} -\bar{\phi}^{0}(x) \\\phi^{-}(x) \end{array}\right)\Bigg]f_{R}(x) + \mathrm{Hermitian\ conjugate\ ,}\nonumber
\end{eqnarray}

\ \\where $f$ represents a fermion and $f'$ represents the corresponding antifermion, $f(x)$ represents a fermion field and $f'(x)$ an antifermion field, the Higgs Mechanism will produce a term to the SM Lagrangian that will give mass both to fermions and antifermions, both to leptons and quarks:

\begin{equation}
m_{f}=m_{f'}=-c_{f}\frac{\lambda}{\sqrt{2}}\mathrm{.}
\end{equation}

\ \\Here the constant $c_{f}$ is also not constrained by theory and is deduced from the measured fermion masses. The mass of a fermion is equal with the mass of the corresponding antifermion.

\section{Physics Beyond the Standard Model}

\ \\The Standard Model of elementary particles and their interactions is the most precise physics theory ever developed. No experiment has shown convincing contradiction with the SM. However, physicists believe that the SM is a low-energy approximation of a higher-energy theory~\cite{SM-MSSM-HiggsReview}~\cite{StandardModelReview}, much the same way that classical physics is a particular case of quantum mechanics. There are various aspects that are not explained yet by the Standard Model: the disappearance of antimatter shortly after Big Bang, the nature of dark matter, the mass of neutrinos, why there are three generations of elementary particles, etc. In addition, in order to have a Higgs boson mass at the electroweak scale, there has to be some mechanism that cancels the radiative corrections of the Higgs spectrum. Elementary particle physicists perform experiments at the Tevatron, the LHC and other particle physics laboratories around the world to test for the validity of the Standard Model, with a hope that new phenomena will be observed. A new theory that would describe correctly these supposedly new phenomena will be called a theory of physics beyond the Standard Model (BSM). 

\subsection{Supersymmetry}

\ \\A popular candidate theory BSM is the theory of supersymmetry (SUSY). It predicts that every SM elementary particle has a corresponding partner that has not yet been observed experimentally. This theory introduces a symmetry between fermions and bosons, predicting that every SM fermion has a bosonic partner and every SM boson has a fermionic partner. This allows for the cancellations of Higgs radiative corrections that would otherwise require a very precise fine tuning that physicists find hard to accept. The supersymmetric partners of SM elementary particles are called ``superpartners''. For example, the superpartner of the top quark, the gluon, the $W^\pm$ and $Z^0$ bosons are called stop ($\tilde{t}$), gluino ($\tilde{g}$), gauginos ($\tilde{\chi}^\pm$) and gaugino ($\tilde{\chi}^0$), respectively.

\ \\If supersymmetry exists, it is a broken symmetry. If it were unbroken, the superpartners would have exactly the same mass as their SM partners and they would have been already observed experimentally. Therefore, if the superpartners exist, they must be more massive than their SM counterparts. 

\ \\The minimal extension to the Standard Model as a supersymmetry is called the Minimal Supersymmetric Standard Model (MSSM). In MSSM there would not be just one Higgs boson, as predicted by the Standard Model, but 5 new particles that play the role of the Higgs boson, three neutral and two electrically charged: $h$, $H$, $A$, $H^+$ and $H^-$~\cite{SymmetryBreakingTheories}. The lightest of the neutral Higgs particles ($h$) has very similar properties 
to those of the Standard Model Higgs boson. This is why if a Standard Model Higgs boson is discovered at the Tevatron or the LHC, precise measurements of its properties would be necessary to check if it is really a Standard Model Higgs boson or a SUSY one. All SUSY models predict the existence of at least two Higgs bosons.

\ \\Although LHC experiments will not have the Tevatron's sensitivity for low mass Standard Model Higgs search within approximately 1-2 years, the LHC experiments will be able to improve upon the Tevatron's results for MSSM Higgs boson searches with 1 $\invfb$ of integrated luminosity at $\sqrt{s}=7~\tev$ that is collected right until the projected major shutdown to prepare the LHC for $\sqrt{s}=14~\tev$ collisions.

\subsection{Dynamic Electroweak Symmetry Breaking}

\ \\The Standard Model and its supersymmetric extension explains the spontaneous symmetry breaking by the introduction of a Higgs mechanism and the prediction of one scalar (spinless) elementary particle recognized as a Higgs boson. However, they do not explain why there  should be in nature a scalar field with a non-zero vacuum expectation value. For this reason, a new theory called ``Technicolor'' providing a dynamic reason for the electroweak symmetry breaking was created in 1979 by Weinberg \cite{DynamicSymmetryBreakingWeinberg1979} and Susskind \cite{DynamicSymmetryBreakingSusskind1979}. Back then, the most massive known fermion was the bottom quark, with a mass of approximately 5 $\gevcc$ \cite{PDG}. The electroweak theory had been developed and it had predicted the existence of the $W$ ($Z$) gauge boson with a mass of approximately 80 (90) $\gevcc$, although they had not yet been observed experimentally. Since the largest fermion mass was negligible with respect to the gauge boson masses, the Technicolor theory explained the mass of the gauge bosons and not the mass of the fermions. The Technicolor theory is very similar to the QCD theory and introduces the dynamic spontaneous symmetry breaking in a similar way as the spontaneous chiral symmetry breaking in QCD. Therefore, the Technicolor theory introduces a new strong interaction, a new gauge group and predicts the existence of new elementary particles called techniquarks. 

\ \\Since the Technicolor theory could not predict the mass of Standard Model fermions, it was an incomplete theory. The theory was extended and models called Extended Technicolor were introduced in order to explain also the fermion masses. However, precise experimental measurements revealed that the predictions of the theory were refuted for quarks as massive as the charm quark. The theory was therefore further refined and Walking Technicolor and Multi-Scale Technicolor emerged. Some combinations between Supersymmetry and Technicolor theories also appeared. 

\ \\When in 1995 the top quark was discovered and was found to have the unexpected large mass of approximately 175 $\gevcc$ \cite{PDG}, which is also the value of the weak scale $v_{weak}=1/\sqrt{2\sqrt{2}G_F}=175\ \gev$, where $G_F$ is the Fermi constant, it was suggested that the top quark could play a special role in the spontaneous symmetry breaking in theories beyond the Standard Model. A theory called Topcolor was created, but since it predicted a top quark mass of 220 $\gevcc$, the theory was immediately refuted. However, when Topcolor and Technicolor theories are combined, a new theory called Topcolor Assisted Technicolor was created, where the top quark is the first of the predicted techniquarks. This theory explains the mass of all gauge bosons and fermions, including the heavy top quark. The predictions of this theory can be checked at the Tevatron or the LHC. 

\ \\In conclusion, there are a variety of dynamic spontaneous symmetry breaking theories that have evolved with time. A well written summary of such theories can be read in Reference \cite{DynamicSymmetryBreakingSummaryHill2003}.

\section{Summary}

\ \\This chapter started by introducing elementary particle physics, the subdomain of physics that studies the elementary particles, the smallest building blocks of matter, and the elementary interactions or forces that allow the elementary particles to form bound states and thus create the structure of matter we see around us. We presented the two experimental methods to study elementary particles: accelerator particle physics and cosmic-ray particle physics. We continued by presenting the advancement of particle physics theories from global gauge theories to local gauge theories. We then presented the current theory of particle physics, a local gauge theory called the Standard Model of elementary particles and their interactions. The Standard Model has been confirmed by all the numerous experiments in elementary particle physics ever conducted. However, the theory by itself does not allow elementary particles to acquire mass. We know experimentally that elementary particles have masses, though, and if they had none, the Universe would be very different than it is today and we would not exist. 

\ \\In 1964 a new theory was proposed by the theorists Higgs, Englert, Brout, Guralnik, Hagen and Kibble, which was later called the Higgs mechanism. The theory explained in an elegant way the spontaneous symmetry breaking of the electro-weak interaction into the electromagnetic interaction and the weak interaction, which is equivalent to splitting an elementary particle into the photon of zero mass and the $Z^0$ boson of non zero mass, which is equivalent to introducing a mechanism to allow the $Z^0$ boson to acquire mass. The Higgs mechanism predicts the existence of a scalar field that pervades the entire Universe, the Higgs field. Each elementary particle couples to the Higgs field with a strength proportional to its mass. The Higgs mechanism is a falsifiable theory, since it predicts the existence of a new elementary particle, the Higgs boson, which is described uniquely by its mass. This thesis will present an experimental search for the existence of the Standard Model Higgs boson and therefore an experimental test of the Higgs Mechanism and the Standard Model of particle physics. We concluded the chapter by discussing several spontaneous symmetry breaking explanations in theories beyond the Standard Model, such as the Supersymmetry or Technicolor theories. 

\ \\In the next chapter we will present the summary of the direct and indirect searches for the Standard Model Higgs boson that have been performed until now at particle accelerators around the world, namely the LEP, the Tevatron and the LHC accelerators. We will also introduce and motivate the Higgs boson direct search presented in this dissertation.

\clearpage{\pagestyle{empty}\cleardoublepage}

\chapter{Standard Model Higgs Boson Experimental Searches\label{chapter:Searches}}

\ \\The Standard Model Higgs boson has not yet been observed experimentally. However, experimental limits on the Standard Model Higgs boson mass have been set both by direct and indirect searches. In this chapter we will introduce the direct searches at the LEP and Tevatron accelerators, as well as the prospects for the LHC accelerator. We will also present the indirect electroweak fits for the Higgs boson. Finally we will present our $WH$ search at CDF at the Tevatron and present our motivation for choosing this search. In this dissertation, by a Higgs boson we mean a Standard Model Higgs boson. 

\begin{figure}[ht]
  \begin{center}
 \includegraphics[width=10.0cm]{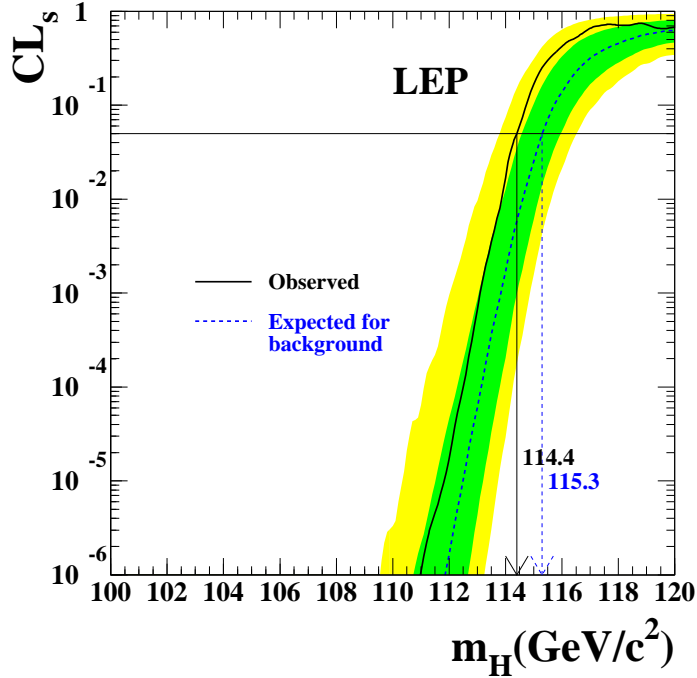}
\caption[Upper limit of Standard Model Higgs boson through direct LEP searches] {The ratio $CL_{s} = CL_{s+b}/CL_{b}$ between the confidence levels for the signal plus background hypothesis and the background only hypothesis as a function of the Higgs boson mass~\cite{LEPCombinationPaper}. The dashed line represents the median expectation. The green and yellow shaded bands around the median expected curve correspond to the $1\ \sigma$ and $2\ \sigma$ probability bands. The observed result is represented by the solid line.}
\label{figure:LepLimit}
\end{center}
\end{figure}

\section{Direct Searches at the LEP Accelerator}

\ \\The Large Electron Positron Collider (LEP) collided electrons and positrons between 1989 and 2000 at centre-of-mass energies $\sqrt{s}$ between 189 and 209 GeV~\cite{LEPCombinationPaper}. A total of 2461 pb$^{-1}$ of integrated luminosity of collision data was analyzed by each of the four detector collaborations at LEP: OPAL, ALEPH, L3 and DELPHI. They looked for Higgs boson production in association with a Z boson, where the Higgs boson decayed to a pair of bottom-antibottom quarks and the Z boson decayed leptonically ($e^{+}e^{-}\to Z^{0}H$, with $Z^{0}\to l^{+}l^{-}$ or $Z^{0}\to \nu \bar{\nu}$ and $H\to b\bar{b}$), or where the Higgs boson decayed to a $\tau$ lepton pair and the Z boson decayed to a generic quark pair ($e^{+}e^{-}\to Z^{0}H$, with $Z^{0}\to q \bar{q}$ and $H\to \tau\bar{\tau}$).

\ \\The results from each experiment and channel were combined by the LEP Electroweak Working Group and a Standard Model Higgs boson was excluded at 95$\%$ confidence level (CL)\footnote{All CDF and Tevatron results presented in this thesis use a Bayesian statistical treatment in setting confidence levels denoted with CL. All other results presented from other experiments use a frequentist approach with confidence level also denoted with CL.}~\cite{PDG} for a mass less than 114.4 $\gevcc$, as we can see in Figure~\ref{figure:LepLimit}. 

\begin{figure}[!ht]
\begin{center}
\includegraphics[width=10.0cm]{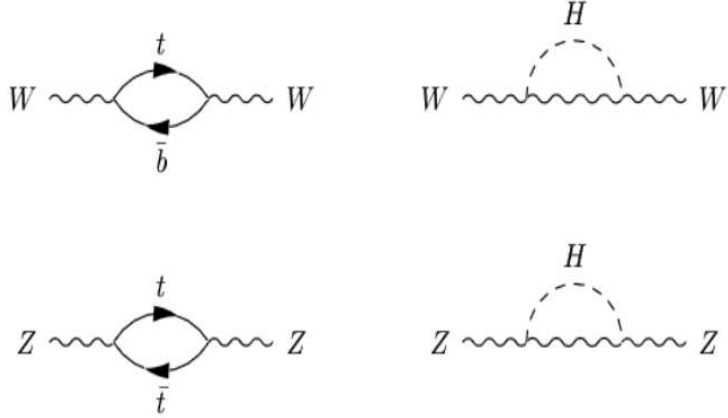}
\caption[Radiative loop contribution to masses of electroweak objects]{Radiative loop contribution to masses of electroweak objects \cite{BarbaraAlvarezPhDThesis}. Precision measurements of the gauge bosons and of the top quark mass can provide a limit on the SM Higgs boson mass.}\label{figure:RadiativeCorrections}
\end{center}
\end{figure}

\section{Electroweak Indirect Fits}

\ \\Although the Higgs boson mass cannot be predicted by theory, it can be constrained through fits on electroweak parameters that are measured by experiment. Radiative corrections due to the Higgs boson loops to the mass of the $W$ and $Z$ bosons or the top quark, as shown in Figure~\ref{figure:RadiativeCorrections}, depend on the Higgs boson mass. Inversely, precise measurements of these masses put an indirect constraint on the Higgs boson mass. 

\begin{figure}[!ht]
\begin{center}
\includegraphics[width=0.55\textwidth]{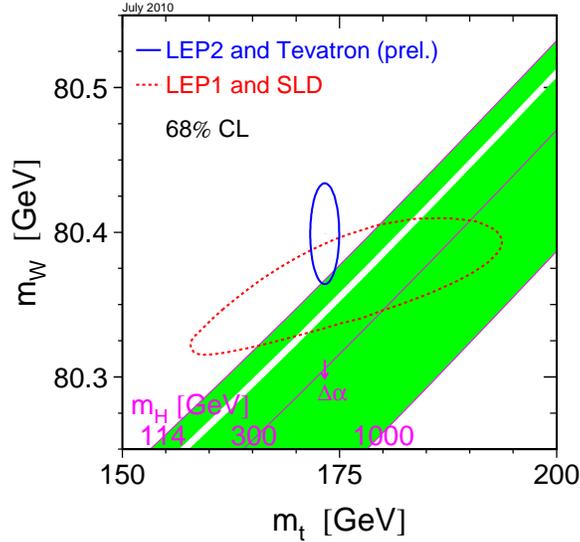}
\caption[Top quark and $W$ boson masses constrain the Higgs boson mass]{SM relationship between $m_{t}$, $m_{W}$ and $m_{H}$~\cite{LEPElectroweakWorkingGroup}. The dashed (solid) line represents indirect constraints on $m_{t}$ and $m_{W}$ from LEP-I and SLD experiments (LEP-II and Tevatron-II) experiments. The white band represents the Higgs boson mass interval excluded at 95\% CL by the Tevatron accelerator in July 2010. Both contours are plotted at 68$\%$ CL. The green contour represents the allowed phase space for $m_{t}$ and $m_{W}$ as a function of the Higgs boson mass. We can see that the two contours agree and they suggest a low mass Higgs boson, just beyond the direct lower mass limit set by the LEP experiments.  The arrow labelled as $\Delta\alpha$ shows the global variation if $\alpha(m_{Z})$ is changed by one standard deviation. This variation gives an additional uncertainty to the SM band shown in the figure.}\label{figure:IndirectElectroweakFits1}
\end{center}
\end{figure}

\ \\The contribution of Higgs boson mass to the mass of gauge bosons can be given by taking into account Feynman diagrams of radiative corrections, such as those in Figure \ref{figure:RadiativeCorrections}, and is given by the following formula:

\begin{eqnarray}
\rho=\frac{M^{2}_{W}}{M^{2}_{Z}(1-\sin^{2}\theta_{W})}=1+\Delta\rho\rm{,}\\
\Delta\rho\equiv\frac{3G_{F}}{8\pi^{2}\sqrt{2}}M_{t}^{2}+\frac{\sqrt{2}G_{F}}{16\pi^{2}}M_{W}^{2}\big[ \frac{11}{3}ln\big( \frac{M_{H}^{2}}{M_{W}^{2}}\big)+...\big]\rm{,}
\end{eqnarray}

\ \\where $G_{F}$ is the Fermi coupling constant, $\theta_{W}$ is the Weinberg angle, 
$M_{t}$, $M_{W}$, $M_{Z}$ and $M_{H}$ are, respectively, the masses of top 
quark, $W$ and $Z$ bosons and Higgs boson. 

\ \\Figure~\ref{figure:IndirectElectroweakFits1}~\cite{IndirectElectroweakFits} shows predictions of Higgs boson mass as a function of the $W$ boson mass $M_{W}$ and the top quark mass $M_{t}$ or {\it vice versa}. The dashed (solid) line represents indirect constraints on $M_{t}$ and $M_{W}$ from LEP-I and SLD experiments (LEP-II and Tevatron-II) experiments. Both contours are plotted at 68\% CL. The green contour represents the allowed phase space for $M_{t}$ and $M_{W}$ as a function of the Higgs boson mass. We can see that the two contours agree and they suggest a low mass Higgs boson, just beyond the direct lower mass limit set by the LEP experiments.  The arrow labelled as $\Delta\alpha$ shows the global variation if $\alpha(M_{Z})$ is changed by one standard deviation. This variation gives an additional uncertainty to the SM band shown in the figure. 

\ \\Figure~\ref{figure:IndirectElectroweakFits2} conveys the same message in a different manner. The plot presents the quality of the Standard Model fit ($\Delta \chi^2$) as a function of the Higgs boson mass. The Higgs mass preferred by the fit is the one that minimizes $\Delta \chi^2$. The latest fit is produced by the LEP Electroweak Working Group~\cite{LEPElectroweakWorkingGroup} using  $m_t = 173.1 \pm 1.3 \gevcc$~\cite{TevatronElectroweakWorkingGroupTopMass} and $m_W = 80.420 \pm 0.031 \gevcc$~\cite{TevatronElectroweakWorkingGroupWMass}:

\begin{figure}[!ht]
\begin{center}
\includegraphics[width=0.55\textwidth]{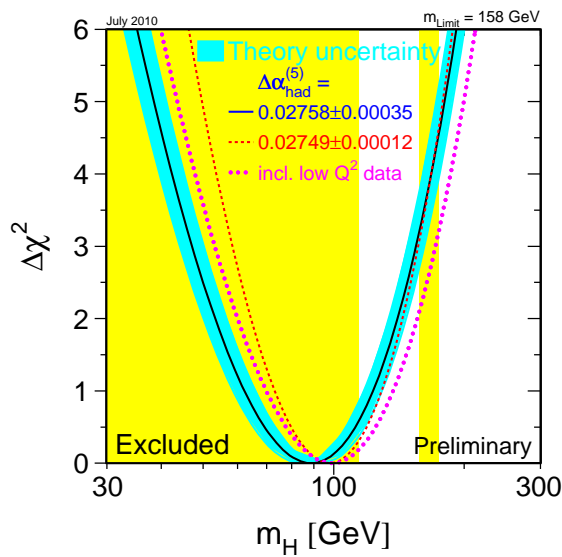}
\caption[Electroweak precision data fit quality versus Higgs boson mass]{The plot presents the quality of fit to electroweak precision data versus Higgs boson mass~\cite{LEPElectroweakWorkingGroup}. The Higgs boson mass range excluded through direct searches at LEP and Tevatron experiments is shown in yellow. The solid dark blue is the nominal fit and the light blue bands represent the theoretical uncertainties on the fit. The 68\% confidence level band is at $\Delta \chi^2=1$ and the 95\% confidence level is at $\Delta \chi^2=2.7$. The dashed and dotted curves represent fit results with slightly different input parameters, such as different theoretical calculations of the vacuum polarization ($\Delta \alpha_{had}^{(5)}$) and values for the $W$ boson mass obtained with low $Q^2$ experiments.}\label{figure:IndirectElectroweakFits2}
\end{center}
\end{figure}

\begin{equation}
m_H = 87^{+35}_{-26}\ \gevcc
\end{equation}

\ \\and the 95\% confidence level upper limit is

\begin{equation}
m_H = 158\ \gevcc ~\cite{LEPElectroweakWorkingGroup}\mathrm{.}
\end{equation}

\ \\We see that the direct preferred Higgs boson mass is excluded by the direct searches at LEP experiments. If this exclusion is taken into account and a new fit is performed, then the 95\% confidence level values increases up to 185 $\gevcc$~\cite{LEPElectroweakWorkingGroup}. 

\section{Direct Searches at the Tevatron}

\ \\The Tevatron accelerator at the Fermi National Accelerator Laboratory (FNAL) in Batavia, Illinois (in the suburbs of Chicago), USA, collides protons and antiprotons at a centre-of-mass energy of 1.96 TeV. For almost two decades, the collisions at the Tevatron were both the most energetic and with the highest instantaneous luminosity in the world. However, in March 2010 the Large Hadron Collider (LHC) at Le Centre Europ\'{e}en de Recherche Nucl\'{e}aire (CERN) in Geneva, Switzerland, has broken Tevatron's record for centre-of-mass energy. As of March 2010, the LHC collides protons and protons at a centre-of-mass energy of 7 TeV, with a predicted 14 TeV energy in about five years. As of May 2011, the LHC has demonstrated instantaneous $pp$ luminosities in excess of $10^{33}\ \rm{cm}^{-2}\rm{s}^{-1}$, more than double the highest $p\bar{p}$ luminosity achieved by the Tevatron. In this dissertation we present a Standard Model Higgs boson search at the Tevatron. 

\begin{figure}[h]
\begin{center}
\includegraphics[angle=270,width=0.7\textwidth,clip=]{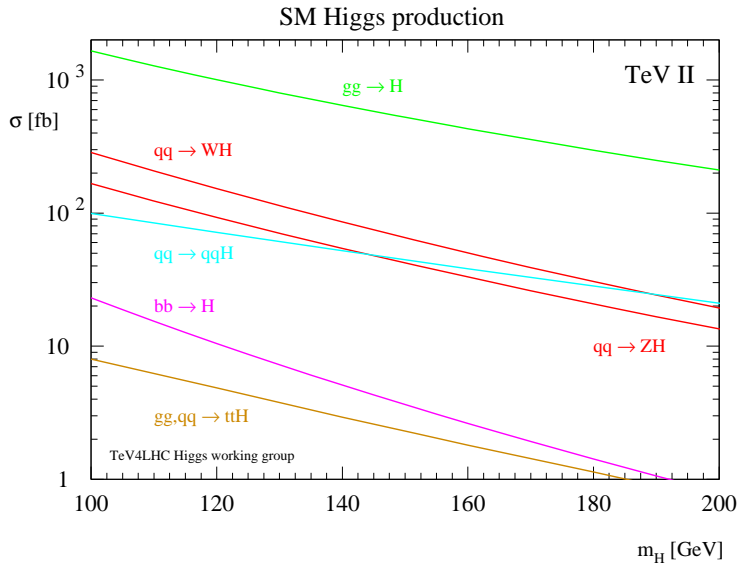}
\caption[SM Higgs production cross sections for $p\pbar$ collisions at 1.96 TeV]
    {SM Higgs production cross sections for $p\pbar$ collisions at a centre-of-mass energy of 1.96 TeV~\cite{HiggsCrossSections}.
      \label{figure:HiggsCrossSectionsTevatron}}
\end{center}
\end{figure}

\subsection{Higgs Boson Production at the Tevatron}

\ \\There are various processes predicted by the Standard Model through which a Higgs boson can be produced at the Tevatron accelerator. The cross sections of these processes vary with the centre-of-mass energy of collisions.

\ \\At the Tevatron, the relative cross sections can be seen in Figure~\ref{figure:HiggsCrossSectionsTevatron}. The most likely is gluon-gluon fusion. Next in line and about ten times less likely is an associated production of a $W$ boson and a Higgs boson. Next comes the associated production between a $Z$ boson and a Higgs boson and is about twice less likely than the previous process. These are followed by vector-boson fusion and the associated production between a top-quark pair and Higgs bosons, which have almost negligible cross sections.

\begin{figure}[h]
\begin{center}
\includegraphics[width=0.7\textwidth,clip=t]{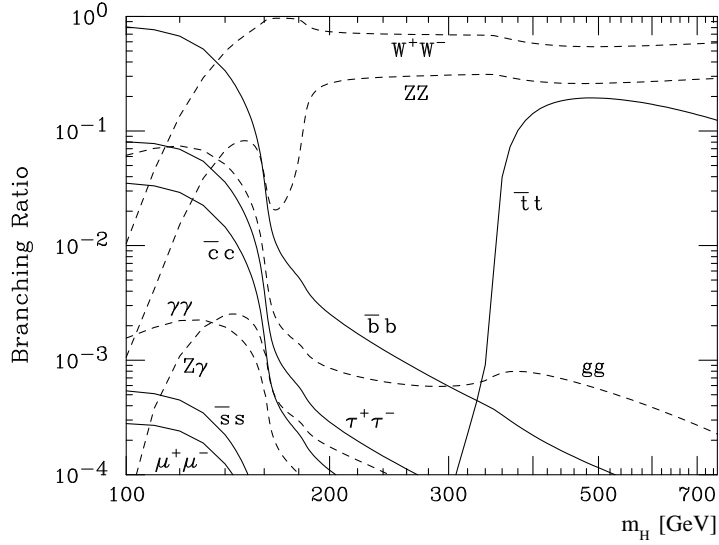}
\caption[Branching ratios for the main decays channels of the SM Higgs boson]
    {Branching ratios for the main decays of the SM Higgs boson~\cite{SM-MSSM-HiggsReview}.
      \label{figure:HiggsBosonDecayModes}}
\end{center}
\end{figure}

\subsection{Higgs Boson Decay}

\ \\The Standard Model predicts the Higgs boson decay modes and their branching ratios independent of the way the Higgs boson was produced\footnote{Be it at the LHC in $pp$ collisions at $\sqrt{s}=7\tev$ or at the Tevatron in $p\pbar$ at $\sqrt{s}=1.96\tev$, the Higgs boson decays in the same way, depending only on its mass. However, the production cross section increases with the centre-of-mass energy.} and depend only on the Higgs boson mass, as we can see in Figure~\ref{figure:HiggsBosonDecayModes}.

\ \\Higgs boson masses up to 114.4 $\gevcc$ have been excluded at 95\% CL by direct searches at experiments at the LEP accelerator. Also, all Higgs boson masses from 186 $\gevcc$ upward have been excluded also at 95\% CL by indirect electroweak fits. The remaining interval of possible Higgs boson masses is divided into values below and above 135 $\gevcc$. If in the interval 114.4 $\gevcc$-135 $\gevcc$ (also called the ``low mass range''), the Higgs boson decays predominantly to a pair of bottom-antibottom quarks ($H\to b\bbar$). On the other hand, if in the interval 135 $\gevcc$-186 $\gevcc$ (also called the ``high mass range''), the Higgs boson decays predominantly to a pair of $W$ bosons ($H\to W^*W$)\footnote{Since both $W$ bosons cannot be on shell for a Higgs boson mass below approximately 160 $\gevcc$, one of the $W$ bosons will be a virtual particle, hence the notation with ``*''.}. 

\ \\A $W$ boson can decay leptonically to a charged lepton and its corresponding neutrino ($W\to e\nu_{e}$, or $W\to \mu\nu_{\mu}$ or $W\to \tau\nu_{\tau}$) or hadronically to a quark-antiquark pair ($W\to q\qbar$).

\subsection{Low Mass Direct Searches at the Tevatron}

\ \\At the Tevatron, the quark pair production through QCD processes has a cross section of about ten orders of magnitude larger than the Higgs boson processes. Quarks hadronize very quickly and they are observed in particle detectors as collimated streams of subatomic particles called jets. The key therefore is that at least one $W$ or $Z$ boson originating from a Higgs boson event is observed in the detector through its leptonic decay products.

\ \\Since an event with a Higgs boson produced through gluon fusion that subsequently decays to a bottom-quark pair ($gg\to H \to b\bbar$) does not present any charged lepton, it is practically impossible to observe a low mass Higgs boson produced through gluon fusion due to the higher backgrounds. 

\ \\However, the process next in line in terms of cross section (WH associated production) can be observed when the $W$ boson is reconstructed through its leptonic decay: $q\qbar \to W^* \to WH \to l\nu_l\ b\bbar$. The charged lepton (an electron or a muon) is reconstructed in the detector and the neutrino is seen as missing transverse energy. Therefore, the compromise is that a ten times smaller signal cross section process brings the advantage of a high QCD background rejection due to the presence in the Higgs boson event of a $W$ boson that is reconstructed through its leptonic decay.

\ \\A similar channel is the $ZH$ associated production. The $Z$ boson is also reconstructed through leptonic decay, either to a pair of charged leptons ($Z \to l^+ l^-$) or to a pair of neutrinos ($Z\to \nu\bar{\nu}$). The sensitivity of this channel is similar but smaller than that of the $WH$ channel.

\ \\It is also possible to search for a Higgs boson produced in gluon fusion that decays to a pair of tau leptons ($gg\to H \to \tau^+ \tau^-$). A tau lepton decays either leptonically or hadronically. By asking for one tau lepton to decay leptonically and another one to decay hadronically this channel is also somewhat sensitive at the Tevatron, although less than the associated production modes.

\begin{figure}[ht]
\begin{center}
\includegraphics[width=12.5cm]{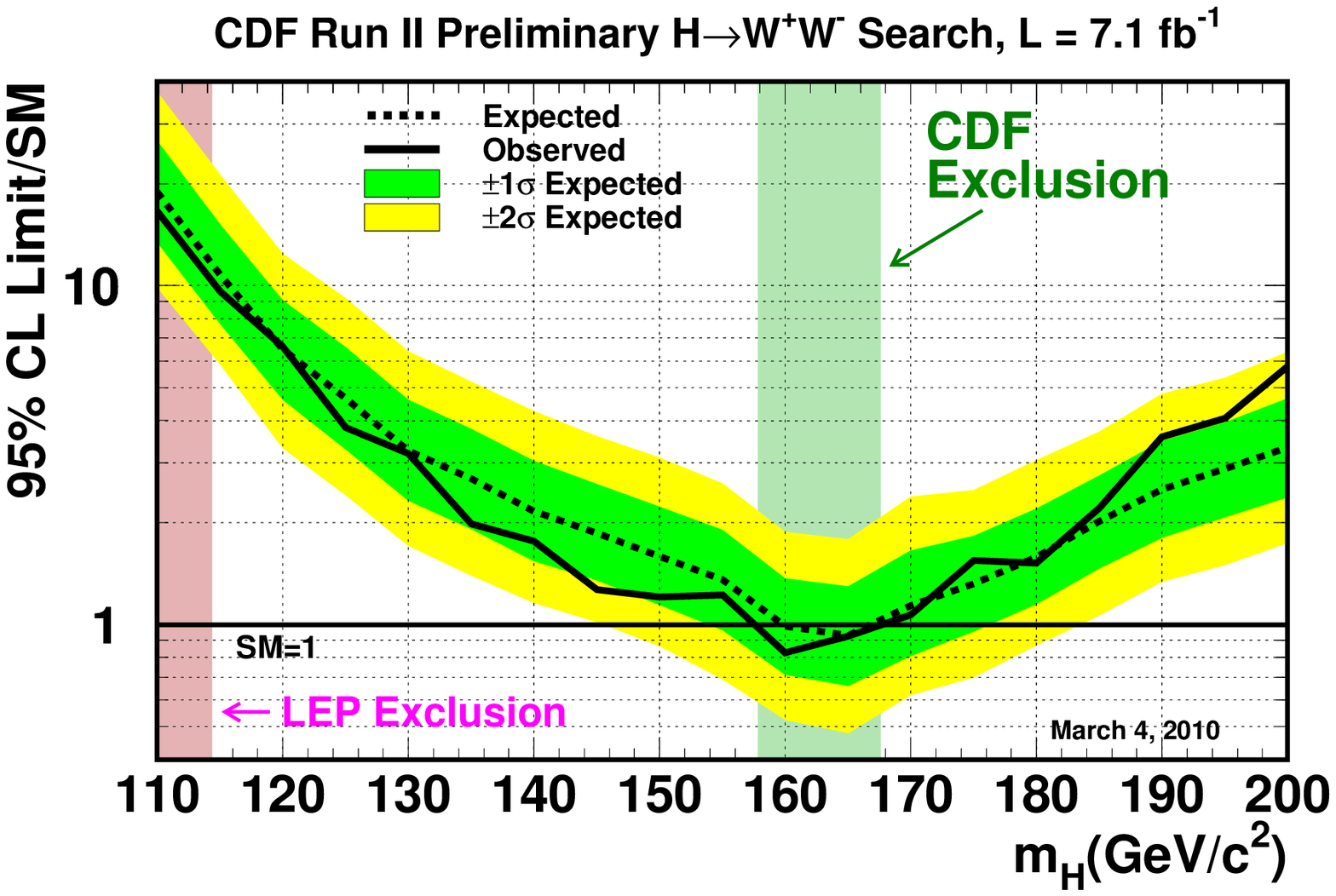}\\
\includegraphics[width=12.5cm]{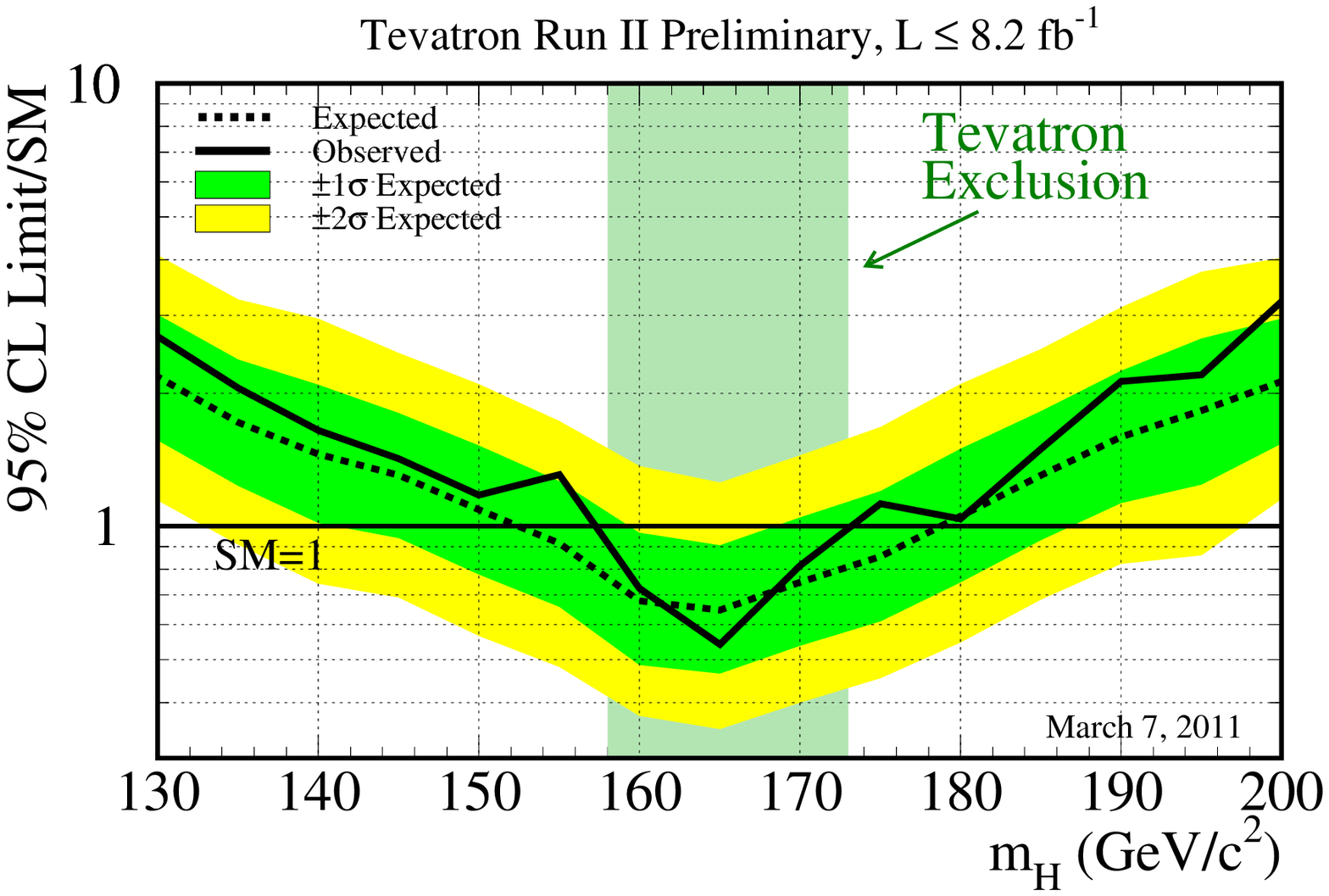}
\caption[CDF and Tevatron combination for high mass Higgs boson as of March 2011.]
{Expected and observed upper limits normalized to the Standard Model (x SM) for a high mass Higgs boson search for CDF (top) and CDF and DZero (bottom), where at most 8.2 $\invfb$ is used, as of March 2011 \cite{HiggsCombinationHighMassMarch2011}. The limits are presented as a function of the Higgs boson mass, between $110\gevcc$ and $200\gevcc$. The horizontal line at 1 represents the Standard Model prediction. The observed upper limits are represented by solid black lines. The expected median upper limits are represented by the dashed black lines. The green (yellow) band represents the 1 (2) standard deviation interval around the expected median upper limit represented by dashed lines.\label{figure:HiggsCombinationHighMassMarch2011}}
\end{center}
\end{figure}

\subsection{High Mass Direct Searches at the Tevatron}

\ \\If in the high mass range, a Higgs boson decays to a pair of $W$ bosons. Reconstructing both $W$ bosons reduces significantly the QCD and electroweak backgrounds. This search is the most sensitive at the Tevatron in the high mass range. 

\ \\The most recent high mass direct search from CDF was performed using an integrated luminosity of 7.2 $\invfb$ and is presented in the top part of Figure \ref{figure:HiggsCombinationHighMassMarch2011}. When combined to the high mass analyses at DZero using up to an integrated luminosity of 8.2 $\invfb$, a Tevatron high mass Higgs boson combination result is achieved, which is presented at the bottom part of Figure \ref{figure:HiggsCombinationHighMassMarch2011}. These results were made public on 7 March 2011 and were presented at the conferences of Winter 2011 \cite{HiggsCombinationHighMassMarch2011}. 

\subsection{This Analysis: SM $WH$ Search at CDF II}

\ \\Our motivation for choosing a Standard Model Higgs boson search for this PhD thesis, although a SUSY Higgs search would have been possible as well\footnote{There are other analyses at CDF and also at the DZero, ATLAS and CMS detectors searching for a SUSY Higgs boson.}, is that we feel that priority should be given to the effort of discovering or excluding a Standard Model Higgs boson. The Tevatron accelerator was chosen because it was the only particle accelerator in the world to have been and still be sensitive to Standard Model Higgs boson on the timescale of the thesis. Given that electroweak indirect fits favour a light Higgs over a more massive one, this thesis performs a low mass Higgs boson search. The most sensitive low mass Higgs boson search at the Tevatron is the associated production $WH$, where the $W$ bosons decay leptonically and the Higgs bosons decay to a pair of bottom quarks. The leading-order Feynman diagram of the $WH \to l\nu b\bbar$ process is presented in Figure~\ref{figure:FeynmanWH}. This is the analysis presented in this dissertation.

\begin{figure}[ht]
\begin{center}
\includegraphics[width=12.0cm]{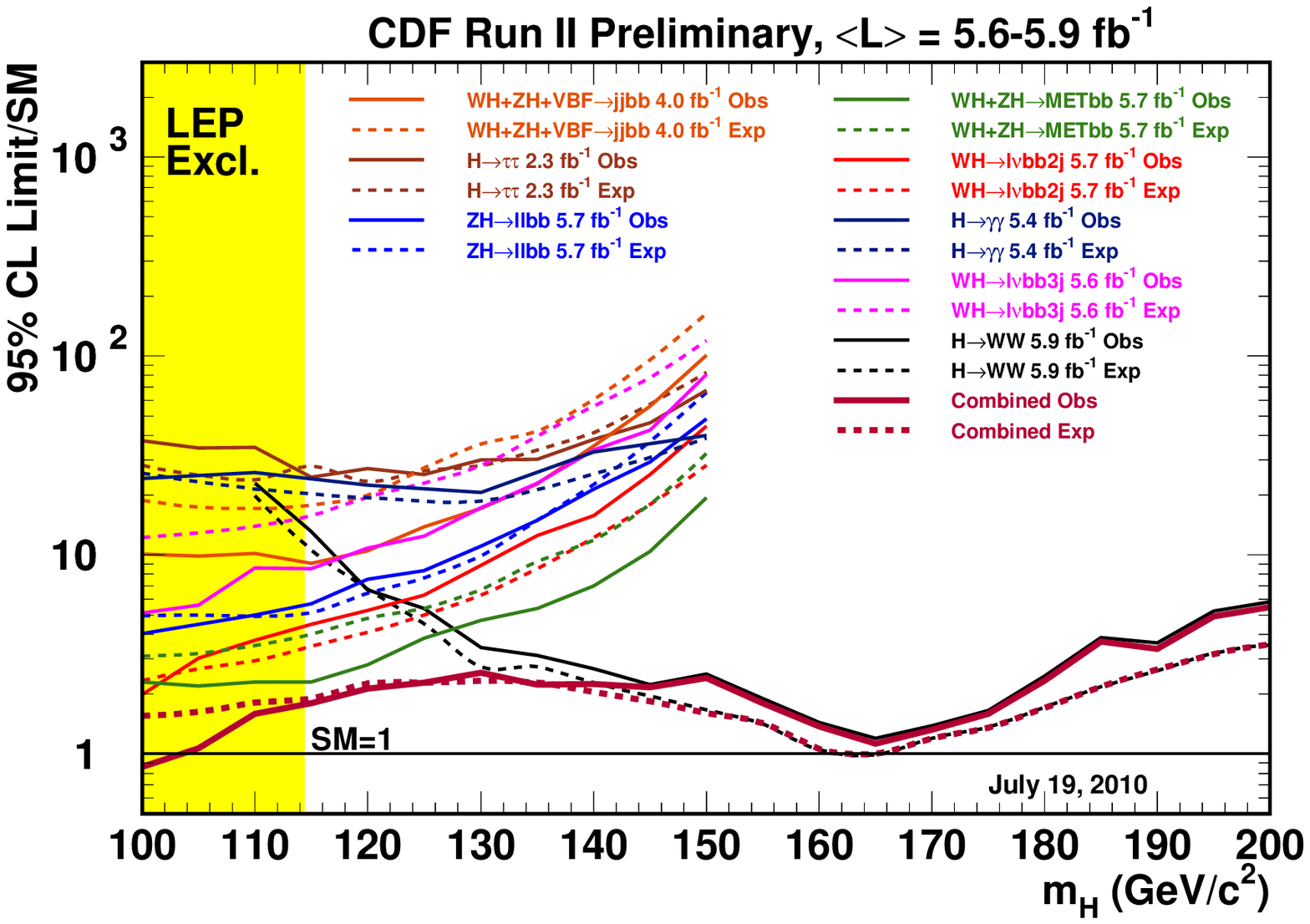}
\includegraphics[width=12.0cm]{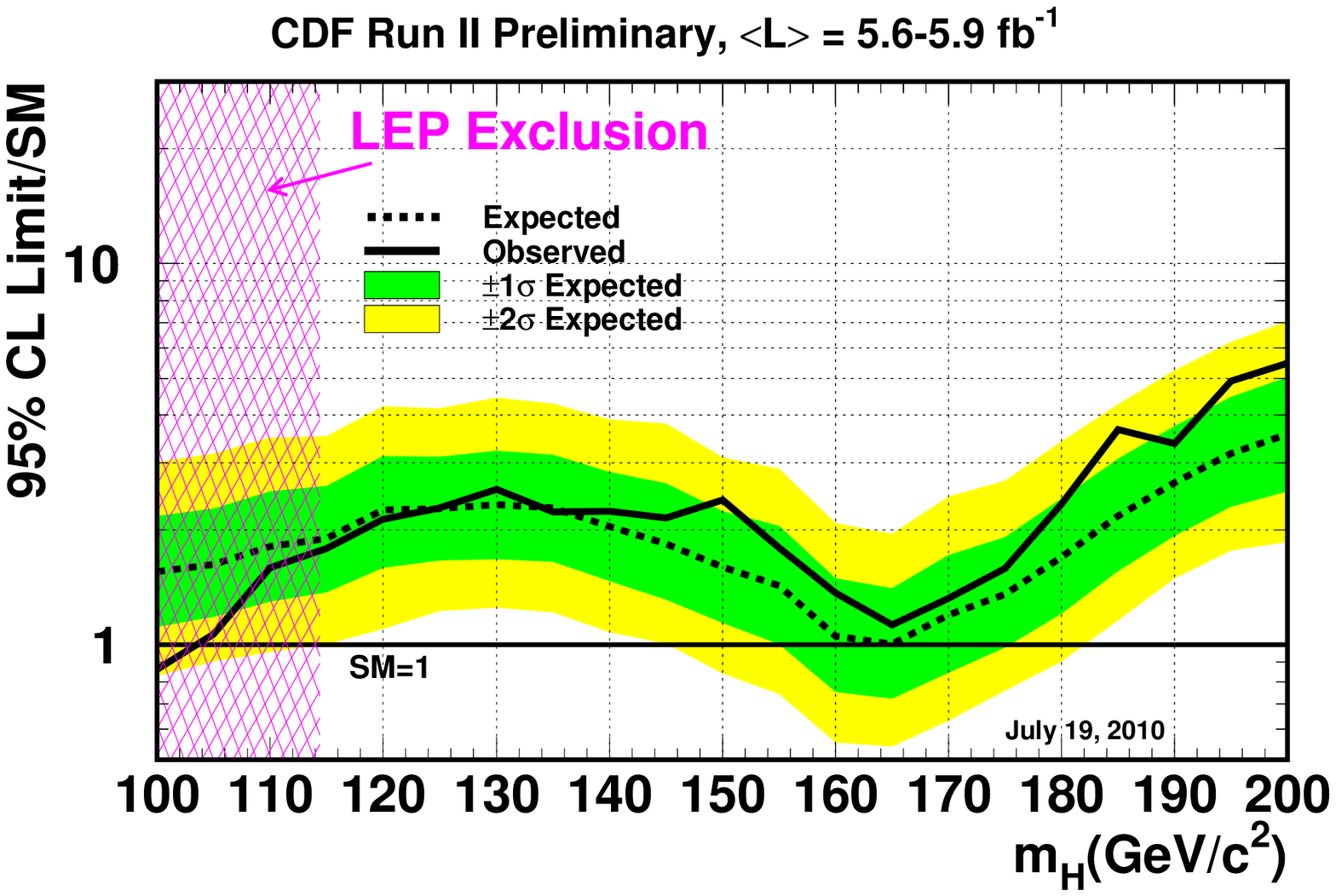}
\caption[CDF Higgs search combination in July 2010.]
{Expected and observed upper limits normalized to the Standard Model (x SM) for a Higgs boson search for CDF individual analyses using between 5.6 and 5.9 $\invfb$ and their combined result, as a function of the Higgs boson mass, between $100\gevcc$ and $200\gevcc$, as of July 2010 \cite{CDFCombinationWebpageJuly2010}. The horizontal line at 1 represents the Standard Model prediction. The yellow (top) or pink (bottom) band is the region excluded through a direct search combining searches at all the experiments at LEP accelerator. The observed upper limits are represented by solid lines. The expected upper limits are represented by the dashed lines. Besides the individual analyses used in this combination, the combined upper limit is represented in dark red in the top plot. In the bottom plot, the green (yellow) band represents the 1 (2) standard deviation interval around the expected upper limit represented in dashed lines.\label{figure:HiggsCombinationAllMassJuly2010CDF}}
\end{center}
\end{figure}

\begin{figure}[ht]
\begin{center}
\includegraphics[width=12.0cm]{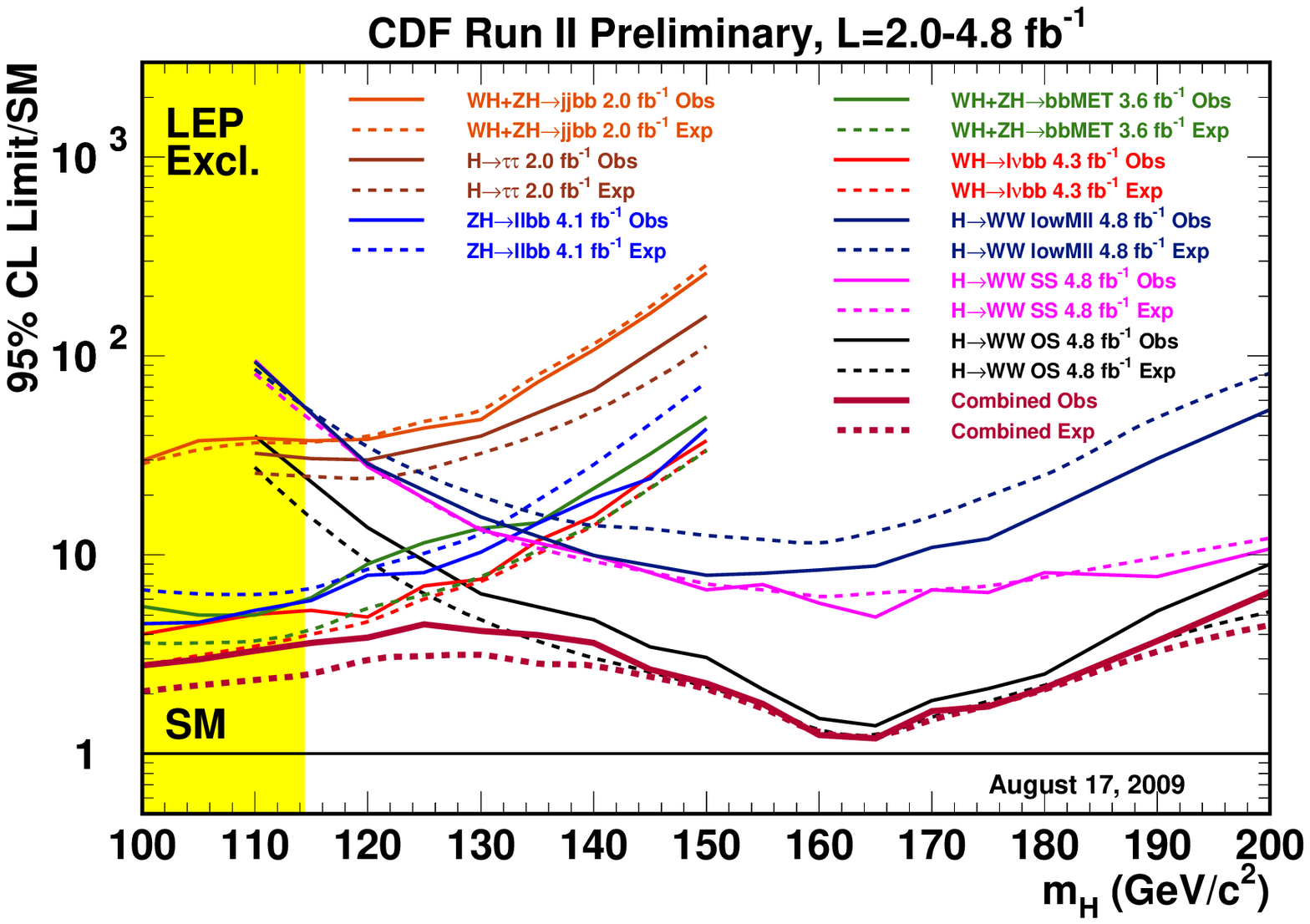}
\includegraphics[width=12.0cm]{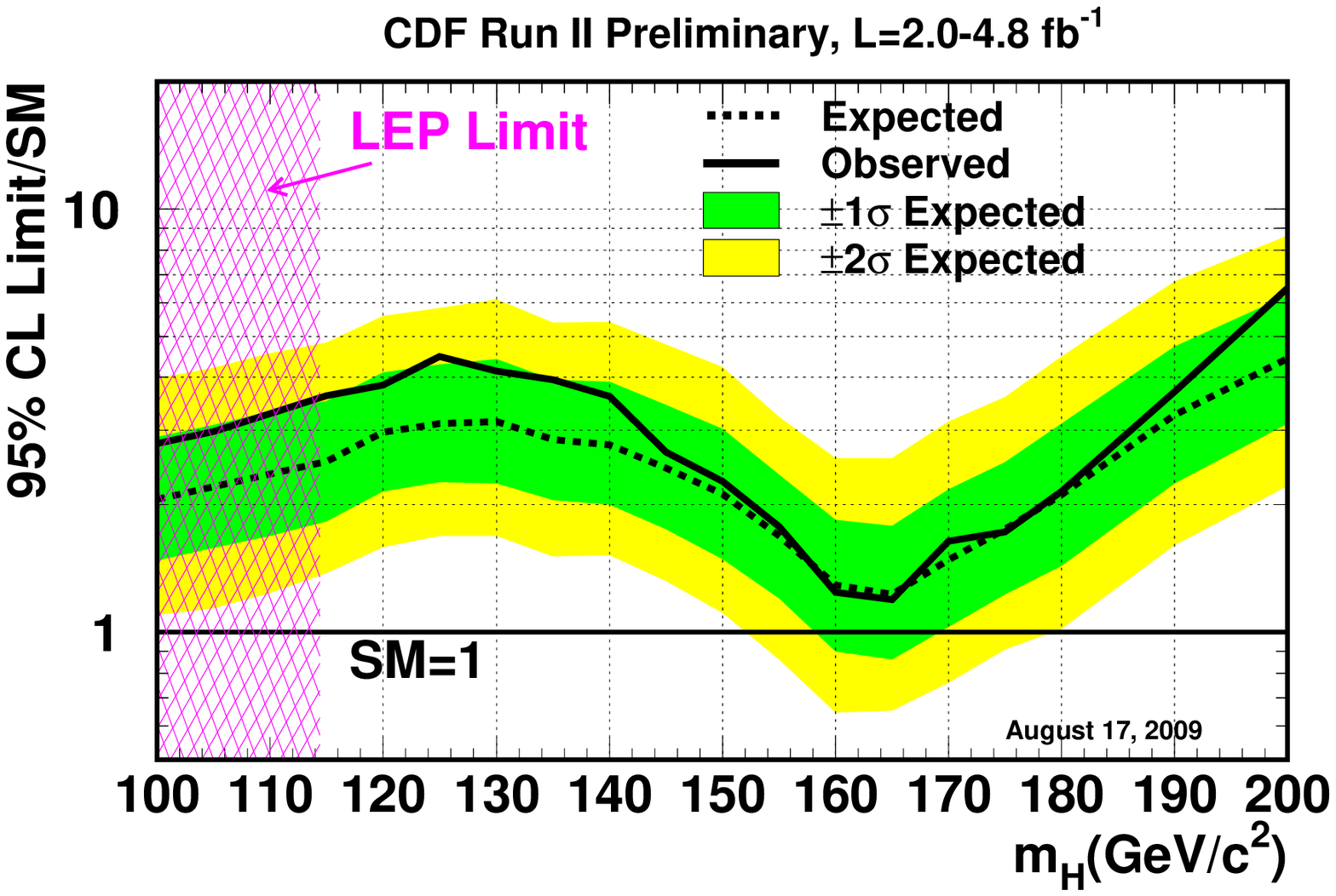}
\caption[CDF Higgs search combination in Aug 2009.]
{Expected and observed upper limits normalized to the Standard Model (x SM) for a Higgs boson search for CDF individual analyses using between 2.0 and 4.8 $\invfb$ and their combined result, as a function of the Higgs boson mass, between $100\gevcc$ and $200\gevcc$, as of August 2009 \cite{CDFCombinationWebpageAug2009}. The horizontal line at 1 represents the Standard Model prediction. The yellow (top) or pink (bottom) band is the region excluded through a direct search combining searches at all the experiments at LEP accelerator. The observed upper limits are represented by solid lines. The expected upper limits are represented by the dashed lines. Besides the individual analyses used in this combination, the combined upper limit is represented in dark red in the top plot. In the bottom plot, the green (yellow) band represents the 1 (2) standard deviation interval around the expected upper limit represented in dashed lines.\label{figure:HiggsCombinationAllMassAug2009CDF}}
\end{center}
\end{figure}

\subsection{Combination of all CDF SM Higgs Analyses}

\ \\The $WH$ search is one of the most sensitive low mass Standard Model Higgs boson modes at CDF, but not the only one. CDF combined in July 2010 all Higgs searches, both at low and high mass, to obtain an improved CDF Standard Model upper limit on the cross section \cite{CDFCombinationWebpageJuly2010}. The individual analyses expected\footnote{Expected limits measure the sensitivity of the analysis and are computed as the median of a distribution of many pseudo-experiments using pseudo-data, as explained in Section~\ref{section:LimitExpectedAndObserved}.} and observed upper limits are presented in the top part of Figure~\ref{figure:HiggsCombinationAllMassJuly2010CDF}. My work contributed directly to the $WH$ search using exactly two tight jets (in red) and therefore contributed also to the combined CDF result, which is also presented in detail in the bottom part of Figure~\ref{figure:HiggsCombinationAllMassJuly2010CDF}. 

\ \\The CDF Higgs combination from July 2010 improves and supersedes the same result from August 2009 \cite{CDFCombinationWebpageAug2009}, to which my work also contributed directly as part of the $WH$ search (in red). The CDF Higgs combination from August 2009 is presented in Figure~\ref{figure:HiggsCombinationAllMassAug2009CDF}. 

\begin{figure}[ht]
\begin{center}
\includegraphics[width=12cm]{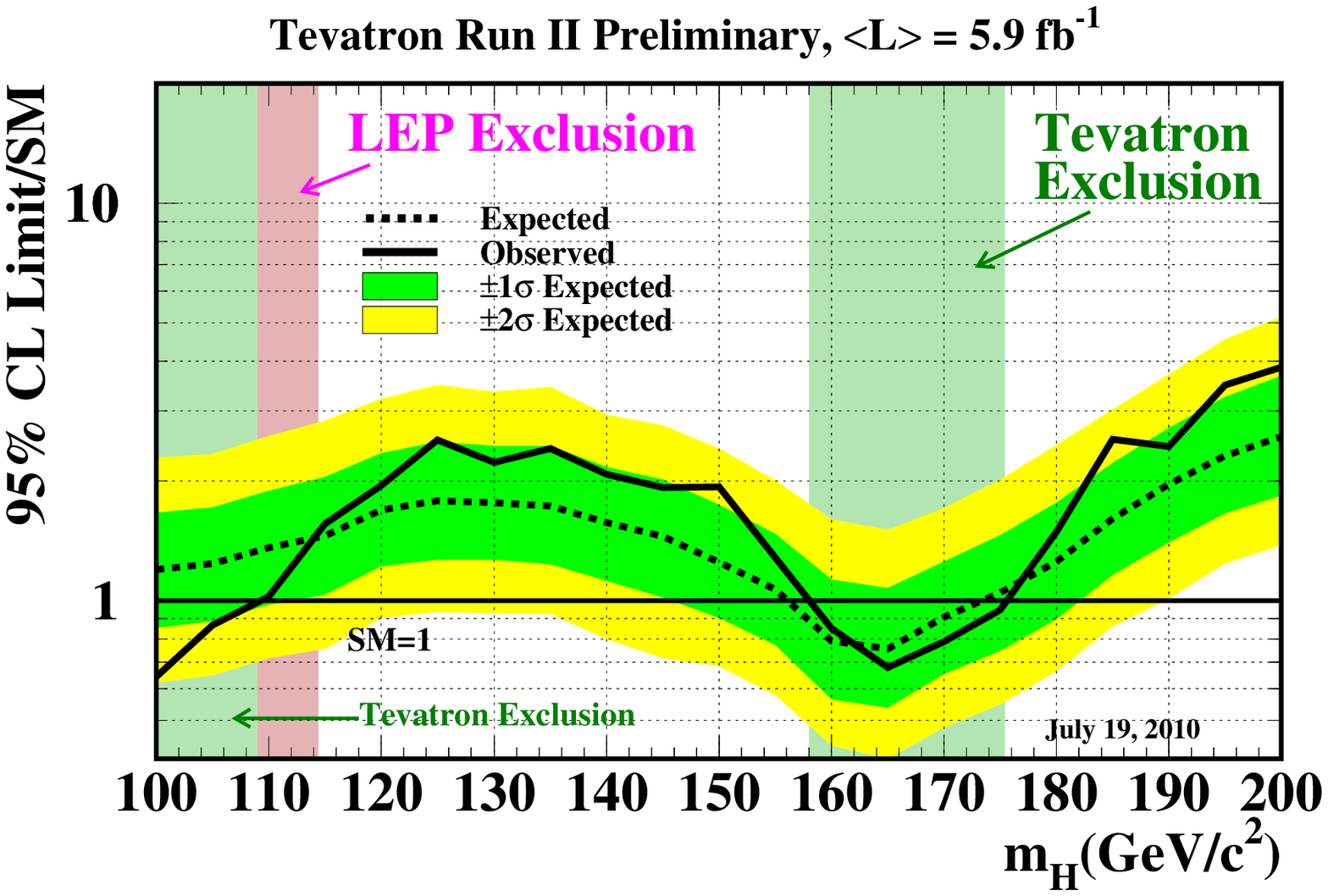}
\includegraphics[width=12cm]{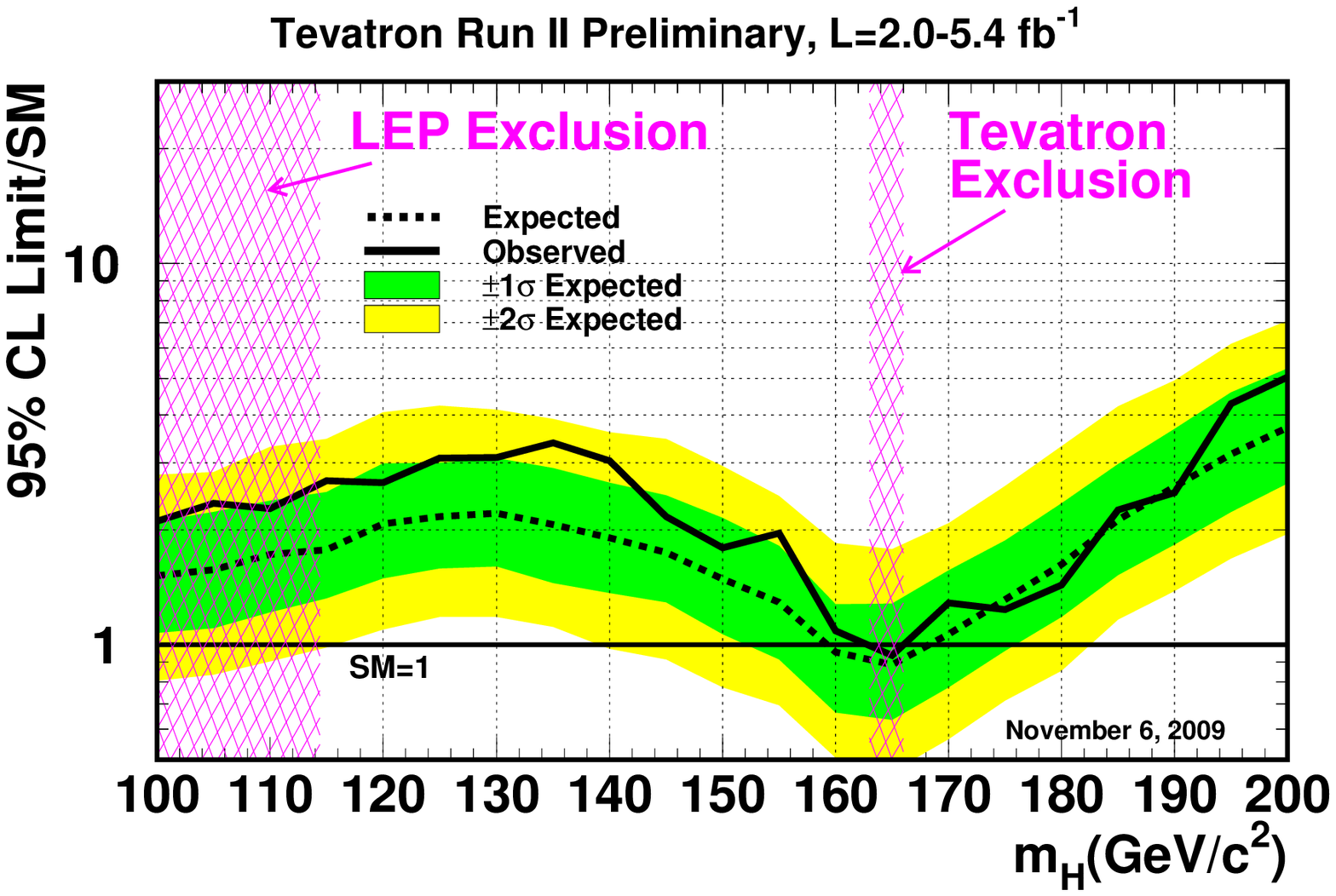}
\caption[Tevatron combination for SM Higgs boson upper limits]
{Expected and observed upper limits normalized to the Standard Model prediction for a Higgs boson search in a combination of searches both at CDF and DZero using between 5.6 and 5.9 $\invfb$, as of July 2010~\cite{TevatronHiggsCombinationJuly2010} (top) and Nov 2009~\cite{TevatronHiggsCombinationNov2009}. The horizontal line at 1 is the Standard Model prediction. The dashed line represents the expected limit. The green (yellow) band represents the one (two) standard deviation interval around the expected limit. The solid line is the measured limit. The 95\% CL Standard Model exclusion intervals are represented by the Higgs boson masses where the measured line goes below the Standard Model line. In July 2010 a SM Higgs boson was excluded at 95\% confidence level in the interval 158-175 $\gevcc$, which supersedes the interval of 163-166 $\gevcc$ of November 2009. 
\label{figure:HiggsCombinationAllMassTevatron}}
\end{center}
\end{figure}

\subsection{Combination of all Tevatron SM Higgs Analyses}

\ \\In order to achieve a higher sensitivity for the Standard Model Higgs boson search, all the low and high mass search channels, both from CDF and DZero, are combined to create one Tevatron cross section times branching ratio plot. The latest Tevatron combination dates from July 2010 and can be seen in the top part of Figure~\ref{figure:HiggsCombinationAllMassTevatron}. It excludes at 95\% CL a SM Higgs boson with a mass in the range 158-175 $\gevcc$~\cite{TevatronHiggsCombinationJuly2010}. In the low mass region, the limit is about 2 times the SM prediction. This result improves upon and supersedes the previous Higgs Tevatron combination result from November 2009, which had excluded the interval $163 \gevcc$ to $166 \gevcc$~\cite{TevatronHiggsCombinationNov2009} and which can be seen in the bottom part of Figure~\ref{figure:HiggsCombinationAllMassTevatron}. My work contributed directly to both of these combined Higgs searches. 

\begin{figure}[h]
\begin{center}
\includegraphics[angle=270,width=0.7\textwidth,clip=]{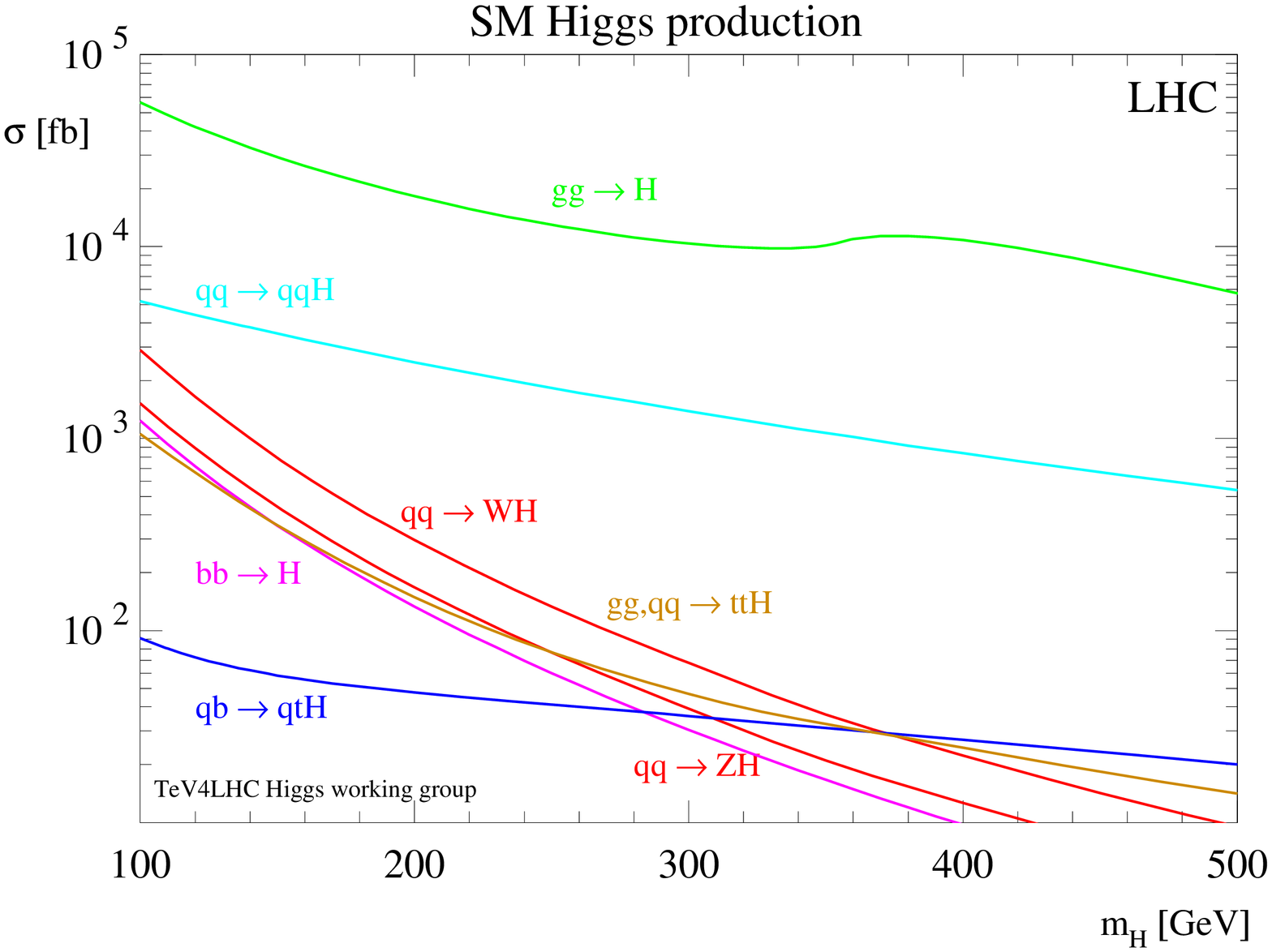}
\caption[SM Higgs production cross sections for $pp$ collisions at 14 TeV ]
    {SM Higgs production cross sections for $pp$ collisions at 14 TeV~\cite{HiggsCrossSections}.
      \label{figure:HiggsCrossSectionsLHC}}
\end{center}
\end{figure}

\section{Expected Higgs Boson Production at the LHC}

\ \\The only other particle accelerator in the world that can search for a Standard Model Higgs boson is the Large Hadron Collider (LHC) at the CERN laboratory in Europe. At the moment, the LHC is colliding protons and protons at centre-of-mass energy of $\sqrt{s}=7\,\tev$. The current run is scheduled to end in 2012, with a ``technical stop'' at the end of 2011. At least $1 \invfb$ of integrated luminosity is expected to be collected during this run by each LHC experiment. This run will be followed by a shutdown lasting about two years when the LHC will be upgraded to be able to run at the design centre-of-mass energy of $\sqrt{s}=14\,\tev$.

\ \\Various processes have different cross sections as a function of the colliding particles and the centre-of-mass energy. Due to similar behaviour for background processes and different behaviour for signal processes, as seen in Figure \ref{figure:HiggsCrossSectionsLHC}, the most sensitive channels for a Higgs boson discovery at the LHC are different than those at the Tevatron. A compilation of the most up-to-date cross section values for SM Higgs boson can be found in Reference~\cite{HiggsCrossSections}.

\ \\The ATLAS \cite{ATLASHiggsResults} and CMS \cite {CMSHiggsResults} experiments have made public their first searches for the Standard Model Higgs boson for the winter conferences of 2011. However, none of those results were as sensitive as the Tevatron combination result from the summer of 2010. 

\section{Summary}

\ \\In this chapter we have presented the history of direct searches for the Standard Model Higgs boson at the LEP, Tevatron and LHC accelerators, as well as the constraints on the Higgs boson mass from fits between precise experimental electroweak measurements and the Standard Model prediction. Since the Standard Model Higgs boson has not yet been observed experimentally and since it is uniquely described by its mass, experimental particle physicists have refuted at 95\% CL the existence of the Higgs boson if its mass is in a certain mass range. For example, in 2000 the combination of several direct searches from four collaborations at the LEP accelerator at CERN have excluded at 95\% CL Higgs bosons with masses smaller than 114.5 $\gevcc$. In the summer of 2010, the combination of several direct searches from the CDF and DZero collaborations at the Tevatron accelerator at Fermilab have excluded at 95\% CL Higgs boson masses in the range of 158-175 $\gevcc$. There were also indirect fits from measurements of the masses of the $W$ boson and top quark to the Standard Model predictions that exclude at 95\% CL Higgs boson masses with values larger than 185 $\gevcc$. 

\ \\ At the winter conferences of 2011 the first searches for the Standard Model Higgs boson from the ATLAS and CMS collaborations at the LHC accelerator at CERN have been made public, but they are not yet more sensitive than the Tevatron results. However, the situation is expected to change in 2012, when the LHC experiments will take the lead from the Tevatron once more and in a few extra years will observe or exclude the Standard Model Higgs boson over its entire available mass range.

\ \\This dissertation presents a direct search for the Standard Model Higgs boson by the Collider Detector at Fermilab collaboration at the Tevatron accelerator at Fermilab. The allowed Higgs mass interval may be divided into a low mass region and a high mass region, depending on the most probable decay channel. Since indirect electroweak fits suggest a Higgs mass in the low mass region, we chose to perform the most sensitive search at the Tevatron for a low mass Higgs, which is the associated production between a $W$ boson and a Higgs boson. The leptonic decay of the $W$ bosons allows us to reconstruct in the detector the electron or muon candidate and thus increase considerably the signal over background ratio and thus the sensitivity of the analysis. 

\ \\In the next chapter we will present the experimental infrastructure used to perform the $WH$ search. 

\clearpage{\pagestyle{empty}\cleardoublepage}

\chapter{Experimental Infrastructure\label{chapter:Experiment}}

\ \\The Higgs boson search presented in this thesis is performed using the Collider Detector at Fermilab (CDF) that surrounds one of the two collision points at the Tevatron accelerator hosted at Fermilab, Batavia, IL, USA. In this chapter we will present the experimental infrastructure of our analysis, namely the accelerator complex at Fermilab and the CDF detector.

\section{The Fermi National Accelerator Laboratory}

\ \\The Fermi National Accelerator Laboratory (also known as Fermilab or FNAL) was founded by Robert W. Wilson (1914-2000) to be the largest particle physics laboratory in the USA\footnote{Besides the scientific aspect, Fermilab offers many other things to the US in general and its local community in particular. Fermilab owns a very large surface of land where the prairie was recovered as it used to be a couple of hundred years ago. In addition, a bison herd is raised in a large farm at Fermilab. Fermilab also hosts numerous lakes where a number of migratory birds take refuge, especially the Canada geese. The public is allowed to walk or bike in the natural environment at Fermilab. Fermilab is also a strong supporter of science communication to the general public. Its second director, Leon Lederman, created a science outreach centre at Fermilab. In addition, there are Saturday morning physics lectures and other public lectures from the latest developments on science and science policy. I had the chance to spend many months at Fermilab and I took great pleasure of enjoying all these various opportunities that Fermilab has to offer.}. As of January 2011, Fermilab remains the only national particle physics laboratory in the US, as other former particle physics laboratories have switched to related fields of science\footnote{For example, the Brookhaven National Laboratory hosts the RHIC accelerator that performs heavy-ion collisions and studies the quark-gluon plasma that existed shortly after the Big Bang, and the Stanford Linear Accelerator Center (SLAC) now uses its electron accelerator to produce free electron lasers.}.

\ \\Fermilab hosts a multi-stage complex accelerator complex, of which the Tevatron is just the most energetic accelerator. The Tevatron performs proton-antiproton collisions that are recorded by two general purpose detectors, CDF and DZero. These collisions are used for fundamental particle physics research. Fermilab also hosts a series of fixed target experiments, as beams of protons are extracted from the Tevatron and collided with fixed targets. In this process secondary beams of mesons, muons or neutrinos are produced. 

\begin{figure}[h]
\begin{center}
\includegraphics[angle=0,width=0.7\textwidth,clip=]{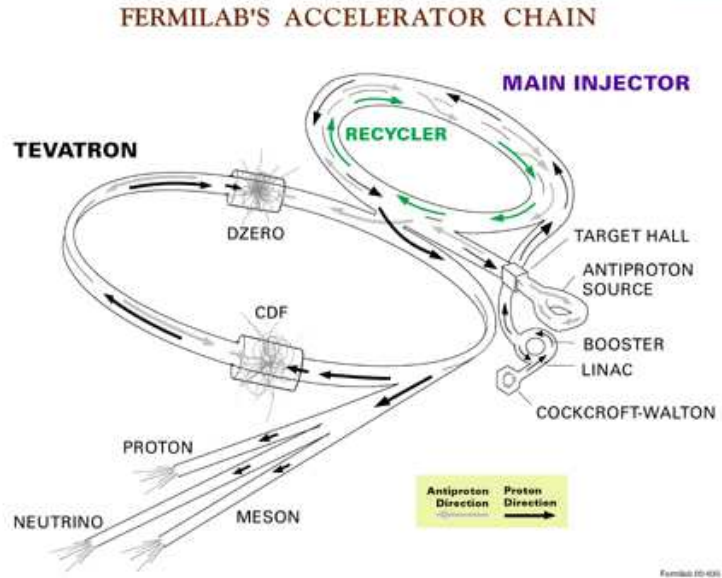}
\caption[Schematic view of the Fermilab Accelerator Complex]
    {Schematic view of the Fermilab Accelerator Complex, where protons and antiprotons are produced, accelerated in subsequent steps and collided in two points around the Tevatron storage ring. Credit image to the CDF collaboration. \label{figure:FermilabAcceleratorComplex}}
\end{center}
\end{figure}

\begin{figure}[h]
\begin{center}
\includegraphics[angle=0,width=0.7\textwidth,clip=]{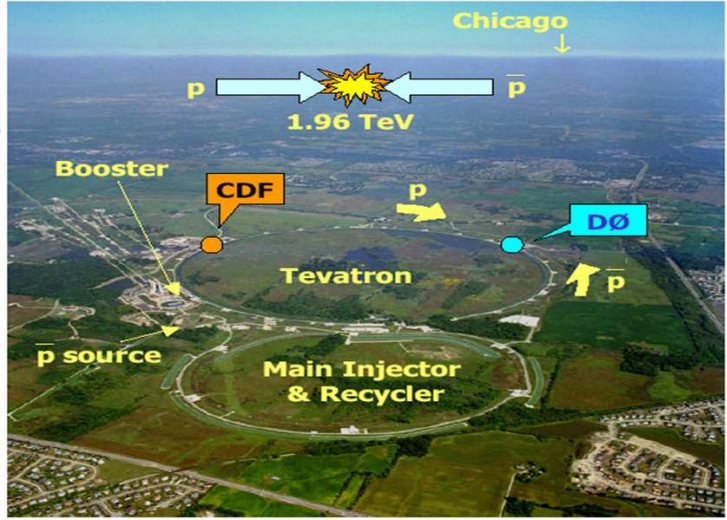}
\caption[Bird's Eye View of the Fermilab Accelerator Complex]
    {Bird's Eye View of the Fermilab Accelerator Complex, looking East. Credit image to the CDF collaboration.\label{figure:TevatronBirdsEyeView}}
\end{center}
\end{figure}

\section{Fermilab Accelerator Complex}

\ \\The Tevatron is a proton-antiproton storage ring where $p\pbar$ collisions are made to occur at a centre-of-mass energy $\sqrt{s}=1.96\,\tev$. Its first collisions were achieved in 1986 and since then a series of improvements allowed for many increases in collision energy and instantaneous luminosity.  The first physics run (Run I) used $\sqrt{s}=1.8\,\tev$ and took place between 1992 and 1996. Between 1997 and 2001, both the accelerator complex and the particle detectors were improved dramatically. While the Tevatron accelerated only 6 bunches of protons or antiprotons in Run I, it accelerates 36 in Run II. Also, while the time interval between two consecutive bunch crossings was 3500 ns in Run I, it decreases to 396 ns in Run II. 

\ \\Run II started in 2001 and as of January 2011 is still in progress. Initially the Tevatron was scheduled to shut down in 2009, when the LHC was supposed to start its physics program at $\sqrt{s}=14\,\tev$ and higher instantaneous luminosity. However, delays in the LHC construction and an incident that stopped the LHC for an extra year motivated prolonging Run II at the Tevatron until September 2011. Furthermore, as the LHC plans to run about 2 years at $\sqrt{s}=7\,\tev$ and then have a one year and a half major shutdown to prepare the LHC for $\sqrt{s}=14\,\tev$ has prompted Fermilab to seek approval and financing for a Run III for the Tevatron from September 2011 through August 2014. 

\ \\The main motivation is that the Tevatron has exhibited excellent performance and is running smoothly, breaking instantaneous luminosity records every month. The Fermilab Accelerator Complex is very well understood after about 25 years of usage. Run II was initially designed to collect 2 $\invfb$ of integrated luminosity and over 10 $\invfb$ have already been delivered, with an expected 12 $\invfb$ before the end of Run II in 2011. 

\ \\Acceleration of protons and antiprotons to collision energies is realized by a complex of eight accelerators, two linear (Cockcroft-Walton and Linac) and six circulating synchrotrons (Booster, Main Injector, Debuncher, Accumulator, Recycler and Tevatron). This huge accelerator complex consumes 30 MW of electric power and stretches over 9 km.

\ \\Proton-antiproton collisions take place at two points around the Tevatron storage ring, in two buildings called BZero and DZero. A general purpose particle detector surrounds each collision point: the Collider Detector at Fermilab (CDF) in the BZero building and the DZero experiment in the DZero building. 

\ \\The accelerator complex at Fermilab~\cite{BuzatuMScThesis} and the CDF and DZero experiments are shown schematically in figure~\ref{figure:FermilabAcceleratorComplex} and from a bird's eye view in figure~\ref{figure:TevatronBirdsEyeView}. The proton and antiproton sources and the various acceleration stages are described in the sections below.

\subsection{Proton Source}

\ \\First, protons have to be produced. A strong electric field ionizes hydrogen atoms at room temperature (0.04 eV/atom) and sends protons and electrons in opposite directions. The protons fall on and stick to a cesium surface. The work needed to free an electron from a cesium surface is smaller than in the case of any other atom, since cesium is the most reactive atom. A falling atom may collide with a group consisting of a proton and two electrons that are temporally together on the cesium surface. The group is thus freed from the surface and it forms a hybrid negative hydrogen ion ($H^{-}$). Thanks to the same electric field, a continuous $H^{-}$ beam of about 25 keV is collected.

\ \\The  $H^{-}$ beam is accelerated by a Cockcroft-Walton accelerator to an energy of 750 keV by a constant electric field. The acceleration voltage is limited by the fact that at high voltages the air creates sparks.

\ \\The $H^{-}$ beam is subsequently accelerated to 400 MeV by a 130 m long linear accelerator called the Linac. The Linac uses alternating current and resonant frequency cavity technology. The continuous beam is therefore bunched up. When outside a cavity, a bunch is accelerated by an electric field. When inside a cavity, a bunch does not see the electric field now in the opposite direction and therefore is not decelerated. As $H^{-}$ particles acquire momentum, cavities and gaps are longer to provide constant acceleration. A typical bunch contains $1.5\cdot10^{9}$ particles. A typical pulse contains 4,000 bunches, a total of $6\cdot10^{12}$ particles and a typical pulse length corresponds to 20 ms. While in the Linac, a particle is accelerated by an electric field of 3MV/m. The beam power is 18 MW when the pulsed hybrid $H^{-}$ ion beam exits the Linac.

\ \\The $H^{-}$ beam is injected into a 475 m circumference circular synchrotron accelerator called the Booster. After the first bending magnet in the Booster, the beam passes through a carbon foil, after which we are left with a $H^{+}$ (proton) beam. Since the 20 ms pulse length for the Linac is much larger than the 2.2 ms Booster circumference, the $H^{-}$ pulse is present for several rotations of the new proton beam inside the Booster. The fact that the incoming pulse is made of $H^{-}$ instead of protons allows the merging of the two beams inside the Booster. The choice of $H^{-}$ for the linear accelerator is therefore not driven by the acceleration process, but by the complex engineering process of transferring the beam from the linear accelerator to a circular accelerator. The same process happens also at the Brookhaven accelerator complex~\cite{RHIC}.

\ \\The Booster synchrotron accelerates charged particles thanks to a resonant frequency cavity. As their momentum increases, particles are kept at a constant radius by a corresponding increase of the magnetic field. The proton beam is accelerated every turn by a 500 kV voltage drop. After completing 16,000 turns in 33 ms, the beam has 8 GeV, exits the Booster and enters the Main Injector synchrotron accelerator. The Main Injector injects 150 GeV protons into the Tevatron synchrotron accelerator and 120 GeV protons into the antiproton production complex.

\subsection{Antiproton Source}

\ \\Antiproton production occurs in the antiproton source. The bunched beam of 120 GeV protons from the Main injector smashes on a 7 cm nickel target every 1.5 s. Particles created in the forward direction are recovered through a lithium lens. A pulsed magnet acting as a charge-mass spectrometer selects only antiprotons. The antiproton beam is pulsed, which means the beam exhibits a large energy spread and a small time spread. To be debunched, the beams passed into another synchrotron accelerator, called the Debuncher. Low (high) energy antiprotons follow the interior (exterior) path, arrive at different times at the resonance frequency cavity. As they see different phases, low (high) energy antiprotons are accelerated (decelerated). After about 100 ms, the antiproton beam is almost continuous, having a small energy spread and a large time spread. After 1.5 seconds in the Debuncher, the beam is injected into a circulating synchrotron called the Accumulator. A new pulsed antiproton beam is then inserted into the Debuncher. It takes 1 million 120 GeV protons to hit the nickel target for 20 antiprotons at 8 GeV each to be injected into the Accumulator.

\ \\The Accumulator uses stochastic cooling to accumulate antiprotons while keeping them at the desired (very small) longitudinal (transverse) momentum for hours, even days. The Accumulator has a shape of a triangle with rounded corners. Stochastic cooling~\cite{StochasticCooling} transforms particles from a hot state, with large spreads in energy, to a cooler state, with smaller spreads in energy, thanks to a feedback technique using pickups and kickers. The triangular shape of the Accumulator is driven by the necessity to have several 16-meter-long straight sections to accommodate the pickups and kickers for the stochastic cooling~\cite{Accumulator}. Van der Meer received in 1984 the Nobel Prize for inventing stochastic cooling.

\ \\The continuous beam of 8 GeV antiprotons from the Accumulator is injected in the Main Injector. The Main Injector replaced in 1998 the Main Ring situated in the same tunnel as the Tevatron. This represents one of the major upgrades from Run I to Run II. The Main Injector accelerates both protons and antiprotons in the same ring, using the same magnetic field for ensuring the circular trajectory for these particles. 150 GeV antiprotons are sent in the Tevatron accelerator where they are accelerated to 980 GeV and collided with the proton beam. When a store ends, almost 75\% of the antiprotons survive. Since creating antiprotons is such a hard task, surviving antiprotons are recuperated in another synchrotron accelerator, called the Recycler.

\ \\The Recycler sits just above the Main Injector and acts as a fixed-energy storage ring thanks to its permanent magnets and stochastic cooling. The Recycler receives antiprotons both from the Accumulator and from the Tevatron at the end of a store. The Recycler acts as an antiproton storage ring until the Tevatron is ready to accept antiprotons in a new store.

\subsection{Tevatron}

\ \\When built in 1983, the Tevatron was the first superconducting synchrotron accelerator. The Tevatron's 1000 superconducting electromagnets can produce a magnetic field as large as 4.2 Tesla. Electromagnet coils are made of 8 mm niobium-titanium alloy wire. One coil contains about 70,000 km of wire. A dipole magnet is about 6.4 m long. Once per turn, particles receive a kick in energy of about 650 kV from a resonance frequency cavity. In about 20 seconds the magnetic field increases gradually from 0.66 Tesla to 3.54 Tesla, while the beam energies increase gradually from 150 GeV to 800 GeV. Meanwhile, the beams turn around the 1 km radius circular accelerator 1 million times. When the beams arrive at 980 GeV, an electric current of more than 4 kA flows through the electromagnet and creates a magnetic field of 4.2 Tesla. For comparison, the superconducting magnets at the LHC will run at 8.4 Tesla when the beam energy will be 7 TeV. Superconducting electromagnet coils kept at liquid helium temperature (4.3 K) have no resistance and therefore dissipate no energy through the Joule effect. Significantly larger currents are able to flow though these coils in order to produce very large magnetic fields. Tevatron's cryogenic system is one of the world's largest, along with HERA's and LHC's. If it absorbs 23 kW of power, it can still maintain the liquid helium temperature. The system can deliver 1000 litres/hour of liquid
helium at 4.2 K.

\ \\Table~\ref{table:FermilabAcceleratorComplex} summarizes the acceleration characteristics of the different stages of the Fermilab $p\pbar$ Accelerator Complex. In this table, $\beta=\frac{v}{c}$ expresses
the speed of the particle as a fraction of the speed of light in vacuum and
$\gamma=\frac{E}{pc}=\frac{1}{\sqrt{1-(\frac{v}{c})^{2}}}$ is the relativistic factor. Also, for highly relativistic particles, kinetic and total energies can be approximated to be the same.

\begin{center}
\begin{table}[h] % [t] puts at top of page
\begin{center}
\caption[Performance of Fermilab Accelerator Complex]
{Performance of Fermilab Accelerator Complex, where C-W=Cockwroft-Walton, L=Linac, B=Booster, Debuncher and Recycler, M=Main Injector, T=Tevatron, A=Accelerator, E=Energy. \label{table:FermilabAcceleratorComplex}}
\begin{tabular}{|c|c|c|c|c|c|c|c|}\hline \hline
A. & H & $H^{-}$ & C-W & L & B & M &
T\\
\hline
E &0.04 eV&25 keV&750 keV&400 MeV&8 GeV& 150 GeV&0.98 TeV\\
\hline
$\beta$ & $9.1\cdot10^{-8}$ & 0.01& 0.04 & 0.71 & 0.99 & $1$ &$1$\\
\hline
$\gamma$ & 1 & 1 & 1 & 1.43 & 9.53 & 161 & 1067\\
\hline\hline
\end{tabular}
\end{center}
\end{table}
\end{center}

\subsection{Collisions and Luminosity}

\subsubsection{Store}

\ \\When 36 new bunches of protons and 36 new bunches of antiprotons enter the Tevatron, it is said that a new store starts. A typical bunch length is 0.43 meters. Both beams have an average energy per accelerated particle of $980\ {\rm GeV}$. A proton bunch contains typically $3.30\cdot 10^{11}$ protons. An antiproton bunch contains typically $3.60\cdot 10^{10}$ antiprotons. Since antiprotons are antimatter, they annihilate with regular matter. This is why antiprotons are accumulated about one order of magnitude less than protons. As the two beams collide head on at a rate of 2.5 million times per second, $p\bar{p}$ scatterings occur at a certain rate per unit of area, which is described by the instantaneous luminosity.

\subsubsection{Luminosity}

\ \\The Tevatron's instantaneous luminosity is given by:

\begin{equation}\label{LuminosityFormula}
\Large{{\cal L}=\frac{fE}{\epsilon_{n}}\cdot\frac{N_{b}N_{p}N_{\bar{p}}}{\beta^{*}}}.
\end{equation}

\ \\The first fraction represents quantities that cannot be easily changed after the experiment is started, such as f, the beam revolution frequency at the Tevatron, which is set by the radius and the speed of light c; E, the beam energy set by the physics goals of the experiment; $\epsilon_{n}$, the beam emittance at injection set by getting the beam into the Tevatron. The second fraction presents quantities that can be changed easily during the period of taking data, such as $N_{b}$, the number of proton or antiproton bunches found at one time in the Tevatron; $\beta^{*}$, the strength of the final focus; $N_{p}$ ($N_{\bar{p}}$), the number of protons (antiprotons) per bunch. 

\ \\However, the instantaneous luminosity delivered by the Tevatron is not calculated by experiments using this formula. It is rather measured by a special detector apparatus described in section~\ref{Section:CherenkovLuminosityCounter}.

\ \\As the store's duration increases, instantaneous luminosity decreases exponentially, in the first few hours due to the intra-beam scattering and later due to antiproton depletion. Instantaneous luminosity is expected to reach 50\% in 7 hours and to reach 1/$e$ in 12 hours. Typically after 24 hours a store is ended as the proton and antiproton bunches are evacuated from the Tevatron. Subsequently new bunches are inserted in the Tevatron and a new store starts. 

\ \\Table~\ref{table:TevatronRunIvsRunII} compares various parameters of Run I and Run II of Tevatron, especially in terms of luminosity~\cite{TevatronAcceleratorRunII}.

\begin{figure}[h]
\begin{center}
\includegraphics[angle=0,width=0.75\textwidth,clip=]{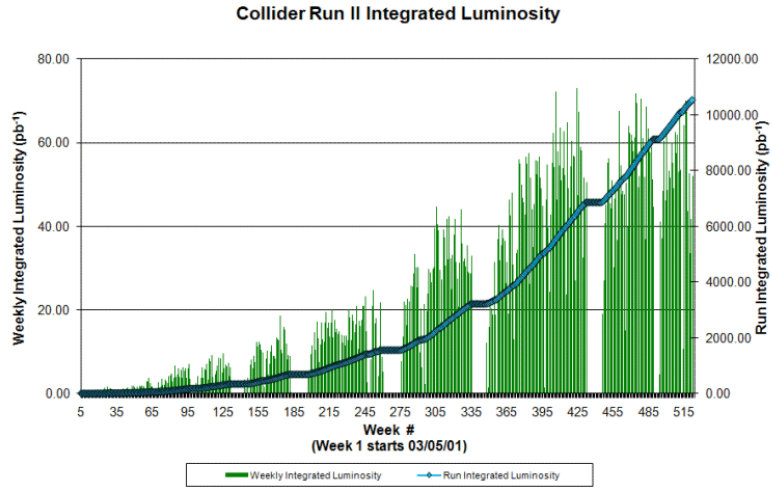}
\caption[Integrated luminosity delivered by the Tevatron during Run II]
    {Tevatron's delivered weekly (left scale) and integrated luminosity (right scale) from Jan 2001 to Apr 2011. The analysis presented in this thesis uses an integrated luminosity of 5.7 $\invfb$. Credit image to the CDF collaboration.\label{figure:TevatronLuminosityDelivered}}
\end{center}
\end{figure}

\begin{center}
\begin{table}[h] % [t] puts at top of page
\begin{center}
\caption[Accelerator parameter comparison between Run I and Run II at the Tevatron]
{Comparison between Run I and Run II at the Tevatron for various accelerator parameters, especially luminosity. \label{table:TevatronRunIvsRunII}}
\begin{tabular}{|l|c|c|}
\hline \hline
Parameter & Run Ib & Run II\\
\hline
Number of bunches ($N_b$) & 6 & 36\\
Number of protons bunch ($N_p$) & $2.3 \cdot 10^{11}$ & $2.7 \cdot 10^{11}$\\
Number of antiprotons per bunch ($N_{\pbar}$) & $5.5 \cdot 10^{10}$ & $3.0 \cdot 10^{10}$\\
Maximum number of antiprotons in a store & $3.3 \cdot 10^{11}$ & $1.1 \cdot 10^{12}$\\
$\beta^*$ [cm] & 35 & 35 \\
Bunch length [m] & 0.6  & 0.37 \\
Time interval between consecutive bunch crossings [ns] & 3500  & 396 \\
Number of $p\pbar$ collisions per bunch crossing & 2.5 & 2.3\\
Energy per $p$~ or $\pbar$ [GeV] & 900  & 980 \\ 
\hline
Integrated Luminosity Delivered by the Tevatron [$\invpb$]& 112 & 8000\\
Peak Instantaneous Luminosity at the Tevatron [$\text{cm}^{\text{-2}} \text{s}^{\text{-1}}$]& $2.0 \cdot 10^{31}$ & $3.6 \cdot 10^{32}$\\
\hline\hline
\end{tabular}
\end{center}
\end{table}
\end{center}

\ \\ A typical integrated luminosity per week is $\int {\cal L}dt=8\,\invpb$. Figure~\ref{figure:TevatronLuminosityDelivered} shows the weekly and integrated luminosity delivered by the Tevatron since the start of Run II. The analysis presented in this thesis uses an integrated luminosity of 5.7$\invfb$. A dataset of about 8.0 $\invfb$ is currently under preparation to be shown at the Summer conferences of 2011. 

\subsubsection{Quench}

\ \\Stores may end prematurely when the beam is lost in a process called quenching. Quenches may happen when a beam hits a superconducting magnet. The magnet is locally not superconducting any more and releases energy by the Joule effect~\cite{JouleEffect}. Soon the whole magnet warms up and is no longer superconducting. Physicists then need to wait for the whole magnet to be cooled down to liquid helium temperature before inserting a new store in the Tevatron.

\subsubsection{Number of Collisions}

\ \\Accelerator based particle physics examines final-state particles created by the initial beam particle collisions. The most interesting processes often have very rare signatures. Counting experiments count the number of events where the particular signature appears; any excess above the estimated background is considered as signal. Per unit time, a signature occurs in a number of events proportional to the physical probability of occurrence (cross section $\sigma$) and to the beam collision conditions in the accelerator complex (instantaneous luminosity ${\cal L}$). However, not all events are reconstructed and identified by the particle physics detector. The experimental efficiency ($\epsilon$) measures the percentage of events that are seen correctly by the detector. Therefore, the observed number of events per unit of time is given by

\begin{equation}\label{NumberEventsInstantaneous}
\frac{dN_{obs}}{dt}=\epsilon \cdot \sigma \cdot {\cal L}. 
\end{equation}

\ \\Integrating the instantaneous luminosity in time provides the integrated luminosity, which gives the total number of observed events:

\begin{equation}\label{NumberEventsInstantaneous}
N_{obs}=\epsilon \cdot \sigma \cdot \int\,{\cal L}dt.
\end{equation}

\ \\Since the physical cross section of many processes increases with the centre-of-mass energy ($\sqrt{s}$), particle physicists try to build accelerators with larger and larger $\sqrt{s}$. Furthermore, particle physicists try to build better detectors with reconstruction efficiencies for various particles very close to one.

\clearpage
\newpage

\begin{figure}[h]
\begin{center}
\includegraphics[angle=0,width=0.7\textwidth,clip=]{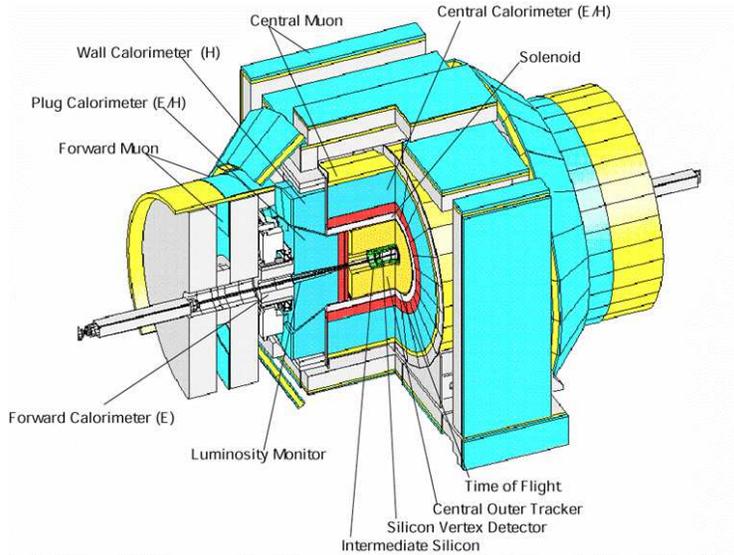}
\caption[A schematic view of the CDF detector]
    {A schematic view of the detector CDF II, with a quadrant taken out to reveal the interior as well. Various subdetector systems are mentioned in the diagram. Proton beams enter from the west side and antiproton beams from the east side. Credit image to the CDF collaboration. \label{figure:CDFIIDetector}}
\end{center}
\end{figure}

\begin{figure}[h]
\begin{center}
\includegraphics[angle=0,width=0.7\textwidth,clip=]{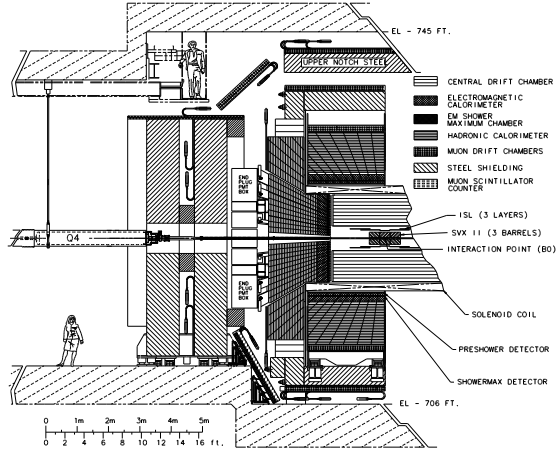}
\caption[An on-scale plane view of the CDF detector]
    {An on-scale plane view of the CDF detector, that also shows the elevation. The west side (from where the proton beams enter CDF) is on the right of the picture. Credit image to the CDF collaboration. \label{figure:CDFIIDetectorElevation}}
\end{center}
\end{figure}

\section{The Collider Detector at Fermilab}

\ \\The Collider Detector at Fermilab II (CDF II or simply CDF) is a general purpose subatomic particle detector that surrounds one of the two $p\pbar$ collision points of the Tevatron accelerator~\footnote{The other general purpose detector at the Tevatron is DZero. There is a friendly competition between CDF and DZero, with each detector collaboration thriving to achieve a physics result first, but with other collaboration needed to confirm the first result before it is fully accepted by the particle physics community. Moreover, only together CDF and DZero have a chance to exclude or discover the Standard Model Higgs boson. Therefore, it is customary that, once or twice a year, CDF and DZero combine their Higgs boson search in what is called a ``Tevatron combination''.} and is described in detail in the Technical Design Report of CDF II Detector~\cite{CDFIIDetector1} and in the CDF paper~\cite{CDFIIDetector2}. During Tevatron Run I there was an older version of CDF, CDF I~\cite{CDFDetector}. The detector underwent a major upgrade in preparation for Run II of Tevatron, thus creating CDF II. 

\ \\The CDF detector has a cylindrical symmetry (i.e. both azimuthal and forward-backward) around the proton-antiproton beam pipe. CDF is formed of three main subdetector systems. The first and innermost one is the tracking system. It is made of a set of silicon strips and an open-cell drift chamber inside a solenoid-produced magnetic field. This system detects momenta of electrically charged particles and displacements of secondary particle vertices with respect to primary vertices\footnote{The primary vertex is the position of the primary hard-scattering interaction.}. Next comes the calorimeter system, which measures energies for both electrically charged and electrically neutral particles as they produce a shower of secondary particles through interaction with dense matter\footnote{The energies of all the secondary particles from the shower are measured as they are absorbed by the calorimeter system. They are all added to obtain the energy of the initial particle.}. Finally, outside the calorimeter system is located the muon subdetector, which measures the momenta of muon candidates\footnote{Muons are minimum ionizing particles and leave only very faint energy deposits in the calorimeter system. This is why we need a dedicated muon detector system on the outer part of the CDF.}. 

\ \\Besides the three main subdetector systems, CDF has a subdetector system called Time of Flight (TOF) that measures the mass of low-momentum charged particles, as well as an instantaneous luminosity counter called the Cherenkov Luminosity Counter (CLC) that uses Cherenkov light emission to measure the number of $p\pbar$ collisions per second inside CDF.

\ \\The CDF detector is built around the ``beam pipe'' , which is a vacuum pipe with a diameter of 2.2 cm through which the proton and antiproton beams circle around the Tevatron. The pipe is made of beryllium because the material provides the lowest particle interaction cross section possible\footnote{The atomic number Z of beryllium is 4. Typically the cross section of interaction of a subatomic particle with a material increases with Z.} while also possessing good mechanical properties. Proton beams circulate clockwise as seen from the top and enter CDF from the west side, while antiproton beams circulate counterclockwise and enter CDF from the east side. 

\ \\This section will cover in detail the different subdetectors of CDF II briefly described above. In figure~\ref{figure:CDFIIDetector} we can see a diagram of CDF II with one quadrant taken out so that we can clearly see the interior. The various subdetector systems are mentioned in the diagram. Figure~\ref{figure:CDFIIDetectorElevation} shows an elevation view of the CDF detector. 

%to update figure%

\subsection{The CDF Coordinate System}

\ \\The coordinate system used by CDF reflects the cylindrical symmetry of the detector. However, depending on the situation, a cylindrical or adapted spherical coordinate system is used.

\ \\The z axis lies along the beamline, with the positive $+z$ side in the direction the protons travel (from west to east). A longitudinal plane is parallel to the z axis, while a transverse plane is perpendicular to the z axis. Although the Cartesian coordinates $x$ and $y$ are not typically used, for the sake of completeness we should note that the $+x$ direction is horizontal towards north and the $+y$ direction is vertical and upward. 

\ \\The radial coordinate $r$ is the radial distance from the beamline and is expressed by the formula:

\begin{equation}\label{r}
r=\sqrt{x^2+y^2}. 
\end{equation}

\ \\The azimuthal angle is noted $\phi$, represents the angle made around the beamline and is expressed by the formula:

\begin{equation}\label{Phi}
\tan \phi = \frac{y}{x}.
\end{equation}

\ \\ The polar angle, denoted $\theta$, represents the angle made with respect to the beamline and is expressed by the formula:

\begin{equation}\label{Theta}
\tan \theta = \frac{r}{z}.
\end{equation} 

\ \\However, since $\theta$ is not a Lorentz-invariant quantity\footnote{Protons and antiprotons are extended objects travelling along the $z$ axis with an energy of 980 GeV. However, not all elementary constituents of protons (generically called partons, such as valance quarks, sea quarks and gluons) have the same momentum along the z axis and the number of particles per unit of $\theta$ angle ($\frac{dN}{d\theta}$)is not Lorentz invariant.}, a derived quantity called rapidity ($y$) is Lorentz invariant and is defined by the formula

\begin{equation}\label{Rapidity}
y \equiv \frac{1}{2} \ln \frac{E+p_zc}{E-p_zc}\rm{,} 
\end{equation}

\ \\ where E is the energy and $p_z$ is the longitudinal momentum of a particle described. 

\ \\Moreover, since in collider physics the particles involved have total energies much larger than their rest energies, we can use the approximation that $E \approx pc$. As such, the rapidity is approximated by a new quantity called pseudorapidity ($\eta$), which is not Lorentz invariant, but is approximately Lorentz invariant, and is defined by

\begin{equation}\label{PseudoRapidity}
\eta \equiv - \ln \tan \frac{\theta}{2}. 
\end{equation}

\ \\The advantage of the $\eta$ quantity is that it does not depend on the mass of the particle and therefore is a pure geometrical quantity and can be used to describe the trajectory of all elementary particles. The cylindrical coordinate system ($r$, $\phi$, $z$) is used to describe the geometry of the detector, while the adapted spherical coordinates ($\eta$, $\phi$) are used to describe the direction of a particle inside the detector. 

\ \\A quantity $R$ is used to measure the distance between two directions of particles inside the detector (in the $\eta$-$\phi$ plane):

\begin{equation}\label{R}
R \equiv \sqrt{(\Delta \eta)^2+(\Delta \phi)^2}. 
\end{equation}

\ \\For a particle with energy $E$ and momentum $p$, we define transverse energy as $E_T=E\sin\theta$ and transverse momentum as $p_T=p\sin\theta$.

\begin{figure}[h]
\begin{center}
\includegraphics[angle=0,width=0.7\textwidth,clip=]{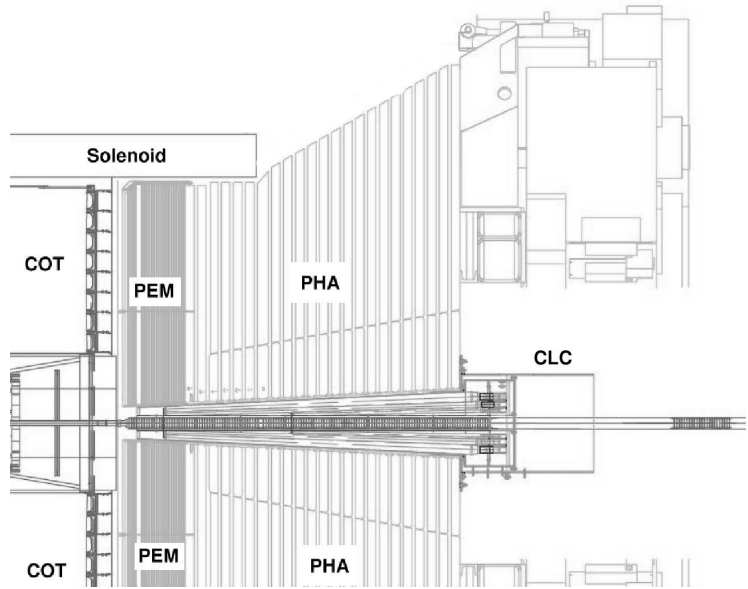}
\caption[Location of the Cherenkov Luminosity Counter at CDF]
    {Location of the Cherenkov Luminosity Counter in CDF. Credit image to the CDF collaboration. \label{figure:CLC}}
\end{center}
\end{figure}

\subsection{The Cherenkov Luminosity Counter}\label{Section:CherenkovLuminosityCounter}

\ \\Although Tevatron accelerator personnel measure the instantaneous luminosity that the Tevatron is delivering, only a fraction of it, albeit one close to unity, is recorded by particle detectors. This is why each detector has its own subdetector that measures instantaneous luminosity as well.

\ \\The general formula of instantaneous luminosity is:

\begin{equation}\label{InstantaneousLuminosityGeneral}
N = L \cdot \sigma \cdot \epsilon, 
\end{equation}

\ \\where $N$ is the number of events (the number of hard (inelastic) scattering interactions) recorded per unit time, $L$ the instantaneous luminosity of the beam, $\sigma$ the cross section of the given event and $\epsilon$ is the efficiency of observing that event in the detector after all selection requirements (the fraction of signal events that pass the selection requirements). In this particular case, $N$ represents the average number of inelastic $p\pbar$ scattering interactions per bunch crossing ($\mu$) times the number of bunch crossings per second ($f_{bc}$), or the rate of bunch crossings):

\begin{equation}\label{NumberInteractions}
N = \mu \cdot f_{bc}. 
\end{equation}

\ \\By plugging equation~\ref{NumberInteractions} into equation~\ref{InstantaneousLuminosityGeneral} we obtain the following equation:

\begin{equation}\label{LuminosityRelation}
\mu \cdot f_{bc} = L \cdot \sigma \cdot \epsilon . 
\end{equation}

\ \\and we can find an expression for the instantaneous luminosity $L$:

\begin{equation}\label{InstantaneousLuminosity}
L = \frac{f_{bc}}{ \sigma}\cdot \frac{\mu}{\epsilon}, 
\end{equation}

\ \\where $f_{bc}$ is the number of bunch crossings per second at the Tevatron Run II ($\frac{1}{396 \rm{ns}}$), and $\sigma$ is the cross section of $p\pbar$ inelastic scattering, which has been determined experimentally~\cite{InelasticCrossSection}. 

\ \\Only $\mu$ and $\epsilon$ are not known. For this purpose, CDF uses long conical gaseous Cherenkov counters to measure the average inelastic interactions per bunch crossing ($\mu$) and the efficiency to detect them ($\epsilon$). The name of this detector subsystem is Cherenkov Luminosity Counter (CLC) \cite{CherenkovLuminosityCounter1} \cite{CherenkovLuminosityCounter2}. It has been designed especially for CDF II in order to cope with increased Tevatron instantaneous luminosity (on the order of $2\cdot 10^{32}\,\rm{cm}^{-2}\rm{s}^{-1}$) and decreased time interval between consecutive bunch crossings (396 ns). 

\ \\This subdetector system is made of two modules located in the forward regions of CDF (both on west and east sides), immediately after the beam pipe, in a 3\degree gap between the beam pipe and the rest of the detector systems, corresponding to a $\eta$ region of $3.7<|\eta|<4.7$, as we can see in Figure~\ref{figure:CLC}. 

\ \\Each CLC module is formed of 3 concentric layers of 18 Cherenkov counters each. Each Cherenkov counter is long and conical, stretching from the interaction region towards one of the forward regions. The Cherenkov cones in the two outer layers (one interior layer) have a length of about 180 (160) cm. Being conical, their diameter increases from 2 cm in the interaction region to 6 cm in the forward region. At the latter end there is a conical mirror that collects light into a 2.5 cm diameter photomultiplier tube (PMT). The light is produced by particles created in $p\pbar$ collisions that travel at large $\eta$ pseudorapidities, thus travelling only inside one cone and emitting Cherenkov light as they travel through the gaseous environment in the cone. Each PMT has a 1 mm thick concave-convex quartz window to collect the ultraviolet light of the Cherenkov spectra with a gain of $2\cdot 10^6$. The gaseous environment is isobutane at atmospheric pressure (1 atm), with the possibility to increase the pressure up to 2 atm if there is a need to increase the yield of Cherenkov light. In this environment, particles emit Cherenkov light with an angle of 3.1\degree if they have a momentum larger than $9.3\mevc$ (electrons) and $2.6\gevc$ (pions). 

\ \\The conical geometry and orientation were chosen so that particles produced in $p\pbar$ collisions close to the centre of the CDF detector can travel a large distance inside of the CLC, producing an important light yield (several hundred photo-electrons), while the particles produced by beam halo or secondary particles travel at smaller $|\eta|$, travel a smaller distance in the CLC and have a smaller light yield. 

\ \\By requiring a certain minimum light yield threshold in each channel, the background from beam halo and secondary particles is rejected and only the signal of particles produced in $p\pbar$ interactions is measured. Also, each module has an excellent time resolution of less than 100 ps. This allows to ask for coincidence hits in the two modules in the forward and rear regions of CDF with respect to the $z$ axis, which improves the signal and background separation. 

\ \\Finally the CLC subdetector measures continuously the parameter $\mu$, while studies on the CLC have measured the efficiency $\epsilon$. Using formula~\ref{InstantaneousLuminosity} CDF measures the instantaneous luminosity it records. The uncertainty on the instantaneous luminosity measurement is 5.9\%~\cite{InstantaneousLuminosityUncertainty} and is quoted as a systematic uncertainty on any measurement at CDF, including the Higgs boson search described in this thesis.

\begin{figure}[h]
\begin{center}
\includegraphics[angle=0,width=0.7\textwidth,clip=]{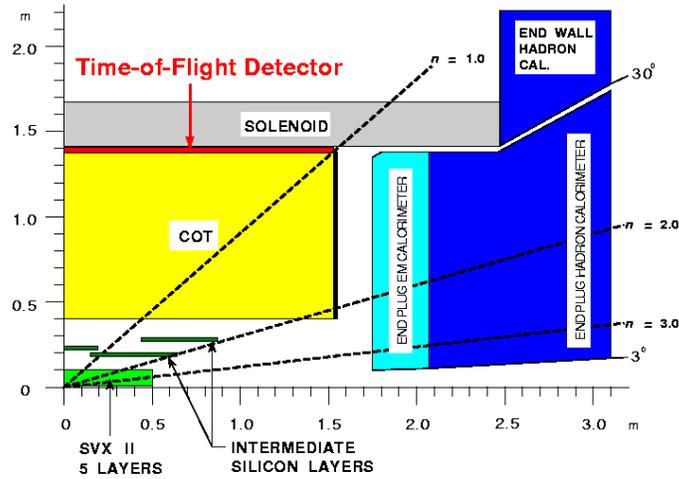}
\caption[Schematic view of the tracking systems at CDF]{Schematic view of the tracking systems at CDF. Credit image to the CDF collaboration. \label{figure:TrackingSystems}}
\end{center}
\end{figure}

\subsection{The Tracking Systems}

\ \\The CDF detector has two tracking systems that provide very precise measurements of charged particle momenta. Both follow the cylindrical geometry of CDF and are embedded in a 1.4 T solenoidal magnetic field. Inside the magnetic field, charged particles follow helical trajectories, whose parameters\footnote{The tracking helical parameters are usually expressed by five non independent variables: $\phi$, $\eta$ (or $\cot \theta$), curvature, $z$ and $d$ (the impact parameter, or the distance of minimum approach between a track and the beam axis).} are measured by a silicon strip vertex detector and a cylindrical open-cell drift chamber.

\ \\The closest to the beam pipe is the silicon detector, which is made up of three subdetectors: Layer 00 (L00), the Silicon Vertex Detector (SVX-II) and the Intermediate Silicon Layers (ISL). It allows precise measurements of the $z$ coordinate for the primary interaction vertex, impact parameters and $\phi$ for tracks, as well as identification of secondary vertices in jets originating from $b$-hadrons.  Next comes the drift chamber, which is called the Central Outer Tracker (COT). A schematic view of the tracking systems in CDF is presented in Figure~\ref{figure:TrackingSystems}.
 
\subsubsection{The Solenoid}

\ \\The superconducting solenoid is 5 m long, is made of Nb-Ti stabilized with Al, is operated at a current of about 5650 A, and generates a uniform 1.4 T magnetic field parallel to the $z$ axis and pointing in the direction of the proton beam. The Lorentz force will bend the trajectories of charged particles. Measuring their curvatures by the tracking systems allows the precise measurements of the momenta of charged particles. 

\subsubsection{The Silicon Vertex Detector}

\ \\The physics principle of a silicon detector is a reversed-biased semiconductor p-n junction. When a charged particle passes through, it deposits energy in the detector material through ionization. This creates electron-hole pairs. Electrons drift towards the anode and holes drift toward the cathode. The energy deposited by the incoming particle is estimated by the amount of charge deposited in the detector, which is to first order proportional to the path length traversed in the material by the charged particle. 

\ \\CDF uses a strip geometry for its silicon microdetectors. This allows to reconstruct the position of the particle as it travels through the detectors. The distance between two silicon microstrips is about 60 $\mu$m. A single incoming charged particle will typically deposit energy in more than one microstrip, forming a charge deposition called a  ``cluster''. CDF employs two types of silicon microstrip detectors: single and double-sided. The single microstrip detectors have only the p side of the junction segmented into strips that are parallel to the $z$ axis, thus allowing $r$-$\phi$ position measurements. The double sided microstrip detectors have in addition the n side segmented in strips at an (stereo) angle with respect to the p sides (that are parallel to the $z$ axis), thus allowing the measurement of the $z$ position as well.

\begin{figure}[h]
\begin{center}
\includegraphics[angle=0,width=0.7\textwidth,clip=]{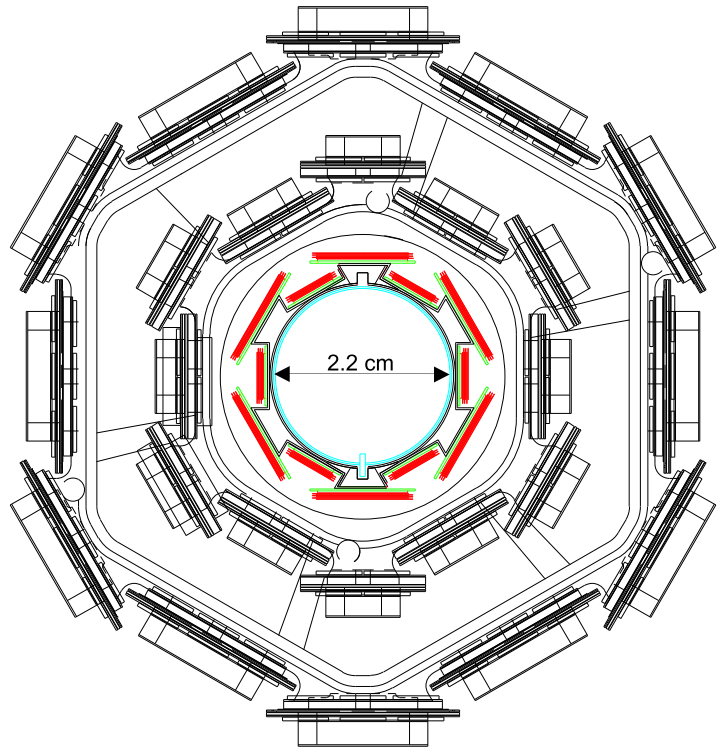}
\caption[Detailed diagram of the Layer 00 silicon detector]{Detailed diagram of the Layer 00 silicon detector (red) mounted directly on the beam pipe at radii between 1.35 cm and 1.62 cm. In the figure we can also see the two innermost layers of SVX-II silicon detector. Credit image to the CDF collaboration. \label{figure:L00}}
\end{center}
\end{figure}

\paragraph{Layer 00}

\ \\\ \\The first layer of the silicon detector, Layer 00 (L00)~\cite{L00}, is mounted directly on the beam pipe between 1.35 cm and 1.62 cm of radius. It has two overlapping hexagonal structures that can be seen in red in Figure~\ref{figure:L00}. These detectors have to be very resistant to radiation from the beam pipe and have only single-sided microstrips. Although they provide only $r$-$\phi$ measurements, they improve the spatial resolution up to 15 $\mu$m per hit, whereas the resolution is 20 $\mu$m per hit without them.

\begin{figure}[h]
\begin{center}
\includegraphics[angle=0,width=0.7\textwidth,clip=]{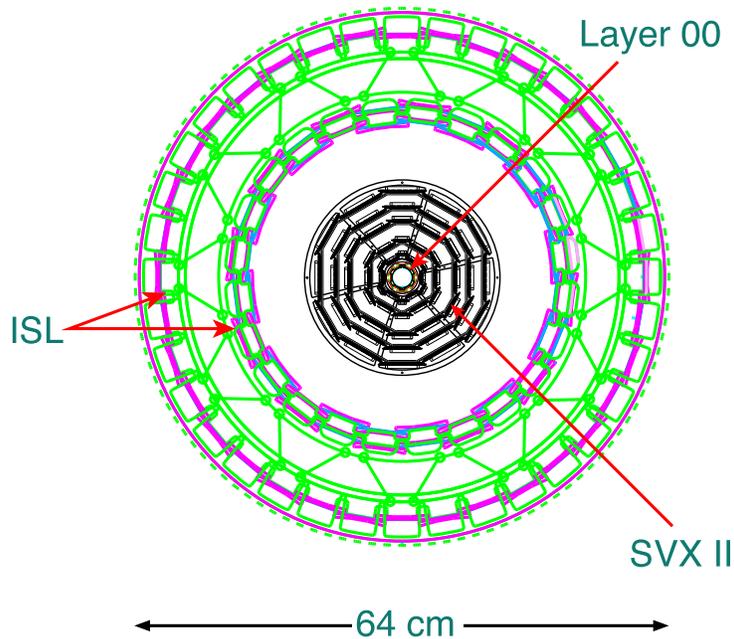}
\caption[Transverse schematic view of the silicon detector systems]{Transversed schematic view of the silicon detector systems: Layer 00 (red), SVX-II (black) and Intermediate Silicon Layers (green). Credit image to the CDF collaboration. \label{figure:SiliconSystems}}
\end{center}
\end{figure}

\paragraph{SVX-II}

\ \\\ \\Next comes the second silicon vertex detector, the SVX-II~\cite{SVX-II}. It extends from a radius of 2.5 cm to a radius of 11 cm and it has a $z$ coverage of $|z|<45$ cm. SVX-II is formed of five concentric layers of double-sided silicon microstrip detectors, which allow for a 3D position measurement with a spacial resolution of about 20 $\mu$m. The first two layers can be seen in black in Figure~\ref{figure:L00} and all the five layers can be seen in black in figure~\ref{figure:SiliconSystems}.

\paragraph{Intermediate Silicon Layers}

\ \\\ \\Further out in radius is the last silicon detector, the Intermediate Silicon Layers (ISL)~\cite{ISL}. ISL is made of three double-sided detector layers. Going from small radius outwards, there is a layer with $1<|\eta|<2$ at $r=20$ cm, then a layer with $|\eta|<1$ at $r=22$ cm, followed finally by a layer with $1<|\eta|<2$ at $r=28$ cm. The three layers can be seen in green in Figure~\ref{figure:SiliconSystems}. 

\ \\Tracks in the central region are reconstructed using both silicon and COT information. However, only silicon information is used to reconstruct tracks in the forward region, up to $1<|\eta|<2.8$.

\begin{figure}[h]
\begin{center}
\includegraphics[angle=0,width=0.3\textwidth,clip=]{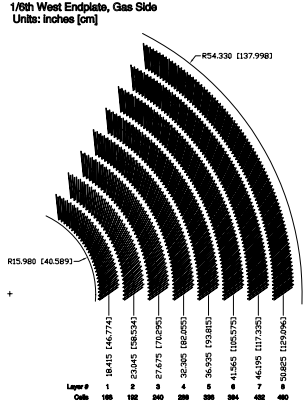}
\caption[Schematic view of sense wire planes in the eight superlayers of COT]{Schematic view of sense wire planes in the eight superlayers of COT. Credit image to the CDF collaboration. \label{figure:COT}}
\end{center}
\end{figure}

\subsubsection{The Central Outer Tracker}

\ \\Besides the silicon detector, the tracking system contains a cylindrical open-cell drift chamber called the Central Outer Tracker (COT)~\cite{COT}. The COT has a cylindrical geometry, extends from a radius of 43.4 cm up to one of 132.2 and has a length of 3.1 m. The COT is made up of eight subdetectors called ``superlayers'', as can be seen in Figure~\ref{figure:COT}. Four superlayers have their sense wires parallel to the $z$ axis (axial superlayers) and therefore can measure trajectories in the $r$-$\phi$ plane. The other four superlayers have their sense wires at a 2\degree angle with respect to the $z$ axis (stereo superlayers), which allows for measurements along the $z$ axis. 

\ \\The 30,240 sense wires in the COT were divided approximately equally between the axial and stereo superlayers. Particles originating from the primary interaction vertex that have $|\eta|<1$ pass through all the eight COT superlayers, but those with $1<|\eta|<1.3$ pass only through four superlayers. The COT is very precise for measurements in the $r$-$\phi$ plane (transverse momentum, $\pt$), but less so for measurements along the $z$ axis (longitudinal momentum, $p_z$). 

\ \\The gas filling the drift cells that constitute the COT is formed by an Argon-Ethane gas mixture and Isopropyl alcohol in the proportions 49.5:49.5:1. The motivation for this choice is that of having a constant drift velocity along the cell width. This produces a drift velocity of 100 $\mu$m/ns and a maximum drift time of 177 ns, which allows for all the electric charges produced by ionization in the drift chamber by the incoming particles to drift away before the next bunch crossing takes places (every 396 ns). The electric field increases exponentially with decreasing distance to the sense wire. This produces an avalanche of electrons very close to the sense wire, which amplifies naturally the measured current and allows for a better measurement. 

\ \\Next the electric currents read by the sense wires are processed by an ASDQ chip~\cite{ASDQ}. The ASDQ amplifies the signal, analyses its shape and height and allows for the measurement of the deposited charge, which is used to measure the ionization along the trail of the charged particle ($dE/dx$), a quantity that is very helpful in discriminating between different types of particles. Next the pulses are digitized by Time to Digital (TDC) boards in the CDF collision hall. 

\ \\Furthermore, pattern recognition algorithms (tracking algorithms) reconstruct helical trajectories of particles in the COT. The position resolution for a track is about 140 $\mu$m and the track $\pt$ resolution measured using cosmic rays varies with the track $\pt$:

\begin{equation}\label{COTResolution}
\frac{\sigma_{\pt}}{(\pt)^2}=0.17\%\frac{1}{\gevc} . 
\end{equation}

\begin{figure}[h]
\begin{center}
\includegraphics[angle=0,width=0.5\textwidth,clip=]{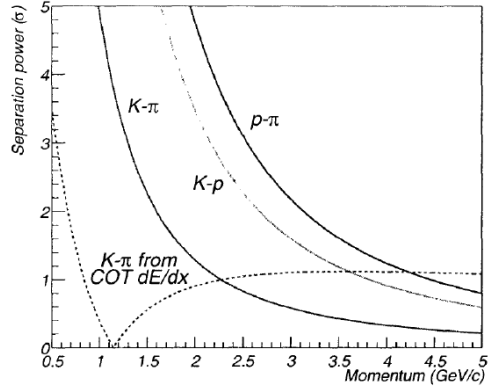}
\caption[Particle identification performance for the Time of Flight subdetector]
    {Time of Flight subdetector performance in distinguishing pions, kaons and protons, superimposed to the COT performance for the same task~\cite{TimeOfFlightImage}. Credit image to the CDF collaboration. \label{figure:TOFPerformance}}
\end{center}
\end{figure}

\subsection{Time of Flight}

\ \\Another improvement of CDF during the upgrade to Tevatron Run II was the addition of a new subdetector aimed at improving particle identification at low momentum. A Time Of Flight (TOF) system~\cite{TimeOfFlight} was built outside the solenoid magnet of the tracking system at a radius of  approximately 1.5 m. The TOF measures the time interval particles travel from the primary vertex of $p\pbar$ collisions to the TOF system. This allows the separation of low momentum protons, kaons and pions. 

\ \\The TOF system consists of 216 bars of scintillating material approximately 3 m in length and with a cross section of $4\times 4 \rm{cm}^2$. These bars are arranged in a cylindrical geometry around the tracking system. The TOF coverage in $\eta$ is $|\eta| < 1$. The scintillating material was chosen to be Zircon 408, because it has a short rise time and a long attenuation length (308 cm). 

\ \\The particle separation principle is the following. All particles are produced at the same time from the same primary interaction. As they pass through the scintillating material, they deposit small flashes of visible light that are detected by photomultiplier tubes (PMT) attached to both ends of each scintillating bar. Next a preamplifier circuit mounted directly on the PMT processes the light signal. Then the readout electronics digitizes both the amplitude and time for the light signal. The time interval is digitized by a Time to Digital Converter (TDC) only when the signal reaches a fixed discrimination threshold. Moreover, the TDC corrects for the effect that a larger amplitude signal reaches the threshold first (time walk effect). If a particle crosses the scintillating bar just in front of a PMT, the time resolution is about 110 ps. If the particle crosses the scintillating bar farther away from a PMT, the resolution worsens~\cite{TimeOfFlightCalibration}.

\ \\At its best performance (110 ps resolution) the TOF can distinguish charged pions and kaons with momenta $p < 1.6\gevc$ with a precision of two standard deviations. This information is complementary to the particle separation due to the $dE/dx$ effect measured by the COT. Figure~\ref{figure:TOFPerformance} shows the effect of TOF and COT superimposed in particle separation.

\begin{figure}[h]
\begin{center}
\includegraphics[angle=0,width=0.7\textwidth,clip=]{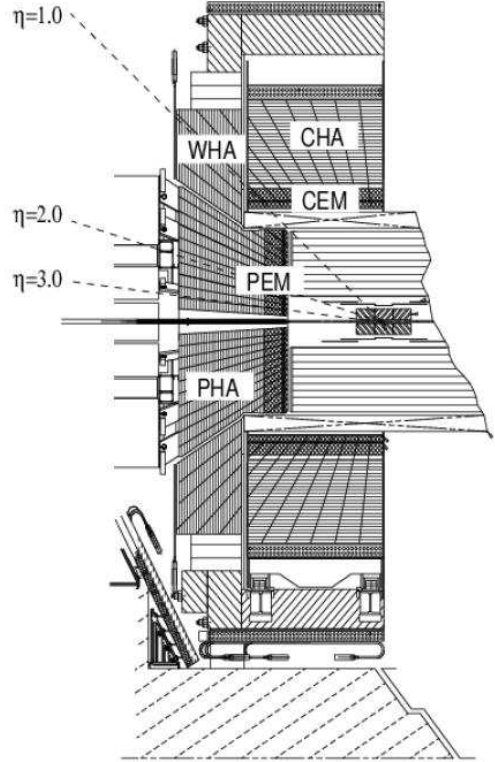}
\caption[Schematic view of the calorimeter detectors in one of the CDF barrels]
{Schematic view of the calorimeter detectors in one of the CDF barrels (PEM, CEM, CHA, WHA, PHA). Credit image to the CDF collaboration. \label{figure:CalorimeterSystem}}
\end{center}
\end{figure}

\subsection{The Calorimeters}

\ \\Besides the trajectory and momentum of a given particle, one needs also to determine its energy in order to reconstruct an event fully. This role is played by the calorimeter systems.

\ \\The calorimeter detectors in CDF II are ``sampling'' calorimeters. They have a sandwich structure, where dense absorbing material alternates with scintillating material. As particles pass through the absorbing material, they develop showers of secondary particles. These secondary particles emit light as they pass through the scintillating material. The light is detected and measured by photomultiplier tubes (PMT). By adding up the energies measured by all the PMT in the calorimeter systems, the energy of the initial particle is measured. 

\ \\Each calorimeter is made of two parts. The innermost part is an electromagnetic calorimeter (EMCAL), which measures the energies of electromagnetic objects (electrons, positrons and photons). EMCAL has a large number of radiation lengths $X_0$ and a small number of interaction lengths. The outermost part is the hadronic calorimeter (HADCAL), which measures the energy of hadrons, such as pions, mesons, $b$ hadrons. HADCAL has a large number of interaction lengths. Electromagnetic particles start to shower immediately as they enter the calorimeter and most of their shower is contained in the EMCAL. On the other hand, hadronic objects deposit very little energy in the EMCAL and most of the energy in the HADCAL. This structure for the calorimeters allows to distinguish between the electromagnetic objects and hadrons. 

\ \\Each calorimeter has a projective geometry, meaning that it is divided in $\eta$ and $\phi$ towers that point towards the interaction region. That way a particle entering a calorimeter tower tends to spend most of its trajectory in that same tower. The calorimeter systems have a $2\pi$ coverage in $\phi$ and $|\eta|<3.6$ in pseudorapidity. 

\ \\The calorimeter system is made up of three parts spanning different geometric regions: the central calorimeter in the barrel region (both electromagnetic, CEM, and hadronic, CHA), the plug calorimeter in the forward region (both electromagnetic, PEM, and hadronic, PHA) and the end wall calorimeter in between the central and plug calorimeters (only hadronic, WHA). 

\ \\In Figure~\ref{figure:CalorimeterSystem} we can see a schematic view of the calorimeter detectors in one of the CDF barrels.  

\subsubsection{Central Electromagnetic Calorimeter (CEM)}

\begin{figure}[h]
\begin{center}
\includegraphics[angle=0,width=0.5\textwidth,clip=]{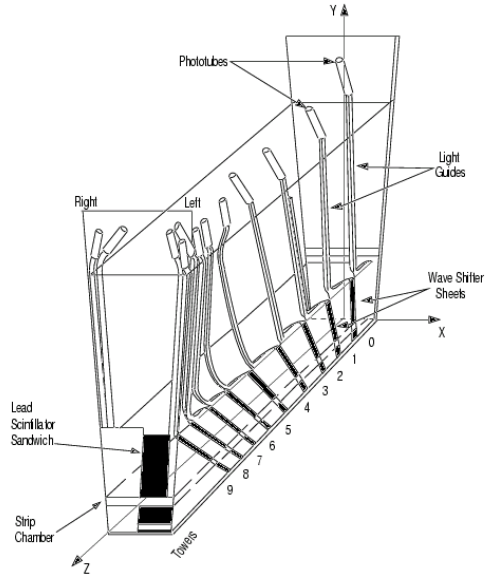}
\caption[Schematic view of a single wedge from the CEM calorimeter]
{Schematic view of a single wedge from the CEM calorimeter. Credit image to the CDF collaboration. \label{figure:CEMWedge}}
\end{center}
\end{figure}

The CEM~\cite{CEMCES} covers the region $|\eta| < 1.1$. The CEM has 24 towers in $\phi$ and 10 towers in $\eta$. It is a sampling calorimeter with lead as the absorbing material alternated with scintillating material. It has 18 radiation lengths ($X_0$). The energy resolution of the CEM is

\begin{equation}
\frac{\sigma_E}{E} = \frac{13.5\%}{\sqrt{E\,(\gev)}} \oplus 2\%\ \rm{,}
\end{equation}

\ \\where the notation $\oplus$ represents the sum in quadrature of the constant and stochastic terms.

\ \\In Figure~\ref{figure:CEMWedge} we can see a schematic view of a single wedge from the CEM calorimeter. 

\subsubsection{Central Electromagnetic Shower Maximum Detector (CES)}

\ \\The goal of CES~\cite{CEMCES} is to improve the position measurement of electromagnetic showers in the CEM. To achieve this goal, CES is a proportional chamber with wire and strip readout located inside the CEM at the position where on average the showers created by electrons, positrons and photons have a maximum number of particles (at a depth of 6 $X_0$) in the CEM. This is also called a shower maximum and it is reflected in the name of the detector. For an electromagnetic object with an energy of 50 GeV, the position resolution achieved by the CES is 0.2 cm. 

\subsubsection{Plug Electromagnetic Calorimeter (PEM)}

\ \\The PEM covers the region $1.1 < |\eta| < 3.6$ and PEM is formed by 24 (48) towers in $\phi$ for the inner (outer) groups and 12 towers in $\eta$ for all groups. As for the CEM, it is also a lead-scintillator sampling calorimeter. Its thickness is slightly larger (21 $X_0$). The energy resolution of the PEM is

\begin{equation}
\frac{\sigma_E}{E} = \frac{14.4\%}{\sqrt{E\,(\gev)}} \oplus 0.7\%\ \rm{.}
\end{equation}

\subsubsection{Plug Electromagnetic Shower Maximum Detector (PES)}

\ \\Similar to CEM, PES~\cite{PES} measures precisely the position of the maximum of electromagnetic showers in the PEM. Also as for the CEM, the PES is located at a depth of 6 $X_0$ radiation lengths inside the PEM. The PES is made of two layers of 5 mm wide scintillator strips and each layer is at a 45\degree angle relative to the other. 

\ \\The properties of the electromagnetic calorimeters at CDF are summarized in Table~\ref{table:CalorimeterSystems}.

\begin{center}
\begin{table}[h] % [t] puts at top of page
\begin{center}
\caption[Summary of electromagnetic calorimeter subsystems at CDF]
{Summary of various properties of the electromagnetic calorimeter subsystems at CDF. \label{table:CalorimeterSystems}}
\begin{tabular}{|l|c|c|c|l|}\hline \hline
System & $\eta$ coverage & Energy resolution (\%) & Thickness & Absorber \\ 
\hline
CEM & $0.0 < |\eta| < 1.1$ & 13.5/$\sqrt{E_T} \oplus 2$   & 18 $X_0$ & 3.2 mm lead \\
PEM & $1.1 < |\eta| < 3.6$ & 14.4/$\sqrt{E_T} \oplus 0.7$ & 21 $X_0$ & 4.5 mm lead \\
\hline 
\hline\hline
\end{tabular}
\end{center}
\end{table}
\end{center}

\subsubsection{Central Hadronic Calorimeter (CHA)}

\ \\The CHA~\cite{CHAWHA} covers the region $|\eta| < 0.9$ and the CHA uses iron as a showering material that alternates with scintillator material. The CHA is segmented in 24 towers in $\phi$ and 8 towers in $\eta$. The CHA is located radially just outward of CEM. Each CHA tower is made of 32 layers, with a total of 4.7 interaction lengths ($\lambda_I$). The energy resolution of CHA is

\begin{equation}
\frac{\sigma_E}{E} = \frac{50\%}{\sqrt{E\,(\gev)}} \oplus 3\%\ \rm{.}
\end{equation}

\begin{figure}[h]
\begin{center}
\includegraphics[angle=0,width=0.7\textwidth,clip=]{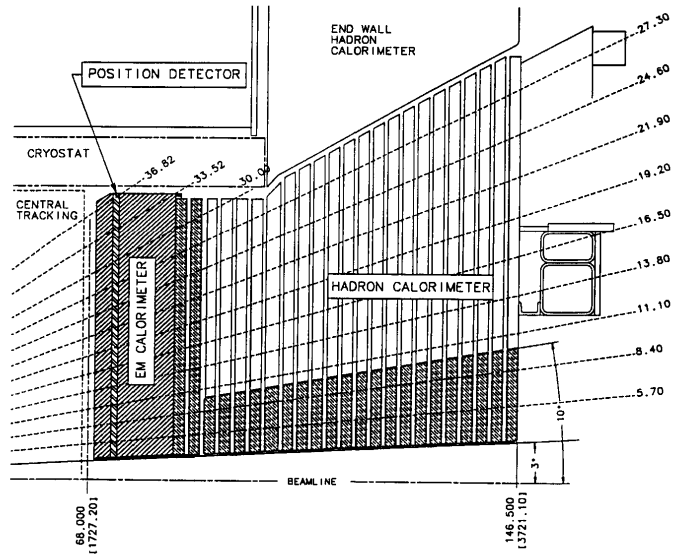}
\caption[Schematic view of one of the two forward PEM and PHA calorimeters]
{Schematic view of one of the two forward (plug) calorimeters (PEM and PHA). Credit image to the CDF collaboration. \label{figure:ForwardCalorimeter}}
\end{center}
\end{figure}

\subsubsection{Plug Hadronic Calorimeter (PHA)}

\ \\The PHA~\cite{PHA} covers the region $1.3 < |\eta| < 3.6$. It is constituted from 23 layers of sandwiched iron (as absorbing material) and scintillating material. The energy resolution for the PHA is:

\begin{equation}
\frac{\sigma_E}{E} = \frac{80\%}{\sqrt{E\,(\gev)}} \oplus 5\%\ \rm{.}
\end{equation}

\ \\In Figure~\ref{figure:ForwardCalorimeter} we can see a schematic view of one of the two forward (plug) calorimeters (PEM and PHA).   

\subsubsection{Wall Hadronic Calorimeter (WHA)}

\ \\The WHA~\cite{CHAWHA} covers the region $0.7 < |\eta| < 1.3$. It extends the $\eta$ coverage between the central and plug hadronic calorimeters. It is also an iron-scintillator sampling calorimeter. There are 15 5.0 cm thick iron layers alternating with 1.0 cm thick scintillator. The energy resolution of WHA is

\begin{equation}
\frac{\sigma_E}{E} = \frac{75\%}{\sqrt{E\,(\gev)}} \oplus 4\%\ \rm{.}
\end{equation}  

\begin{figure}[h]
\begin{center}
\includegraphics[angle=0,width=0.8\textwidth,clip=]{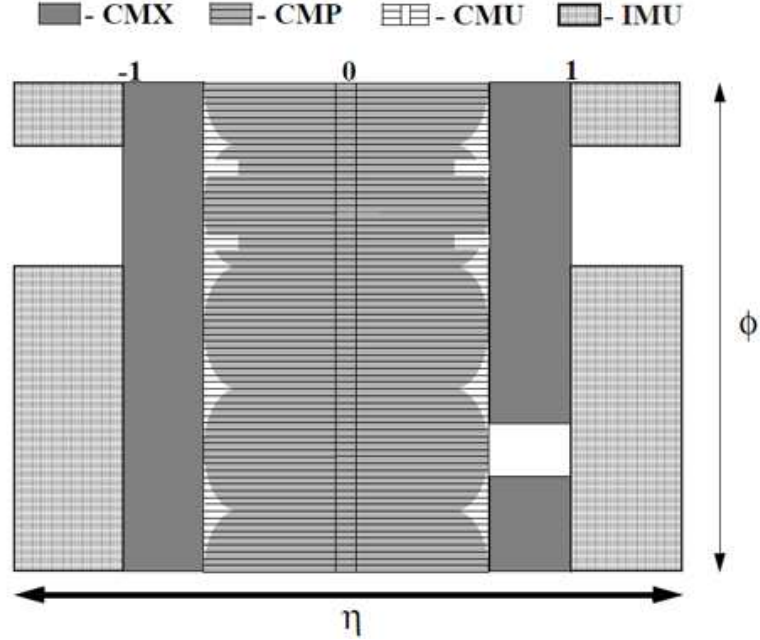}
\caption[Diagram in the $\eta$-$\phi$ plane of the muon systems at CDF]
{Diagram in the $\eta$-$\phi$ plane of the muon systems at CDF: CMU, CMP, CMX and BMU muon detectors. The BMU detector is referred in this diagram as IMU. Credit image to the CDF collaboration. \label{figure:MuonCoverage}}
\end{center}
\end{figure}

\subsection{The Muon Chambers}

\ \\Although nearly all particles are absorbed by the calorimeter system, muons pass through the calorimeters as minimum ionizing particles and then exit the calorimeter system\footnote{Neutrinos are the only known particles that leave the detector completely undetected.}. The outermost subdetector of CDF is a muon system. It is made out of single wire drift chambers and scintillator counters for fast timing, located radially just outside the calorimeter system.

\ \\There are various muon subsystems as follows: the Central Muon Detector (CMU), the Central Muon uPgrade Detector (CMP), the Central Scintillator uPgrade (CSP), the Central Muon eXtension Detector (CMX), the Central Scintillator eXtension (CSX), the Toroid Scintillator Upgrade (TSU), the Barrel Muon Upgrade (BMU) and the Barrel Scintillator Upgrade (BSU). 

\ \\The CMU, CMP and CSP systems cover an $|\eta|$ range of $|\eta|<0.6$, the CMX and CSX systems cover an $|\eta|$ range of $0.6<|\eta|<1.0$ and the TSU, BMU and BSU subsystems cover an $|\eta|$ range of $1.0<|\eta|<2.0$. The muon subsystems can be seen in Figure~\ref{figure:MuonCoverage}.

\ \\The innermost muon system is CMU~\cite{MuonSystemsRunI}. It was built for CDF I and is located just outside the CHA calorimeter, at a radius of 350 cm and arranged in 12.6\degree wedges in $\phi$. Each wedge has three layers (stacks) and each stack has four rectangular drift chambers. Each drift chamber is filled with the same gaseous composition as the COT (argon-ethane-alcohol 49.5:49.5:1) and has a 50 $\mu$m sense wire in the middle of the cell, parallel to the $z$ axis. 

\ \\ The CMU is followed by another muon system, the CMP~\cite{MuonSystemsRunII}. The CMP has rectangular geometry and is formed of four layers of drift chambers of the same constitution as above. In preparation for Tevatron Run II, a 60 cm thick layer of steel was added. The $\pt$ threshold for the CMU (CMP) is 1.4 (2.2) $\gevc$. 

\ \\On the outer side of CMP there is the CSP~\cite{MuonSystemsRunII}, formed of a single rectangular layer of scintillating material with a wave guide to transport the light to a PMT. The CSP fast response is used in triggering. 

\ \\The CMX muon system is located at each edge between the CDF barrel and forward regions. It has a conical geometry with drift chambers similar to the CMP. Also, it has a scintillating system called the CSX, similar to the CSP. The CMX system covers 360\degree with 15 wedges in $\phi$. Each wedge is formed of eight layers of drift chambers in the radial direction. 

\ \\Various properties of the muon subsystems are summarized in Table~\ref{table:MuonSubsystems}.

\begin{center}
\begin{table}[h] % [t] puts at top of page
\begin{center}
\caption[Summary of various properties of the muon subsystems at CDF]
{Summary of various properties of the muon subsystems at CDF. \label{table:MuonSubsystems}}
\begin{tabular}{|l|c|c|c|}\hline \hline
 & CMU & CMP/CSP & CMX/CSX\\ 
\hline
$\eta$ coverage & 0-0.6 & 0-0.6   & 0.6-1.0\\
Minimum $\pt$ [$\gevc$] & 1.4 & 2.2 & 1.4\\
\hline \hline
Drift Tubes \\
\hline 
Thickness [cm] & 2.68 & 2.5 & 2.5 \\
Width [cm] & 6.35 & 15 & 15 \\
Length [cm] & 226 & 640 & 180 \\
Maximum drift time [$\mu$s] & 0.8 & 1.4 & 1.4 \\
\hline \hline
Scintillators \\
\hline
Thickness [cm] & N/A & 2.5 & 1.5 \\
Width [cm] & N/A & 30 & 30-40 \\
Length [cm] & N/A & 320 & 180 \\
\hline\hline
\end{tabular}
\end{center}
\end{table}
\end{center}

\ \\Based on the timing information provided by the individual drift chambers, short ``tracks'' of ionization in the drift chambers (called ``stubs'') are reconstructed in the muon chambers. At CDF a muon candidate is formed when a track reconstructed by the COT is matched to a stub in the muon system. In the reconstruction process, a $\chi^2$ in the $\phi$ coordinate ($\chi_\phi^2$) is computed for the match between a COT track and a muon stub. To ensure good quality of muon candidates, an upper limit cut is set on $\chi_\phi^2$. 

\begin{figure}[h]
\begin{center}
\includegraphics[angle=0,width=0.8\textwidth,clip=]{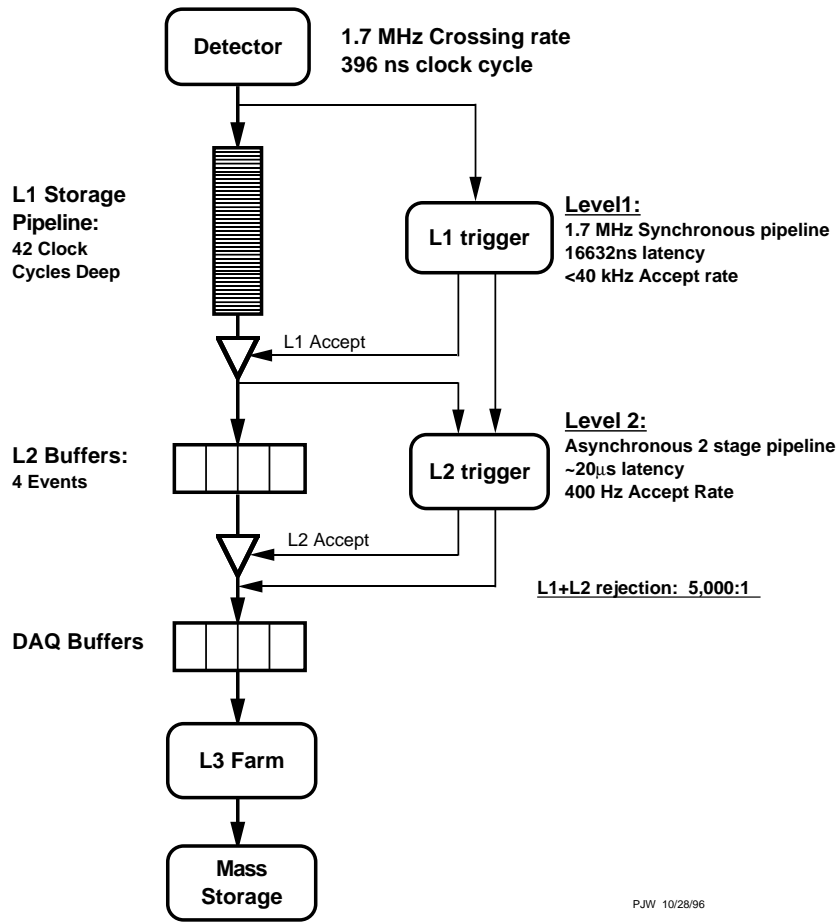}
\caption[Diagram of the CDF trigger system]
{Diagram of the CDF trigger system (Level 1, Level 2, Level 3, Mass storage). Credit image to the CDF collaboration. \label{figure:CDFDataFlow}}
\end{center}
\end{figure}

\subsection{The Trigger System and Data Acquisition}

\ \\The trigger system is indispensable at each collider physics experiment because the collision rate is much larger than the maximum rate of data events stored for analysis. At the Tevatron the theoretical bunch crossing (collision) rate is 2.5 MHz for 36 bunches of protons and 36 bunches of antiprotons. In practice, there are 1.7 MHz $p\pbar$ collisions per second. Even so, this rate is a lot larger than the 50 Hz rate to save data events\footnote{In CDF, an event is defined as all the recorded particle activity in the detector during a bunch crossing. On average, more than one $p\pbar$ hard scattering happens during an event, with multiple primary vertices present in the event (pile-up).} to magnetic tape\footnote{One might be surprised that in the era of CDs, DVDs and USB storage keys, collision data is saved on old fashioned magnetic tapes. The motivation comes from the fact that the latter perserve the data for more years and with lower costs than the former.}. Also, on average a data event needs about 250 kB of data storage. Even if the rate to copy these to magnetic tape would be sufficient, there would not be enough storage space to save all these events. Nor would it be easy to analyze all these events afterwards. 

\ \\There are three essential conditions that the CDF trigger system has to meet. First, the trigger has to be quick enough to make a decision about an event (whether to save it or reject it) before the next event comes in (in other words, there should be zero dead time). Second, the Tevatron collider imposes that a new event comes in every 396 ns. Third, the magnetic tape can only save about 50 events per second. 

\ \\In conclusion, the goal of the CDF trigger system is to analyze, every second, 1.7 million events and decide also every second which 50 of these events will be saved to tape.

\ \\Figure~\ref{figure:CDFDataFlow} represents a diagram of the three trigger levels in the CDF trigger system. 

\subsubsection{CDF Trigger Levels}

\ \\There are three trigger levels at CDF, and each needs a certain amount of time to reach a decision whether to reject an event or send it to the next trigger level. 

\ \\The first trigger level (L1) uses hardware-based custom electronics to try to reconstruct physics objects using only a subset of CDF information. L1 makes a decision by object count and energy values. The second trigger level (L2) also uses custom designed hardware to reconstruct better physics objects. This information is passed to programmable processors to make decisions. The third trigger level (L3) uses all CDF information and a PC farm of about 500 CPUs to do a full event reconstruction and decide if the event is kept. 

\subsubsection{Trigger Paths}

\ \\A trigger path represents a sequence of requirements that an event has to pass at L1, L2 and then at L3. About 100 trigger paths are implemented by the CDF II trigger system. An event will be saved to tape if it passes the requirements of at least one trigger path. 

\subsubsection{L1 Trigger Level}

\ \\L1 reduces the event rate from 1.7 MHz to about 40 KHz, thus discarding about 98\% of events. In order to have enough time to analyze each event, L1 uses a pipeline and 42 buffers so that L1 has 2 $\mu$s to analyze each event. 

\ \\The input for L1 comes from the tracking, calorimeter and muon systems. The decision to send an event to L2 is taken using the number and energy values of electron, muon and jet candidates, as well as the value of missing transverse energy~\footnote{As neutrinos escape the detector without being detected, the energy they carry appears as missing energy in the event. Since the partons in protons or antiprotons have a well measured energy in the transverse plane but a distribution of energy along the $z$ axis, we can only say that the vector sum of energies should be zero in the transverse plane, but we are not able to say the same for the $z$ axis. This is why we refer only to missing transverse energy. Also note that this happens at all hadron colliders, LHC included, and that at lepton colliders, such as the PEP-II collider (for the BABAR experiment) and the KEKB collider (for the Belle experiment), the energy is well measured in all directions and they can refer to missing energy, not only missing transverse energy.}, or the kinematic properties of a pair of tracks. 

\ \\A subsystem called eXtremly Fast Tracker (XFT)~\cite{XFT} reconstructs at L1 high-$\pt$ tracks ($\pt > 1.5\,\gevc$) using COT information. XFT uses a digitized readout of ionization (hits) in the COT. The hits from COT's superlayers are combined into segments. Pattern recognition algorithms group segments into tracks that cross the entire COT. The tracks reconstructed by the XFT are combined with information about energy deposits in calorimeter (muon) systems to produce L1 electron (muon) candidates. 

\ \\Energies of jets, electron and photon candidates, as well as missing transverse energy and sums of jet energies in an event are approximated using clusters of energy in the calorimeter systems. 

\ \\The only muon system used at L1 is CMU. 

\begin{figure}[h]
\begin{center}
\includegraphics[angle=0,width=0.4\textwidth,clip=]{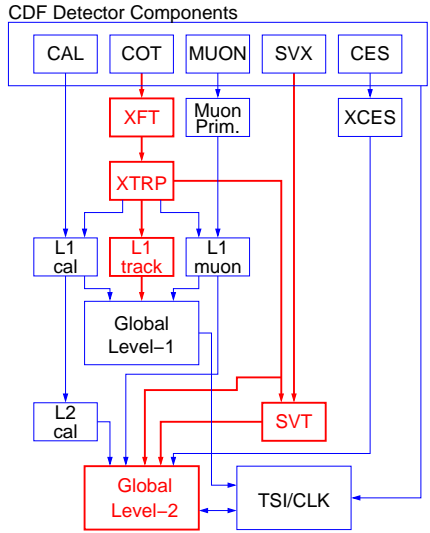}
\caption[Diagram of the first and second trigger levels at CDF]
{Diagram of the first (L1) and second (L2) trigger levels at CDF. Credit image to the CDF collaboration. \label{figure:CDFTriggerL1AndL2}}
\end{center}
\end{figure}

\subsubsection{L2 Trigger Level}

\ \\Events accepted by L1 are sent to L2, where the event rate is reduced from 40 KHz to about 500, thus discarding about 99\% of events passed to L2. L2 uses 4 buffers in order to have enough time to analyze each event (about 20 $\mu$s). 

\ \\ At L2 a better event reconstruction is performed. The tracking is improved by taking into account the silicon tracking information as well. Also, better track reconstruction and calorimeter clustering~\footnote{Jets are reconstructed at L2 with the help of L2 cluster finder (L2CAL), which starts from a seed of 3 GeV calorimeter tower and adds adjacent towers with energies larger than 1 GeV.} for jet finding algorithms are used. The reconstruction of electron and photon candidates is improved by taking into account as well the information from the central calorimeter shower maximum subdetector (CES)~\footnote{This way the resolution for electron and photon showers is better than the cluster location. This information is combined with the tracking information to reconstruct better electron candidates.}. Also, secondary vertices~\footnote{Jets originating from a $b$ quark contain a secondary vertex displaced by about 3 mm from the primary $p\pbar$ interaction vertex due to the fact that $b$ quarks live longer than other quarks before decaying. The decay products appear emerging from the same vertex. This is called secondary vertex.} are reconstructed at L2 inside certain jets using the Silicon Vertex Tracker (SVT) trigger subsystem~\cite{SVT}~\footnote{The information from SVXII subdetector is combined with the L1 XFT reconstructed track to reconstruct both a more precise track and reconstruct a secondary vertex inside a jet that originates from a $b$ quark.}. Also, all muon systems are used to combine hits in the muon chambers with the L1 XFT reconstructed track to produce L2 muon candidates.

\ \\Figure~\ref{figure:CDFTriggerL1AndL2} represents a block diagram of the L1 and L2 trigger levels at CDF.  

\subsubsection{L3 Trigger Level}

\ \\If an event is accepted at L2, it is sent to L3, where a full detector readout is done and a full reconstruction is done using the computer farm. If an event passes L3 requirements then it is sent to the Consumer Server/Logger (CSL) that is the final component of the CDF data acquisition. 

\subsubsection{Consumer Server/Logger}

\begin{figure}[h]
\begin{center}
\includegraphics[angle=0,width=1.0\textwidth,clip=]{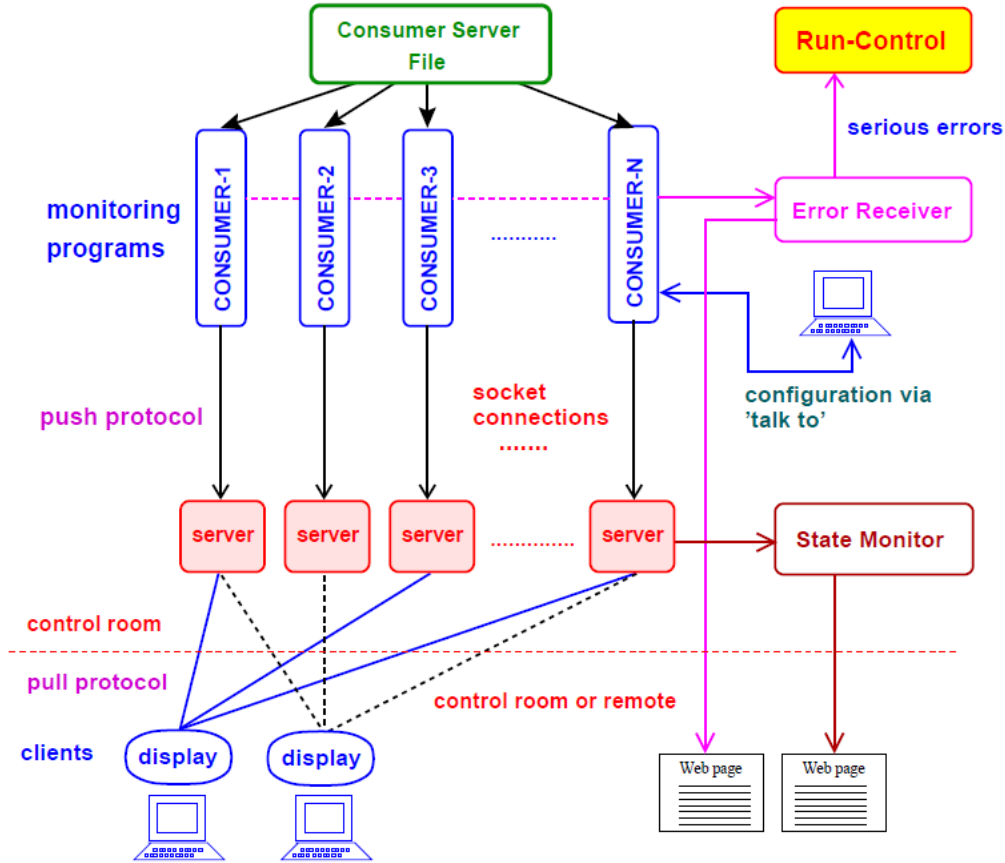}
\caption[Design diagram of the Consumer Server/Logger]
{Design diagram of the Consumer Server/Logger. Credit image to the CDF collaboration. \label{figure:CDFTriggerAndDAQDiagram}}
\end{center}
\end{figure}

The CSL categorizes events by trigger path, writes to disk those that pass at least one of the trigger paths and sends a fraction of these events to online processors for online monitoring of data quality.  Figure~\ref{figure:CDFTriggerAndDAQDiagram} represents a diagram of Consumer server/Logger. 

\ \\As a graduate student on CDF, I did three one-week online data quality monitoring shifts~\footnote{At CDF these shifts are called ``Consumer Operator shifts``.} and one three-month online data acquisition and detector control shift~\footnote{At CDF these shifts are called ``ACE shifts``.}. 

\section{Summary}

\ \\In this chapter we have presented the experimental infrastructure used for the $WH$ associated production direct search presented in this thesis. We started by introducing the US national particle physics laboratory, Fermilab. We then presented in detail the Fermilab accelerator complex that accelerates and collides protons and antiprotons at a centre-of-mass energy of $\sqrt{s}=1.96\ \tev$ with the help of a chain of particle accelerators formed by the proton source, the Cockwroft-Walton, Linac, Booster, Debuncher, Recycler, Main Injector and the Tevatron accelerator. We continued with the detailed description of the Collider Detector at Fermilab, the apparatus that records the elementary particles produced in the proton-antiprotons collisions delivered by the Tevatron accelerator. We first introduced the CDF coordinate system and the Cherenkov Luminosity Counter. We then described the several subdetectors structured as layers of CDF which measure several properties of the elementary particles. The first layer is formed by the tracking system, which measures precisely the momenta of elementary particles. The second layer is the calorimeter system, which measures precisely the energy of elementary particles. The third layer is dedicated to the muons. From a total of 2.5 million collisions per second, only about 100 collisions per second are chosen to be stored by the trigger system, which is deployed by way of three levels.

\ \\In the next chapter we will present how the several detector systems are used in order to reconstruct the elementary particles used in our $WH$ analysis.

\clearpage{\pagestyle{empty}\cleardoublepage}

\chapter{Object Identification\label{chapter:Object}}

\ \\The energy deposits in the CDF subdetectors are digitized and transformed into electronic signals that are then reconstructed to high-level objects such as primary vertices, tracks, calorimeter clusters and muon stubs. Physics objects such as electrons, muons, jet candidates and missing transverse energy are reconstructed by applying cuts, or selection criteria, on high level objects. In this analysis we present a $WH\to l\nu b\bbar$ search. The final state contains therefore a charged lepton (an electron or a muon~\footnote{We do not consider the tau lepton final state directly. However, tau leptons that decay leptonically to electrons or muons plus neutrinos contribute to our signal, background and data selection.}), missing transverse energy due to the undetected neutrino, and two jets originating from bottom ($b$) quarks.

\ \\A primary vertex represents the reconstructed position of the primary $p\pbar$ collision that produced primary particles. In the case of our signal, these are the $W$ boson and the Higgs boson. 

\ \\A track represents the reconstructed trajectory of a charged particle in the tracking systems from the particle's electromagnetic interactions in these detector systems.

\ \\A calorimeter cluster is a collection of adjacent calorimeter towers where energy is deposited due to an incoming particle, either electrically charged or electrically neutral.

\ \\A muon stub is a collection of energy deposits in adjacent muon chambers in the muon systems. 

\ \\This chapter presents the reconstruction techniques for the basic physics objects used in this analysis, such as tracks, primary vertices and calorimeter clusters. Next comes the reconstruction of high-level physics objects such as electron, muon and jet candidates, as well as missing transverse energy, which are used by the event selection. Finally, two algorithms used to identify jets originating from $b$ quarks are described. 

\section{Track Identification}

\ \\Tracks are reconstructed at CDF using information from the tracking systems. They are used in primary vertex reconstruction, charged lepton identification and identification of jets originating from $b$ quarks. A track represents a reconstructed three dimensional helical trajectory of a charged particle passing through the solenoidal magnetic field in the tracking systems and is described by the following parameters.

\ \\The half-curvature of the trajectory ($C$) is defined as 
\begin{equation}\label{TrackCurvature}
C=\frac{1}{2Q\rho}\,\rm{,}
\end{equation}

\ \\where $Q$ is the electric charge of the track, and $\rho$ is the radius of the circle formed by the projection of the helical trajectory on the transverse plane. $C$ carries the same sign as the electrical charge of the particle and is inversely proportional to the transverse momentum ($\pt$) of the track.

\ \\$\cot \theta$ is the cotangent of the polar angle of the trajectory at the closest approach to the primary interaction vertex.

\ \\The impact parameter ($d_0$) is the minimal distance in the transverse plane between the trajectory and primary interaction vertex. It is given by the expression

\begin{equation}\label{ImpactParameter}
d_0=Q\cdot(\sqrt{x_0^2+y_0^2}-\rho)\,\rm{,}
\end{equation}

\ \\where $x_0$ and $y_0$ are the coordinates in the transverse plane of the centre of the circle of the helix.

\ \\$\phi_0$ is the azimuthal angle of the trajectory at the closest approach to the primary interaction vertex.

\ \\$z_0$ is the $z$ coordinate position of the trajectory at the closest approach to the primary interaction vertex.

\subsection{Tracking Algorithms}

\ \\There are three tracking algorithms that we use in this analysis: COT stand-alone tracking, Outside-In (OI) tracking and Inside-Out (IO) tracking.

\subsubsection{COT stand-alone tracking}

\ \\The COT stand-alone tracking algorithm uses only COT information (with no information from the silicon detectors). Electromagnetic interactions inside the COT cells or silicon detectors are called hits. First, hits in each superlayer are fit together to reconstruct short tracks. Then, short tracks from all the superlayers are fit together to form a COT stand-alone track. The details of this latter fit are the following. Since axial and stereo superlayers alternate, first a fit is performed where only the axial superlayers are considered in the order from the most outer one to the most inner one. Then, stereo layers are added and a new fit is performed. The final COT stand-alone track needs to have hits in at least 2 axial and 2 stereo superlayers. This tracking algorithm is used in the central region of the detector, corresponding to $|\eta|<1.1$.

\subsubsection{Outside-In tracking}

\ \\The Outside-In (OI) tracking algorithm starts with a COT stand-alone track (called a seed track) and adds high-resolution hits from silicon detector information. First the axial silicon hits are added to the COT stand-alone track. Then, for each silicon wafer, every hit on a stereo silicon strip is added to a different copy of the current track. After the last silicon wafer has been processed, there are a multitude of track candidates. The Outside-In track is chosen to be that with the largest number of silicon hits and with the lowest $\chi^2$ over the number of degrees of freedom. This tracking algorithm is used in the central region of the detector ($|\eta|<1.1$). 

\subsubsection{Inside-Out tracking}

\ \\The Inside-Out (IO) tracking algorithm is needed for tracking in the forward regions of CDF. The algorithm starts with hits in at least three layers of the silicon detector. Then hits in the COT that have not already been used by the COT stand-alone and OI algorithms are added to produce the final IO track.

\section{Primary Vertex Identification}

\ \\CDF uses two main algorithms to reconstruct primary interaction vertices (PV). The locus of all PVs represents the beamline, or the luminous region of the detector.

\subsection{Primary Vertex Reconstruction Algorithms}

\ \\The ZVertexFinder algorithm~\cite{ZVertexFinder} takes as an input a set of tracks passing minimum quality requirements based on the number of silicon and COT hits. The algorithm computes an error weighted average ($z_{0}$) of $z$ coordinates of these tracks, which is given  by

\begin{equation}\label{ImpactParameter}
z_{0}=\frac{\sum_{i}\left(z_{i}^{0}/\delta_{i}^{2}\right)}
{\sum_{i}\left(1/\delta_{i}^{2}\right)} \,\rm{.}
\end{equation}

\ \\The algorithm outputs a collection of PVs characterized by their own quality, track multiplicity, $z$ position, $z$ position error and $\pt$. However, PVs output by the ZVertexFinder algorithm present no $x$ and $y$ position information. Each reconstructed PV corresponds either to a hard scattering or to an underlying event of a hard scattering. It may also happen that a physical PV gets reconstructed into two PVs due to tracking resolution. The PV transverse momentum $\left(p_{T, PV}\right)$ is defined as the sum of the transverse momenta of its tracks $\left(\sum_{tracks} p_{T}\right)$ and conveys the information of how energetic a PV is. Typical $WH$ PV candidates have
$\left(\sum_{tracks} p_{T}\right)$ on the order of 100 GeV. The PV quality conveys the information of how well the PVs are reconstructed. PV quality is based on the track multiplicity, as shown in Table~\ref{table:PrimaryVertex}.

\begin{center}
\begin{table}[h] % [t] puts at top of page
\begin{center}
\caption[Primary Vertex Quality Criteria]
{Primary Vertex Quality Criteria. \label{table:PrimaryVertex}}
\begin{tabular}{|c|c|}
\hline \hline Criterion & Quality Value\\
\hline Number Si -tracks$\ge$3 & 1\\
\hline Number Si -tracks$\ge$6 & 3\\
\hline Number COT-tracks$\ge$1 & 4\\
\hline Number COT-tracks$\ge$2 & 12\\
\hline Number COT-tracks$\ge$4 & 28\\
\hline Number COT-tracks$\ge$6 & 60\\
\hline\hline
\end{tabular}
\end{center}
\end{table}
\end{center}

\ \\The PV with the best chances to be the PV of the interaction triggered on is considered the event PV. The CDF collaboration used to use a run-averaged beamline position as an event PV. CDF developed in 2003 an algorithm called the
PrimeVertexFinder~\cite{PrimeVtx1}~\cite{PrimeVtx2} that reconstructs a 3D event PV on an event-by-event basis. This algorithm allowed CDF to improve the efficiency of identifying jets as originating from a bottom quark (b-tagging) for shorter secondary vertex displacements and to reduce the systematic uncertainties due to the run-dependent beam position variation. PrimeVertexFinder takes as an input a set
of good quality tracks in good agreement ($\chi^{2}<10$) with a  seed vertex (usually the beamline position or one of the PVs output by ZVertexFinder). These tracks are reconstructed to a new 3D PV and checked if they are still in good agreement with the new PV. Tracks with $\chi^{2}>10$ are rejected. The remaining tracks are reconstructed to a new 3D PV. The procedure is iterated until all remaining tracks have a $\chi^{2}<10$ with respect to the latest PV. The last 3D PV becomes the event PV. 3D position information is crucial for b-tagging techniques that use information about the bottom quark lifetime.

\ \\A PV position is represented by $\left(x_{PV},y_{PV},z_{PV}\right)$.  A typical longitudinal width is $\sigma_{z}=29$ cm. A typical transverse width is circular, smaller at the centre of the detector, $\sigma_{\perp, z=0\ \rm{cm}}=30\ \mu \rm{m}$ and larger at the extremities, $\sigma_{\perp, z=40\ \rm{cm}} \simeq 50\ \mu \rm{m}$. Typical $x_{PV}$ and $y_{PV}$ are very small, on the order of tens of microns. Event PV reconstruction is trusted only in the luminous region ($|z_{PV}|\leq 60$ cm). Events with the event PV outside the luminous region are rejected (luminous cut).

\subsection{Primary Vertex Definition Studies}

\ \\In my Master of Science thesis~\cite{BuzatuMScThesis} I compared two possible definitions for the primary interaction vertex for high-$p_T$ events with the charged lepton+missing transverse energy+jets signature (top quark pair production) at CDF. The analysis presented in this PhD thesis also uses a signature of a charge lepton + missing transverse energy + jets (the $WH$ search). The only difference is that there are two jets in our case and there are four jets in that for $t\tbar$. The two possible primary vertex definitions were:

\begin{itemize}
\item The primary vertex with the closest $z$ coordinate to the $z_0$ coordinate of the charged lepton (electron or muon) reconstructed track.
\item The primary vertex with the largest transverse momentum of all the primary vertices with the $z$ coordinate within 5 cm of the $z_0$ coordinate of the charged lepton (electron or muon) reconstructed track.
\end{itemize}

\ \\The study concluded that both definitions are equally efficient and therefore confirmed that the definition used by CDF was the best possible one. 

\subsection{Primary Vertex Reconstruction at L1 Trigger Level Study \label{TriggerPrimaryVertexStudy}}

\ \\During my first year of PhD studies I performed a study within the Higgs Trigger Task Force (HTTF)~\cite{HTTF} at CDF. The goal of HTTF was to design new triggers to increase the acceptance of Higgs boson events. As part of that effort, I performed a study to evaluate if the triggers would benefit from the ability to know at L1 Trigger Level if the primary interaction vertex was in the east or west side of CDF. If the answer were yes, then a hardware based hit count would have been implemented to evaluate in which half of CDF more particle activity was recorded. However, the study showed that this information would not have changed the trigger efficiency significantly and therefore primary vertex identification was not introduced at L1. However, other efforts were proven to be worthwhile. For example, a new missing transverse energy + jets trigger was designed, and we currently use this in our analysis. 

\section{Calorimeter Clustering Algorithm}

\ \\High transverse momentum electrons, photons and jets leave energy deposits in the calorimeter systems in adjacent calorimeter towers. Together they form a calorimeter cluster, which is reconstructed using a clustering algorithm. The first step is finding a seed cluster that has an energy deposit larger than a certain threshold value. Then neighbouring clusters with energy deposits larger than another lower threshold are also added to the cluster. The total energy of the cluster is the sum of the energy deposits in each calorimeter tower. The position of the cluster is computed as an energy-weighted average of the positions of each tower in the cluster. Then, in order to improve the precision for the cluster position, the calorimeter cluster is matched with a cluster in the shower maximum detector. The latter type of cluster is built with a similar algorithm, but one that is optimized to achieve a better position resolution. 

\section{Charged Lepton Identification}

\ \\In this section we will discuss charged lepton identification. Except when otherwise mentioned, a process described for a particle is also true for its antiparticle. In this analysis we use electron, muon and isolated track candidates. 

\subsection{Electron Identification}

\ \\An electron typically deposits most of its energy in the electromagnetic calorimeters. The basic selection for an electron candidate is a high $\pt$ track, isolated from other activity in the tracking systems, which is matched to an electromagnetic calorimeter cluster. The isolation requirement ensures that the charged lepton candidate originates in the primary interaction vertex and therefore is the daughter particle of the $W$ boson, and does not originate in a $B$ hadron semi-leptonic decay, as in the case of a jet originating in a $b$ quark. Further selection criteria summarized in Table~\ref{table:ElectronIdentification} are required to reconstruct the tight central electron candidates in the region $|\eta|<1.1$ (CEM calorimeter) and therefore noted in this thesis as CEM.  

\begin{center}
\begin{table}[h] % [t] puts at top of page
\begin{center}
\caption[Criteria for central electron candidate (CEM) identification]
{Criteria for central electron candidate (CEM) identification. \label{table:ElectronIdentification}}
\begin{tabular}{|l|c|c|}
\hline \hline Criterion & Central Electron (CEM)\\
\hline
$\et$ & $> 20 \gev$\\
$E_{HAD}/E_{EM}$ & $\le (0.055 + (0.00045 \cdot E))$\\ 
Isolation & $\le 0.1$\\
Track $z_0$& $\le$ 60 cm\\
Track $p_T$& $\ge 10 \gevc$\\
COT Axial Segments & $\ge 3$\\
COT Stereo Segments & $\ge 2$\\
$L_{shr}$ & $\le$ 0.2\\
E/p & $\le$ 2.0 for $\pt$ $\le$ 50 $ \gevc$\\
$\chi^2$ & $\le$ 10.0\\
$Q \cdot \Delta x$ & $-3.0\le Q \cdot \Delta x \le 1.5$\\
$|\Delta z|$ & $\le 3.0$ cm\\
\hline\hline
\end{tabular}
\end{center}
\end{table}
\end{center}

\ \\Here is the explanation of the notations from Table~\ref{table:ElectronIdentification}:

\begin{itemize}
\item The transverse energy of the calorimeter cluster is $\et$. The trigger requires $\et > 18 \gev$, but the analysis requires $\et > 20 \gev$ to make sure that the trigger is fully efficient. 
\item $E_{HAD}/E_{EM}$ is the ratio between the energy deposited in the hadronic calorimeters (CHA, WHA or PHA) and the energy deposited in the electromagnetic calorimeters (CEM). This ratio is very small for electron candidates, on the order of 5\%.
\item Isolation is defined as the ratio between energy deposited in all the additional towers located in a cone of radius $R=\sqrt{(\Delta \phi)^2+(\Delta \eta)^2}=0.4$ around the calorimeter cluster and the energy of the calorimeter cluster itself. Isolation is required to be smaller than 0.1, which means that the calorimeter cluster is required to be isolated from other significant energy deposits in the calorimeter.
\item Track $z_0$ is the $z$ coordinate position where the isolated track intersects the beamline.
\item Track $\pt$ is the transverse momentum of the electron candidate that is explicitly measured by the track curvature.
\item COT Axial (Stereo) Segments is the number of axial (stereo) COT layers that have at least 5 hits each associated with the track. 
\item $L_{shr}$ is a quantity that measures how well the theoretical electron shower profile matches the distribution of energy in the calorimeter cluster.
\item E/p is the ratio between the energy of the calorimeter cluster and the track momentum.
\item $\chi^2$ is the $\chi^2$ of the fit of the shower-maximum profile measured by the shower maximum detector (CES or PES) with respect to the electron test beam data.
\item $\Delta x$ is the signed difference in $x$ coordinate between the track and the calorimeter cluster when the track is extrapolated to the position of the shower maximum. 
\item Q is the measured electric charge of the electron candidate (negative for electron and positive for positron). 
\item $|\Delta z|$ is the absolute value in the $z$ coordinate between the position of the calorimeter cluster and the position of the track that is extrapolated to the position of the shower maximum. 

\end{itemize}

\subsection{Muon Identification\label{subsection:MuonIdentification}}

\ \\A muon behaves like a minimum ionizing particle due to its rest mass, which is about 200 times larger than that of the electron~\cite{PDG}. Therefore muons deposit very little energy in the calorimeter systems. The outer layer of the CDF detector is instrumented by muon chambers. Various types of muon candidates are reconstructed that bear the name of the muon detector that records them. Ionization deposits from a muon candidate in a given muon detector constitute a ``stub''. 

\ \\The basic selection for a muon candidate is a well reconstructed high-$\pt$ track isolated from other activity in the detector that is matched to a muon stub. Other selection cuts required to refine the selection above are summarized in Table~\ref{table:MuonIdentification} for two types of tight muon candidates (CMUP and CMX). 

\ \\CMUP muon candidates are reconstructed in the region $|\eta|<0.6$ and a track is required to match stubs in both CMU and CMP muon detectors. Some very energetic hadrons are able to deposit some of their energy outside the calorimeter systems, in the muon detectors, and therefore fake the muon signature. The ``fake muon'' fraction in the collected sample is decreased considerably by requiring both CMU and CMP stubs.

\ \\CMX muon candidates are reconstructed in the region $0.65<|\eta|<1.0$ and require a stub in the CMX muon detector. 

\begin{center}
\begin{table}[h] % [t] puts at top of page
\begin{center}
\caption[Criteria for muon candidate (CMUP and CMX) identification]
{Criteria for muon candidate (CMUP and CMX) identification. \label{table:MuonIdentification}}
\begin{tabular}{|l|c|}
\hline \hline Criterion & CMUP and CMX\\
\hline
$\pt$ & $> 20 \gev$ \\
$E_{HAD}$ & $< 6 + \max (0,(p-100)\cdot 0.0280)\gev$ \\
$E_{EM}$  & $< 2 + \max (0,(p-100)\cdot 0.0115)\gev$ \\
Isolation & $\le 0.1$ \\
Track $z_0$& $\le$ 60 cm\\
Track $p_T$& $\ge 10 \gevc$ \\
COT Axial Segments & $\ge 3$\\
COT Stereo Segments & $\ge 2$\\
Impact Parameter $d_0$ & $<$ 0.2 cm (0.02 with silicon hits) \\
$\chi^2$ & $<$ 2.3\\
\hline
CMU $\Delta x$ & $<$ 3 cm\\
CMP $\Delta x$ & $<$ 5 cm\\
CMX $\Delta x$ & $<$ 6 cm\\
CMX $\rho_{\rm{COT}}$ & $>$ 140 cm\\
\hline\hline
\end{tabular}
\end{center}
\end{table}
\end{center}

\subsection{Isolated Track Identification\label{subsection:IsolatedTrackIdentification}}

\ \\Our original contribution to the $WH$ analysis is the introduction of a new charged lepton reconstruction method based on a high-$\pt$ track isolated from energy deposits in the tracking systems and with no requirements about energy deposits in the calorimeter or muon detectors. We call these ``isolated tracks''. The isolation requirement is necessary in order to ensure that the track corresponds to a charged lepton produced in a decay of a $W$ boson and not part of a jet of hadrons that originate in quarks. The fact that the isolated track is not required to match a calorimeter cluster or a muon stub allows to recover real charged leptons that arrive in non-instrumented regions of the calorimeter or muon detectors, as seen in Figure \ref{figure:ScatterEtaPhi}. Thus, the signal acceptance of the $WH$ search is increased and this has the potential to increase the $WH$ search sensitivity.

\ \\The first analysis that used isolated tracks at CDF \cite{CDFTopLeptonPlusTrack} was a top quark cross section measurement in top quark pair production where each top quark decays to one $W$ boson and a $b$ quark and where each $W$ boson decays leptonically to a charged lepton and a corresponding neutrino, using an integrated luminosity of $1.1\ \invfb$. In this analysis one charged lepton was either an electron or a muon candidate and the second charged lepton was an isolated track candidate. The events used in this analysis were selected using an electron-inclusive trigger or a muon-inclusive trigger.

\ \\In our analysis we have exactly one charged lepton candidate and in this channel the charged lepton is an isolated track candidate. However, we do not have an isolated track trigger at CDF. This is why we use three triggers with requirements on MET and jets, but not on charged leptons. One of my original contributions consisted in parameterizing the trigger efficiency turnon curves and measuring the systematic uncertainty of this procedure.

\ \\My work contributed directly to the neural network $WH$ searches that used one (two) MET-based triggers in $2.7\ \invfb$ ($4.3\ \invfb$), as described in the Ph.D. thesis \cite{JasonSlaunwhiteThesis} (\cite{YoshikazuNagaiThesis}) and in their corresponding publications and CDF and Tevatron combinations. As my work evolved in time, this thesis presents for the first time the addition of a third MET-based trigger and a novel method to combine an unlimited number of triggers short of having an ``OR'' between triggers, as described in detail in Subsection \ref{section:ISOTRKTriggers}. 

\ \\In order to reconstruct isolated tracks, we select on an event by event basis a set of good quality tracks with criteria that meet the requirements detailed in Table \ref{table:GoodQualityTracks}. 

\begin{center}
\begin{table}[h] % [t] puts at top of page
\begin{center}
\caption[Good quality tracks criteria]
{Good quality tracks criteria. \label{table:GoodQualityTracks}}
\begin{tabular}{|l|l|}
\hline \hline Variable & Cut\\
\hline
$\pt$ & $> 0.5 \gev$ \\
$\Delta R$ (track, candidate) & $<$ 0.4\\
$\Delta z_0$ (track, candidate) & $<$ 5 cm\\
COT Axial Hits & $\ge 20$\\
COT Stereo Hits & $\ge 10$\\
\hline\hline
\end{tabular}
\end{center}
\end{table}
\end{center}

\ \\Next, for each good quality track we define and compute a quantity called ``track isolation'' as 

\begin{equation}\label{TrackIsolation}
\rm{Track\ Isolation}=\frac{\pt({\rm{track\ candidate}})}{\pt({\rm{track\ candidate}})+\sum \pt({\rm{other\ tracks)}}}\,\rm{,}
\end{equation}

\ \\where $\pt({\rm{track\ candidate}}$ is the transverse momentum of the specific track we analyze (candidate) and $\sum \pt({\rm{other\ tracks)}}$ is the sum of the transverse momenta of all good quality tracks within a cone radius of 0.4 of the candidate track. Given this definition, a track is fully isolated if it has a track isolation of 1.0. However, very seldom a track is fully isolated in a hadronic collision. In this analysis we consider a track to be isolated if it has an isolation larger than 0.9, which means that at least 90\% of the $\pt$ in the vicinity of the track corresponds to the track itself.

\ \\From the sample of good quality tracks with isolation larger than 0.9 we select the sample of ``isolated track'' candidates by requiring further tighter track reconstruction criteria, as described Table \ref{table:IsolatedTrackIdentification}.

\begin{center}
\begin{table}[h] % [t] puts at top of page
\begin{center}
\caption[Criteria for isolated track candidates identification]
{Criteria for isolated track candidates identification. \label{table:IsolatedTrackIdentification}}
\begin{tabular}{|l|l|}
\hline \hline Criterion & ISOTRK\\
\hline
$\pt$ & $> 20 \gev$ \\
$\eta$ & $<$ 1.2 \\
$z_0$ & $\le$ 60 cm\\
Isolation & $\ge 0.9$ \\
COT Axial Hits & $\ge 24$\\
COT Stereo Hits & $\ge 20$\\
$\chi^2$ & $> 10^{-8}$\\
Impact Parameter $d_0$ & $<$ 0.2 cm (0.02 with silicon hits) \\
\hline\hline
\end{tabular}
\end{center}
\end{table}
\end{center}

\ \\The purity for isolated tracks is about 80\%, whereas the purity for TIGHT charged lepton candidates is about 90\%, as shown in Figure~\ref{figure:Pretag_QCD}.

\subsection{Charged Lepton Reconstruction Scale Factor\label{subsection:ChargedLeptonScaleFactors}}

\ \\Our detector is not as efficient to reconstruct charged lepton candidates in Monte-Carlo-simulated events as in real data events. For this reason we correct each simulated event by a scale factor for its specific charged lepton reconstructed category, which is defined as the ratio between the reconstruction efficiencies in data and simulated events.

\begin{figure}[ht]
  \begin{center}
    \includegraphics[width=13cm]{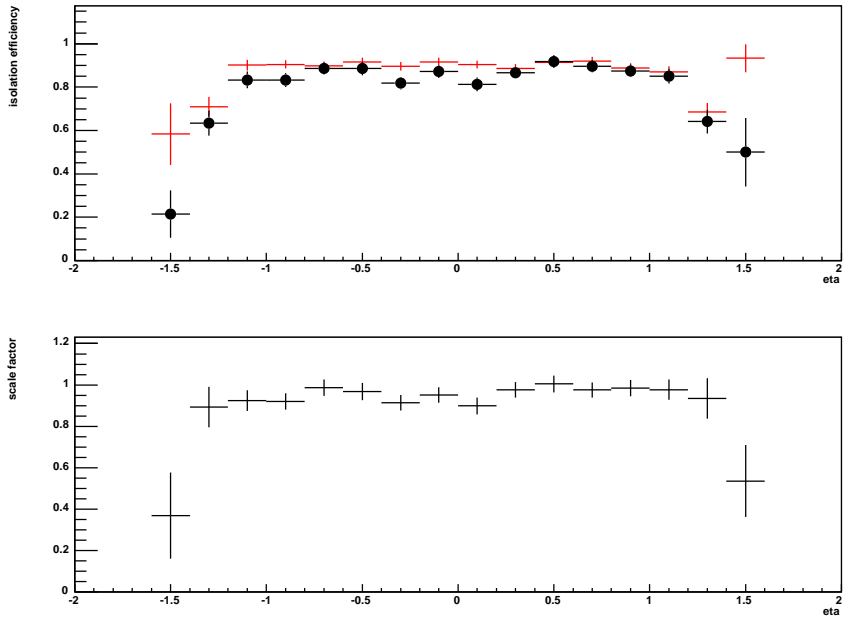}                                          
    \caption[ISOTRK reconstruction scale factor as a function of lepton $\eta$]{Isolated track reconstruction efficiency and scale factor as a function of lepton $\eta$. The upper plot shows the reconstruction efficiency in both data (black) and simulated (red) events, while the lower plot shows the resulting scale factor as the ratio of the histograms from the top plot.}
    \label{figure:ISOTRKScaleFactorsEta}
  \end{center}
\end{figure}

\begin{figure}[ht]
  \begin{center}
    \includegraphics[width=13cm]{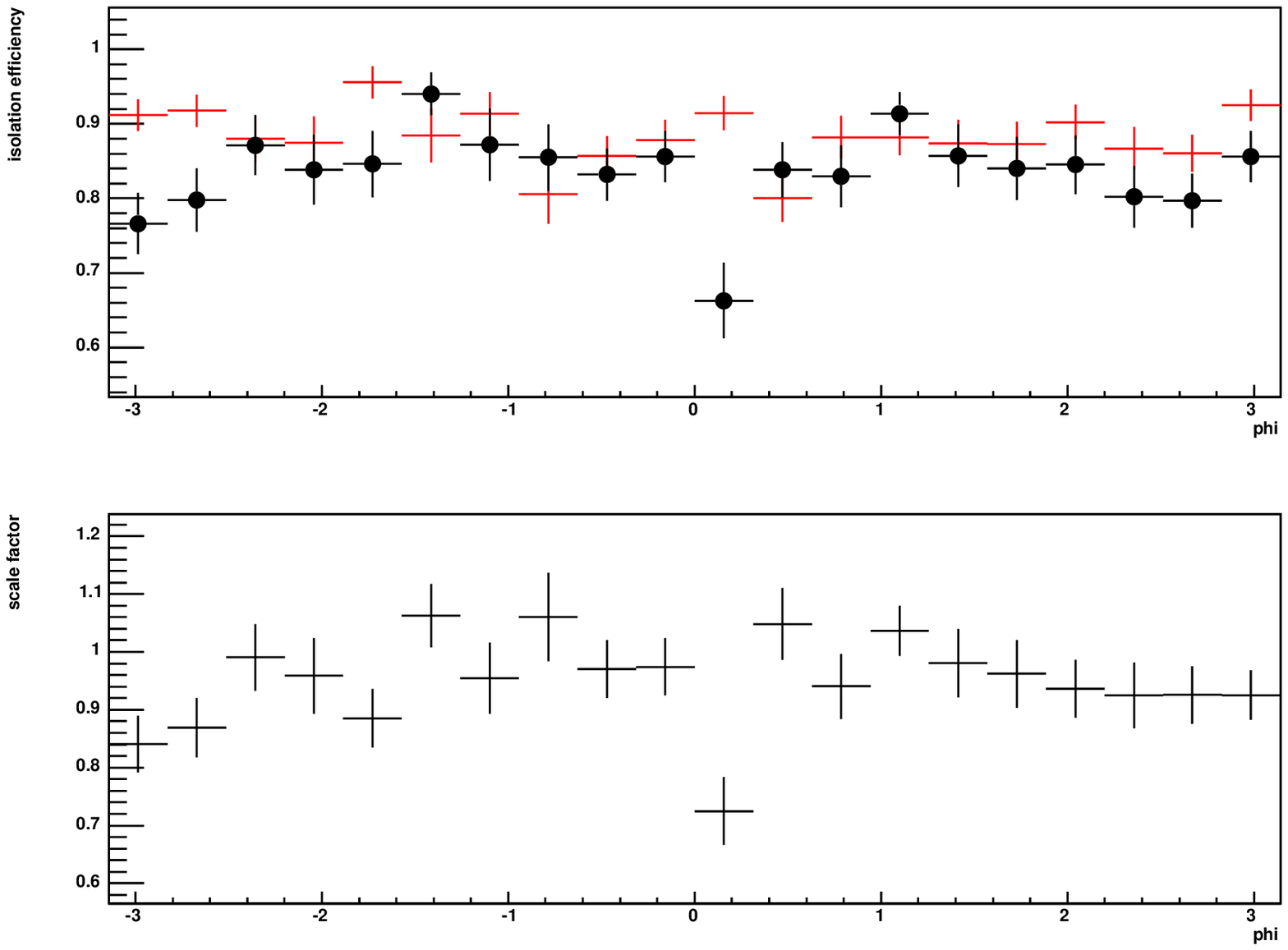}                                          
    \caption[ISOTRK reconstruction scale factor as a function of lepton $\phi$]{Isolated track reconstruction efficiency and scale factor as a function of lepton $\phi$. The upper plot shows the reconstruction efficiency in both data (black) and simulated (red) events, while the lower plot shows the resulting scale factor as the ratio of the histograms from the top plot.}
    \label{figure:ISOTRKScaleFactorsPhi}
  \end{center}
\end{figure}

\begin{figure}[ht]
  \begin{center}
    \includegraphics[width=13cm]{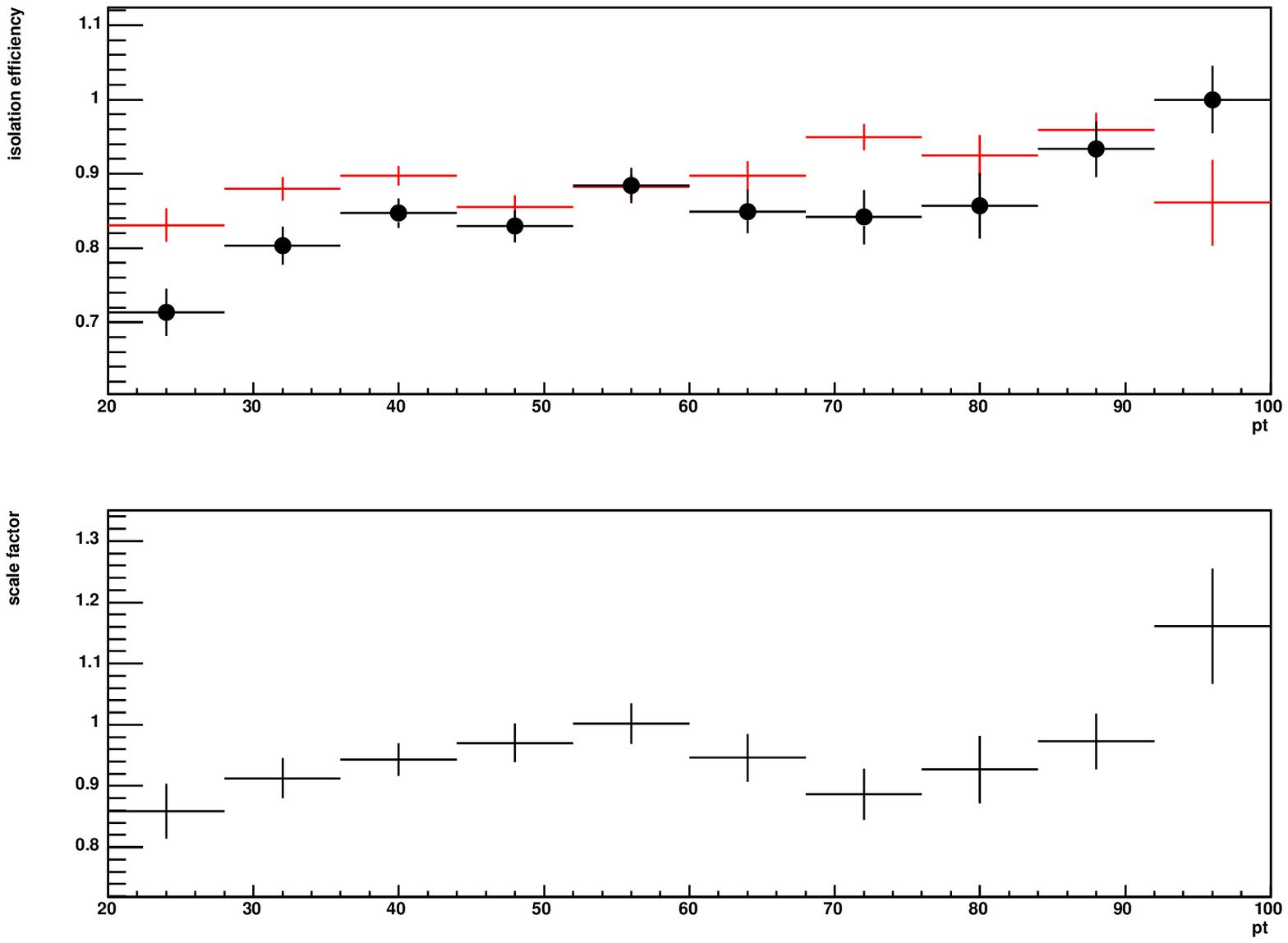}                                          
    \caption[ISOTRK reconstruction scale factor as a function of lepton $\pt$]{Isolated track reconstruction efficiency and scale factor as a function of lepton $\pt$. The upper plot shows the reconstruction efficiency in both data (black) and simulated (red) events, while the lower plot shows the resulting scale factor as the ratio of the histograms from the top plot.}
    \label{figure:ISOTRKScaleFactorsPt}
  \end{center}
\end{figure}

\subsubsection{Scale Factor Measurement for Isolated Tracks\label{subsubsection:ISOTRKScaleFactor}}

\ \\ We studied $WH$ Monte Carlo simulated events and concluded that ISOTRK charged lepton candidates are muon candidates in 85\% of cases, electron candidates in 7\% of cases and tau lepton candidates in 8\% of cases \cite{JasonSlaunwhiteThesis}. We measure the scale factor for muon ISOTRK charged leptons and correct the systematic uncertainty on that value for the fact that in 15\% of the cases ISOTRK events are not muon candidates.

\ \\We select a sample of events where a $Z$ boson decays to a muon-antimuon pair and use the generic method called ``tag and probe'', which is also used to measure the scale factor for the reconstruction of CMUP muons in Reference \cite{FirstWZCDFII}. We select events with a well reconstructed tight muon (CMUP or CMX), which is considered the tag leg, and a high-$\pt$ track isolated from other activity in the detector, which is called the probe leg. We ask further selection requirements to improve the purity of the $Z$ boson sample: the invariant mass of the tag and prob legs should be in the $Z$ boson mass window ($81-101\ \gevcc$); the absolute value of the $z$ position between the two candidates has to be smaller than 4 cm; the legs must have opposite electric charges; for data events, the tag charged lepton must fire the CMUP or the CMX muon trigger; the event must pass the cosmic veto to ensure it is not produced by cosmic rays\footnote{Cosmic rays that reach the CDF detector situated 10 meters underground and shielded with thick concrete walls are mostly muons. Such a muon from cosmic rays is typically very energetic and therefore its trajectory is not curved much by the solenoid magnetic field. Thus, it will produce an almost straight track in the tracking systems. On the other hand, the software reconstruction is looking for tracks starting from the centre of the detector and will reconstruct this one muon as two back-to-back muon candidates.}; the probe charged lepton must have $\pt > 20 \gevc$ and be matched to a muon stub. After these cuts, we have a very pure $Z$ boson sample and we are confident that also the probe muon candidate is a real muon. We measure the efficiency that this probe muon candidate is indeed reconstructed as an ISOTRK charged lepton. We divide the efficiencies of data and Monte Carlo simulated event to obtain the ISOTRK reconstruction scale factor for each jet bin. For events with exactly two tight jets, as in our main analysis, we obtain the average value of $0.937 \pm 0.009$. Figures \ref{figure:ISOTRKScaleFactorsEta}, \ref{figure:ISOTRKScaleFactorsPhi} and \ref{figure:ISOTRKScaleFactorsPt} show the simulated and data efficiencies and scale factor as a function of the ISOTRK $\phi$, $\eta$ and $\pt$. 
                                                                                                                                                
\ \\However, we do not quote a systematic uncertainty of 1\%. We take into account that in 15\% of cases the ISOTRK charged lepton is either an electron or a tau lepton candidate. We assign a 25\% uncertainty for these cases \cite{JasonSlaunwhiteThesis} \cite{YoshikazuNagaiThesis}. The total ISOTRK scale factor systematic uncertainty is computed as a weighted average, namely $0.85\cdot 1\% + 0.15 \cdot 25\% = 4.5\%$. Therefore, the ISOTRK scale factor for our analysis is $0.937 \pm 0.042$. 

\ \\The code for this procedure was written by a postdoctoral researcher (Nils Krumnack). For the past two years, I maintained the code after he left the collaboration and I used it to measure the ISOTRK scale factors for the $WH$ analyses of 2009, 2010 and 2011. 

\subsubsection{Scale Factor Measurement for Tight Charged Leptons\label{subsubsection:TightScaleFactor}}

\ \\We use a very similar method to measure the scale factor for the reconstruction of tight muon candidates, CMUP and CMX. For the dataset of 5.7 $\invfb$ used in this analysis, we measure for CMUP candidates a scale factor of $0.892 \pm 0.002$ and for CMX candidates a scale factor of $0.948 \pm 0.002$.

\ \\A similar method with the exception that a $Z$ boson sample decaying to electron pairs is selected is used to measure the scale factor for the CEM tight electron to be $0.977 \pm 0.001$.  

\section{Missing Transverse Energy Identification\label{section:METObject}}

\ \\Neutrinos are the only subatomic particles that leave the detector completely undetected. Their momentum and energy appears to be missing. Since the longitudinal energies of the colliding partons are unknown and not necessarily equal, we can only say that the total transverse energy of the $p\pbar$ collision is zero. Therefore we observe a missing transverse energy (MET or $\met$) due to the neutrino. 

\ \\The vector missing transverse energy ($\vec{\met}$) is the opposite of the vector sum of all energy deposits in calorimeter towers, with the $z$ position of the neutrino closest to the beamline considered to be the $z$ vertex position. The missing transverse energy is the absolute value of the vector missing transverse energy ($\met=|\vec{\met}|$). From its definition we see that $\met$ is formed by missing transverse energy from all undetected particles, such as one or more neutrinos, but also subatomic particles predicted by theories beyond the Standard Model, such as the lightest supersymmetric particle or a signature of particles travelling in extra spatial dimensions. In our analysis, real $\met$ is produced by only one neutrino from the $W$ boson decay.

\ \\Real missing transverse energy can also be faked by mismeasured energy deposits in the calorimeter, especially due to jets. Typically jet energies are underestimated, as seen in the next section, which translates into an overestimation of the $\met$. Events that produce jets in pure QCD processes tend to have fake $\met$ as we will see in the background chapter.

\ \\At trigger level, $\met$ is computed using only calorimeter information and assuming the primary interaction vertex is located in the centre of the detector. As shown in subsection~\ref{TriggerPrimaryVertexStudy},I performed a study in the context of the Higgs Trigger Task Force that showed that improving the primary vertex position measurement at L1 trigger level by noting in which side of the detector it is located does not improve significantly the offline $z$ vertex position reconstruction. This study confirmed this approach to be optimal at trigger level. However, the offline reconstructed $\met$ is corrected for the true $z$ vertex position, for the corrected jet energies, and by subtracting the momenta of minimum ionizing high-$\pt$ muons and adding back the transverse energy of the calorimeter towers crossed by the muon. These corrections can be summarized in the following equation:

\begin{equation}
\met = \met^{\text{raw}} - \sum_{\text{muon}}\pt + \sum_{\text{muon}}\et(\text{EM} + \text{HAD}) - \sum_{\text{jet}}\et(\text{Jet Energy Correction})
\end{equation}

\begin{figure}[h]
\begin{center}
\includegraphics[angle=0,width=0.4\textwidth,clip=]{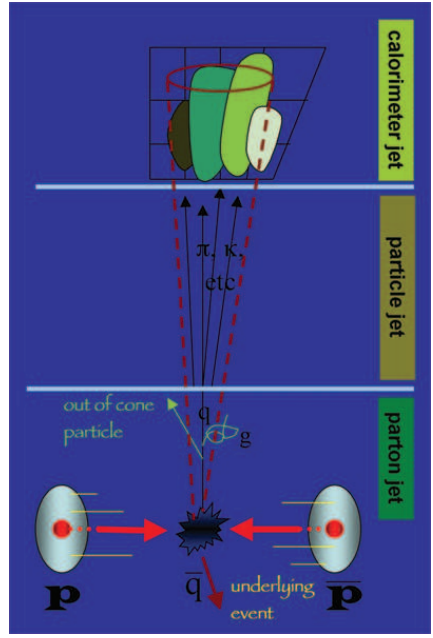}
\caption[Diagram of jet production at CDF]
{Diagram of jet production at CDF, from parton-level jet to particle-level jet and then to calorimeter-level jet. Credit image to the CDF collaboration. \label{figure:JetProduction}}
\end{center}
\end{figure}

\section{Jet Identification}

\ \\A jet consists of a collimated shower of secondary particles produced in the hadronization of a quark or gluon produced in the primary $p\pbar$ interaction. There are various possible algorithms to reconstruct jets from calorimeter towers. The algorithm used in this analysis is called JETCLU~\cite{JETCLU}. JETCLU is a cone algorithm that searches for towers with energy deposits within a cone of radius $\Delta R = \sqrt{(\Delta \eta)^2+(\Delta \phi)^2} = 0.4$ in the $\eta$-$\phi$ plane around the seed cluster. If a charged lepton is reconstructed within this cone, its energy is neglected in the calculation of the jet energy.

\ \\A diagram of the jet production at CDF, passing from parton jet to particle jet and then to calorimeter jet, is shown in Figure~\ref{figure:JetProduction}.

\begin{figure}[h]
\begin{center}
\includegraphics[angle=0,width=1.0\textwidth,clip=]{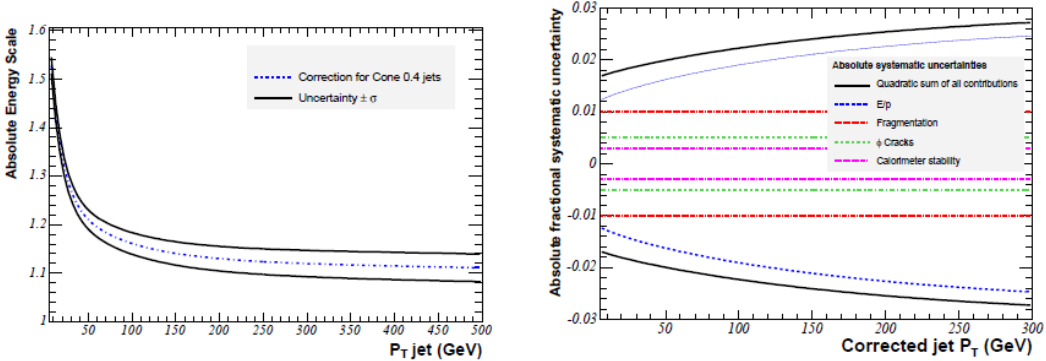}
\caption[Absolute energy jet corrections and their systematic uncertainty]
{The left hand side plot shows the absolute jet corrections for a cone size of 0.4 as a function of the calorimeter-level jet $\pt$. The right hand side plot represents the jet energy scale uncertainty due to detector calibration and simulation. The solid lines represent the total systematic uncertainty and the dotted lines the partial contributions. Credit image to the CDF collaboration. \label{figure:AbsoluteEnergyCorrections}}
\end{center}
\end{figure}

\ \\The energy of the jet is corrected for various effects ~\cite{JetCorrections}:
\begin{itemize}

\item Relative Energy Corrections take into account the fact that the detector is not uniform in $\eta$, since the plug and central calorimeters have different geometries and because there are cracks between calorimeter towers. Central calorimeters are better calibrated than the plug ones. This is why plug calorimeters are corrected with respect to the central calorimeters using Pythia Monte Carlo (MC) simulated events and di-jet data events. In our analysis, both data and MC events are corrected with respect to $\eta$ to make sure there is a uniform jet energy response across the detector. 

\item Multiple Interaction Corrections take into account the effect that typically there are more than one $p\pbar$ interactions in a bunch crossing. For an instantaneous luminosity on the order of $10^{32}\rm{cm}^{-2}s^{-1}$ the average number of primary interactions is three. This is why CDF uses algorithms to select the correct primary interaction vertex. But also the energy of the jets is corrected for the effect that energy deposits from particles produced in $p\pbar$ interactions other than the one that interests us happen in the same calorimeter cluster as the one for the selected jet. These corrections are derived from minimum bias data events and are parametrized as a function of the number of primary vertices in the bunch crossing (event). 

\item Absolute Energy Corrections take into account the effect of the non linearities and the uninstrumented regions of the detector. These corrections map the hadron-level jet after its hadronization from a quark or gluon to the calorimeter level-jet. 

\item Underlying Event Corrections subtract the energy deposited in the calorimeter towers by the underlying event. The underlying event is represented by soft energy depositions due to the spectator quarks and gluons in the protons and antiprotons.

\item Out-of-Cone Corrections add the energy deposited in calorimeter towers that have not been reconstructed by the JETCLU algorithm to be part of the jet. 
\end{itemize}

\ \\The corrected transverse jet energy can be summarized by the following equation:

\begin{equation}
\et^{\mbox{\footnotesize{parton}}} = (E_{T,\,{\mbox{\footnotesize jet}}}^{\mbox{\footnotesize measured}} 
\times f_{\mbox{\footnotesize rel}} - \et^{\mbox{\footnotesize MI}} \times N_{\mbox{\footnotesize vtx}})
\times f_{\mbox{\footnotesize abs}} - \et^{\mbox{\footnotesize UE}} + \et^{\mbox{\footnotesize OOC}} 
\label{eq:JEcorr}
\end{equation}

\ \\where the correction factors are $f_{\mbox{\footnotesize rel}}$ (the scale factor that makes the jet energy measurement uniform for various $\eta$, $\et^{\mbox{\footnotesize MI}} \times N_{\mbox{\footnotesize vtx}}$ (the correction factor for multiple $p\pbar$ interactions per bunch crossings), $f_{\mbox{\footnotesize abs}}$ (the absolute energy correction determined by matching the parton energy to the jet energy), $\et^{\mbox{\footnotesize UE}}$ (the scale factor that corrects for the underlying event) and $\et^{\mbox{\footnotesize OOC}}$ (the out-of-cone correction for the energy of the initial parton that is not reconstructed in the jet cone). 

\ \\The corrections and their systematic uncertainty can be seen in Figure~\ref{figure:AbsoluteEnergyCorrections}. The transverse jet energy resolution~\cite{JetCorrections} is given approximately by

\begin{equation}
\frac{\sigma(\et)}{\et} = \frac{1.0}{\et [\gev]}  \oplus 0.1\,.
\end{equation}

\ \\In our analysis we classify events by the number of tight jets. Tight jets are jets with the following tight selection criteria: $\et > 20 \gev$ and $|\eta|<2.0$. Events can also have loose jets. Loose jets are jets with looser selection criteria and they are exclusive to the tight jets. Loose jets have $12 \gev < \et < 20 \gev$ and $|\eta|<2.0$ or $\et > 12 \gev$ and $2.0<|\eta|<2.4$.

\section{$b$-jet Identification\label{section:bTagging}}

\ \\In this analysis we search for an associated production of a $W$ boson and a Higgs boson, where the latter decays to a pair of bottom-antibottom quarks ($b\bbar$ pair). Each quark hadronizes and is seen in the detector as a jet. An essential point of our analysis is to identify for a given jet if it is produced by a $b$ quark or not. After hadronization, $b$ hadrons (mesons or baryons) travel for a relatively long lifetime, on the order of a few picoseconds, before decaying. Therefore, we can see in the silicon detectors a secondary vertex displaced by about three millimetres from the primary $p\pbar$ interaction vertex. The tracks originating in the secondary displaced vertex have on average larger values for the $d_0$ parameter. 

\ \\In this analysis we use two $b$-tagging algorithms: Secondary Vertex Tagger (\secvtx) and Jet Probability Tagger (\jetprob).

\begin{figure}[h]
\begin{center}
\includegraphics[angle=0,width=0.4\textwidth,clip=]{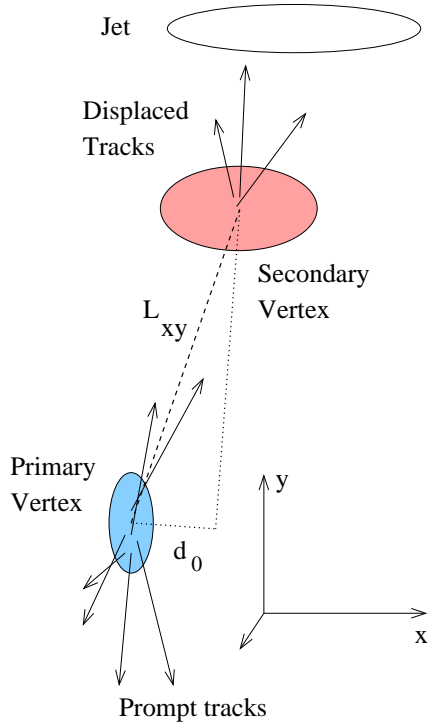}
\caption[Diagram of a secondary (displaced) vertex specific to a $b$ quark]
{Diagram of a secondary vertex displaced with respect to a primary interaction vertex, typical for a jet originating from a $b$ quark. Credit image to the CDF collaboration. \label{figure:SecVtxDiagram}}
\end{center}
\end{figure}

\subsection{Secondary Vertex Algorithm}

\ \\The Secondary Vertex algorithm (\secvtx) \cite{SecVtx} reconstructs secondary vertices displaced with respect to the primary interaction vertex. \secvtx~operates not on an event by event basis, but on a jet by jet basis, which means that one event can have two or more jets identified by \secvtx~as originating from a $b$ hadron, as shown in Figure~\ref{figure:SecVtxDiagram}.

\begin{figure}[h]
\begin{center}
\includegraphics[width=6.5cm]{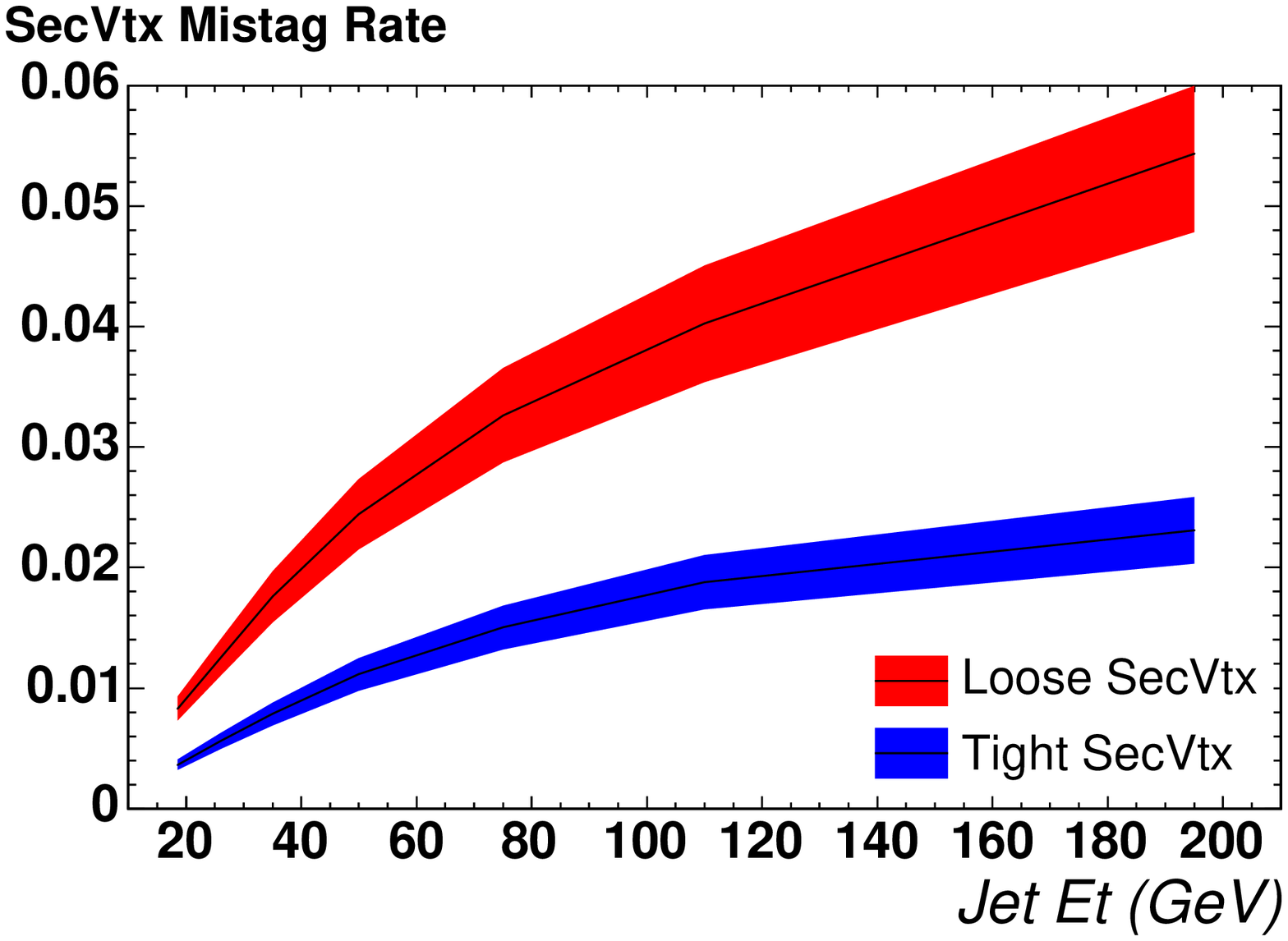}
\includegraphics[width=6.5cm]{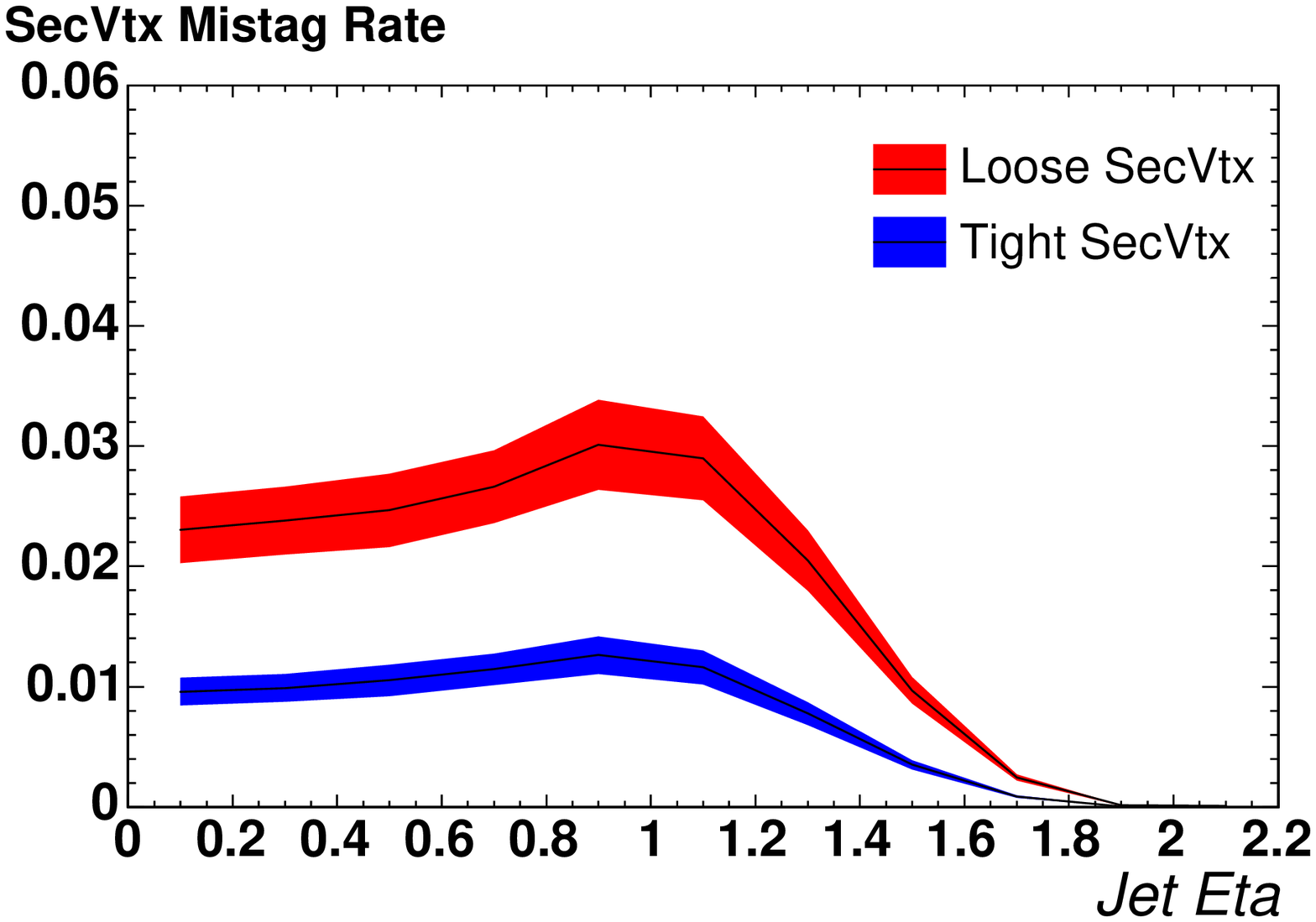}
\caption[Mistag rates of \secvtx as a function of jet $\et$ and $\eta$]
{Mistag rates of \secvtx~as a function of jet $\et$ (labelled here ``Jet Et'') and $|\eta|$ (improperly labelled here ``Jet Eta''). The rate is measured using inclusive jet data. Credit image to the CDF collaboration. \label{figure:SecVtxMistag}}
\end{center}
\end{figure}

\ \\\secvtx~starts by looking at well reconstructed tracks by the Inside-Out tracking algorithm in the cone of radius 0.4 around the calorimeter cluster. That means that silicon tracks are required first and then a match to a COT track is required. In order to reject poorly reconstructed tracks, all tracks used by \secvtx~must have $\pt > 0.5 \gevc$, their impact parameter $d_0$ is corrected for the primary interaction vertex and meets the criterion of $|d_0| < 0.3\,\rm{cm}$, and the distance in the $z$ coordinate between the track and the primary vertex should be less than 5 cm ($|z_{\rm{track}}-z_{\rm{primary \ vertex}}|<5\,\rm{cm}$). In addition, all tracks have to pass a certain number of hits in the silicon detector and the COT and track fit have $\chi^2$ criteria.

\ \\Once all the tracks inside the jet cone are reconstructed, \secvtx~has a two pass approach. 

\ \\First, \secvtx~tries to fit three loosely defined tracks to a common secondary vertex. These tracks have to have $\pt > 0.5 \gevc$, $|\frac{d_0}{\sigma_{d_0}}|>2.5$ and at least one of the tracks needs to have $\pt > 1.0 \gevc$.

\ \\If no secondary vertex is found in the first pass, then in the second pass only two tracks are required, but they have to pass tighter selection criteria of $\pt > 1.0 \gevc$ and $|\frac{d_0}{\sigma_{d_0}}|>3.0$.

\ \\If a secondary vertex is found by the first or second pass, then its two dimensional decay length is measured with respect to the primary vertex $\Lxy$, together with the uncertainty $\sigma_{\Lxy}$. From these we obtain the decay length significance $S_\Lxy=\frac{\Lxy}{\sigma_{\Lxy}}$~. $\Lxy$ is positive (negative) when the tracks emerging the secondary vertex are heading in the same (opposite) direction as the jet. If $|S_\Lxy| \ge 7.5$ the jet is \secvtx~tagged, with a positive tag if $S_\Lxy \ge 7.5$ and negative tag if $S_\Lxy \le -7.5$. 

\ \\A positive \secvtx~tag means that the jet has been identified as originating from a $b$ quark. A negative \secvtx tag means that the tracks have not been properly identified and that the primary vertex lies in front of the jet. The negatively tagged jets are unlikely to be produced by a $b$ hadron and are used to estimate the percentage of positively tagged jets that actually originate from light quarks (such as $u$, $d$, $s$ and $g$), also called the mistag rate. The mistag rate is parameterized as a function of various variables as seen in Figure~\ref{figure:SecVtxMistag}. 

\begin{figure}[h]
\begin{center}
\includegraphics[width=6.5cm]{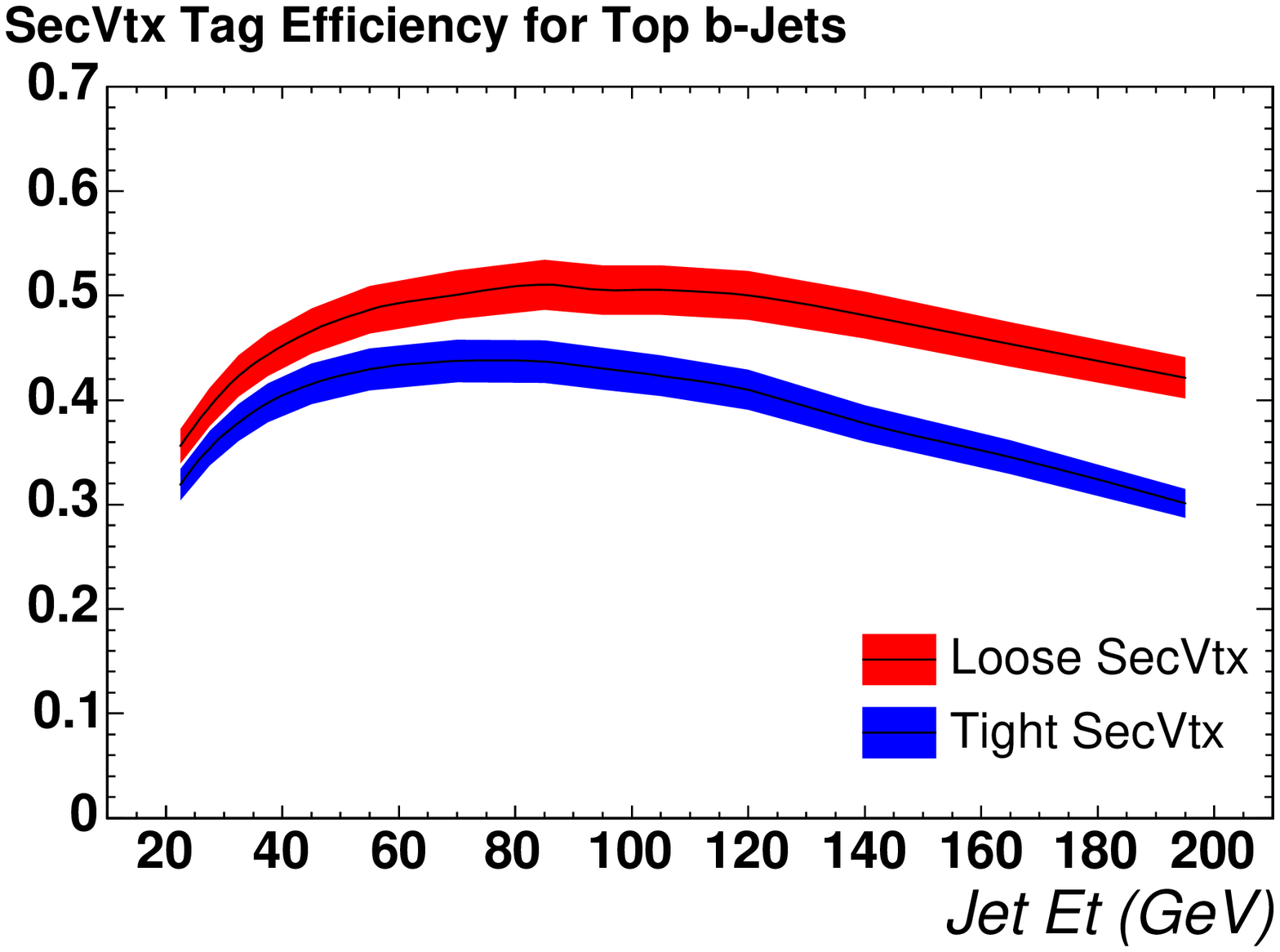}
\includegraphics[width=6.5cm]{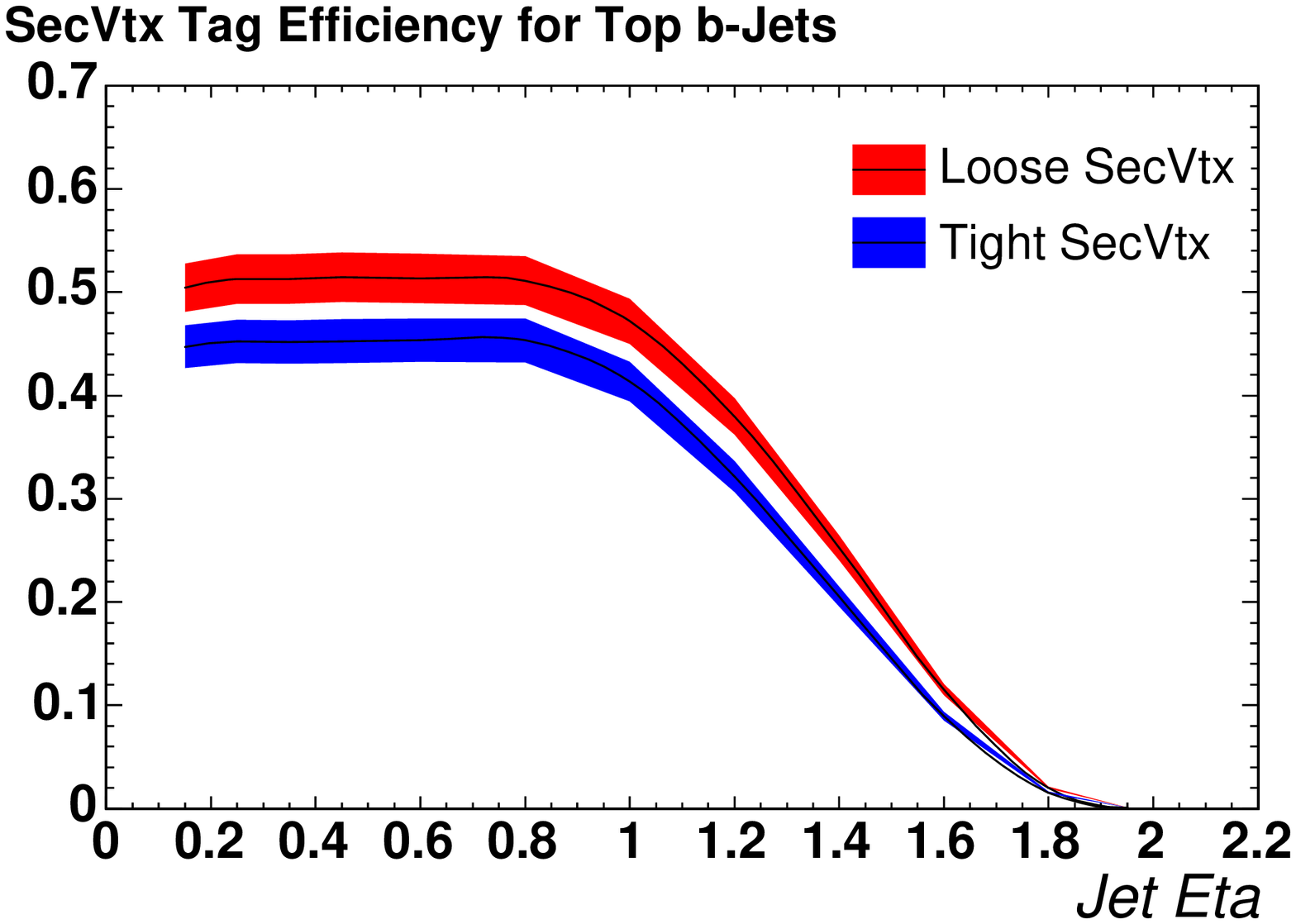}\\
\includegraphics[width=6.5cm]{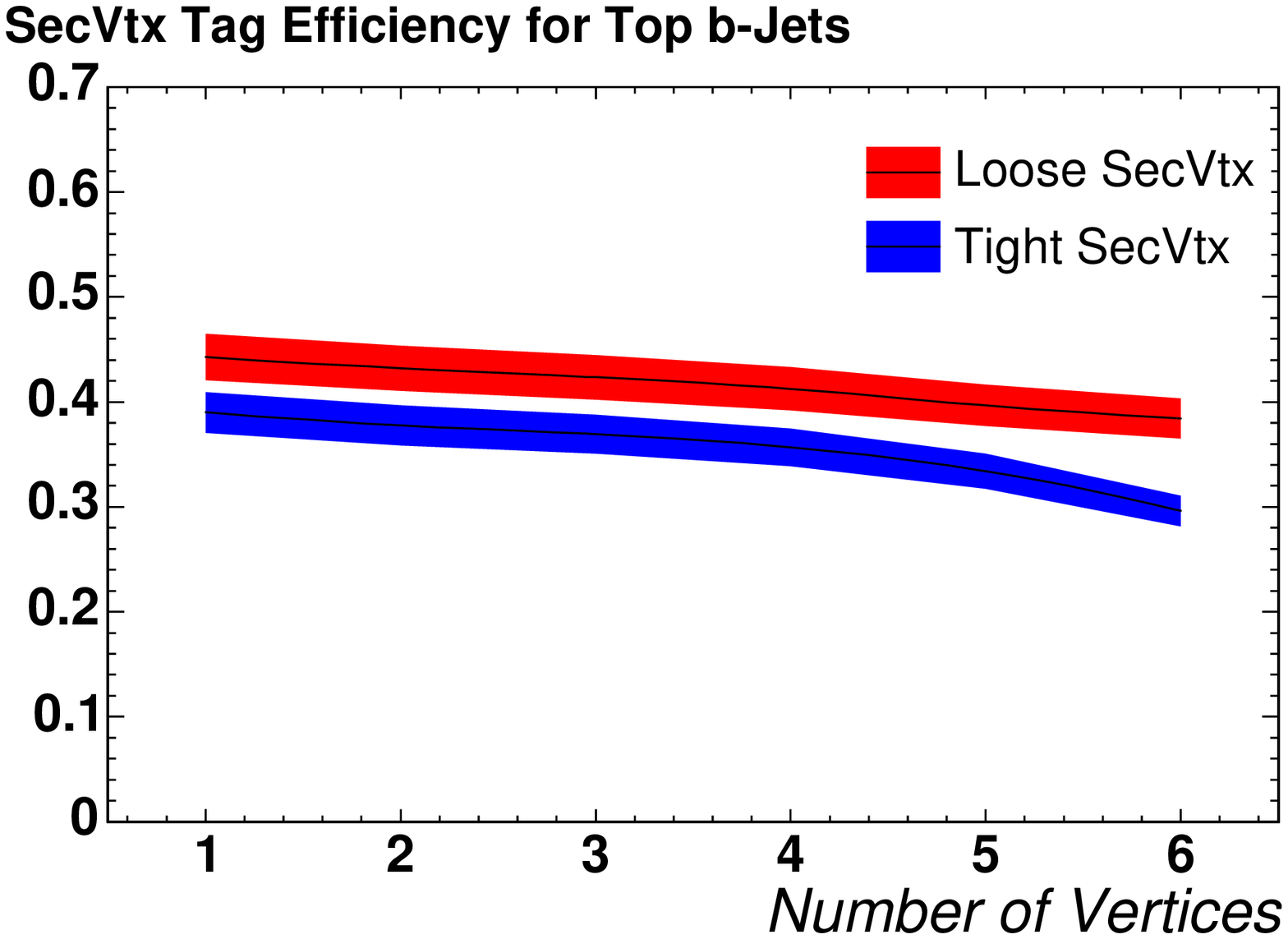}
\caption[Parametrization for $b$-tagging efficiency for the \secvtx~tagger]
{$b$-tagging efficiency of \secvtx~as a function of jet $\et$ (labelled here ``Jet Et''), jet $|\eta|$ (improperly labelled here ``Jet Eta''), and number of primary interaction vertices. The operation point called ``tight'' is used for this analysis. Credit image to the CDF collaboration. \label{figure:SecVtxEff}}
\end{center}
\end{figure}

\ \\The \secvtx~algorithm is tuned to accept only a very low mistag rate, on the order of 1-2\%, and this translates to a trade off in efficiency of only 40\% as seen in Figure~\ref{figure:SecVtxEff}. What it means is that 60\% of jets originating from a $b$ quark will not be positively tagged by \secvtx . 

\ \\Also, out of the remaining positively tagged jets, about half originate from the $c$ quark, which is also relatively massive and long lived. In fact, in general $b$-taggers are actually heavy flavour taggers, where the flavour quarks are the $b$ and $c$ quarks.

\ \\Furthermore, the efficiency for tagging a jet with \secvtx~is different between data and Monte Carlo simulated events. We define a $b$-tagging efficiency scale factor on a jet per jet basis for \secvtx~defined as the ratio between the efficiencies for data and simulated events. We selected a jet sample enriched in jets originating in $b$ quarks and we measure a $b$-tagging scale factor of $0.96 \pm 0.05$.

\begin{figure}[h]
  \begin{center}
 \includegraphics[width=6.5cm]{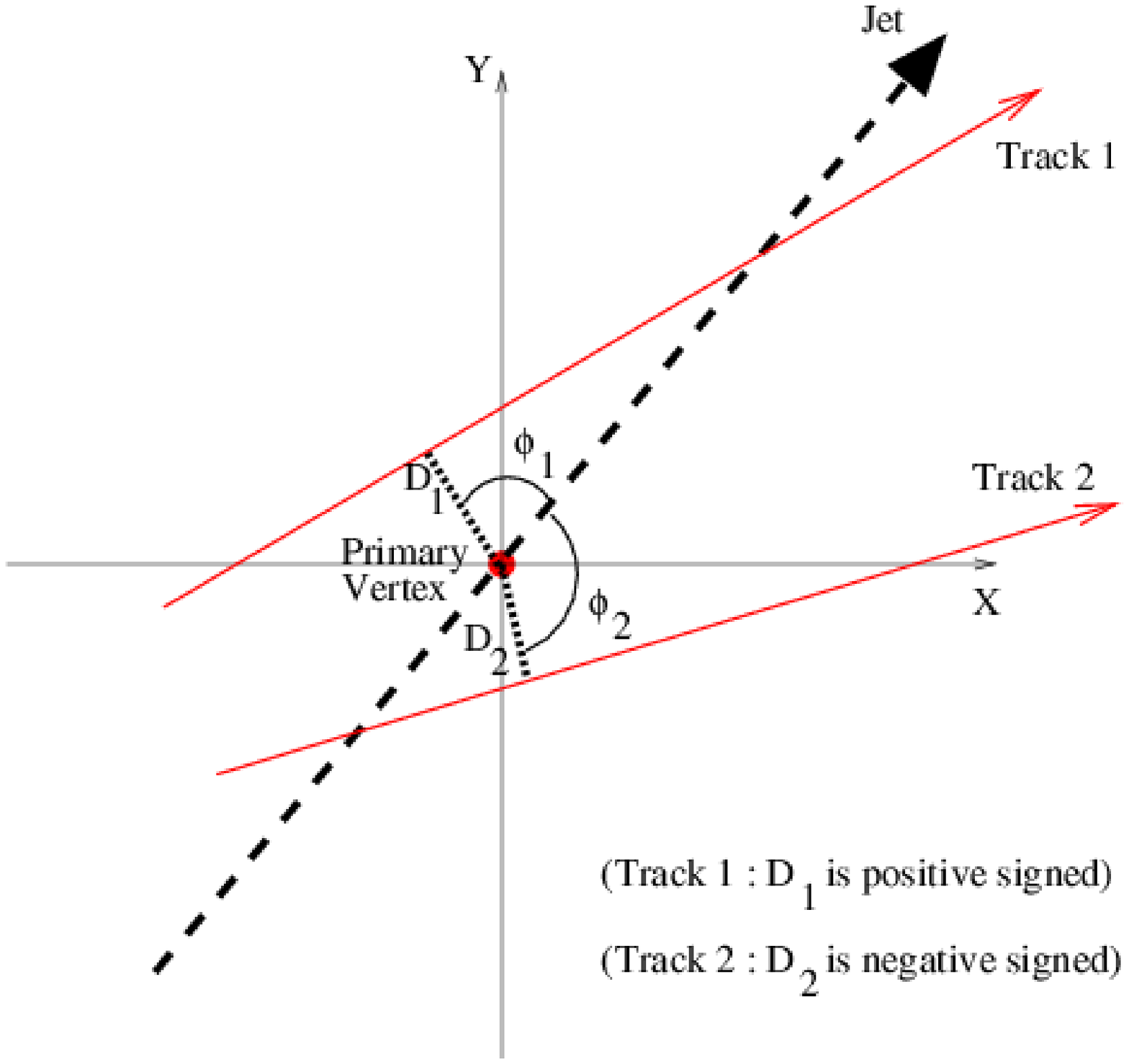}
 \includegraphics[width=6.5cm]{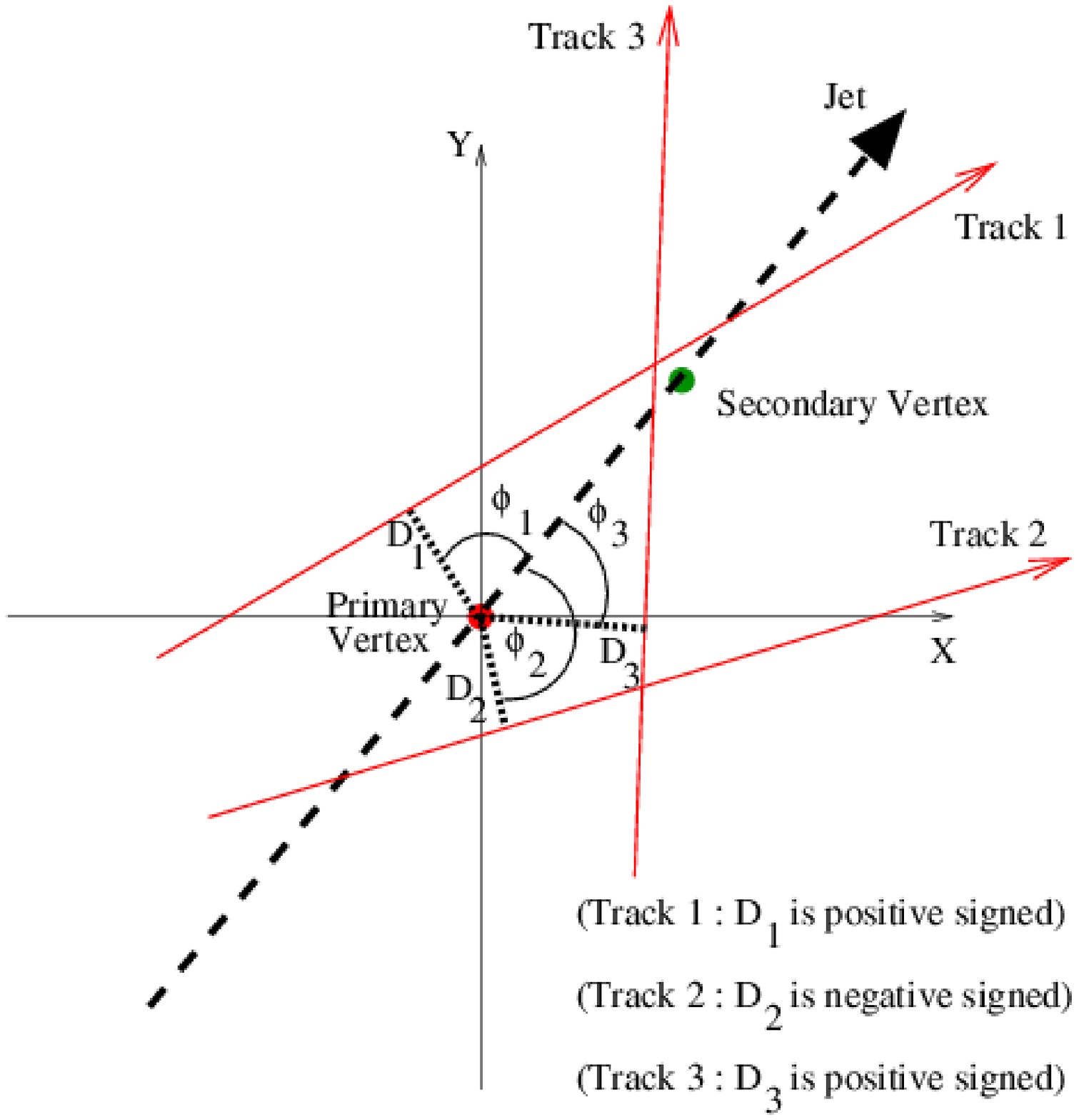}
\caption[Track impact parameter used by the \jetprob~$b$-tagging algorithm]{Tracks from a primary vertex (left) and from a secondary vertex (right). Illustration of the track impact parameter definition that is used by the \jetprob~$b$-tagging algorithm. Credit image to the CDF collaboration. \label{figure:ImpactParameterDefinition}}
\end{center}
\end{figure}

\subsection{Jet Probability Algorithm}

\ \\Another jet by jet basis $b$-tagging algorithm employed in this analysis is Jet Probability (\jetprob) \cite{JetProb1}~\cite{JetProb2}. \jetprob~looks at the distribution of impact parameters for the tracks reconstructed in the cone of the jet to estimate a probability that the ensemble of all these tracks is consistent with originating from the primary interaction vertex.  

\ \\The impact parameter of a track is considered positive (negative) if the angle between the track and the jet it belongs to is smaller (larger) than 90 degrees. More precisely, an impact parameter in the $\eta$-$\phi$ plane is positive (negative) if $\cos \phi > 0$ ($\cos \phi < 0$), where $\phi$ is the angle between the direction of the jet and the distance of closest approach between the track and the primary vertex, as seen in Figure~\ref{figure:ImpactParameterDefinition}. 

\begin{figure}[h]
  \begin{center}
 \includegraphics[width=6.5cm]{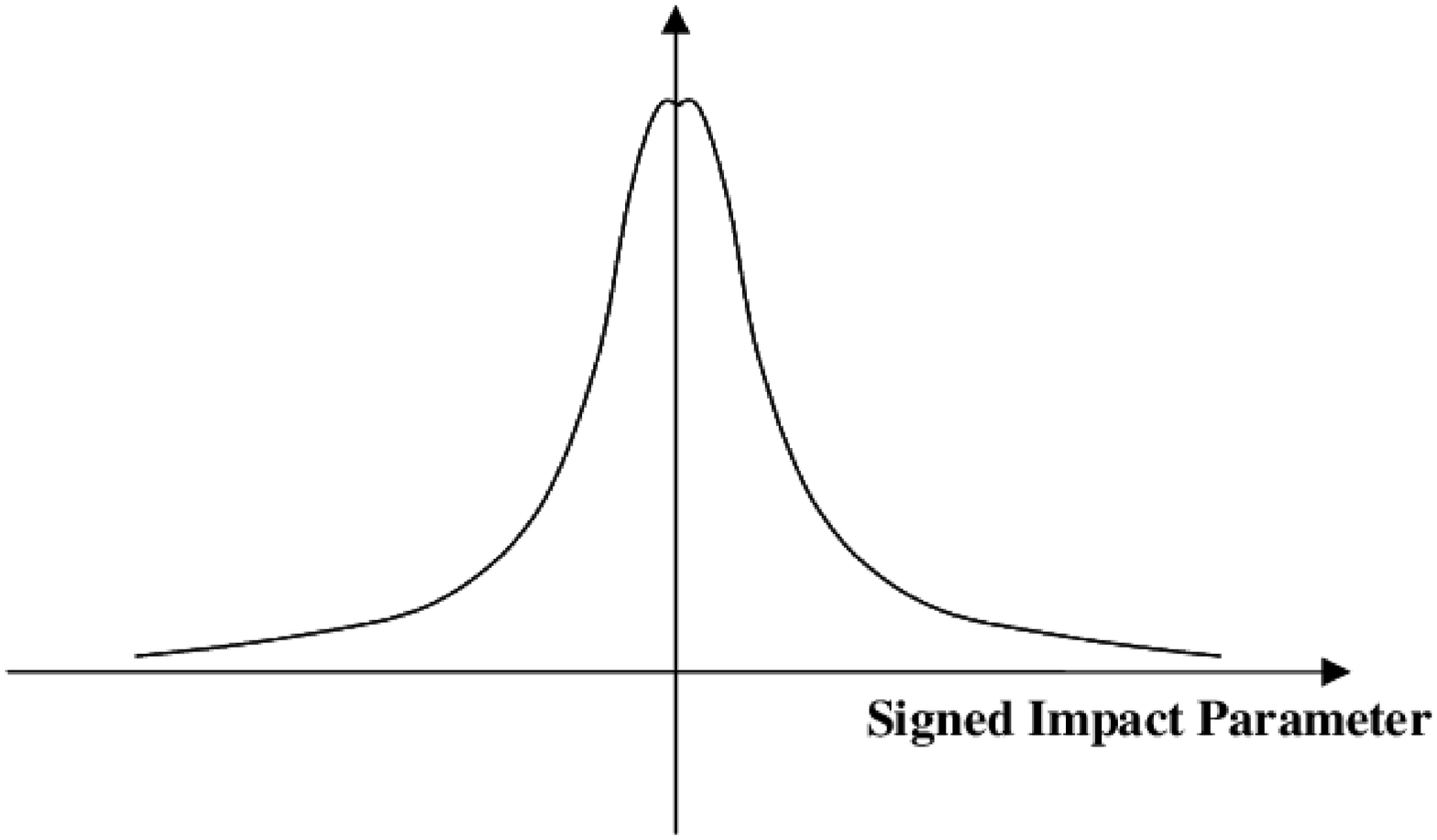}
 \includegraphics[width=6.5cm]{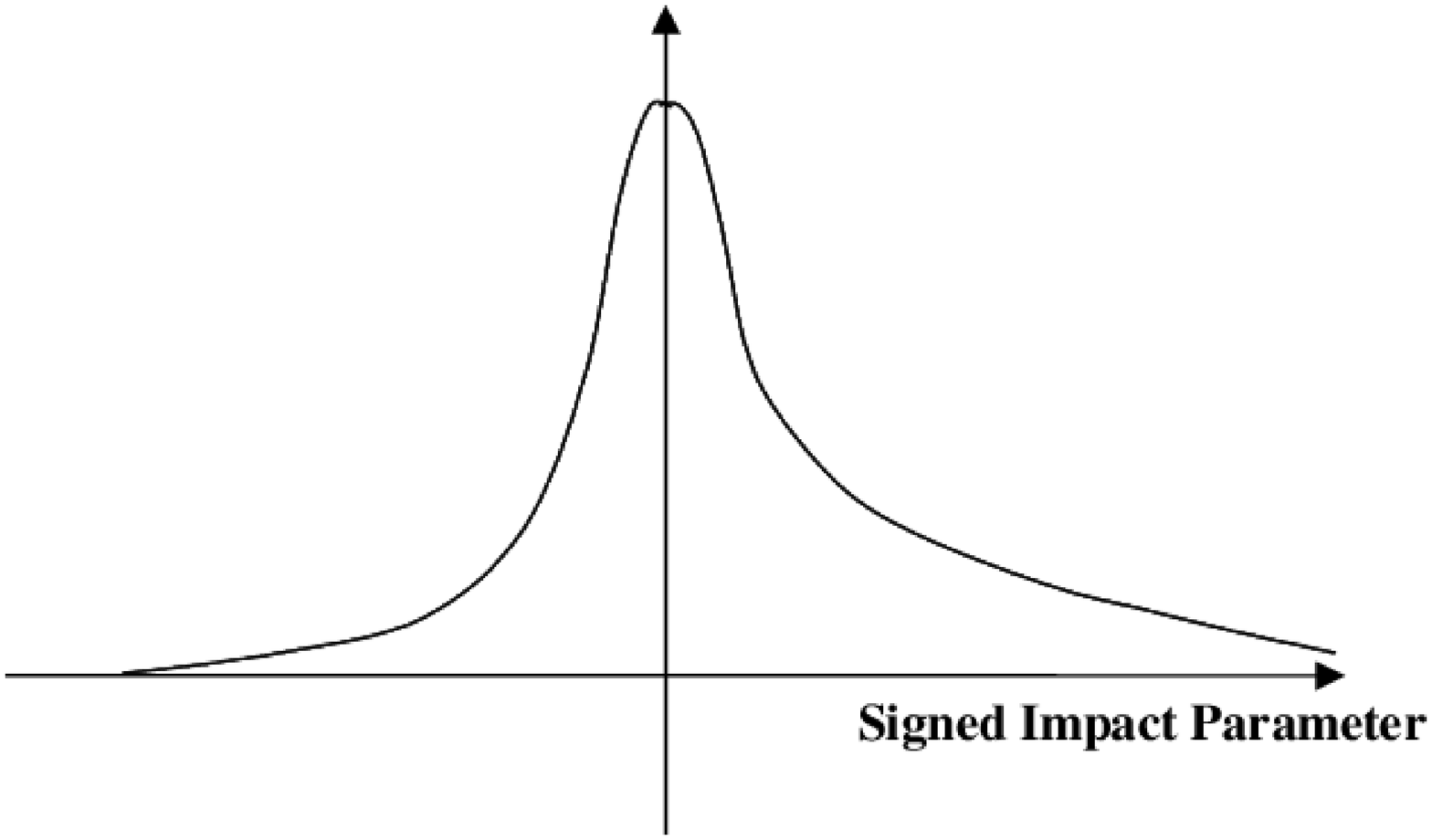}
\caption[Signed track impact parameter distribution]{Tracks from  primary vertex (left) and from a secondary vertex (right). Illustration of signed track impact parameter distribution that is used by the \jetprob~$b$-tagging algorithm. Credit image to the CDF collaboration. \label{figure:ImpactParameterDistribution}}
\end{center}
\end{figure}

\ \\Jets originating from light quarks ($u$, $d$, $s$) and gluons typically decay very close to the primary interaction vertex and their tracks should appear to emerge from the primary interaction vertex. However, due to finite tracking resolution, these reconstructed tracks have non zero impact parameter values, as seen in the top of Figure~\ref{figure:ImpactParameterDefinition}. A track has equal chances to have a positive or negative impact parameter. Therefore, the distribution of impact parameter values for tracks from light flavour jets is symmetric around zero, as seen in the top of Figure~\ref{figure:ImpactParameterDistribution}. 

\begin{figure}[h]
\begin{center}
\includegraphics[width=6.5cm]{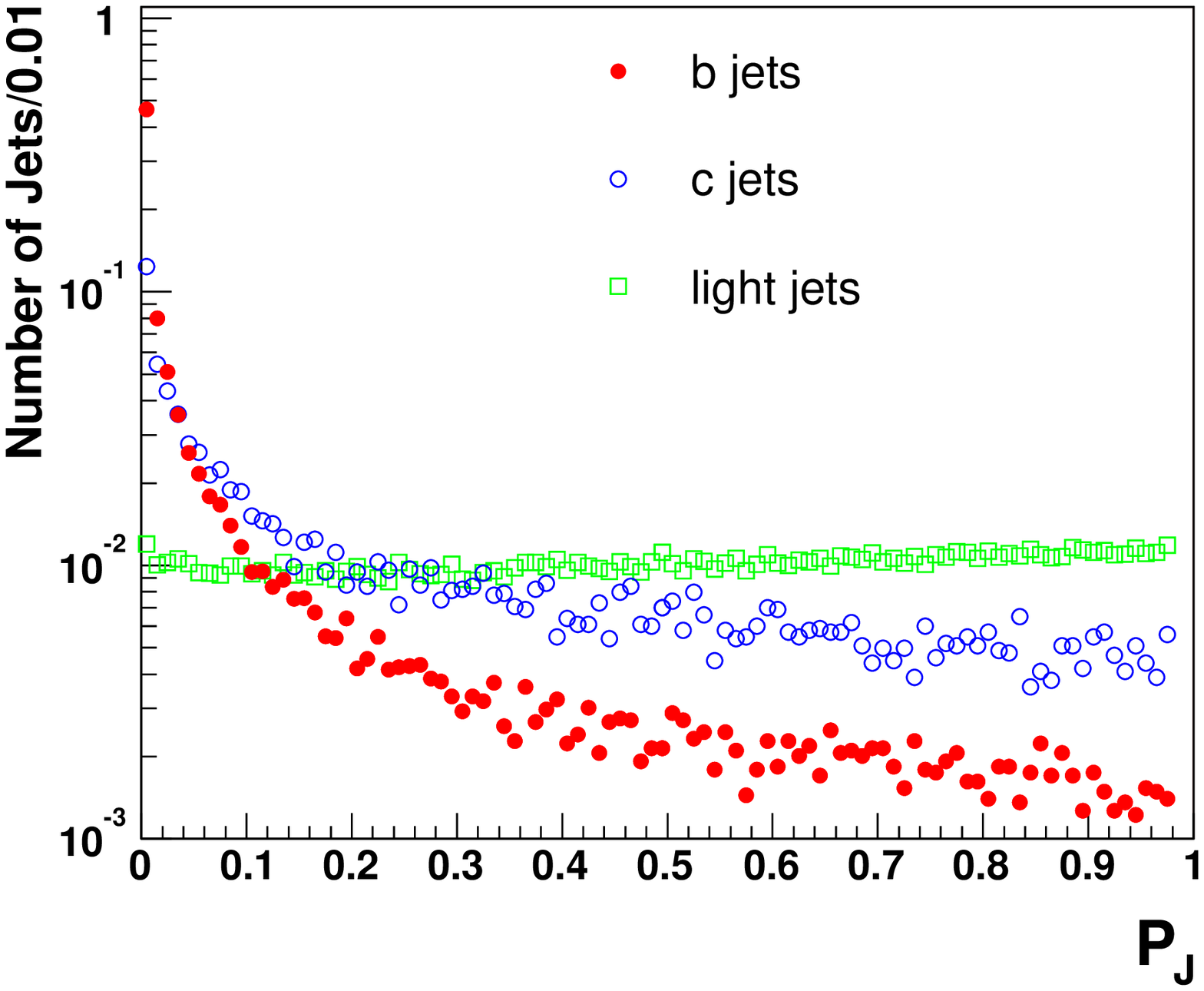}
\includegraphics[width=6.5cm]{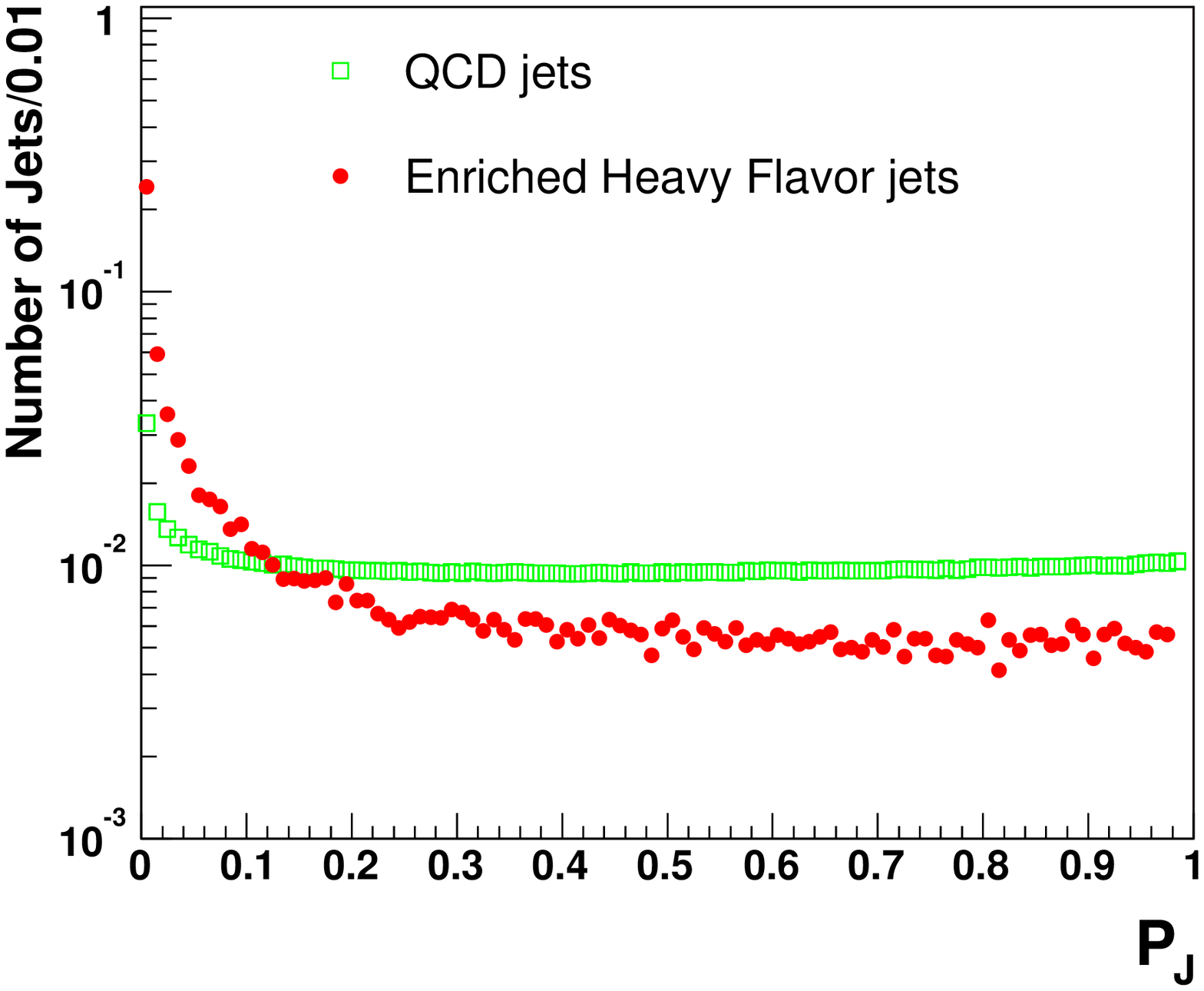}
\caption[Jet Probability $b$-tagger distributions for simulated and data events]{At left, Jet Probability $b$-tagger distributions from Monte Carlo Simulated events for jets originating from $b$ quarks (full red circles), $c$ quarks (empty blue circles) and light quarks or gluons (empty green squares). At right, Jet Probability $b$-tagger distributions from data events: inclusive electron data enriched in heavy flavour jets (full red circles) and generic QCD events selected with a trigger requiring jets with energy larger than 50 GeV (empty green squares). Credit image to the CDF collaboration. \label{figure:JetProbDistributions}}
\end{center}
\end{figure}

\ \\However, jets originating from heavy quarks ($c$ and $b$) will have tracks that emerge from a secondary displaced vertex, which translates to impact parameter values that are on average positive and larger in absolute values, as seen in the right hand side of Figure~\ref{figure:ImpactParameterDefinition}. Therefore, the distribution of impact parameter values for tracks from heavy flavour jets is elongated toward positive values, as seen in the bottom of Figure ~\ref{figure:ImpactParameterDistribution}.

\ \\The \jetprob~algorithm uses only good quality tracks that have $\pt > 0.5 \gevc$, $|d_0| < 0.1\,\rm{cm}$, more than 3 hits in the silicon detector, more than 20 COT axial segment hits and more than 17 COT stereo segment hits, as well as less than 5 cm in the $z$ coordinate between the track and the primary interaction vertex.

\ \\The first step is to quantify a probability for every track with a positive impact parameter that it originates from the primary interaction vertex (we note $P_k$ the probability of the k$^{th}$ such track). Tracks with negative impact parameters are only used to quantify the uncertainty on the impact parameters, which depends on tracking detector resolution, beam spot size and multiple scattering. The impact parameter significance $S_{d_0}$ is defined as the ratio of the impact parameter and its uncertainty

\begin{equation}
S_{d_0} = \frac{d_0}{\sigma_{d_0}}\,\rm{.} 
\end{equation}

\ \\The individual track probability is parametrized as a function of its impact parameter significance:

\begin{equation}
P_k\left(S_{d_0}\right) = \frac{\int_{-\infty}^{-|S_{d_0}|}\,R(S)dS}{\int_{-\infty}^{0}\,R(S)dS}\,\rm{.} 
\end{equation}

\ \\The next step is to quantify a probability on a jet by jet basis ($P_{\rm{jet}}$) that the jet assumed to be made up of N well reconstructed tracks with positive impact parameter is consistent with originating from a primary interaction vertex (i.e. that the jet is originating from a light flavour quark or gluon): 

\begin{equation}
P_{\text{jet}} = \Pi \sum_{k=0}^{N-1} \frac{(-\text{ln}\Pi)^k}{k!}
\end{equation}

\ \\ where

\begin{equation}
\Pi = \prod_{k=1}^{N} P_k \, \rm{.}
\end{equation}

\ \\By the definition of $P_{\rm{jet}}$ and all the things explained above, we can deduce that the distribution of $P_{\rm{jet}}$ is uniformly distributed between 0 and 1 for light flavour jets and is peaking at 0 for heavy flavour jets, as seen in Figure~\ref{figure:JetProbDistributions}. In this analysis we ask for $P_{\rm{jet}} < 0.05$ where the \jetprob~$b$-tagging efficiency is approximately 33\% and the \jetprob~$b$-tagging scale factor is $0.78 \pm 0.05$. A mistag matrix is also measured for \jetprob.

\section{Summary}

\ \\This chapter presented the reconstruction techniques for the physics objects used in this analysis. First the basic physics objects such as tracks, primary vertices and calorimeter clusters were introduced. Later, we described the reconstruction of high-level physics objects such as electrons, muons, isolated tracks and jet candidates, as well as missing transverse energy, which are used by the event selection. Finally, two algorithms used to identify jets originating from $b$ quarks were described. We also present the scale factors for charged lepton identification and $b$-tagging algorithms. All these physics objects are used to reconstruct all the event, background and data events for the $WH$ search presented in this thesis. 

\ \\Since all signals and all but one background processes are simulated using Monte Carlo generators, in the next chapter we present how an event is generated in a Monte Carlo simulation and what are the particularities of simulation for each physics process. 

\clearpage{\pagestyle{empty}\cleardoublepage}

\chapter{Monte Carlo Simulated Events \label{chapter:Simulation}}

\ \\In this chapter we describe the Monte Carlo simulated events used as signal and background. In this analysis, the signal is represented by the associated production of a Standard Model Higgs boson and a $W$ boson ($WH$ process). We also consider a small contribution to the signal represented by the associated production of the Higgs boson and a $Z$ boson, where the $Z$ boson decays to a pair of charged leptons and one of the charged leptons is not reconstructed in the detector ($ZH$ process). The following processes have identical or very similar final state signatures and therefore can mimic the signal to constitute background processes for this analysis: $W$+jets, top quark pair ($t\bar{t}$), single top, $Z$+jets, diboson and non-W (QCD) production. We use Monte Carlo generators to simulate signal and background processes, except for the non-W (QCD) background, which is estimated using a data sample. 

\section{Monte Carlo Simulation}

\ \\Simulation is needed to predict the distribution of signal and background events in the data event sample, as well as signal efficiencies and scale factors between data and Monte Carlo simulated events. 

\ \\A typical particle physics event has the following steps from initial $p\pbar$ interaction up to detection of final state particles in CDF:
\begin{itemize}
\item Parton Distribution Functions. In a $p\pbar$ collision there is actually a parton-parton collision that takes place, whereas the other partons are spectator partons. Parton is a generic term used for a constituent of a proton or neutron and represents physically either a quark or a gluon\footnote{Historically speaking, first came evidence of a structure in protons and neutrons and only decades later came the confirmation of quark and gluon existence.}. A parton has a certain probability to carry a certain fraction of the total momentum of the proton or antiproton. This probability is called a parton distribution function. They are measured in particle physics experiments and then are used as inputs for theory or Monte Carlo simulated events. 
\item As a proton and antiproton approach each other at momenta of $0.98\tevc$, a shower of partons appears from each parton in the proton and antiproton. This is called generation of initial state partons.
\item As two of these partons collide, they transfer momentum to each other and they can even change flavour. This process is called the hard scattering event. 
\item Just as the incoming partons were branching, the final state partons may also branch. This produces the final state partons.
\item Because the strong force described by QCD, which mediates the interaction between partons, does not allow the existence of neutral coloured particles, an outgoing parton will transform part of its energy to produce the mass of new partons that together form neutral hadrons travelling in the same direction as the initial parton. This process is called fragmentation.
\item Most hadrons produced are unstable and decay to other particles, thus producing the final state particles that deposit energy in the detector.
\end{itemize}

\ \\In following subsections we will describe the various tools used for Monte Carlo simulated events, both for signal and background processes: event generators, parton shower and hadronization generators, detector simulation.

\subsection{Event Generators}

\ \\ The signal Monte Carlo simulated events are generated with \pythia~v6.2~\cite{PYTHIA}, a general-purpose event generator. The hard parton scattering processes are computed at leading order matrix elements. \pythia~uses the parton distribution functions (PDF) provided by CTEQ5L~\cite{CTEQ}. In this manner, a full particle physics event is generated, with parton shower and hadronization included.

\ \\The $W$+jets and $Z$+jets background events are simulated using \alpgen~\cite{ALPGEN}. It is an event generator specialized in electroweak bosons ($W$ and $Z$ bosons) produced in association with a desired number of jets coming from either quarks or gluons. \pythia~is also used to generate simulated events for the following background processes: diboson ($WW$, $WZ$ and $ZZ$) production and top quark pair ($t\tbar$) production. The single-top background events are simulated using \madevent~\cite{MADEVENT} as event generator. As it produces events at parton level, it is sensitive to the top quark polarization that plays a role in the distribution of kinematic quantities in the events.  The top quark mass is assumed to be 172.5$\gevcc$ when modelling $t\bar{t}$ and single top production. The pure QCD (non-W) background events are not simulated at all. Instead, the contribution of this background process is measured directly from data events.

\subsection{Parton Showering and Hadronization}

\ \\Irrespective of the event generator used, all simulated events use \pythia~to model the parton showering, gluon radiation and then hadronization. The parton showering process allows for initial and final state gluon radiation. These gluons then decay to quark pairs and thus increase the number of jets detected in CDF. Beside the initial hard scattering events, more particles may be detected in CDF due to effects of multiple interactions and beam remnants. Once these particles are produced, they are all passed to the hadronization stage of the simulation. In the case of $p\pbar$ collisions at CDF, the hadronization takes place at small $Q^2$ and large $\alpha_s$ and therefore perturbation theory calculations cannot be used. Instead, phenomenological models that depend on the Monte Carlo generator are employed.

\ \\The table~\ref{table:MonteCarloGenerators} summarizes the event generators used for different background and signal processes used in this analysis.

\begin{center}
\begin{table}[h] % [t] puts at top of page
\begin{center}
\begin{tabular}{|c|c|c|}
\hline \hline
Process & Event Generator & Parton Showering\\
\hline
$WH$ & \pythia & \pythia \\
$ZH$ & \pythia & \pythia \\
$W$+jets & \alpgen & \pythia \\
$Z$+jets & \alpgen & \pythia \\
Diboson & \pythia & \pythia \\
$t\tbar$ & \pythia & \pythia \\
Single top & \madevent & \pythia \\
non-W (QCD) & data & data \\
\hline \hline
\end{tabular}
\caption[Monte Carlo generators used for signal and background event simulation]{Monte Carlo event generators and parton showering software programs used for Monte Carlo simulated events for signal and background processes used in this analysis.}
\label{table:MonteCarloGenerators}
\end{center}
\end{table}
\end{center}

\subsection{Detector Simulation}

\ \\Once the final state particles are generated, their propagation through the detector (their interaction with matter in the detector) is simulated using \geant~\cite{GEANT}. Interaction with the silicon detectors is simulated using an unrestricted Landau distribution and a simple ionizing particle path length geometrical model. Interaction with the tracking chamber system is simulated using \garfield~\cite{GARFIELD}, whose parameters are tuned to match CDF COT data~\cite{COT}. Interaction with the calorimeter system uses the \gflash~\cite{GFLASH} package. No special parametrization is used for interaction with the muon system. A detailed description of the CDF simulation can be seen in Reference~\cite{CDFSimulation}.

\subsection{Monte Carlo Validation}

\ \\Since the number of events due to background processes is estimated using Monte Carlo simulated events, it is essential to validate that the Monte Carlo simulation models the data correctly. This is why we compare background estimation and data distributions for all the kinematic quantities used in this analysis, but also for other quantities as well. 

\begin{figure}[h]
\begin{center}
\includegraphics[width=10.0cm]{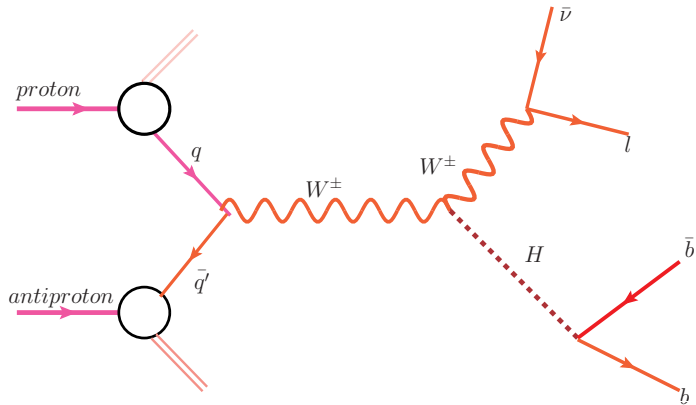}
\caption[Feynman diagram for the $WH$ associated production and decay]
    {Leading order Feynman diagram for the $WH$ associated production and decay, the signal process of our search. Credit image to the CDF collaboration.
      \label{figure:FeynmanWH}}
\end{center}
\end{figure}

\section{Signal Samples}

\ \\In this analysis the signal process is the associated production of a $W$ boson and a Higgs boson, where a $W$ boson decays leptonically to a charged lepton and a neutrino and the Higgs boson decays to a pair of bottom-antibottom quarks. The leading-order Feynman diagram of the $WH \to l\nu b\bbar$ process is presented in Figure~\ref{figure:FeynmanWH}.

\ \\The signature of this process consists of a reconstructed charged lepton candidate (an electron or a muon directly and a tau candidate indirectly through its decay to an electron or a muon), missing transverse energy (due to the fact that the neutrino escapes the detector without being detected) and two jets (due to the bottom and antibottom quarks). This final state signature is often called "lepton + jets", but it would be more correct to call it ``charged lepton + missing transverse energy + jets''. 

\ \\Since the same signature is obtained also for the associated production of a Standard Model Higgs boson and a $Z$ boson, where the $Z$ boson decays to a pair of charged leptons and one of them is not reconstructed in the detector, we also consider this small contribution to the signal of the analysis (the $ZH$ channel). 

\ \\Since the mass is unknown for the Standard Model Higgs boson, we have produced several $WH$ and $ZH$ samples assuming several values of Higgs boson masses that start at $100 \gevcc$, end at $150 \gevcc$ and increment in steps of $5 \gevcc$. We have chosen this mass range because we are looking for a Higgs boson in the "low mass" region. The Monte Carlo event generator is \pythia, which treats the Higgs boson as a resonance. The production cross section and decay branching fraction depend on the assumed Higgs boson mass and include all the latest higher order QCD and electroweak corrections~\cite{CrossSectionBranchingRatio}, as illustrated in Table~\ref{table:SignalCrossSectionsAndBranchingFraction}.

\begin{table}[h] % [t] puts at top of page
\begin{center}
\begin{tabular}{|c|c|c|c|}
\hline \hline
$M(H)$ & $\sigma(p\bar{p} \rightarrow W{^\pm}H)$ & $\sigma(p\bar{p} \rightarrow ZH)$ & $\branchingratio(H \rightarrow b\bar{b})\
$ \\
\hline
100 $\gevcc$ &  0.2919 pb & 0.1698 pb & 0.8033\\
105 $\gevcc$ &  0.2484 pb & 0.1459 pb & 0.7857\\
110 $\gevcc$ &  0.2120 pb & 0.1257 pb & 0.7590\\
115 $\gevcc$ &  0.1819 pb & 0.1089 pb & 0.7195\\
120 $\gevcc$ &  0.1564 pb & 0.0944 pb & 0.6649\\
125 $\gevcc$ &  0.1351 pb & 0.0823 pb & 0.5948\\
130 $\gevcc$ &  0.1169 pb & 0.0719 pb & 0.5118\\
135 $\gevcc$ &  0.1015 pb & 0.0630 pb & 0.4215\\
140 $\gevcc$ &  0.0883 pb & 0.0553 pb & 0.3304\\
145 $\gevcc$ &  0.0770 pb & 0.0487 pb & 0.2445\\
150 $\gevcc$ &  0.0673 pb & 0.0429 pb & 0.1671\\
\hline \hline	
\end{tabular}
\caption[Signal cross section and branching ratio values for each Higgs boson mass]{Cross section values for the $WH$ and $ZH$ production in proton-antiproton collisions at the centre-of-mass of $\sqrt{s}=1.98\ \tev$ at the Tevatron accelerator, in units of picobarns (pb), as well as branching ratio of the Higgs boson decay to bottom quark pairs, as a function of the Higgs boson mass~\cite{CrossSectionBranchingRatio}.}
\label{table:SignalCrossSectionsAndBranchingFraction}
\end{center}
\end{table}

\ \\The Monte Carlo samples model a different number of primary interaction vertices per event in order to simulate the relatively low, medium and high instantaneous luminosity that the Tevatron accelerator provided during its current Run II. 

\section{Background Samples}

\ \\In general there are two types of background: reducible and irreducible. The reducible backgrounds can be reduced given enough data sets and/or improved analysis techniques. They typically have different signatures than the signal processes or they have the same signature but are made up of one or more objects incorrectly reconstructed (fake objects)\footnote{As examples of fake object reconstruction we quote a real electron being reconstructed as a jet and vice versa, or mismeasured jet, photon and electron energies that appear to produce missing transverse energy in the event.} or by one particle not being reconstructed at all. The irreducible backgrounds have the same final states and therefore, even if all high level objects are correctly reconstructed, those backgrounds will not go away. All one can do is measure them correctly.

\ \\In this analysis we use data or Monte Carlo simulated events to measure the contribution of the background processes to our sample. Each of them will be described in more detail below.

\begin{figure}[h]
\begin{center}
\includegraphics[width=13cm]{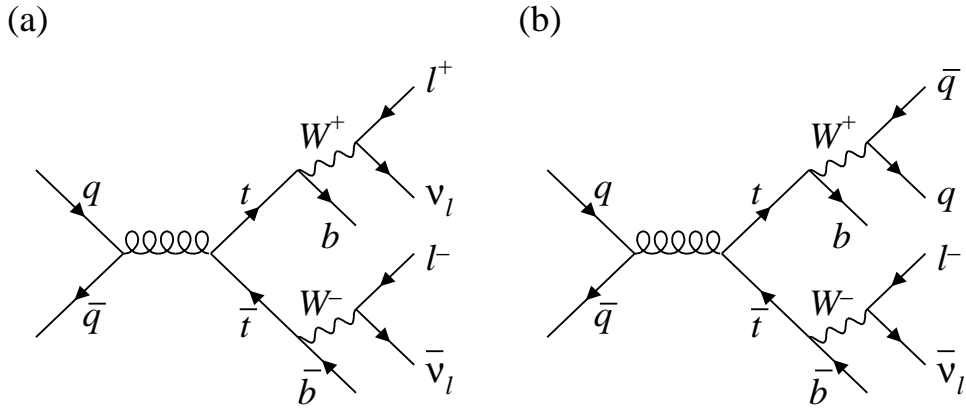}
\caption[Feynman diagram for the top quark pair background]
    {Leading order Feynman diagrams for the top quark pair production and decay \cite{SingleTopObservationPRDCDF}, reducible background processes for our $WH$ search, in the case the reconstruction misses a charged lepton in the process on the left or two jets in the process on the right.
      \label{figure:FeynmanTopPair}}
\end{center}
\end{figure}

\subsection{Top Quark Pair Production}

\ \\The first reducible background is due to top quark pair production ($t\tbar$). A top quark decays almost 100\% of the time to a $W$ boson and a bottom quark. For this analysis, the background process happens when one $W$ boson decays leptonically and some of the second $W$ boson decay daughter particles products are not reconstructed properly. Since in principle better detector and analysis techniques can reduce the fraction of events reconstructed incorrectly, this background is reducible. The leading order Feynman diagrams for the $t\tbar$ production and decay are illustrated in Figure~\ref{figure:FeynmanTopPair}, where the process mimics our $WH$ signal if one of the charged leptons (process in the left diagram), or two of the jets (process in the right diagram) are not reconstructed.

%\ \\If the second $W$ boson is reconstructed through its hadronic decay, the $t\tbar$ signature is still "lepton+missing transverse energy+jets", but the number of jets here is four, whereas the number of jets for our signal is two. Even so, not all jets are reconstructed and therefore the $t\tbar$ events appear not only in the four-jet bin, but also in the three-jet and two-jet bins, as seen in Figure~\ref{figure:TopPairJetBins}. In this analysis we use only the two-jet bin. The two-jet bin $t\tbar$ process where one $W$ boson decays leptonically and another one either is not detected or decays hadronically represents a reducible background, since in principle with more data and better analysis techniques the $W$ boson should be reconstructed completely.

%\begin{figure}[h]
%\begin{center}
%\includegraphics[width=7.0cm]{./simulation/TopPairJetBins.eps}
%\caption[Top-quark-pair event prediction as a function of jet multiplicity]
%    {Top quark pair production reconstructed event yield as a function of reconstructed jet multiplicity. A considerable fraction of these events appear in the two-jet bin and therefore constitute a reducible background for our $WH$ search. Credit image to the CDF collaboration.
%      \label{figure:TopPairJetBins}}
%\end{center}
%\end{figure}

\subsection{$Z$ Boson + Jets Production}

\ \\The second reducible background is the associated production of a $Z$ boson and a gluon, where the $Z$ boson decays to two charged leptons (and one is not reconstructed at all) and the gluon decays to a $b\bbar$ pair (and the mismeasured jet energies produce a fake missing transverse energy). The leading order Feynman diagram for this process is shown in the Figure~\ref{figure:FeynmanZjetsQCD} (a).

\begin{figure}[h]
\begin{center}
\includegraphics[width=13cm]{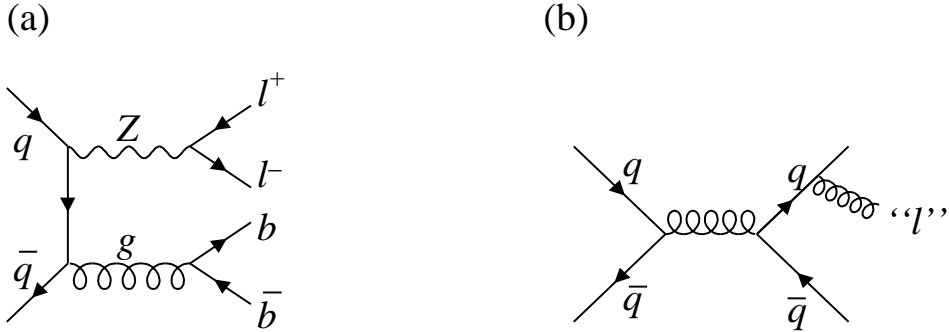}
\caption[Feynman diagram for the $Z$+jets background]
    {Leading order Feynman diagrams for the $Z$+jets reducible and non-W (QCD) processes production and decay, reducible backgrounds for the $WH$ search \cite{SingleTopObservationPRDCDF}. The $Z$+jets process can mimic our signal if a charged lepton is not reconstructed and jet energies are mismeasured. The non-W (QCD) jet production can mimic our signal if one jet is incorrectly reconstructed as a charged lepton and if jet energies are mismeasured. \label{figure:FeynmanZjetsQCD}}
\end{center}
\end{figure}

\subsection{Non-W (QCD) Multi-jet Production}

\ \\The third reducible background of this analysis is the multijet production described by the QCD theory, where one jet fakes an electron candidate and the mismeasured energy of all the jets fakes the missing transverse energy. Since semi-leptonic decays of hadrons containing $b$ or $c$ quarks produce muons, a jet could in principle also fake a muon candidate. However, we reduce most of these cases by requiring the isolation requirement on the charged lepton candidates. Thus a fake $W$ boson is reconstructed in the event, whereas the physical process contains none, as seen in the leading order of the non-W (QCD) background is illustrated in the Figure~\ref{figure:FeynmanZjetsQCD} (b). It is the only background for which the event yield and kinematic shapes are estimated using a data sample, and not one of Monte Carlo simulated events. As we analyze an increasingly larger integrated luminosity, we gain a better understanding of the process where jets fake real electrons and the jet energy resolution. This is why the non-W (QCD) is a reducible background.

\begin{figure}[h]
\begin{center}
\includegraphics[width=13cm]{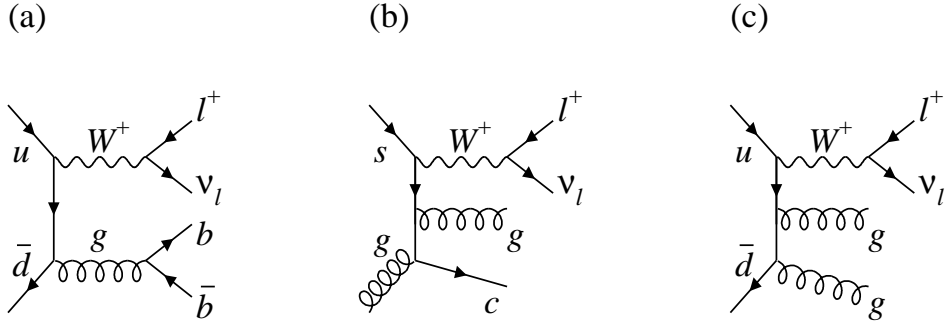}
\caption[Feynman diagrams for the $W$ + jets backgrounds]
    {Leading order Feynman diagrams for several $W$ + jets associated production and decay processes \cite{SingleTopObservationPRDCDF}. Diagram (a) represents the irreducible Wbb background, diagram (b) represents the reducible Wcc background and diagram (c) represents the reducible Wlf background. \label{figure:FeynmanWjets}}
\end{center}
\end{figure}

\subsection{$W$ Boson + Jets Production}

\ \\There are several $W$+jets processes. Just as in the case of the signal, these processes present a real $W$ bosons and two jets. The jets may originate from light flavour partons (up, down, strange quarks and any type of gluons) or from heavy flavour partons (charm and bottom quarks). For the signal process, both jets originate from bottom quarks. After we employ algorithms to identify jets that originate from bottom quarks ($b$-tagging algorithms), the signal over background ratio increases. 

\ \\However, these algorithms are not perfect. On one side, light flavour jets may be incorrectly identified as heavy flavour jets and thus become a background for our signal, which we denote $W$ + Light Flavour jets, or shortly, W+LF or Mistags. Such processes are shown in Figure \ref{figure:FeynmanWjets} (c) where two jets are produced from two gluons and also in Figure \ref{figure:FeynmanWjets} (a) if we replace the two bottom quarks with two light quarks. On another side, these algorithms also identify charm quarks as bottom quarks. In other words, although they are denoted $b$-tagging algorithms, in reality they tag heavy flavour quarks, with about half tagged jets originating either from a bottom or charm quark. Such a process is presented in Figure \ref{figure:FeynmanWjets} (b) and is called $W$ + cj. The process of Figure \ref{figure:FeynmanWjets} (a) where we replace the two bottom quarks with two charm quarks is called Wcc. In this thesis we add presented these processed together under the common name of Wcc. 

\ \\Since the $b$-tagging algorithms may be improved if more time is available and calibrated better in Monte Carlo and data simulated events as we collected a data sample that corresponds to a larger integrated luminosity, these W+LF and Wcc processes are reducible backgrounds for our $WH$ signal.

\ \\The first irreducible background is the associated production of a $W$ boson and two jets that originate from bottom quarks, as seen in \ref{figure:FeynmanWjets} (a). It is called irreducible since even if we had perfect reconstruction algorithms, the process has exactly the same final state as our signal. In order to reduce an irreducible background event prediction, one has to look at subtle kinematic differences between the two processes, as just reconstructing correctly the final state is not enough any more. We call this process Wbb background. We also use the denomination of $W$ + Heavy Flavour, W+HF, for the sum of Wcc and Wbb. 

\ \\For each process, the $W$ boson decay is produced leptonically to either electron, muon or tau leptons plus neutrinos. Also, each of these processes is simulated with various numbers of extra generic partons (W+HF, namely $W+bb+0p$, $W+bb+1p$,$W+bb+2p$). All the subsamples have to be added in order to obtain the generic W+HF background event yield. Some extra jets can be produced by \pythia~to account for the initial state radiation (ISR) and final state radiation (FSR). In order not to count incorrectly the number of jets in these events, we use the MLM prescription~\cite{ALPGEN}.

\ \\Each of these samples is simulated using \alpgen~for matrix element generation and \pythia~for parton showering. After a jet-based flavour overlap removal algorithm \cite{HeavyFlavourRemoval} is applied to both W+LF and W+HF samples, they are all added together, weighting each sample by its production cross section.

\begin{figure}[h]
\begin{center}
\includegraphics[width=13cm]{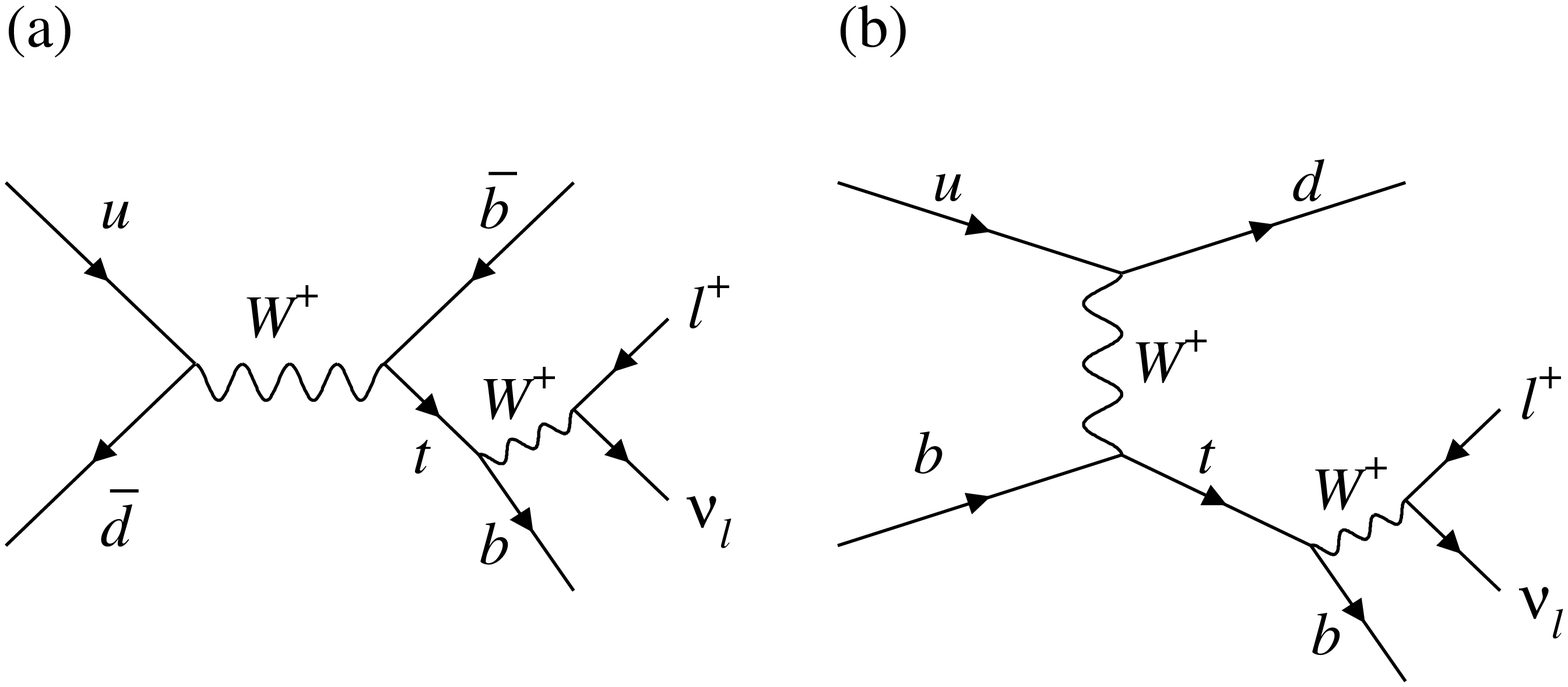}
\caption[Feynman diagrams for the single top background]
    {Feynman diagrams for the single top associated production and decay \cite{SingleTopObservationPRDCDF}, background processes for our $WH$ search. The the s-channel (a) is an irreducible background because of the presence of two bottom quarks in the final state, while the t-channel (b) is in principle a reducible background due to the presence of a generic get in the final state.
      \label{figure:FeynmanSingleTop}}
\end{center}
\end{figure}

\subsection{Single Top Quark Production}

\ \\The second irreducible background process for this analysis is the single top production, where a top quark is produced by the electroweak force in association with a bottom quark (s-channel, which has the exact final state as our $WH$ signal and therefore is an irreducible background) and in association with a bottom quark and a generic quark, where the generic quark escapes detection (t-channel, in principle a reducible background due to the presence of the generic jet). The s-channel (t-channel) leading order Feynman diagrams are illustrated in the left (right) side of Figure~\ref{figure:FeynmanSingleTop}. The single top production has been observed (discovered) experimentally at CDF \cite{SingleTopObservationPRDCDF}. 

\ \\Single top events are generated with \madevent~and the parton showering is done with \pythia. 

\begin{figure}[h]
\begin{center}
\includegraphics[width=13cm]{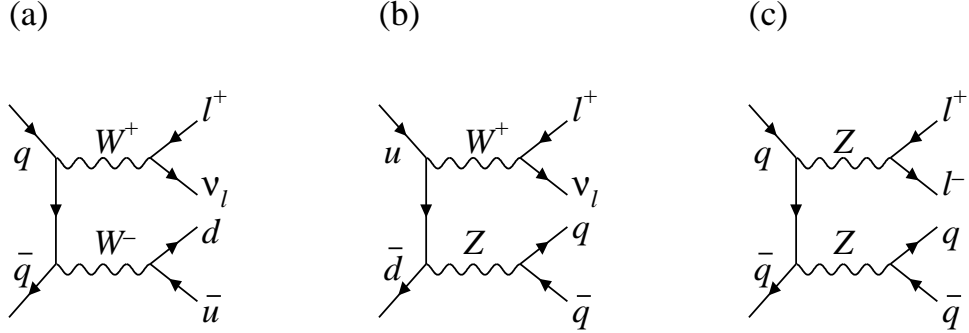}
\caption[Feynman diagrams for the electroweak diboson backgrounds]
    {Feynman diagrams for the electroweak diboson production and decay \cite{SingleTopObservationPRDCDF}. The $WW$ (a) and $WZ$ (b) are irreducible backgrounds, whereas $ZZ$ (c) is a reducible background processes for our $WH$ search.
      \label{figure:FeynmanDiboson}}
\end{center}
\end{figure}

\subsection{Diboson Production}

\ \\The electroweak diboson production processes are the associated production of two $W$ bosons ($WW$), two Z bosons ($ZZ$) or the associated production of a $W$ boson and a $Z$ boson ($WZ$). The $WW$ and $WZ$ processes constitute an irreducible background when one $W$ boson decays leptonically and the second $W$ boson or the $Z$ boson decay hadronically to two quarks. The $ZZ$ process constitutes a reducible background when one $Z$ boson decays leptonically to a pair of charged leptons, one of which is not reconstructed at all, and the second $Z$ boson decays hadronically. For these processes, both the matrix elements and the parton showering are simulated using \pythia. The Feynman diagrams of these processes can be seen in Figure~\ref{figure:FeynmanDiboson}. The non-resonant production of diboson processes predicted by the Standard Model have been observed at the Tevatron~\cite{CDFWZObservation}~\cite{CDFDibosonNonResonantMeasurement1}~\cite{CDFDibosonNonResonantMeasurement2}. 

\section{Summary}

\ \\We started this chapter by describing a three-step methodology to simulate $p\bar{p}$ collisions that produce elementary particles that are recorded by the CDF detector. The first step is to model the $p\bar{p}$ interaction using Monte Carlo event generators, such as \pythia, \alpgen~and \madevent. If quarks are produced, they hadronize and form a shower, which is modelled by \pythia. Finally, the propagation of such showers and other particles through the CDF detector is modelled using \geant. 

\ \\In the second part of the chapter, we described the signal processes for the $WH$ search. The main signal process is the $WH$ associated production, where the $W$ boson decays leptonically and the Higgs boson decays to a $b\bar{b}$ pair. The second signal process is the $ZH$ associated production, where the Higgs boson decays to a $b\bar{b}$ pair and the $Z$ boson decays to two charged leptons, but one of them is not reconstructed in the CDF detector. The second process is only a small contribution to the main $WH$ process. We presented the Feynman diagrams for the signal processes, as well as their cross sections and branching ratios for the Higgs boson masses studies in this analysis. 

\ \\We continued by presenting all the main background processes to our $WH$ search. We divided the backgrounds into two main categories. The reducible backgrounds (top-quark-pair production, $Z$-boson-plus-jets production, Non-$W$-QCD-multi-jet production) are those processes that can be separated from the signal in principle, with datasets with large integrated luminosity and with better analysis techniques. The irreducible background processes ($W$-boson-plus-jets production, diboson production, single-top-quark production) are those processes that cannot be separated further from the signal even with larger datasets and improved analysis techniques because they contain the same final state as the signal. We also presented the Feynman diagrams for the background processes.

\ \\In the next chapter we will present the event selection, both online (at the trigger level) and offline (at analysis level) to select from the many $p\bar{p}$ collisions stored by the CDF those that have the signature presented in this chapter, namely one charged lepton, missing transverse energy and two jets that originate from bottom quarks. 

\clearpage{\pagestyle{empty}\cleardoublepage}

\chapter{Event Selection\label{chapter:Selection}}

\ \\In this analysis we select candidate events consistent with the signature of one charged lepton (electron or muon), missing transverse energy and two tight jets (at least one required to be $b$-tagged by the \secvtx~algorithm). We analyze a data sample that corresponds to an integrated luminosity of $5.7 \pm 0.3\ \invfb$.

\ \\In this chapter we will describe the online and offline event selection. The online selection is performed by the three levels of the trigger system that selects, every second, about 100 bunch crossings out of the about 2 million bunch crossings that take place per second. The total information recorded by the CDF detector during a bunch crossing is called an event. These selected events are saved on magnetic tape and later are analyzed in detail in the offline analysis. 

\section{Online Event Selection}

\ \\Events for the tight lepton categories are selected using the charged lepton information. Events with CEM charged leptons are required to pass the high transverse momentum CEM trigger. Events with CMUP charged leptons are required to pass the trigger that asks for ionization both in the CMU and CMP detector. Events with CMX charged leptons have to pass the trigger requirements of CMX ionization deposits. 

\ \\One of our contributions to this analysis is the addition of a loose charged lepton category that uses at trigger level the orthogonal information to the charged lepton, namely the missing transverse energy and jets. We use three missing transverse energy + jets triggers to select isolated track events, thereby increasing the signal acceptance. 

\ \\Each trigger has three levels of selections, each more stringent than the previous. The details of selection for all these triggers at all the trigger levels are presented in detail below. 

\subsection{CEM Trigger\label{section:CEMTrigger}}

\ \\The trigger used for selecting tight central electron candidate events (CEM) is called ``ELECTRON\_CENTRAL\_18''. At L1, it requires a track with $\pt > 8 \gevc$, a calorimeter tower with $\et > 8 \gev$ and the ratio between the energy deposited in the hadronic calorimeter to that in the electromagnetic calorimeter $E_{HAD}/E_{EM} < 0.125$. At L2, it requires a calorimeter cluster with $\et > 16 \gev$ matched to a track of $\pt > 8 \gevc$. At L3, it requires an electron candidate with $\et > 18 \gev$ matched to a track of $\pt > 9 \gevc$. We note that as we move up the trigger levels, object reconstruction becomes more sophisticated and the selection cuts are tighter\footnote{We see that only towers and tracks are required at L1, but clusters reconstructed around towers and matching between clusters and tracks are asked for at L2 and full electron reconstruction is performed at L3. We also note that the cut values increase as we move from L1 to L3.}. This is a general feature of triggers.

\ \\We use a data sample of $W$ boson events decaying leptonically to an electron and a neutrino to measure the efficiency of this trigger. The efficiency of a trigger is the percentage of signal Monte Carlo simulated events that meet all of the trigger requirements. We find an average efficiency of $0.961\pm 0.001$ for the CEM trigger.

\subsection{CMUP Trigger\label{section:CMUPTrigger}}

\ \\The trigger used for selecting tight central muon candidate events (CMUP) has the name of ``MUON\_CMUP18''. At L1, it requires a track with $\pt > 4 \gevc$ that is matched to ionization in both the CMU and CMP muon chamber that is consistent a muon candidate with $\pt > 6 \gevc$. At L2, it requires a track with $\pt > 15 \gevc$ and that the calorimeter cluster in the direction of the track and muon stubs have ionization deposits consistent with a minimum ionizing particle. At L3, it requires a fully reconstructed COT track with $\pt > 18 \gevc$ that, if extrapolated to the CMU (CMP) detector, matches hits in this detector within $\Delta x_{\rm{CMU}}< 10\ \rm{cm}$ ($\Delta x_{\rm{CMP}}< 20\ \rm{cm}$).

\ \\We measure the efficiency of the CMUP trigger by collecting a data sample of $Z$ bosons that decay to a muon-antimuon pair, where one muon passes the trigger requirements and the second muon is checked if it passes the trigger selection or not. We find a CMUP trigger efficiency of $0.877 \pm 0.002$.

\subsection{CMX Trigger\label{section:CMXTrigger}}

\ \\The trigger used for selecting another category of tight central muon candidate events, more forward than the CMUP muons, is called ``MUON\_CMX18\_DPS''. At L1, it requires a track with $\pt > 8 \gevc$ that is matched to ionization in the CMX muon chamber. At L2 and L3, the criteria are identical to that of CMUP trigger, only that they refer to the CMX detector. 

\ \\In a similar way to CMUP trigger, we measure a trigger efficiency of $0.902 \pm 0.002$ for the CMX trigger. 

\subsection{Triggers for ISOTRK\label{section:ISOTRKTriggers}}

\ \\Our main original contribution to this analysis is the use of three missing transverse energy + jets triggers that are used to select ISOTRK events by using information orthogonal to the charged lepton ones, namely the missing transverse energy and the jets. In order to use these three MET-based triggers, we had to parameterize the trigger efficiency turnon curve as a function of trigger MET for each of the triggers and each trigger level. We also introduced a novel method to combine the three different triggers in order to maximize the event yield and yet do not have a logical ``OR'' between the triggers in order to avoid correlations and measure easier and correctly the systematic uncertainty related to the combined trigger efficiency. Due to the long description necessary for this work, we present it in detail in Appendix \ref{chapter:METTriggers} and Appendix \ref{chapter:CombineTriggers}.

\section{Offline Event Selection}

\ \\In this section we will describe the baseline offline event selection, followed by the description of the $b$-tagging categories used and the QCD veto used for every charged lepton category.

\subsection{Baseline Event Selection}

\ \\The offline event selection makes sure selected events pass a series of criteria that help discriminate the $WH$ signal against a large background. 

\begin{itemize}
\item Events must fire the trigger specific to their charged lepton category. There is a special treatment for ISOTRK channel which, constitutes one of our original contributions to this analysis.

\item The $z$ coordinate of the primary interaction vertex must be within 60 cm of the centre of the detector, the so-called fiducial region. 

\item One and only one high transverse momentum isolated charged lepton candidate must be reconstructed in the event. We require muon (electron) candidates to have $\pt > 20 \gevc$ ($\et > 20 \gev$). We note that the trigger requirement was of  $\pt > 20 \gevc$ (($\et > 20 \gev$), but we require higher values in the offline selection to make sure that the event is in the plateau region of an possible trigger turnon.

\item The $z$ coordinate difference between the charged lepton track and the primary interaction vertex has to be smaller than 5 cm ($|z_{\rm{track}}-z_0|<5$ cm). 

\item Photon conversion events are vetoed. High energy electrons can emit photons primarily as they pass through the detector due to bremsstrahlung radiation. As these photons are also very energetic, they decay to electron-positron pairs. Events with such secondary electrons are identified and removed thanks to their two tracks with a small opening angle emerging from a vertex far away from the primary interaction vertex.

\item Cosmic ray events are vetoed, as discussed in the muon identification subsection \ref{subsection:MuonIdentification}.

\item Events are vetoed if the invariant mass between the reconstructed charged lepton and any other track with opposite charge in the event falls in the $Z$ boson mass window ($76-106 \gevcc$).

\item The event must present large missing transverse energy $\met > 20 \gev$ for all charged lepton categories. The missing transverse energy is corrected for the position of the primary interaction vertex, for jet energies and the presence of muons, as described in detail in the object reconstruction chapter.

\item The event must contain exactly two tight jets reconstructed using a JETCLU algorithm with a radius $\Delta R=0.4$. Each jet is required to have $\et > 20 \gev$ and be in the central region of the detector ($|\eta|<2.0$) \footnote{There are two other WH analyses at CDF \cite{BarbaraAlvarezPhDThesis} \cite{MartinFrankPhDThesis} that studied in addition the sample of events with exactly three tight jets. The third jet may appear from the underlying event, the initial state radiation or the final state radiation.}. 

\item Each event is required to pass some selection criteria designed to reject a significant part of the non-W (QCD) background events (QCD veto). The QCD veto is specific to every charged lepton category and will be discussed below.

\item Each event must have at least one jet that is $b$-tagged using \secvtx~algorithm in order to discriminate against events with light flavour jets. The detailed tag categories are descried below.  
\end{itemize}

\subsection{$b$-tagging Categories\label{bTaggingCategories}}

\ \\Identifying jets that originate in $b$ quarks is essential for this analysis, helping discriminate against the W+LF background. Since for our $WH$ signal events both jets originate from a $b$ quark, we ask all events to have at least one jet tagged by \secvtx. Since \secvtx~is more efficient than \jetprob~ and we want to maximize the number of events that have two tagged jets, we use the following $b$-tagging categories:

\begin{itemize}
\item \secvtx~tight + \secvtx~tight (SVTSVT): Both jets in the event are tagged with \secvtx at the tight operating point.
\item \secvtx~tight + \jetprob~5\% (SVTJP05): One jet in the event is tagged by \secvtx at the tight operating point and another one is not tagged by \secvtx, but is tagged by \jetprob~algorithm at 5\% operating point. A jet is considered $b$-tagged by \jetprob~if it has a probability of less than 5\% to emerge from a primary interaction vertex.
\item \secvtx~tight (SVTnoJP05): One jet in the event is tagged by \secvtx~at the tight operating point and another one is neither tagged by \secvtx, nor by \jetprob. 
\end{itemize}

\ \\By the construction of the $b$-tagging categories, we can see that they are orthogonal. Our analysis is divided in several channels based on the charged lepton category and the $b$-tagging categories. 

\subsection{QCD Veto\label{subsection:QCDTemplates}}

\ \\In order to evaluate this background event yield, we select events based on a non-W model specific for each charged lepton category in the analysis. These selection criteria are very close to the charged lepton ones, with one or two cuts required to fail. Therefore samples of enriched non-W events are selected, which give the shape for the non-W background for different kinematic quantities. The normalization of these shapes (the exact yield of background events) is calculated later on using a fit in data for missing transverse energy distributions, as we will see in detail in the background estimation chapter. 

\ \\For the tight central electron (CEM) channel of the analysis, the non-W model is called ``anti-electron'' and it tries to model the jets that are incorrectly reconstructed as an electron candidate. The electron trigger is required to have fired as well as passing all but two of the following kinematic quantities: $E_{HAD}/E_{EM}$, $\chi^2$, $L_{shr}$, $Q \cdot \Delta x$ and $|\Delta z|$.

\ \\For the tight forward electron (PHX) channel of the analysis, the non-W model is called ``jet-electron''. It selects jet candidate events with a transverse energy $\et > 20 \gev$, $0.80<E_{HAD}/E_{EM}<0.95$ and at least four tracks in the jet. This makes sure that no real electrons are selected in this sample.

\ \\For the central muon (CMUP and CMX) channels of the analysis, the non-W model is called ``non-isolated muons''. It selects muon candidates that pass all the selection criteria but fail the ``isolation'' one. That means that the charged lepton track is required to be surrounded by other activity in the COT, which is typical for jets. 

\ \\The same criteria are required for the central loose muon candidates (ISOTRK) channel of this analysis, where the non-W model is called ``non-isolated loose muons''. We select loose muon candidates that pass all the selection criteria but fail the ``isolation'' cut.

\ \\In order to reduce the non-W (QCD) background in data samples before asking any $b$-tag requirement (Pretag samples) and in the SVTnoJP05 samples, we apply further selection cuts that are generically called ``QCD veto''. The QCD veto will not be applied for the SVTSVT and SVTJP05 tagging categories. The QCD veto applied is specific for each charged lepton category.

\ \\We first define $m_T^W$ as the reconstructed transverse mass of the W boson:

\begin{equation}\label{TransverseMass}
m_T^W = \sqrt{2 \cdot (p_{T,{\mbox{lep}}}\cdot \met - \vec{p_{T,{\mbox{lep}}}} \cdot \vec{\met})} \rm{,}
\end{equation}

\ \\We then define $\phi_{\met,jet2}$ as the azimuthal ($\phi$) angle between the missing transverse energy vector ($\vmet$) and the second most energetic jet in the event (also called the second-leading jet) and $\metsig$ as the missing transverse energy significance given by the following formula:

% to make met, jet1 to be uniform in these formulas

\begin{equation}\label{METSignificance}
\metsig = \frac{\met}{\sqrt{\sum_{\mbox{jets}}C_{JES}^{2}\cos^{2}(\Delta\phi_{\met,\,jet1})\et + \cos^{2}(\Delta\phi_{vtx\met,\,corr\met})(\met - \sum_{\mbox{jets}}\et)  }}\rm{,}
\end{equation}

\ \\where $C_{JES}$ is the jet energy correction factor and $\Delta\phi_{vtx\,\met,\,corr\,\met}$ is the azimuthal angle difference between the uncorrected and corrected missing transverse energy directions. 

\ \\For the CEM category we require that:
\begin{itemize}
\item [-] $m_T^W > 20 \gev$
\item [-] $\metsig \le -0.05\cdot m_T^W + 3.5$
\item [-] $\metsig \le 2.5-3.125\cdot \Delta \phi_{\met,\,jet2}$
\end{itemize}

\ \\For the CMUP, CMX and ISOTRKcategories we require that:
\begin{itemize}
\item [-] $m_T^W > 20 \gev$\rm{.}
\end{itemize}

\subsection{Further Event Selection for ISOTRK}

\subsubsection{Ensuring Exactly One Charged Lepton}

\ \\For ISOTRK category we require a high transverse momentum isolated track in the central region of the detector ($|\eta|<1.2$), as it was described in Subsection \ref{subsection:IsolatedTrackIdentification}. Since this is a loose charged lepton category, we apply further cuts to ensure that the isolated track is genuinely a charged lepton produced in a $W$ boson decay and that the event is not counted already in another charged lepton category, since the signature of our analysis contains exactly one charged lepton. Therefore we apply the following veto cuts:
\begin{itemize}
\item Tight Lepton Veto: If the event has a reconstructed CEM, CMUP or CMX lepton candidate, that event can not be an ISOTRK event.
\item Tight Jet Veto: If an isolated track is within a cone radius $\Delta R=0.04$ of a tight jet, then most likely the track belongs to a particle produced in a quark decay in a secondary vertex. We reject the event to avoid that the track does not originate from the primary interaction vertex.
\item Two or More Isolated Track Veto: If two or more isolated tracks have been reconstructed in the event, we veto the event.
\end{itemize}

\subsubsection{Ensuring Trigger-Specific Jet Requirements}

\ \\Also, the jet selection is specific for each of the three MET-based triggers. As  described in Appendix \ref{chapter:CombineTriggers}, in my new method of combining three MET-based triggers, we compute on an event by event basis the trigger efficiency for each trigger and then we set it to zero if the event fails the jet selection specific to that trigger. Then we choose the largest of the weights and we require that it is strictly larger than zero, thus ensuring that the jet selection is passed for at least one of the triggers. We weight simulated events by this final weight. For data events, we check that the trigger that gave the largest weight fires for the particular event. If it does, the event is kept. If it does not, the event is rejected, even if it could fire other triggers. We do not even check if other triggers fire and it is this specific handling of triggers that allows us to select the maximum signal acceptance without having correlations (OR) between the triggers. 

\section{Summary}

\ \\In this chapter we have presented the online (trigger) and offline event selection. We started by describing in detail the charged-lepton inclusive triggers used in this analysis, namely the CEM, CMUP and CMX triggers. Our original contribution is the isolated track charged lepton category, for which we introduce a novel method to combine three different MET-plus-jets-based triggers. We continued by describing the baseline event selection for all charged lepton categories. We then introduced the three different $b$-tagging categories employed in this analysis with the help of two different $b$-tagging algorithms. We then presented the selection used to remove a large fraction of the Non-W (QCD) background, which is specific to each charged lepton category. The novel charged lepton category we introduced needs further specific event selection, such as vetoing events with two or more charged leptons or isolated tracks and ensuring the event passes a jet selection specific to the chosen MET-based trigger. 

\ \\The event selection criteria described in this chapter are applied both to data and Monte-Carlo-simulated events. In the next chapter we will present the methodology and the result for the calculation of signal event yield prediction. In the following chapter we will present the background estimation method. 

\clearpage{\pagestyle{empty}\cleardoublepage}

\chapter{Higgs Boson Signal Estimation\label{chapter:Signal}}

\ \\In this chapter we use Monte Carlo calculations to estimate the number of $WH$ ($ZH$) signal events in our sample and the systematic uncertainty on this number.

\begin{figure}[h]
\begin{center}
%ISOTRK at the top layer, looks more impressive, but less correct
%\includegraphics[angle=0,width=0.45\textwidth,clip=]{./signal/ScatterPlots/scatter_muon1.eps}
%\includegraphics[angle=0,width=0.45\textwidth,clip=]{./signal/ScatterPlots/scatter_electron1.eps}
%ISOTRKat the bottom layer, looks less impressive, but more correct
\includegraphics[angle=0,width=0.45\textwidth,clip=]{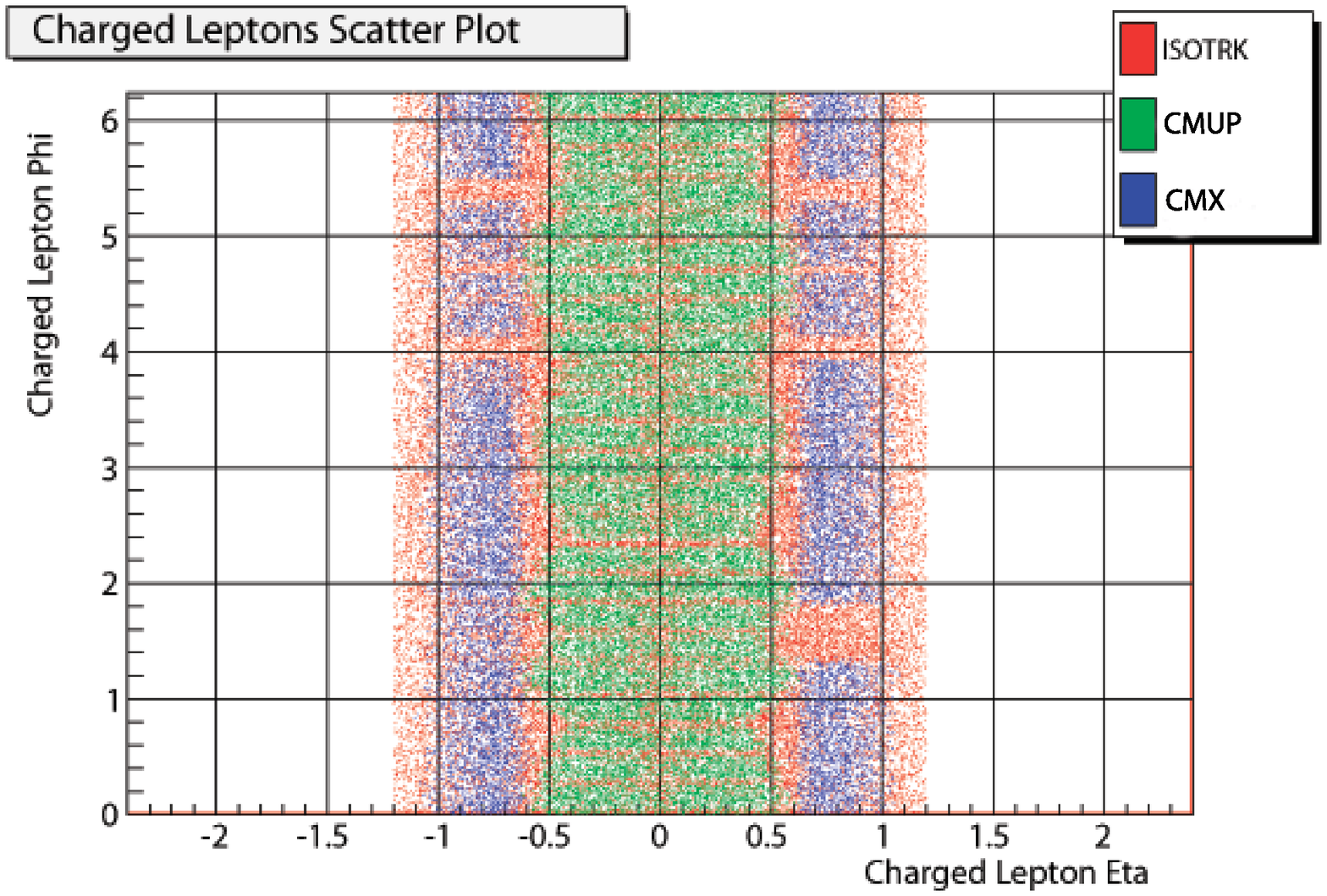}
\includegraphics[angle=0,width=0.45\textwidth,clip=]{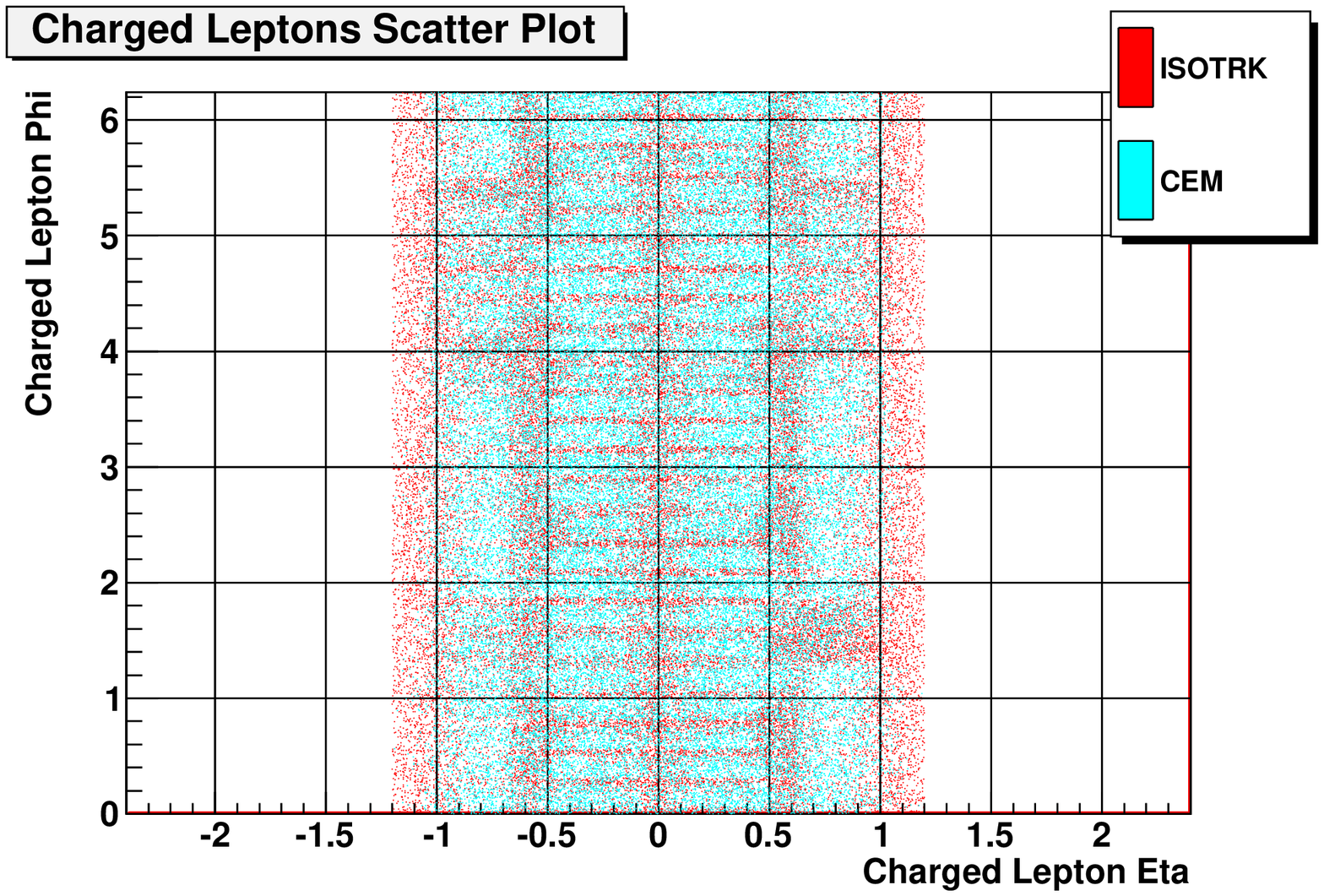}
\caption[Charged lepton scatter plot for eta and phi.]
{Scatter plot for charged lepton in our analysis after full event selection in a Pretag sample for Monte Carlo simulated events for the $WH$ signal, where the Higgs boson has a mass of 115 $\gevcc$. The left plot shows gaps in the $\eta$-$\phi$ coverage for these CMUP and CMX muon candidates due to non instrumented regions in the muon detectors are filled by the ISOTRK charged lepton candidates. The right plot shows gaps in the $\eta$-$\phi$ distribution of the electron candidate CEM. Since the CEM electron candidates present a smooth distribution, ISOTRK candidates are mostly muon candidates. \label{figure:ScatterEtaPhi}}
\end{center}
\end{figure}

\section{Signal Prediction Estimation}

\ \\We calculate the number of predicted signal events from by the main signal channel $N_{WH\to l\nu b\bbar}$ using Equation \ref{formula:WHEstimation}. The supplementary contribution from $N_{ZH\to l(l) b\bbar}$ is calculated using Equation \ref{formula:ZHEstimation}.  

\begin{equation} 
N_{WH\to l\nu b\bbar} = \sigma (p\pbar \to WH) \cdot  \left( \sum_{l=e,\mu,\tau} \branchingratio(W \to l\nu) \right) \cdot \branchingratio(H \to b\bbar) \cdot \left(\int {\cal{L}}\,dt\right) \cdot \epsilon_{W(Z)H\to l\nu b\bbar}\, \rm{.}
\label{formula:WHEstimation}
\end{equation}

\begin{equation} 
N_{ZH\to l\nu b\bbar} = \sigma (p\pbar \to ZH) \cdot  \left( \sum_{l=e,\mu,\tau} \branchingratio(Z \to ll) \right) \cdot \branchingratio(H \to b\bbar) \cdot \left(\int {\cal{L}}\,dt\right) \cdot \epsilon_{W(Z)H\to l\nu b\bbar}\, \rm{.}
\label{formula:ZHEstimation}
\end{equation}

\ \\In these equations, $\sigma (p\pbar \to W(Z)H)$ represents the $W(Z)H$ production cross section in $p\pbar$ collisions at a centre-of-mass energy of $\sqrt{s}=1.98\tev$. These values are a function of the Higgs boson mass and are presented in Table~\ref{table:SignalCrossSectionsAndBranchingFraction}. The $\sum_{l=e,\mu,\tau} \branchingratio(W \to l\nu)$ is $0.324\pm 0.003$ \cite{PDG} and represents the branching ratio of a $W$ boson leptonic decay and is equal to the sum of the branching ratio of a $W$ boson decay to an electron and an electron neutrino plus the branching ratio of a $W$ boson decay to a muon and a muon neutrino plus the branching ratio of a $W$ boson decay to a tau lepton and a tau neutrino. The $\sum_{l=e,\mu,\tau} \branchingratio(Z \to ll)$ is $0.104\pm 0.001$ \cite{PDG} and represents the branching ratio of a $Z$ boson decay charged leptons and is equal to the sum of the branching ratio of a $Z$ boson decay to an electron-antielectron pair plus the branching ratio of a $Z$ boson decay to a muon-antimuon pair plus the branching ratio of a $Z$ boson decay to a tau-antitau pair. The $\branchingratio(H \to b\bbar))$ is the branching ratio of the Higgs boson decay to a $b\bbar$ pair. These values are a function of the Higgs boson mass and are presented in Table \ref{table:SignalCrossSectionsAndBranchingFraction}. The $\int {\cal{L}}\,dt$ is the integrated luminosity. For this analysis, it has a value of $5.70 \pm 0.34$ $\invfb$. The $\epsilon_{W(Z)H\to l\nu b\bbar}$ is the efficiency of the signal selection and is given by equation

\begin{equation} 
\epsilon_{W(Z)H\to l\nu b\bbar} = \epsilon_{z_0} \cdot \epsilon_{\rm{trigger}} \cdot \epsilon_{\rm{lepton ID}} \cdot \epsilon_{WH\to l\nu b\bbar}^{\rm{MC}}\,\rm{.}
\label{formula:EfficiencyEstimation}
\end{equation} 

\ \\ In Equation \ref{formula:EfficiencyEstimation}, $\epsilon_{z_0}$ is the efficiency of the cut that requires the primary vertex position to be situated within 60 cm of the centre of the detector ($|z_0|<60$ cm). The $\epsilon_{\rm{trigger}}$ is the efficiency of the requirement that the event fires the required trigger. It is measured in data for each trigger and therefore is specific to each charged lepton category. The $\epsilon_{\rm{lepton ID}}$ is the ratio of the lepton identification efficiencies for data and for simulated signal events, also called lepton identification scale factor. The $\epsilon_{WH\to l\nu b\bbar}^{\rm{MC}}$ the fraction of signal events, after the requirement of $|z_0|<60$ cm, that pass all the other kinematic selections of the analysis. For the various $b$-tagging categories, this term also takes into account the $b$-tagging scale factor between data and signal simulated events.

\ \\For a Higgs boson mass of $115 \gevcc$, the computed $WH$ and $ZH$ signal event predictions for each charged lepton and $b$-tagging category are presented in Table~\ref{table:WHZHAPrediction115}. My original contribution to this analysis, the ISOTRK charged lepton category, increases the $WH$ ($ZH$) signal prediction by 33\% (66\%) over the TIGHT charged lepton category alone. 

\ \\This increase is easily understood from Figure \ref{figure:ScatterEtaPhi}, which represents the $\eta$-$\phi$ distribution of a sample of Monte Carlo simulated events for the $WH$ signal, where the Higgs must have a mass of 115 $\gevcc$, after the full event selection in the Pretag sample. My original contribution to this analysis, the addition of ISOTRK charged lepton candidates, fills the gaps, as we see in red. This increases the number of selected events containing charged leptons and thus the signal event prediction. Unlike the case of muon candidates, the distribution for electron candidates is smooth. This is why ISOTRK candidates are mostly muon candidates. The calorimeter detectors still have very small non instrumented regions, such as between the wedges of the calorimeter towers. A detailed study \cite{JasonSlaunwhiteThesis} identified that ISOTRK candidates are muon candidates in 85\% of cases, electron candidates in 6\% of cases and tau candidates in 7\% of cases. 

\begin{table}[h] % [t] puts at top of page
\begin{center}
\begin{tabular}{|c|c|c|c|c|c|}
\multicolumn{6}{c}{}\\
\hline
\multicolumn{6}{|c|}{CDF Run II Preliminary 5.7 fb$^{-1}$}\\
\multicolumn{6}{|c|}{Number of Expected $WH$ ($ZH$) Events at $M(H)$ = 115 GeV }\\
\hline Tag Sample & CEM & CMUP & CMX & ISOTRK & \% Increase \\
\hline Pretag    & 12.07 (0.27) & 6.47 (0.48) & 3.20 (0.23) & 7.34 (0.64) & 34 (65)\\
\hline SVTSVT    & 1.72 (0.04)  & 0.86 (0.06) & 0.44 (0.03) & 1.01 (0.09) & 33 (69)\\
\hline SVTJP05   & 1.24 (0.03)  & 0.64 (0.04) & 0.32 (0.02) & 0.74 (0.06) & 34 (67)\\
\hline SVTnoJP05 & 4.17 (0.09)  & 2.23 (0.16) & 1.12 (0.08) & 2.46 (0.22) & 33 (67)\\
\hline \hline
\end{tabular}
\caption[Expected number of signal events for a Higgs boson mass of 115 $\gevcc$]{Expected number of $WH$ and $ZH$ signal events for an assumed Higgs boson mass of $115 \gevcc$ and an integrated luminosity of 5.7 $\invfb$ as a function of charged lepton and $b$-tagging information. The last column represents the percentage increase in signal prediction due to our original contribution ISOTRK charged leptons over TIGHT (CEM+CMUP+CMX) charged leptons alone. In each column, the first values represent the $WH$ signal and the numbers in brackets represent the $ZH$ signal.}
\label{table:WHZHAPrediction115}
\end{center}
\end{table}

\section{Systematic Uncertainty on Signal Event Prediction}

\ \\In this section we describe the various contributions to the systematic uncertainty we quote on the $WH$ and $ZH$ signal event prediction.

\subsection{Trigger}

\ \\The systematic uncertainty on the trigger used is measured by selecting data events with an orthogonal trigger and then asking the fraction of events that fire our trigger of interest. This process is also done for each charged lepton category. The systematic uncertainties are measured to be $< 1.0\%$ for the TIGHT charged leptons. For the ISOTRK charged lepton the analysis employs the new trigger parametrization described in Appendix \ref{chapter:METTriggers} and the novel method to combine triggers described in Appendix \ref{chapter:CombineTriggers}, with the method to compute systematic uncertainty described in Section \ref{section:METTriggerSystematics}. We measured a trigger systematic uncertainty for the ISOTRK category of 3\%.

\subsection{Lepton Identification}

\ \\The lepton identification systematic uncertainty is measured by comparing a data sample highly enriched in $Z$ boson events with a $Z$ boson sample of Monte Carlo simulated events using \pythia as event generator. $Z$ boson decay to charged leptons are used to evaluate the lepton identification systematic uncertainty for each charged lepton category. Details are presented in Subsection \ref{subsection:ChargedLeptonScaleFactors}.

\subsection{Initial and Final State Radiation}

\ \\The systematic uncertainty due to the effect of the initial state radiation (ISR) and final state radiation (FSR) is calculated by changing in the Monte Carlo simulation the parameters related to ISR and FSR to their half and double values. The systematic uncertainty is quoted as half of the difference between the signal event prediction with the two changes \cite{YoshikazuNagaiThesis}.

\subsection{Parton Distribution Functions}

\ \\Another systematic uncertainty source is the fact that the parton distribution functions (PDFs) are not perfectly known, neither for the protons, nor the antiprotons. We first compute three systematic uncertainties that we use to compute the final PDF systematic uncertainty.

\ \\The PDFs for the simulations used in this analysis use CTEQ5L \cite{CTEQ}, which is parameterized using 20 eigenvectors. We weight the nominal Monte Carlo simulated events for each of the 20 eigenvectors and for each we compute a signal event prediction. The first PDF systematic uncertainty is quoted as the quadrature sum between the differences between the nominal and weighted event prediction.

\ \\We also compute the signal event prediction using MRST72~\cite{MRST72} as PDF generator. The absolute value of the difference between the CTEQ5L and MRST72 signal prediction is quoted as the second PDF systematic uncertainty.

\ \\In addition we compute the signal event prediction using PDFs that are generated using different values of the coupling constant of the strong force, i.e. different QCD energy scales. We use MRST72 ($\Lambda_{\rm{QCD}}=228 \mev$) and MRST75 ($\Lambda_{\rm{QCD}}=300 \mev$). The absolute value of the difference between the CTEQ72 and MRST75 signal predictions is quoted as the third PDF systematic uncertainty.

\ \\The final PDF systematic uncertainty is computed by adding in quadrature the maximum between the first and the second with the third PDF systematic uncertainty \cite{YoshikazuNagaiThesis}.

\begin{figure}[ht]
  \begin{center}
    \includegraphics[width=0.32\textwidth]{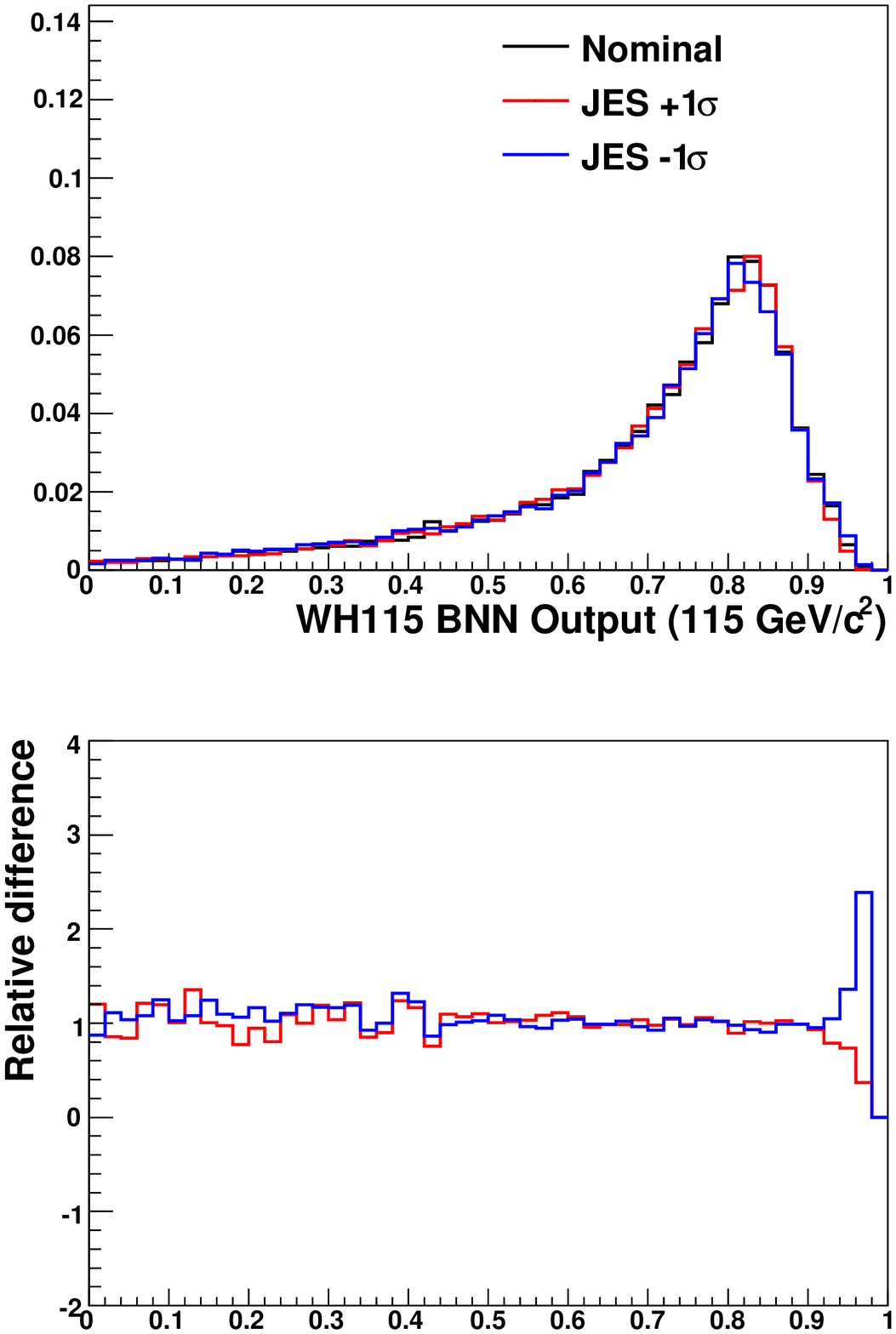}
    \includegraphics[width=0.32\textwidth]{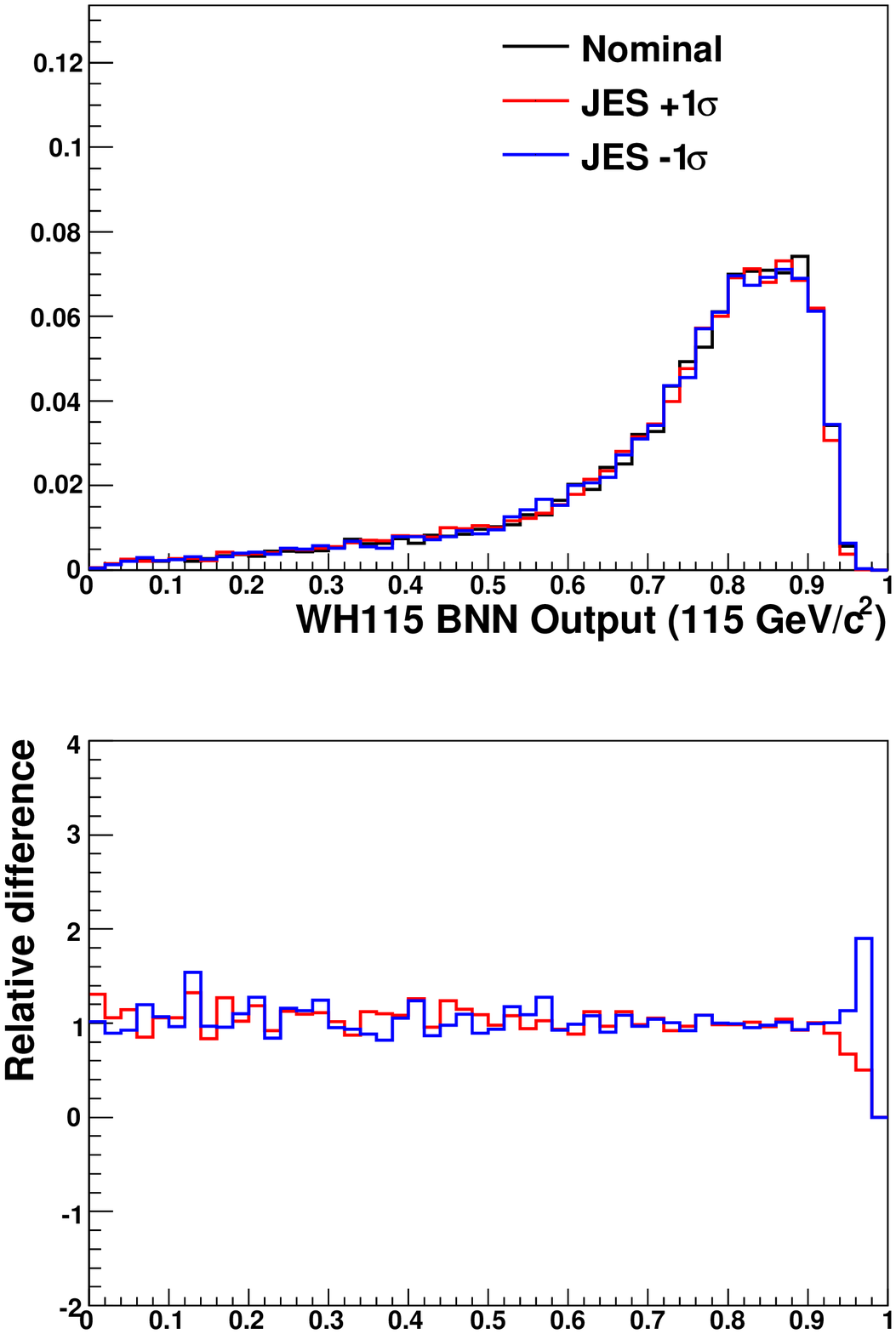}
    \includegraphics[width=0.32\textwidth]{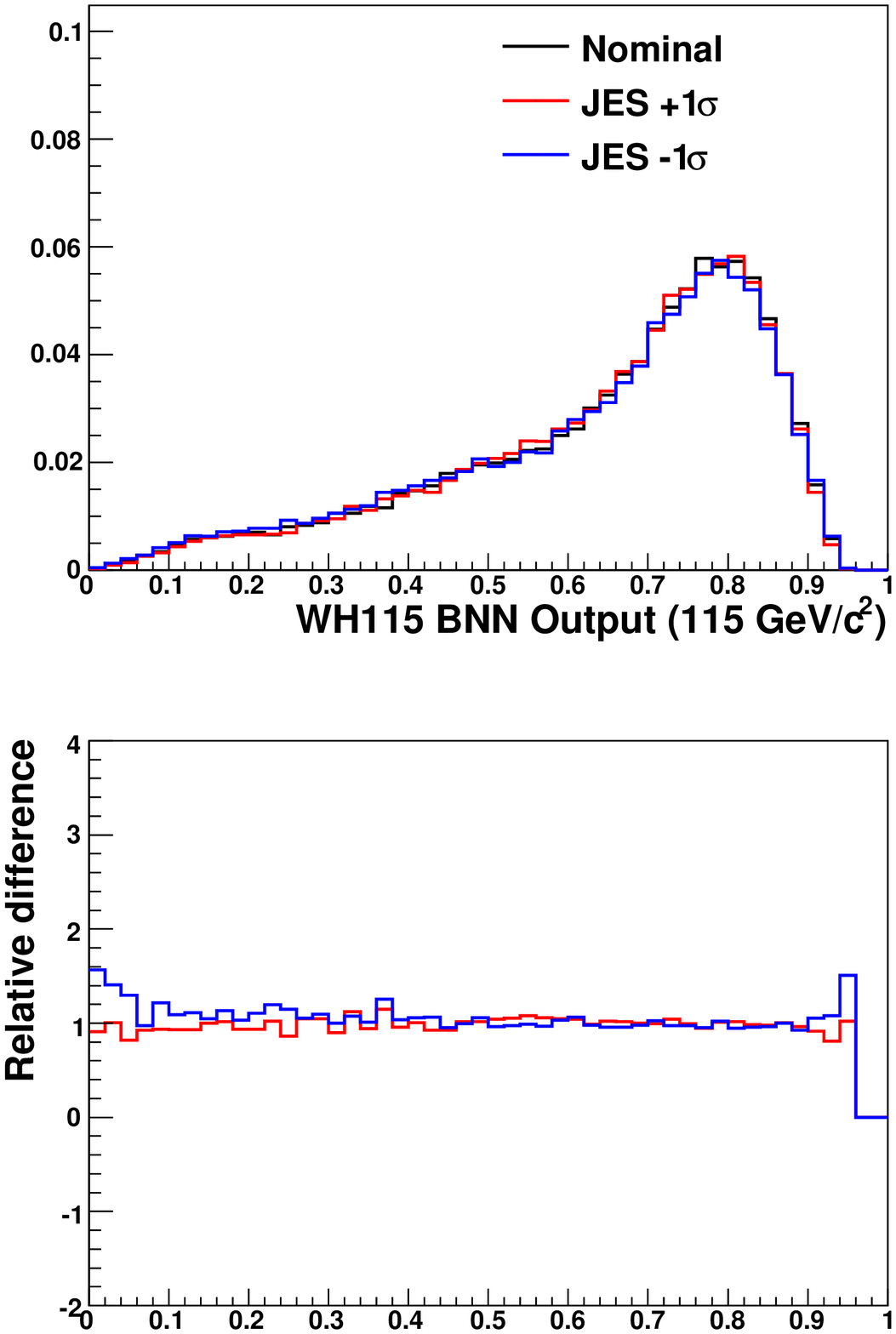}
    \caption[TIGHT BNN output shapes overlaid for JES variations]{TIGHT BNN output shape for default and $\pm$1~sigma JES for $WH$ signal ($m_H = 115~\text{GeV}/c^2$). From left to right SVTSVT, SVTJP05 and SVTnoJP05, respectively. The horizontal axis represents the value of the BNN output, which is the final analysis discriminant.}
    \label{figure:shapeSystematicTIGHTWH}
  \end{center}
\end{figure}

\begin{figure}[ht]
  \begin{center}
    \includegraphics[width=0.32\textwidth]{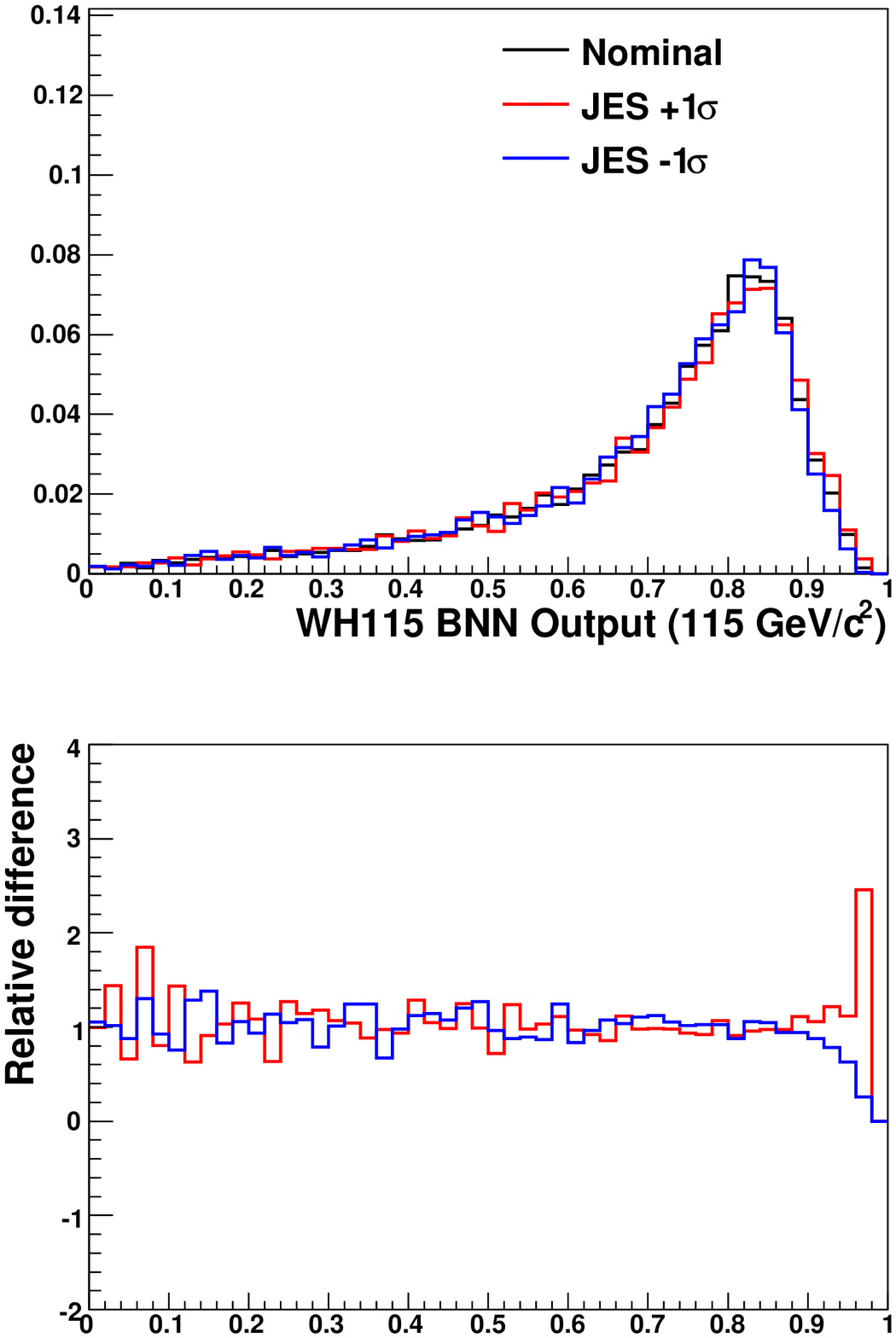}
    \includegraphics[width=0.32\textwidth]{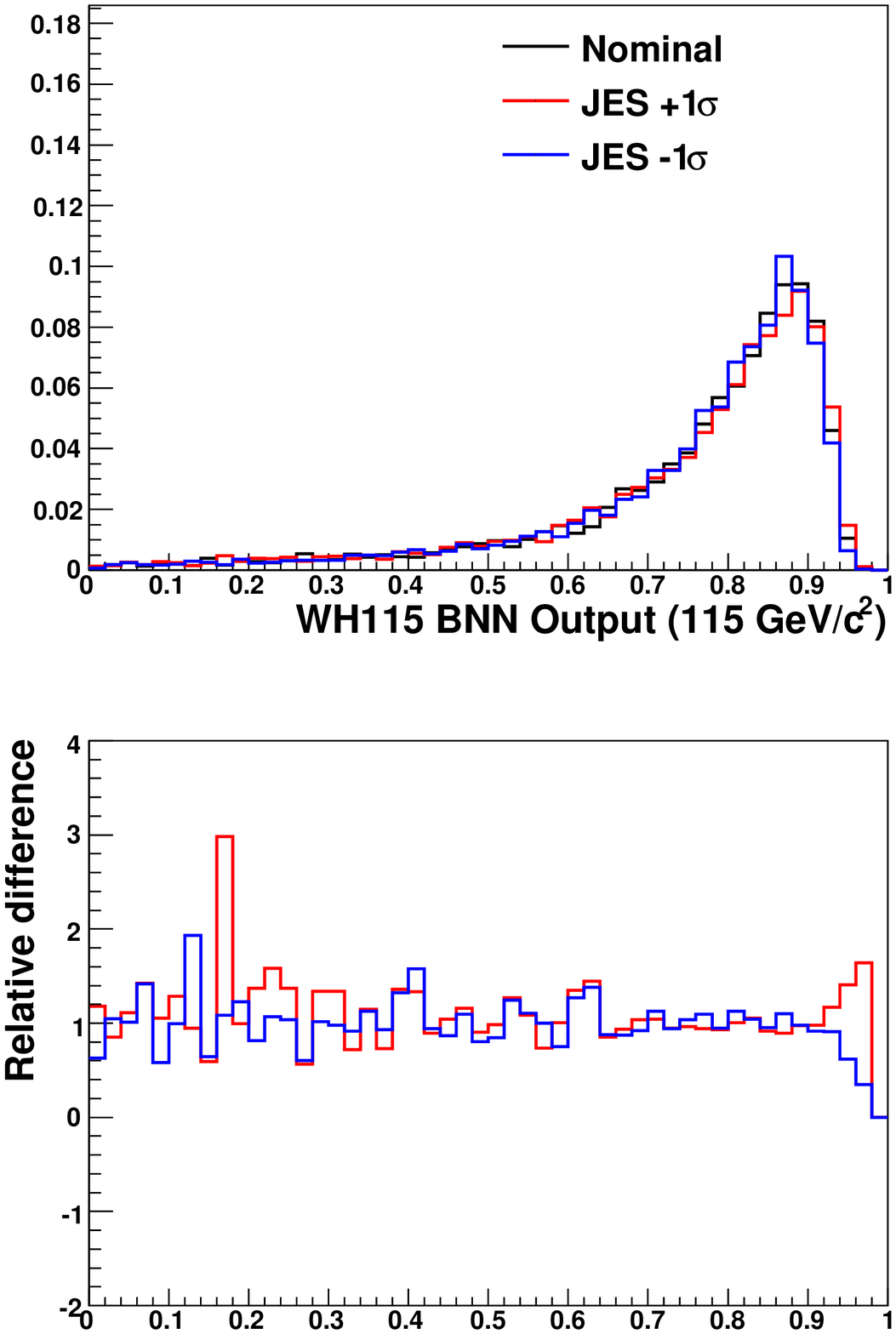}
    \includegraphics[width=0.32\textwidth]{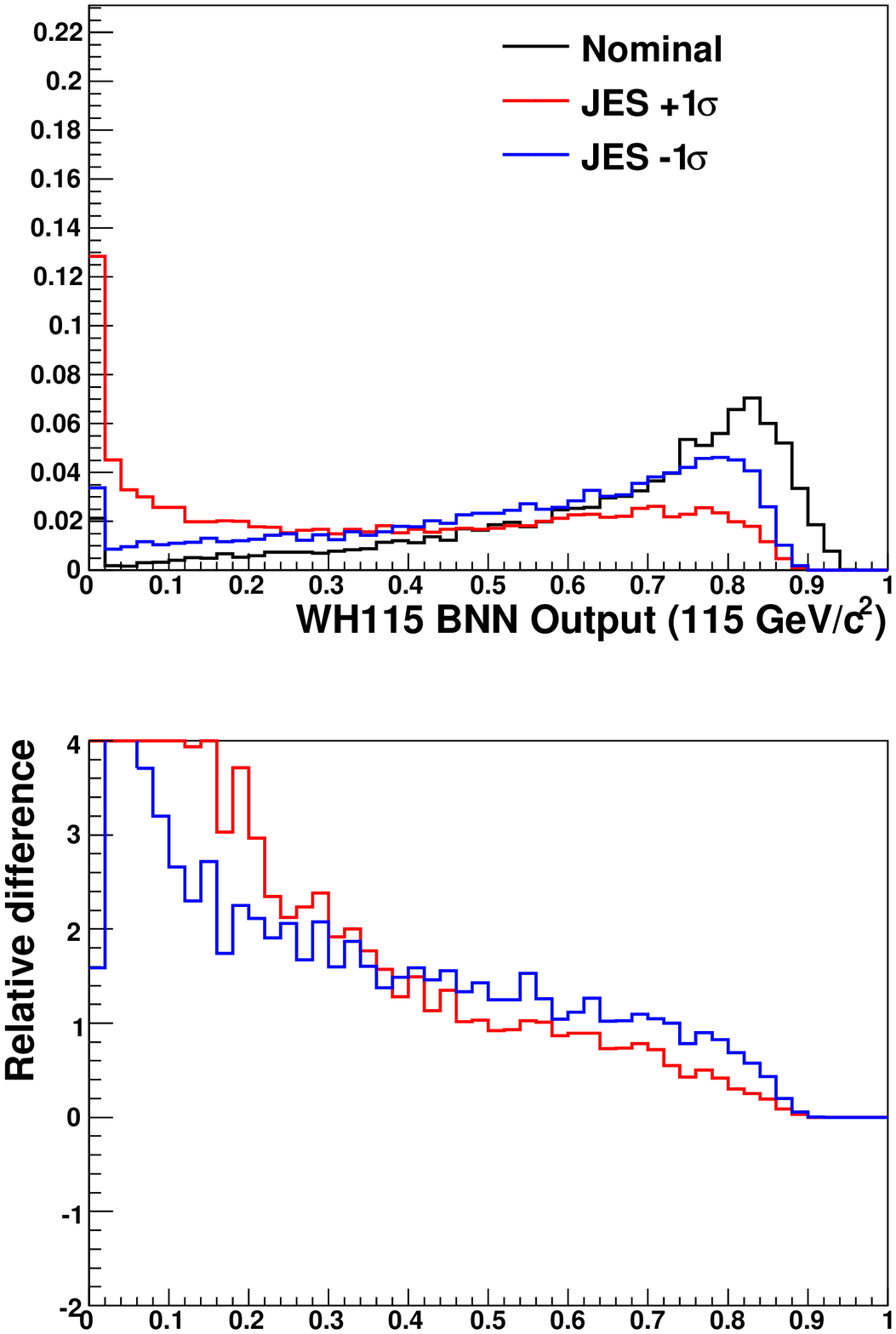}
    \caption[ISOTRK BNN output shapes overlaid for JES variations]{ISOTRK BNN output shape for default and $\pm$1~sigma JES for $WH$ signal ($m_H = 115~\text{GeV}/c^2$). From left to right SVTSVT, SVTJP05 and SVTnoJP05, respectively. The horizontal axis represents the value of the BNN output, which is the final analysis discriminant.}
    \label{figure:shapeSystematicISOTRKWH}
  \end{center}
\end{figure}

\subsection{Jet Energy Scale}

\ \\First we compute the nominal $WH$ signal event prediction using a Higgs boson mass of $115 \gevcc$. Then, for the same Higgs boson mass, we scale the jet energy scale \cite{JetCorrections} up and down by one standard deviation. We compute the signal event prediction for these two cases. We take the largest deviation from the nominal signal predictions as the jet energy scale rate systematic uncertainty. 

\ \\In this analysis we also consider one jet energy scale shape systematic. We scale the jet energies up and down by one standard deviation and then we propagate this to all reconstructed variables, including the final analysis discriminant, described in detail in Chapter~\ref{chapter:Discriminant} and denoted the BNN output. Both the central and the alternate plus and minus BNN shapes are used in the limit calculation, as described in Section \ref{section:LimitAllBinsOneChannel}. The top plots in Figure \ref{figure:shapeSystematicTIGHTWH} (\ref{figure:shapeSystematicISOTRKWH}) presents the BNN shapes for JES zero, JES plus, JES minus for each of the $b$-tagging categories (SVTSVT, SVTJP05 and SVTnoJP05, as described in Subsection~\ref{bTaggingCategories} and the TIGHT (ISOTRK) charged lepton category. The bottom plots in the same figures represent the ratio of the JES plus and JES minus to JES Zero. 

\subsection{$b$-tagging Scale Factor}

\ \\The systematic uncertainty on the $b$-tagging scale factor between data and Monte Carlo simulated events is computed in Section \ref{section:bTagging}. 

\subsection{Integrated Luminosity}

\ \\The integrated luminosity of 5.7 $\invfb$ used in this analysis has been measured with the Cherenkov Luminosity Counter, which as been described in detail in ~\ref{Section:CherenkovLuminosityCounter}. The measurement has a systematic uncertainty of 6\% \cite{InstantaneousLuminosityUncertainty}.

\subsection{Systematic Uncertainty Values}

\ \\The computed values for the total systematic uncertainties are presented in Table~\ref{Table:SystematicUncertaintyCentral} for tight central charged lepton categories (CEM, CMUP and CMX) and in Table~\ref{Table:SystematicUncertaintyISOTRK} for the loose central charged lepton category (ISOTRK).

\newpage

\begin{center}
\begin{table}[htb] % [t] puts at top of page
\begin{center}
\begin{tabular}{|l|c|c|c|c|c|c|}
\hline \hline
$b$-tagging category & Lepton ID & Trigger & ISR/FSR/PDF & JES & b-tagging & Total\\
\hline \hline
SVTSVT    & 2\% & $<$1\% & 4.9\% & 2.0\% & 13.6\% & 15.9\%\\
SVTJP05   & 2\% & $<$1\% & 4.9\% & 2.8\% & 8.1\%  & 10.1\%\\
SVTnoJP05 & 2\% & $<$1\% & 3.0\% & 2.3\% & 4.3\%  & 6.1\%\\
\hline \hline
\end{tabular}
\caption[Systematic uncertainties for TIGHT charged lepton candidates]{Systematic uncertainty values on the signal event prediction for the tight central charged lepton categories (CEM, CMUP, CMX) for each b-tagging category.}
\label{Table:SystematicUncertaintyCentral}
\end{center}
\end{table}
\end{center}

\begin{center}
\begin{table}[htb] % [t] puts at top of page
\begin{center}
\begin{tabular}{|l|c|c|c|c|c|c|}
\hline \hline
$b$-tagging category & Lepton ID & Trigger & ISR/FSR/PDF & JES & b-tagging & Total\\
\hline \hline
SVTSVT    & 4.5\% & 3\% & 7.1\%  & 1.7\% & 8.6\%  & 12.5\%\\
SVTJP05   & 4.5\% & 3\% & 6.4\%  & 2.4\% & 8.1\%  & 11.9\%\\
SVTnoJP05 & 4.5\% & 3\% & 8.4\%  & 4.7\% & 4.3\%  & 11.8\%\\
\hline \hline
\end{tabular}
\caption[Systematic uncertainties for ISOTRK charged lepton candidates]{Systematic uncertainty values on the signal event prediction for the loose central charged lepton category (ISOTRK) for each b-tagging category.}
\label{Table:SystematicUncertaintyISOTRK}
\end{center}
\end{table}
\end{center}

\section{Summary}

\ \\The first part of this chapter has presented the calculation of event yield predictions for the $WH$ and $ZH$ signal processes, which is the multiplication of the integrated luminosity with the signal cross section, decay branching ratios and efficiencies for the various selection cuts. For the novel charged lepton category we introduced, we explained how real charged leptons that go towards non-instrumented regions of the detector are recovered and thus increase the signal event yield. 

\ \\In the second part of the chapter we presented the signal systematic uncertainties that we take into account in this analysis, namely the trigger, lepton identification, initial and final state radiation, parton distribution functions, jet energy scale, $b$-tagging scale factor and integrated luminosity.

\clearpage{\pagestyle{empty}\cleardoublepage}

\chapter{Background Estimation\label{chapter:Background}}

\ \\Our background estimation method assumes that we have correctly identified all the processes that mimic our $WH$ signal: top quark pair, single top (both s- and t-channel), diboson ($WW$, $WZ$ and $ZZ$), $Z$+jets (both $Z$ + Heavy Flavour (HF) jets and $Z$ + Light Flavour (LF) jets), $W$+jets (both W+LF and W+HF) and non-W (QCD) production. In Chapter~\ref{chapter:Simulation} we described in detail our signal and background processes, especially their Feynman diagrams, signatures and Monte Carlo generators used to simulate the events. In this chapter we will describe how data and Monte Carlo generated events are used to compute the contribution of each background type in our data sample.

\ \\We recall that we call the ``pretag'' sample the events that pass all the event selection requirements, but are not required to pass any $b$-tagging information. The $b$-tagging categories (SVTSVT, SVTJP05, SVTnoJP05) are orthogonal to each other and each is a sub-sample of the pretag sample. The Pretag sample is dominated by background events, with very little signal contribution. This is why we use it as a ``control region'' or ``sideband'' to check that the background modelling is correct and in agreement with the data events. Table \ref{table:PretagObservedEvents} gives the observed number of events for the Pretag data sample. Also, we use the pretag sample to measure the fraction of the data events that are non-W (QCD) and $W$+jets (both LF and HF) events, which is then extrapolated to the ``tagged'' categories signal region, as we will discuss in the next sections. 

\begin{table}[hbt]
  \begin{center}
    \begin{tabular}{|c|c|}
    \hline \hline
    Category & Pretag Data Observation\\
    \hline \hline
    TIGHT    & 83788\\
    ISOTRK   & 21486 \\
    \hline \hline
  \end{tabular}
  \caption[Observed number of events for the Pretag data sample]{Observed number of events for the Pretag data sample for the TIGHT and ISOTRK charged lepton categories. }
  \label{table:PretagObservedEvents}
  \end{center}
\end{table}

\section{Top Quark and Other Electroweak Backgrounds}

\ \\We first compute the expected background events in our data sample background processes that are well modelled at tree level by Monte Carlo simulations (top quark pair production, single top, diboson and $Z$+jets). We use the same procedure and formulae as those used for the signal acceptance calculation, described in detail in Chapter~\ref{chapter:Signal} and summarized by the following formula:

\begin{equation}\label{N_EQK_TOP}
N_{p\pbar \to X}=\epsilon^{event} \cdot \epsilon^{tag} \cdot \sigma_{p\pbar \to X} \cdot \int \! {\cal{L}}\, dt \,\rm{.}
\end{equation}

\ \\We denote the estimated number of events due to electroweak processes (diboson or $Z$+jets) as $N_{EWK}$ and the estimated number of events due to top quark processes (top quark pair and single top) as $N_{TOP}$. They will be used in the background estimation of QCD and $W$+jets processes, as described below.

\ \\We use the cross sections and branching ratios specific for each background process, as seen in Table~\ref{table:BackgroundCrossSections}. 

\begin{table}[hbt]
  \begin{center}
    \begin{tabular}{|c|c|}
    \hline \hline
    Process & Theoretical Cross Sections\\
    \hline \hline
    $WW$   & 11.34 $\pm$ 0.70 pb \\
    $WZ$   & 3.22  $\pm$ 0.30 pb \\
    $ZZ$   & 1.20  $\pm$ 0.20 pb \\
    Single Top  s-channel &  1.05 $\pm$ 0.07 pb\\
    Single Top  t-channel &  2.10 $\pm$ 0.19 pb\\
    $Z+\mbox{jets}$   &  787.4 $\pm$ 85.0 pb\\
    $t\bar{t}$ & 7.04 $\pm$ 0.44  pb \\
    \hline \hline
  \end{tabular}
  \caption[NLO theoretical cross sections and uncertainties for background events]{NLO theoretical cross sections and uncertainties used in the computation of predicted background events for processes that are well modelled at tree level (top quark pair production, single top s-channel, single top t-channel, diboson ($WW$, $WZ$, $ZZ$) and $Z$+jets). The simulated events use a top quark mass ($m_t = 172.5~\gevcc$).}
  \label{table:BackgroundCrossSections}
  \end{center}
\end{table}

\ \\The efficiency term $\epsilon^{event}$ multiplies all the individual efficiencies, except the $b$-tagging scale factor. The efficiency term $\epsilon^{tag}$ is the $b$-tagging scale factor for the event. For the pretag sample, $\epsilon^{tag}=1$. 

\ \\We measure the systematic uncertainty on the background normalization values due to the $b$-tagging efficiency by varying the scale factor and mistag probabilities within one standard deviation of their values and then reproducing this entire procedure. 

\section{Pretag: $W$+Jets and QCD Events Faking a W boson}

\ \\As described in detail in Subsection~\ref{subsection:QCDTemplates}, we developed a QCD template (modelling) for each of the charged lepton categories\footnote{Each charged lepton category is susceptible to different kinds of faking due to the different reconstruction criteria.} by selecting events that fail one or two relevant cuts for charged lepton selection. These samples are enriched in fake $W$ boson candidates and therefore represent a model for the non-W (QCD) processes. However, the QCD background is the most poorly predicted and the least understood, which requires us to assign a large systematic uncertainty on its normalization estimate.

\ \\We fit the $\met$ distribution in pretag data to a sum of $\met$ background shapes. In the fit the electroweak ($N_{EWQ}$) and top ($N_{TOP}$) normalization values are fixed, but the normalization of QCD ($N_{QCD}$) and $W$+jets ($N_{W+jets}$) are allowed to float. The fit is performed in the range $0 \gev <\et < 120 \gev$ using the $\met$ distributions
of the QCD and W+jets samples, which are obtained after removing the $\met > 20 \gev$ from the standard event selection. The normalization of the QCD and W+jets samples are then fixed by the fit.

\ \\We define and extract from the fit the fraction of QCD events ($F_{QCD}^{pretag}$) from the total number of events in the pretag data sample after the standard selection $\met > 20 \gev$ cut is applied  - equation~\ref{F_QCD_pretag}), which we call $N^{pretag}$:

\begin{equation}\label{F_QCD_pretag}
F_{QCD}^{pretag}=\frac{N_{QCD}^{pretag}(\met > 20 \gev)}{N^{pretag}(\met > 20 \gev)}\ \ \rm{.}
\end{equation}

\ \\Then the number of QCD events in the pretag sample ($N_{QCD}^{pretag}$) is given by:

\begin{equation}\label{N_QCD_pretag}
N_{QCD}^{pretag}=N^{pretag}\cdot F_{QCD}^{pretag}\,\rm{.}
\end{equation}

\ \\We are careful to check how the QCD normalization changes when we modify the histogram binning, the $\met$ fit interval, the $\met$ cut for definition of $F_{QCD}$, as well as the non-W models used for CEM, CMUP, CMX, ISOTRK and PHX. As a conclusion of these studies, we assign a conservative 40\% systematic uncertainty on the QCD normalization. Despite this very large uncertainty, the total number of QCD events is relatively small in the final sample, thanks to the $\met$ cut we apply in the final analysis. This permits the analysis to remain sensitive to the $WH$ process. 

\ \\At this stage we have estimated the number of electroweak, top quark and QCD processes in the pretag data set. The remaining events in the sample are therefore $W$+jets, as resulting from the following formula:

\begin{equation}\label{N_Wjets_pretag}
N_{W+jets}^{pretag}=N^{pretag}\cdot(1-F_{QCD}^{pretag}) - N_{EWK}^{pretag} - N_{TOP}^{pretag} \,\rm{.}
\end{equation}

\ \\The $W$+jets sample is divided in two: $W$+heavy flavour (W+HF) and $W$+light flavour (W+LF). 

\ \\The W+LF background contains light flavour jets that are incorrectly tagged to originate from a heavy flavour quark ($b$ or $c$) by the $b$-tagging algorithm. This phenomenon is called ``mistag'', as discussed in Section \ref{section:bTagging}. 

\section{Tag: $W$+Jets and QCD Events Faking a W boson}

\ \\Now we reproduce the procedure described above for each of the $b$-tagged categories. The background templates are all weighed by $\epsilon^{tag}$, as described in the section above.

\subsection{QCD Background}

%\ \\The crucial point of the background estimation method and the benefit of the fit in the background enriched region is that we assume that the QCD fraction measured in the pretag sample is equal to the one in each of the $b$-tagging categories. Then f

\ \\For a given category the QCD normalization is given by Equation~\ref{N_QCD_tag} (to be compared with Equation~\ref{N_QCD_pretag} for pretag) and the $W$+jets normalization is given by equation~\ref{N_Wjets_tag} (to be compared with Equation~\ref{N_Wjets_pretag} for pretag).

\begin{equation}\label{N_QCD_tag}
N_{QCD}^{tag}=N^{tag}\cdot F_{QCD}^{pretag}\,\rm{.}
\end{equation}

\begin{equation}\label{N_Wjets_tag}
N_{W+jets}^{tag}=N^{tag}\cdot(1-F_{QCD}^{pretag}) - N_{EWK}^{tag} - N_{TOP}^{tag} \,\rm{.}
\end{equation}

\subsection{W+HF}

\ \\The normalization of W+HF in a tagged sample ($N_{W+HF}^{tag}$) is computed starting from the normalization of $W$+jets in the pretag sample, that is multiplied by the fraction of pretag events with jets matched to heavy flavour quarks ($F_{HF}$), by the scale factor between data and Monte Carlo for the heavy flavour fraction ($K = 1.4 \pm 0.4$) and by the $b$-tagging efficiency:

\begin{equation}\label{N_WHF_tag}
N_{W+HF}^{tag}=N_{W+jets}^{pretag} \cdot (F_{HF} \cdot K) \cdot \epsilon^{tag} \,\rm{.}
\end{equation}

\ \\The heavy flavour fraction $F_{HF}$ is measured from Monte Carlo simulated events of all the processes that produce one and only one real $W$ boson. This quantity does not agree with the data prediction exactly and their ratio is represented by the scale factor $K$, which acts as a correction for the heavy flavour fraction $F_{HF}$ that is applied to Monte Carlo simulated events from our analysis. The $K$ factor is measured in the 1-jet bin of the analysis, that is a ``sideband'' and not a signal region and has the largest statistics of all the jet bins. We then assume that K has the same value across all the jet bins. 

\subsection{W+LF}

\ \\The normalization of W+LF in a tagged sample ($N_{W+HF}^{tag}$) is computed starting from the normalization of $W$+jets in the pretag sample, that is multiplied by the fraction of pretag events that is not matched to heavy flavour ($1-(F_{HF}\cdot K)$) and by the overall fake tag rate ($\epsilon^{mistag}$), also called the mistag rate:

\begin{equation}\label{N_WHF_tag}
N_{W+LF}^{tag}=N_{W+jets}^{pretag} \cdot (1- F_{HF} \cdot K) \cdot \epsilon^{mistag} \,\rm{.}
\end{equation}

\ \\The mistag rate is measured for each $b$-tagging algorithm in generic light jet data samples. The reason that sometimes light flavour jets are tagged as heavy flavour jets is due to finite tracking resolution. We measure the mistag rate using the negatively tagged jets, as explained in Section~\ref{section:bTagging}. In the end we produce a function that inputs various jet quantities and outputs the mistag probability for that jet. We call this function a mistag matrix. The mistag matrix for \secvtx~is a function of jet $\eta$, jet $\et$, number of interaction vertices in the event, jet track multiplicity and the scalar sum of all the transverse energy in the event. The mistag matrix for \jetprob~uses the same information and, in addition, uses the $z$ coordinate of the primary interaction vertex. 

\ \\Once we obtain the mistag rate on a jet-per-jet basis, we need to do the same thing for the entire event. We add the mistag jet rates to obtain an event mistag rate. We sum all the event mistag probabilities to obtain the total mistag probability, $\epsilon^{mistag}$.

\ \\We measure the systematic uncertainty on the normalization of the W+LF background by fluctuating the per-jet tag rates by one standard deviation and then reproducing the entire procedure. 

\begin{figure}[htbp]
  \begin{center}
    \includegraphics[width=5.0cm]{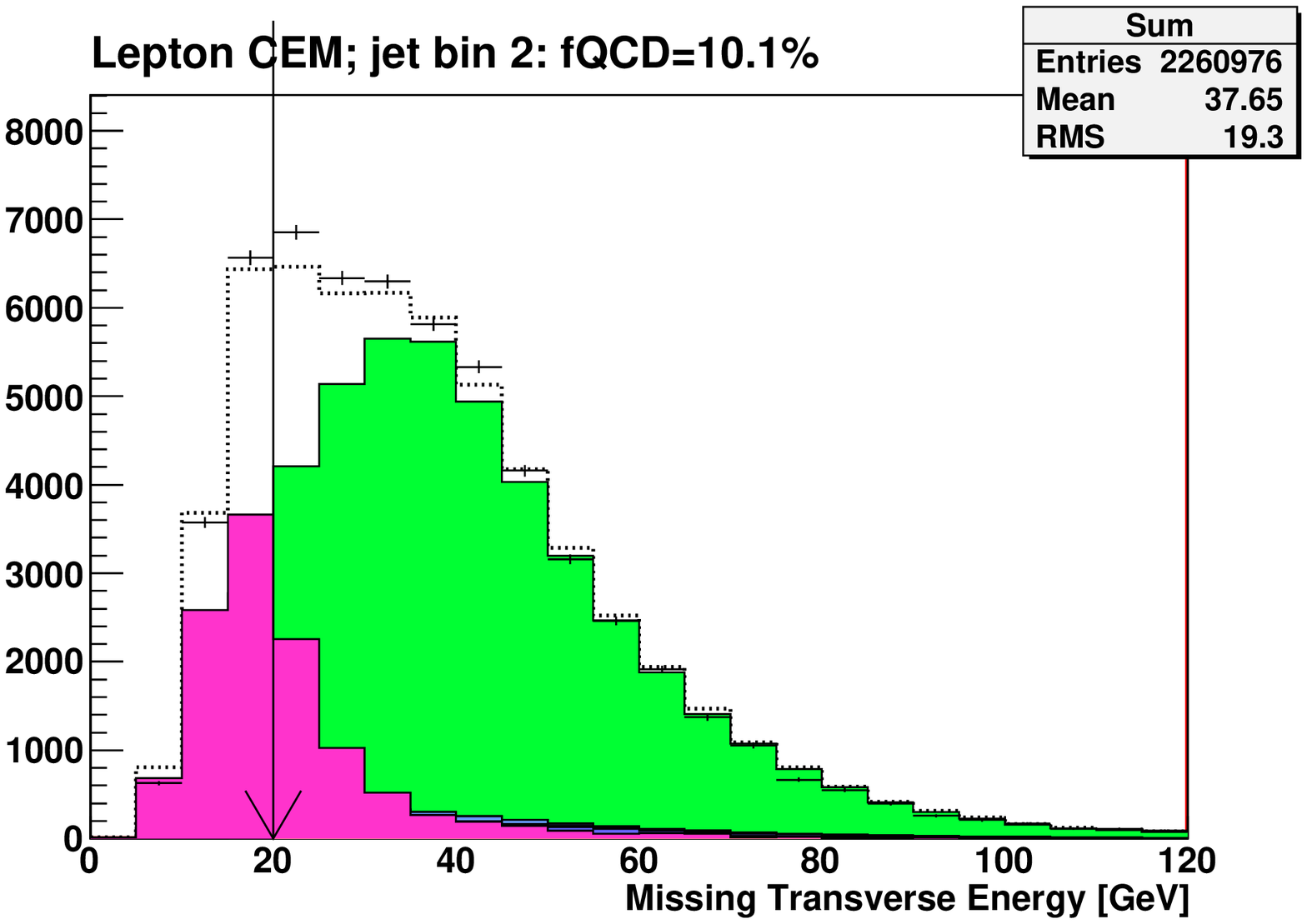}
    \includegraphics[width=5.0cm]{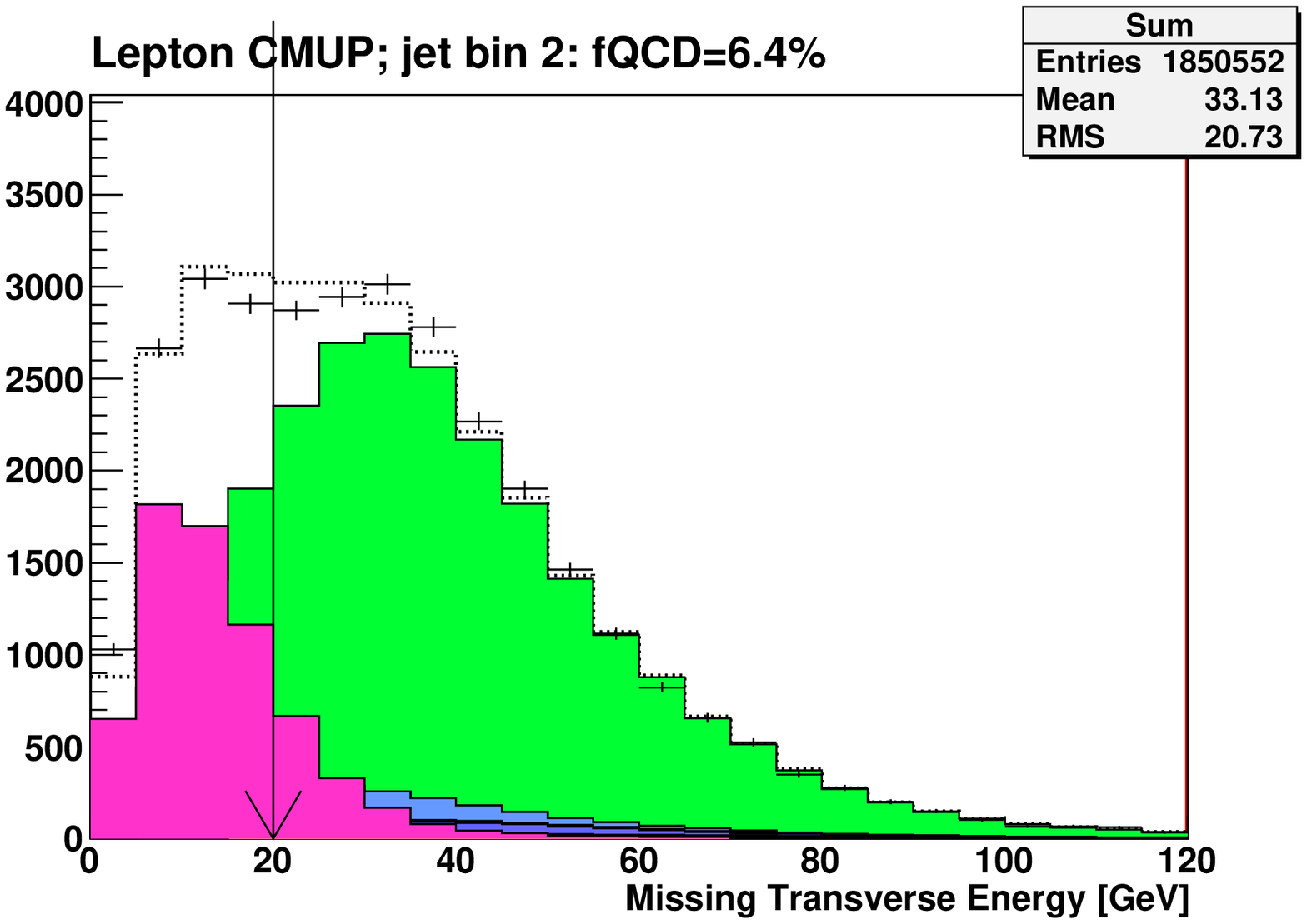}
    \includegraphics[width=5.0cm]{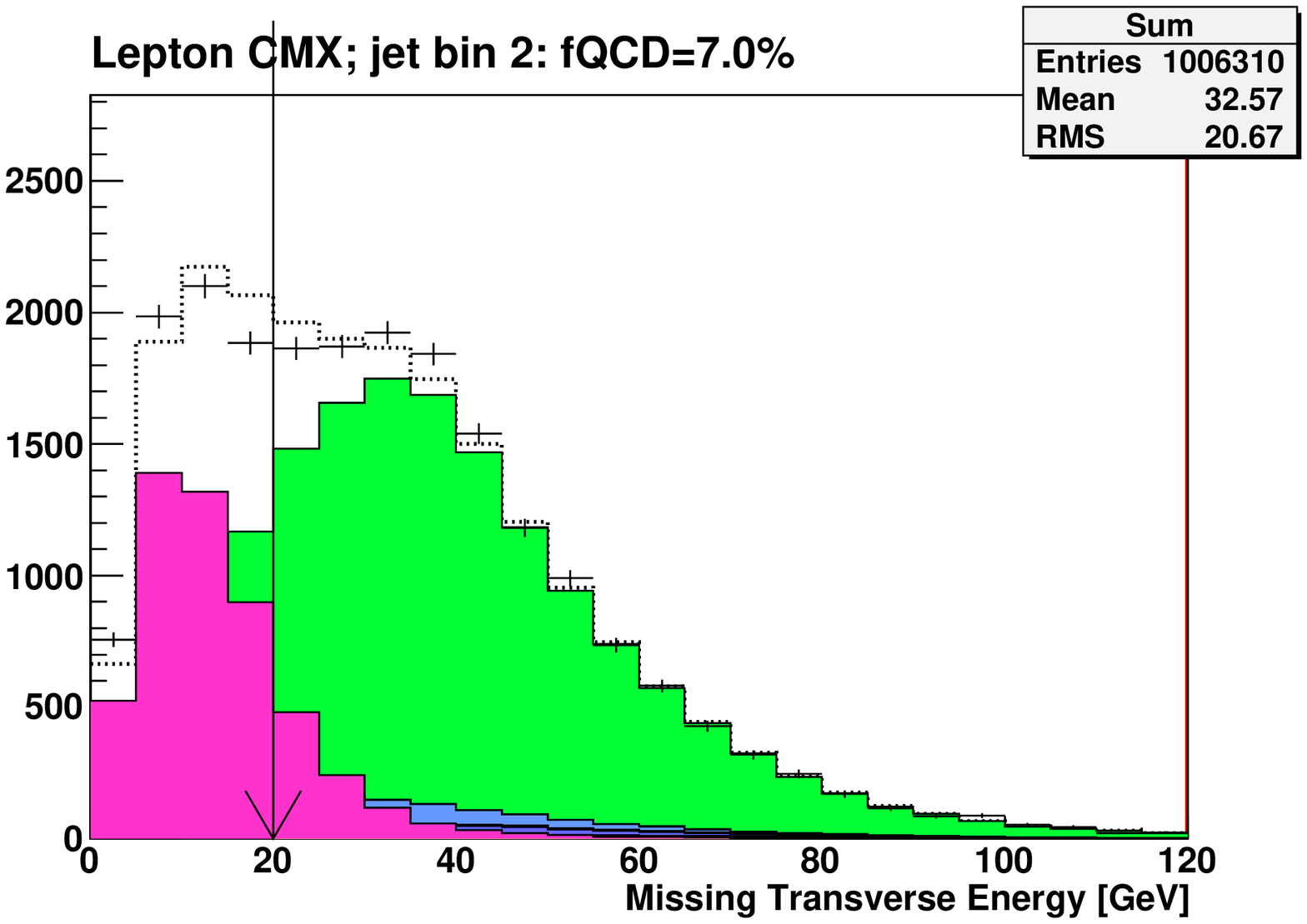}
    \includegraphics[width=5.0cm]{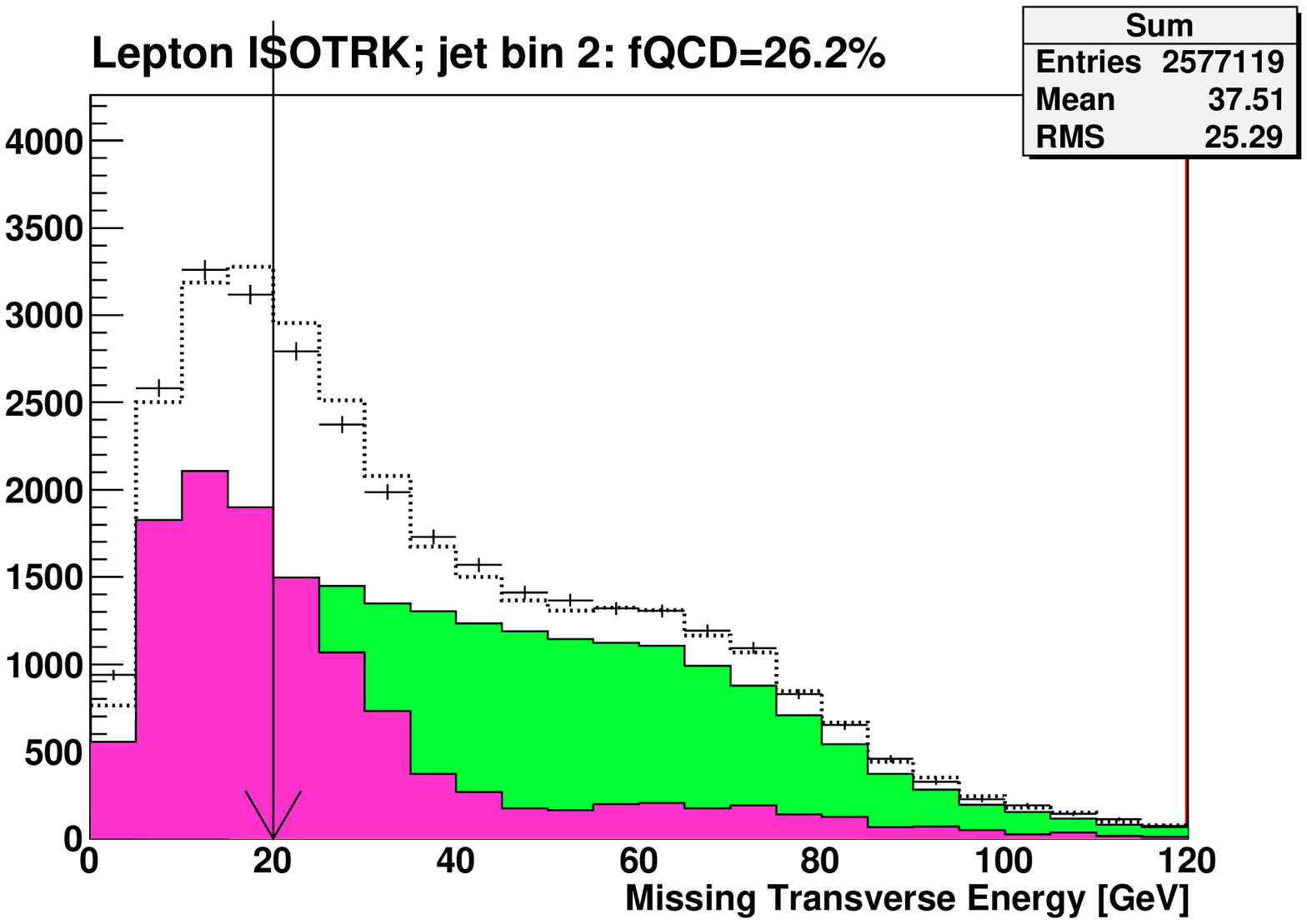}
    \caption[QCD (Non-W) fraction estimate for the Pretag sample]{QCD (Non-W) fraction estimate for the Pretag sample. The horizontal axis represents the fully corrected MET. The QCD background is represented in pink, the remainder of backgrounds in green. The dashed line represents the sum of all the backgrounds and the points represent the data. The figures represent (left to right and top to bottom) the CEM, CMUP, CMX and ISOTRK charged lepton categories.}
    \label{figure:Pretag_QCD}
  \end{center}
\end{figure}

\begin{figure}[htbp]
  \begin{center}
    \includegraphics[width=5.0cm]{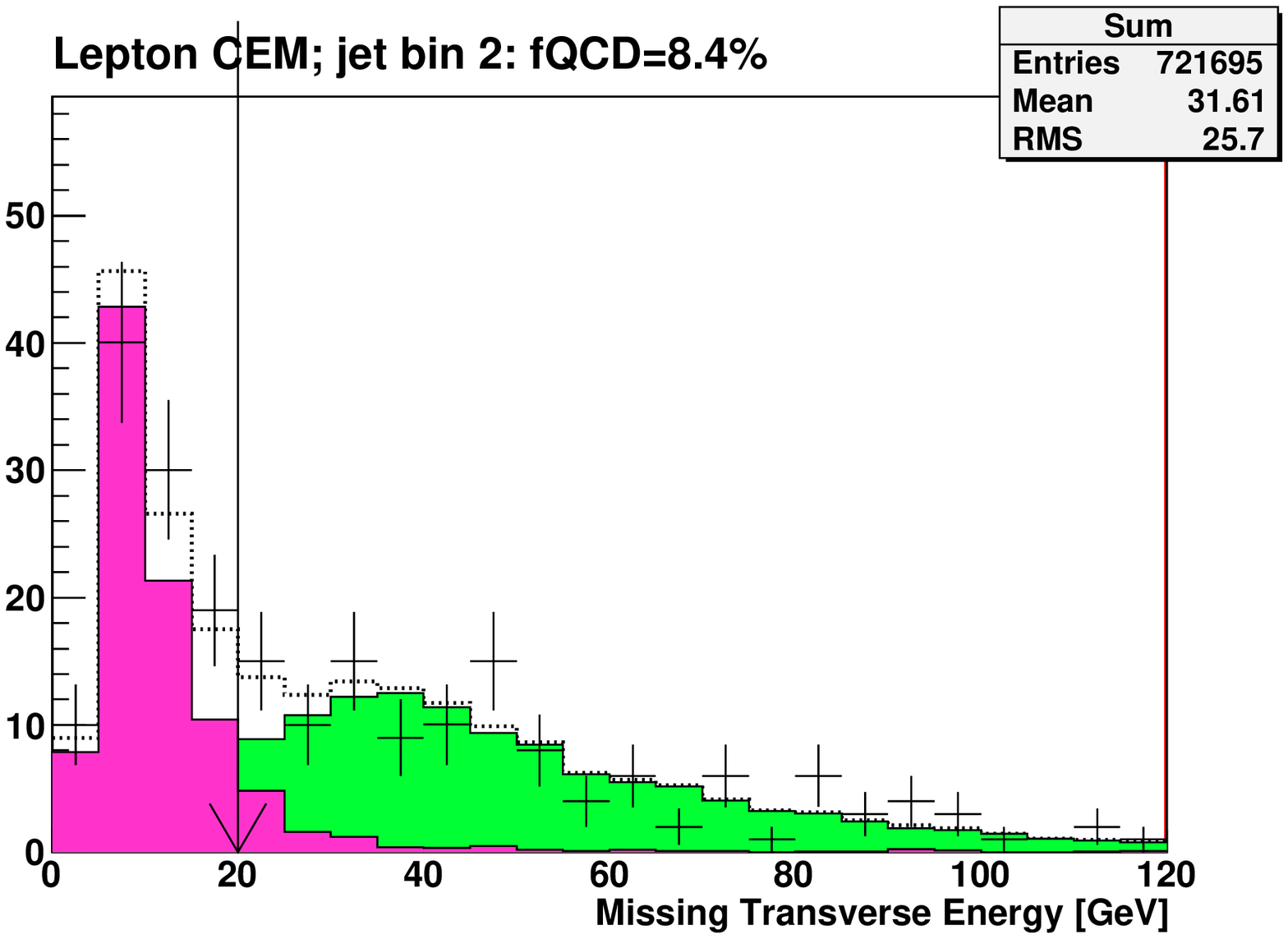}
    \includegraphics[width=5.0cm]{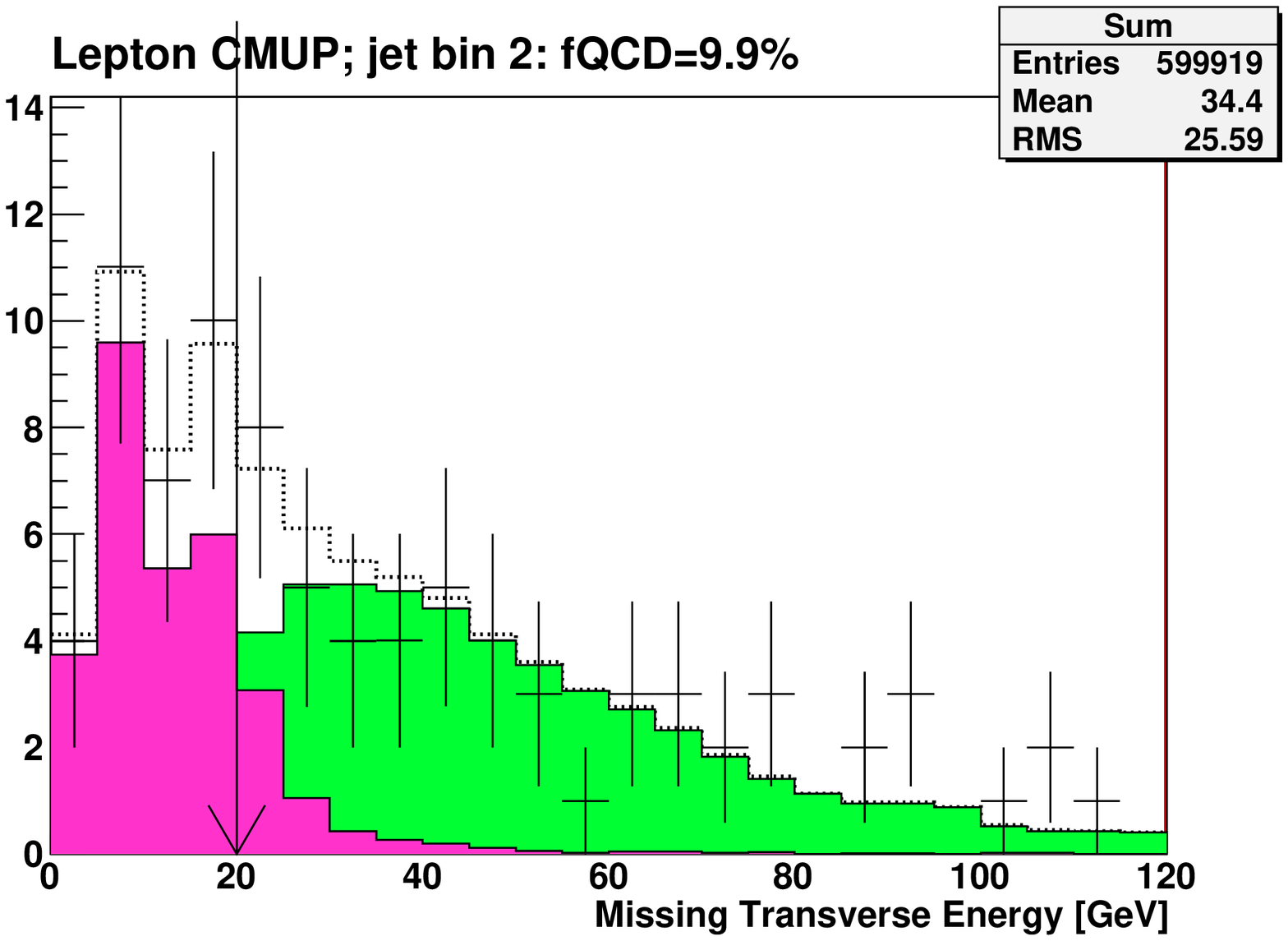}
    \includegraphics[width=5.0cm]{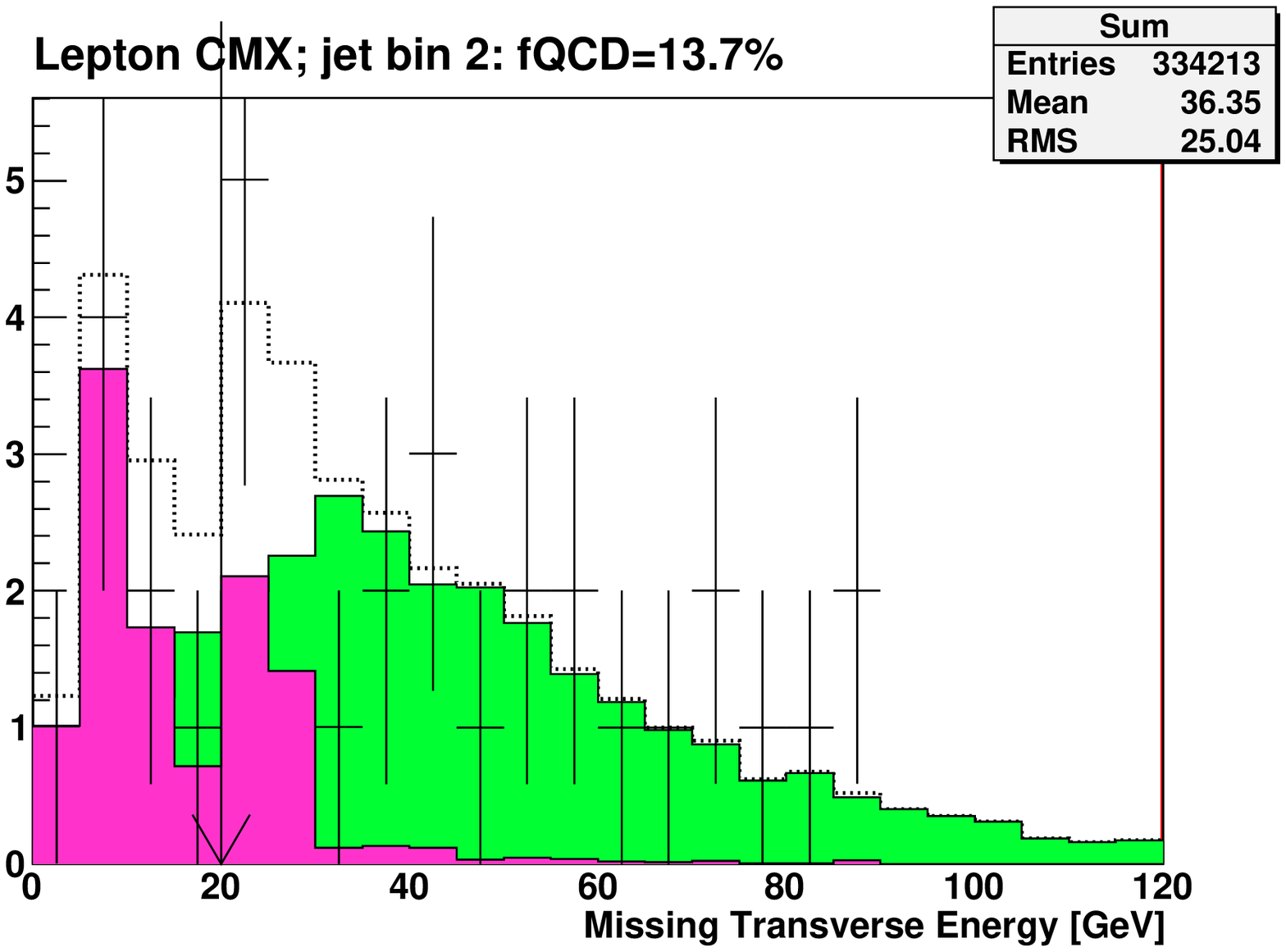}
    \includegraphics[width=5.0cm]{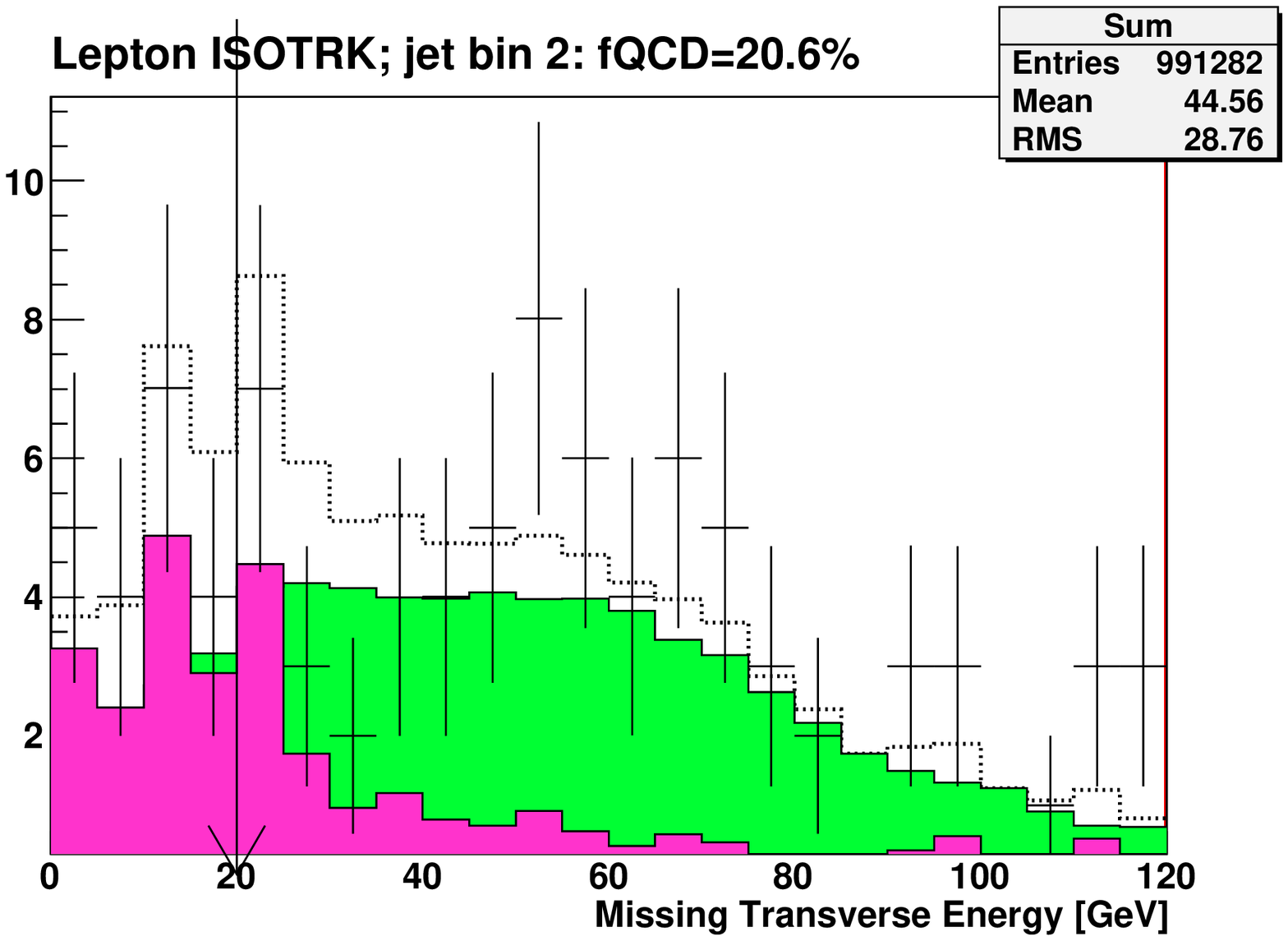}
    \caption[QCD (Non-W) fraction estimate for the SVTSVT sample]{QCD (Non-W) fraction estimate for the SVTSVT sample. The horizontal axis represents the fully corrected MET. The QCD background is represented in pink, the remainder of backgrounds in green. The dashed line represents the sum of all the backgrounds and the points represent the data. The figures represent (left to right and top to bottom) the CEM, CMUP, CMX and ISOTRK charged lepton categories.}
    \label{figure:SVTSVT_QCD}
  \end{center}
\end{figure}

\begin{figure}[htbp]
  \begin{center}
    \includegraphics[width=5.0cm]{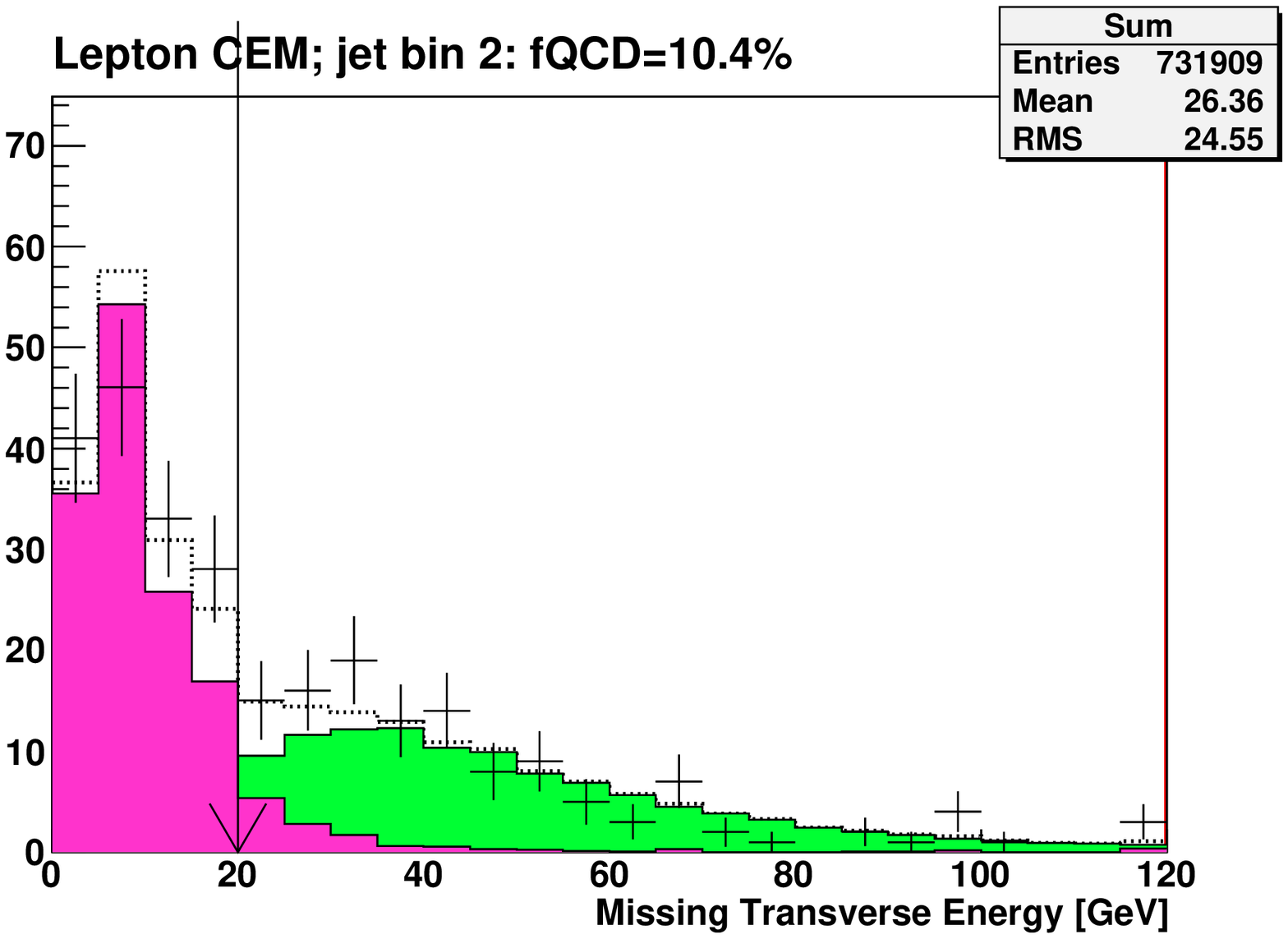}
    \includegraphics[width=5.0cm]{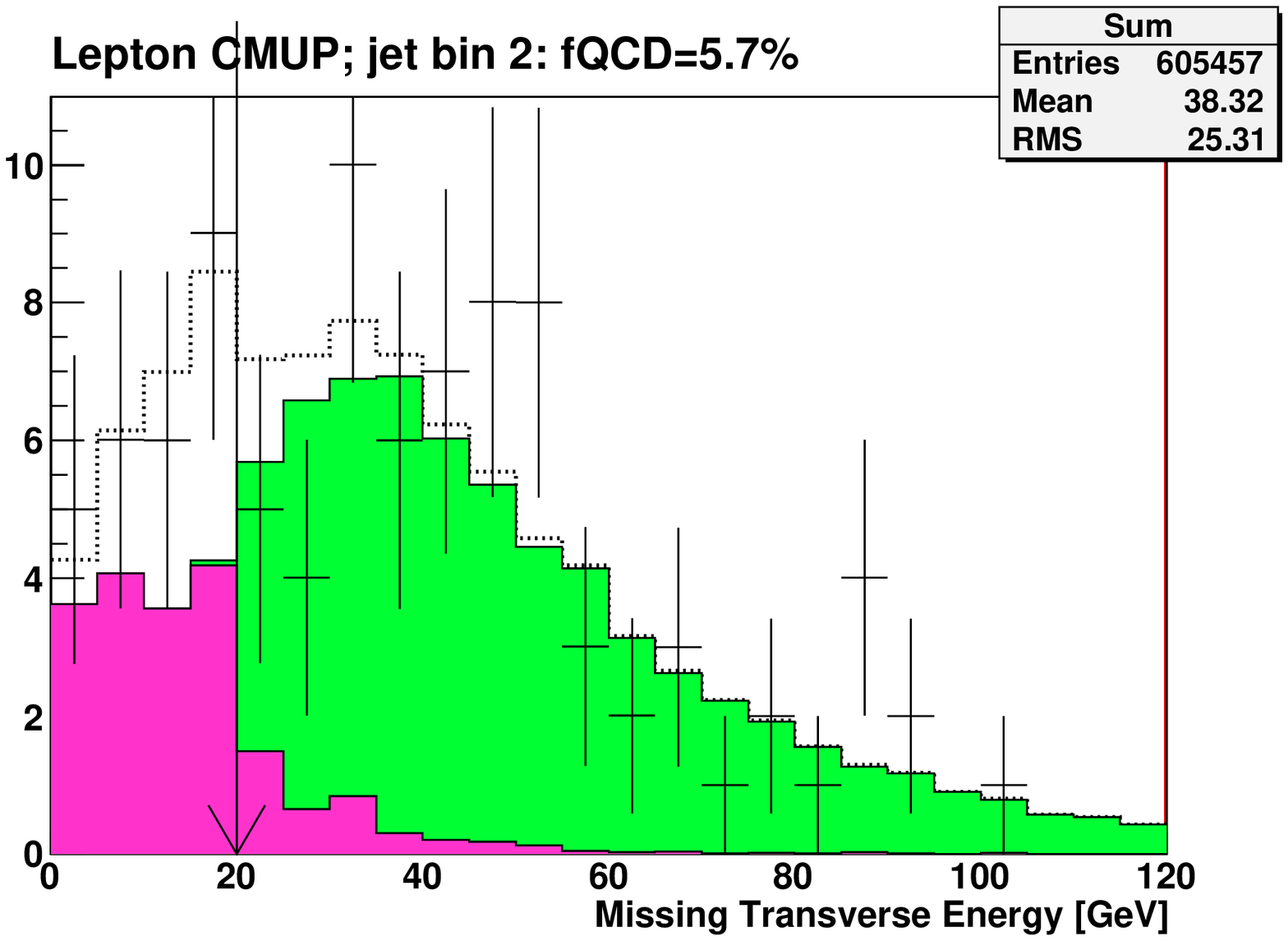}
    \includegraphics[width=5.0cm]{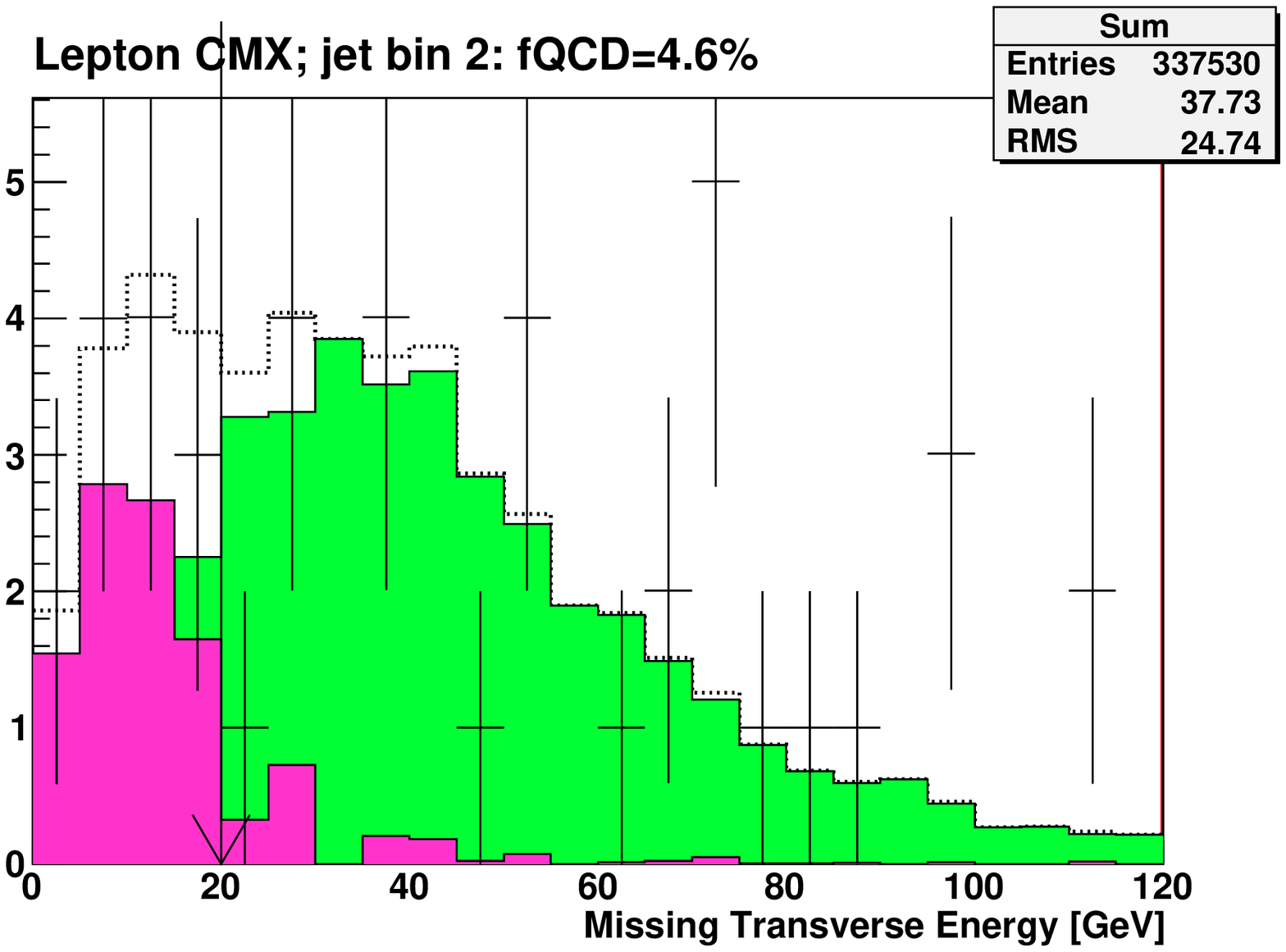}
    \includegraphics[width=5.0cm]{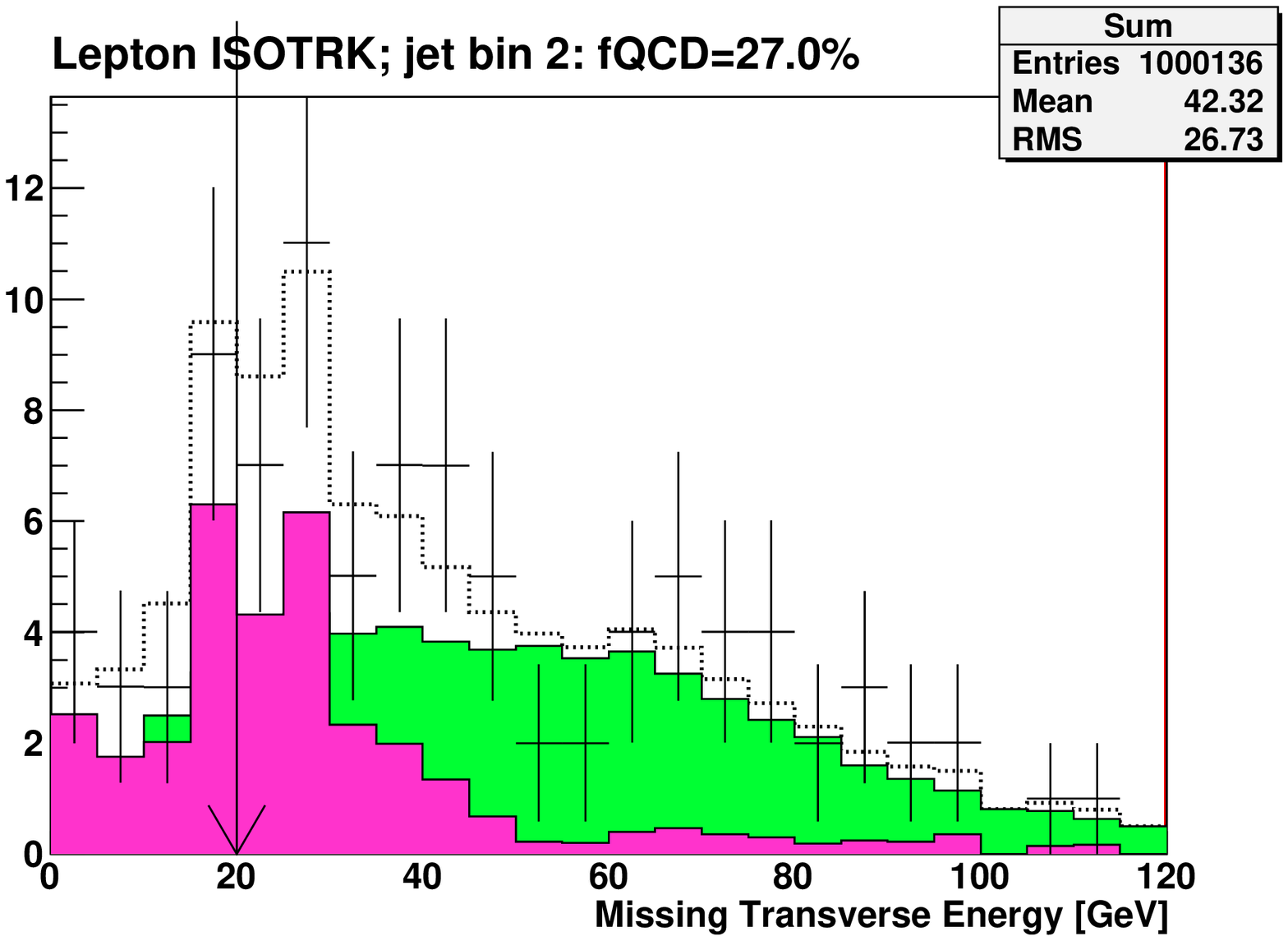}
    \caption[QCD (Non-W) fraction estimate for the SVTJP05 sample]{QCD (Non-W) fraction estimate for the SVTJP05 sample. The horizontal axis represents the fully corrected MET. The QCD background is represented in pink, the remainder of backgrounds in green. The dashed line represents the sum of all the backgrounds and the points represent the data. The figures represent (left to right and top to bottom) the CEM, CMUP, CMX and ISOTRK charged lepton categories.}
    \label{figure:SVTJP05_QCD}
  \end{center}
\end{figure}

\begin{figure}[htbp]
  \begin{center}
    \includegraphics[width=5.0cm]{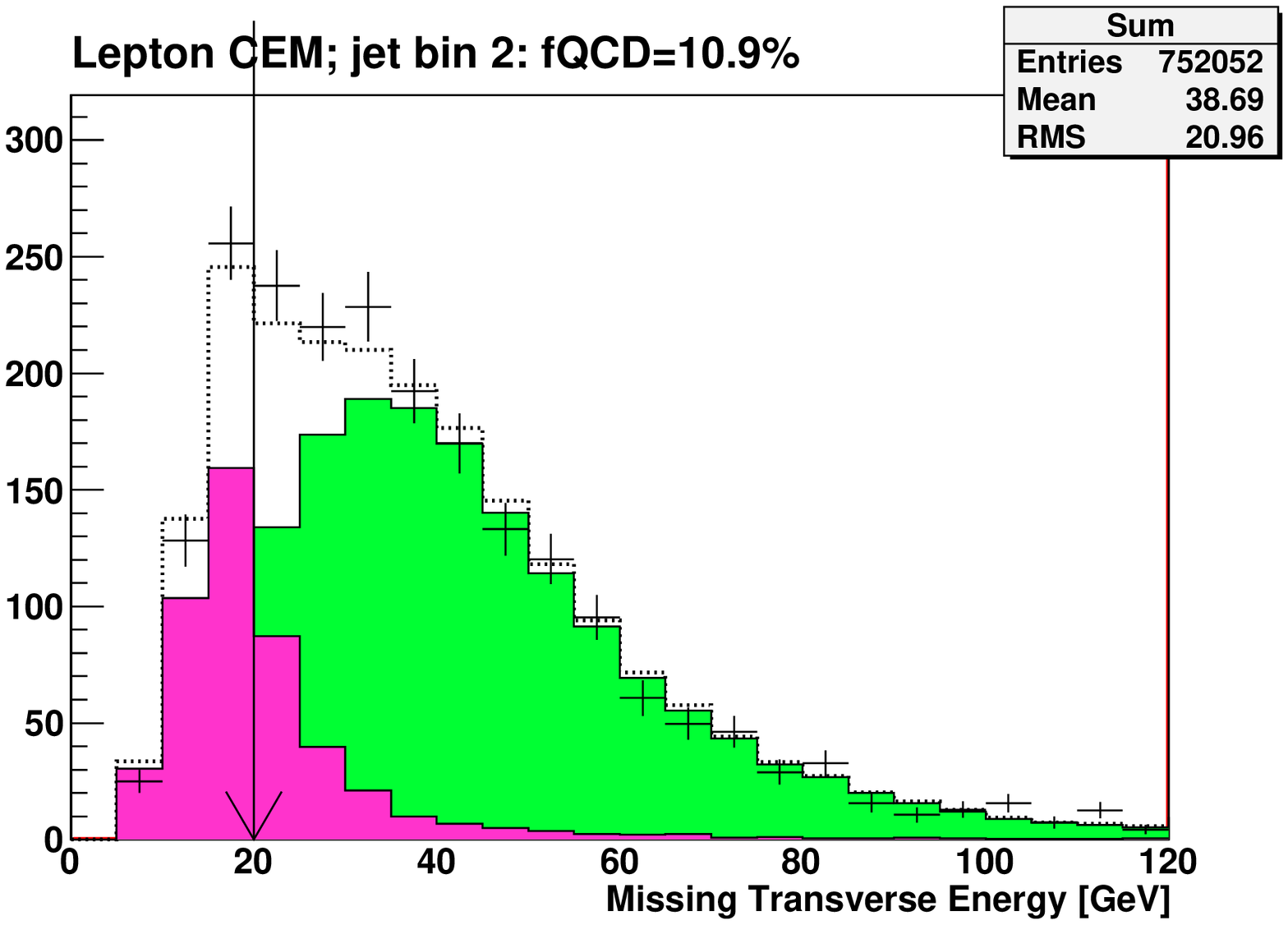}
    \includegraphics[width=5.0cm]{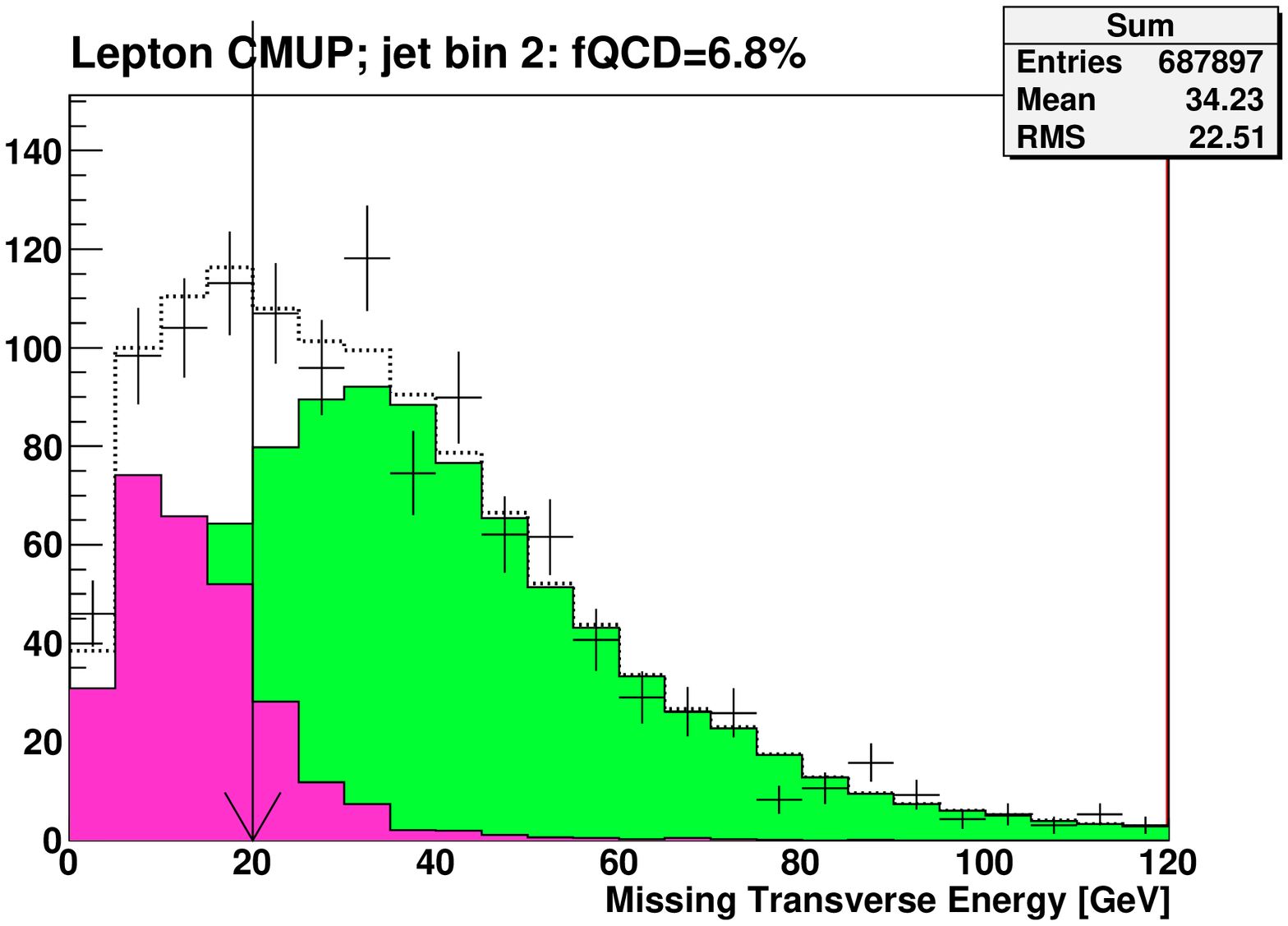}
    \includegraphics[width=5.0cm]{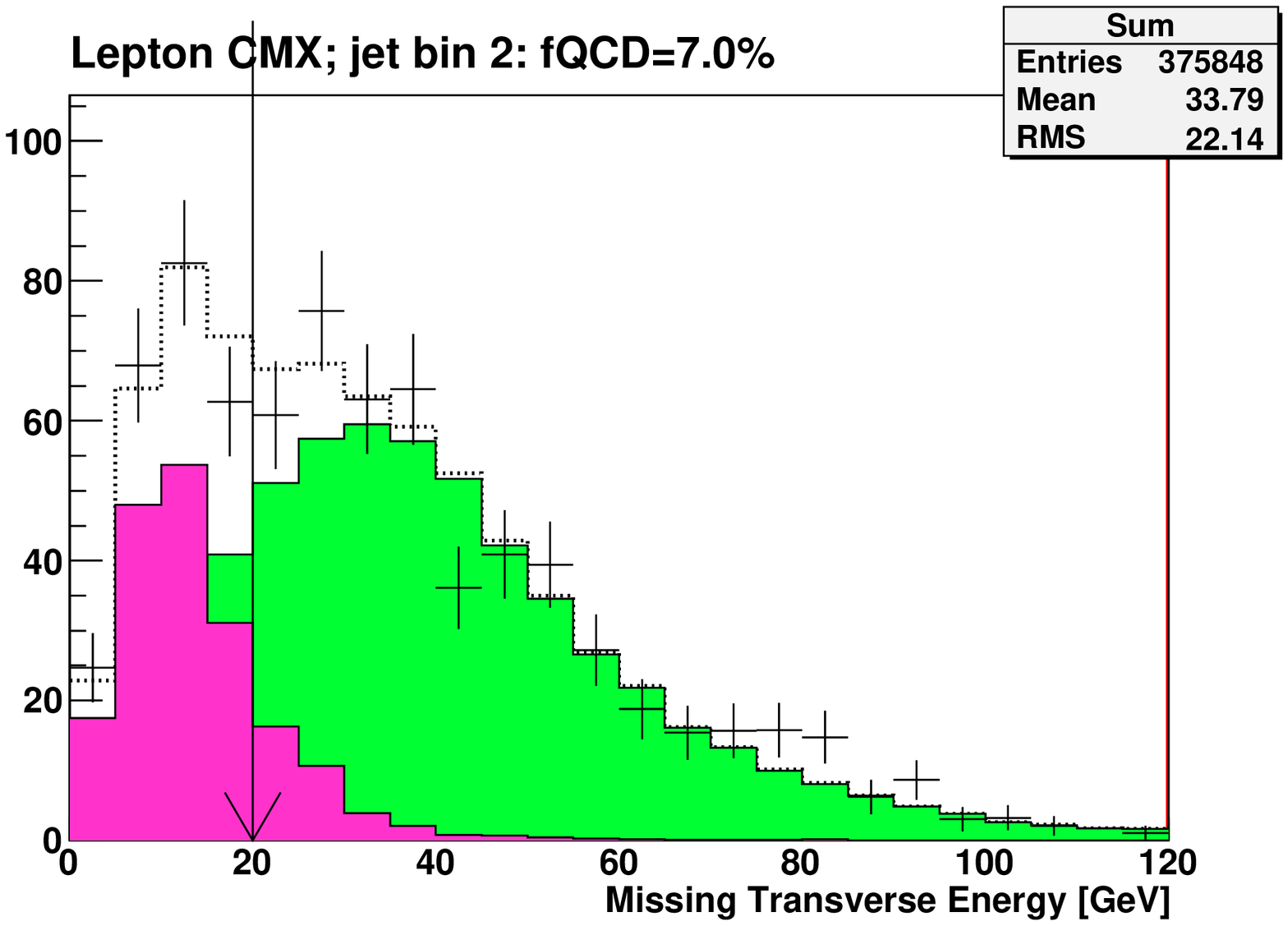}
    \includegraphics[width=5.0cm]{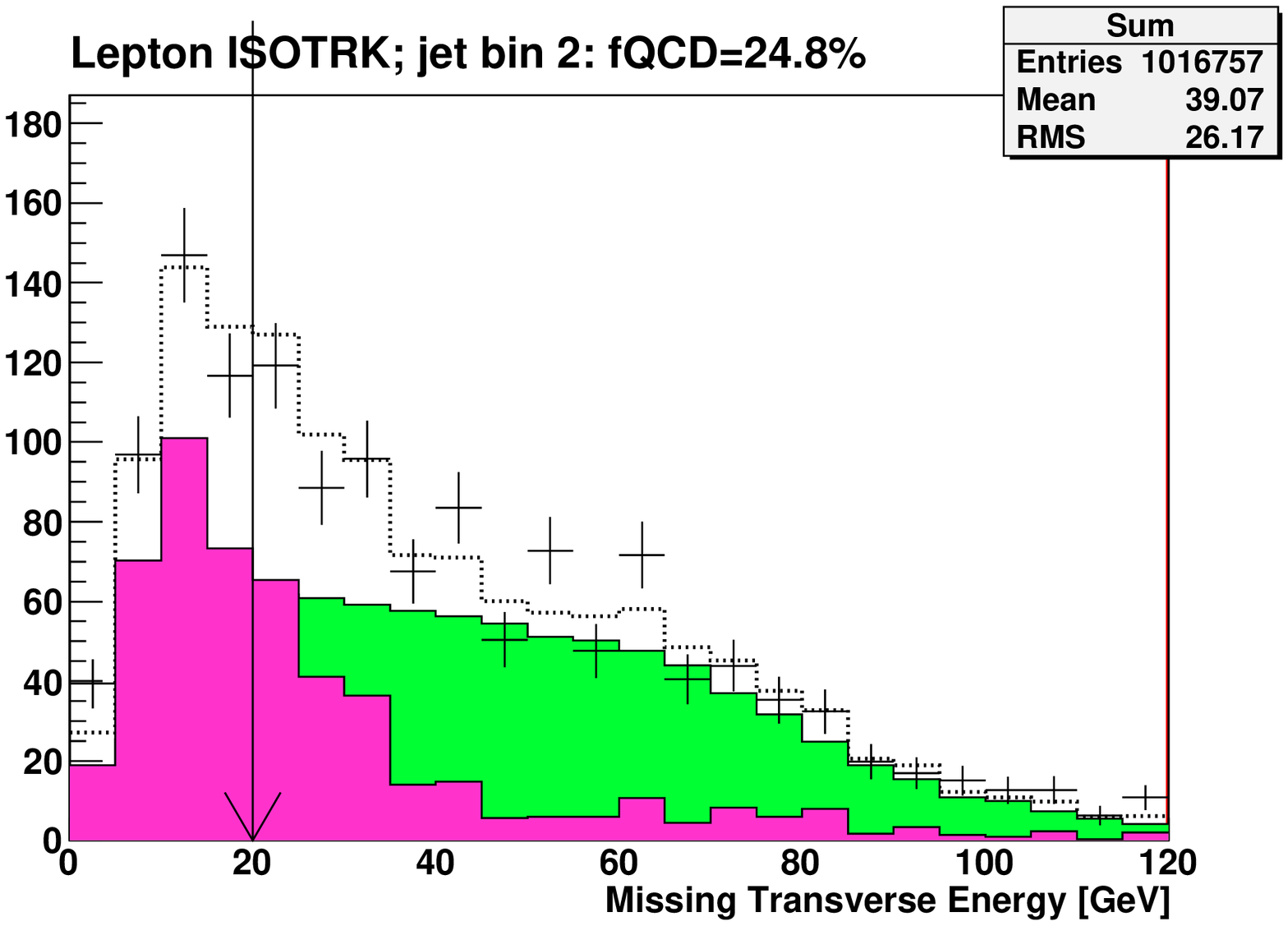}
    \caption[QCD (Non-W) fraction estimate for the SVTnoJP05 sample]{QCD (Non-W) fraction estimate for the SVTnoJP05 sample. The horizontal axis represents the fully corrected MET. The QCD background is represented in pink, the remainder of backgrounds in green. The dashed line represents the sum of all the backgrounds and the points represent the data. The figures represent (left to right and top to bottom) the CEM, CMUP, CMX and ISOTRK charged lepton categories.}
    \label{figure:SVTnoJP05_QCD}
  \end{center}
\end{figure}

\section{Background Fits and Event Counts}

\ \\Figures \ref{figure:Pretag_QCD} through \ref{figure:SVTnoJP05_QCD} show the results fitting the $\MET$ distribution in the Pretag and SVTSVT, SVTJP05 and SVTnoJP05 tag regions for each charged lepton category. We note that we fit separately for CEM, CMUP, CMX charged lepton categories and only later add the templates and event counts. We set a default 40\% systematic uncertainty on the QCD (non-W) background normalization. Due to low statistics, for ISOTRK SVTSVT and SVTJP05 we use a 100\% systematic uncertainty.

%\section{Event Counts for Each Analysis Channel}

\ \\Table \ref{table:EventCounts} shows the event counts for each of the six analysis channels given by the two charged lepton categories (TIGHT and ISOTRK) and the three $b$-tagging categories (SVTSVT, SVTJP05 and SVTnoJP05). The event counts are estimated using Monte-Carlo simulated samples for the signal and background processes and the data sample for the real observed events.  

\begin{table}
  \begin{center}
\begin{tabular}{|c|cc|cc|cc|}
\multicolumn{7}{c}{}\\
\hline
\multicolumn{7}{|c|}{CDF Run II Preliminary 5.7 fb$^{-1}$}\\
\multicolumn{7}{|c|}{Predicted number of events, Uncertainty and Percent of Uncertainty}\\
\hline \hline
   Sample Type  & SVTSVT & Err & SVTJP05 & Err & SVTnoJP05 & Err \\
\hline \hline
\multicolumn{7}{|c|}{TIGHT Charged Lepton}\\
\hline
   DiTop & 57.43 $\pm$ 8.46 & 0.15 & 46.2 $\pm$ 5.05 & 0.11 & 212.6 $\pm$ 21.5 & 0.1\\
   STopS & 22.04 $\pm$ 2.62 & 0.12 & 16.32 $\pm$ 1.42 & 0.087 & 56.33 $\pm$ 4.78 & 0.085\\
   STopT & 5.96 $\pm$ 0.8 & 0.13 & 5.98 $\pm$ 0.77 & 0.13 & 92.16 $\pm$ 9.42 & 0.1\\
   WW & 0.8 $\pm$ 0.2 & 0.25 & 2.97 $\pm$ 0.96 & 0.32 & 84.57 $\pm$ 11.8 & 0.14\\
   WZ & 5.17 $\pm$ 0.77 & 0.15 & 4.15 $\pm$ 0.49 & 0.12 & 23.38 $\pm$ 2.64 & 0.11\\
   ZZ & 0.19 $\pm$ 0.03 & 0.16 & 0.18 $\pm$ 0.03 & 0.17 & 0.83 $\pm$ 0.08 & 0.096\\
   Zjets & 3.19 $\pm$ 0.44 & 0.14 & 4.08 $\pm$ 0.56 & 0.14 & 62.48 $\pm$ 8.47 & 0.14\\
   Wbb & 117.98 $\pm$ 48.1 & 0.41 & 104.35 $\pm$ 42.1 & 0.4 & 694.19 $\pm$ 279 & 0.4\\
   Wcc & 13.54 $\pm$ 5.61 & 0.41 & 44.14 $\pm$ 18.3 & 0.42 & 793.33 $\pm$ 321 & 0.41\\
   Wlf & 6.54 $\pm$ 1.66 & 0.25 & 25.81 $\pm$ 8.86 & 0.34 & 913.58 $\pm$ 120 & 0.13\\
   QCD & 20.37 $\pm$ 8.14 & 0.4 & 18.92 $\pm$ 7.57 & 0.4 & 271.08 $\pm$ 108 & 0.4\\
\hline
   Bkg & 253.21 $\pm$ 76.8 & 0.3 & 273.1 $\pm$ 86.1 & 0.32 & 3204.53 $\pm$ 888 & 0.28\\
\hline
   WH115 & 3.02 $\pm$ 0.39 & 0.13 & 2.2 $\pm$ 0.15 & 0.068 & 7.52 $\pm$ 0.53 & 0.07\\
   ZH115 & 0.13 $\pm$ 0.01 & 0.077 & 0.09 $\pm$ 0 & 0 & 0.33 $\pm$ 0.03 & 0.091\\
\hline
   Obs & 213 & 0 & 234 & 0 & 2952 & 0\\

\hline \hline
\multicolumn{7}{|c|}{ISOTRK Charged Lepton}\\
\hline
\hline
   DiTop & 22.54 $\pm$ 3.49 & 0.15 & 18.48 $\pm$ 2.23 & 0.12 & 85.07 $\pm$ 9.48 & 0.11 \\
   STopS & 7.98 $\pm$ 1.02 & 0.13 & 5.95 $\pm$ 0.59 & 0.099 & 20.39 $\pm$ 1.98 & 0.097 \\
   STopT & 2.2 $\pm$ 0.31 & 0.14 & 2.17 $\pm$ 0.3 & 0.14 & 31.47 $\pm$ 3.54 & 0.11 \\
   WW & 0.24 $\pm$ 0.06 & 0.25 & 0.85 $\pm$ 0.3 & 0.35 & 23.17 $\pm$ 3.4 & 0.15 \\
   WZ & 1.49 $\pm$ 0.24 & 0.16 & 1.23 $\pm$ 0.15 & 0.12 & 6.97 $\pm$ 0.86 & 0.12 \\
   ZZ & 0.11 $\pm$ 0.02 & 0.18 & 0.07 $\pm$ 0.01 & 0.14 & 0.39 $\pm$ 0.04 & 0.1 \\
   Zjets & 2.05 $\pm$ 0.3 & 0.15 & 2.64 $\pm$ 0.39 & 0.15 & 37.01 $\pm$ 5.22 & 0.14 \\
   Wbb & 29.97 $\pm$ 13.1 & 0.44 & 25.89 $\pm$ 11.2 & 0.43 & 164.47 $\pm$ 71.2 & 0.43 \\
   Wcc & 3.36 $\pm$ 1.48 & 0.44 & 10.21 $\pm$ 4.54 & 0.44 & 162.51 $\pm$ 70.6 & 0.43 \\
   Wlf & 2.09 $\pm$ 0.64 & 0.31 & 7.91 $\pm$ 3.15 & 0.4 & 246.06 $\pm$ 56.7 & 0.23 \\
   QCD & 15.42 $\pm$ 6.17 & 0.4 & 20.22 $\pm$ 8.09 & 0.4 & 230.14 $\pm$ 92.1 & 0.4 \\
\hline
   Bkg & 87.45 $\pm$ 26.8 & 0.31 & 95.62 $\pm$ 31 & 0.32 & 1007.65 $\pm$ 315 & 0.31 \\
\hline
   WH115 & 1.01 $\pm$ 0.14 & 0.14 & 0.74 $\pm$ 0.07 & 0.095 & 2.46 $\pm$ 0.21 & 0.085\\
   ZH115 & 0.09 $\pm$ 0.01 & 0.11 & 0.06 $\pm$ 0.01 & 0.17 & 0.22 $\pm$ 0.02 & 0.091\\
\hline
   Obs & 75 & 0 & 75 & 0 & 929 & 0 \\
\hline
\hline
\end{tabular}
  \caption[Summary of background, signal and data event counts]{Summary of background and signal predicted event number and data observed event number for each of the six analysis channels. The table uses the following notations for backgrounds and signals: DiTop - top quark pair; STopS - single top s channel; STopT - single top t channel; WW - $WW$; WZ - $WZ$; ZZ - $ZZ$; Zjets - $Z$ plus jets; Wbb - $W$ plus $b\bar{b}$; Wcc - $W$ plus $c\bar{c}$ or $W$ plus $cj$; Wlf - $W$ plus light flavour jets incorrectly tagged as heavy flavour jets (mistags); QCD - non-$W$ (QCD); WH115 (ZH115) - $WH$ ($ZH$) assuming a $115 \gevcc$ mass for the Higgs boson. The systematic uncertainties are added linearly in a conservative but realistic approach, since most of the systematic uncertainties are correlated between all background processes.}
  \label{table:EventCounts}
  \end{center}
\end{table}

\section{Summary}

\ \\This chapter has presented the methodology to calculate the event yield prediction for the various background processes. Since any blind analysis compares a control sample with an analysis sample, we define the control sample as the Pretag sample and the signal samples as each of the three $b$-tagging categories. Although the signal samples are included in the Pretag samples, since the Pretag sample is much larger this is a very good approximation to the orthogonality of the control sample and the pretag sample. We then presented the computation of the top quark and electroweak background event yield by using the same methodology used for the signal processes. It is only the $W$+jets and the QCD background yields that are determined from a missing transverse energy fit to the data distribution. A slightly more complex procedure is used for the $b$-tagging category than for the Pretag sample. We presented the fit plots in order to demonstrate the quality of the fits. We concluded with the presentation of the table with the event yield for the signal and background processes, as well as the measured event counts, for each charged lepton and $b$-tagging category, as well as the uncertainty on these values due to the systematic uncertainty. In all categories, the background prediction and data agree within the systematic uncertainty for the background. 

\ \\Since we do not see a signal excess, we employ multivariate techniques to separate the signal and background even more. We detail these in the following chapter. 

\clearpage{\pagestyle{empty}\cleardoublepage}

\chapter{Neural Network Discriminant\label{chapter:Discriminant}}

\ \\The event selection is optimized to separate as much as possible the signal and background processes. Requiring $b$-tagging separates them even more. For a given $b$-tagging category, a $WH$ search based on a counting experiment would be a search for a bump in the distribution of the invariant mass of the two jets (dijet invariant mass). However, since the expected signal is about two orders of magnitude smaller than the background prediction, as seen in Chapter \ref{chapter:Background}, such a bump would not be visible. The main reason why a counting experiment $WH$ analysis is not sensitive enough is that the dijet invariant mass describes just the Higgs candidate system and ignores any information about the reconstructed $W$ boson candidate (charged lepton and missing transverse energy), as well as any correlation between the $W$ and Higgs systems. 

\ \\For improved sensitivity, in this analysis we use a multivariate technique. In general, a multivariate technique uses several kinematic distributions to discriminate further the signal events from the background events. The multivariate technique chosen for this search is an artificial neural network. 

\section{Artificial Neural Networks Overview}

\ \\Artificial Neural Networks (ANN) are multivariate technique functions that are produced through an iterative training process. An ANN is formed of several interconnected nodes. Each connection is weighted by a sigmoid function. The first layer contains input nodes, the last layer contains the output nodes, and the rest of the nodes are organized into intermediate hidden layers. Using a feed-forward process, an ANN allows the information to flow from the input nodes to the output nodes. For a given event, the input nodes receive values of certain kinematic quantities and the output nodes give the neural network output values for that event. 

\begin{figure}[ht]
\begin{center}
\includegraphics[width=7.0cm]{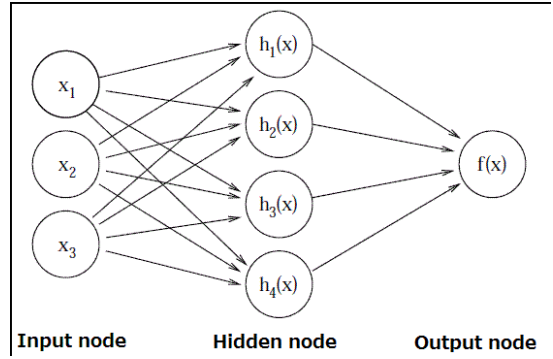}
\caption[Schematic view for an artificial neural network]
{The schematic view of an artificial neural network used in this analysis to produce the final discriminant between signal and background \cite{YoshikazuNagaiThesis}.\label{figure:SchematicNeuralNetwork}}
\end{center}
\end{figure}

\section{Neural Network Structure}

\ \\In this analysis we use as a final discriminant the output of an ANN with only one hidden intermediate layer and with only one node in the output layer, as seen in Figure~\ref{figure:SchematicNeuralNetwork}. The ANN is basically a function that takes various quantities from the event as an input and returns only one number. 

\ \\If there are $N_{inputs}$ nodes in the input layer, every node $j$ in the hidden layer is described by a sigmoid function that depends on the neural network input values $x(x_1, x_2, \cdots, x_i, \cdots, x_{N_{inputs}})$:

\begin{equation} 
h_j(x) = \frac{1}{1+\text{exp}(-\displaystyle{\sum_{i}u_{ij}x_i})}\,\rm{.}
\label{NeuralNetworkWeighting}
\end{equation}

\ \\The weights $u_{ij}$ are determined by training the artificial neural network. Once trained, the ANN output from the only node in the third layer is computed using a linear combination between the hidden layer values:

\begin{equation} 
f(x) = \sum_{j}v_jh_j(x)\,\rm{,}
\label{NeuralNetworkOutput}
\end{equation}

\ \\where the weights $v_j$ are also determined by training the neural network. 

\section{Neural Network Training}

\ \\Before the training starts, all the weights  ($h_{ij}$ and $v_j$) have some initial values. At the end of the training process, these weights would be such that the ANN output is as close as possible to the chosen target value ($t$), which means we optimize the quantity $E$:

\begin{equation} 
E = \frac{1}{2}(f(x)-t)^2\,\rm{.}
\label{NeuralNetworkTraining}
\end{equation}

\ \\In this analysis, we choose for the ANN a target value of 1 for signal samples and 0 for background samples. Therefore we use signal and background samples for the training process. We divide these samples in half. One half is used for training and the other half is used for validation of the training on independent dataset samples.

\ \\For each event from the training sample, we minimize the quantity $E$. We then use back propagation to change the values of the weights ($h_{ij}$ and $v_j$) with an amount proportional to the ANN output to the target. The training is finished when new events would not change any more these weights in a significant way.

\ \\This is not the first search that uses an ANN at CDF. It is now a standard multivariate technique that is used in high energy physics experiments. 

\ \\There are many types of ANN. In this search we use a Bayesian Neural Network (BNN) algorithm~\cite{BayesianNeuralNetwork}~\cite{BayesianNeuralNetworkBook}. The advantage of BNN over other artificial neural network algorithms is that it is less prone to over training because of the Bayesian statistical interpretation where each weight is considered as a posterior probability in Bayes' theorem. 

\ \\For this analysis, we employ distinct BNN discriminant functions which were optimized for one of the three $b$-tagging categories: SVTSVT, SVTJP05 and SVTnoJP05. 

\begin{figure}[htbp]
  \begin{center}
    \includegraphics[width=6.5cm]{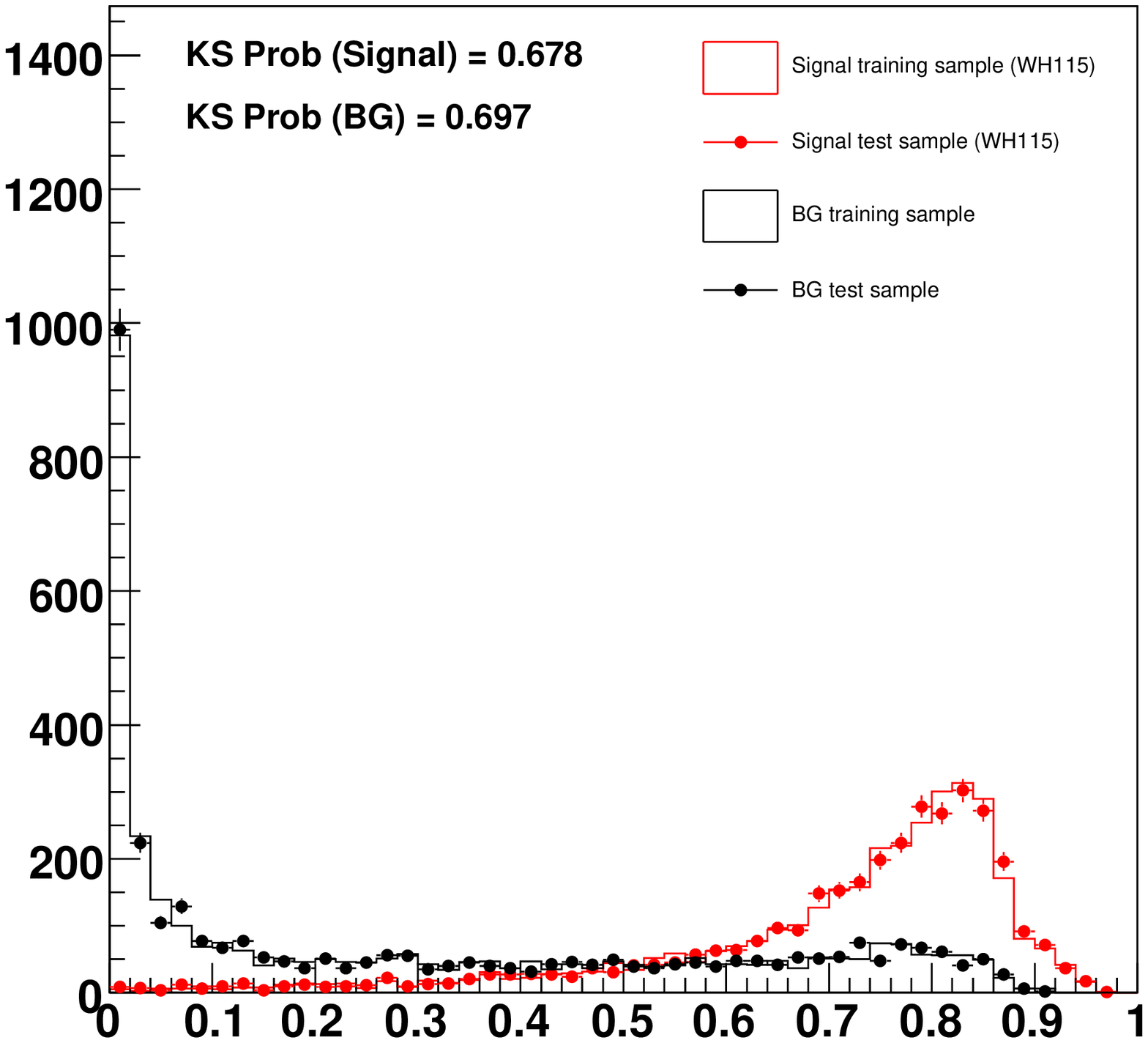}
    \includegraphics[width=6.5cm]{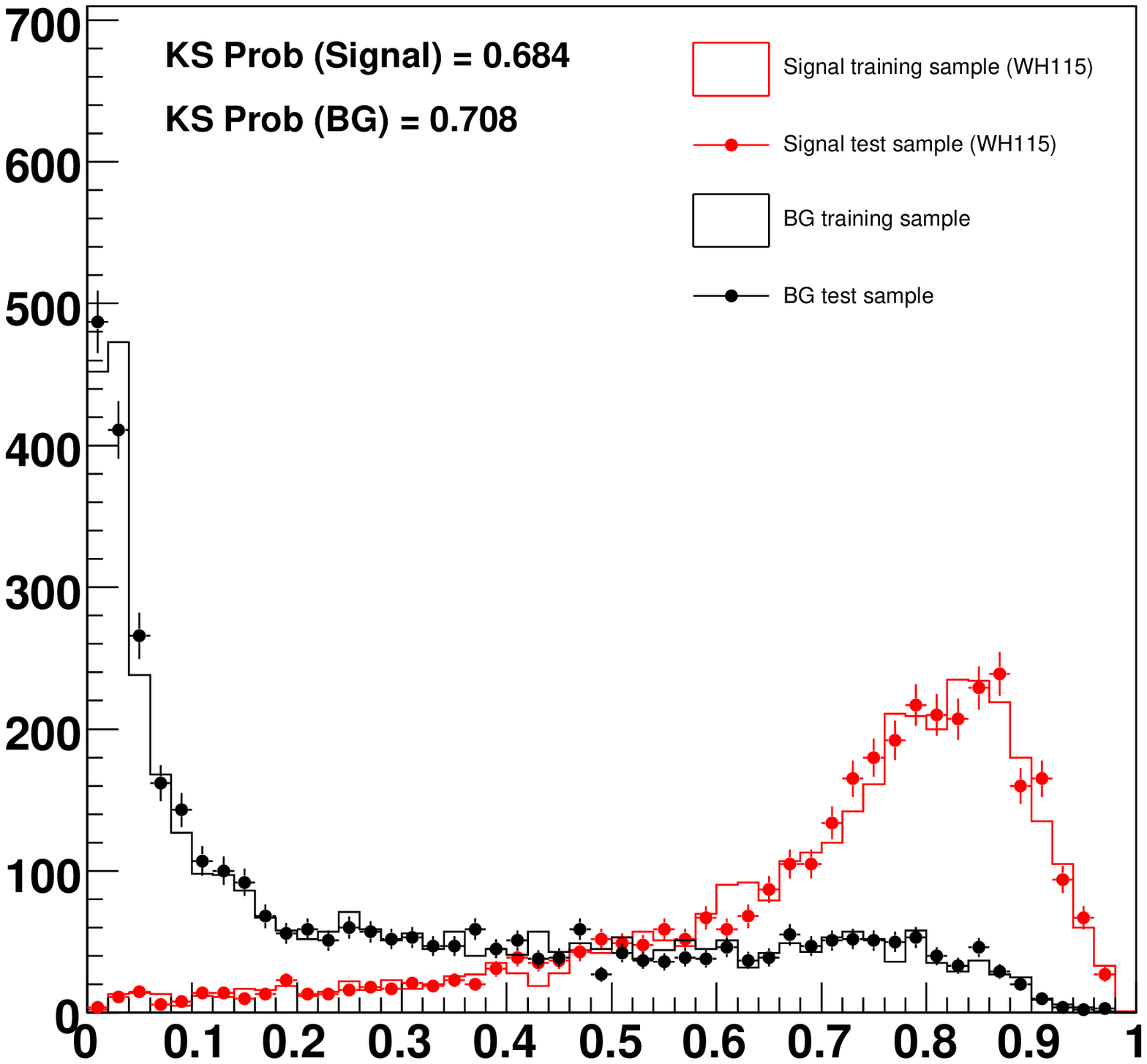}\\
    \includegraphics[width=6.5cm]{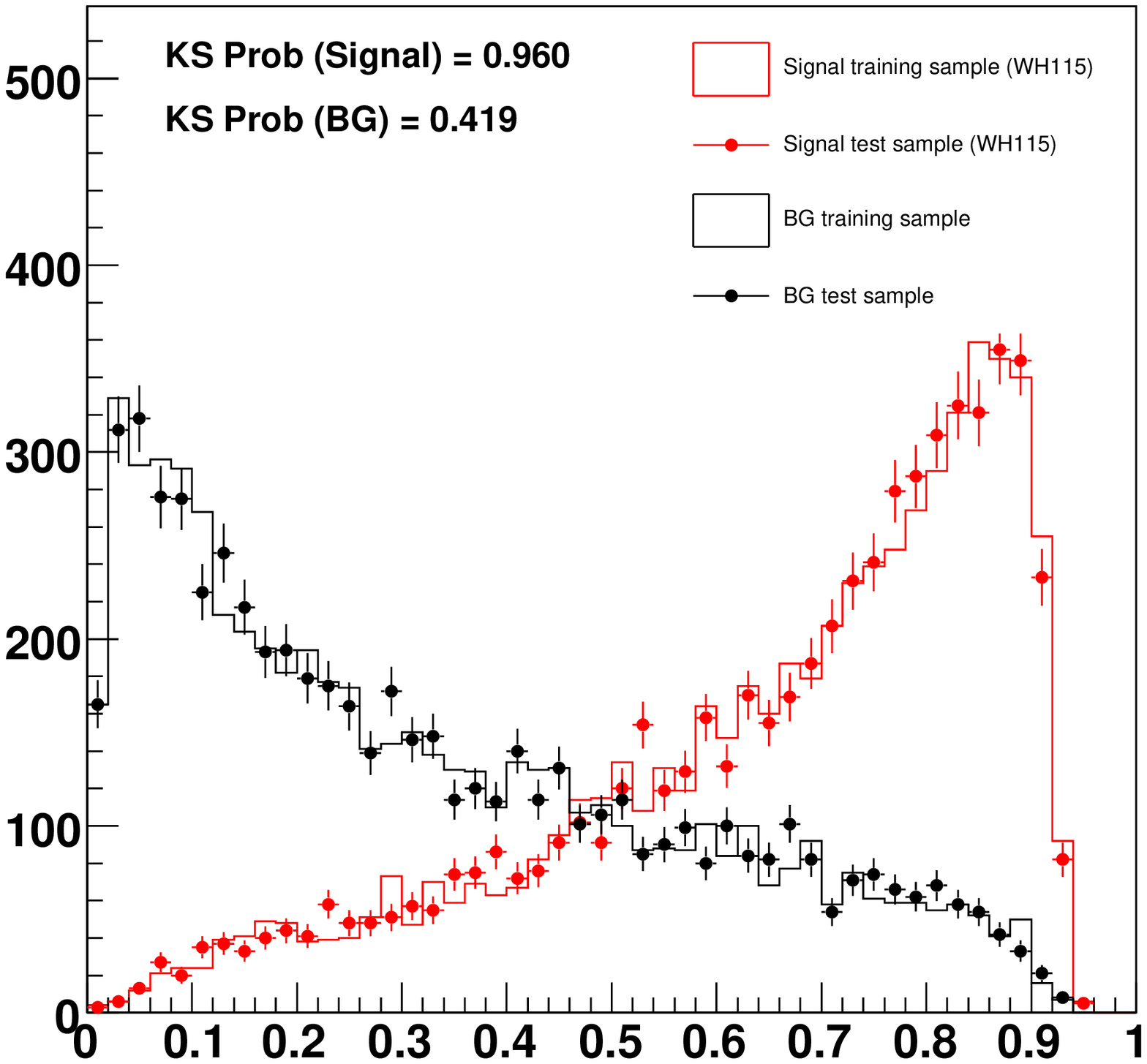}
\caption[Overtraining check for our Bayesian artificial neural network]
{Bayesian Neural Network output for $WH$ signal (red) and background (black) simulated events with a Higgs boson mass of $115 \gevcc$ for the training sample (solid lines) and orthogonal test sample (circles) \cite{YoshikazuNagaiThesis}. We can see that indeed the signal peaks at output values of 1 and background at output values of 0. We also see that the two samples give the same shape and normalization, which confirms that the BNN training is to be trusted. The top left plot refers to the SVTSVT $b$-tagging category, the top right plot refers to the SVTJP05 $b$-tagging category and the bottom plot refers to the SVTnoJP05 $b$-tagging category. \label{figure:NeuralNetworkOvertrainingCheck115}}
  \end{center}
\end{figure}

\section{Neural Network Training Check}

\ \\Once training is done, we perform a check for overtraining. Overtraining may happen when the neural network ''learns'' the training sample instead of ``generalizing'' a model from the training sample. In overtraining the smallest statistical fluctuation is believed to be part of the real model. For this reason, an overtrained neural network has very little predictive power. If overtrained, the neural network output distribution for an independent test sample will not be the same as the one for the training sample.

\ \\We check that our neural networks are not overtrained by comparing the training sample shape to that for a test sample which was not used in training. Figure \ref{figure:NeuralNetworkOvertrainingCheck115} shows examples of an overtraining check for a Higgs mass of 115 $\gevcc$. We conclude that our BNN discriminant is not overtrained, since the response of the training sample is in good agreement with that of the test sample.

\section{Neural Network Inputs}

\ \\Each BNN is optimized independently to separate the $WH$ signal from the various background processes (W+HF, W+LF, top quark pair, single top s-channel production). We use distinct BNN input quantities for each tagging category.

\ \\For the SVTSVT $b$-tagging category, seven kinematic quantities are used as inputs. The first one is the invariant mass of the two jets in the event ($M_{jj}$). This quantity is computed after both jets have their energies corrected using another type of artificial neural network, as described in Section~\ref{Section:JetNeuralNetwork}. The second quantity is the scalar sum of the transverse momenta of the charged lepton and the jets, from which the missing transverse energy is subtracted ($\pt$ imbalance). The third one is the invariant mass of the charged lepton, missing transverse energy and one of the two jets, where we choose the jet that produces the largest invariant mass ($M_{l\nu j}^{max}$). The fourth one is the charge of the charged lepton multiplied by its $\eta$ coordinate ($Q_{lep}\cdot \eta_{lep}$). The fifth one is the scalar sum of the transverse energy of the loose jets of the events ($\sum \et (\rm{loose\ jets)}$). The loose jets are orthogonal to the tight jets and have $\et > 12 \gev$ and $|\eta|<2.4$. The tight jets are explicitly not included in this summation. The sixth one is the transverse momentum of the reconstructed $W$ boson candidate, computed as the vector sum of the transverse momentum of the charged lepton and missing transverse energy ($\pt (W)$). The seventh one is the scalar sum of the transverse energies of all the objects in the event, such as jets, charged lepton and missing transverse energy ($H_T$).

\ \\For the SVTJP05 $b$-tagging category, we also use seven kinematic quantities as inputs. Five of them are the same as for the SVTSVT category, namely $M_{jj}$, $Q_{lep}\cdot \eta_{lep}$, $\sum \et (\rm{loose\ jets)}$, $\pt (W)$ and $H_T$. Instead of $M_{l\nu j}^{max}$ we use $M_{l\nu j}^{min}$, which means we pick the jet that minimizes the quantity. We also use missing transverse energy ($\met$) instead of $\pt$ imbalance. Both changes are motivated by the fact that for each $b$-tagging category we have tested a large number of input parameter combinations and we have chosen the inputs that give the largest sensitivity in that category.

\ \\For the SVTnoJP05 $b$-tagging category, we also use seven kinematic quantities as inputs, namely $M_{jj}$, $Q_{lep}\cdot \eta_{lep}$, $\sum \et (\rm{loose\ jets)}$, $\pt (W)$, $H_T$, $\met$ and $\pt$ imbalance. 

\section{Background Modelling Check}

\ \\We plot the distribution of each of the inputs and outputs for BNN to check that the total background prediction agrees with the observed data distribution, as seen in Figures~\ref{fig:ControlPlots1} ~\ref{fig:ControlPlots2} and in those of Appendix \ref{chapter:ControlPlots}.  This procedure cross checks that our various background processes are modelled well. As all distributions agree in normalization and shape with the data distributions, we are safe to use the BNN output in the final limit calculation.

\begin{figure}[ht]
  \begin{center}
    \includegraphics[width=6.7cm]{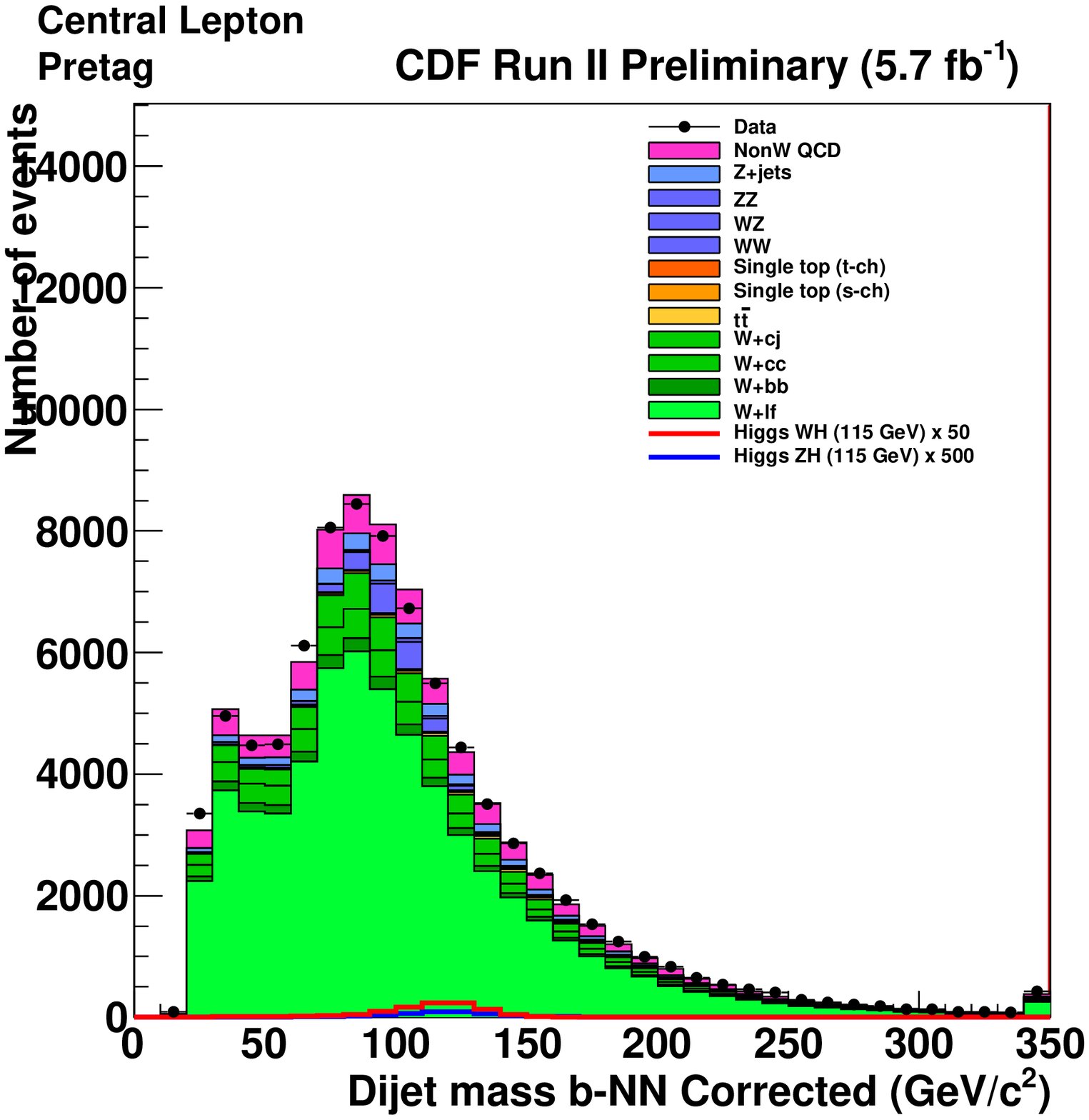}
    \includegraphics[width=6.7cm]{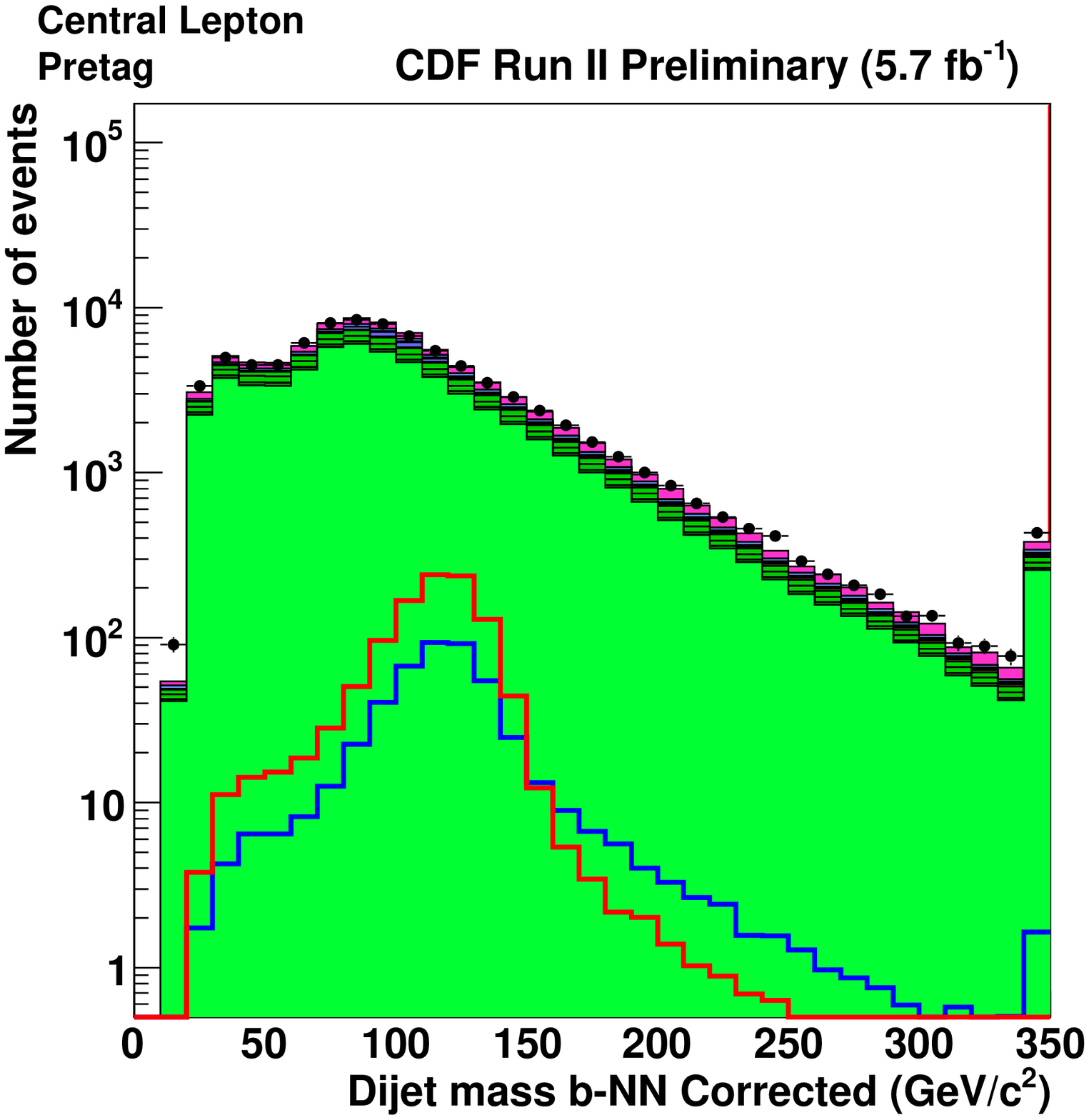}
    \includegraphics[width=6.7cm]{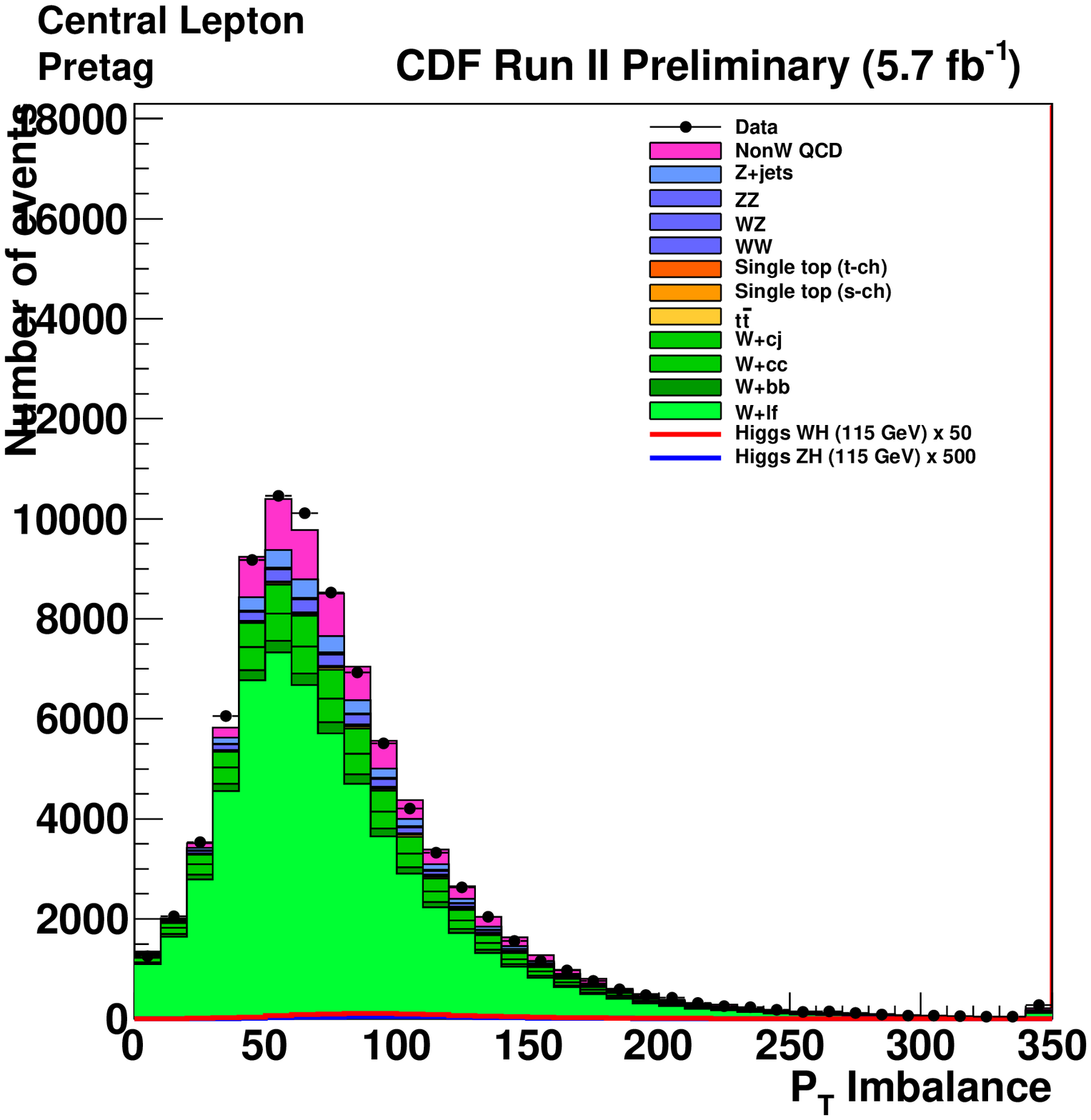}
    \includegraphics[width=6.7cm]{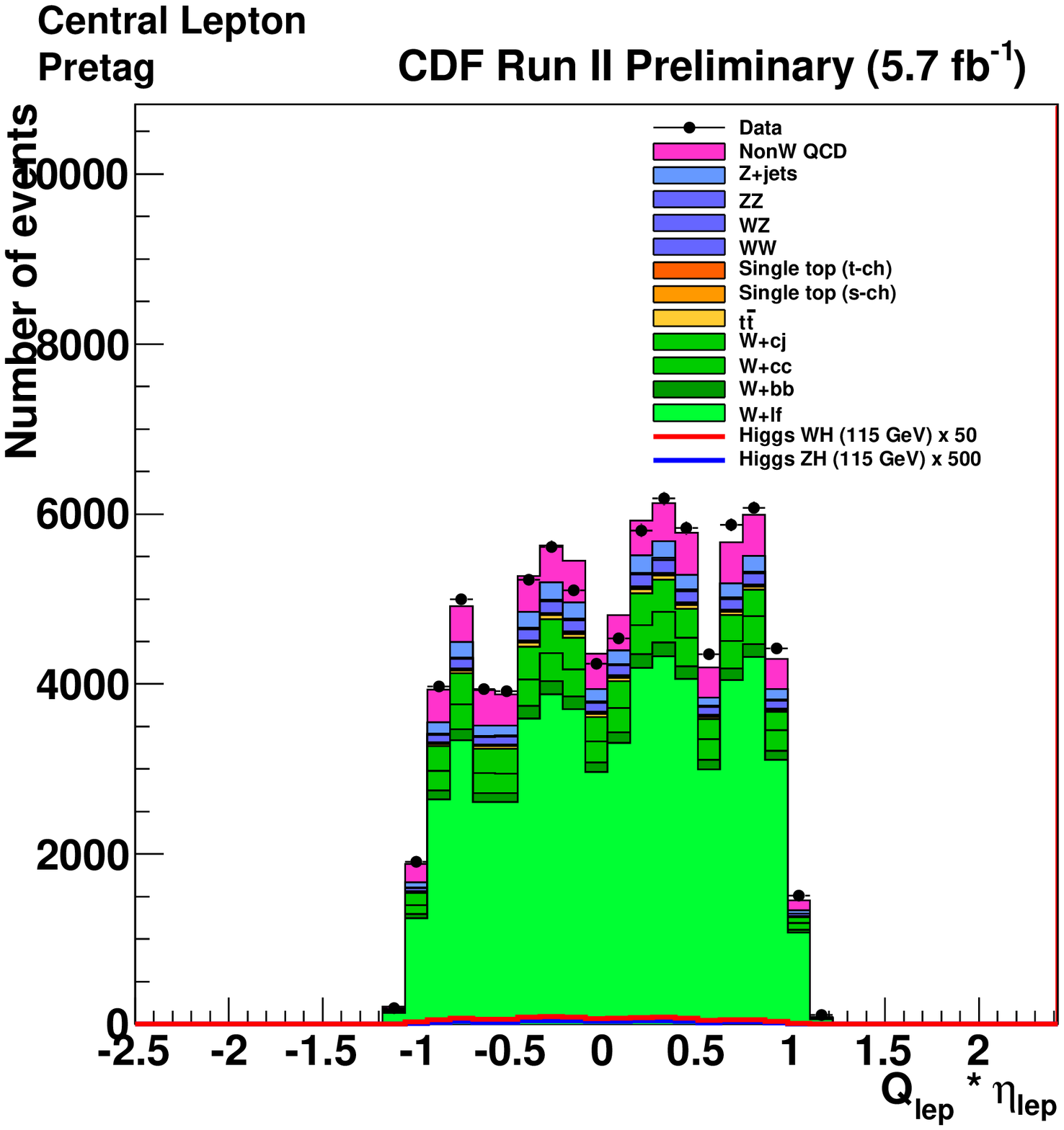}
    \includegraphics[width=6.7cm]{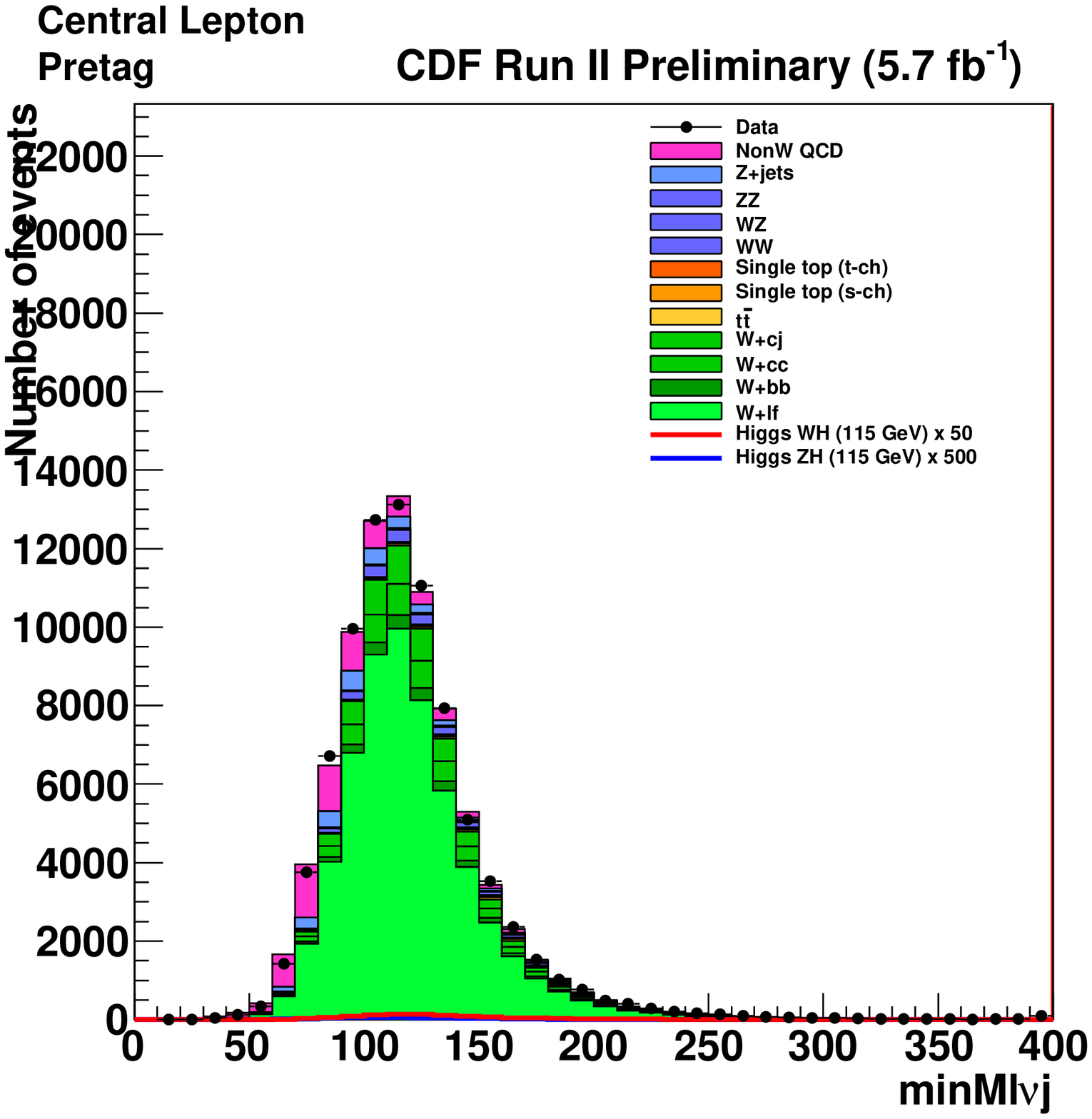}
    \includegraphics[width=6.7cm]{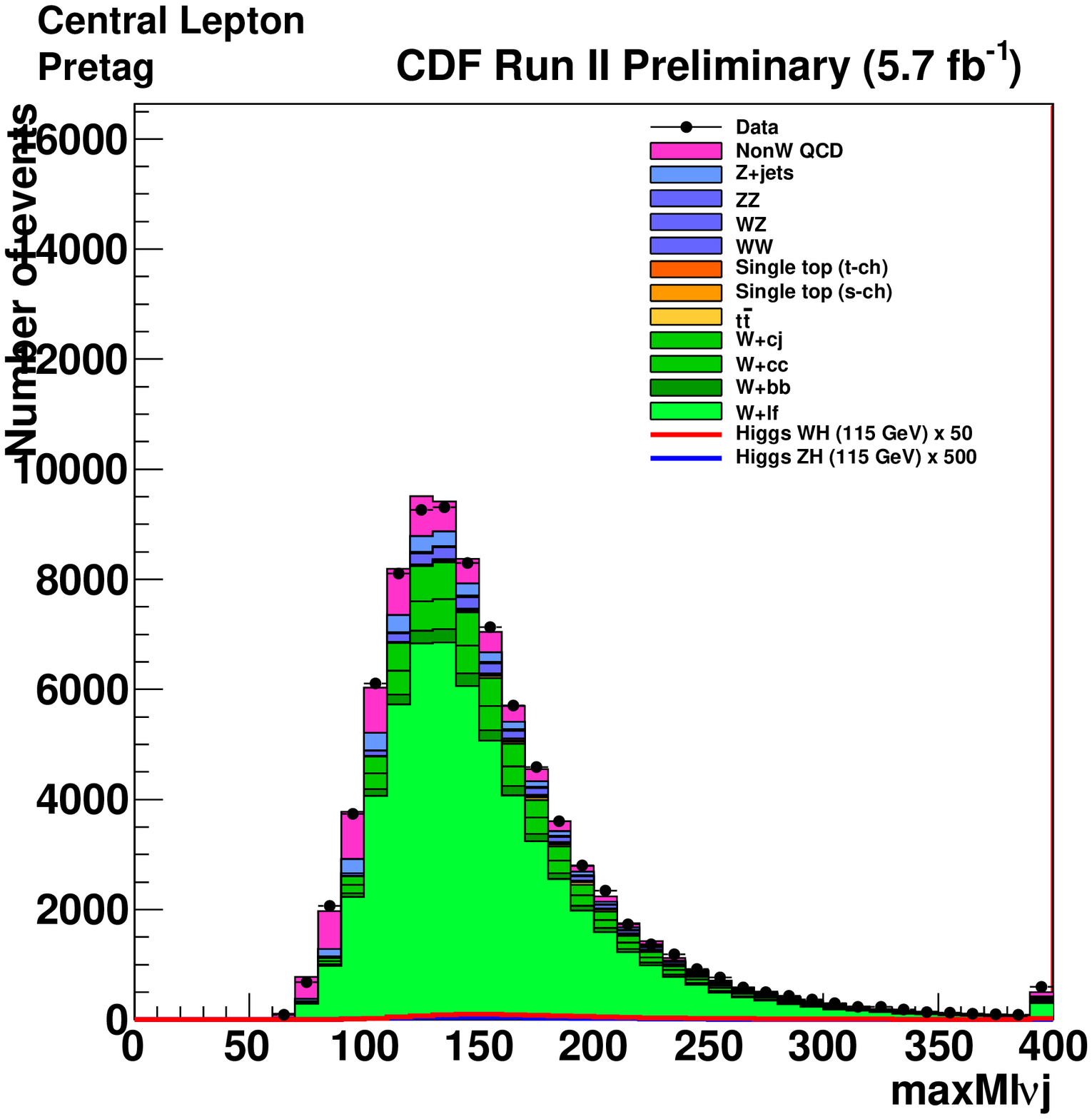}
    \caption [Control plots for TIGHT Pretag BNN Input and Output 1/1]{First half of the control plots for TIGHT charged lepton Pretag BNN Input Variables. \label{fig:ControlPlots1}}
  \end {center}
\end {figure}

\begin{figure}[ht]
  \begin{center}\includegraphics[width=6.7cm]{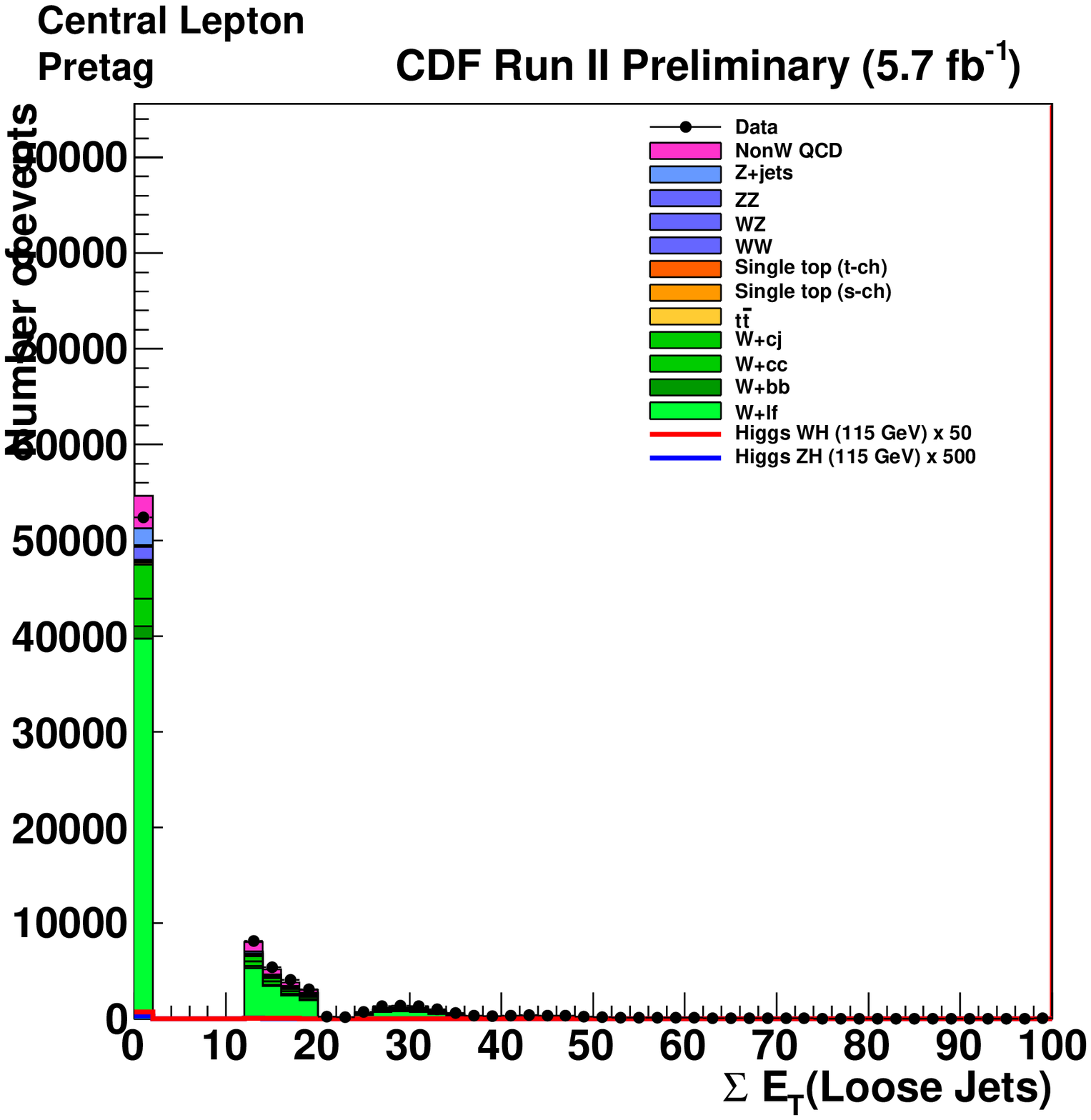}
    \includegraphics[width=6.7cm]{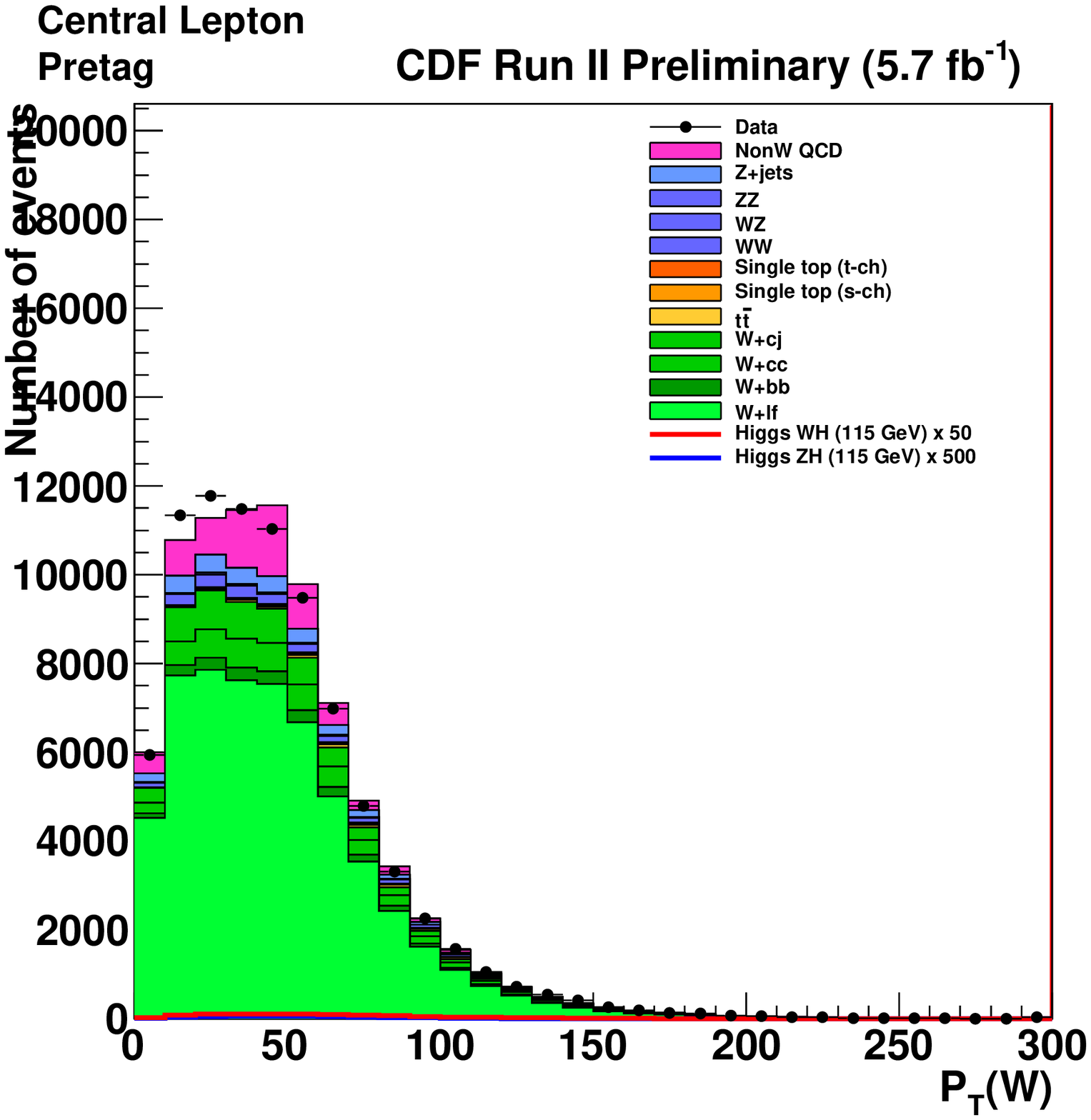}
    \includegraphics[width=6.7cm]{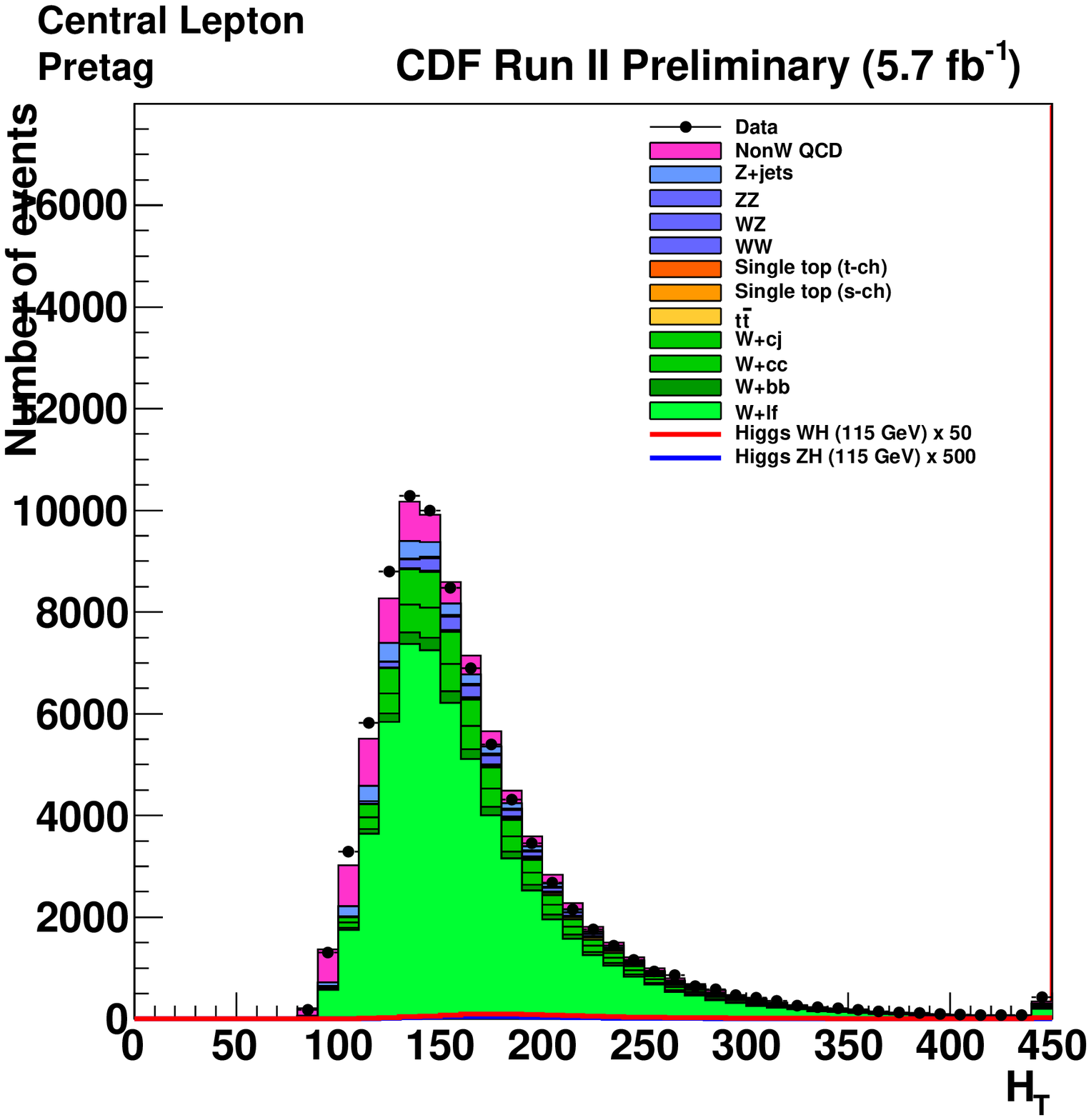}
    \includegraphics[width=6.7cm]{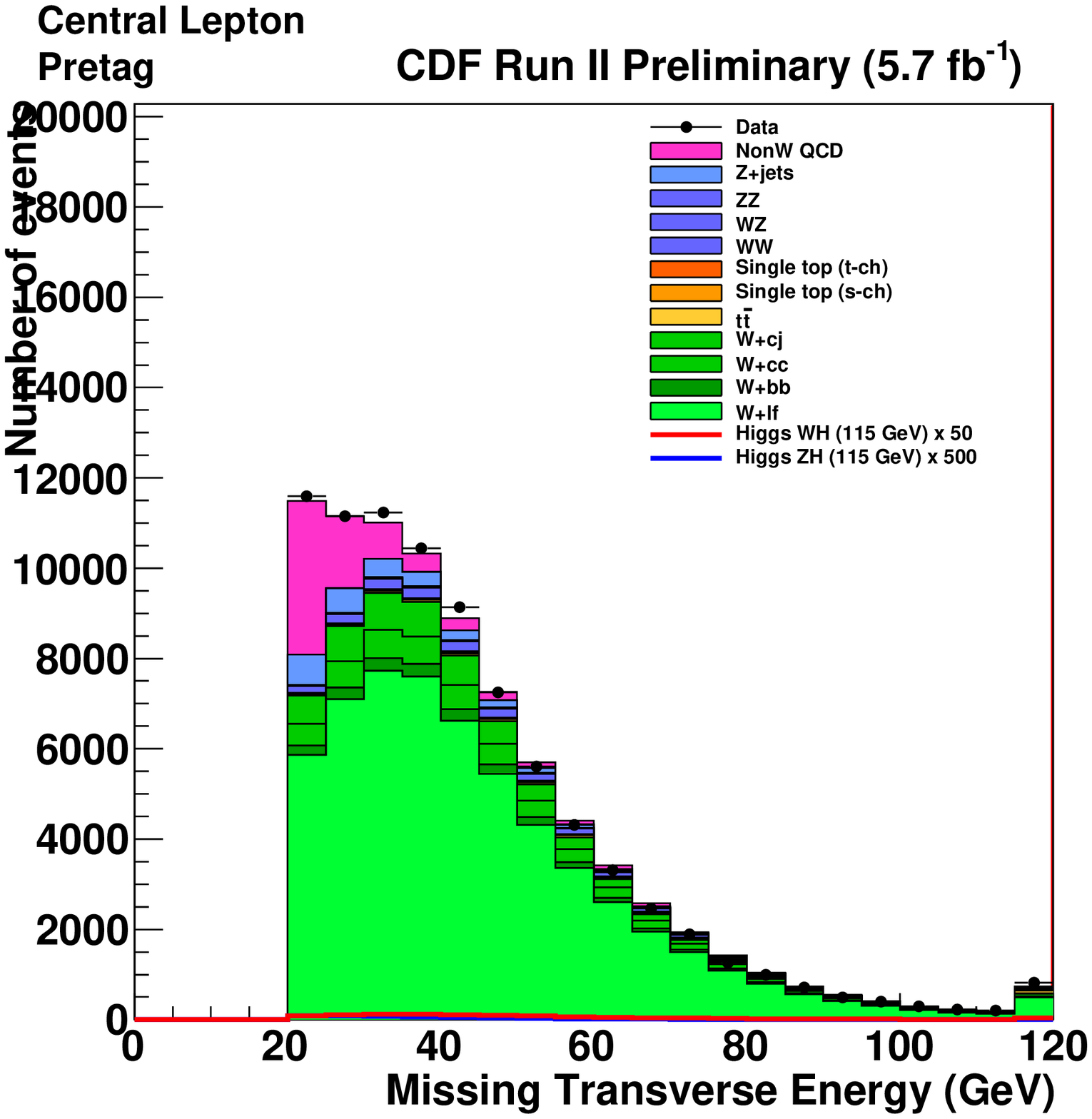}
    \includegraphics[width=6.7cm]{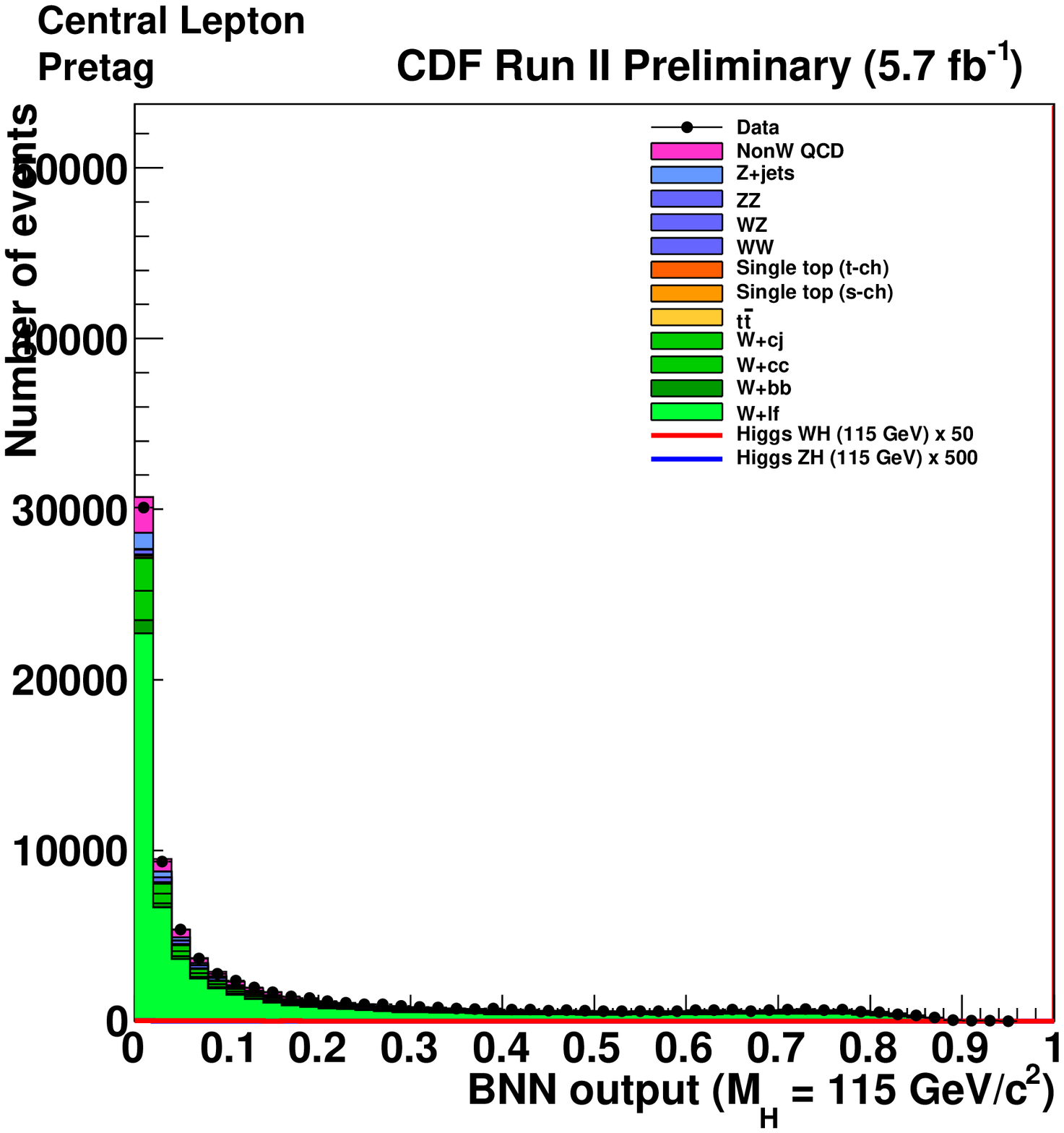}
    \includegraphics[width=6.7cm]{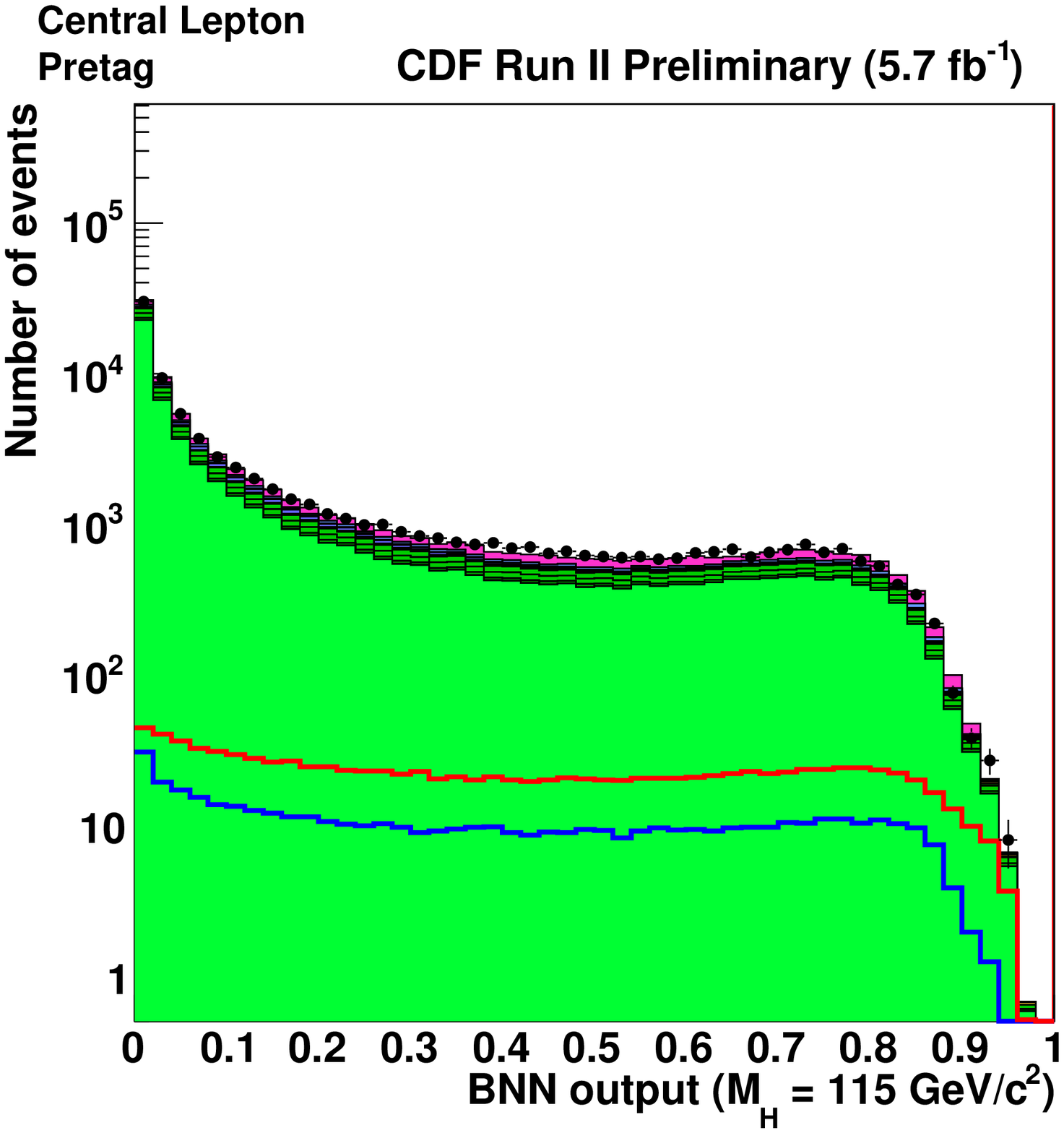}
    \caption [Control plots for TIGHT Pretag BNN Input and Output 1/2]{Second half of the control plots for TIGHT charged lepton Pretag BNN Input and the BNN Output Variable for the Higgs mass of 115 $\gevcc$. \label{fig:ControlPlots2}}
  \end {center}
\end {figure}

\begin{figure}[ht]
  \begin{center}
    \includegraphics[width=6.5cm]{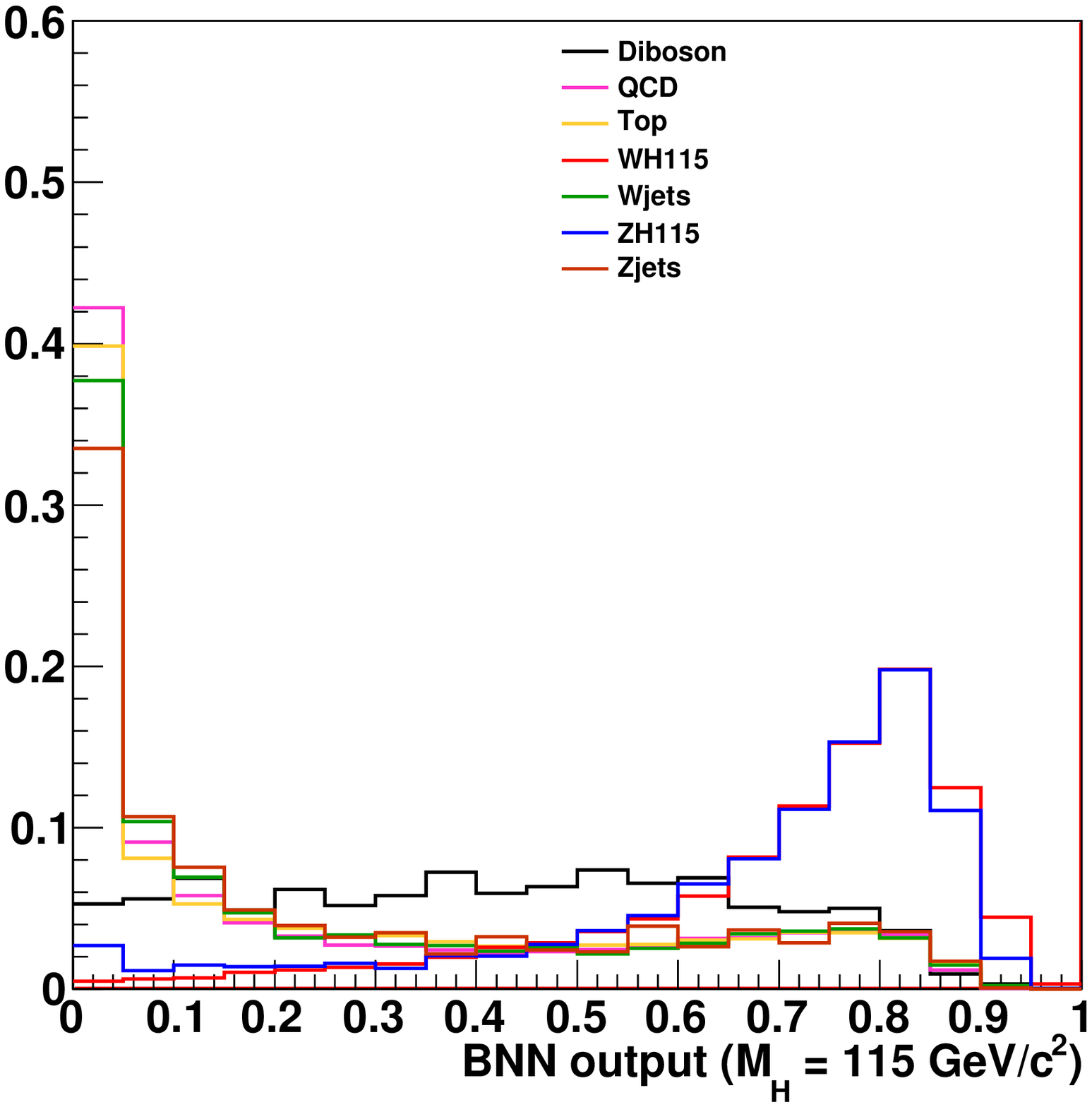}
    \includegraphics[width=6.5cm]{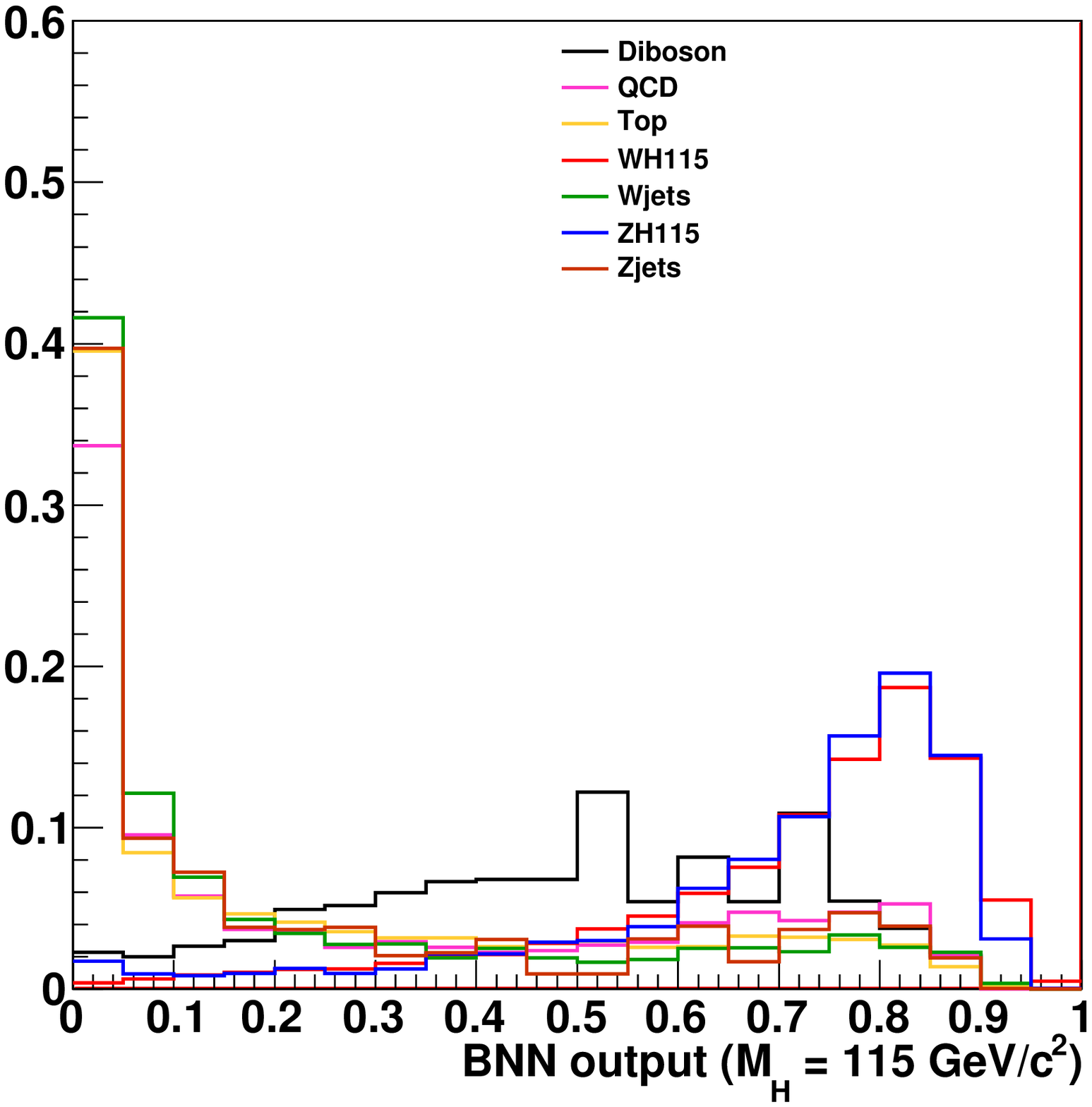}\\
    \includegraphics[width=6.5cm]{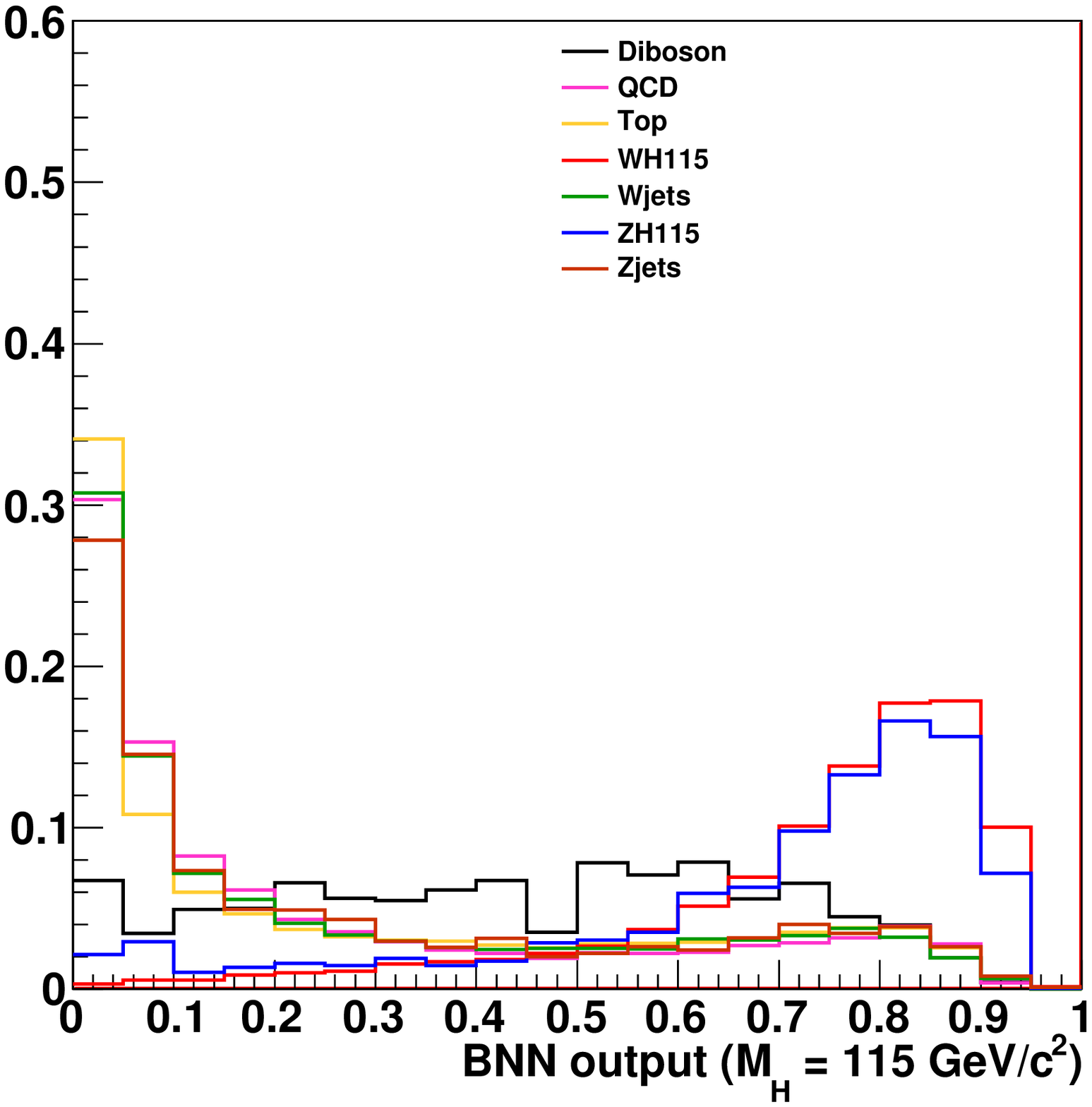}
    \includegraphics[width=6.5cm]{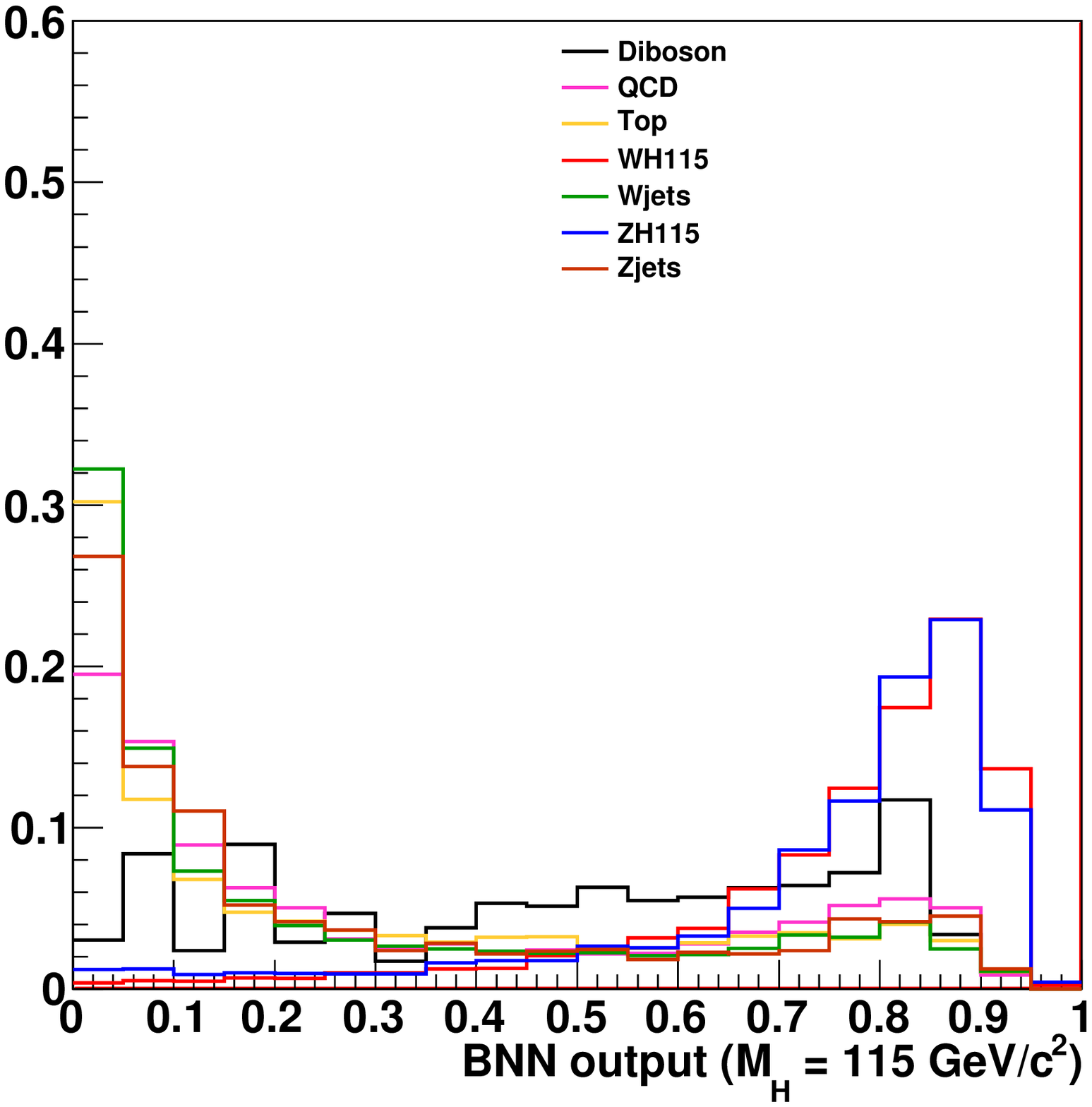}\\
    \includegraphics[width=6.5cm]{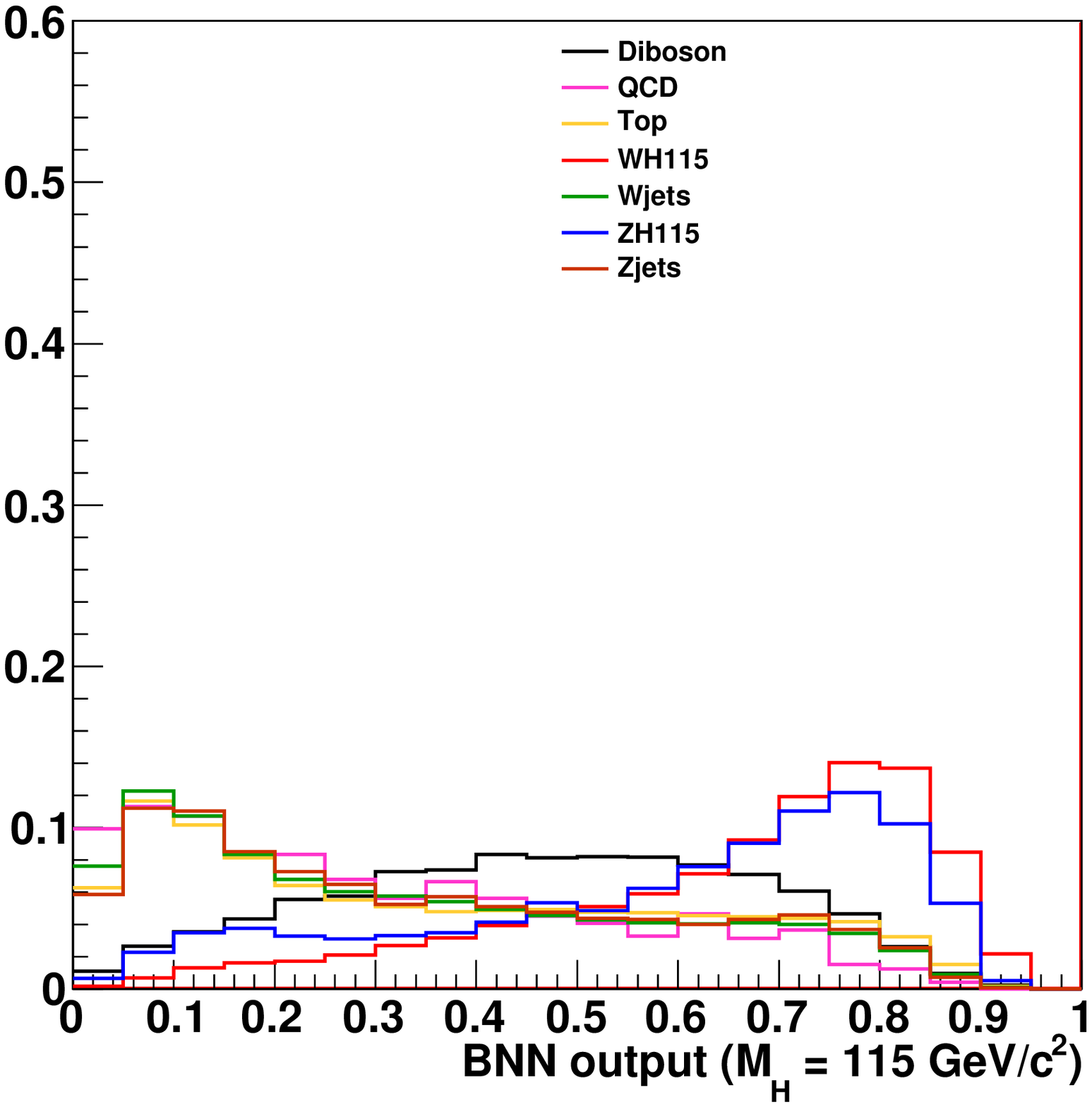}
    \includegraphics[width=6.5cm]{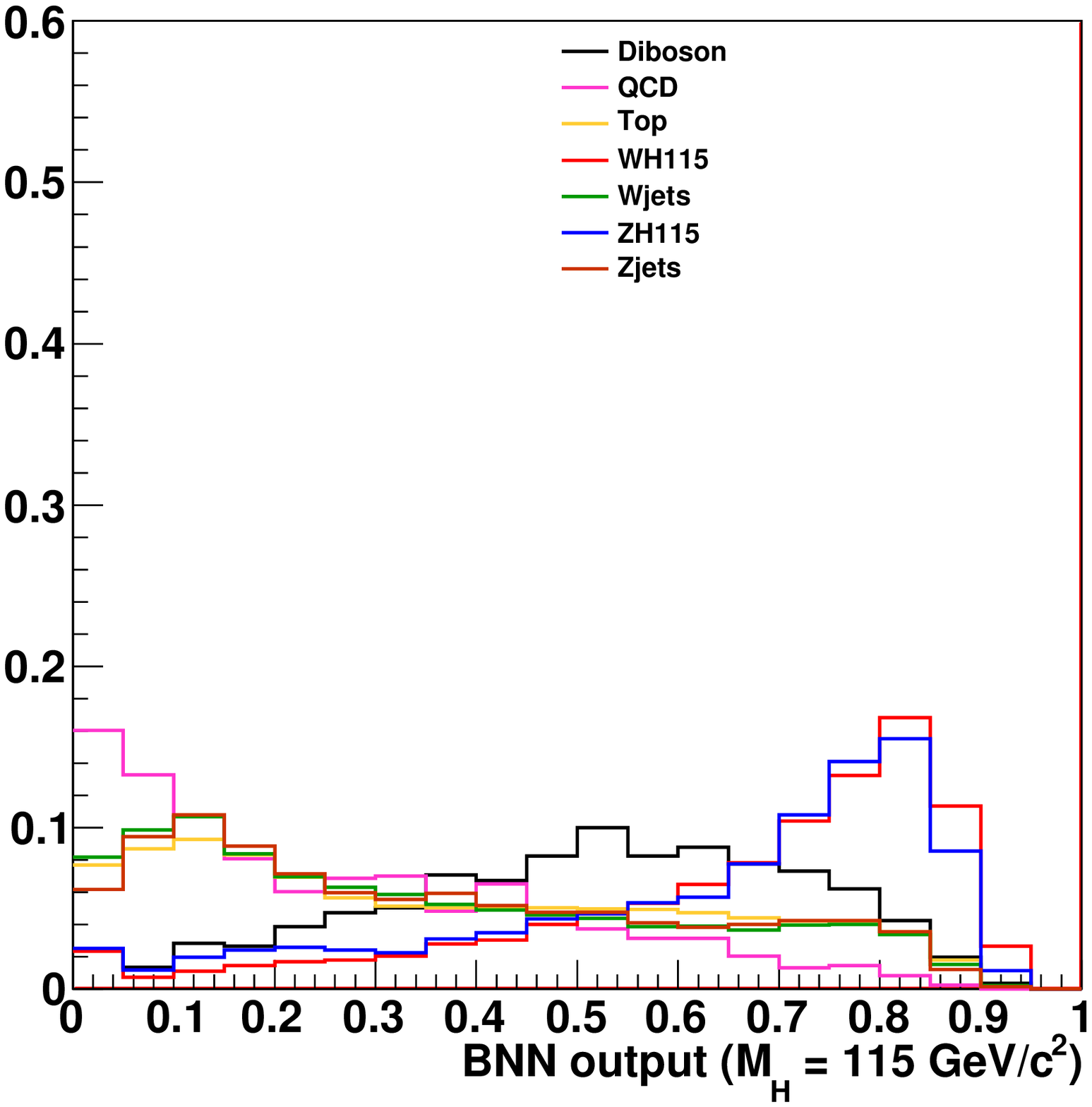}\\
\caption[Normalized BNN output distributions for signal and background]
{Normalized-to-unit-area BNN output distributions for signal and various background types for the various $b$-tagging categories and a Higgs boson mass of $115 \gevcc$. From top to bottom, the plots are specific to the SVTSVT, SVTJP05 and SVTnoJP05 $b$-tagging categories, respectively. The left (right) plots correspond to the TIGHT (ISOTRK) charged lepton categories, respectively. \label{figure:NeuralNetworkOutput115}}
  \end{center}
\end{figure}

\section{Neural Network Output}

\ \\From a physics point of view, the BNN takes as an input an entire event (from which it selects the information required by the input nodes) and gives as an output only one value, which represents the probability the event is signal or background. The BNN is trained for signal to peak at output values of 1 and background at output values of 0. The training sample is obtained using the same event selection described in Chapter~\ref{chapter:Selection}. The training is done independently for Higgs boson masses between 100 and 150 $\gevcc$ with increments of 5 $\gevcc$. Indeed, this is confirmed in the normalized-to-unit-area BNN output distributions for signal and background for the various $b$-tagging categories and the TIGHT and ISOTRK charged lepton categories, where the Higgs boson is assumed to have a mass of $115 \gevcc$, which can be seen in Figure~\ref{figure:NeuralNetworkOutput115}.

\section{Neural Network for Jet Energy Correction \label{Section:JetNeuralNetwork}}

\ \\All quantities in the analysis are computed using the jet energies corrected for various effects, as described in Chapter \ref{chapter:Object}. However, as the dijet invariant mass is the most sensitive observable to distinguish between the $WH$ signal and various types of background, we make an extra effort to reduce its uncertainty. This improves directly the $WH$ search sensitivity. 

\ \\We have designed a method that corrects the jet energies once more, depending on the information if they have been $b$-tagged or not \cite{b-NN_NIM}. Each correction is implemented on a per-jet basis. Typically the jet energies measured by the calorimeter are underestimated and even more so for jets originating in $b$ quarks due to the semi-leptonic decays of the $b$ quark producing a muon that is a minimum ionizing particle in the calorimeter. We add vertex and tracking information about the jet in order to improve the jet energy resolution.

\ \\To achieve this goal, we use a multivariate regression technique in the form of a second artificial neural network algorithm. All neural network algorithms provided by ROOT have been tested and the best results were obtained with the Broyden-Fletcher-Goldfarb-Shanno (BFGS) method \cite{BFGS}. 

\ \\We train three BFGS neural networks, one for jets tagged by \secvtx, one for jets not tagged by \secvtx~ but tagged by \jetprob,  and one for jets that are neither tagged by \secvtx~ nor by \jetprob. We use only $WH$ signal samples, that are divided in two. One half is used as training sample and another one as test sample. We do not use background samples because we are focused on improving the dijet mass resolution for signal. This quantity would anyway have a broad distribution for background processes. The input nodes use quantities that describe the jet and will be enumerated below. There is only one output value and its target value has a specific value for each jet taken as an input by the neural network. In principle this method works for any type of jet, but it must be trained for the specific type of jet. 

\subsection{BFGS Inputs} 

\ \\For the neural network trained on \secvtx-tagged jets, we use the following nine quantities as inputs: jet $\et$; jet $\pt$; uncorrected jet $\et$; jet transverse mass; jet decay length ($\Lxy$); uncertainty on the jet decay length ($\sigma(\Lxy)$); sum of transverse momenta of tracks originating from the secondary vertex identified by the \secvtx algorithm; the maximum $\pt$ of tracks inside the jet; the scalar sum of transverse momenta of tracks inside the jet. For the jets not tagged by \secvtx, we use the same input variables except $\Lxy$ and $\sigma(\Lxy)$.

\subsection{BFGS Output}

\ \\The neural network is designed to have as an output ($NN_{output}$) the extra correction factor by which the standard-corrected jet transverse energies ($\et$) have to be multiplied in order to achieve the newly corrected values (${\et}_{corr}$).

\begin{equation} 
{\et}_{corr} = NN_{\rm{output}} \cdot \et\,\rm{.}
\label{NeuralNetworkOutputJet}
\end{equation}

\ \\For each jet, the target value ($t$) for $NN_{output}$ is represented by the ratio between the Monte Carlo generator level transverse energy (${\et}_{gen}$) and the standard-corrected transverse energy ($\et$).

\begin{equation} 
t = \frac{{\et}_{gen}}{\et}\,\rm{.}
\label{NeuralNetworkTargetJet}
\end{equation}

\ \\Following this procedure, after the training, ${\et}_{corr}$ will be closer to ${\et}_{gen}$ than $\et$. 

\begin{figure}[ht]
\begin{center}

\includegraphics[width=5.0cm,clip=]{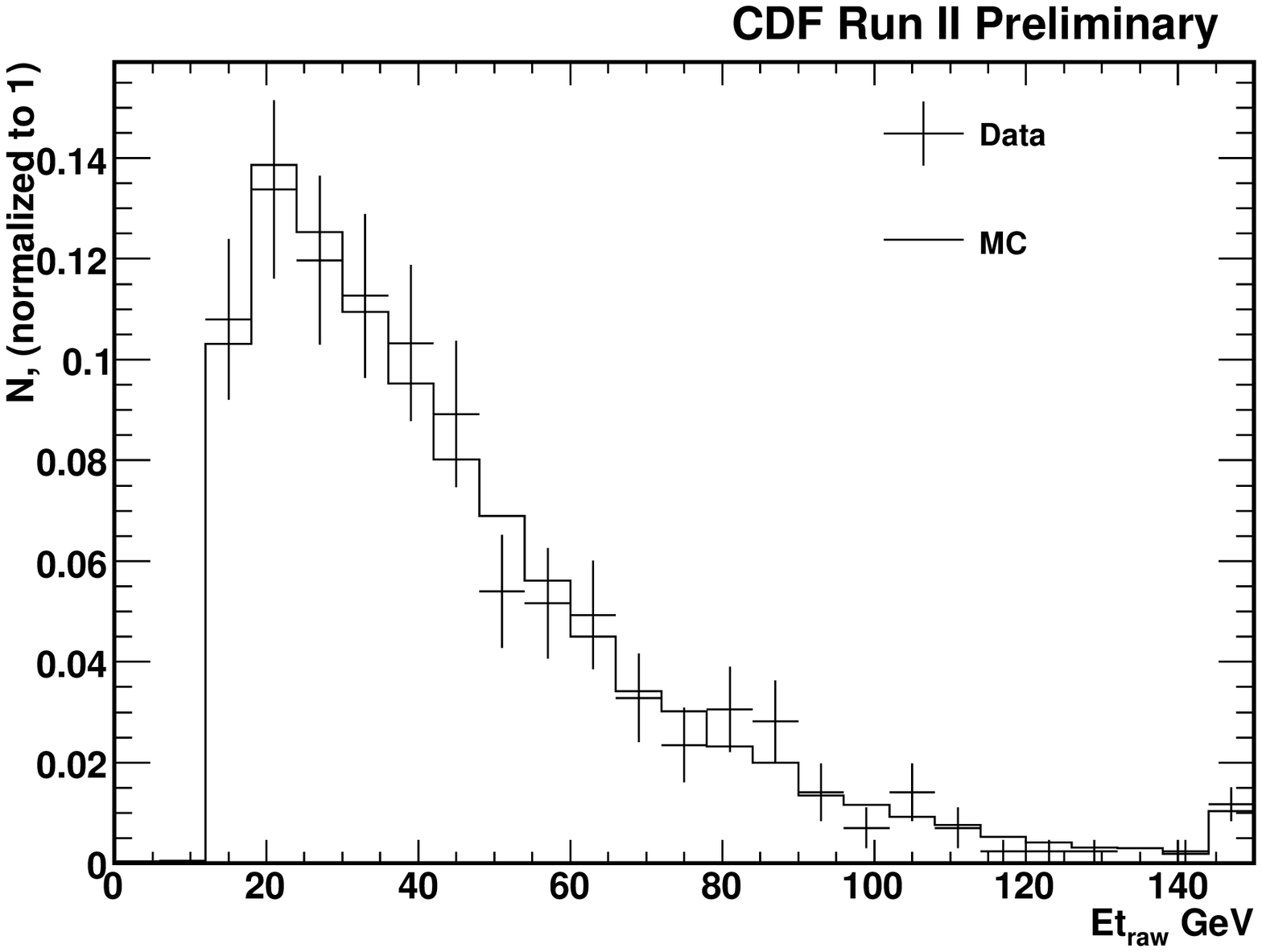}
\includegraphics[width=5.0cm,clip=]{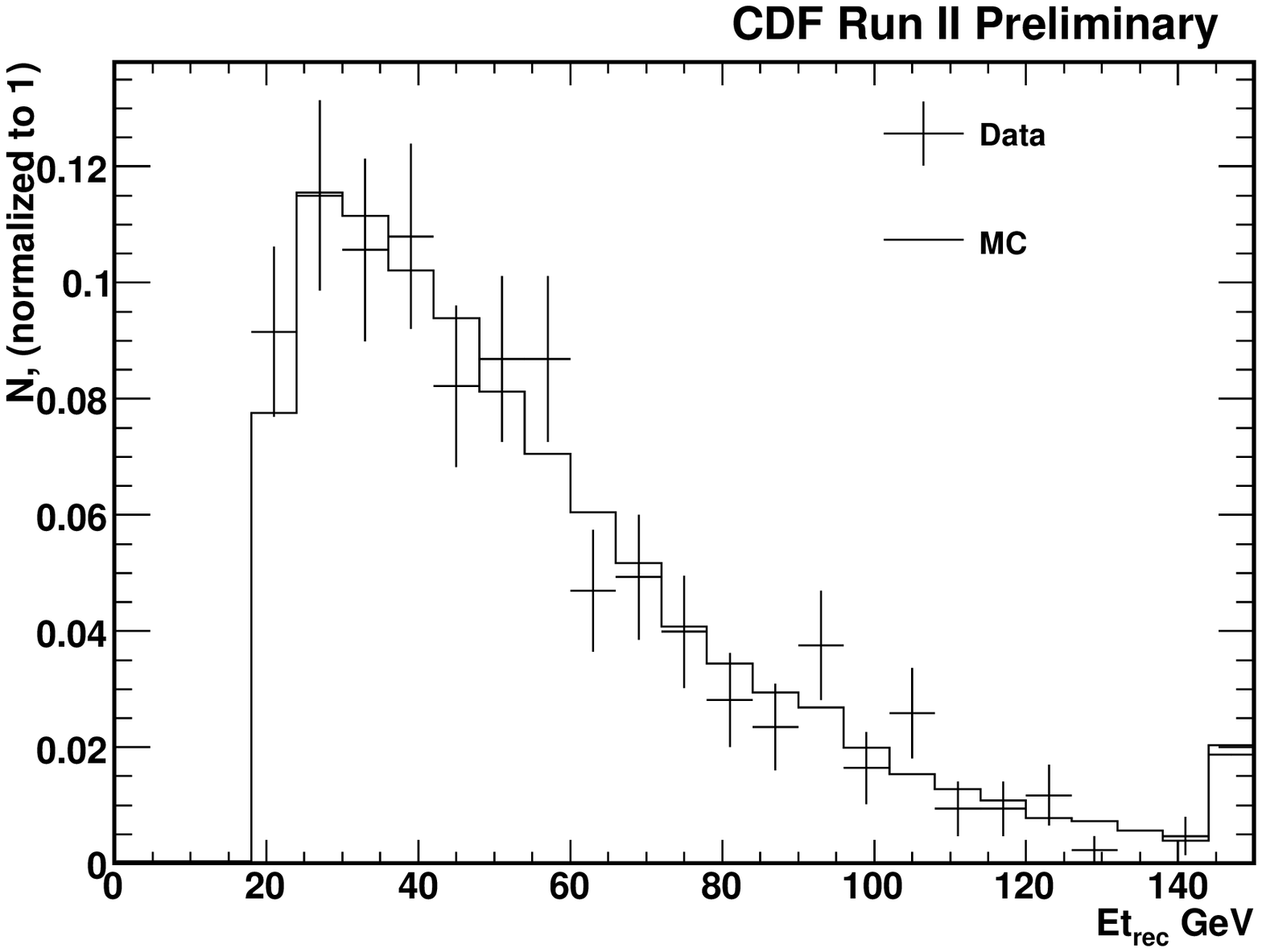}
\includegraphics[width=5.0cm,clip=]{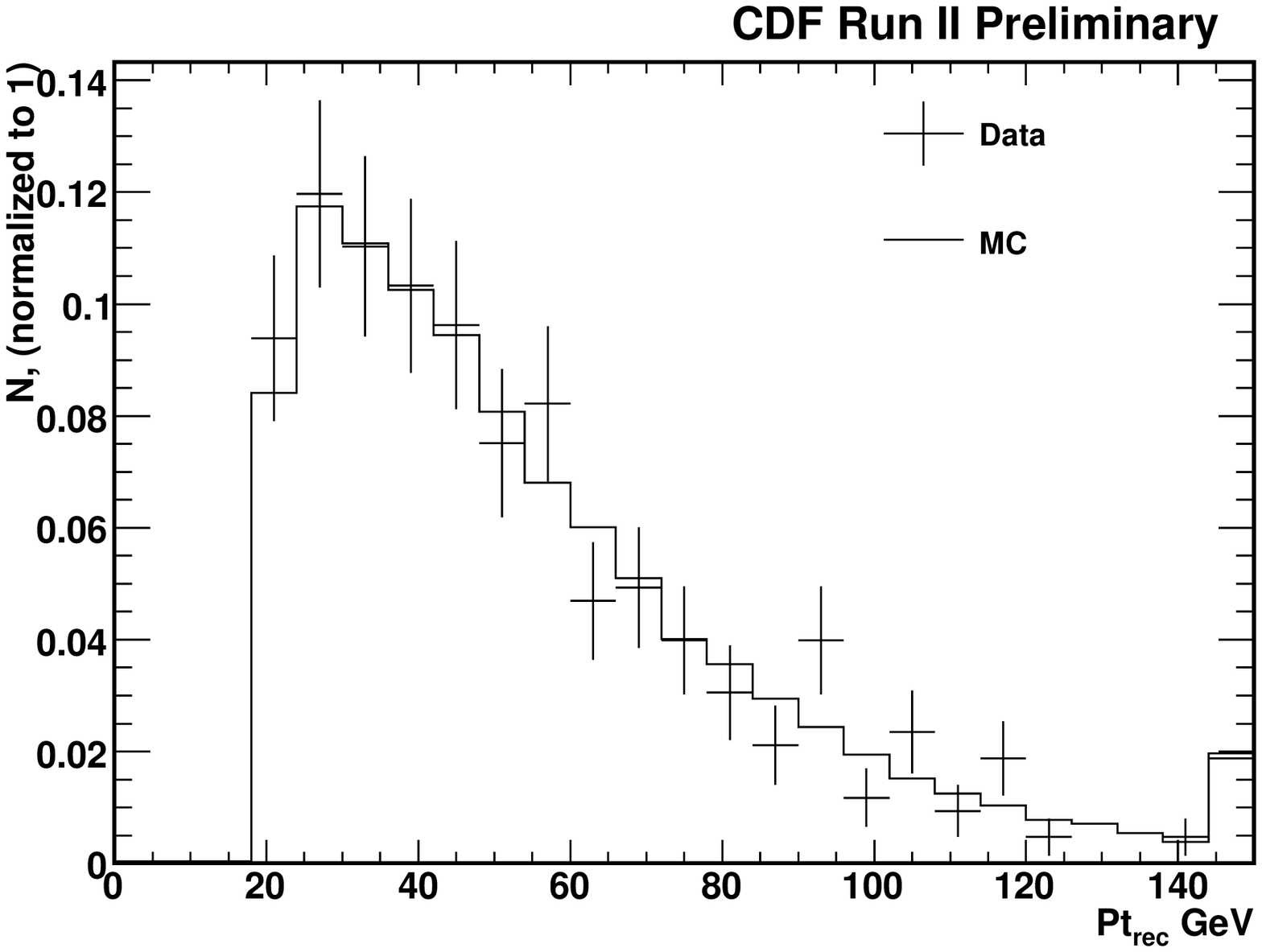}
\includegraphics[width=5.0cm,clip=]{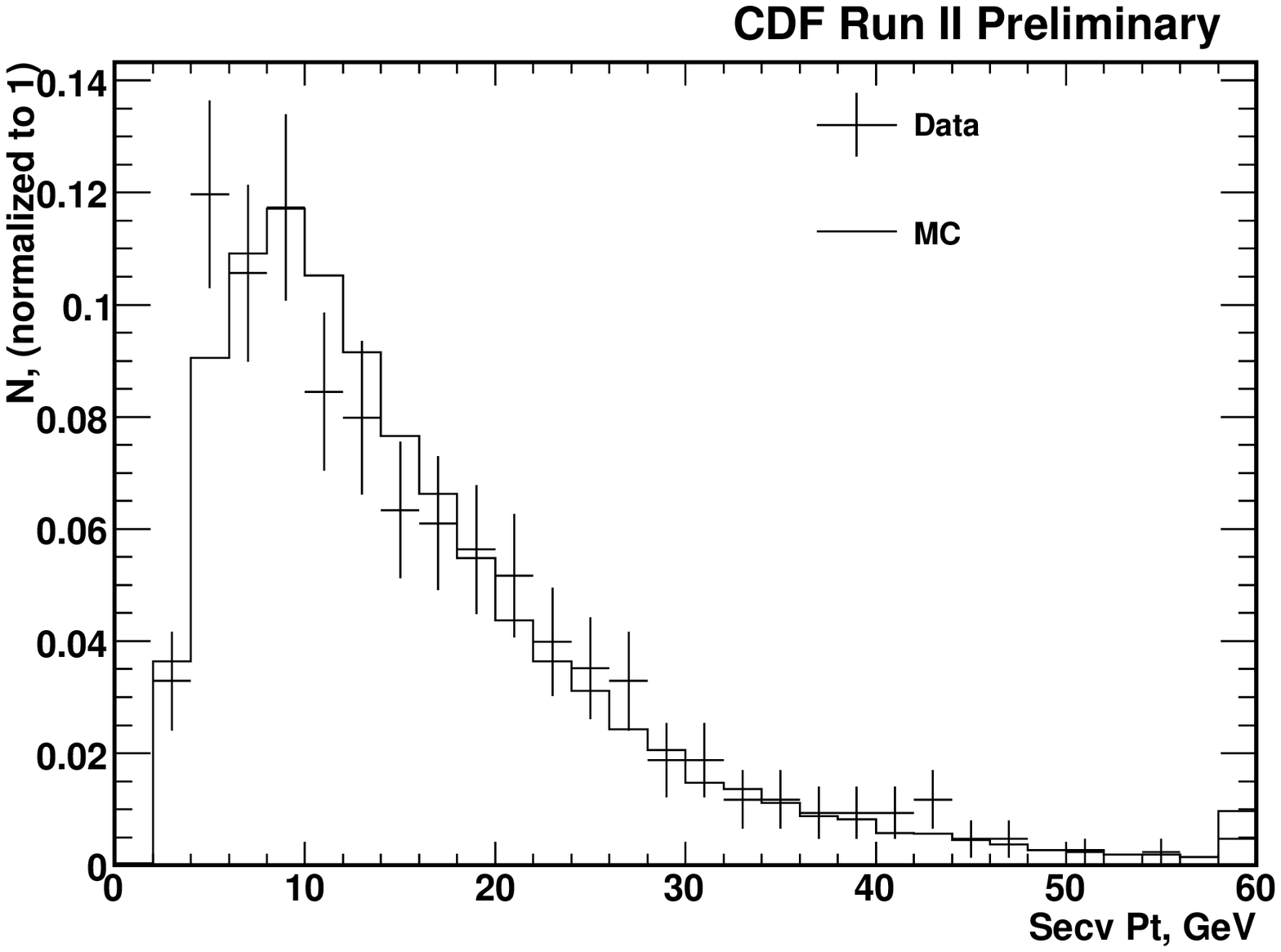}
\includegraphics[width=5.0cm,clip=]{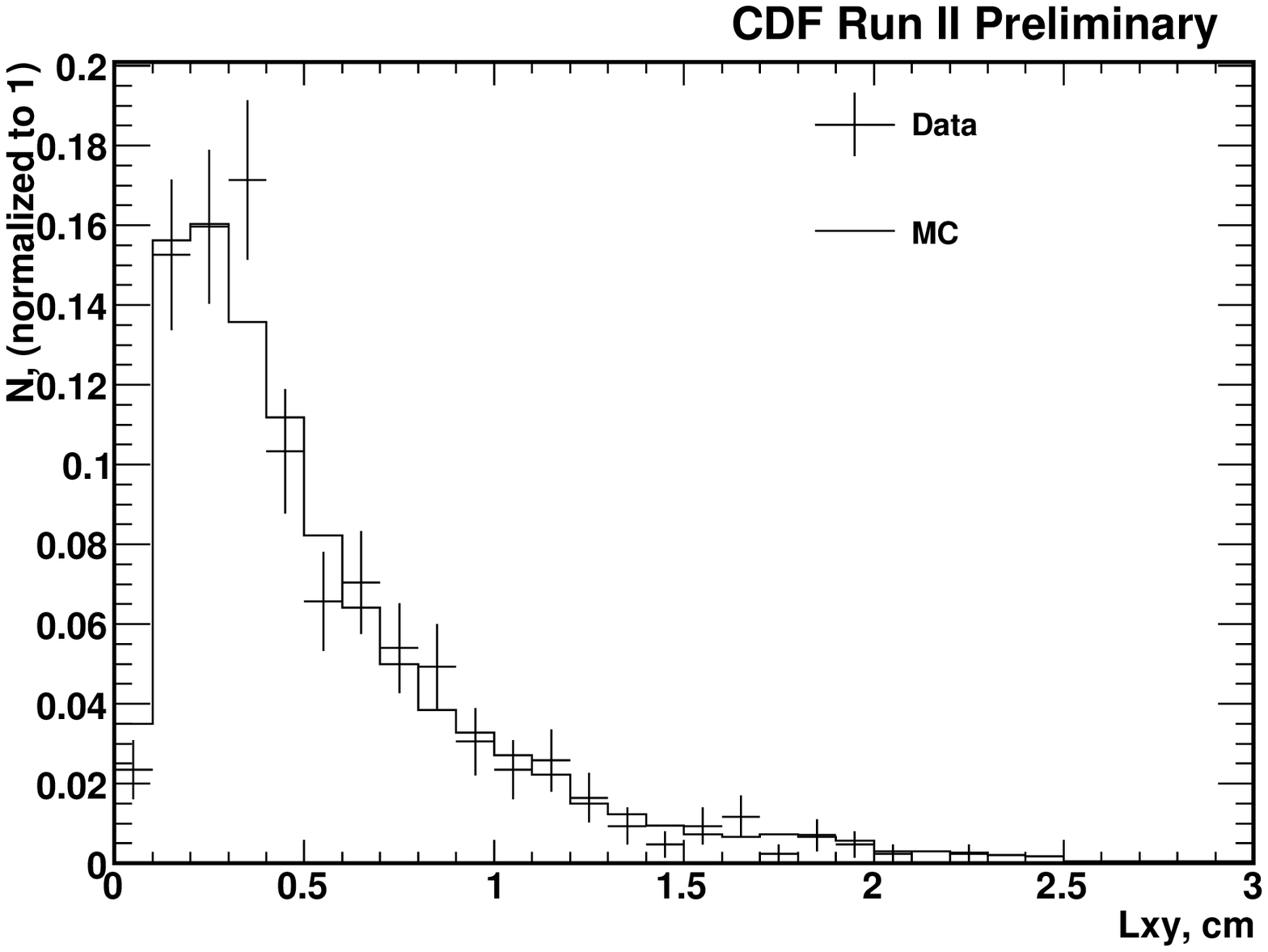}
\includegraphics[width=5.0cm,clip=]{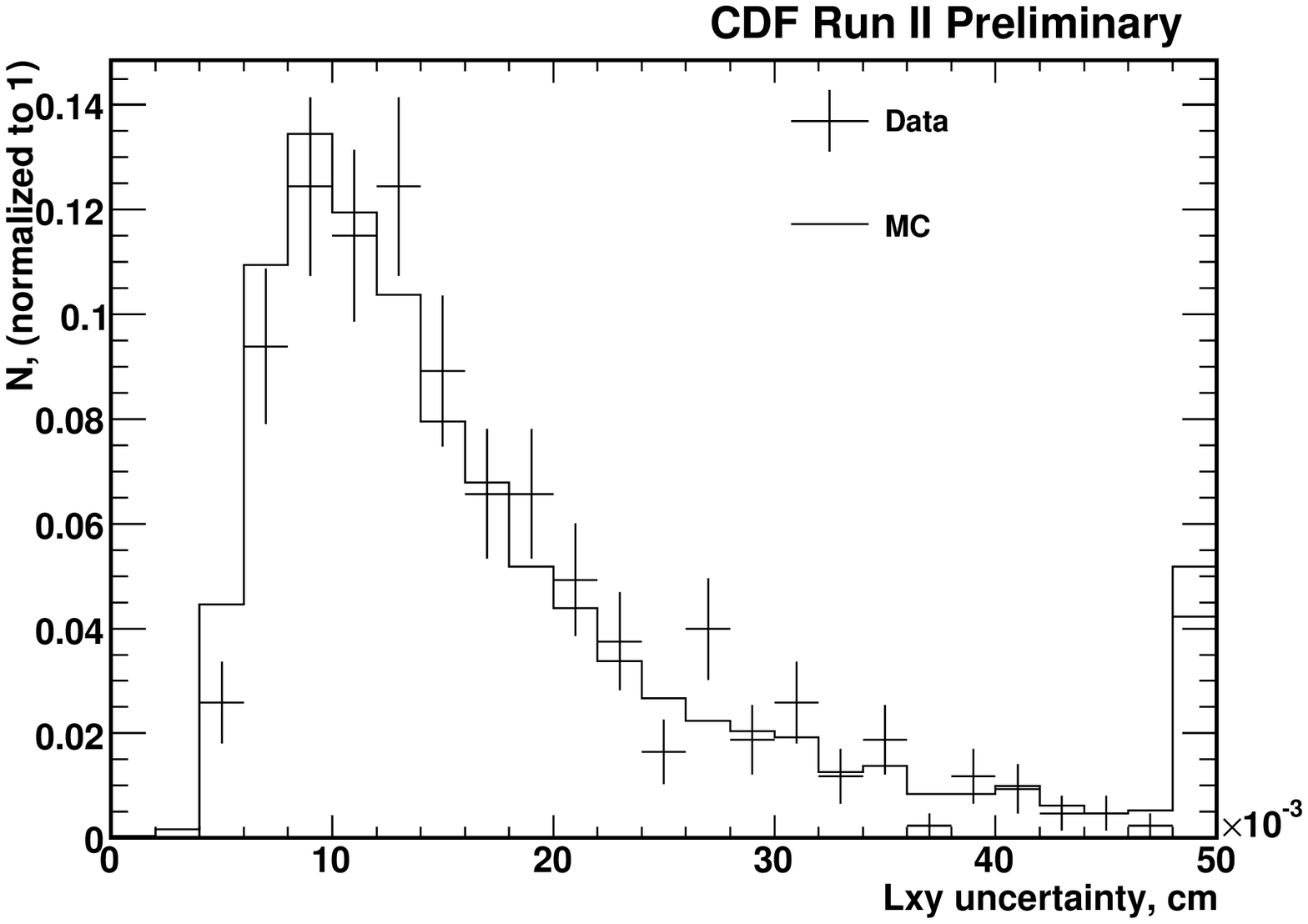}
\includegraphics[width=5.0cm,clip=]{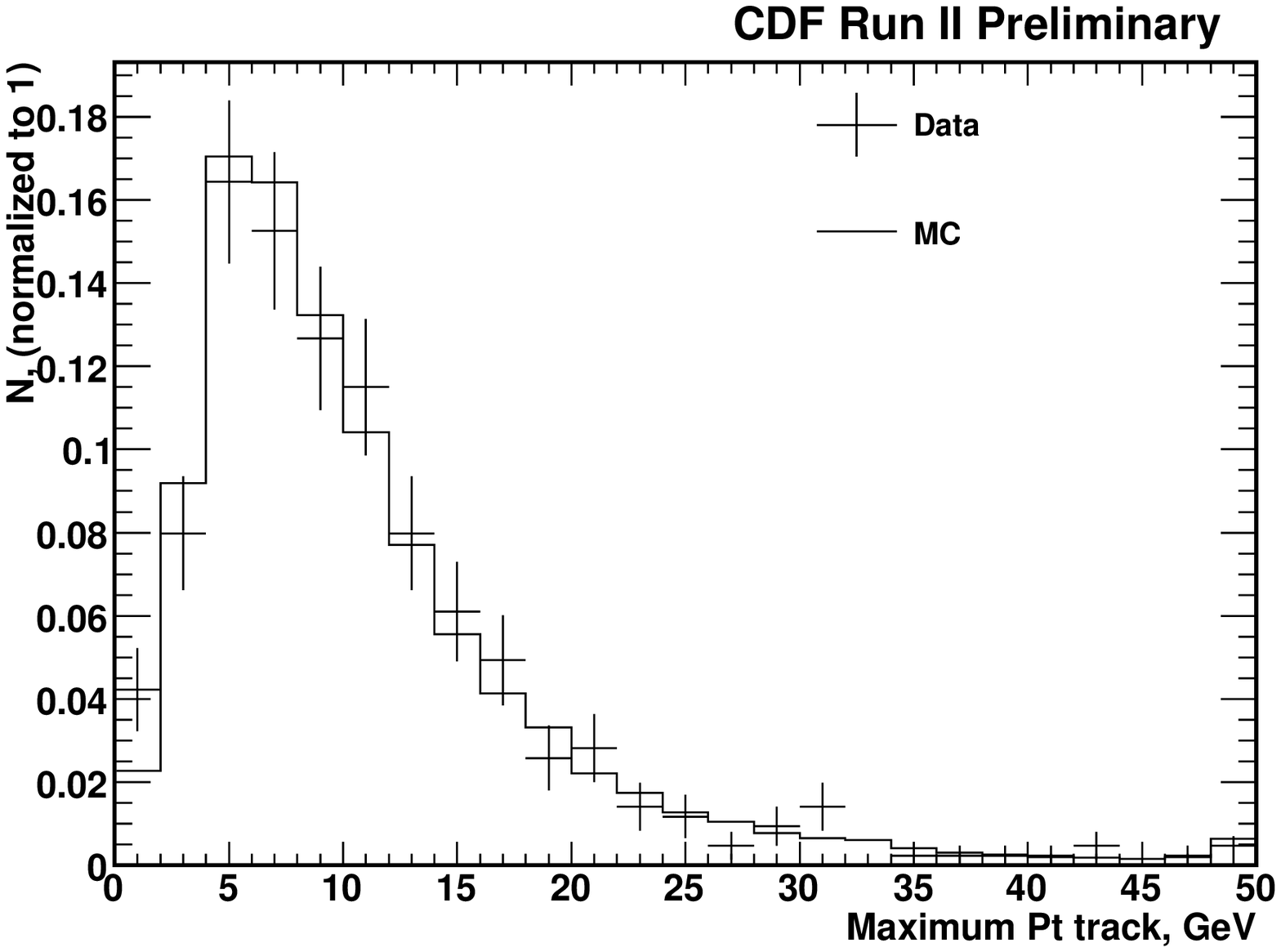}
\includegraphics[width=5.0cm,clip=]{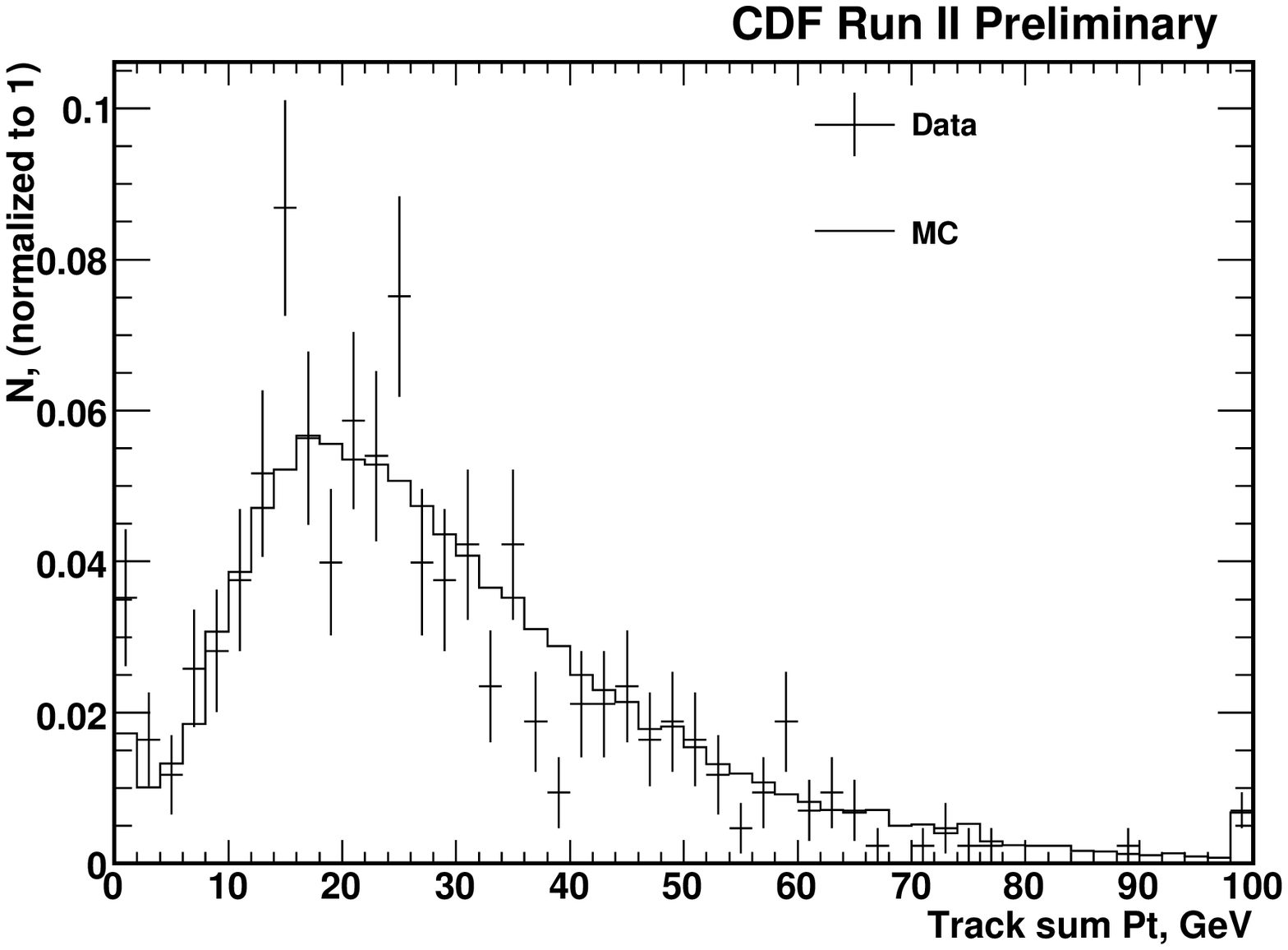}
\includegraphics[width=5.0cm,clip=]{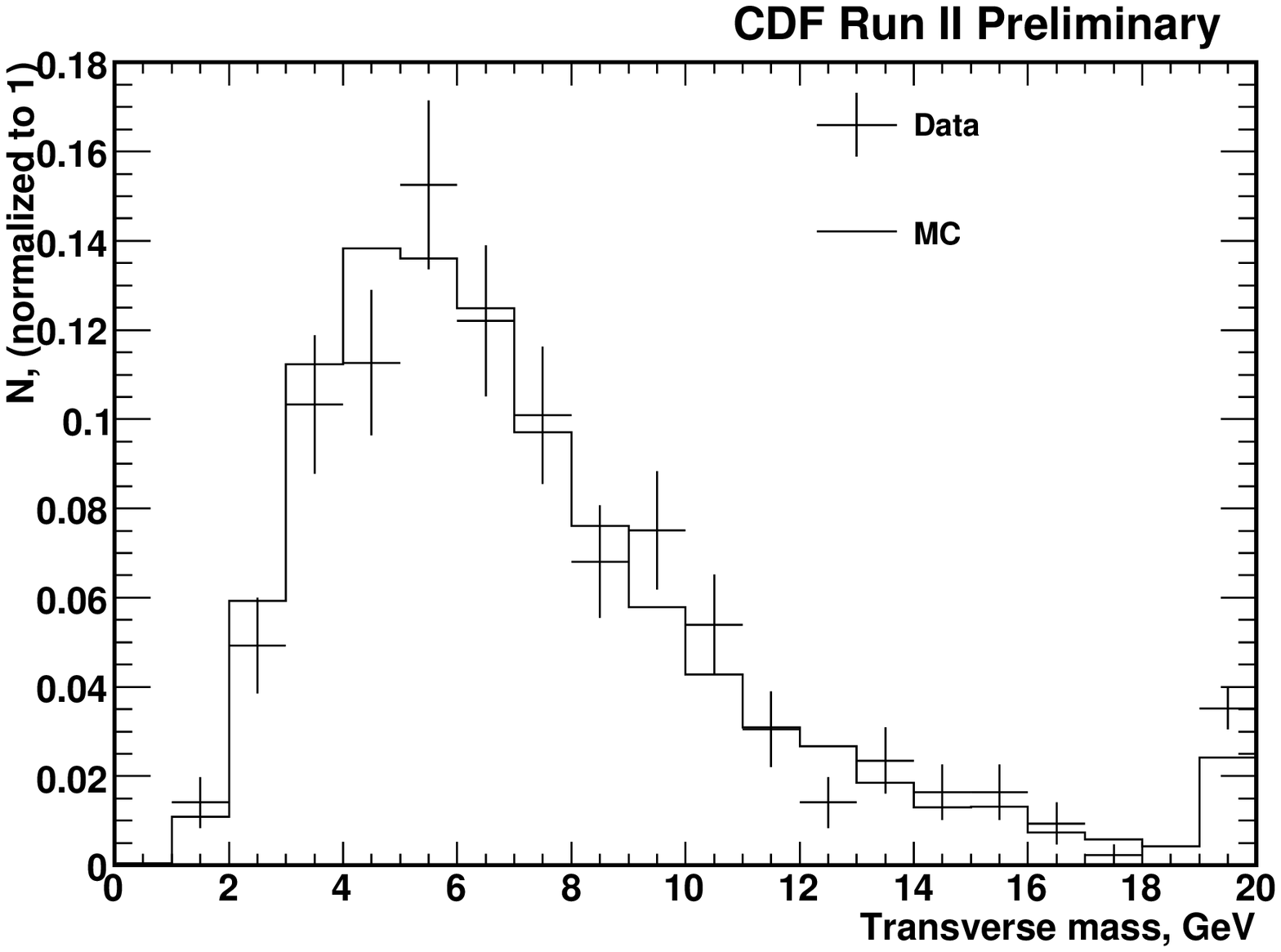}
\includegraphics[width=5.0cm,clip=]{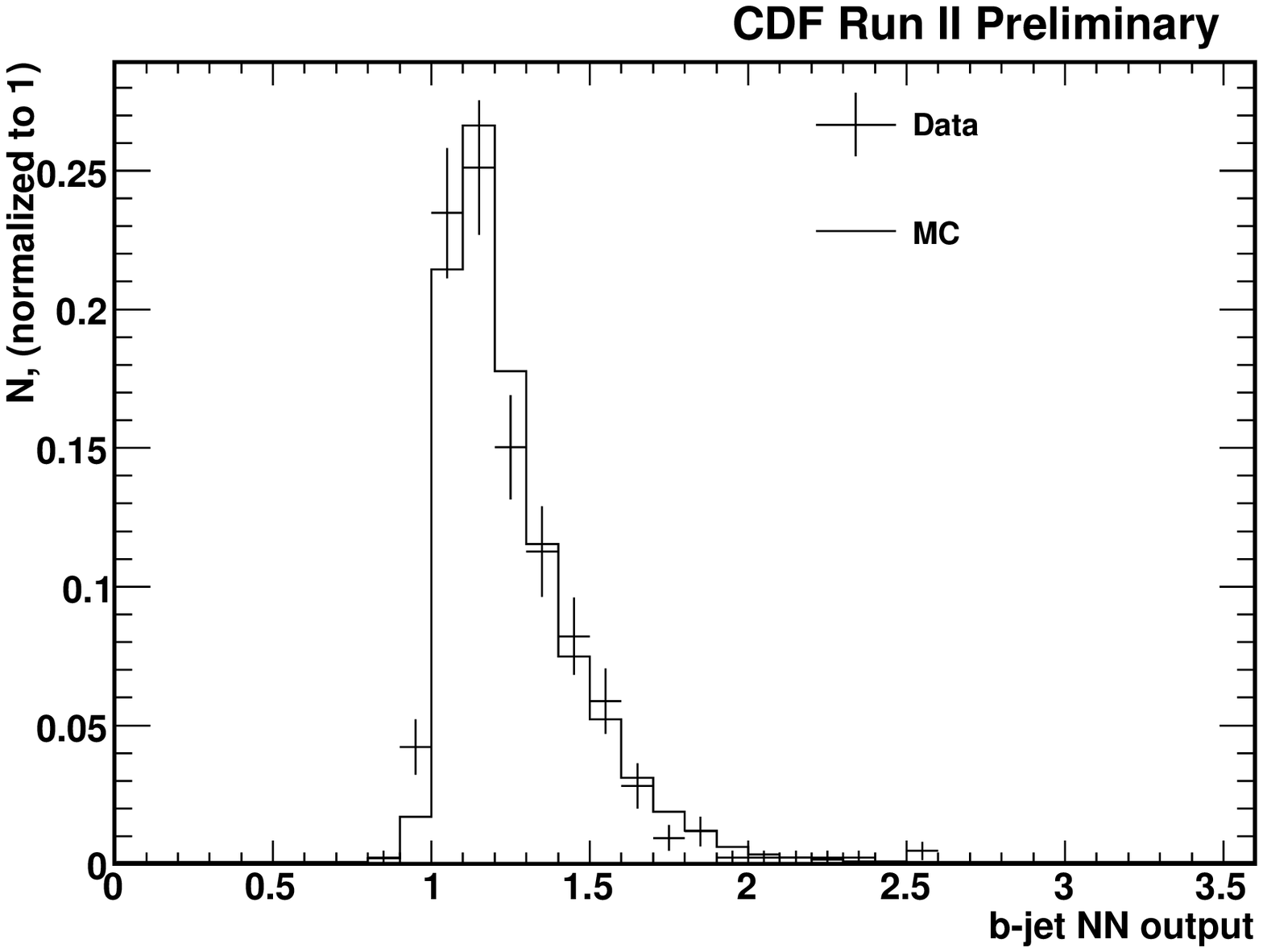}

\caption [Neural-network-based $b$-jet energy correction variables]{NN $b$-jet energy correction input variables for {\secvtx} tight-tagged jets and the output of $b$-jet neural network correction. The b-NN jet energy correction was implemented by Timo Aaltonen and these plots were produced by me. They are shown in a paper submitted to Nuclear Instruments and Methods, for which I am one of the four authors \cite{b-NN_NIM}. The figures represent a sample of TIGHT charged leptons in the SVTSVT tag category events where both tight jets in the events are shown on the same histogram in order to double the statistics.}
\label{figure:NNbCrorrST}
\end{center}
\end {figure}

\subsection{Background Modelling Check}

\ \\Also for this second neutral network type used for $b$-jet energy corrections, we check that the data reproduce well the background modelling for all the kinematic distributions of the inputs and output of these two jet energy correction neural networks. For SVTSVT events, we sum up the histograms for each of the two tight jets in the events so that we double the statistical power. Figure \ref{figure:NNbCrorrST} shows the input and the output of the neural network based SVT-correction, on a jet by jet basis, proving that this correction works both for Mote-Carlo-simulated and data events.

\begin{figure}[ht]
\begin{center}
\includegraphics[width=7.0cm]{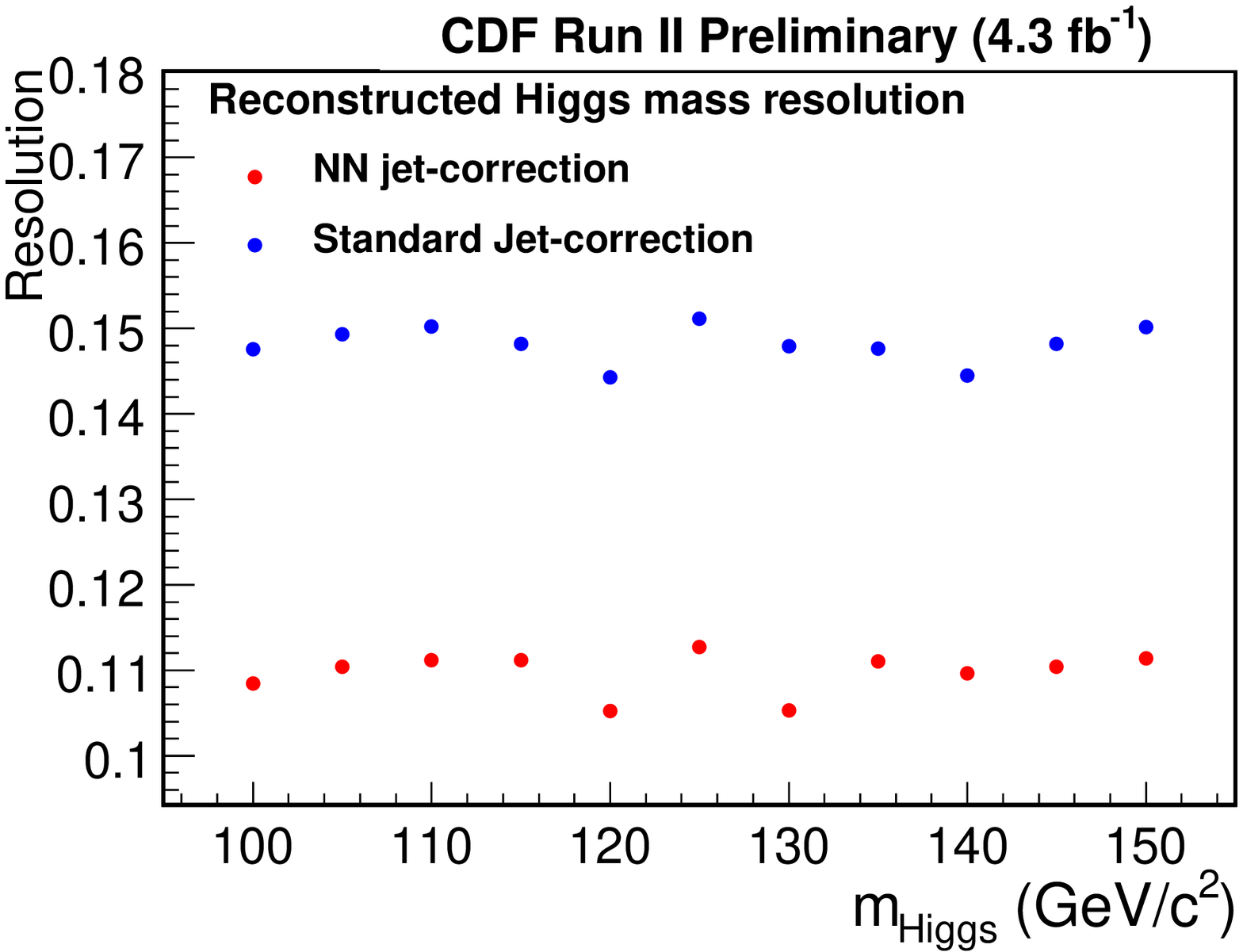}\\
\includegraphics[width=7.0cm]{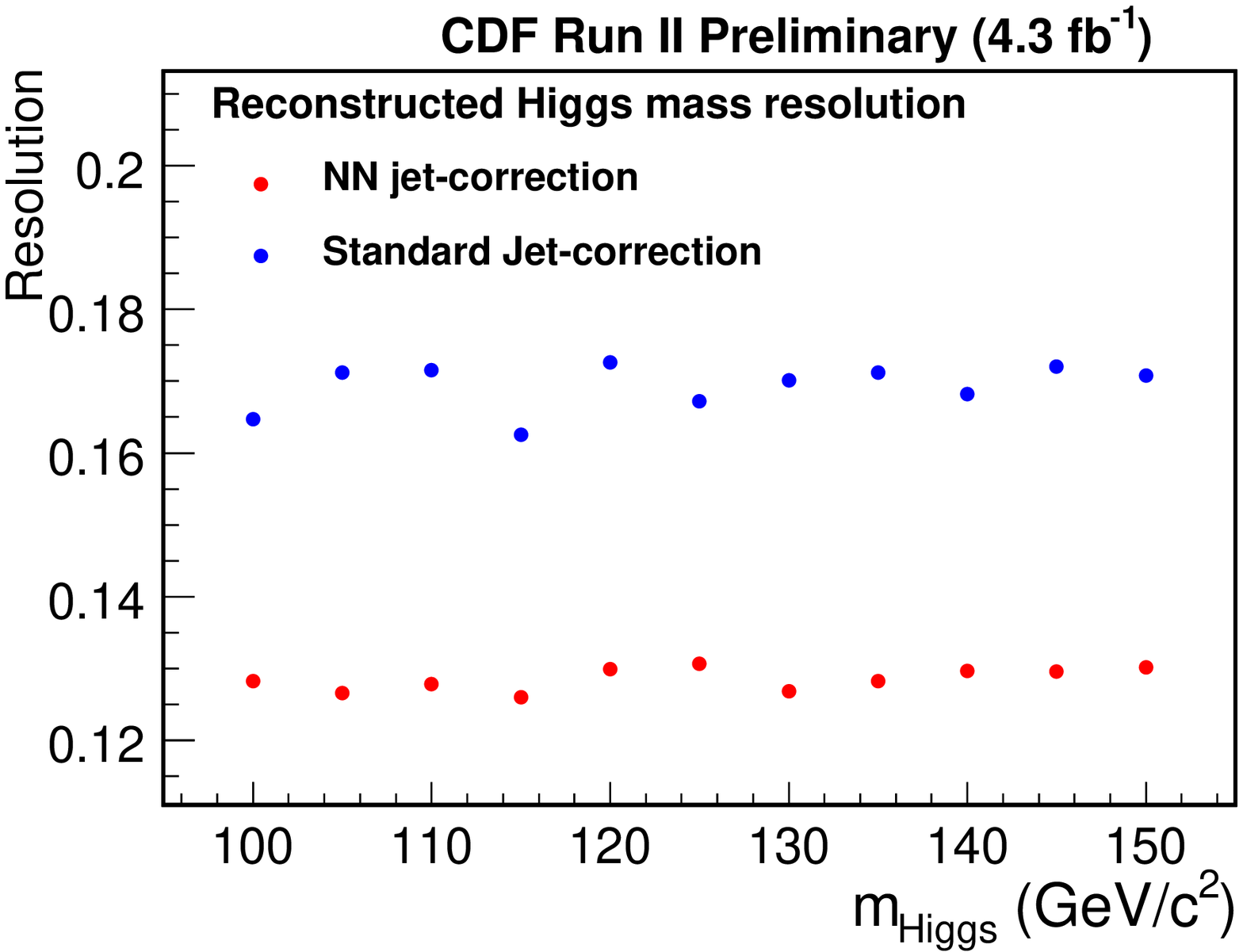}
\caption[Dijet invariant mass resolution before and after $b$-NN corrections]
{Reconstructed Higgs boson invariant mass resolution in a $WH$ sample using the standard jet energy correction and the neural network jet energy correction. The top (bottom) plot presents the SVTSVT (SVTnoJP05) $b$-tagging category \cite{YoshikazuNagaiThesis}. Although not exemplified in this plot, we also apply the jet energy correction to the dijet invariant mass in the Pretag and SVTJP05 samples. We see an improvement of dijet mass resolution across the entire Higgs boson mass range when the neural network corrections are used. As this quantity is the most sensitive input to the final discriminant neural  network, our $WH$ exclusion limits improve due to this correction. Credit image to Timo Aaltonen. \label{figure:MjjResolution}}
\end{center}
\end{figure}

\subsection{Dijet Invariant Mass Resolution}

\ \\We now compute the dijet invariant mass using ${\et}_{corr}$ instead of $\et$ for all the samples used in our analysis: Pretag, SVTSVT, SVTJP05 and SVTnoJP05. This specific correction improves the resolution of the dijet invariant mass from about 15\% to about 11\% in the SVTSVT category and from about 17\% to about 14\% in the SVTnoJP05 category across all the Higgs boson mass range, as we can see Figure~\ref{figure:MjjResolution}.

\section{Summary}

\ \\In conclusion, in this chapter we have presented two artificial neural networks that we use in our analysis. One corrects the energy of the jets based on the information of whether the jet is tagged or not by the $b$-taggers used in our analysis. The corrected energies are used to compute the dijet invariant mass variable, which in turn is the main input variable to the final analysis discriminant, another artificial neural network (BNN) trained to separate the signal and the various background processes. In the next chapter we will present how the BNN output is used in order to compute an upper limit on the Higgs boson cross section times branching ratio. 

\clearpage{\pagestyle{empty}\cleardoublepage}

\chapter{Upper Limits on Higgs Boson Production\label{chapter:Limit}}

\ \\In this thesis we present a search for the existence of the Standard Model Higgs boson. In the absence of an observation, as a final result we present an upper limit on the Higgs boson cross section times branching ratio at 95\% credibility level (CL), as is typical in Bayesian statistical inference in experimental particle physics. We compute the upper limit using the BNN neural network output distributions for signal, background and data processes. Since the BNN distributions are binned, we use a binned likelihood technique. In Section \ref{section:LimitOneBinNoSystematics} we introduce an unbinned Bayesian likelihood technique, i.e. the treatment of one specific bin in the distribution, which is equivalent to a simple counting experiment. In Section \ref{section:LimitOneBinWithSystematics} we explain how each independent systematic uncertainty is taken into account as a nuisance parameter that is convolved with the likelihood. In Section \ref{section:LimitAllBinsOneChannel} we present how the method is generalized to a given number of bins for one analysis channel, i.e. a charged lepton and $b$-tagging category pair. In Section \ref{section:LimitAllChannels} we explain how combining all analysis channels is equivalent to adding new bins to only one distribution. In Section \ref{section:LimitExpectedAndObserved} we explain how we use pseudo-experiments to compute the expected limit (the sensitivity of the analysis) and read data to compute the observed limit (our result). In Section \ref{section:LimitWH} we present the expected and observed upper limits in the $WH$ search in the TIGHT, ISOTRK, and TIGHT+ISOTRK channels, with all $b$-tagging categories combined. 

\section{Bayesian Upper Limit Calculation\label{section:LimitOneBinNoSystematics}}

\ \\One a bin-by-bin basis, we denote with $\mu$ the expected number of events and with $n$ the measured number of events. Our goal is to evaluate the probability that we expect $\mu$ events when we measure $n$ events, given that the probability that we measure $n$ events given we expect $\mu$ events is described by the Poisson distribution expressed by Formula \ref{formula:PoissonStatistics}. This process is called statistical inference.

\begin{equation} 
P(n|\mu)=\frac{\mu^{n}e^{-\mu}}{n!}\ \rm{.}
\label{formula:PoissonStatistics}
\end{equation}

\ \\There are two principal methodologies for statistical inference. The frequentist approach assumes no prior knowledge about the generic probabilities $P(\mu)$ and $P(n)$ and introduces a test statistic variable that is used to test if the data distribution agrees more with the background-plus-signal hypothesis, or with the background-only hypothesis. The frequentist approach is used to set upper limits on the Higgs boson at the DZero experiment at Fermilab. At CDF we use the Bayesian approach, where prior information is taken into account as well. We start with Bayes' theorem:

\begin{equation} 
P(\mu|n)=\frac{L(n|\mu)\pi(\mu)}{P(n)}\ \rm{,}
\label{BayesTheorem}
\end{equation}

\ \\where $P(\mu|n)$ is the posterior probability on the expected number of events $\mu$, i.e. the probability distribution of $\mu$ after the experiment is performed; $L(n|\mu)$ is the probability to measure n events given the expected number of events $\mu$, has a Poisson distribution and is given by formula \ref{formula:Likelihood}; $\pi(\mu)$ is the prior probability on the expected number of events $\mu$, i.e. the probability distribution of $\mu$ before the experiment is even performed; $P(n)$ is the probability of distribution for the observed number of events n, which is evaluated as a normalization constant given by the condition that the sum of all probabilities should be exactly 1, as shown in Equation \ref{equation:SumOfProbabiltiesIsUnity}.

\begin{equation}
L(n|\mu)=\frac{\mu^{n}e^{-\mu}}{n!}\ \rm{.}
\label{formula:Likelihood}
\end{equation}

\begin{equation}
\int_{\mu_{\rm{min}}}^{\mu_{\rm{max}}}\, P(\mu|n)\, d\mu= 1\ \rm{.}
\label{equation:SumOfProbabiltiesIsUnity}
\end{equation}

\ \\By combining Equations \ref{BayesTheorem} and \ref{equation:SumOfProbabiltiesIsUnity} we obtain

\begin{equation}
\int_{\mu_{\rm{min}}}^{\mu_{\rm{max}}}\, \frac{L(n|\mu)\pi(\mu)}{P(n)}\, d\mu= 1\ \rm{,}
\label{NormalizationFactor1}
\end{equation}

\ \\which leads to the formula for $P(n)$ 

\begin{equation}
P(n)=\int_{\mu_{\rm{min}}}^{\mu_{\rm{max}}}\, L(n|\mu)\pi(\mu)\, d\mu\ \rm{.}
\label{NormalizationFactor2}
\end{equation}

\ \\A decision has to be made about what prior information should be assumed about the expected number of events distribution. We assume a flat prior with $\pi(\mu)=\rm{constant}$ in order not to bias us towards a certain distribution of the signal and yet help us deal with systematic uncertainties. Given that $P(n)$ is also a constant with respect to $\mu$ we define $c_{n}$ as a constant that depends only on $n$, and given by the formula

\begin{equation}
c(n)=\frac{\pi(\mu)}{P(n)} \rm{.}
\label{Formula_cn}
\end{equation}

\ \\Combining Equations \ref{BayesTheorem} and \ref{Formula_cn}, we can express the posterior probability

\begin{equation} 
P(\mu|n)=c_{n}\frac{\mu^{n}e^{-\mu}}{n!}\ \rm{.}
\label{Posterior_mu_n}
\end{equation}

\ \\In our analysis, the number of expected events $\mu$ is represented by the sum of the expected number of background events ($B$) and the expected number of signal events ($S$), as seen in equation

\begin{equation} 
\mu=B+S \rm{.}
\label{ExpectedValue}
\end{equation}

\ \\ Also, the observed number of events $n$ is the number of measured data events ($D$), as in
\begin{equation} 
n=D \rm{.}
\label{ObservedValue}
\end{equation}

\ \\By combining Equations \ref{Posterior_mu_n}, \ref{ExpectedValue} and \ref{ObservedValue}, the equation for the likelihood becomes 

\begin{equation} 
P(B+S|D)=c_{D}\frac{(B+S)^{D} \cdot e^{-(B+S)}}{D!}\ \rm{.}
\label{Posterior_BSD}
\end{equation}

\ \\In our analysis, we want to set an upper limit on the value of the number of Higgs signal events $S$, given not only the measured number of data events $N$, but also the number of expected background events $B$. All the predictions of the Standard Model have been confirmed, except the existence of the Higgs boson. This is why we are sure, within uncertainties, of the existence of the background processes and of their predicted number of events $B$. We then reinterpret the posterior probability as the posterior probability of having $S$ expected signal events given $B$ expected background events and $D$ measured data events and we note it as $P(S|B,D)$. After the change of variable from $B+S$ to $S$, Equation \ref{Posterior_BSD} becomes 

\begin{equation} 
P(S|D,B)=c_{D}^{'}\frac{(B+S)^{D} \cdot e^{-(B+S)}}{D!}\ \rm{.}
\label{Posterior_S}
\end{equation}

\ \\We denote with $s$ the number of signal events predicted by the Standard Model. We express the true value for the signal predicted events $S$ by introducing the ratio $f$ as $S=f\cdot s$. We also denote that the total background event prediction is formed of several background processes. We therefore note $B=\sum_k b_k$, where k is the index of a given background sample. The expected value is now

\begin{equation} 
\mu = \sum_k b_k+f\cdot s\ \rm{.}
\label{ExpectedValue2}
\end{equation}

\ \\ After a change of variable from $S$ to $f$, Equation \ref{Posterior_S} becomes

\begin{equation} 
P(f|D,\sum b_k,s)=c_{D}^{''}\frac{(\sum_k b_k+f\cdot s)^{D} \cdot e^{-(\sum_k b_k+f\cdot s)}}{D!}\ \rm{.}
\label{Posterior_f}
\end{equation}

\ \\By definition, the minimum value that $f$ can take is zero. This case corresponds to the background-only hypothesis. Any positive value of $f$ corresponds to the background-plus-signal hypothesis. In this analysis, we set a 95\% credibility level upper limit on $f$, as is typical in particle physics experimental searches for new particles (the Standard Model Higgs boson search or new elementary particles predicted by theories beyond the Standard Model). In other words, we want to find the upper value $l$ for the ratio $f$ between the true number of expected events $S$ and the Standard Model prediction $s$, such that the probability that $f$ lies in the interval [0,$\,l$] is 0.95. Therefore, we have to solve for $l$ in the equation

\begin{equation} 
\int_{0}^{l} P(f|D,\sum b_k,s)\, df=0.95 \rm{.}
\label{UpperLimitEquation}
\end{equation}

\ \\By combining Equations \ref{Posterior_f} and \ref{UpperLimitEquation}, we have to solve for $l$ in the equation

\begin{equation} 
\int_{0}^{l} c_{D}^{''}\frac{(\sum_k b_k+f\cdot s)^{D} \cdot e^{-(\sum_k b_k+f\cdot s)}}{D!}\, df=0.95 \rm{.}
\label{UpperLimitEquation_f}
\end{equation}

\ \\Similarly with equation \ref{NormalizationFactor1}, the constant term $c_{D}$ is given by the condition

\begin{equation} 
\int_{0}^{\infty} c_{D}^{''}\frac{(\sum_k b_k+f\cdot s)^{D} \cdot e^{-(\sum_k b_k+f\cdot s)}}{D!}\, df=1
\label{Equation_c_D}
\end{equation}

\ \\to be 

\begin{equation} 
c_{D}^{''}=\frac{1}{\int_{0}^{\infty} \frac{(\sum_k b_k+f\cdot s)^{D} \cdot e^{-(\sum_k b_k+f\cdot s)}}{D!}} \rm{.}
\label{c_D}
\end{equation}

\ \\By combining Equations \ref{UpperLimitEquation} and \ref{c_D} we obtain the equation for $l$:

\begin{equation}
\displaystyle 
\frac{\int_{0}^{l} \frac{(\sum_k b_k+f\cdot s)^{D} \cdot e^{-(\sum_k b_k+f\cdot s)}}{D!}\, df}{\int_{0}^{\infty} \frac{(\sum_k b_k+f\cdot s)^{D} \cdot e^{-(\sum_k b_k+f\cdot s)}}{D!}\, df}=0.95 \rm{.}
\label{UpperLimitEquation2}
\end{equation}

\ \\Equation \ref{UpperLimitEquation2} cannot be solved analytically, so numerical integrators have to be used. The strategy is to consider increasingly higher values of $l$, compute the integral and stop immediately after the value of 0.95 is reached. 

\ \\In conclusion, this is the Bayesian approach to computing the upper limit $l$ for an analysis with only one bin in our final discriminant (for a simple counting experiment) and without taking any systematic uncertainties into account. In the following sections we will present how to take into account the systematic uncertainties, several bins and several analysis channels. 

\section{Taking Systematic Uncertainties Into Account\label{section:LimitOneBinWithSystematics}}

\ \\The procedure described in the previous section does not take into account the systematic uncertainties on the Standard Model predicted events for several background channels ($b_k$) and for the Higgs boson process ($s$). 

\ \\There are two types of systematic uncertainties. Rate systematic uncertainties apply to all the bins in the distribution. Shape systematic uncertainties are rate systematic uncertainties applied on a bin-by-bin basis. In this section we will describe how rate systematic uncertainties are introduced in a one-bin analysis. Shape systematics for binned discriminants will be described in the following section. In our analysis we employ only symmetric systematic uncertainties characterized by the standard deviation $\sigma$. 

\ \\For each independent systematic uncertainty we introduce a nuisance parameter $\nu_j$ as a coefficient to the expected number of events for a particular process. Most systematic uncertainties affect multiple physics processes. If we consider $\Delta_{\rm{sig}}$ and $\Delta_{\rm{bkg,k}}$ the sets of nuisance parameters that apply to the signal and to the $k^{\rm{th}}$ background process, respectively, then we can express the expected number of events $\mu$ as

\begin{equation} 
\mu \left(b_k, s, \sum_{\nu_j}\right)= \sum_k \left(\left(\prod_{j \in \Delta_{\rm{bkg,k}}}\nu_j\right)\right)\cdot b_k+f \cdot\left(\prod_{j \in \Delta_{\rm{sig}}}\nu_j \right)\cdot s\ \rm{.}
\label{ExpectedValue3}
\end{equation}

\ \\We model each nuisance parameter $\nu_j$ as a truncated Gaussian distribution. We recall that a Gaussian distribution of a variable $x$ with a mean $m$ and a variance $\sigma$ is given by

\begin{equation}
G(x|m,\sigma)=\frac{1}{\sqrt{2\pi\sigma^2}}\cdot e^{-\frac{(x-m)^2}{2\sigma^2}}\ \rm{.}
\label{Gaussian}
\end{equation}

\ \\Since nuisance parameters are used as coefficients for the predicted number of events for each background and signal process, we set $\sigma$ to the value of the chosen systematic uncertainty for that particular process expressed as a ratio of the absolute systematic uncertainty and the absolute expected number of events. For example, if a systematic uncertainty is 3\%, we set $\sigma=0.03$. This implies that the expected value $m$ is set to 1.0. Therefore, nuisance parameters $\nu_j$ are described by the following Gaussian distribution:

\begin{equation}
G(\nu_j|1.0,\sigma)=\frac{1}{\sqrt{2\pi\sigma^2}}\cdot e^{-\frac{(\nu_j-1.0)^2}{2\sigma^2}}\ \rm{.}
\label{Gaussian}
\end{equation}

\ \\We truncate the Gaussian distribution for each nuisance parameter in order to keep only those values that make physical sense, such as positive values. 

\ \\As per the instructions in the statistics section of the Particle Data Group review \cite{PDG}, we take the systematic uncertainties into account by convolving each nuisance parameter into the likelihood function, integrating over the nuisance parameters and then reproducing the reasoning in the previous section using the new likelihood. The likelihood is a function of the nuisance parameters $\nu$ since it is a function of the expected number of events $\mu$ which is a function of the nuisance parameters, as described in Equation \ref{ExpectedValue3}. The likelihood is given by

\begin{equation} 
L(D|f,\sum b_k,s, \sum \nu_j)= \frac{\mu(\nu_j)^{D} \cdot e^{-\mu(\nu_j)}}{D!}\rm{.}
\label{Likelihood_explicit_before}
\end{equation}

\ \\The new likelihood is a convolution of the old likelihood from Equation \ref{Likelihood_explicit_before} with all nuisance parameters modelled by Gaussian distributions and is given by 

\begin{equation} 
L(D|f,\sum b_k,s)=\int ... \int \frac{\mu(\nu_j)^{D} \cdot e^{-\mu(\nu_j)}}{D!} \cdot \prod_j G(\nu_j|1.0,\ \sigma_j) \prod_j d\nu_j\rm{.}
\label{Likelihood_explicit_after}
\end{equation} 

\ \\Each integral is performed between a lower range, chosen such that the quantity is positive, and infinity. This process is called integrating out the nuisance parameters and the result is that the new likelihood does not depend any more on the nuisance parameters. Since such an integral over many parameters is very difficult to compute with normal Monte Carlo methods, a Markov Chain Monte Carlo (MCMC) \cite{PDG} method is used in our analysis.

\ \\Since the likelihood depends now only on $f$, $s$ and $b_k$, as if the systematic uncertainties did not exist, we have reduced our problem to the simpler problem solved in the previous section. The upper limit $l$ is computed  by yet another integration given by the generic formula

\begin{equation}
\displaystyle 
\frac{\int_{0}^{l} L(D|f,\sum b_k,s)\, df}{\int_{0}^{\infty}L(D|f,\sum b_k,s)\, df}=0.95 \rm{.}
\label{UpperLimitEquation3}
\end{equation}

\section{Taking All The Bins Into Account\label{section:LimitAllBinsOneChannel}}

\ \\In the previous two sections we have described the Bayesian statistical inference of an upper limit assuming there is only one bin in our chosen distribution, as is the case in a simple counting experiment. We increase the sensitivity of the analysis if we take advantage of the shapes of the distributions, i.e. we consider each bin of the BNN output distribution separately. On a bin-by-bin basis the signal over background ratio (S/B) changes. By construction of our BNN output, the lower S/B is achieved for low values of BNN and higher S/B is achieved for high values of BNN. 

\ \\Rate systematics such as described above apply the same way to all the bins. However, shape systematics must now be taken into account as well. A shape change is actually a change in values on a bin-by-bin basis. A shape systematic can be understood as a rate systematic that is bin-specific. For each shape systematic we introduce a new nuisance parameter modelled by a Gaussian distribution. We integrate all the rate and shape nuisance parameters in the new binned likelihood in order to obtain the analysis likelihood. The only difference with respect to the previous section is the integration range for the shape nuisance parameters. For rate systematics, they were the value where the parameter became positive and infinity. For a shape systematic, the value is computed from the templates for the shape upper fluctuation, lower fluctuation and central value. In our analysis we use only one shape systematic, namely the one due to jet energy scale for the jet energy. 

\ \\The likelihood of the $i^{\rm{th}}$ bin is given by

\begin{equation} 
L_i(D_i|f,\sum b_{ik},s_i, \sum \nu_{ij})= \frac{\mu_i(\nu_{ij})^{D_i} \cdot e^{-\mu(\nu_{ij})}}{D_i!}\ \rm{,}
\label{Likelihood_explicit_before_onebin}
\end{equation}

\ \\where $\nu_{ij}$ represents both the rate and shape nuisance parameters for the $i^{\rm{th}}$ bin. Since all bins are statistically independent, the likelihood for all the binned BNN distribution is given by the product of the likelihoods for each bin, namely

\begin{equation} 
L(D|f,\sum b_k,s, \sum \nu_j)= \prod_{i \in \rm{bins}} L_i(D_i|f,\sum b_{ik},s_i, \sum \nu_{ij})\ \rm{.}
\label{Likelihood_explicit_before_allbin}
\end{equation} 

\ \\We have now reduced the problem to that of a one bin distribution. We integrate out the nuisance parameters as in Equation \ref{Likelihood_explicit_after} and we compute the upper limit as in Equation \ref{UpperLimitEquation2}. 

\section{Taking Into Account All Analysis Channels\label{section:LimitAllChannels}}

\ \\In the previous section we have presented the limit calculation for one analysis channel with a binned distribution. However, our analysis has six independent channels, each channel given by a pair of charged lepton and $b$-tagging categories. In this analysis we have two charged lepton (TIGHT and ISOTRK) and three $b$-tagging categories (SVTSVT, SVTJP05 and SVTnoJP05). 

\ \\We first perform the analysis in each category separately, which means computing a likelihood for each category. Since all channels are statistically independent, we combine all these categories by multiplying all these likelihoods together, which is equivalent to considering all the bins in the discriminant output from the six channels juxtaposed in only one histogram. We have reduced the problem of multiple channels to the simpler problem of only one channel and now we proceed as in the previous section. 

\section{Expected Limits and Observed Limits\label{section:LimitExpectedAndObserved}}

\ \\Before computing the limits using the real data distribution (an unblind analysis) we evaluate the sensitivity of our analysis to a $WH$ (plus $ZH$) signal (a blind analysis). We simulate the number of data events (pseudo-data events) by picking randomly according to a Poisson probability density function a real expected number of events in the interval generated by the background prediction allowed to fluctuate smoothly in its one sigma interval. Therefore, we do not use data, but pseudo-data. We do not use real events, but pseudo-events. Each pseudo-event is characterized by a different random number so that the value $D$ is specific for each event. Also, in the pseudo-events we assume there is no signal at all ($S$=0). The median of the distribution of upper limits is considered the median expected limit and it characterizes the sensitivity of our analysis. To ensure enough pseudo-experiment statistics to compute the lower and upper one and two standard deviation bounds on the median expected limit, and yet limit the CPU power consumption, we perform 3000 pseudo-experiments\footnote{The computation of 3000 pseudo-experiments for each mass point, for each charged lepton category and b-tagging category pair, and for all the categories combined takes on the order of 3 days using the parallel computing facilities of the CDF collaboration.}.

\ \\Contrary to the expected limit that uses pseudo-data in many pseudo-experiments, the observed limit (the real result of our analysis) uses real data in only one experiment. Both upper limits using pseudo-experiments and the real experiment use the methodology described in the sections above. When we present our expected and observed limits, we check that the observed limit is within the two standard deviation interval around the median expected value for each Higgs boson mass point. 

\begin{figure}[ht]
\begin{center}
   \includegraphics[width=.45\textwidth]{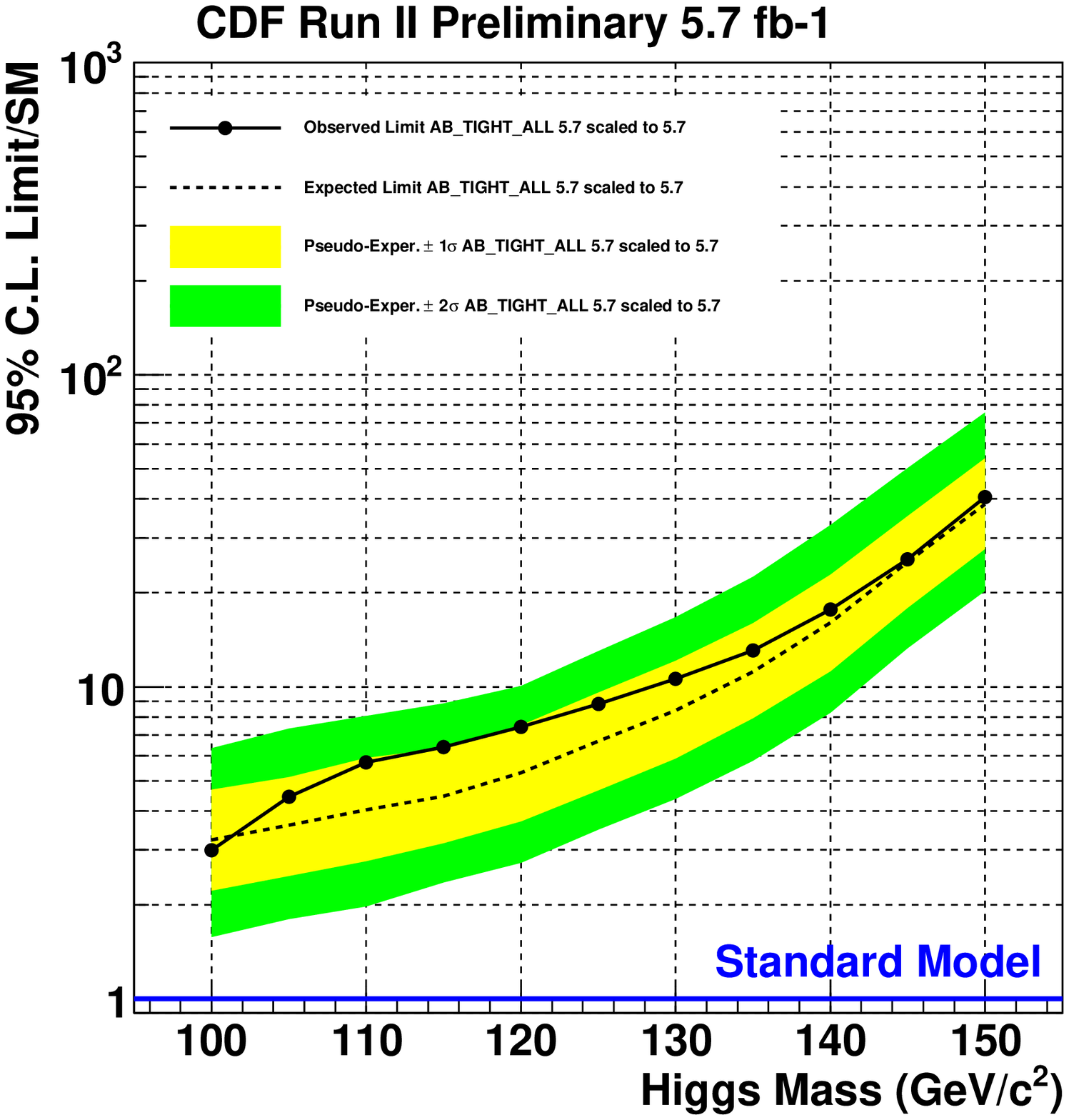}
   \includegraphics[width=.45\textwidth]{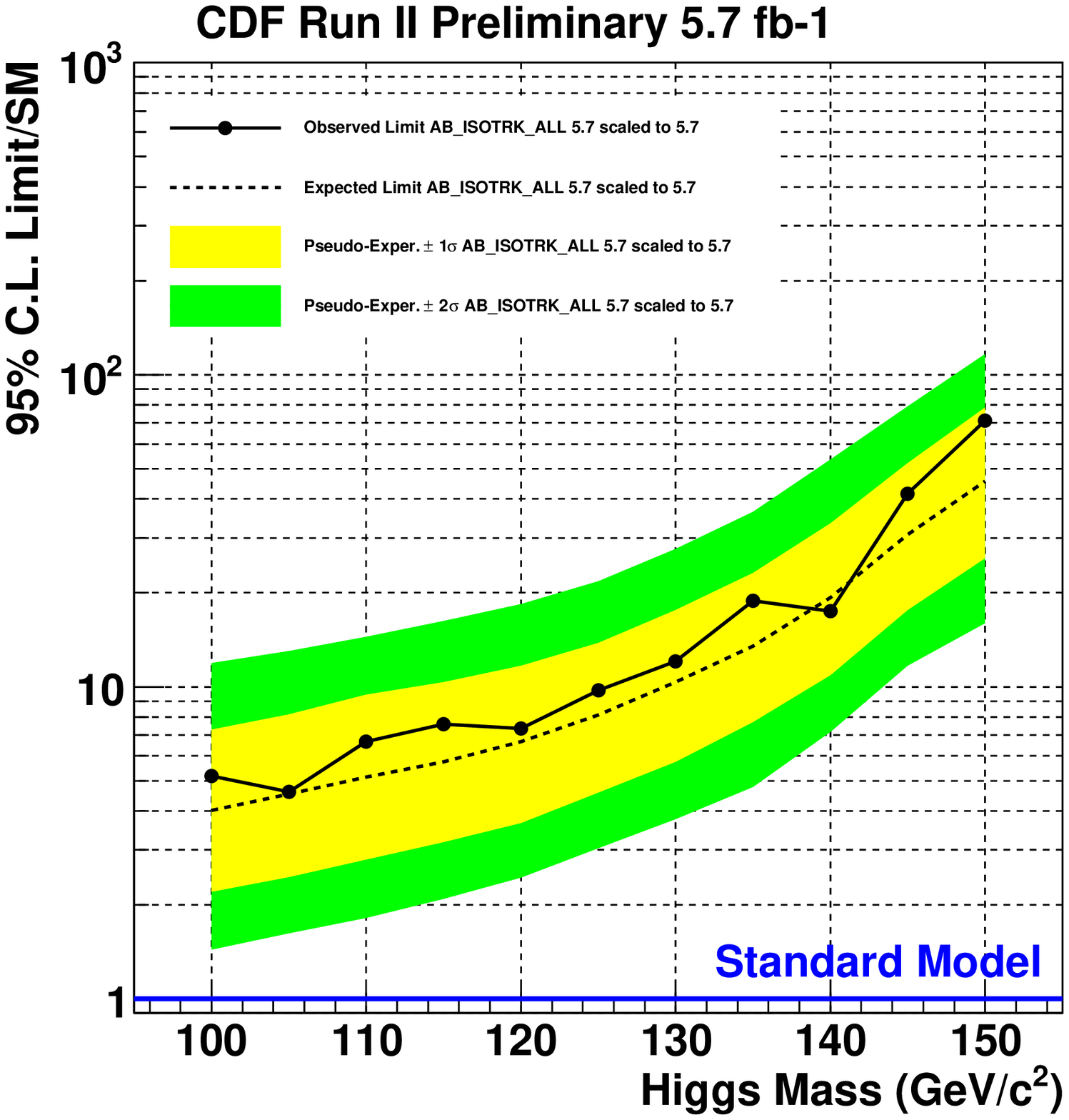}\\
    \includegraphics[width=.9\textwidth]{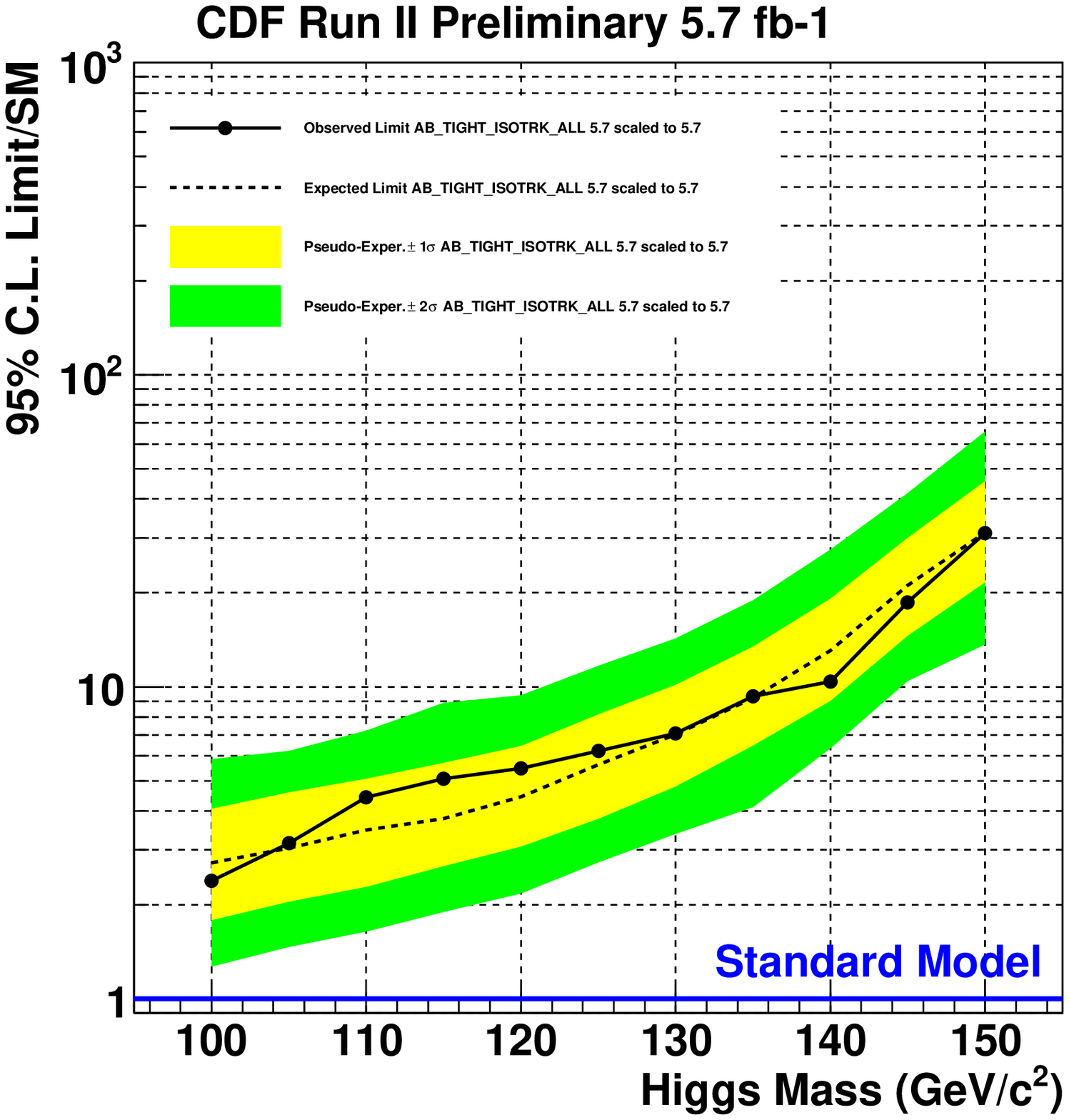}
\caption[Our result: upper limits on Standard Model Higgs boson]
{Expected and observed cross-section upper limits for a $WH$ search for the TIGHT (top left), ISOTRK (top right) and TIGHT+ISOTRK (bottom) charged lepton category and all $b$-tagging categories combined at CDF using 5.7 $\invfb$, as a function of the Higgs boson mass, between $100\gevcc$ and $150\gevcc$. The horizontal line at 1 represents the Standard Model prediction. The expected upper limits are represented by the dashed line. The yellow (green) band represents the 1 (2) standard deviation interval around the expected upper limit. The observed upper limits are represented by the solid line. \label{figure:Limits}}
\end{center}
\end{figure}

\section {WH Neural Network Upper Limits\label{section:LimitWH}}

\ \\At CDF a binned likelihood technique such as described in the previous sections is implemented in the MCLIMIT \cite{ConfidenceLevel} \cite{TevatronHiggsCombinationMarch2011} package that we used for our analysis as well. We search for a Higgs signal excess in our BNN neural network output distributions. Since we find no evidence for such an excess, we set upper limits on the $WH$ production cross section times the branching ratio: $\sigma(p\pbar \to WH) \cdot \branchingratio(H \to b\bar{b})$. We present the upper limits as ratios (normalized) to the Standard Model predicted ones (x SM) for TIGHT, ISOTRK and TIGHT combined with ISOTRK, each with all $b$-tagging categories combined in Table \ref{table:Limits} and in Figure \ref{figure:Limits}.

\begin{table}
\begin{center}
\begin{tabular}{|l|l|l|l|l|l|l|l|l|l|l|l|l|}
\multicolumn{12}{l}{}\\
\hline \hline
\multicolumn{12}{|c|}{CDF II Preliminary 5.7 $\invfb$}\\
\hline M(H) GeV/$c^2$ & 100 & 105 & 110 & 115 & 120 & 125 & 130 & 135 & 140 & 145 & 150 \\
\hline \hline
\multicolumn{12}{|c|}{TIGHT}\\
\hline
\hline
Exp2sP & 6.37 & 7.34 & 8.06 & 8.84 & 10.1 & 13.0 & 16.7 & 22.5 & 33.0 & 50.1 & 75.8 \\
Exp1sP & 4.69 & 5.13 & 5.91 & 6.36 & 7.53 & 9.59 & 12.1 & 16.0 & 22.9 & 35.2 & 54.0 \\
Exp    & 3.23 & 3.6 & 4.03 & 4.46 & 5.29 & 6.69 & 8.41 & 11.2 & 16.1 & 25.1 & 38.5 \\
Exp1sM & 2.22 & 2.48 & 2.76 & 3.15 & 3.71 & 4.67 & 5.90 & 7.92 & 11.2 & 17.9 & 27.6 \\
Exp2sM & 1.58 & 1.80 & 1.98 & 2.36 & 2.73 & 3.49 & 4.38 & 5.8 & 8.27 & 13.3 & 20.3 \\
\hline
Obs    & 3.00 & 4.43 & 5.73 & 6.4 & 7.43 & 8.8 & 10.6 & 13.1 & 17.7 & 25.6 & 40.5 \\
\hline
\hline
\multicolumn{12}{|c|}{ISOTRK}\\
\hline
Exp2sP & 11.9 & 13.0 & 14.4 & 16.2 & 18.4 & 21.8 & 27.7 & 36.4 & 53.6 & 79.2 & 117 \\
Exp1sP & 7.29 & 8.17 & 9.45 & 10.3 & 11.7 & 13.8 & 17.6 & 23.2 & 33.5 & 52.0 & 78.6 \\
Exp    & 4.02 & 4.53 & 5.13 & 5.76 & 6.67 & 8.13 & 10.3 & 13.5 & 19.3 & 30.7 & 45.6 \\
Exp1sM & 2.20 & 2.46 & 2.80 & 3.18 & 3.66 & 4.58 & 5.74 & 7.72 & 10.9 & 17.6 & 25.8 \\
Exp2sM & 1.44 & 1.62 & 1.82 & 2.09 & 2.45 & 3.04 & 3.76 & 4.78 & 7.18 & 11.7 & 16.00 \\
\hline
Obs    & 5.19 & 4.61 & 6.68 & 7.6 & 7.34 & 9.74 & 12.1 & 18.8 & 17.5 & 41.5 & 71.20 \\
\hline
\hline
\multicolumn{12}{|c|}{TIGHT and ISOTRK}\\
\hline
Exp2sP & 5.86 & 6.23 & 7.22 & 8.87 & 9.38 & 11.7 & 14.3 & 19.0 & 27.5 & 41.6 & 65.7 \\
Exp1sP & 4.06 & 4.59 & 5.07 & 5.70 & 6.46 & 8.16 & 10.2 & 13.4 & 19.2 & 30.0 & 45.6 \\
Exp    & 2.73 & 3.04 & 3.47 & 3.79 & 4.44 & 5.62 & 7.04 & 9.20 & 13.1 & 21.1 & 31.2 \\
Exp1sM & 1.79 & 2.05 & 2.29 & 2.66 & 3.08 & 3.78 & 4.80 & 6.48 & 9.04 & 14.6 & 21.6 \\
Exp2sM & 1.27 & 1.47 & 1.64 & 1.90 & 2.18 & 2.74 & 3.38 & 4.12 & 6.37 & 10.4 & 13.7 \\
\hline
Obs    & 2.39 & 3.15 & 4.42 & 5.08 & 5.48 & 6.24 & 7.09 & 9.32 & 10.4 & 18.6 & 31.1 \\
\hline \hline
\end{tabular}
\caption[Our result: upper limits, on Standard Model Higgs boson]{Upper limits, expressed as multiples of the Standard Model prediction, for TIGHT charged leptons only, ISOTRK charged leptons only, TIGHT and ISOTRK charged leptons combined, in all cases with all $b$-tagging categories combined, as a function of the SM Higgs boson mass using 5.7 $\invfb$ at CDF. $m_H$ represents the hypothetical mass of the Higgs boson and is expressed in $\gevcc$. Exp represents the expected median limit using 3000 pseudo-experiments. Exp2sP (Exp2sM) represents the upper (lower) bound on the two standard deviations around Exp. Exp1sP (Exp1sM) represents the upper (lower) bound on one standard deviation around Exp. Obs represents the observed upper limit using data.}
\label{table:Limits}
\end{center}
\end{table}

\begin{figure}[b]
\begin{center}
    \includegraphics[width=.70\textwidth]{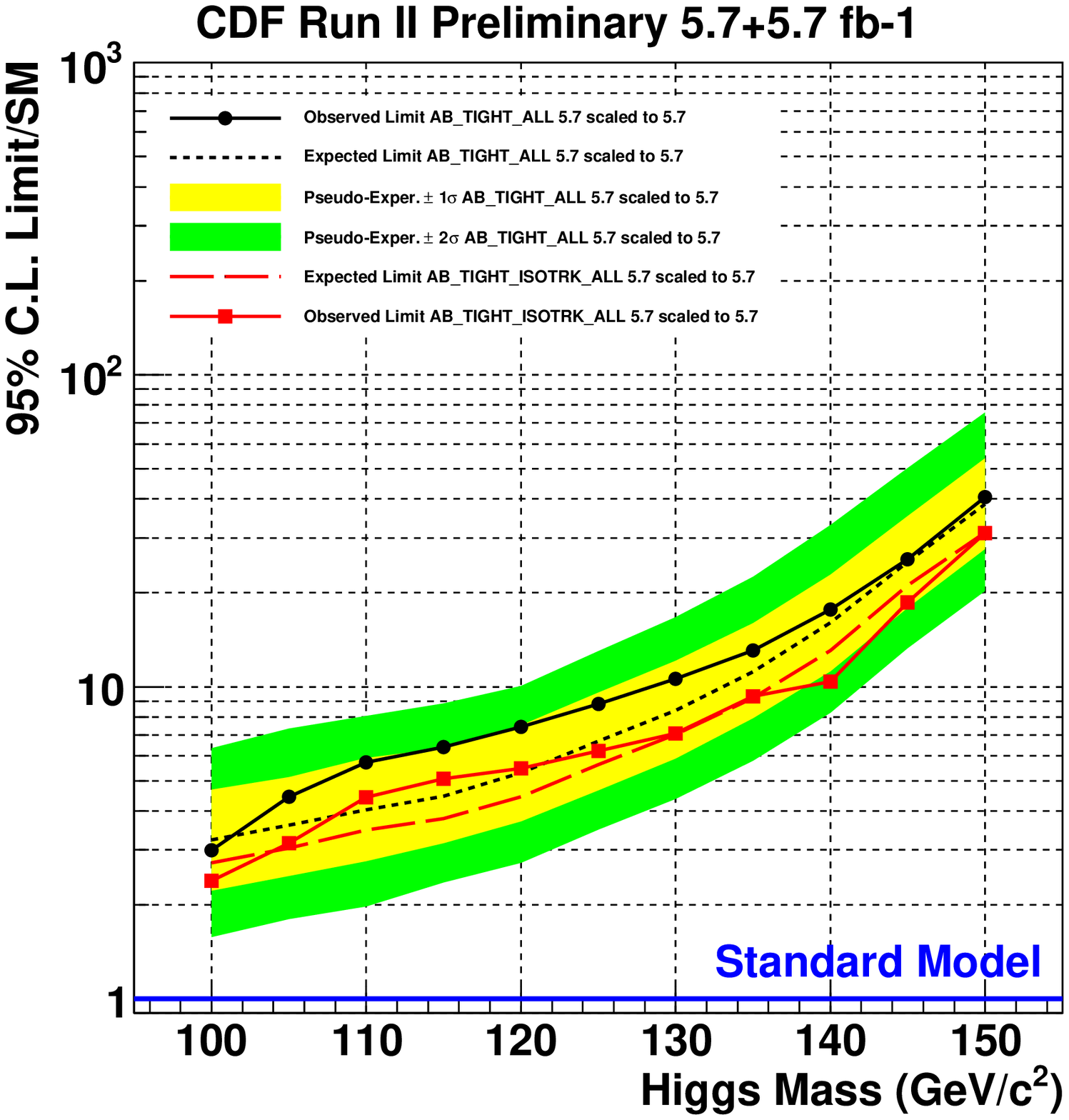}
\caption[Improvement due to ISOTRK on the top of TIGHT category]
{Expected and observed cross-section upper limits for a $WH$ search overlaid for the TIGHT category and TIGHT combined with ISOTRK category and all $b$-tagging categories combined at CDF using 5.7 $\invfb$, as a function of the Higgs boson mass, between $100\gevcc$ and $150\gevcc$. The horizontal line at 1 represents the Standard Model prediction. The expected upper limits are represented by the dashed lines in black (TIGHT) and red (TIGHT+ISOTRK). The yellow (green) band represents the 1 (2) standard deviation interval around the expected upper limit for TIGHT. The observed upper limits are represented by the solid lines in black (TIGHT) and red (TIGHT+ISOTRK). \label{figure:LimitTIGHTvsTIGHTISOTRK}}
\end{center}
\end{figure}

\section{Impact of Our Original Contribution}

\ \\Since my original contribution to the $WH$ analysis is the addition of the ISOTRK charged lepton category with respect to the TIGHT charged lepton category, it is shown in Table \ref{table:LimitTIGHTvsTIGHTISOTRK} for each Higgs boson mass point the expected and observed upper limits for the TIGHT category only and the TIGHT combined with ISOTRK, as well as the percentage by which the limits become smaller and thus better. In Figure \ref{figure:LimitTIGHTvsTIGHTISOTRK} it is shown in black the expected and observed limits for the TIGHT charged lepton category only and in red the ones from the combination of the TIGHT and ISOTRK categories. The improvement both for the expected limits and observed limits visible in Table \ref{table:LimitTIGHTvsTIGHTISOTRK} is now also visible in a visual form. 

\begin{table}[f]
\begin{center}
\begin{tabular}{|l|l|l|l|l|l|l|l|l|l|l|l|l|}
\multicolumn{12}{l}{}\\
\hline \hline
\multicolumn{12}{|c|}{CDF II Preliminary 5.7 $\invfb$}\\
\hline 
\multicolumn{12}{|c|}{\% Improvement due to ISOTRK over TIGHT}\\
\hline \hline
\multicolumn{12}{|c|}{Expected limits}\\
\hline
M(H) in $\gevcc$ & 100 & 105 & 110 & 115 & 120 & 125 & 130 & 135 & 140 & 145 & 150 \\
\hline
TIGHT     & 3.23 & 3.60 & 4.03 & 4.46 & 5.29 & 6.69 & 8.41 & 11.2 & 16.1 & 25.1 & 38.5 \\
\hline
TIGHT+ISOTRK   & 2.73 & 3.04 & 3.47 & 3.79 & 4.44 & 5.62 & 7.04 & 9.20 & 13.1 & 21.1 & 31.2 \\
\hline
\% Improvement   & 15.5 & 18.4 & 16.1 & 17.7 & 16.1 & 16.0 & 16.3 & 17.9 & 18.6 & 15.9 & 19.0 \\
\hline \hline
\multicolumn{12}{|c|}{Observed limits}\\
\hline 
TIGHT   & 3.00 & 4.43 & 5.73 & 6.40 & 7.43 & 8.80 & 10.6 & 13.1 & 17.7 & 25.6 & 40.5 \\
\hline
TIGHT+ISOTRK  & 2.39 & 3.15 & 4.42 & 5.08 & 5.48 & 6.24 & 7.09 & 9.32 & 10.4 & 18.6 & 31.1 \\
\hline \hline
\end{tabular}
\caption[Improvement due to ISOTRK on the top of TIGHT category]{Expected and observed cross-section upper limits for TIGHT and TIGHT+ISOTRK analysis, as well as percentage improvement in the expected limit when ISOTRK is combined with TIGHT.}
\label{table:LimitTIGHTvsTIGHTISOTRK}
\end{center}
\end{table}

\section{Summary}

\ \\In this chapter we have presented the computation methodology and results for the upper limit on the Higgs boson cross section times branching ratio as a function of the Higgs boson mass for the TIGHT, ISOTRK and TIGHT+ISOTRK cases, when all $b$-tagging categories are combined. In the first part of the chapter we introduce the methodology of the limit calculation for a one-bin analysis without and with systematic uncertainties. We then presented the methodology for a multi-bin analysis. Finally, we presented how several channels are combined, such as our $b$-tagging categories. We then introduced the pseudo-experiments, which use data simulation by allowing the background prediction to fluctuate within its uncertainty, in order to measure the sensitivity of the analysis. We then presented the real measurement using real data. In the last part of the chapter we presented both in plots and tables that adding the ISOTRK charged lepton category on the top of the TIGHT charged lepton category improves both the expected and observed upper limits significantly. 

\clearpage{\pagestyle{empty}\cleardoublepage}

\chapter{Conclusions and Discussions\label{chapter:Conclusion}}

\section{Summary}

\ \\In this dissertation we have presented an experimental test of the current theory of particle physics and their interactions, the Standard Model (SM). All the predictions of the SM have been observed experimentally, except one, namely the existence of a new elementary particle called the Higgs boson. If the particle is discovered, the SM is confirmed experimentally. If the particle is excluded, then the SM is refuted, which means that the true model of nature is not the SM, but some other theory beyond the SM. The mass of the SM Higgs boson is unconstrained by the theory, but direct and indirect experimental searches constrain it at 95\% confidence level between 114.4 and 158 or between 175 and 185 $\gevcc$. The preferred Higgs boson mass from SM indirect fits is towards the lower edge of the allowed mass ranges. We therefore performed an experimental search of the Higgs boson that was most sensitive to possible low Higgs boson masses.

\ \\We study a sample of $p\bar{p}$ collisions at the Tevatron at the centre-of-mass energy $\sqrt{s}=1.96\ \tev$ that corresponds to an integrated luminosity of 5.7 $\invfb$ collected by the Collider Detector at Fermilab. There are many ways a Higgs boson is hypothetically produced at the Tevatron, but independently of the production mode, once produced, a Higgs boson would decay the same way, depending just on its mass. For masses below 135 $\gevcc$, the Higgs boson is expected to decay predominantly to a $b\bar{b}$ pair, whereas for higher mass it decays predominantly to a $W^+W^-$ pair. Given our preference for a low mass Higgs boson, we choose the most promising channel to identify a Higgs boson that decays to $b\bar{b}$ pairs: the associated production between the Higgs boson and the $W$ boson, where the $W$ boson decays leptonically. Our analysis channel is therefore $WH\rightarrow l\nu b\bar{b}$. 

\ \\We select events consistent with a signature of a high-$\pt$ charged lepton (electron or muon) candidate, large $\met$ and exactly two jets. In order to improve the signal over background ratio, we require that at least one of the jets is identified to originate in a $b$ quark. We use a sample of events without this requirement as a Pretag control sample. In order to discriminate further between signal and background events, we employ an artificial neural network. Finally, using a Bayesian statistical inference technique, we compute expected and upper limits on the cross section times branching ratio with respect to the SM prediction for Higgs masses between 100 and 150 $\gevcc$. 

\ \\The main charged lepton categories at CDF are electron and muon candidates with stringent reconstruction criteria. An electron (muon) candidate is typically a high-$\pt$ isolated track that is matched to a calorimeter cluster (muon stub). We reconstruct tight electron (muon) candidates using an electron(muon)-inclusive trigger. We add together all the $WH$ events selected using tight charged leptons into the TIGHT sample. 

\ \\Our detector has uninstrumented regions both at the calorimeter level, such as the small space between calorimeter towers, and at the muon detector level, where the eta-phi coverage is not uniform. We introduce a novel charged-lepton category with looser reconstruction criteria, namely a high-$\pt$ isolated track that is freed from the requirement to match a calorimeter cluster or a muon stub. Such charged-lepton candidates recover also real charged leptons that would have been otherwise lost in the non-instrumented regions of the detector. We call the $WH$ sample collected in this way the ISOTRK sample. We make sure that the ISOTRK and TIGHT samples are orthogonal.

\ \\As there is no ISOTRK-dedicated trigger at CDF, we use triggers that make use of the orthogonal information in the event, namely the $\met$ and the jets. We have three such MET-based triggers at CDF. We parameterized at each of the three trigger levels the trigger efficiency turnon curves as a function of trigger $\met$ and identified the appropriate jet kinematic selection so that the efficiency is flat with respect to jets and only varies with respect to trigger $\met$. We also measured the prescale of one of the triggers that is prescaled. Since not all triggers were used for all the runs in our dataset, we measured the fraction of luminosity where each of the possible combinations of the triggers were used.

\ \\There are many possible ways to combine the triggers and indeed in the $WH$ search our contribution on this topic was a work in progress. As such, the $WH$ analysis using 2.7 $\invfb$ from the summer of 2008 used only one MET-based trigger. The $WH$ analyses using 4.3 $\invfb$ from the summer of 2009 and using $5.7 \invfb$ from the summer of 2010 used two different MET-based triggers. The ISOTRK channel was an original contribution to these analyses. These results went into the CDF and Tevatron combinations from those years and were presented at the major summer conferences. 

\ \\In this latest analysis we use the same integrated luminosity as the $WH$ analysis of the summer of 2010. The motivation is that we focused on two main issues. The $WH$ group at CDF of which I am an active member decided to develop a new data analysis software framework called WHAM, completely independent of the one used for the 2010 analysis. This new framework is more modular, more flexible and dedicated to the single charged lepton plus missing transverse energy plus jets. Several analyses such as $WH$, $WZ$, single top, technicolor, $t\bar{t}H$ are currently produced in this framework. WHAM allows the sharing of almost all the tools between the analyses and as such avoids redundancy, allows an analysis improvement to be instantly propagated to the other related analyses, and allows greater scrutiny of a common piece of code and thus identify bugs easier. I am one of the three main authors of the software framework. My first task was therefore to reproduce in the new framework the 2010 $WH$ analysis in the TIGHT and ISOTRK category, while my second task was to improve the MET-plus-jets parameterization in order to improve the ISOTRK category, the results of which are embodied in this dissertation.  

%\ \\Our second major contribution is the improvement of the ISOTRK sample event count through the introduction of the third MET-based trigger, as well as a novel method to combine in an efficient way an unlimited number of triggers in order to maximize the signal acceptance and yet avoid correlations between triggers such as is the case when one takes ``OR'' between triggers. Taking an ``OR'' between triggers is not wrong, but one has to be very careful to evaluate correctly the systematic uncertainty of the trigger combination. The method we introduce allows to easily compute correctly the systematic uncertainty, which a priory is also smaller than in the case of correlated triggers. Moreover, it is by construction the most efficient method short when no correlations are used. It is such a method that is presented in this thesis for the ISOTRK category after having been reviewed by the Higgs group at CDF.

\ \\Our final result is represented by the expected and observed 95\% CL upper limits on the Higgs boson cross section times branching ratio with respect to the SM when all categories are combined. The expected upper limits vary between 2.73 $\times$~SM and 31.2 $\times$~SM for a mass range of 100-150 $\gevcc$. The improvement in sensitivity due to addition of the ISOTRK charged lepton category is between 16 and 19\% for the entire mass range. The observed upper limits vary between 2.39 $\times$~SM and 31.1 $\times$~SM for the entire mass range. Since the upper limit set by the LEP experiments is 114.4 $\gevcc$, it is interesting to note that for a 115 $\gevcc$ Higgs boson, we compute an expected upper limit of 3.79 $\times$~SM and an observed upper limit of 5.08 $\times$~SM. 

\section{Future Prospects}

\ \\This section presents the future prospects and potential improvements of the $WH$ search at CDF, of the Higgs combined searches at the Tevatron and of the original method to combine an unlimited numbers of MET-plus-jets triggers that I introduce in this analysis. 

\subsection{$WH$ search at CDF}

\ \\By the time this thesis is submitted, there are about two more months remaining until the summer conferences of 2011. I will lead the $WH$ effort through the internal review process and a possible subsequent publication for approximately 7 months thanks to a Universities Research Association\footnote{Universities Research Association (URA) is the association of universities that manages Fermilab. McGill University and the University of Toronto are the only universities in Canada that are members of the URA.} Visiting Scholar Postdoctoral Fellowship and a grant for travel to Fermilab. It is both the goal of the CDF experiment and my personal goal to improve this analysis with the newly available datasets and as many of the following potential improvements in the analysis technique: add a forward electron charged lepton category; migrate some of the current ISOTRK events in a separate loose muon category and use triggers dedicated to them; replace the current $b$-tagging algorithms with a newer one that has been produced by the CDF collaboration. These improvements are expected to sharpen the analysis sensitivity significantly more than just the addition of more integrated luminosity would allow us to, as seen in the following section. 

\begin{figure}[htb]
\begin{center}
\includegraphics[width=7.0cm]{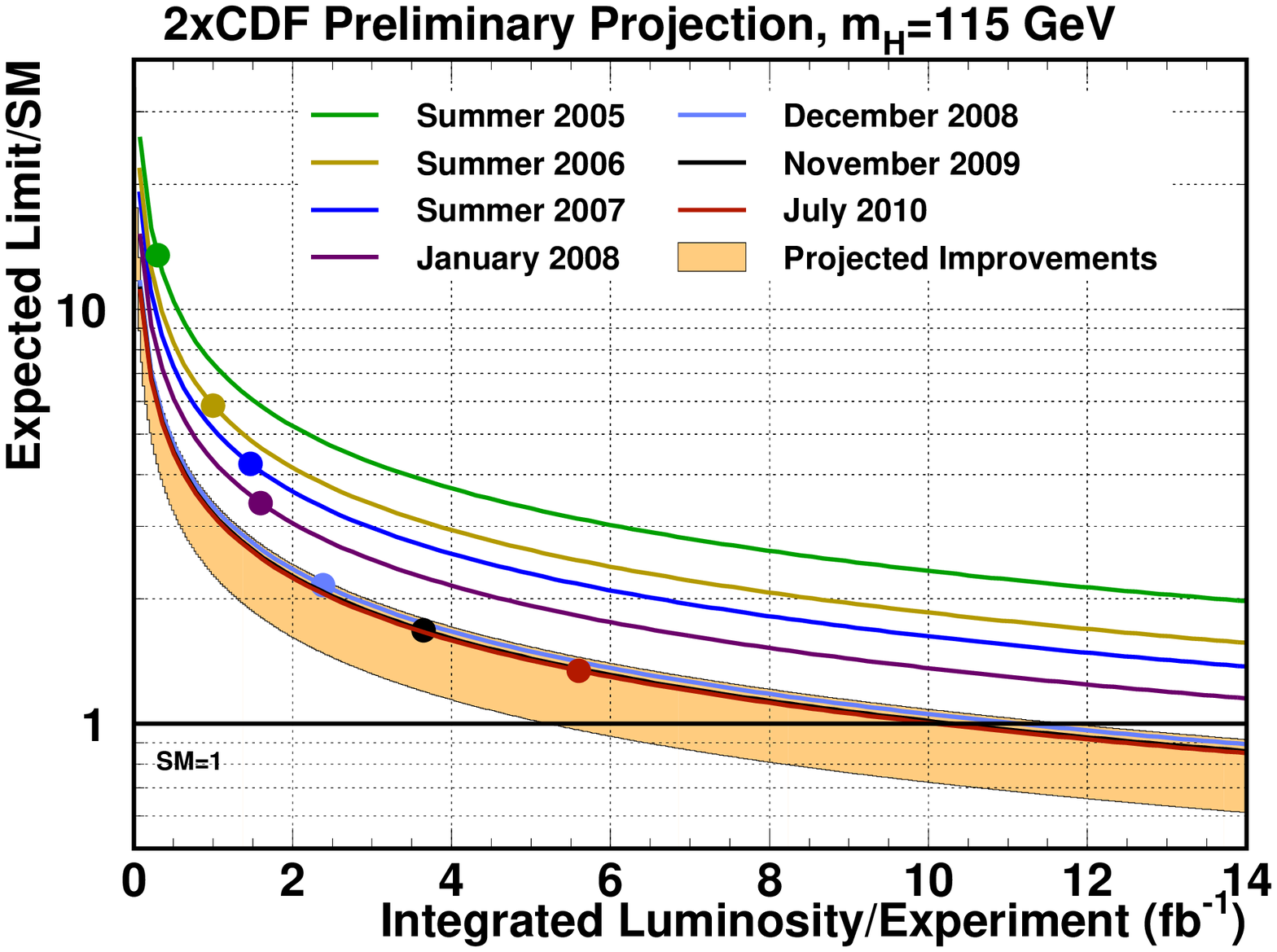}
\includegraphics[width=7.0cm]{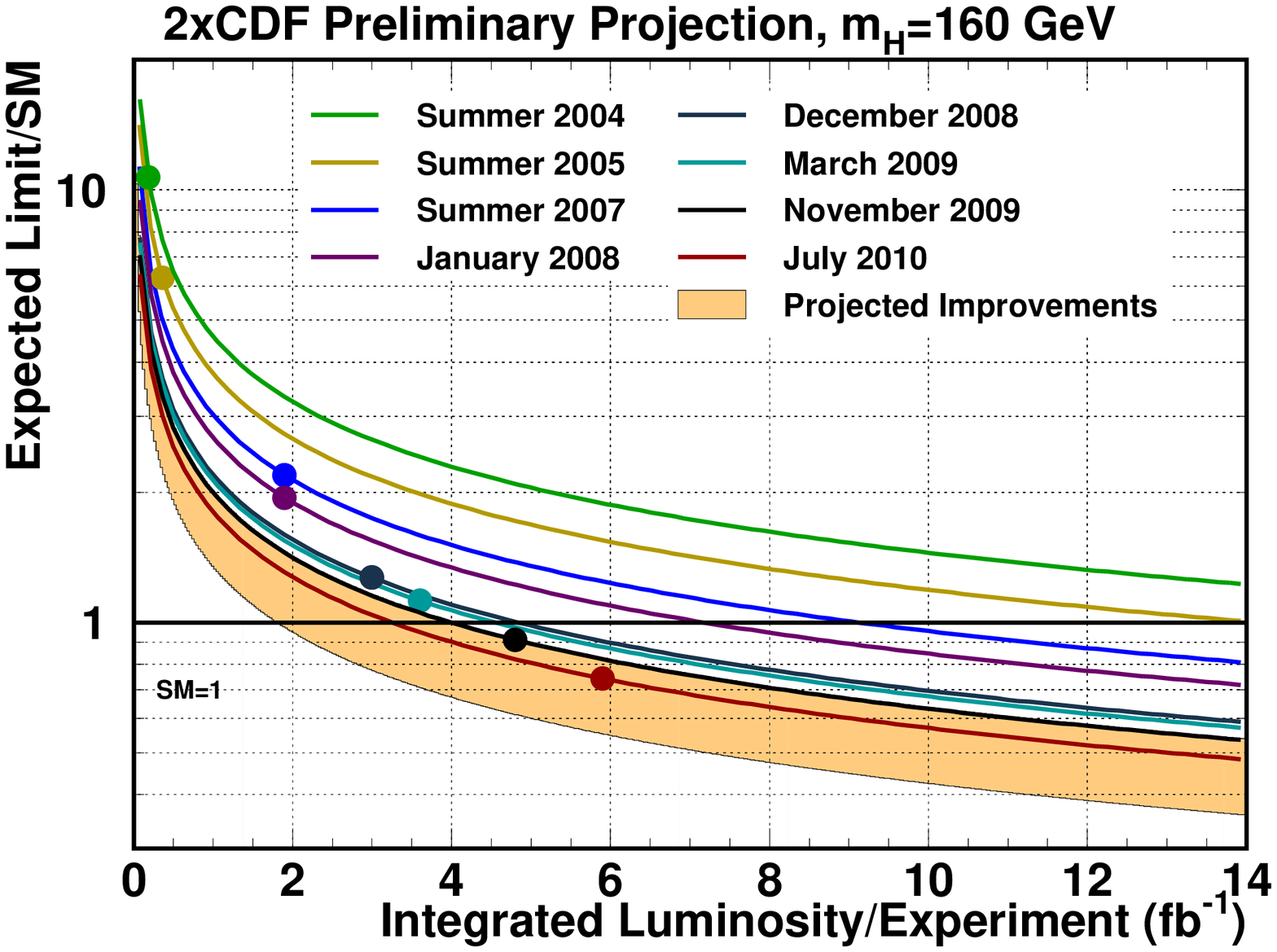}
\caption[Projected expected median sensitivity scaling with integrated luminosity]
{Projected expected median sensitivity scaling with total integrated luminosity used by CDF, assuming a combination of two CDF experiments, as of July 2010. The solid lines represent the sensitivity projections as a function of integrated luminosity with analysis improvement, while the dots represent the actual limits set for a Higgs boson mass of $115 \gevcc$ (top) and $160 \gevcc$ (bottom). The brown band represents the addition of analysis improvements that are expected to be added to the CDF analyses in the future and that could allow to reach Standard Model sensitivity with 9 $\invfb$ that is expected to be collected by the end of Run-II of the Tevatron accelerator in the summer of 2011. Credit image to the CDF collaboration. \label{figure:SensitivityImprovementsx2CDF}}
\end{center}
\end{figure}

\subsection{Higgs search at the Tevatron}

\ \\The cross section times branching ratio sensitivity of Higgs searches at the Tevatron typically improve continuously due to two reasons: using larger integrated luminosity data sets, as Tevatron accelerator performs excellently and is scheduled to run until 30 September 2011; improving the analysis techniques. If analysis improvements are ignored, the sensitivity scales as $1/\sqrt{\int \! {\cal{L}} dt}$. Over the past few years, CDF has managed to improve always the expected sensitivity more than just could be achieved by using larger data sets, as can be seen in Figure~\ref{figure:SensitivityImprovementsx2CDF} for a Higgs boson mass of $115 \gevcc$ (top) and $160 \gevcc$ (bottom), where twice larger datasets than CDF only are assumed, simulating at first order an expected Tevatron combination between the CDF and DZero experiments, that have almost similar data sets and analysis sensitivity. Also, Figure~\ref{figure:SensitivityImprovementsx2CDF} suggests that with about 9 $\invfb$ of integrated luminosity collected until the expected end of Run-II at the Tevatron in summer 2011, if the potential analysis improvements identified are implemented, then the Tevatron combination would approach a Standard Model sensitivity at both the low and high Higgs boson mass.

\subsection{Trigger Combination Method}

\ \\Our original method to combine triggers is already in use by other CDF analyses due to the fact that the method is implemented by a user friendly software package called ABCDF that we designed. The $WZ\rightarrow l\nu b\bar{b}$ and $ZH \rightarrow \nu\bar{\nu}b\bar{b}$ will use the three triggers combined by the novel method in the searches they prepare for the summer of 2011. The method has potential applications in searches of physics beyond the Standard Model, such as supersymmetry, that have as key signatures large missing transverse energy and jets. As such, it is already applied at other CDF analyses and can be applied to other analyses at DZero at the Tevatron or at ATLAS and CMS at the Large Hadron Collider at CERN. The method has the advantage that it incorporates an unlimited number of triggers and each trigger is allowed to have its own jet selection, prescale and interval of applicability. 

\subsection{Higgs searches at the LHC}

\ \\Both ATLAS and CMS experiments at the Large Hadron Collider have started to present results on SM Higgs searches. As of April 2011, none of these searches has become more sensitive than those of CDF and DZero at the Tevatron. However, the LHC has broken the instantaneous luminosity record of the Tevatron and as more data is rapidly collected by the well-performing ATLAS and CMS detectors, they will surpass in sensitivity finally the Tevatron searches. For now, the jury is still out and the friendly competition between the Tevatron and LHC is ongoing. Being in experimental particle physics is indeed living in interesting times. 

\section{Conclusions}

\ \\We have presented a $WH$ search at CDF and we introduced a novel charged lepton category that improved the sensitivity of the search by 16-19\% across a Higgs boson mass of 100-150 $\gevcc$. In the process we developed a novel method to combine an unlimited number of MET-plus-jets triggers, which is already being used by other CDF analysis and has the potential to be useful for other experiments as well, since triggers based on this signature are key to some physics beyond the Standard Model scenarios. 

\clearpage{\pagestyle{empty}\cleardoublepage}

\appendix
\chapter{MET-based Trigger Parametrization\label{chapter:METTriggers}}

\ \\Events that do not present a tight charged lepton, but have a high-$\pt$ track isolated from other activity in the tracking system are called isolated track (ISOTRK) events. This new charged lepton category has looser reconstruction criteria than the tight charged leptons and represents my original contribution to the $WH$ analysis. At CDF the tight charged lepton events are collected using dedicated inclusive electron or muon triggers. However, there is no trigger dedicated to ISOTRK events. For this reason, we use triggers based on the orthogonal information to the charged lepton, namely the missing transverse energy (MET) and jets. We call them generically MET-based triggers. At CDF we have three such triggers. We denote them with MET2J, MET45 and METDI. For Monte Carlo simulated events we have to model the trigger selection by applying on an event-by-event basis a weight that represents the trigger efficiency. This chapter describes the parametrization that we measured for these MET-based triggers used in the analysis for the ISOTRK category. The novel method we introduced to combine these three MET-based triggers is described in Appendix \ref{chapter:CombineTriggers}. 

\section{Three MET-based Triggers at CDF}

\ \\There are three trigger levels at CDF, which we denote L1, L2 and L3. At each trigger level, quantities are reconstructed more correctly, using successively greater computing resources, than at the previous trigger level. As a general rule, trigger requirements become more stringent as the trigger level is higher. The offline event selection is then even tighter. In this section we will describe the trigger requirements for each of the three MET-based triggers employed in this analysis.

\subsection{MET + 2 Jets Trigger}

\ \\The MET + 2 jets trigger (MET2J) has been active since the beginning of Run II at the Tevatron. A data event fires the L1 of MET2J if it has a MET larger than 28 GeV ($\met > 28\ \gev$), in which case it is sent automatically to be studied by L2. In order to pass the L2 requirements, a data event must have $\met > 30\ \gev$ and at least two jets, one of them with a transverse energy $\et > 20\ \gev$ and reconstructed in the central region of the detector ($|\eta|<1.1$) and the other jet with $\et > 15\ \gev$ and $|\eta|<2.0$. Not all events that pass requirements at L2 are sent to be analyzed by the L3, which means that this trigger is prescaled. The prescale is done in an automatic way as a function of the instantaneous luminosity. We measure the prescale of this trigger, as seen in Section \ref{section:TriggerPrescales}. Events that reach L3 and meet the requirement of $\met > 35\ \gev$ fire the full MET2J trigger and are saved on tape to be used in our analysis. Since there are real data events that do not meet these trigger criteria, they are not stored and therefore not used in the analysis. For this reason, not all Monte Carlo simulated events should be used, or rather they should all be used to preserve the statistics, but they should be weighted to simulate the trigger selection. 

\ \\In time, there were four major versions of the MET2J trigger used at CDF. The trigger evolved as the instantaneous luminosity increased and as the requirements of the specific physics groups changed. For example, previous versions required a $\met > 25\ \gev$ at L1, did not require that one of the jets be central at L2, and there was no prescale at L2. For a given run, only one version of the trigger was used. In this analysis we parameterize the trigger efficiency averaged out over the several historical versions, as if there were only one version. We make sure our offline requirements are tighter than the most recent and stringent requirements of the MET2J trigger. 

\subsection{MET Trigger}

\ \\The MET-only trigger has requirements only on $\met$, but not on jets. This trigger has also been in existence since the beginning of Run II. It comes in two historical versions. The first version was used for the first 2.3 $\invfb$ of the integrated luminosity and required at L1 $\met > 25\ \gev$, at L2 $\met > 25\ \gev$ and $\met > 45\ \gev$. The second version started being used afterwards for the remaining of 3.4 $\invfb$ of the integrated luminosity used in this analysis. The physics desire was to decrease the $\met$ value in order to select more events from rare processes, such as Higgs or physics beyond the Standard Model. As such, it is required at L1 that $\met > 28\ \gev$, at L2 $\met > 35\ \gev$ and at L3 $\met > 40\ \gev$. Just as in the case of the MET2J trigger, we parametrize the trigger as if it has only one version which we call MET45. This trigger was never prescaled. 

\ \\One caveat is that somewhere in the early data taking there was a bug in the MET45 trigger and although the event fired and the information was stored, the information is not to be trusted. Therefore, in the run range 178637-192363, which approximates 3\% of the total integrated luminosity used in this analysis, we treat data events as if the trigger MET45 was not defined and therefore could not have fired. We have to simulate this in Monte Carlo events as well. The novel method we introduce to combine triggers takes this into account easily, as seen in Appendix \ref{chapter:CombineTriggers}. 

\subsection{MET + Dijet Trigger}

\ \\The third and last MET-based trigger at CDF is the MET + dijet trigger. As its name suggests, it is very similar to the MET2J trigger and in order to avoid confusion it is denoted in this thesis as METDI. This trigger was first introduced when about 2.4 $\invfb$ of integrated luminosity have already been collected and has never been changed since. This trigger was designed by the Higgs Trigger Task Force and was optimized for the Higgs boson search. This trigger was never prescaled. Since this trigger was applied only in about 42\% of the integrated luminosity, we also have to simulate that in Monte Carlo events. The novel method we implemented does that easily, as described in Appendix \ref{chapter:CombineTriggers}. 

\section{Variable Choice for Trigger Parametrization}

\ \\We want to parametrize the trigger efficiency turnon curves as a function of only one variable that is common for all the three triggers and apply cuts so that we are in the plateau regions with respect to all the other variables. For the MET-based triggers, the naturally quantity for the parametrization is the missing transverse energy and the trigger specific selections are based on the kinematic distributions of the two jets in the event. 

\ \\Since the parametrization is performed using an offline data sample and since we also apply the parametrization in the analysis online, we need to choose one $\met$ quantity computed offline that is as close as possible to the $\met$ quantities used at trigger level. As discussed in Section \ref{section:METObject}, the fully corrected $\met$ on which we apply a cut at 20 GeV for all events in our analysis is raw $\met$ corrected for the position of the primary interaction vertex, for the jet energies in the event and for the energy deposited by the muon in the calorimeter (which is relevant in the case of ISOTRK charged lepton events, which are muons in 85\% of the time). From a physics point of view, $\met$ represents the missing transverse energy due to the neutrino in the final state. However, these corrections are not performed at trigger level and therefore the physical meaning of the trigger $\met$ is the missing transverse energy of the $W$ boson (and not of the neutrino!). 

\ \\Ideally we should choose raw $\met$ for our parametrization. However, studies have shown that this variable is not well modelled in the control sample (Pretag) of the analysis. Therefore we correct this quantity for the position of the primary interaction vertex and the energy of the jets, but not the energy of the muon. Its physics meaning remains the missing transverse energy of the $W$ boson, but it is now modelled better. We denote this quantity trigMET and we use it for the trigger turnon curve efficiency parametrization.

\section{Trigger-Specific Jet Selection}

\ \\The next step is to identify for each trigger the specific jet cuts that allow for the remaining data events that the turnon curve parametrization is indeed flat in any jet quantities and depends only on trigMeET. All events in our analysis must have exactly two jets with $\et > 20\ \gev$ and $|\eta|<2.0$. 

\ \\The jet selection specific for the MET2J trigger is the following: both jets need to have $\et > 25\ \gev$, one of them must be in the central region of the detector ($|\eta|<0.9$), while the second jet must have $|\eta|<2.0$, and also the two jets have to specially separated ($\Delta R > 1.0$). 

\ \\The MET45 trigger does not have any jet requirements at trigger level. Therefore, for this trigger the jet selection is the same as for the non-ISOTRK charged lepton categories, namely exactly two jets with $\et > 20\ \gev$ and $|\eta|<2.0$. 

\ \\For the METDI trigger we studied that the optimum specific jet selection requires the most energetic jet to have $\et > 40\ \gev$, the second most energetic jet to have $\et > 25\ \gev$ and both jets to have $|\eta|<2.0$. 

\ \\We can already see that each trigger must be applied only in its specific jet kinematic region, which is equivalent to assuming the trigger is not defined in the other kinematic regions. The method we introduce in Chapter \ref{chapter:CombineTriggers} also takes this easily into account.

\begin{figure}[htbp]
  \begin{center}
    \includegraphics[width=9.5cm]{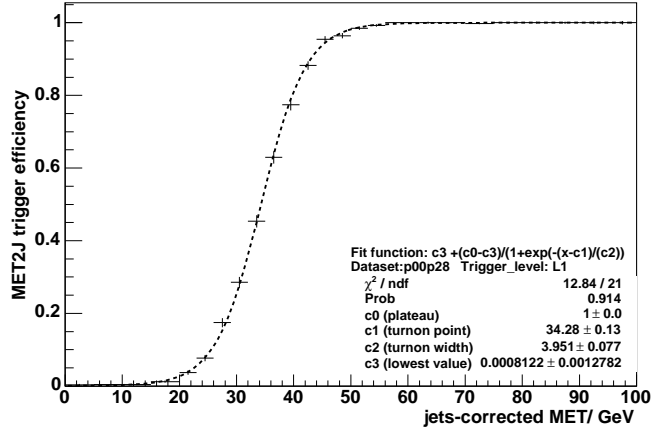}
    \includegraphics[width=9.5cm]{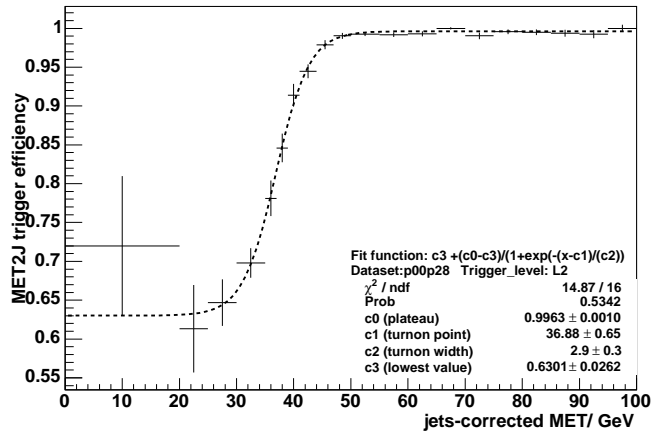}
    \includegraphics[width=9.5cm]{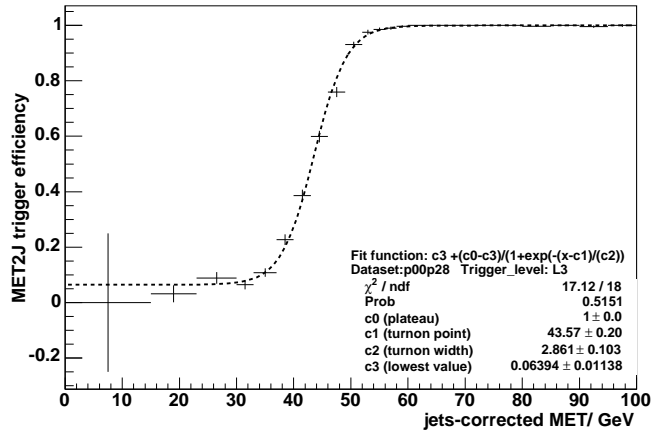}
    \caption[MET2J trigger efficiency turn-on curve parametrization]{MET2J trigger turn-on curves, parameterized as a function of trigMET. The figures show, from top to bottom, the L1, L2 and L3 turn-on curves, respectively. The turn-on curves were measured in the full dataset used in this analysis, and do not include the effect of the prescale for MET2J trigger.}
    \label{figure:ParametrizationMET2J}
  \end{center}
\end{figure}

\begin{figure}[htbp]
  \begin{center}
    \includegraphics[width=9.5cm]{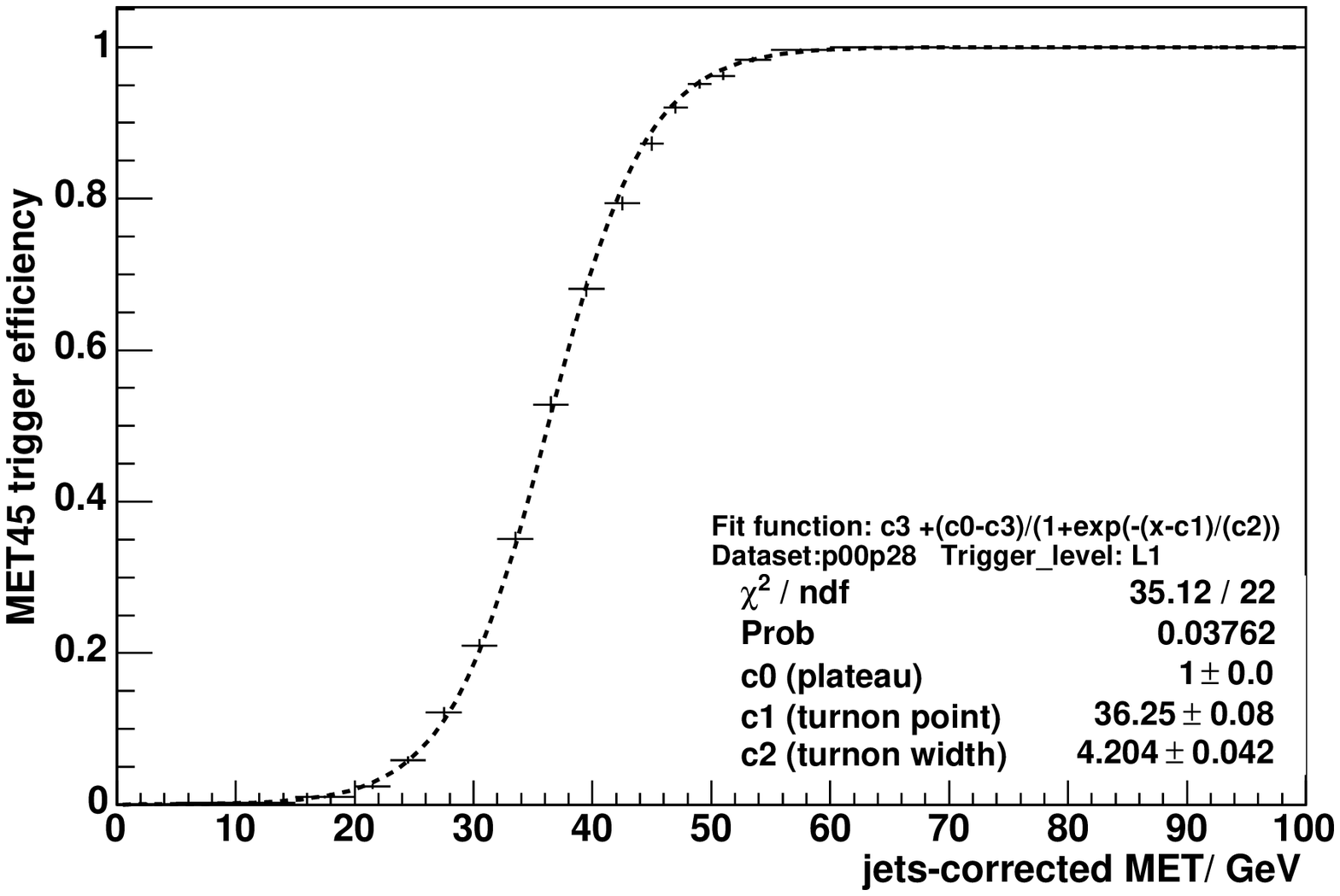}
    \includegraphics[width=9.5cm]{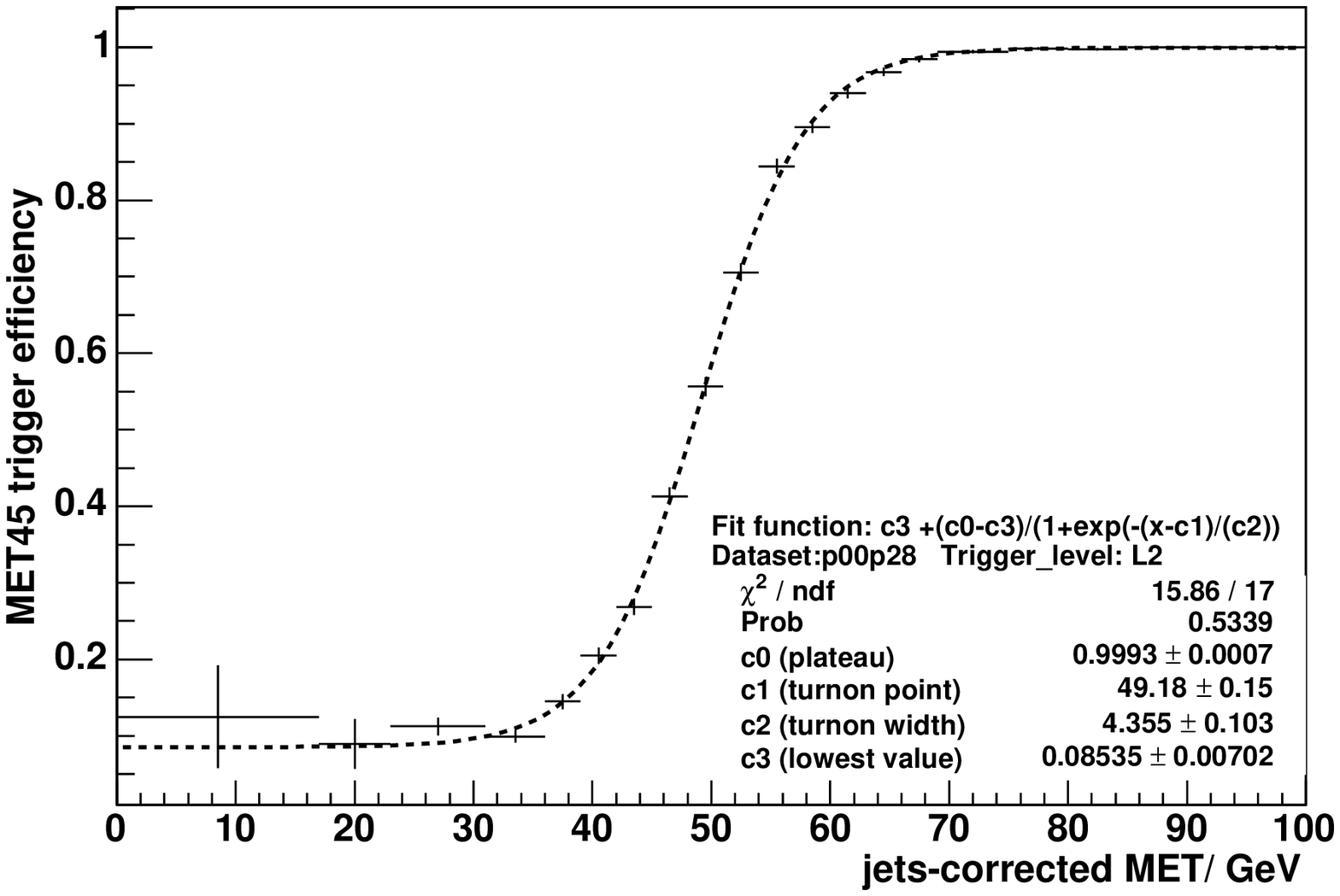}
    \includegraphics[width=9.5cm]{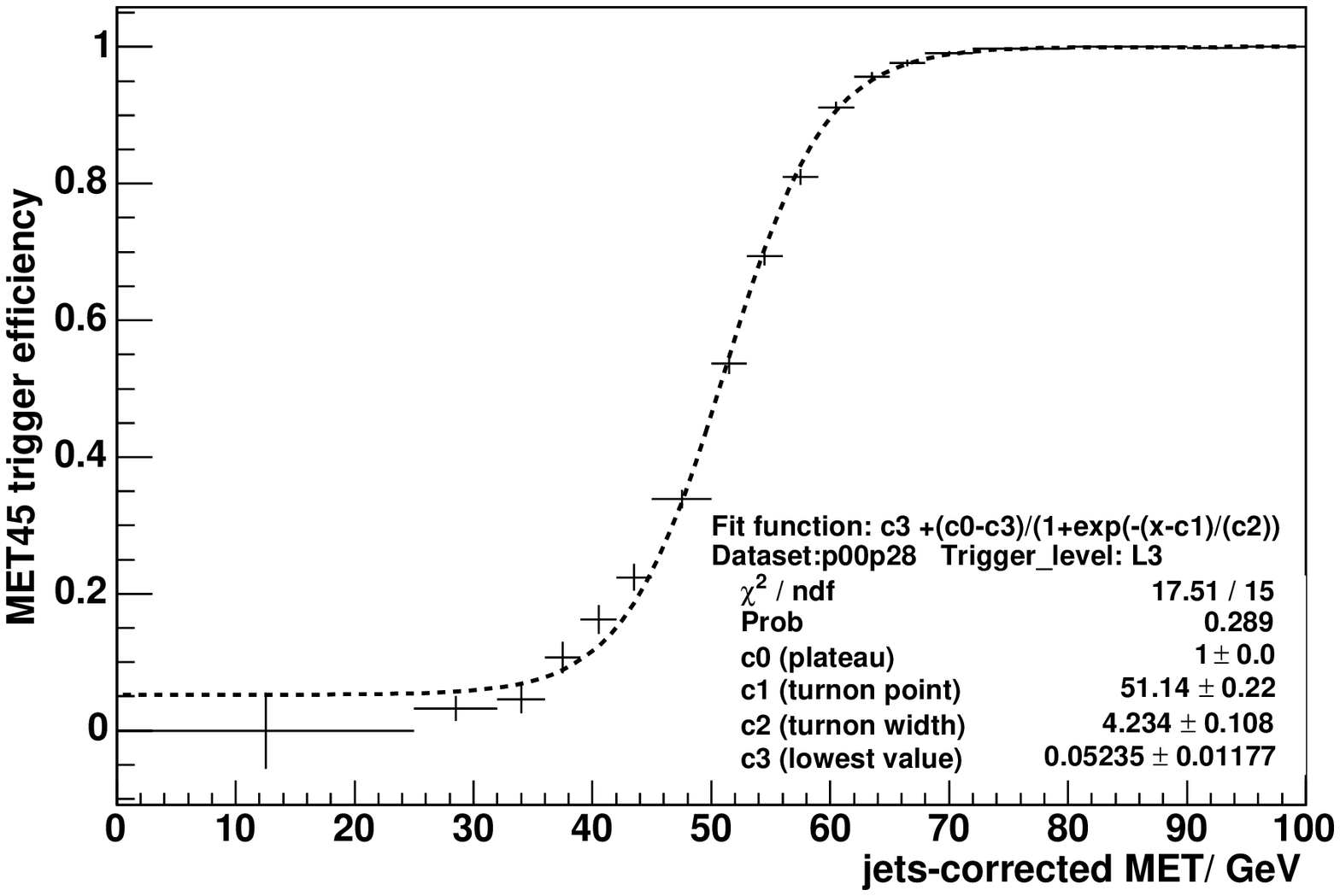}
    \caption[MET45 trigger efficiency turn-on curve parametrization]{MET45 trigger turn-on curves, parameterized as a function of trigMMET. The figures show, from top to bottom, the L1, L2 and L3 turn-on curves, respectively. The turn-on curves were measured in the full dataset used in this analysis.}
    \label{figure:ParametrizationMET45}
  \end{center}
\end{figure}

\begin{figure}[htbp]
  \begin{center}
    \includegraphics[width=9.5cm]{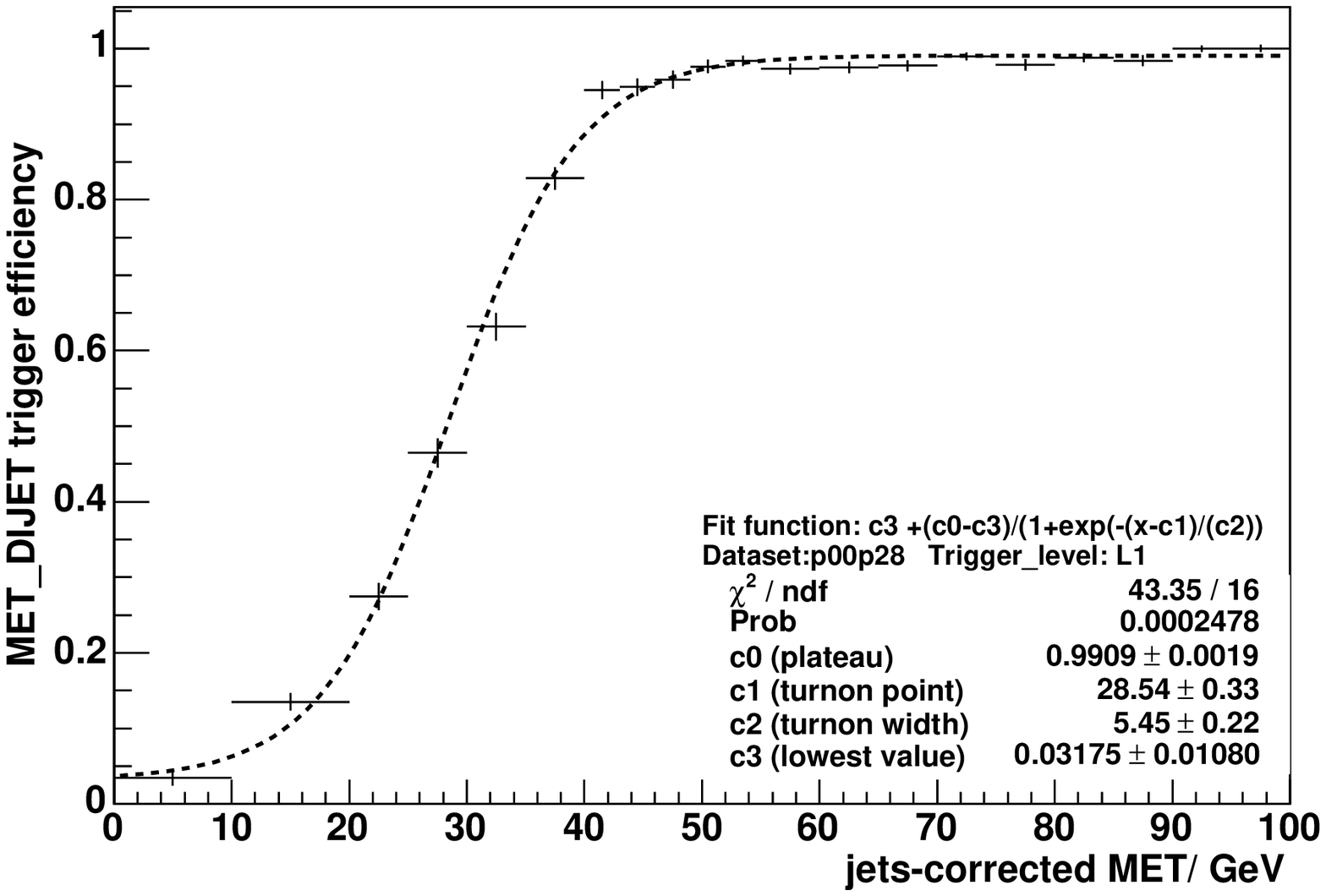}
    \includegraphics[width=9.5cm]{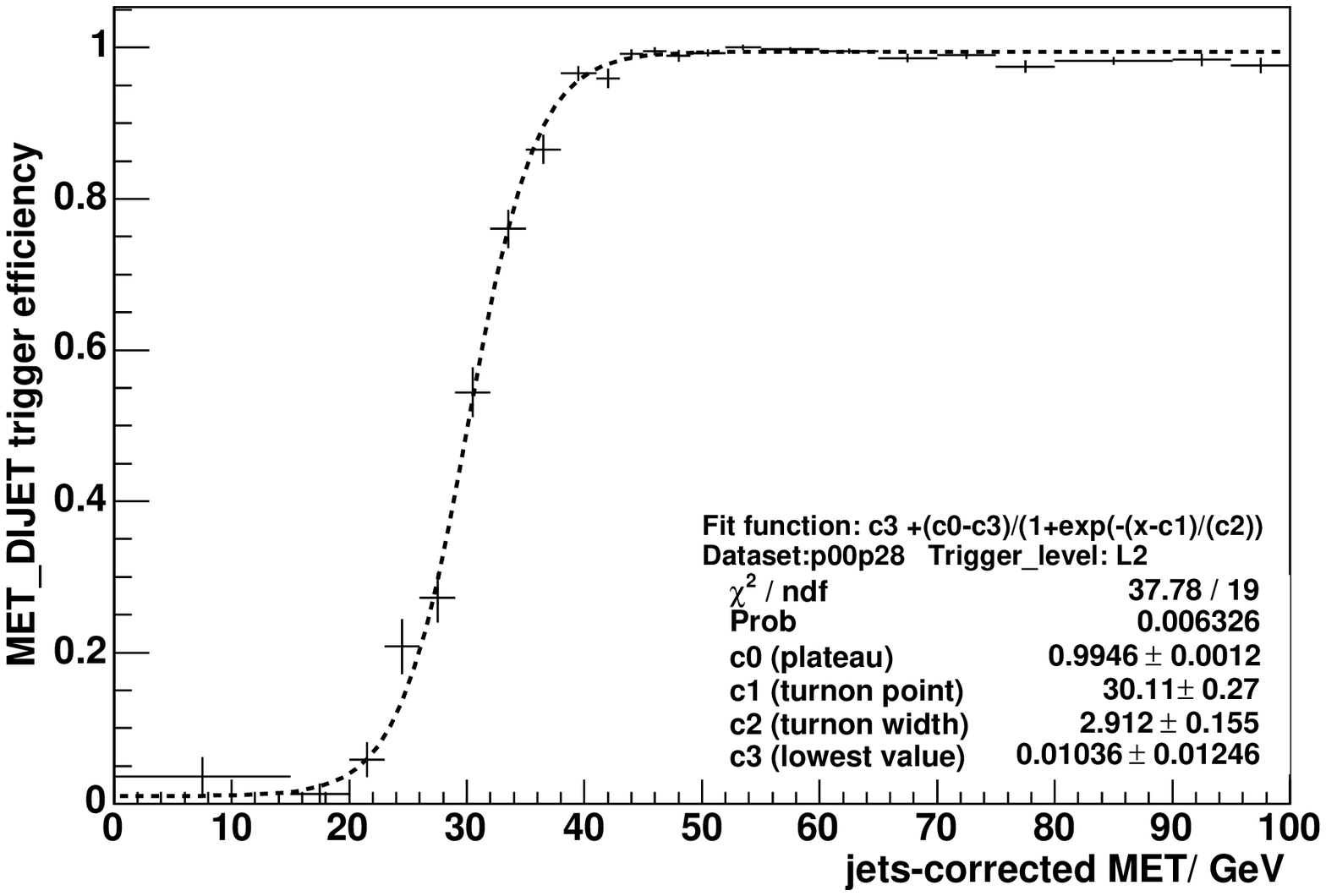}
    \includegraphics[width=9.5cm]{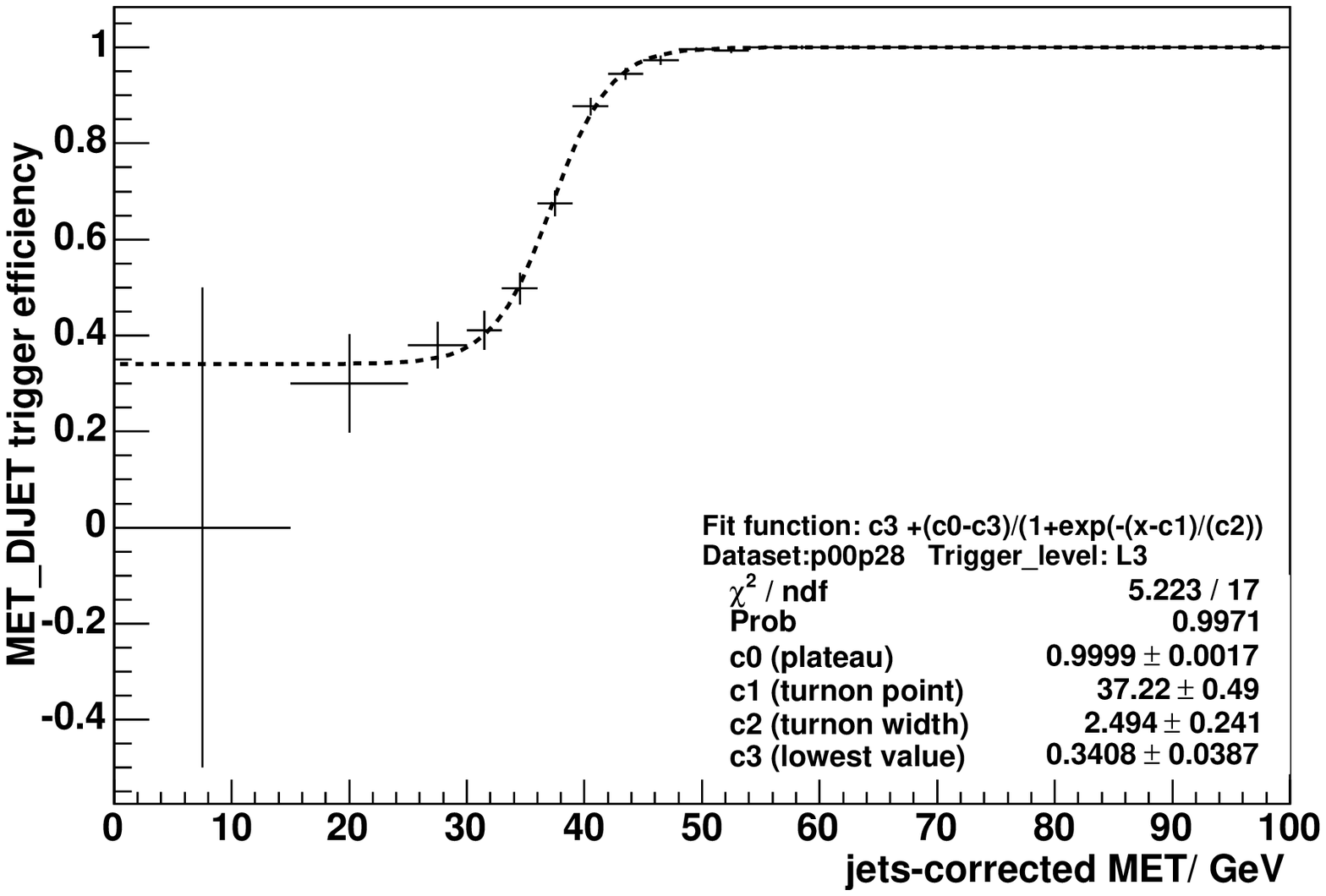}
    \caption[METDI trigger efficiency turn-on curve parametrization]{METDI trigger turn-on curves, parameterized as a function of trigMMET. The figures show, from top to bottom, the L1, L2 and L3 turn-on curves, respectively. The turn-on curves were measured in the full dataset used in this analysis.}
    \label{figure:ParametrizationMETDI}
  \end{center}
\end{figure}

\section{Parametrization for MET-based Triggers}

\ \\Since we use these MET-based triggers for the ISOTRK charged lepton category and previous studies have shown that ISOTRK candidates are in 85\% of cases muon candidates, we measure the trigger turnon curves using a data sample collected with a muon inclusive trigger. We require exactly one reconstructed CMUP muon candidate which fires the CMUP trigger. We compute trigger efficiency turnon curves for each of the three MET-based triggers and for each of the three trigger levels in such a way that none of these turnon curves takes into account the prescale of the trigger, if any. In the following paragraph we describe the procedure for one generic case.

\ \\We select the subset of events that pass the jet selections specific for this trigger. For these events we fill a histogram for the variable trigMET. This is the denominator histogram. We then fill another histogram for the same variable, but only for the events that also fired the chosen MET-based trigger. This is the numerator histogram. Since the CMUP and the MET-based trigger are uncorrelated for all practical purposes, we divide the numerator and denominator histograms to obtain the efficiency turnon histogram for the chosen trigger. We fit the efficiency histogram to a sigmoid function with four parameters and as a function of trigMET, given by

\begin{equation} 
\rm{Eff(trigMET)}=c_3+\frac{c_0-c_3}{1+e^{-\frac{trigMET-c_1}{c2}}}\ \rm{,}
\label{formula:SigmoidFunction}
\end{equation}

\ \\where $c_0$ represents the highest plateau efficiency, $c_1$ represents the central value of the turnon region (and is measured in $\gev$), $c_2$ represents the width of the turnon region (and is also measured in $\gev$) and $c_3$ represents the lowest efficiency value. The fit returns the four parameters which uniquely defines the efficiency as a function of trigMET only for a given trigger and trigger level. 

\ \\In order to ensure that the turnon curves do not include the effect of the eventual trigger prescale (the method is general, although we know that only the trigger MET2J is prescaled only at L2), for L2 we include in the denominator the requirement that the event was sent from L1 to L2 and for L3 that the event was sent from L2 to L3. 

\ \\The nine turnon curves are presented in the Figure \ref{figure:ParametrizationMET2J} (MET2J trigger), the Figure \ref{figure:ParametrizationMET45} (MET45 trigger), and the Figure \ref{figure:ParametrizationMETDI} (METDI trigger).

\section{Parametrization for Systematic Uncertainty Evaluation\label{section:METTriggerSystematics}}

\ \\We repeat the procedure above in bins of the following kinematic quantities: $\et$, $\eta$ and $\phi$ of both jets, absolute value of the $\Delta \eta$, $\Delta \phi$, $\Delta R$, $\Delta \et$, as well as fully corrected $\met$ of the analysis and the fraction of total luminosity that corresponds to each run. The number of bins are chosen automatically by the code in order to have a minimum specified number of events in the turnon region of the distribution so that the fit is performed correctly. 

\ \\Whereas the standard trigger weight is the value of the central turnon curve for the event trigMET, the systematic weight that corresponds to the variable $\eta$ is given by the same trigMET applied to a different turnon curve specific to the bin of the particular event $\eta$. The same is repeated for all systematic uncertainties. The total weighted average is compared between the central turnon curve and each of the systematic values and the largest percentage difference is quoted as the systematic uncertainty of the analysis. This procedure is general and works to all trigger combination methods, including the one we introduce in this analysis and we present in the next chapter.

\section{Prescales for MET-based Triggers\label{section:TriggerPrescales}}

\ \\Since the MET45 and METDI triggers are unprescaled, their prescales are $1.000 \pm 0.000$. However, the trigger MET2J is prescaled at L2 for a large fraction of the integrated luminosity. We computed the total integrated luminosity for MET2J and we divided it to the total integrated of another trigger (which requires a jet with an energy larger than 100 $\gev$) which has also been continuously active since the beginning of Run II and was never prescaled. Therefore we computed the prescale for MET2J to be $0.92 \pm 0.05$. 

\section{Integrated Luminosities for MET-based Triggers}

\ \\Not all triggers were defined at the same time. In order to properly simulate that in Monte Carlo events, we measured the fraction of integrated luminosity for each combination of MET-based triggers were defined, as shown in Table \ref{table:IntegratedLuminositiesMETTriggers}. 

\begin{table}[h]
\begin{center}
\begin{tabular}{|c|c|c|l|}
\hline
\hline
MET2J & MET45 &  METDI & Fraction\\
\hline
No & No & No & 0 \\
No & No & Yes & 0 \\
No & Yes & No & 0.13\% \\
No & Yes & Yes & 0 \\
Yes & No & No & 2.53\% \\
Yes & No & Yes & 0 \\
Yes & Yes & No & 36.52\% \\
Yes & Yes & Yes & 60.81\% \\
\hline
\hline
\end{tabular}
\caption[Fraction of integrated luminosity where MET-based triggers are defined]{Fraction of integrated luminosity where MET-based triggers are defined for each possible combination of the MET-based triggers.}
\label{table:IntegratedLuminositiesMETTriggers}
\end{center}
\end{table}

\ \\We have now all the information needed about triggers in order to combine them in our analysis. We introduce a novel trigger combination technique that we present in the next chapter.

\clearpage{\pagestyle{empty}\cleardoublepage}
\chapter{Novel Method to Combine Triggers\label{chapter:CombineTriggers}}

\ \\Once the MET-based triggers are parametrized, there are many ways they can be combined in an analysis. Our trigger parameterizations were used for the $WH$ analyses for the summers of 2008, 2009 and 2010, which were presented in two Ph.D. theses, CDF and Tevatron Higgs combinations, one PRL paper and two PRD drafts under preparation, as presented in detail in the section entitled ''Original Contributions''. 

\section{The Method for the 2.7 $\invfb$ $WH$ search}

\ \\The analysis for the summer of 2008 used 2.7 $\invfb$ of integrated luminosity. It used only one MET-based trigger, namely MET2J, and therefore there were no complications. For both Monte-Carlo-simulated and data events only the same trigger would be used. For data events, the event is checked if it has fired the trigger. If it has, the event is kept and is given a weight of exactly 1.0. If not, the event is rejected, which is equivalent to receiving a weight of exactly 0.0.

\ \\As a side note, this analysis used a simplified and less precise trigger parameterization than the one we described in Appendix \ref{chapter:METTriggers}. The analysis used only one trigger efficiency turnon curve measured across all trigger levels and which included the trigger prescale. The trigger weight was applied to all Monte-Carlo-simulated events without exception. 

\section{The Method for the 4.3 and 5.7 $\invfb$ $WH$ searches}

\ \\The analysis for the summer of 2009 (2010) used 4.3 (5.7) $\invfb$ of integrated luminosity. Both analyses used two MET-based triggers, namely MET2J and MET45. Since MET2J could be applied only for a subset of the jet kinematic phase space where MET45 could be applied and since efficiency studies suggest that MET2J is more efficient than MET45, it was decided that for events with two jets with $\et > 25 \ \gev$, one jet central with $|\eta|<0.9$ and the other jet with $|\eta|<2.0$, and with non overlapping jets with $\Delta R > 1.0$, only the MET2J would be used, whereas for the remaining phase space up to two jets with $\et > 20 \ \gev$ and $|\eta|<2.0$ only MET45 would be used. This method is very clean and brings no complications, since for a kinematic region, both the Monte-Carlo-simulated and data events will use the same trigger. 

\ \\For a given event, the jet kinematic region was checked. If the event was in the tight kinematic region and it was a data event, then the event was checked if it fired the MET2J trigger. If it did, the event was kept. Otherwise, it was rejected. If the event was simulated, then a weight given by the multiplication of the weights given by the L1, L2, L3 and turnon curves for MET2J and by the prescale of MET2J was used for that event. If the event was not in the tight jet kinematic region, but it was in the looser kinematic region, then the procedure described above was done for the MET45 trigger, with the caveat that if the data event was in the particular run range where the MET45 trigger had a bug, then the event was rejected even if the trigger fired. But how to model that correctly in Monte Carlo simulation? Since the effect was small, on the order of 2.6\% (Table \ref{table:IntegratedLuminositiesMETTriggers}), the effect was ignored and included in the systematic uncertainty of the procedure.  

\ \\The method described in the previous section does not use a local ``OR'' between the MET2J and MET45 triggers in order to avoid the correlations between the triggers. The price to pay is a smaller event acceptance. The advantage is that systematic uncertainty is easier to calculate correctly and is smaller than in the case of correlated triggers. The main feature of the method described in the previous section is that in the kinematic phase space of jet selection, a trigger is assigned to one region and another trigger is assigned to another region. We stress that this is done before the data events are checked if the trigger fire. The choice of kinematic regions and the chosen trigger for each kinematic region is based on an \emph{a priori} study. In the case of the particular method described in the previous section, the turnon curves for the MET2J (Figure \ref{figure:ParametrizationMET2J}) and the MET45 (Figure \ref{figure:ParametrizationMET2J}) triggers were compared and it was observed that the MET2J trigger has a turnon region on smaller trigMET values than MET45 and therefore has a potential larger efficiency than MET45. In the kinematic region that was not tight, but it was loose, only MET45 trigger could be used, so there was no dilemma. For the tight jet region, though, since correlation between triggers was to be avoided, only one trigger had to be chosen and based on the reasoning above this was MET2J. 

\ \\This procedure increased the signal acceptance over MET2J only, but it was clear that does not provide the maximum signal acceptance. For example, for some events in the tight jet region it is possible that the trigger weight is larger for MET45 than for MET2J, or that the event was not selected by the MET2J stream due to a prescaling or other effect, but was selected by the MET45 trigger. Why not divide this kinematic region further into smaller kinematic regions and for each of them reevaluate if to use the MET2J or the MET45 triggers? But how should this division be done? And how to take into account the fact that for 2.6\% of data events the trigger MET45 was not defined? And what about using also the METDI trigger, which is defined only for about 60.8\% of the integrated luminosity (Table \ref{table:IntegratedLuminositiesMETTriggers})? It seems there would be very many kinematic regions and there would be a very complicated bookkeeping. 

\ \\In the next section we propose a general method that solves all these problems for any number of triggers combining. We apply this method to optimize the signal acceptance for our $WH$ analysis. 

\section{The Novel Method for Combining Triggers\label{NovelMethodCombineTriggers}}

\ \\The solution we propose in order to maximize the event selection while using only one trigger per kinematic region with minimal bookkeeping is that of considering the largest possible number of kinematic regions, i.e. the number of events in that particular Monte-Carlo-simulated or data sample. Our idea is to consider each event as an independent kinematic region. Just as in the case of the method above, we study \emph{a priori} which is the most efficient trigger in the kinematic region, choose that trigger and ignore the other triggers. Since the kinematic region means that unique event, it means we choose the trigger that has a priory the largest probability to fire that event. Below we will go into the details of how such a probability is computed for every trigger for the particular event. Once they are computed, though, we choose the trigger as the one that has the largest a priory probability. If all probabilities are strictly zero, then the event is rejected both if it is a data event or a MC event. This is equivalent to a study done for a particular kinematic region and choosing that all events in that kinematic region are assigned to only one trigger, such as MET2J was preferred to MET45 in the tight jet region of the previous method. For a data event, the chosen trigger is checked if it fired the event. If it has, the event is kept, or equivalently is assigned a weight of strictly 1.0. If the event did not fire the trigger, the event is thrown away, or equivalently is assigned a weight of strictly 0.0, and the other triggers are not checked at all. Ignoring the other triggers is the main feature that allows the orthogonality between triggers. It is crucial that the trigger is chosen before the trigger is checked if it fired. For a Monte Carlo event it is assumed automatically that the trigger has fired the event and the probability of the chosen trigger is returned as the trigger weight. 

\ \\The a priori probability that an event fires a particular trigger is given by the product between the weight at L1 for that particular trigger, the weight at L2 for that particular trigger, the weight at L3 for that particular trigger, the prescale for that particular trigger.

\begin{equation} 
P=w_{L1}\cdot w_{L2}\cdot w_{L3} \cdot PS \cdot JS \cdot TD  \ \rm{,}
\label{formula:TriggerProbability}
\end{equation}

\ \\where $w_{L1}$, $w_{L2}$, $w_{L3}$ are weights of that particular trigger at L1, L2 and L3, respectively, are given by Formula \ref{formula:SigmoidFunction}, and vary as a function of the event trigMET; $PS$ represents the average prescale of the evaluated trigger, with the same value for all events, given by Formula \ref{formula:Prescale}; $JS$ represents the jet selection and is 1 if the jet selection for that particular trigger is met by the event and zero if it is not, given by Formula \ref{formula:JetSelection}; $TD$ represents the condition if the chosen trigger is defined for the event and is 1 if this happens and 0 if it does not. 

\begin{equation} 
PS = 
\begin{cases} 
0.92, & \mbox{if MET2J}\\
1.00, & \mbox{if MET45}\\
1.00, & \mbox{if METDI}\\
\end{cases}
\label{formula:Prescale}
\end{equation}

\begin{equation} 
JS = 
\begin{cases} 
1.00, & \mbox{if chosen trigger jet selection is passed}\\
0.00, & \mbox{if chosen trigger jet selection is failed}\\
\end{cases}
\label{formula:JetSelection}
\end{equation}

\begin{equation} 
TD = 
\begin{cases} 
1.00, & \mbox{if chosen trigger is defined}\\
0.00, & \mbox{if chosen trigger is not defined}\\
\end{cases}
\label{formula:JetSelection}
\end{equation}

\ \\For data events, $TD$ is evaluated easily as each event contains the information if the trigger is defined or not. For MET45 we also consider the trigger is not defined for the 2.6\% of the integrated luminosity where the trigger has a bug. As seen in Table \ref{table:IntegratedLuminositiesMETTriggers}, for 0.13\% of the integrated luminosity no MET-based trigger is defined, for 2.53\% of the integrated luminosity only MET2J is defined, for 36.52\% of the integrated luminosity only MET2J and MET45 are defined and for 60.81\% of the integrated luminosity all three MET-based triggers are defined. The simulation of this in Monte Carlo was the really tricky part with the previous methods. 

\ \\In our approach, for every Monte Carlo simulated event a random number is chosen from a uniform distribution between 0 and 1. It represents the probability that the event falls in any of the integrated luminosity intervals based on which it is decided which triggers are defined or not for the MC event. If the random number is in the interval [0.0000-0.0013] then all triggers are assumed to be undefined and TD=0.0 for all triggers. For the interval [0.0013-0.0266] TD=1.0 for MET2J and TD=0.0 for MET45 and METDI. For the interval [0.0266-0.3918] TD=1.0 for MET2J and MET45 and TD=0.0 for METDI. For the interval [0.3918,1.000] TD=1.0 for all three MET-based triggers.

\ \\Formula \ref{formula:TriggerProbability} ensures that each trigger is allowed in the competition with the other triggers to decide which is more likely to fire for the particular event only if the trigger is meaningful for that particular event, i.e. if the event passes the jet selection specific and necessary for the trigger to be considered and if the trigger was actually defined for that particular run to which the event belongs. If this does not happen, the trigger ends up with a Probability of zero, which makes sure the trigger does not win over the other triggers. If the probability is different than zero, than it is the probability given by the turnon curves at each of the three trigger values as well as the prescale of that trigger. 

\ \\We implemented this method into a software package called ABCDF\footnote{The name of the package comprises the author's initials followed by those of the Collider Detector at Fermilab experiment} that allows for a user-friendly inclusion of the method into an analysis. The ABCDF package will take in the event information and return a probability between 0 and 1 for Monte Carlo simulated events and either 0 or 1 for data events. A selection cut of probability strictly larger than 0.001 is necessary to make sure the data events for which the chosen trigger did not fire are rejected from the analysis, as they would have a weight of strictly 0.0. ABCDF is part of the CDF software repository and can be used by any analysis. In fact, it is being used already by several analysis at CDF for the three MET-based triggers. 

\ \\The method is a general one and can be used with any number of MET-based triggers. Each trigger would come with its own specific kinematic selection. Information other than jet information can be included in the kinematic selection. This method was used in the analysis presented in this dissertation and is currently being used by other analyses at CDF. It also has potential applications at other experiments, as MET+jet triggers are used for new physics searches such as supersymmetry at the LHC experiments. The method can be used on other trigger types and even the parametrization could be done as a function of another variable for each different trigger, i.e. $w_Li$ could be computed by different formulae specific for each trigger. 

\ \\The total Monte Carlo event count is the count of events that pass the event selection weighted by the total Probability of the chosen trigger, i.e. the largest Probability amongst the available triggers. The systematic uncertainty is calculated easily by considering other turnon curves in bins of several kinematic quantities enumerated in Section \ref{section:METTriggerSystematics} which changes on an event by event basis the $w_L$ values and may change not only the Probability of each trigger, but also the chosen trigger. The largest percentage difference for all systematic variations with respect to the standard event count is typically considered to constitute a sensible systematic uncertainty.

\ \\An assumption of our method is that the triggers are very efficient. The assumption is a very good one in the case of our analysis. However, if the triggers were very inefficient, an OR method would provide a significantly higher even yield. One would have to model correctly the correlation between triggers and compute the systematic uncertainty for the OR trigger combination.

\clearpage{\pagestyle{empty}\cleardoublepage}

\chapter{WHAM package in CDF\label{chapter:WHAM}}

\section{Introduction}

\ \\Associated WH production search Analysis Modules (WHAM) is a new data analysis framework for the CDF collaboration. It builds on a previous framework developed by the top quark study group inside CDF and adds functionality through its improved modular structure. WHAM performs all the analysis stages from data and Monte Carlo ntuples produced by the CDF production group up to the final measurement, such as limits or cross sections. I am one of the main contributors to WHAM code development and validation. Also, WHAM has been used to produce the results presented in this thesis.

\section{Motivation}

\ \\The main motivation of WHAM is to perform a combination analysis between two WH searches inside CDF: the one using as a discriminant an artificial neural network (WHNN, the search presented in this thesis) and the one using a boosted decision tree and matrix elements (WHME). Studies inside CDF has shown that there is no 100$\%$ correlations between the two discriminants and therefore more information could be extracted by combining the two searches. This can be achieved if a superdiscriminant is built that takes as inputs the event by event basis values for discriminants of each analyses. CDF latest combination between WHNN and WHME showed a 10$\%$ increase in sensitivity up to the best performing analysis~\cite{WHCombination2.7fb-1}. 

\ \\The key words here are "on an event by event basis". That means we need to make sure both analysis have the exact event selection. Also we have to make sure both analyses reconstruct the same way various kinematic quantities such as jet energies, missing transverse energies, dijet invariant mass and have the same values for these on an event by event basis. So far each analysis used its own framework, its own definition of loose charged leptons and its own way for corrected various kinematic quantities for various effects. This makes a combination between WHNN and WHME very time and resource consuming. The combination cited above is the only one achieved so far.

\ \\The goal of WHAM is to allow an easy combination of these two analyses at any desired moment. This will allow CDF collaboration to present combined and therefore improved results for the WH search at each conference cycle.

\ \\But that is not all. WHAM allows very easily to perform searches that have the same or very close signature as the WH one: Technicolour and $t\tbar H$ searches, as well as $WZ$ and single top measurements. So far these analyses have had their own framework. 

\ \\The fact that all these analyses could be done in a common framework allows that a given improvement by one collaborator once integrated in WHAM and validated for one analysis can be used automatically and thus help improve all the other analyses as well.

\ \\One last point is that WHAM does all the steps from original data and Monte Carlo simulations up to the final result with a minimal number of actions from a user. In the context as the number of CDF collaborators is decreasing, WHAM is very helpful in last years of CDF data taking and data analyzing because one postdoctoral student will be able to update very easily all these analyses with new data.

\section{Implementation}

\ \\The WHAM code is contained in various folders with suggestive names and where code with very specific goal is placed: setup, inputs, commands, modules and results. 

\ \\The folder "setup" allows for the easy setup of the entire analysis framework.

\ \\The folder "inputs" contains all the inputs needed for the analyses: 

\begin{itemize}
\item [-] Lists of files with events to be processed, either data or Monte Carlo signal or background simulated events;
\item [-] Lists of data runs with good quality data;
\item [-] Text files with information needed for the analysis such as analysis cuts, tasks to be performed, luminosities and data-simulation scale factors. This allows to change input parameters in the analysis without recompiling the code. Also, these input parameters are always saved with the results and they can be retrieved if in doubt on the input parameters of a given result. 
\end{itemize}

\ \\The folder "modules" contains in a modular way most of the code that performs tasks in WHAM:

\begin{itemize}
\item [-] The folder "dep" contains the dependencies files with extension .o and .d produced during the compilation of code;
\item [-] The folder "shlib" contains the shared objects with extension .so produced during the compilation of code
\item [-] The folder "inc" is the include directory that contains symbolic links to all the packages inside the "modules" folder;
\item [-] The folder "external" contains all the packages already existing in the CDF software archive that WHAM uses, such as high level object reconstruction code, $b$-tagging algorithm and their mistag matrices, background calculation methods, limit calculation method, manipulating .root files, boost\footnote{Boost provides free peer-reviewed portable C++ source libraries at \url{http://www.boost.org}.}, NKRoot\footnote{NKRoot is a collection of tools for ROOT file manipulation needed for particle physics data analysis, developed by Nils Krumnack while a postdoc on CDF and still being developed as he is a researcher on ATLAS. The package is available for free to anyone at \url{http://www-cdf.fnal.gov/~nils/root/}.} and ABCDF\footnote{ABCDF is the software package I developed in order to model the trigger simulation for missing energy + jets triggers and it is placed in the external package as it can be used by other CDF frameworks and analyses as well.}.
\end{itemize} 

\ \\The folder "native" contains all the packages produced only for WHAM, such as event reconstruction, various computations, such as the discriminant output on an event by event basis, making analysis trees from the original data or simulated events from cdf production group, making histogram root files that are used for background estimation.

\ \\The folder "commands" contains various tasks such as submitting computing jobs to the CDF's Central Analysis Farm (CAF), the limit calculation code, various data analysis ROOT macros and various scripts that read log files and compute things with that information or merge existing root files.

\ \\The folder "results" contains all the results of tasks from "modules". Here we store the ROOT trees produced with WHAM using CAF that are used then for background estimation and limit calculation, the limit results, text files with enumerations of events that pass our event selection and their kinematic properties (event dumps), histogram files used for background calculation, various smaller trees for various studies such as signal acceptance improvement or jet energy resolution improvement. 

\section{Future Plans}

\ \\At the moment that this thesis is submitted, only the WHNN analysis has been performed and validated using the WHAM framework. Other analyses are in progress, such as single top measurements and technicolour and $WZ$ searches.

%Next the WHME will do the same thing, followed by a combination of the two analyses in WHAM. Still, the two analyses have each their own ROOT tree filled separately but for the same events and with the same quantities. A natural step is to produce only one tree with all the information needed by both analyses. That way it is even easier and less error prone to combine the two analyses. 

%\ \\Also there is a work in progress to perform the single top measurement and Technicolour search using WHAM.

\clearpage{\pagestyle{empty}\cleardoublepage}
\chapter{Control Plots\label{chapter:ControlPlots}}

\ \\This appendix presents a selection of relevant plots for our analysis. In each plot, all backgrounds, properly normalized, are stacked. Data points are overlaid in order to show the good modelling of the backgrounds. The two signal processes are also overlaid after having been multiplied by a factor in order to be visible on the plots. Only the most sensitive analysis channel (TIGHT SVTSVT) is shown in order not to make the thesis too long. Since the Pretag category is a control sample for each analysis channel, also the TIGHT Pretag category is shown. 

\ \\For each category, a collection of plots relevant to the kinematic distribution of the event is shown, such as the transverse energy, $\eta$ and $\phi$ of the two jets and the charged lepton, as well as the $\met$, the transverse mass of the $W$ boson, the $\Delta \phi$ between the $\met$ and each jet and the charged lepton, followed by the $\Delta R$ between the two jets. Also for each category a collection of plots showing the variables used as input to the BNN final discriminant, as well as the BNN output for a Higgs mass of 115 $\gevcc$, are shown. 

%\clearpage

%%% For Kinematic Variables PretagSVTSVT TIGHT charged lepton %%%

\begin{figure}[ht]
  \begin{center}
    \includegraphics[width=6.7cm]{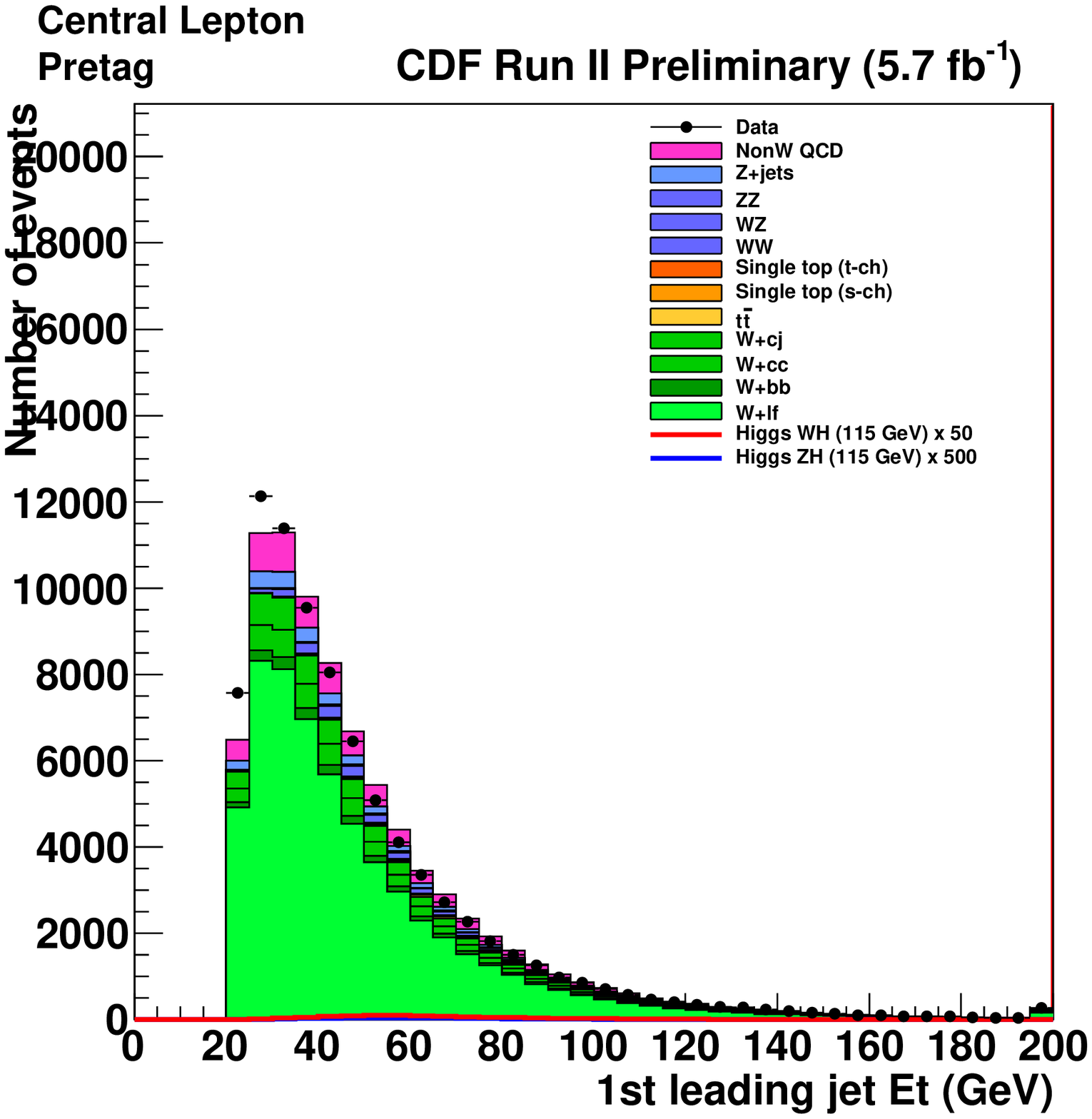}
    \includegraphics[width=6.7cm]{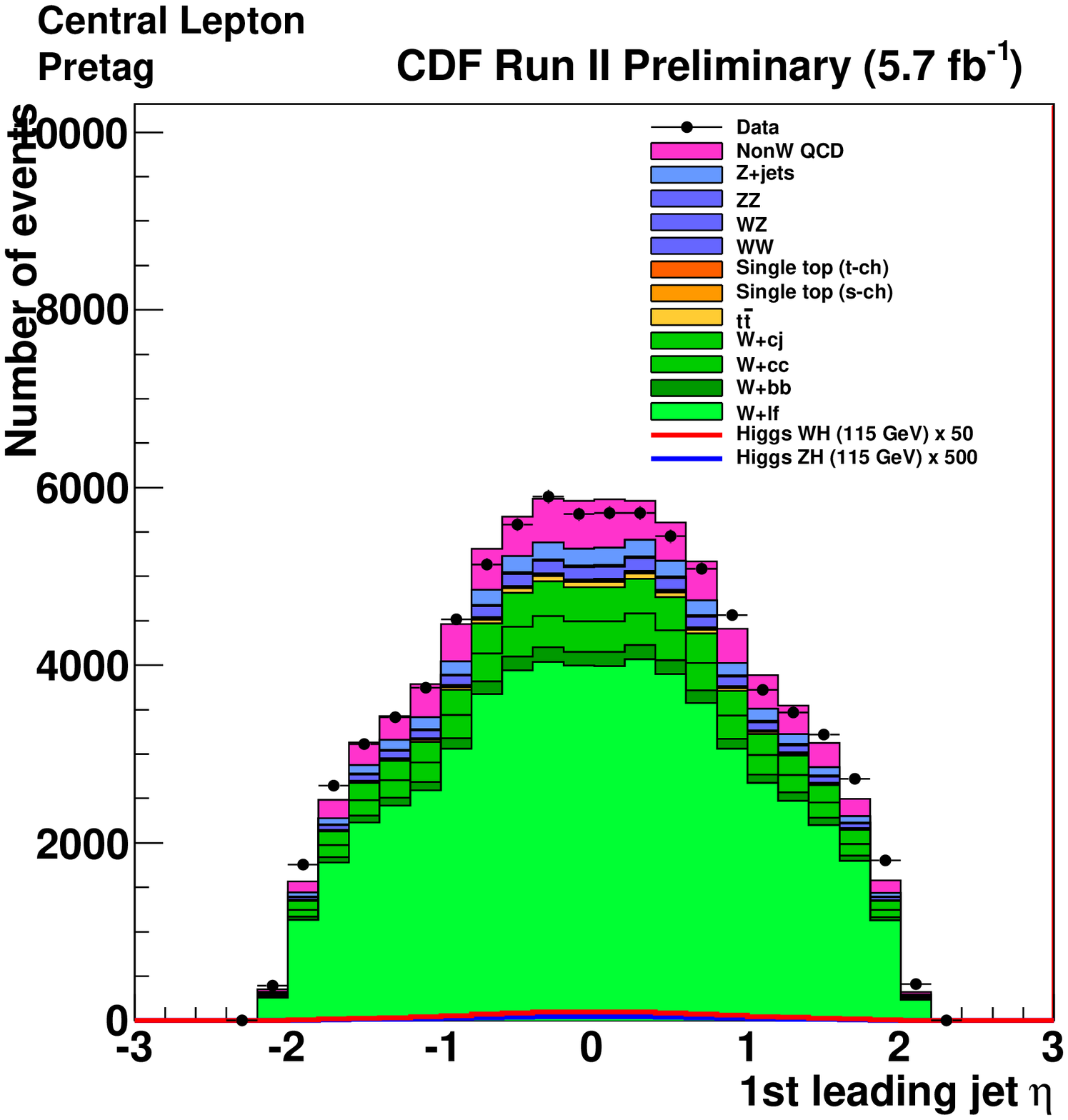}
    \includegraphics[width=6.7cm]{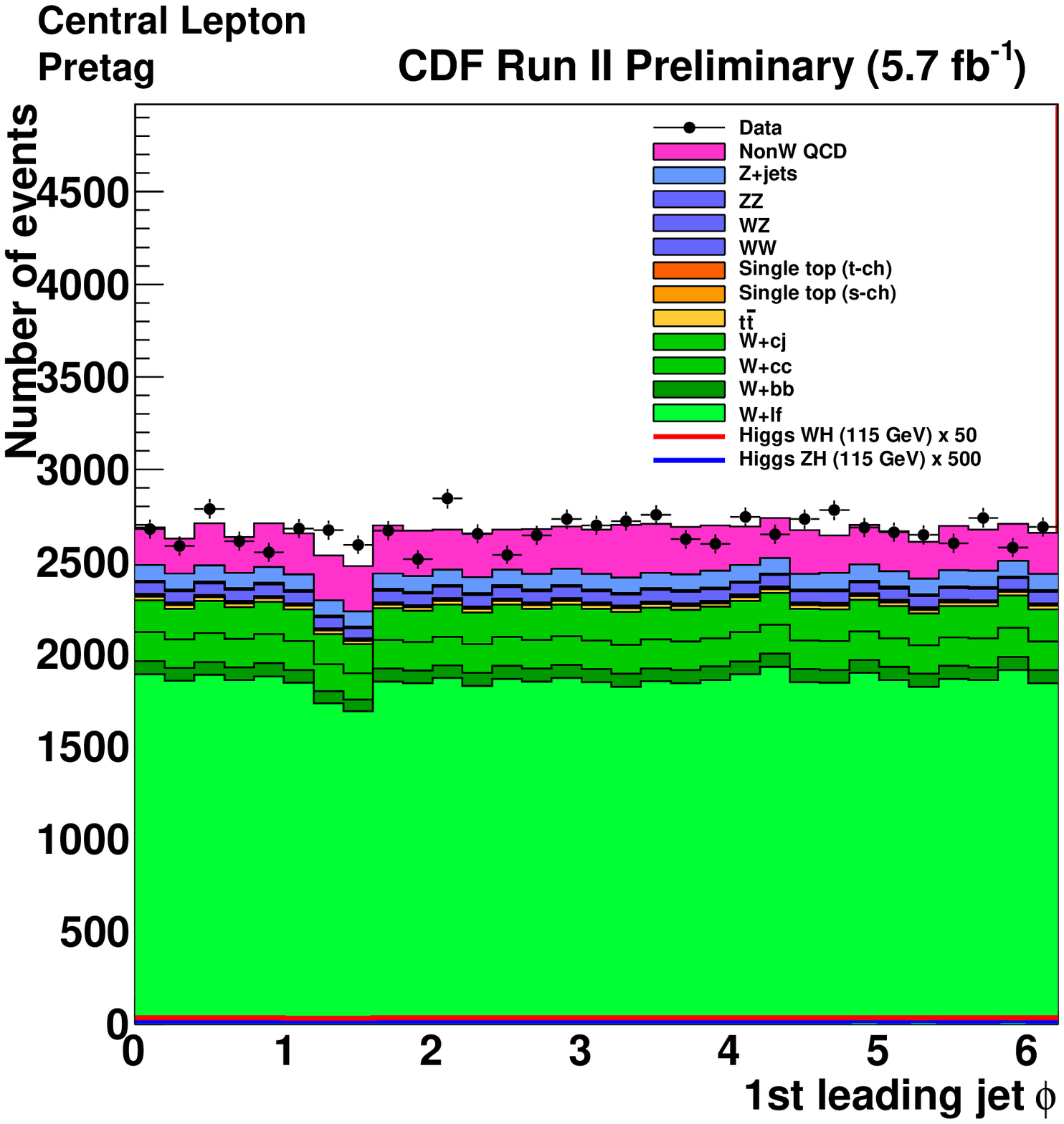}
    \includegraphics[width=6.7cm]{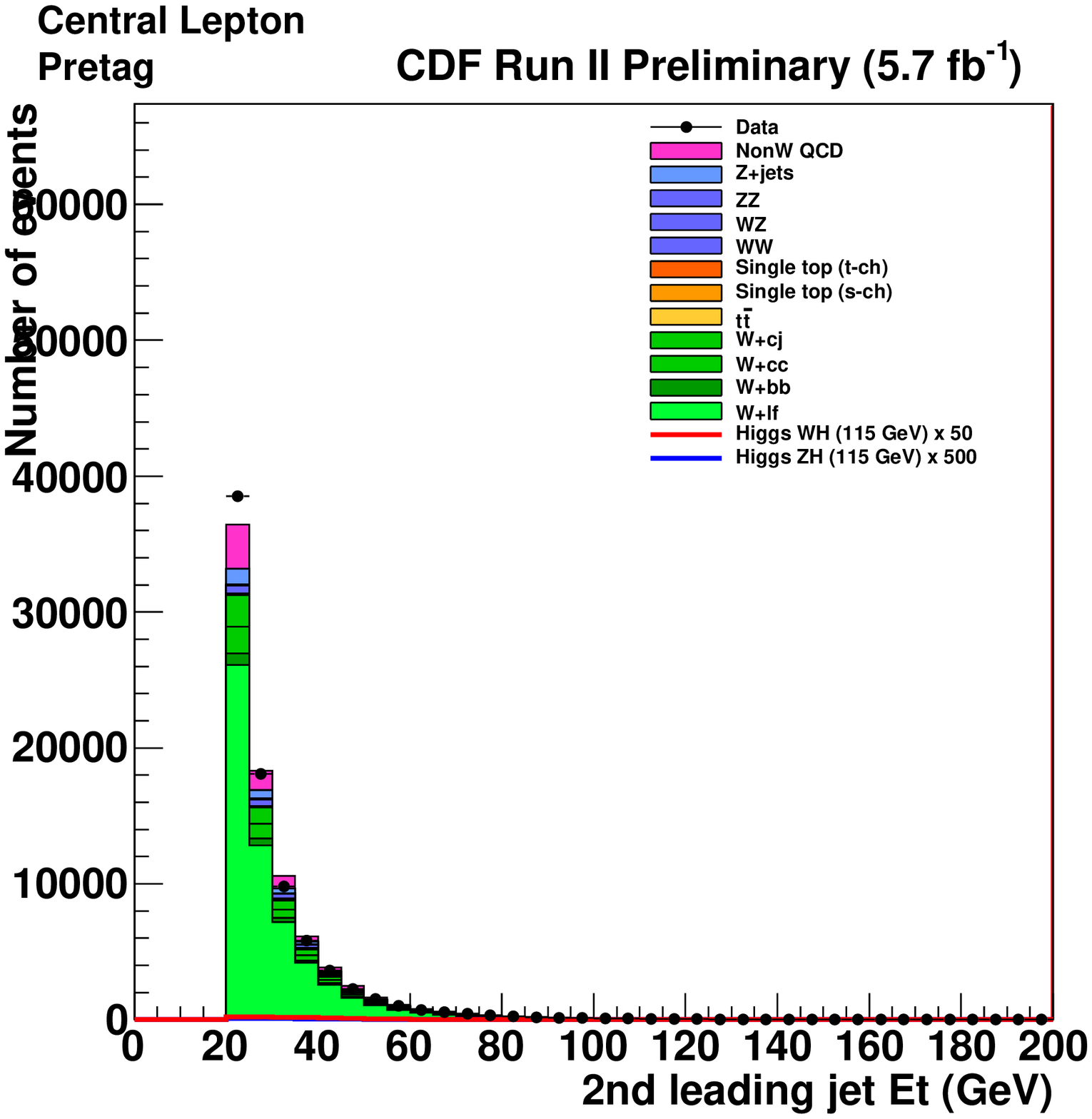}
    \includegraphics[width=6.7cm]{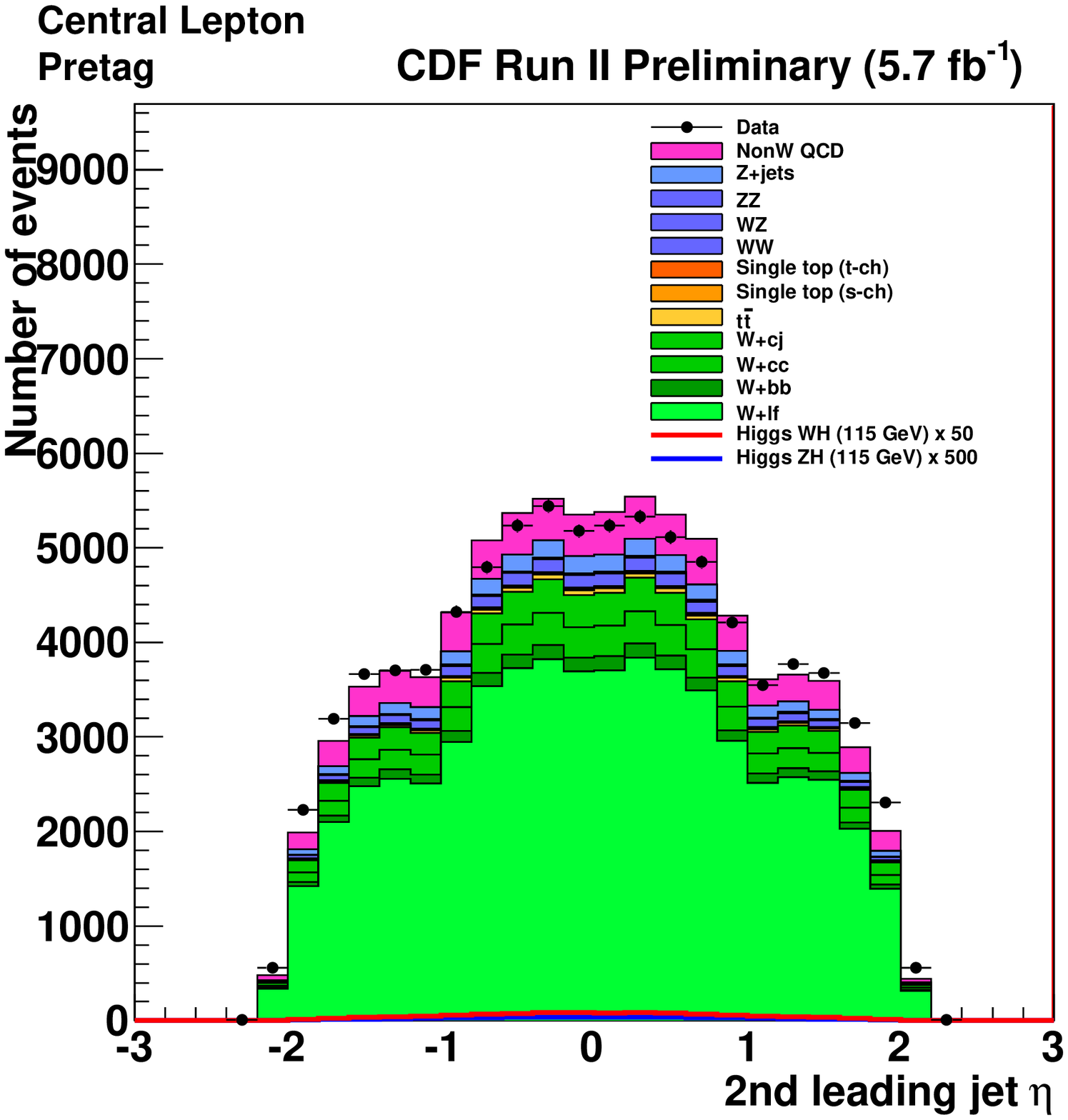}
    \includegraphics[width=6.7cm]{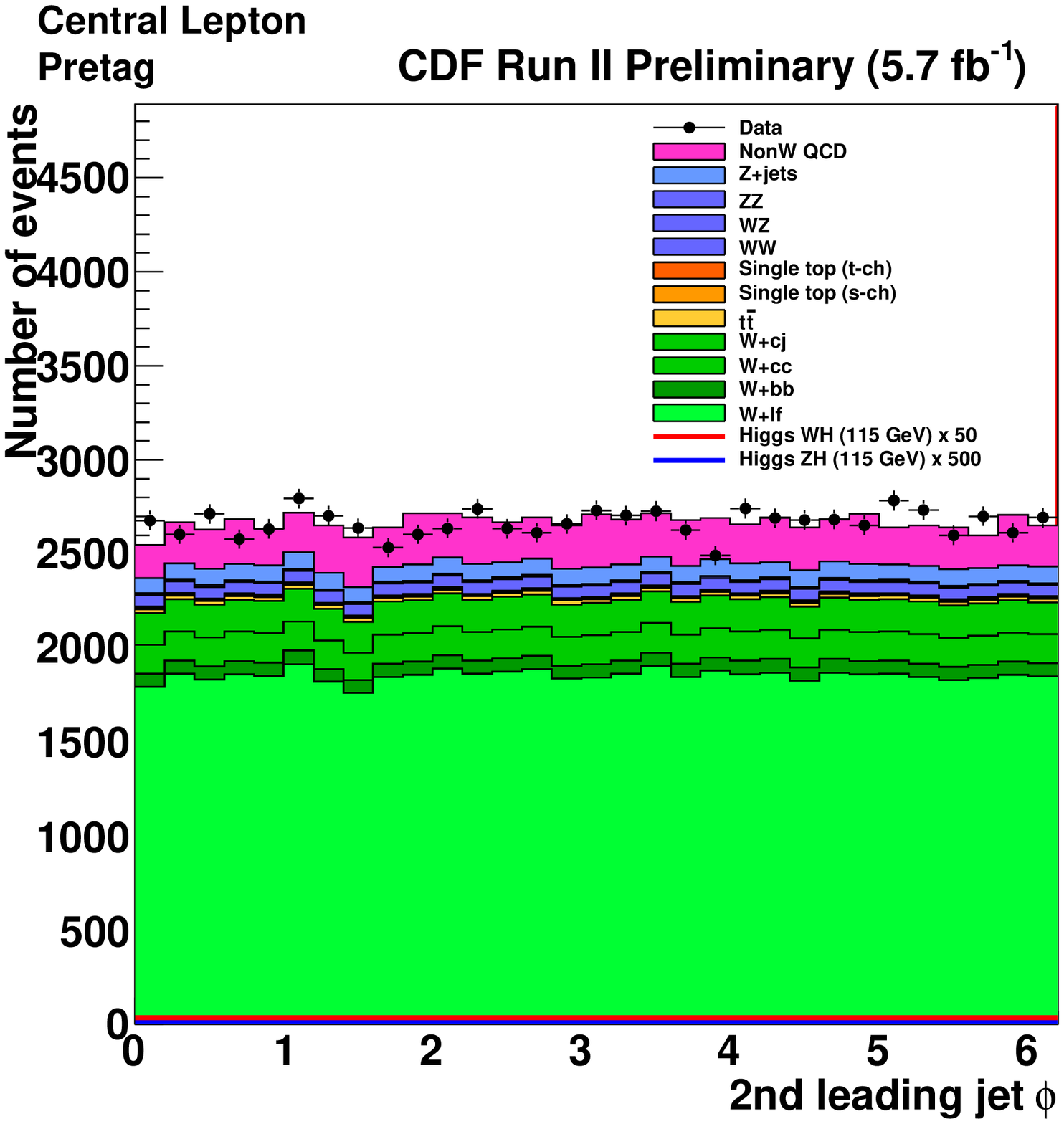}
    \caption [Control plots for TIGHT Pretag Kinematic Variables 1/3]{First part of the control plots for TIGHT charged lepton Pretag kinematic variables.}
  \end {center}
\end {figure}

\begin{figure}[ht]
  \begin{center}
    \includegraphics[width=6.7cm]{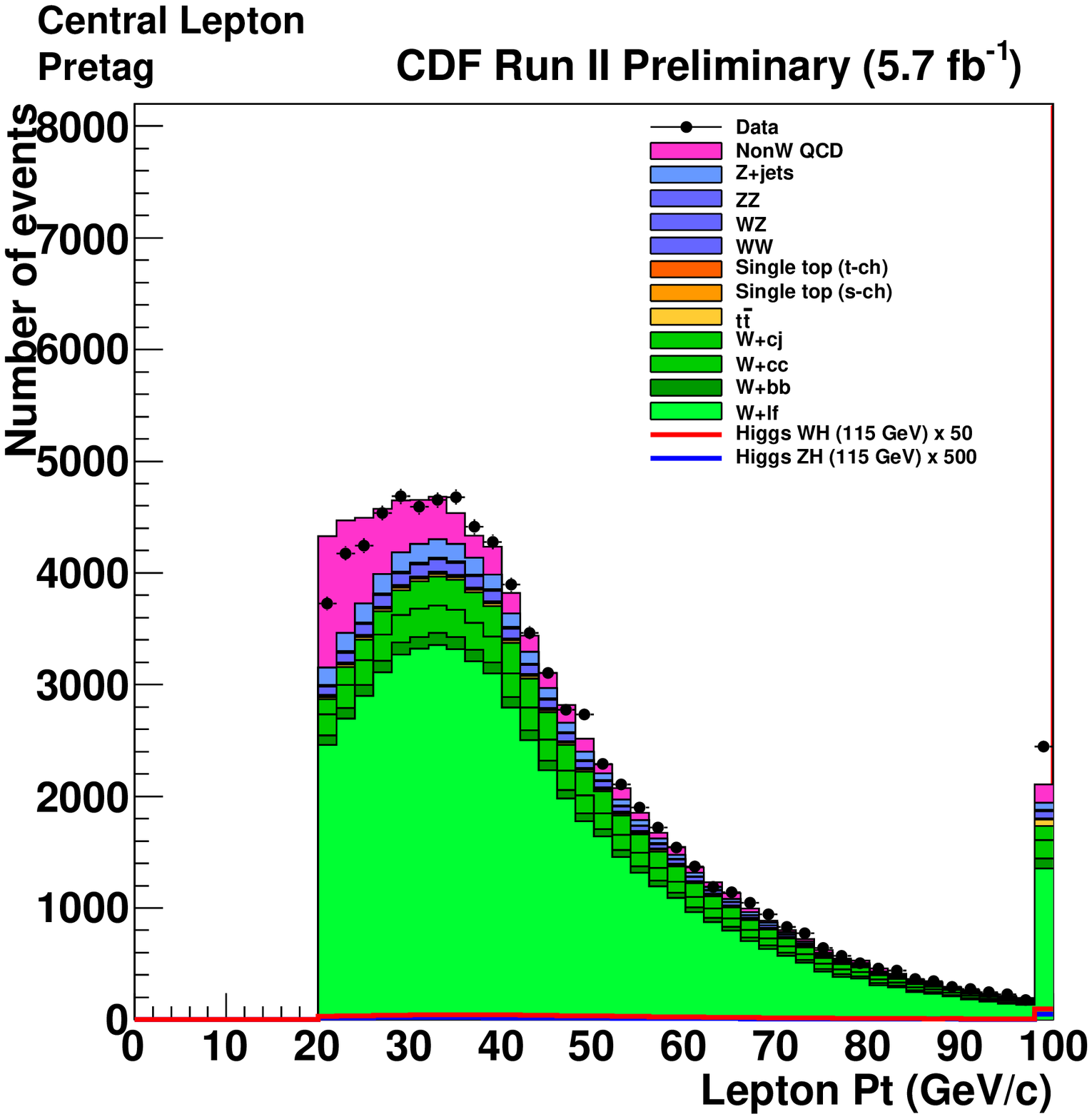}
    \includegraphics[width=6.7cm]{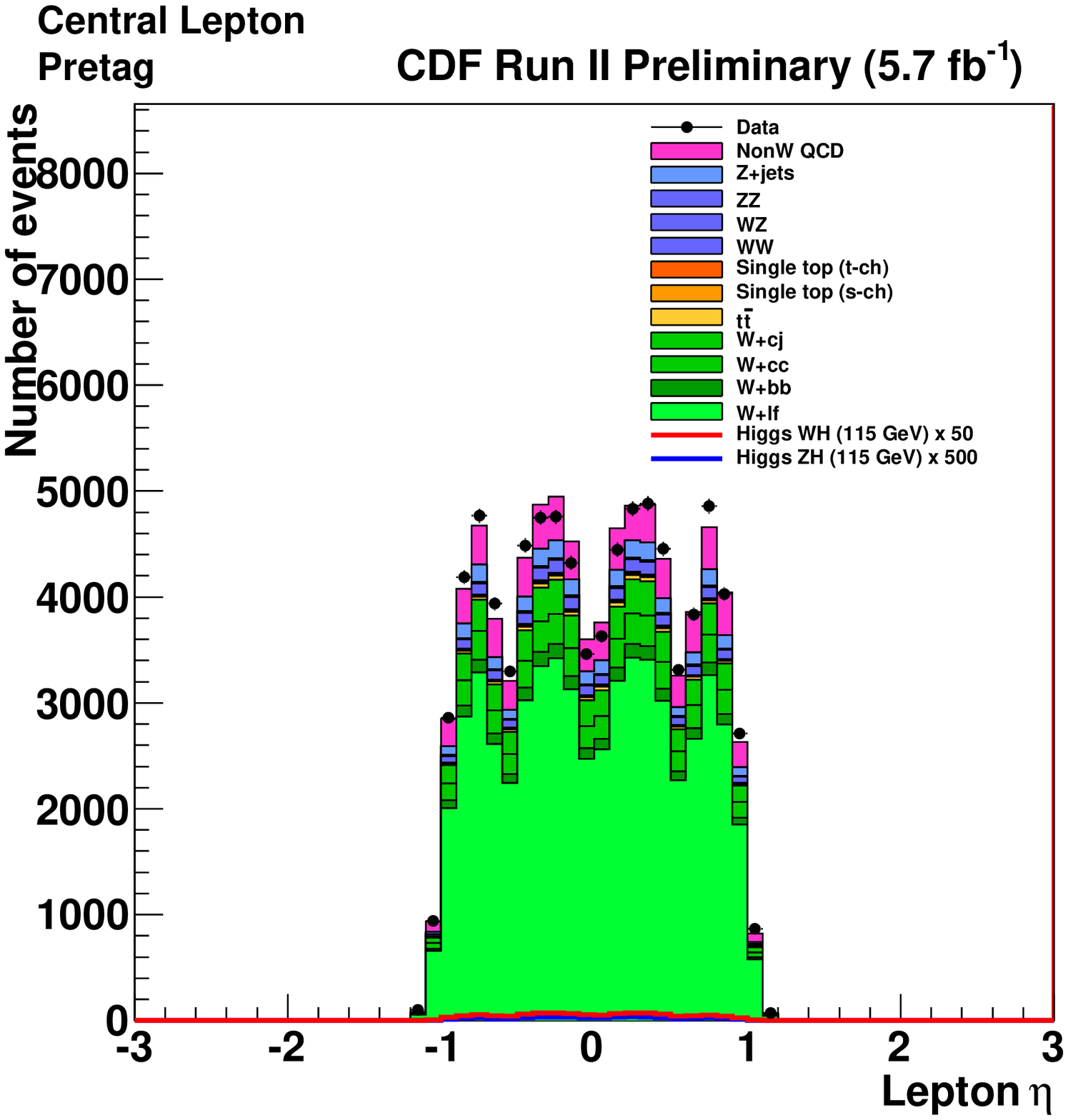}
    \includegraphics[width=6.7cm]{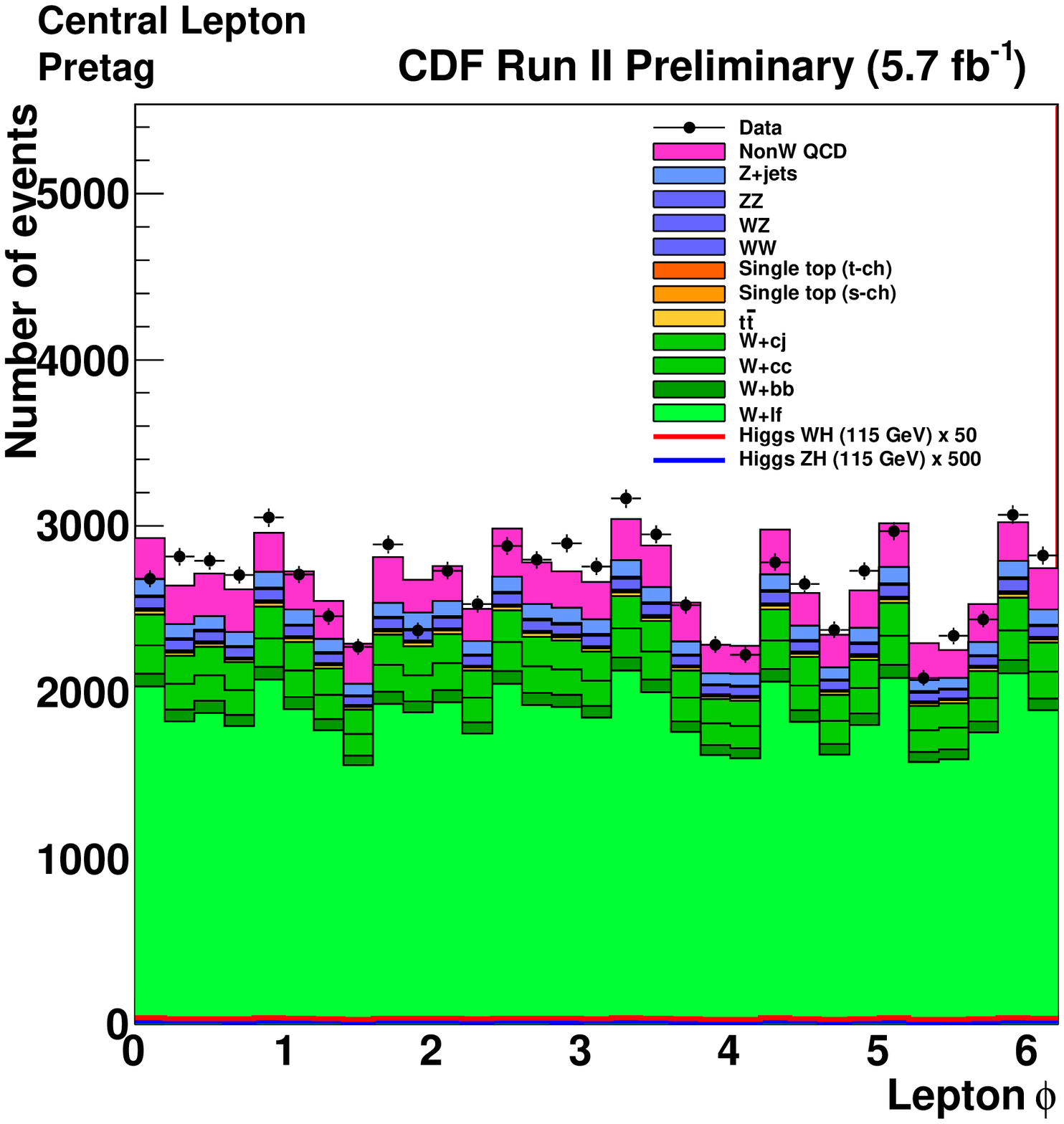}
    \includegraphics[width=6.7cm]{./appendix/ControlPlots/Kinematics/WH_PretagSVTSVT_CEMCMUPCMX_met.eps}
    \includegraphics[width=6.7cm]{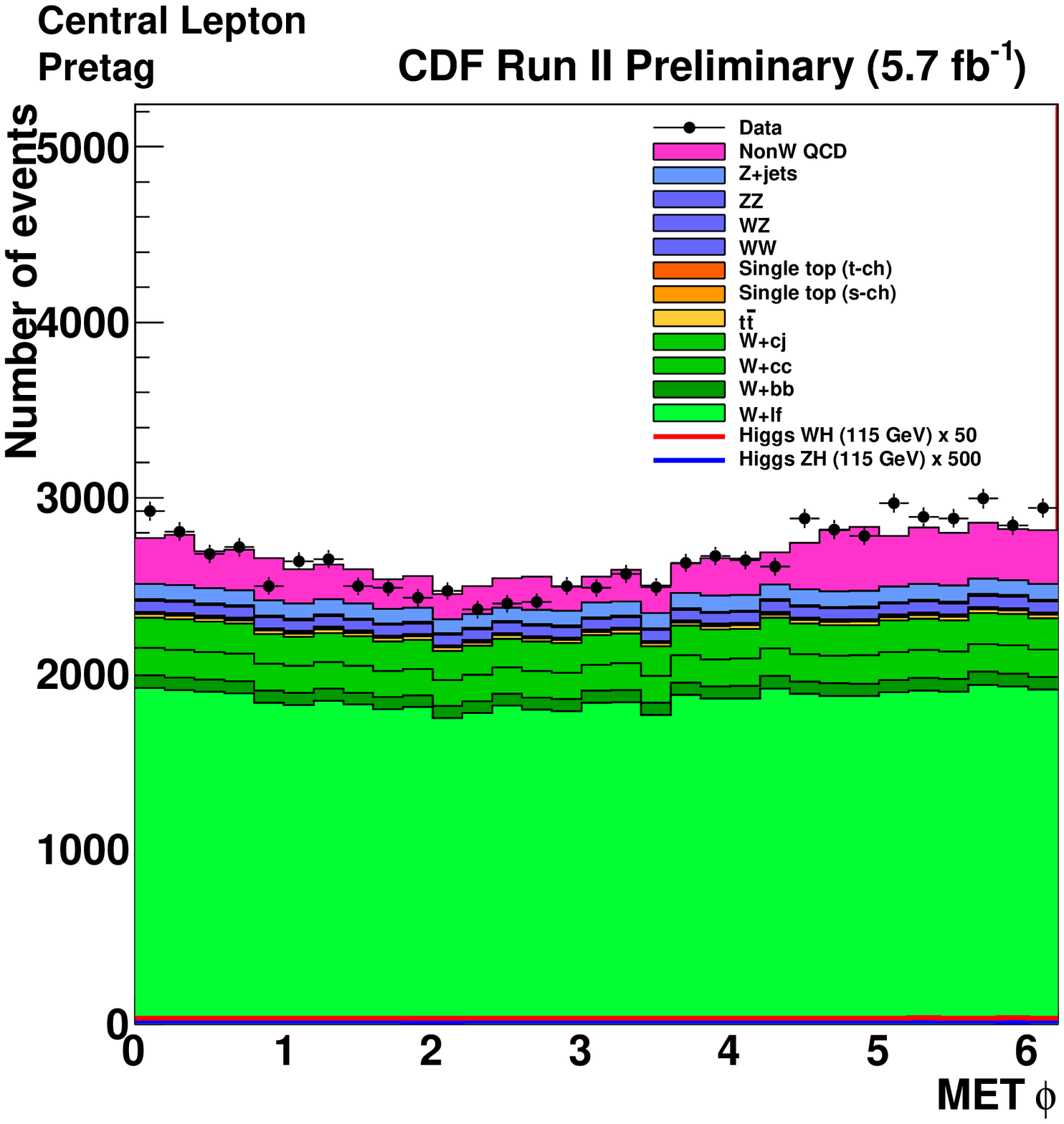}
    \includegraphics[width=6.7cm]{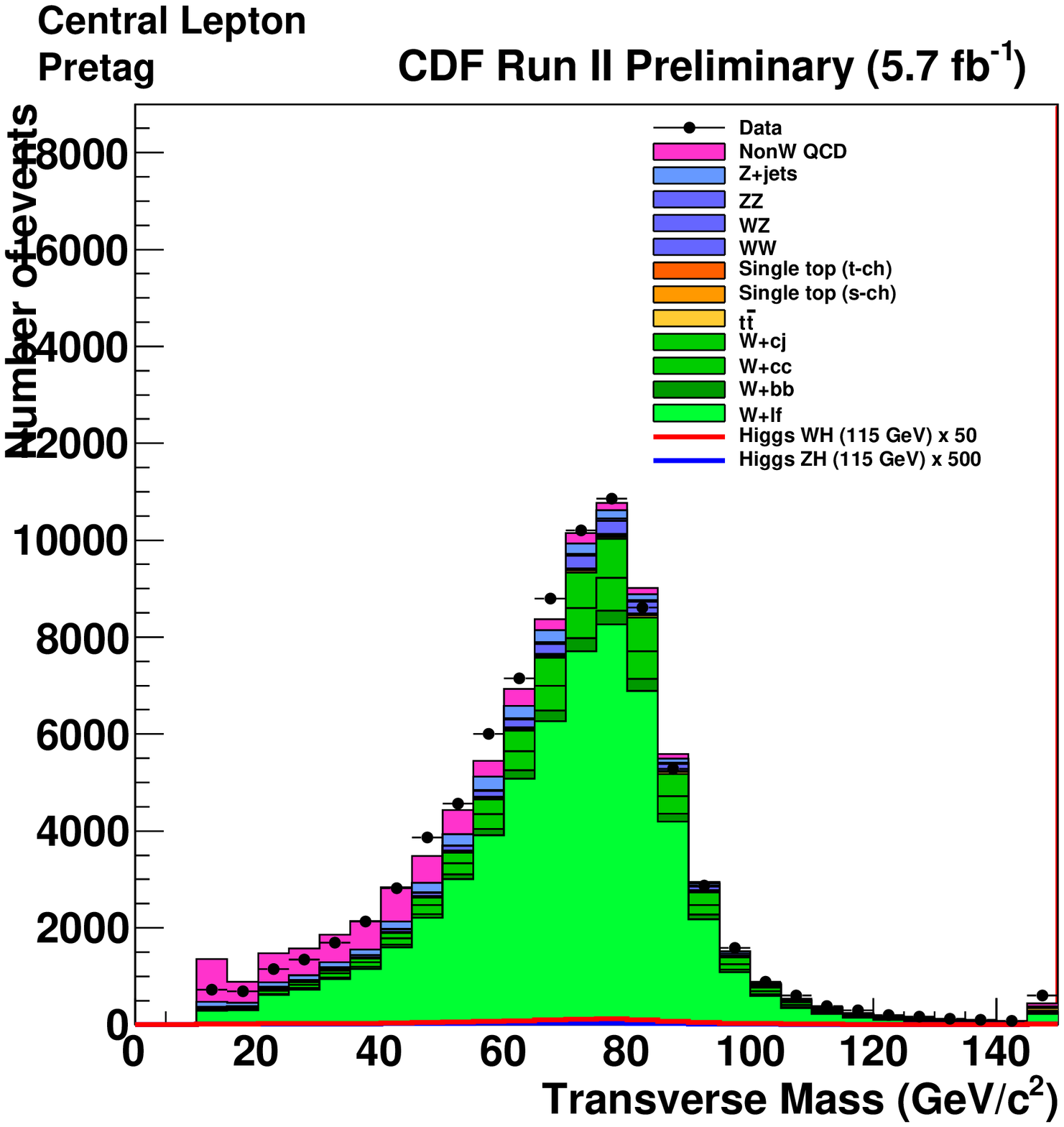}
    \caption [Control plots for TIGHT Pretag Kinematic Variables 2/3]{Second part of the control plots for TIGHT charged lepton Pretag kinematic variables.}
  \end {center}
\end {figure}

\begin{figure}[ht]
  \begin{center}
    \includegraphics[width=6.7cm]{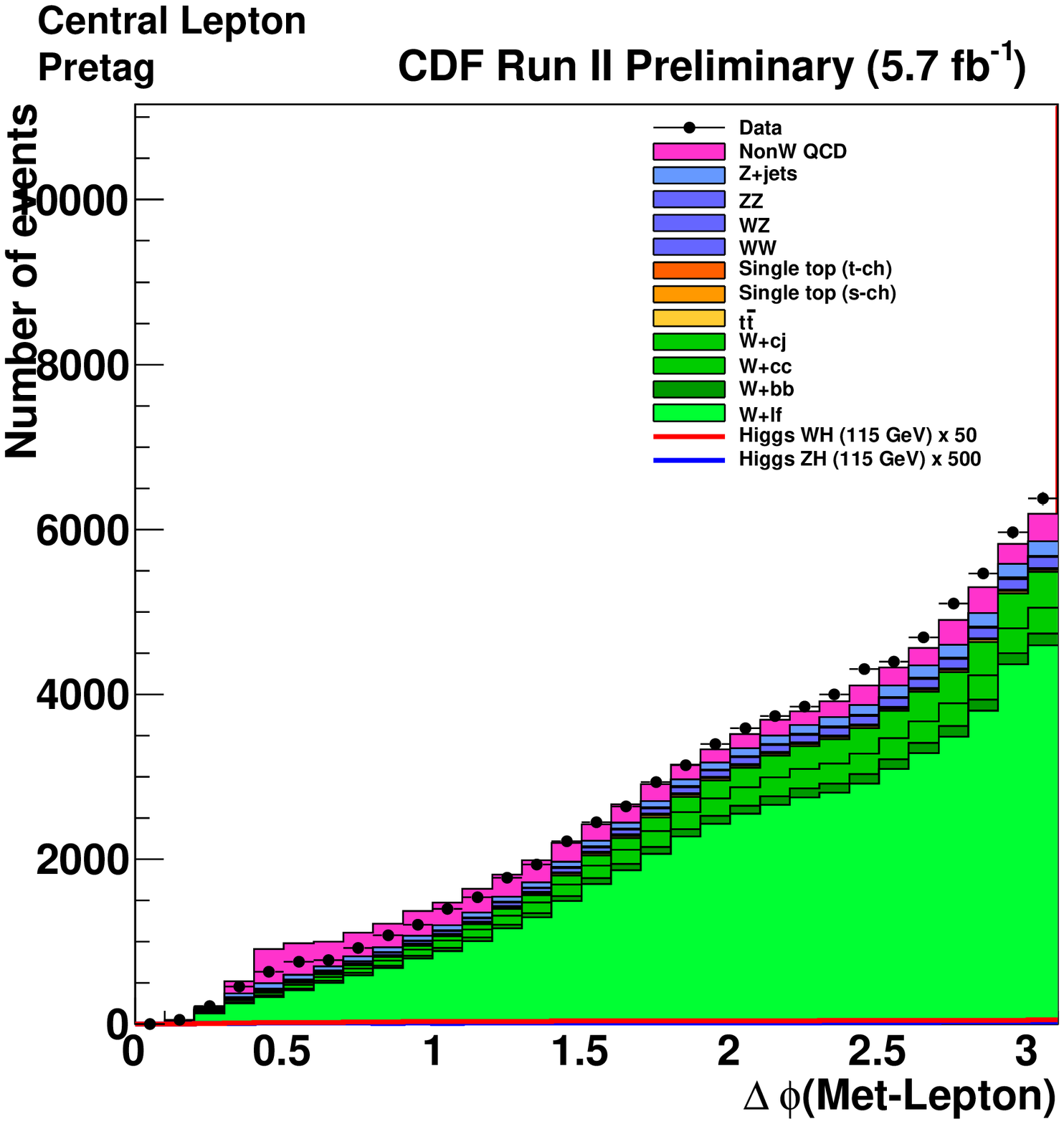}
    \includegraphics[width=6.7cm]{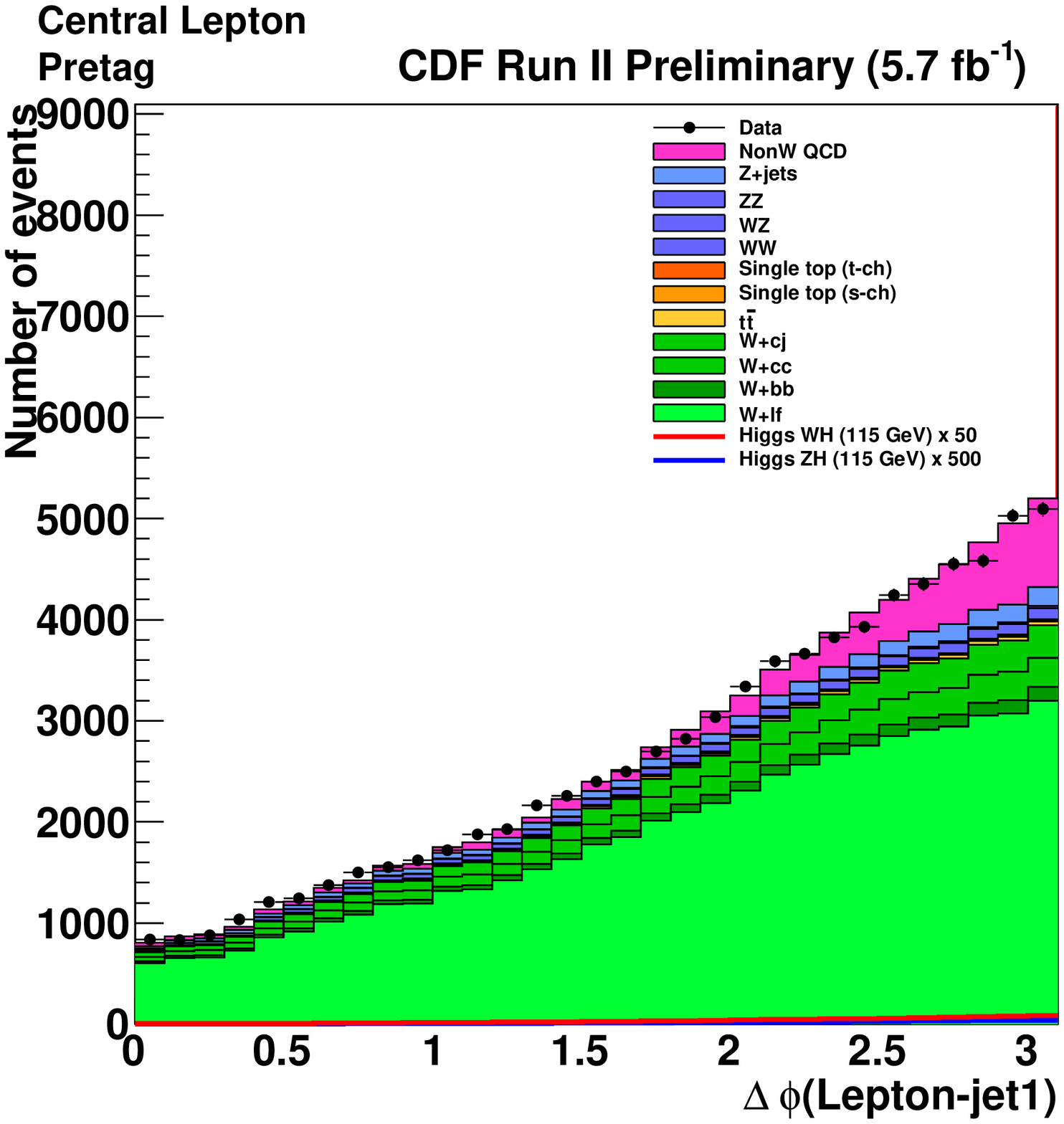}
    \includegraphics[width=6.7cm]{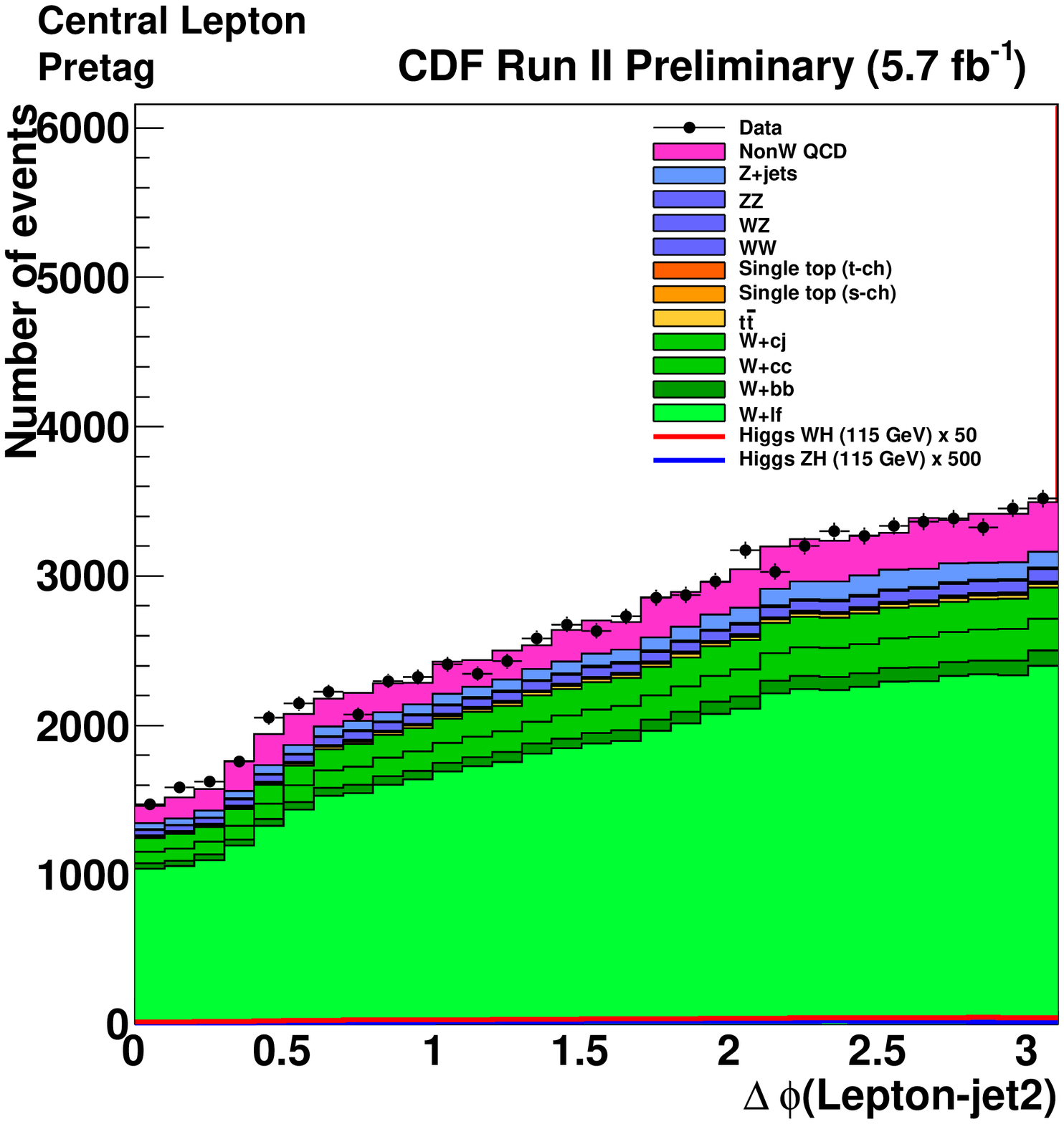}
    \includegraphics[width=6.7cm]{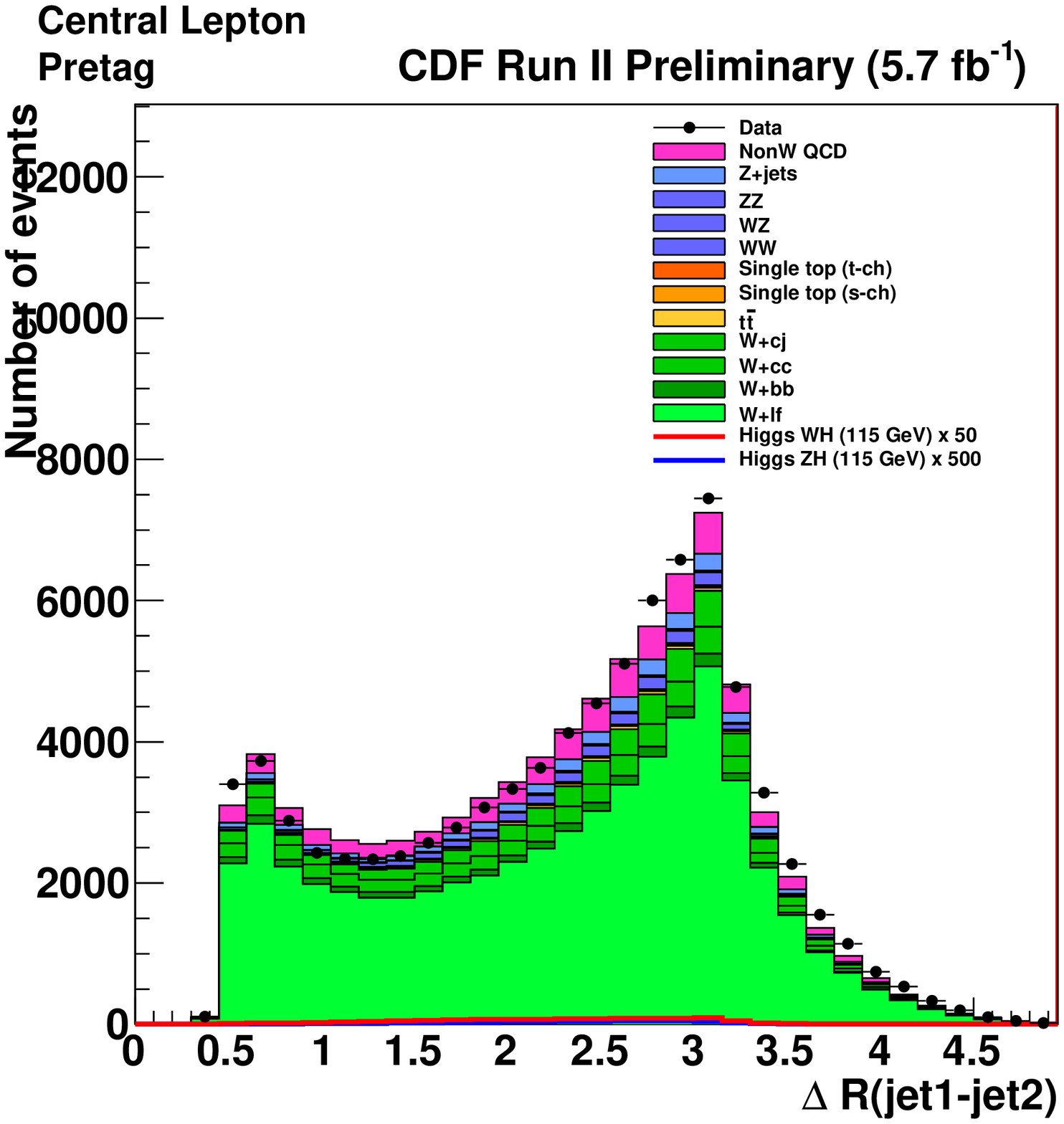}
    \caption [Control plots for TIGHT Pretag Kinematic Variables 3/3]{Third part of the control plots for TIGHT charged lepton Pretag kinematic variables.}
  \end {center}
\end {figure}

\clearpage

%\input{./appendix/ControlPlots/controlPlots_BNN_PretagSVTSVT_CEMCMUPCMX.tex}
%\clearpage

%\clearpage

%%% For Kinematic Variables SVTSVT TIGHT charged lepton %%%

\begin{figure}[ht]
  \begin{center}
    \includegraphics[width=6.7cm]{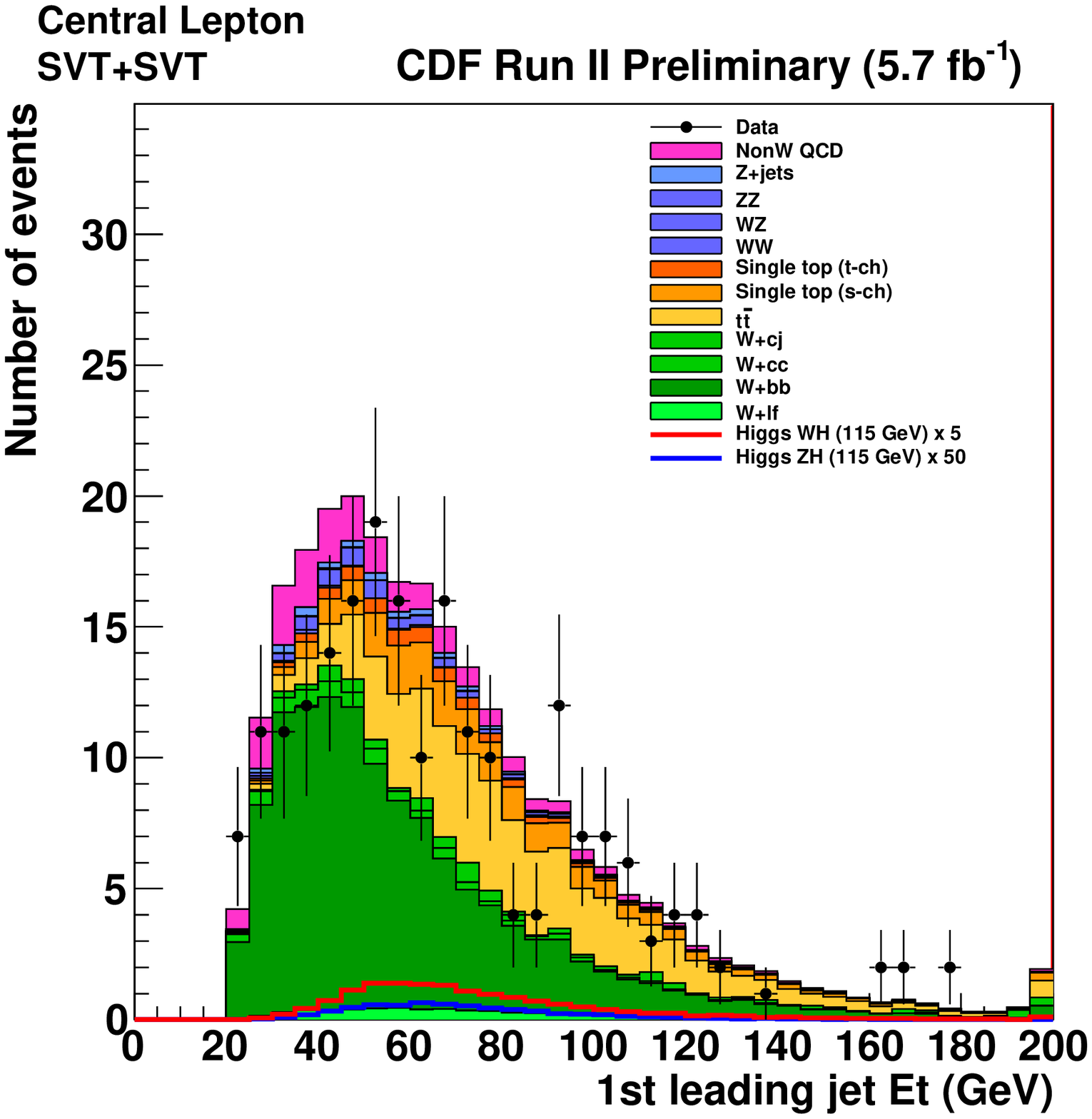}
    \includegraphics[width=6.7cm]{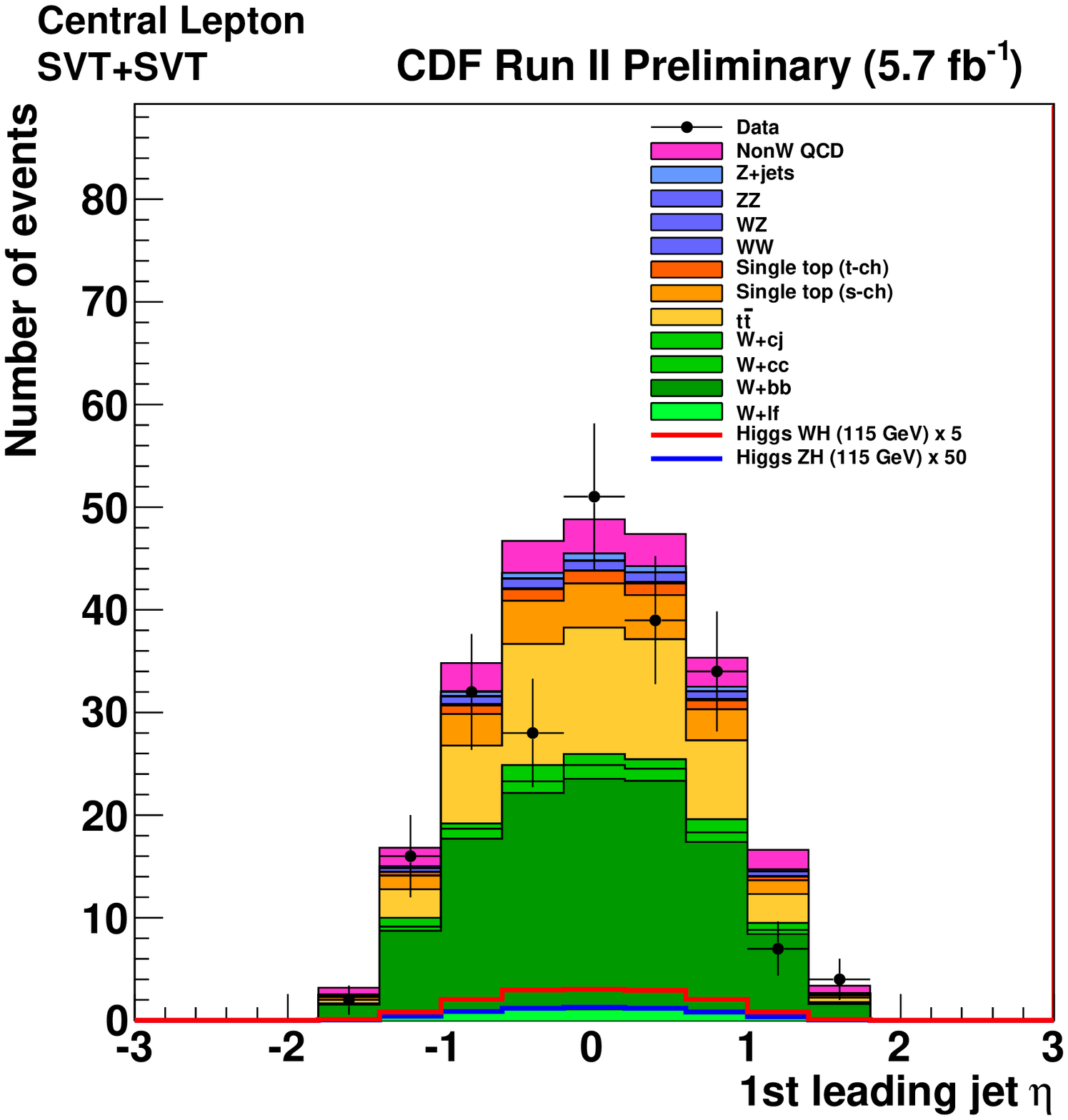}
    \includegraphics[width=6.7cm]{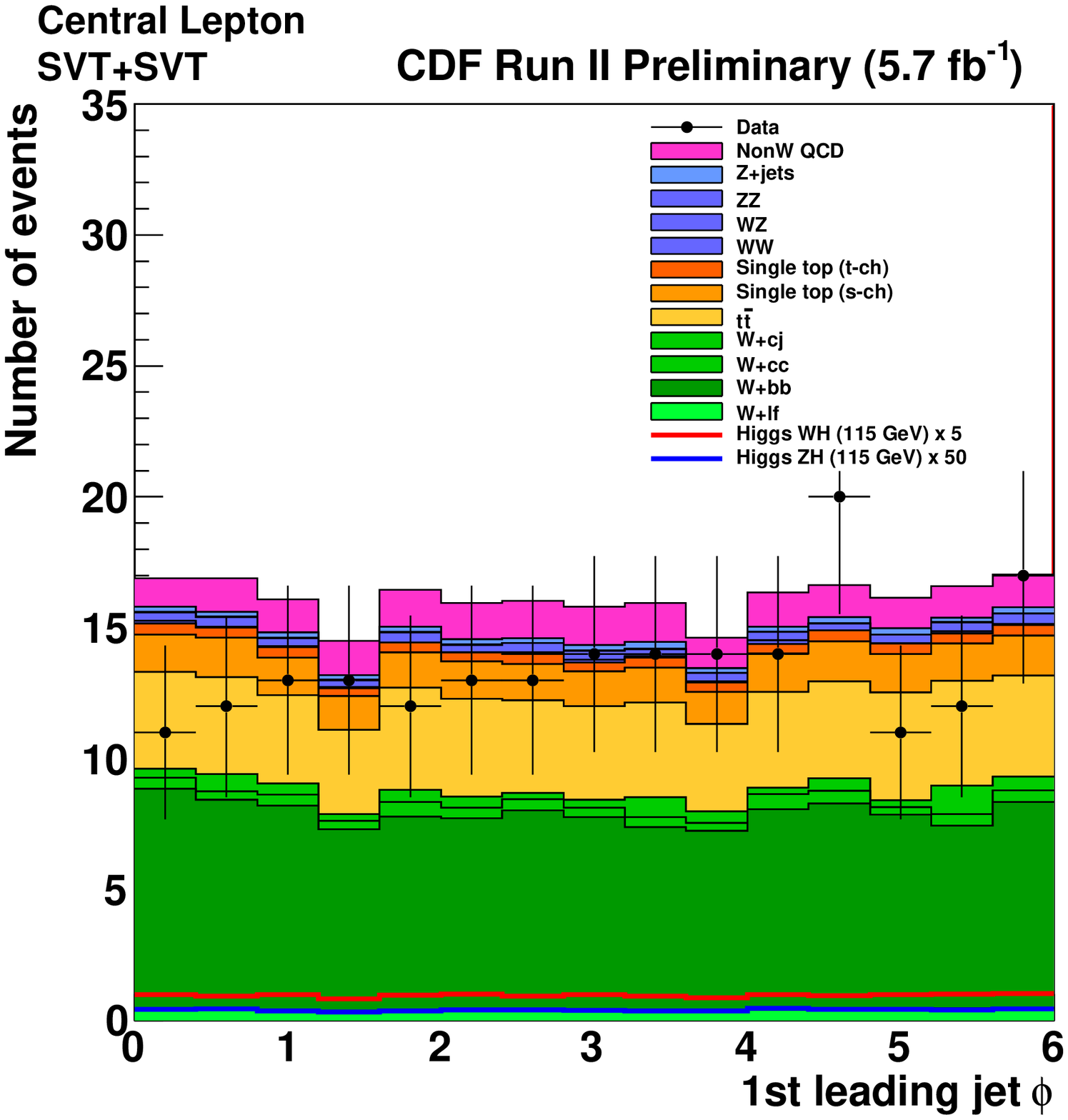}
    \includegraphics[width=6.7cm]{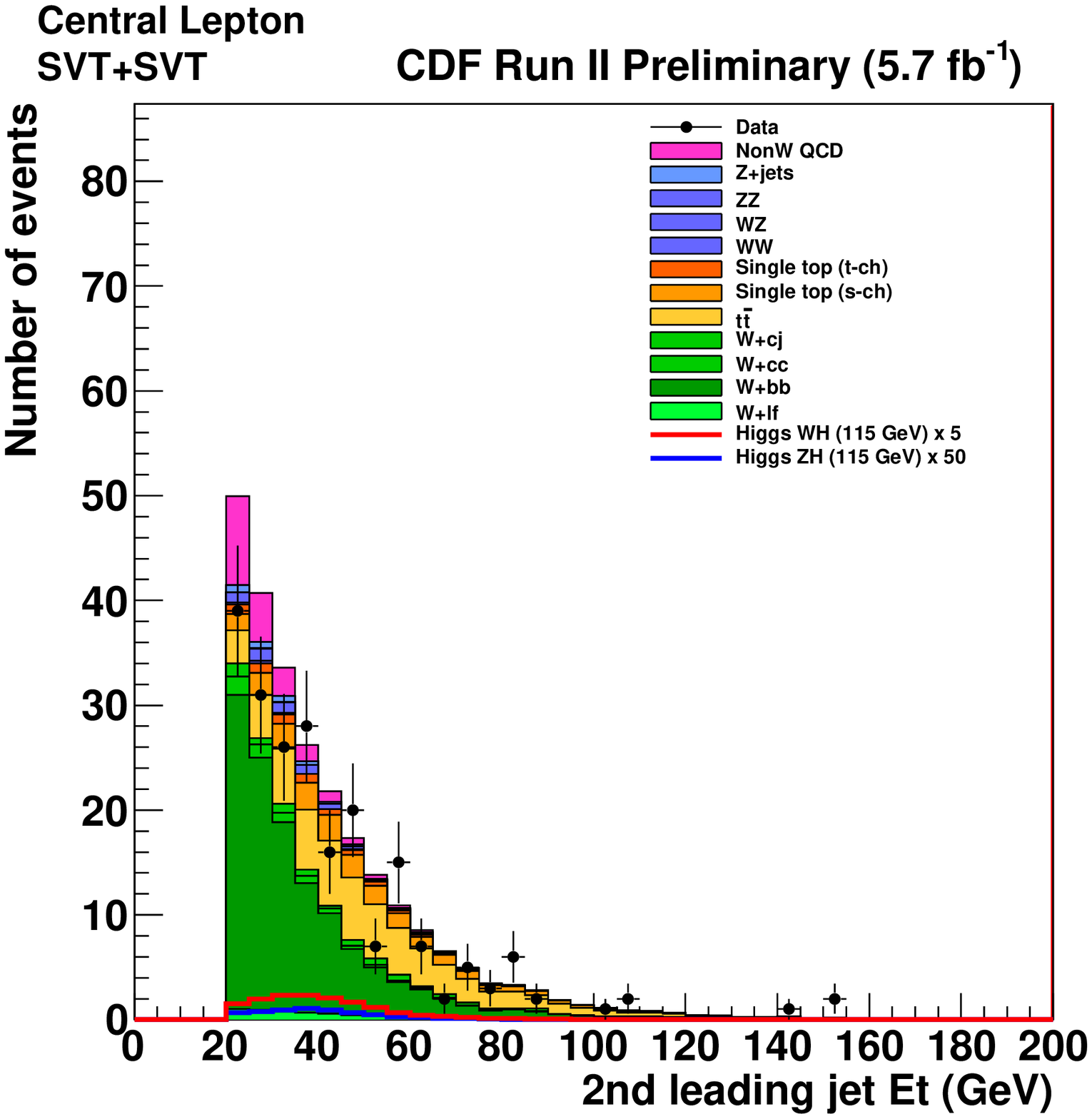}
    \includegraphics[width=6.7cm]{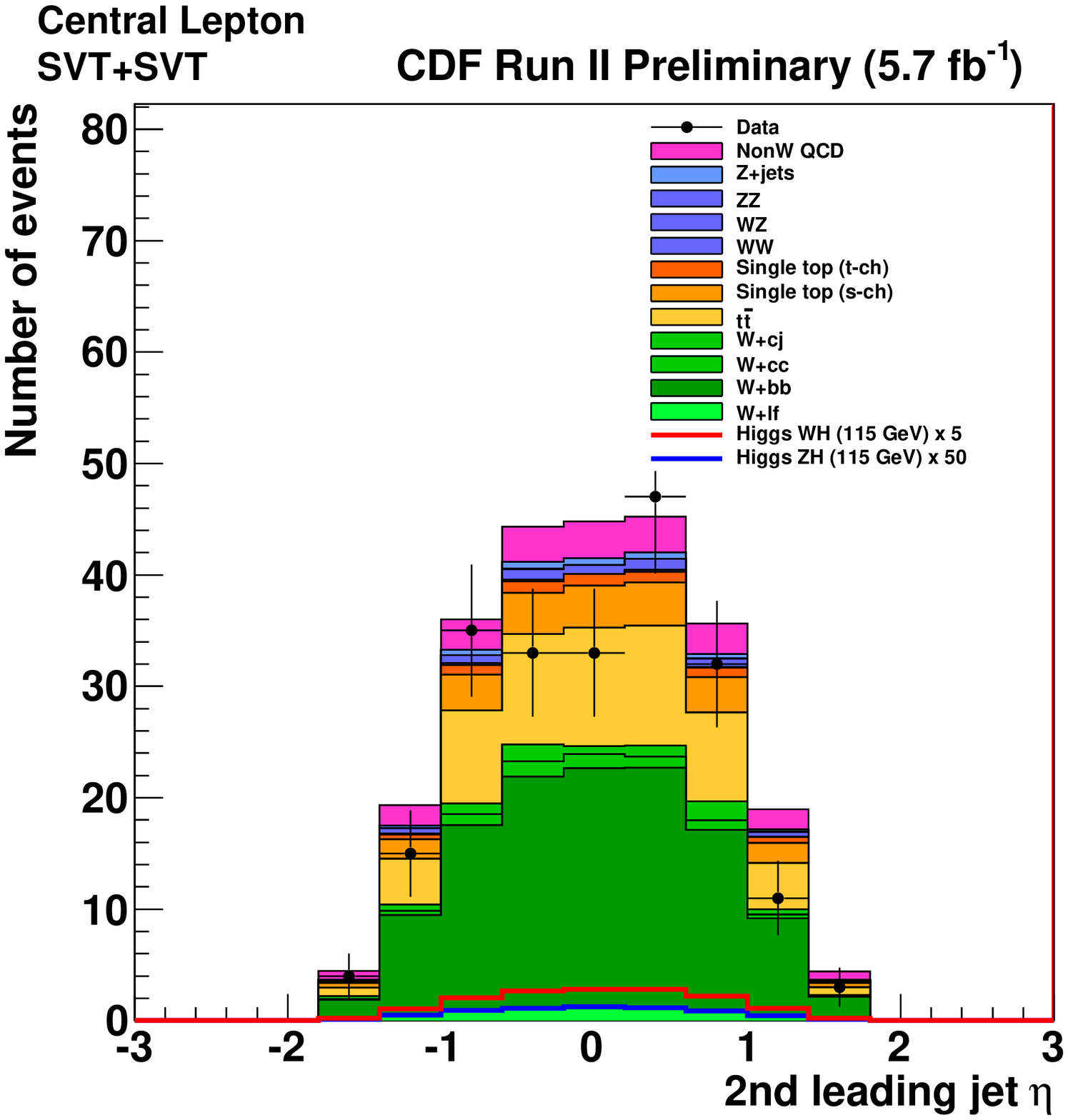}
    \includegraphics[width=6.7cm]{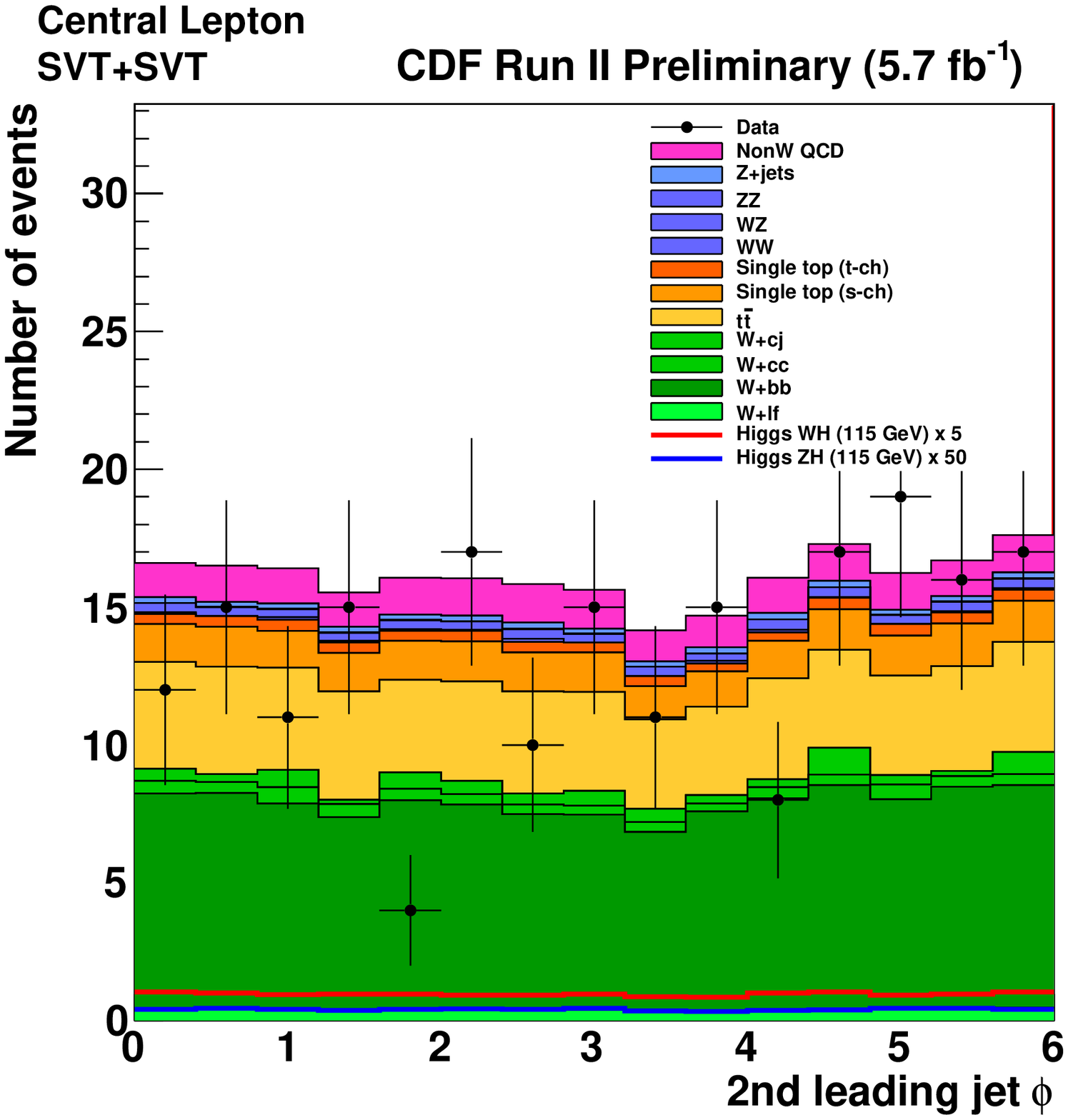}
    \caption [Control plots for TIGHT SVTSVT Kinematic Variables 1/3]{First part of the control plots for TIGHT charged lepton SVTSVT kinematic variables.}
  \end {center}
\end {figure}

\begin{figure}[ht]
  \begin{center}
    \includegraphics[width=6.7cm]{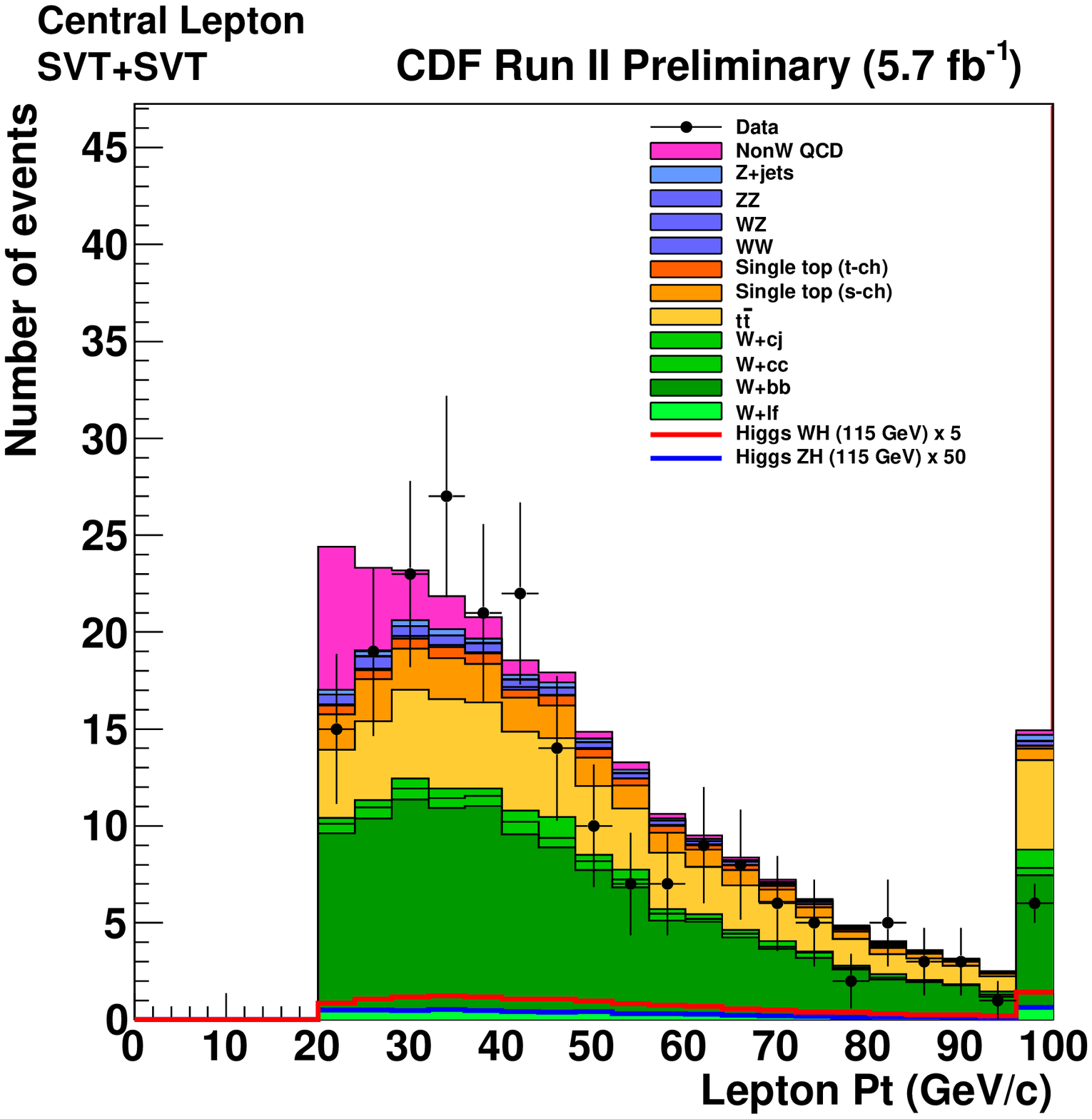}
    \includegraphics[width=6.7cm]{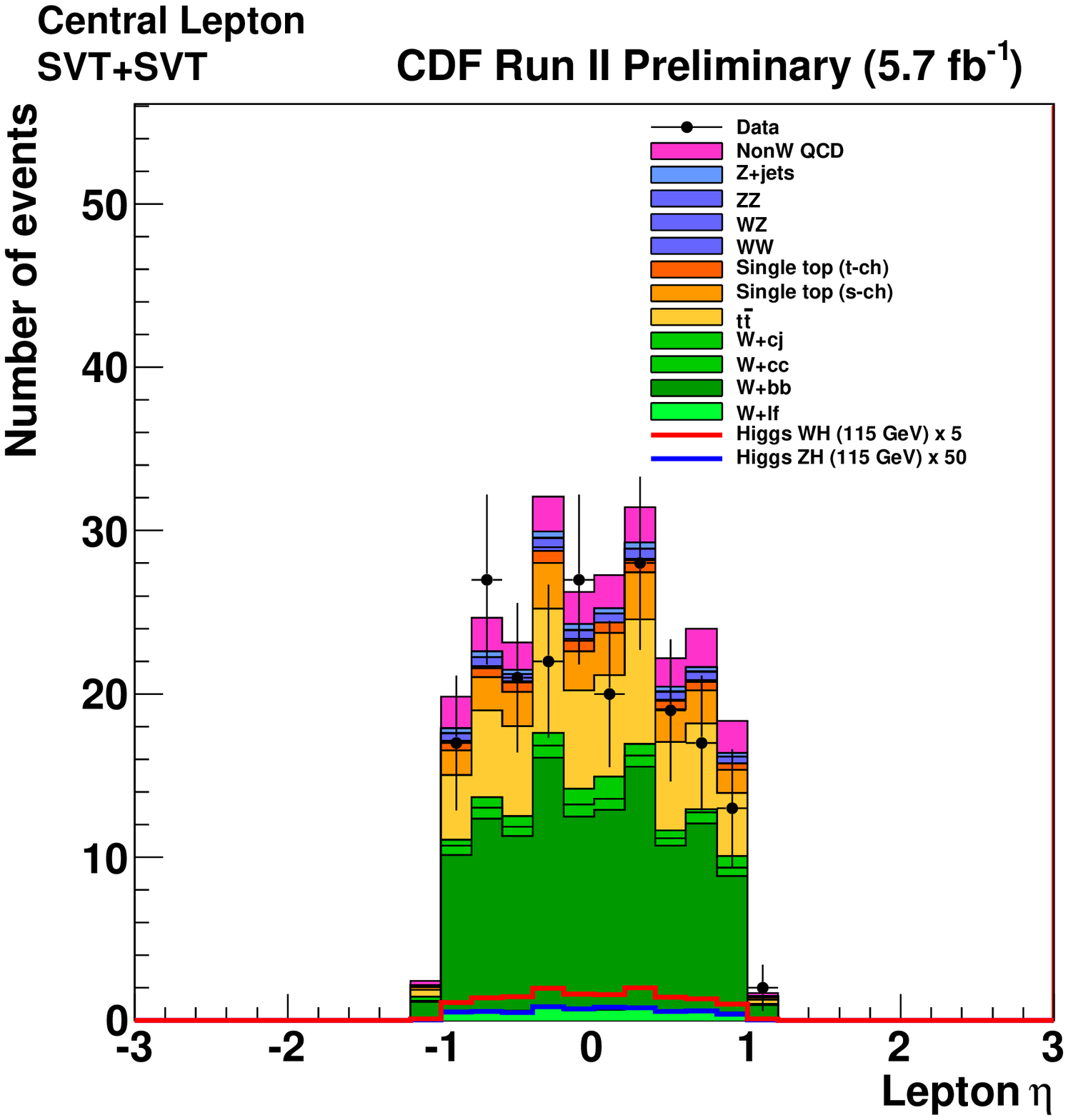}
    \includegraphics[width=6.7cm]{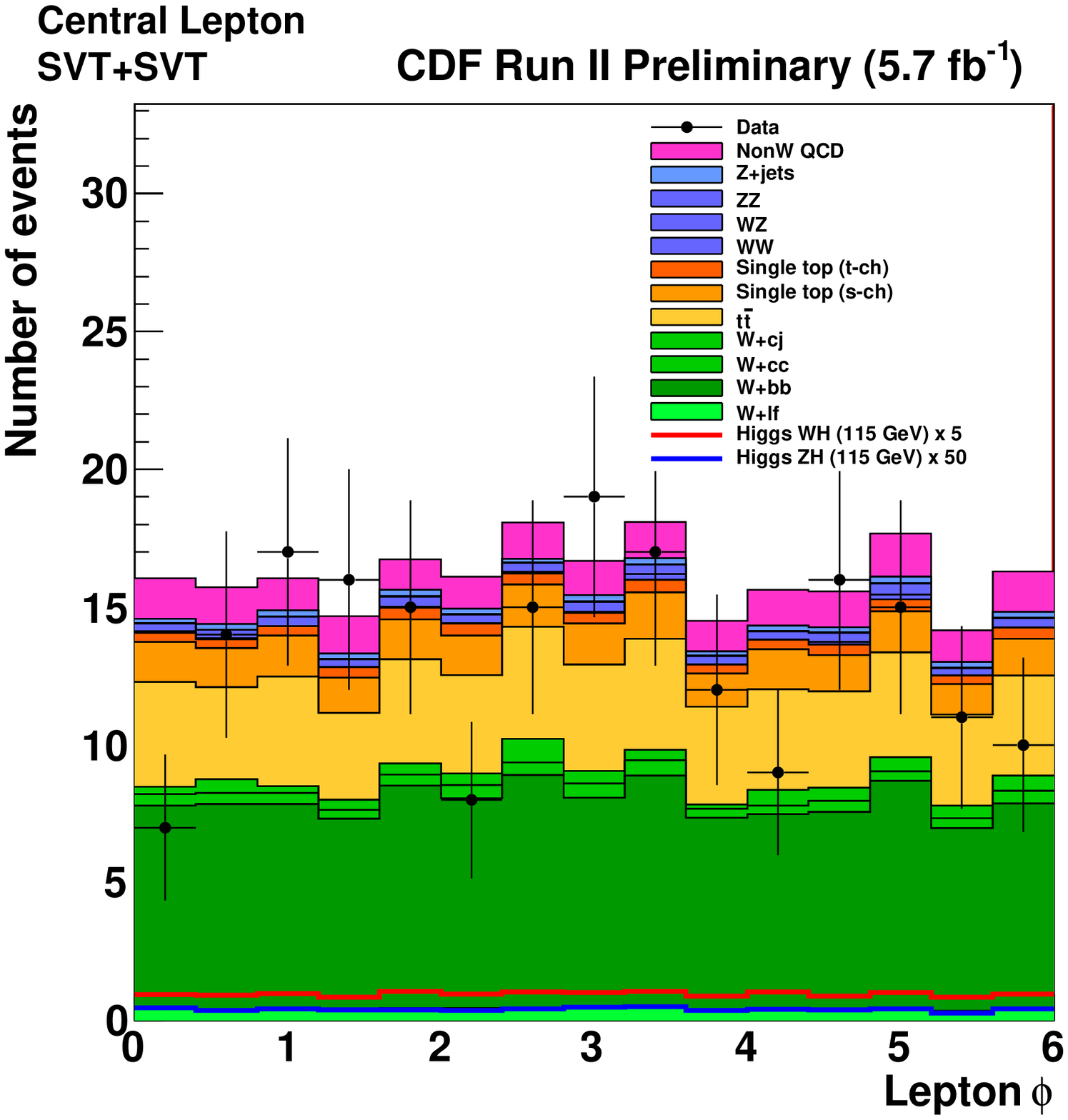}
    \includegraphics[width=6.7cm]{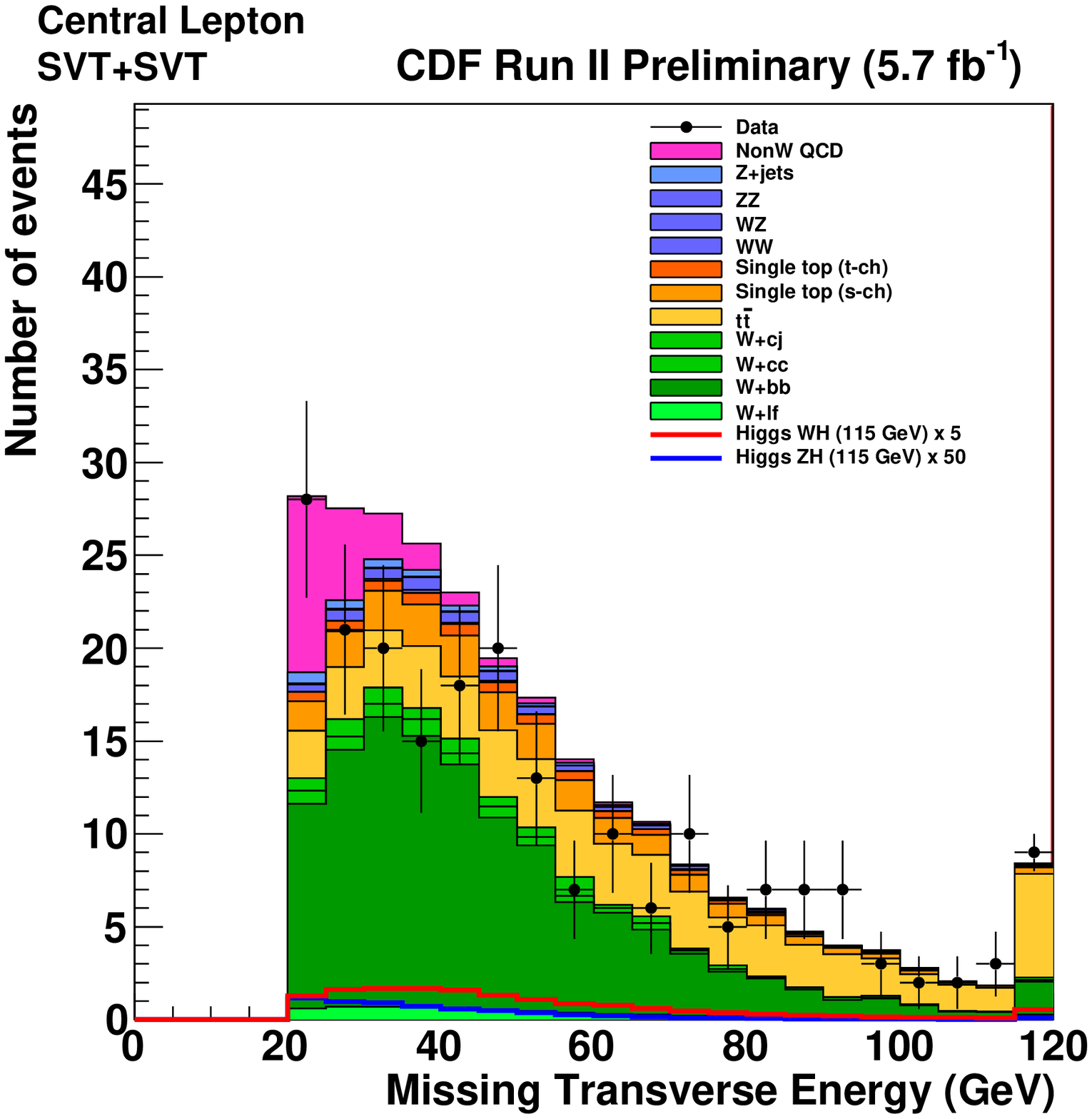}
    \includegraphics[width=6.7cm]{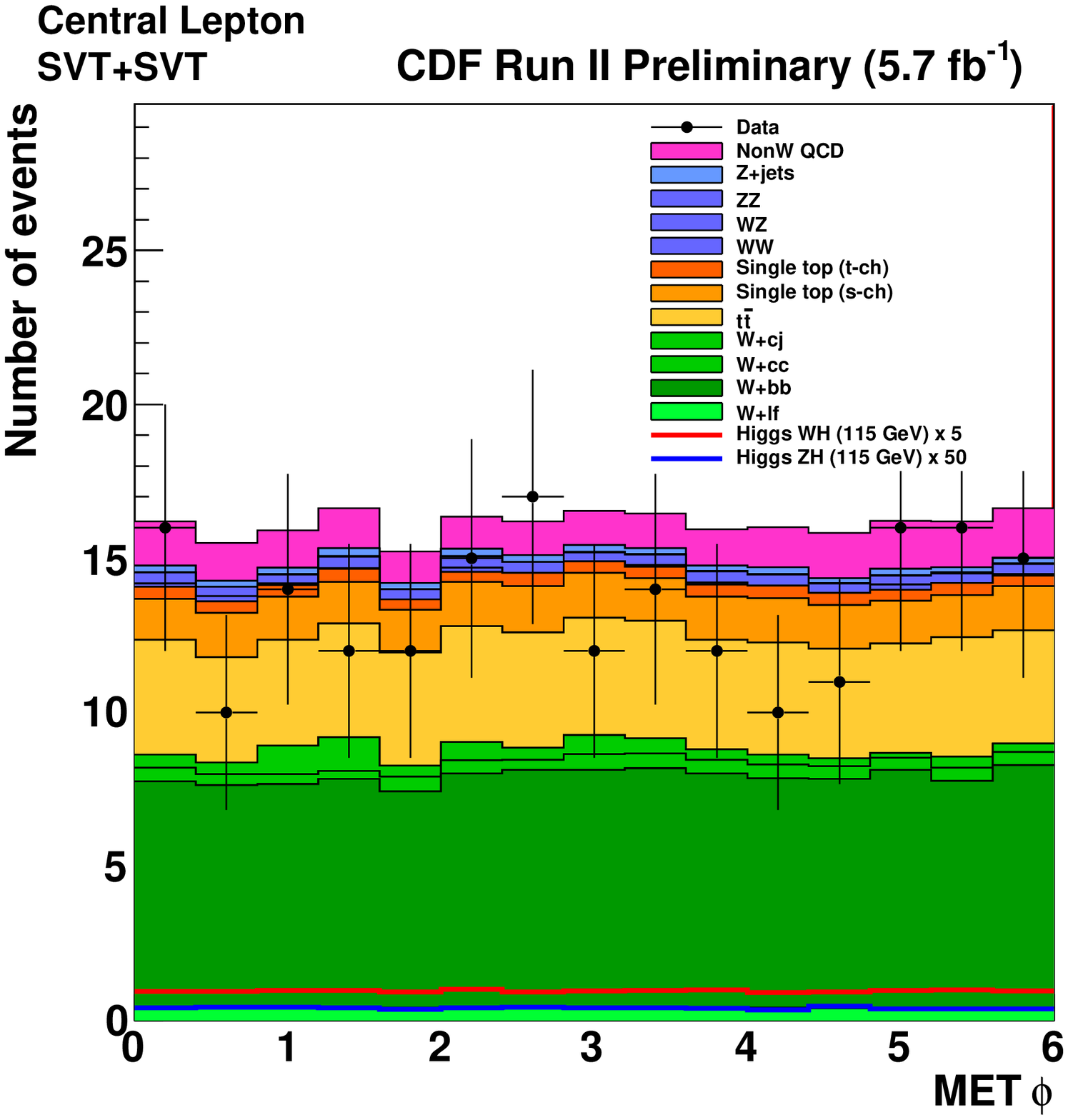}
    \includegraphics[width=6.7cm]{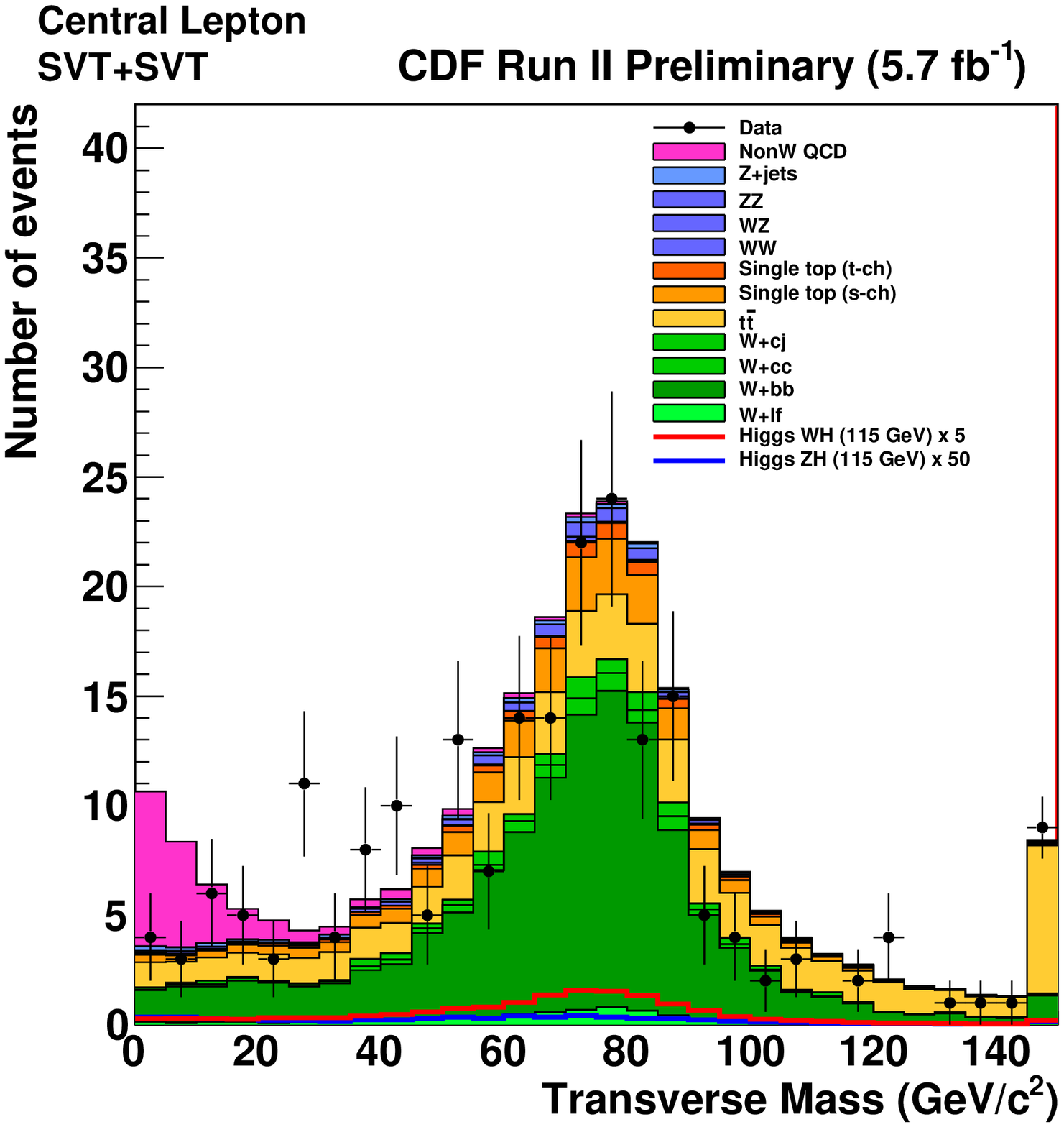}
    \caption [Control plots for TIGHT SVTSVT Kinematic Variables 2/3]{Second part of the control plots for TIGHT charged lepton SVTSVT kinematic variables.}
  \end {center}
\end {figure}

\begin{figure}[ht]
  \begin{center}
    \includegraphics[width=6.7cm]{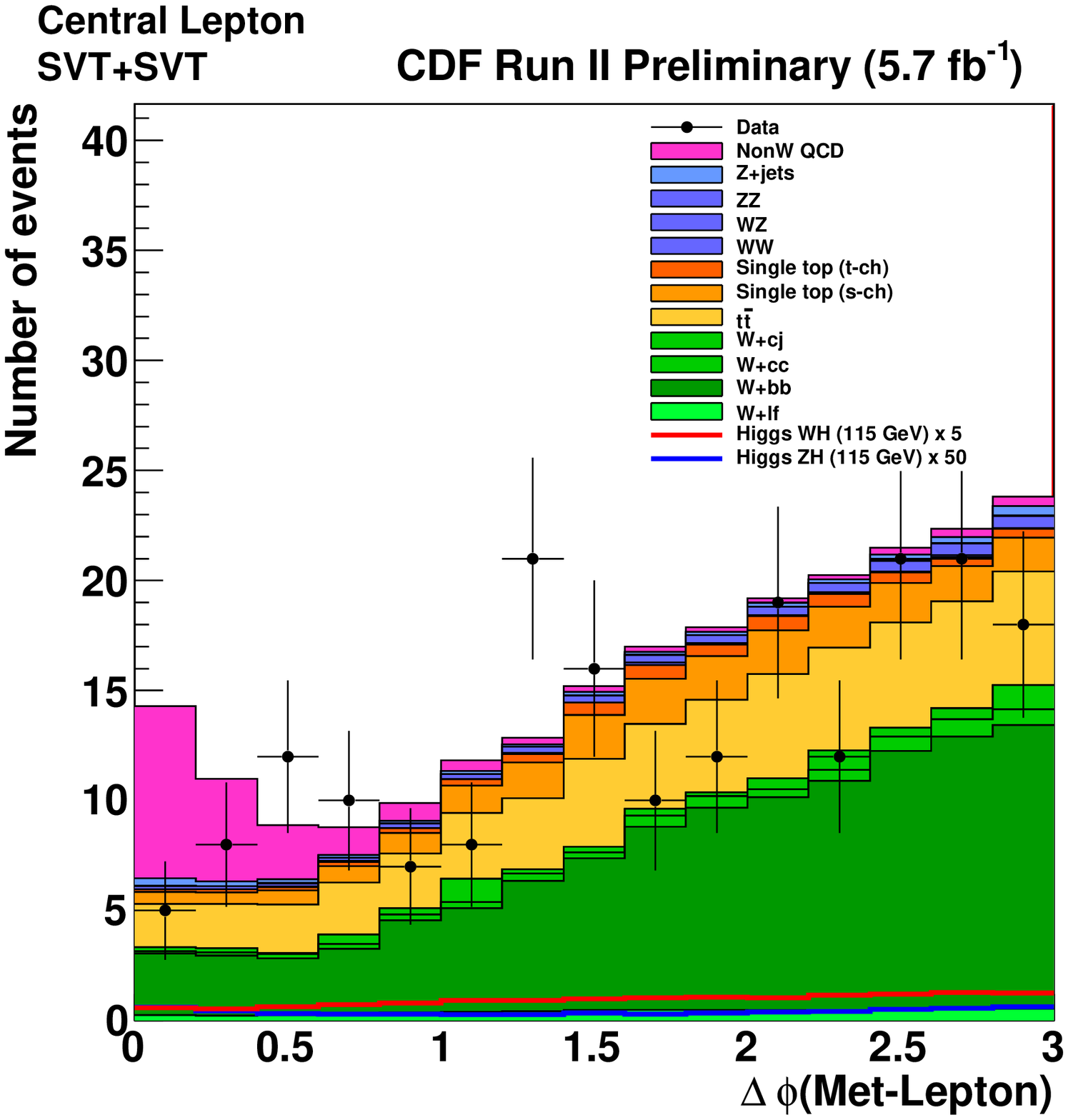}
    \includegraphics[width=6.7cm]{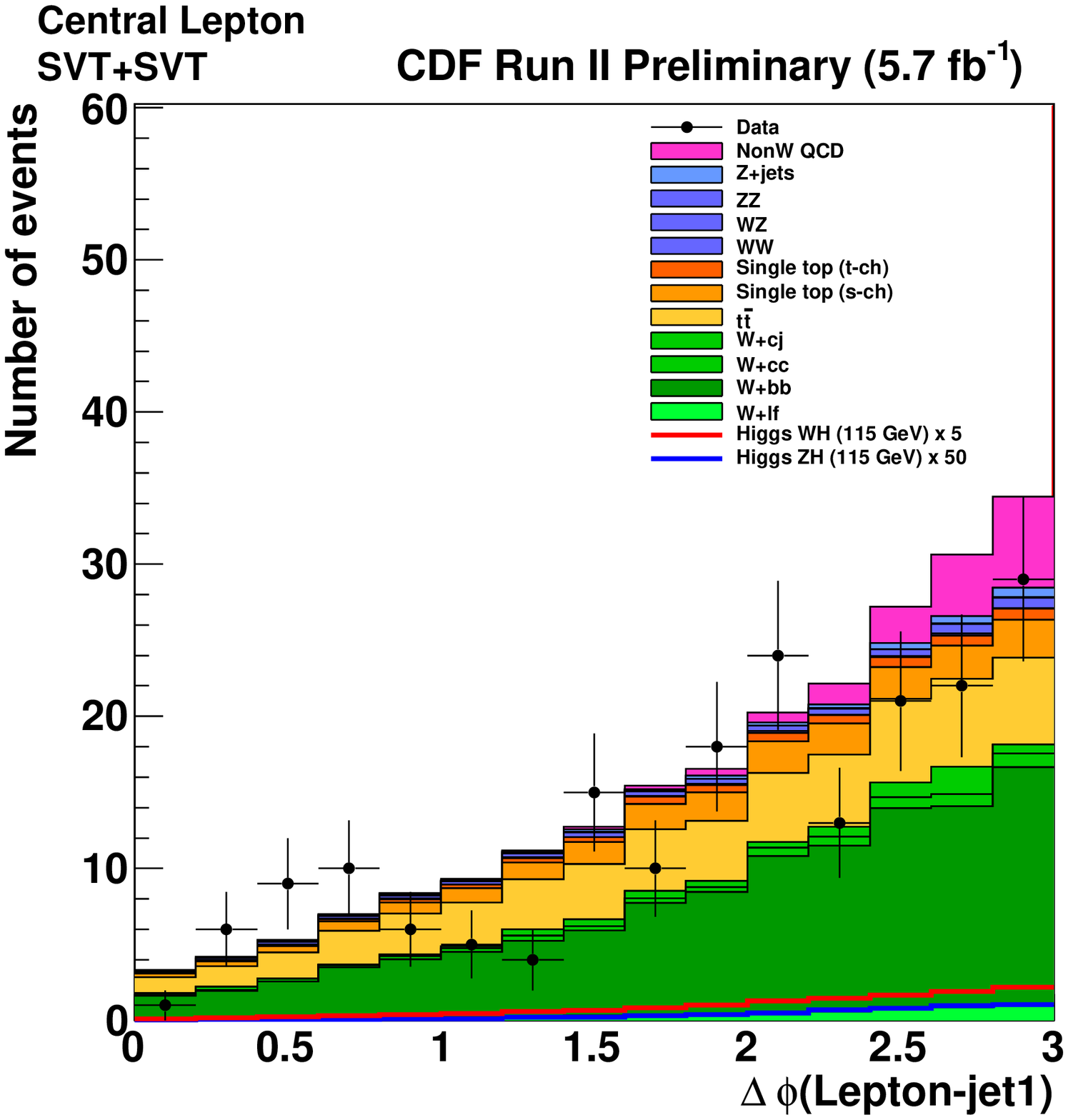}
    \includegraphics[width=6.7cm]{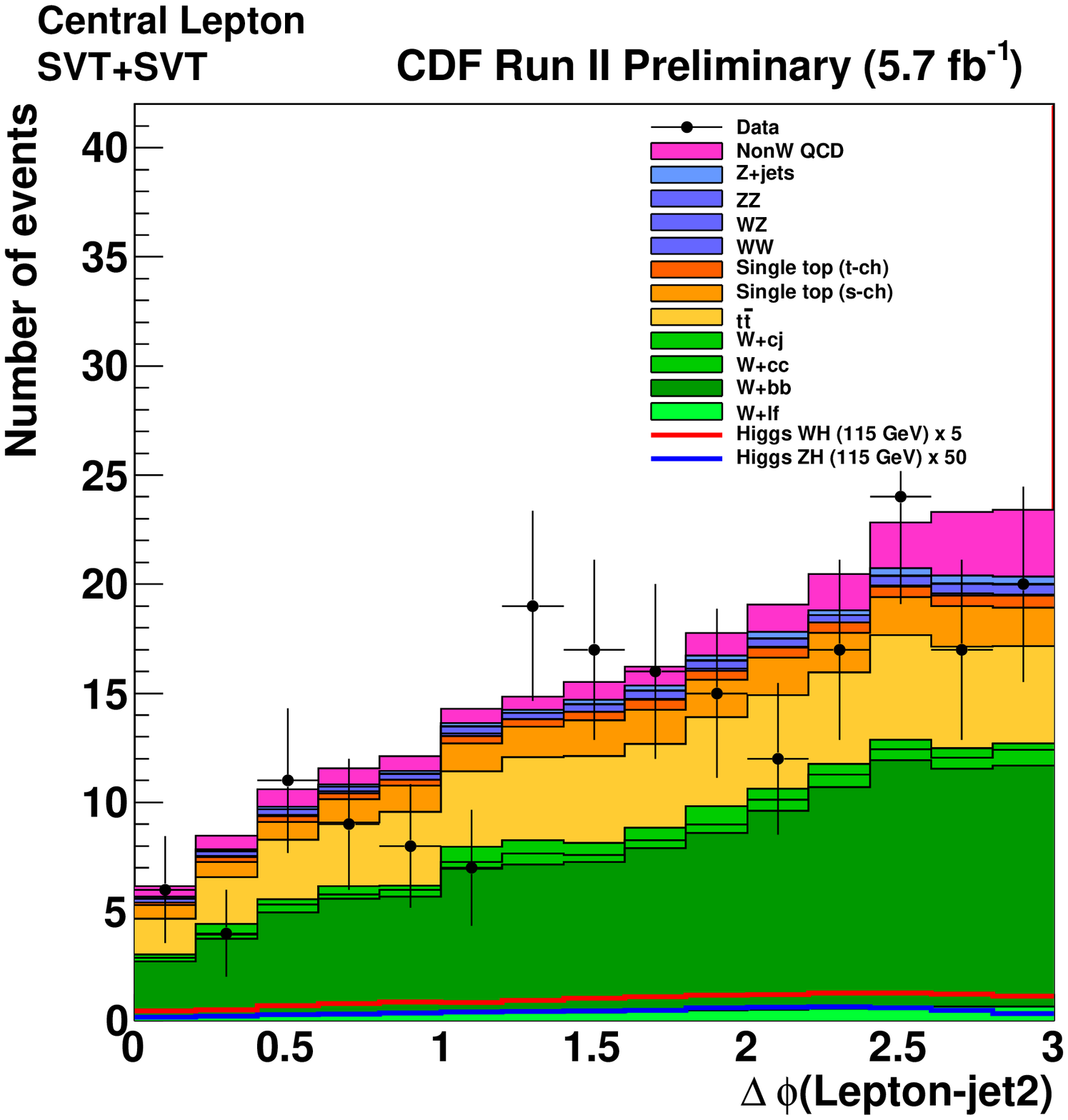}
    \includegraphics[width=6.7cm]{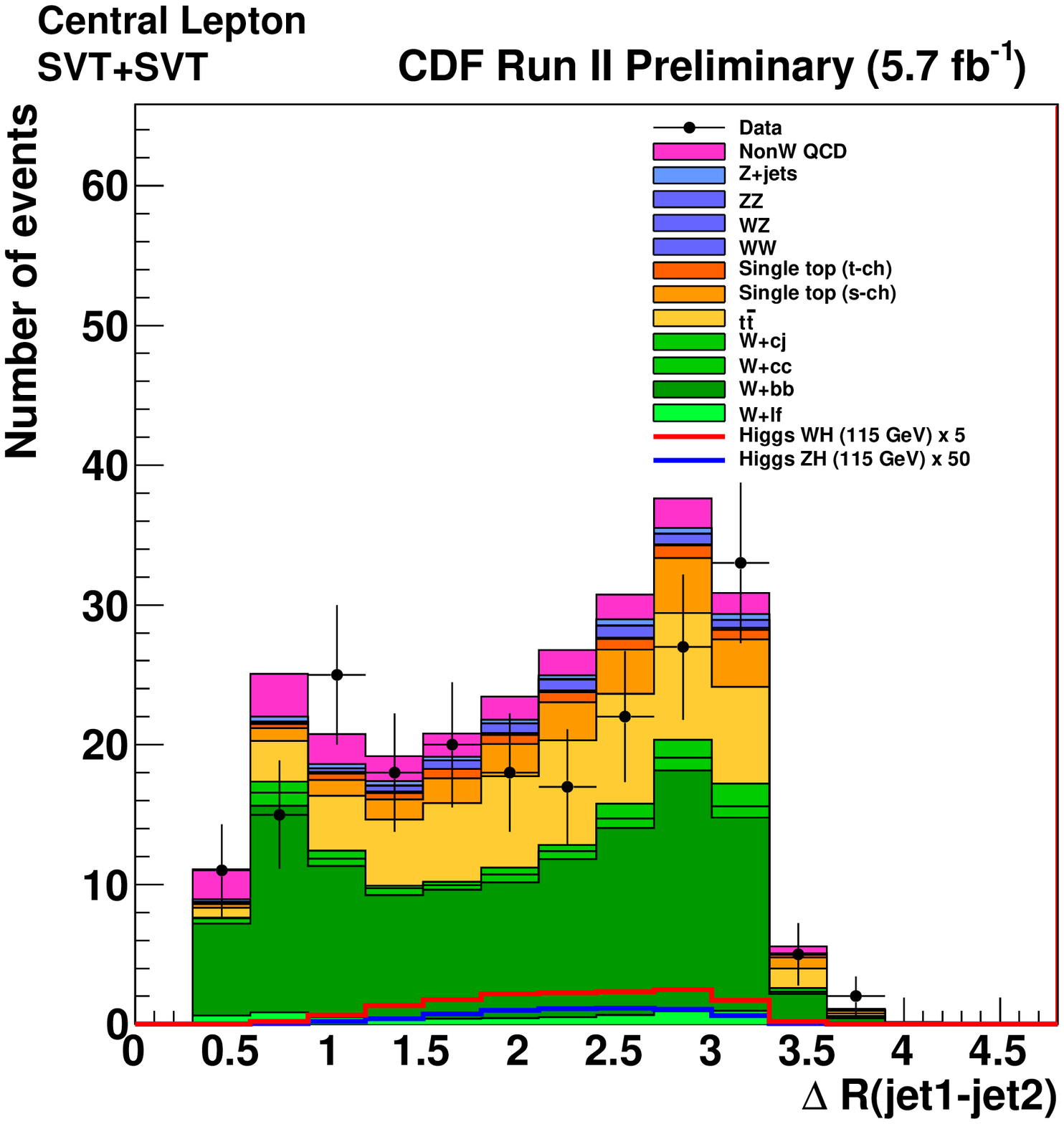}
    \caption [Control plots for TIGHT SVTSVT Kinematic Variables 3/3]{Third part of the control plots for TIGHT charged lepton SVTSVT kinematic variables.}
  \end {center}
\end {figure}

\clearpage

%\clearpage

%%% For BNN inputs and outputs SVTSVT TIGHT charged lepton %%%

\begin{figure}[ht]
  \begin{center}
    \includegraphics[width=6.7cm]{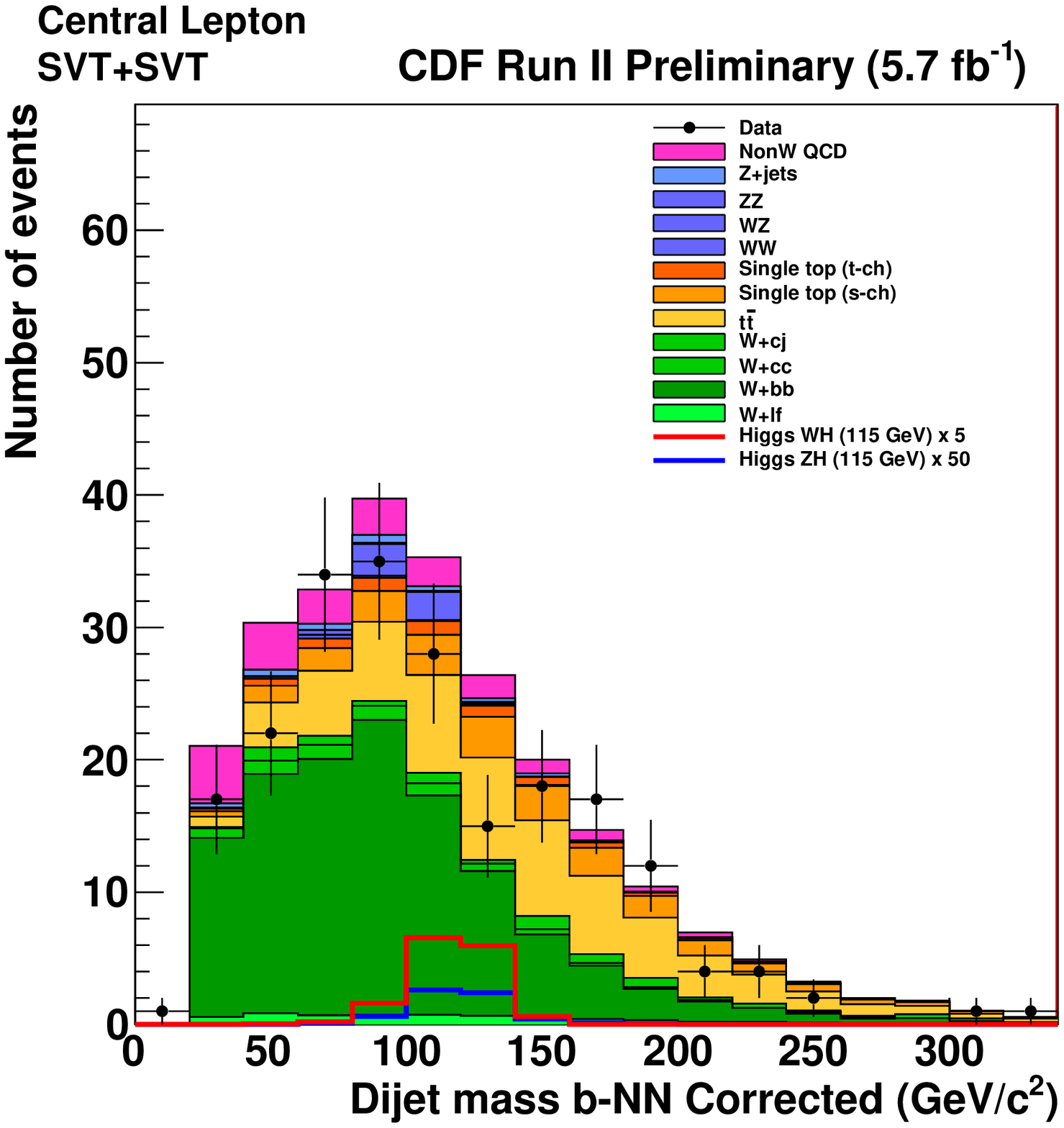}
    \includegraphics[width=6.7cm]{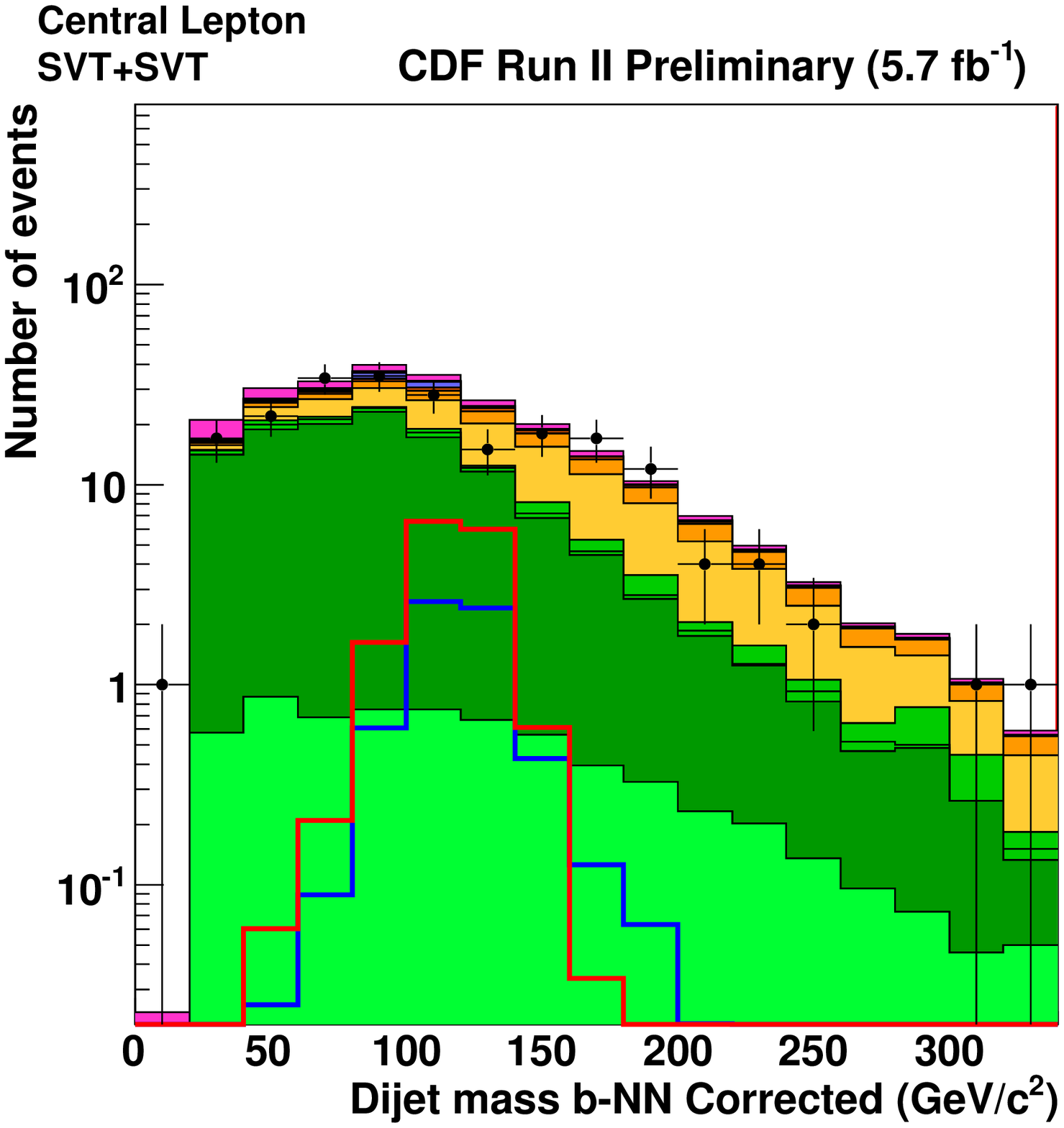}
    \includegraphics[width=6.7cm]{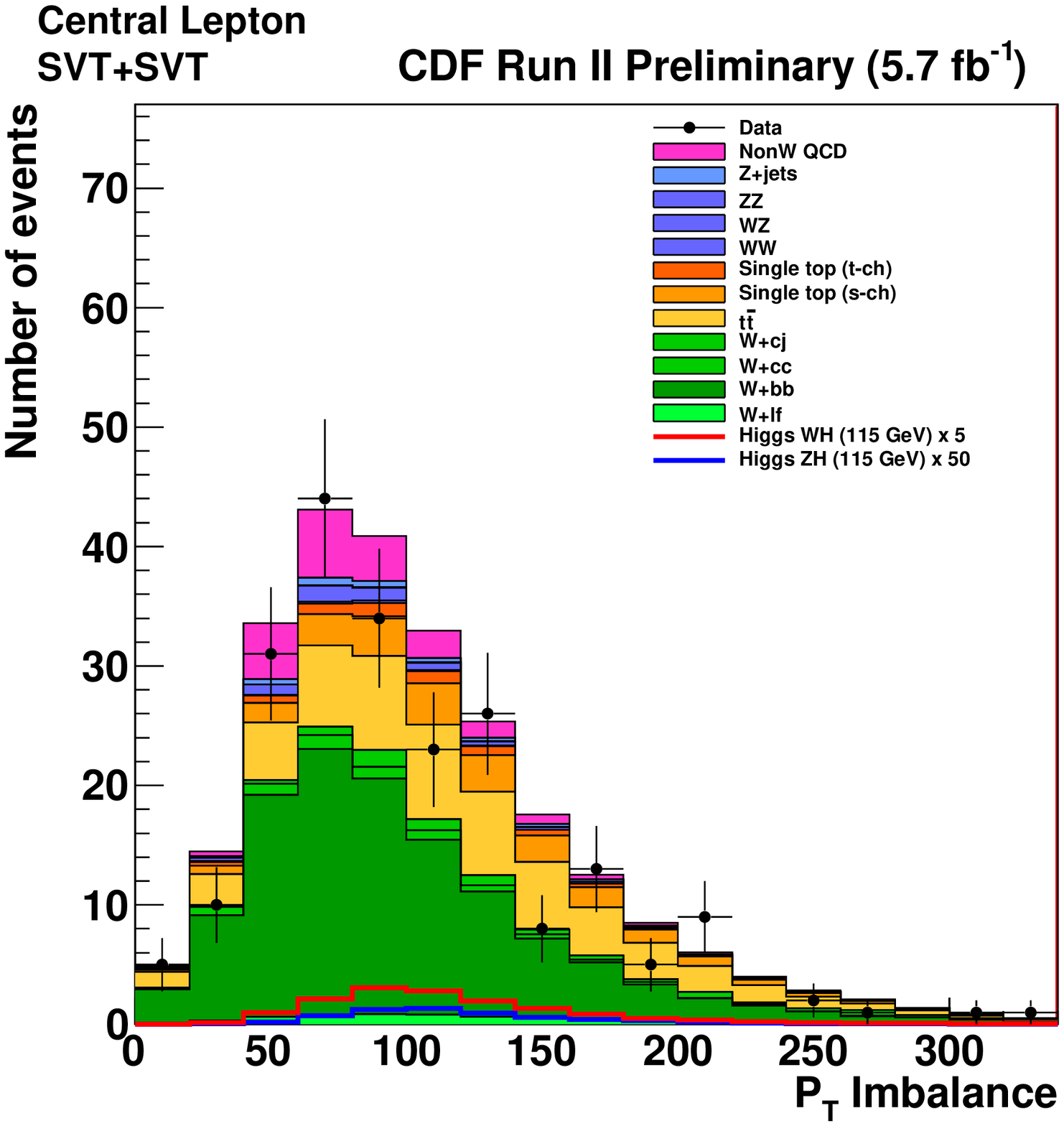}
    \includegraphics[width=6.7cm]{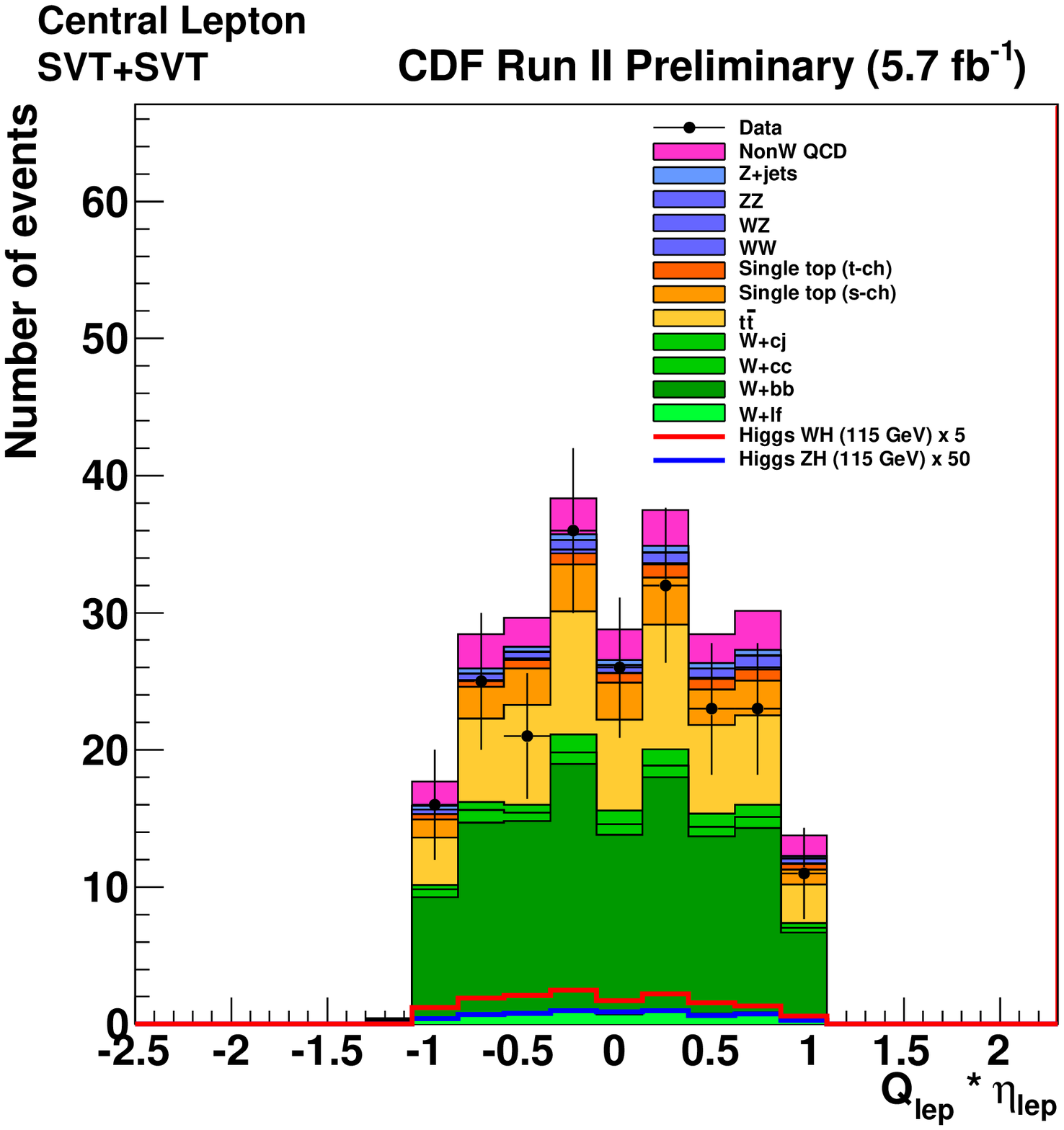}
    \includegraphics[width=6.7cm]{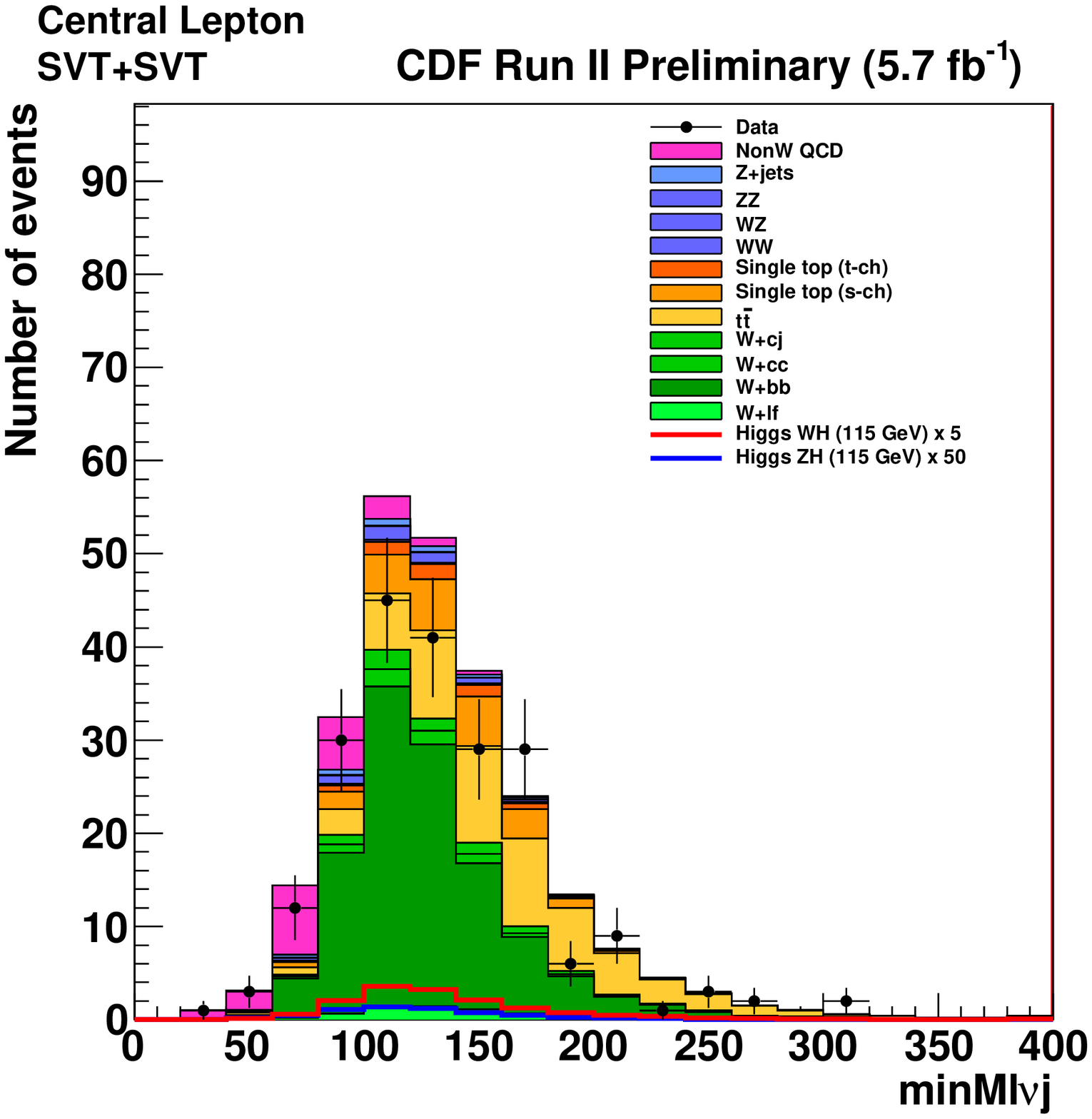}
    \includegraphics[width=6.7cm]{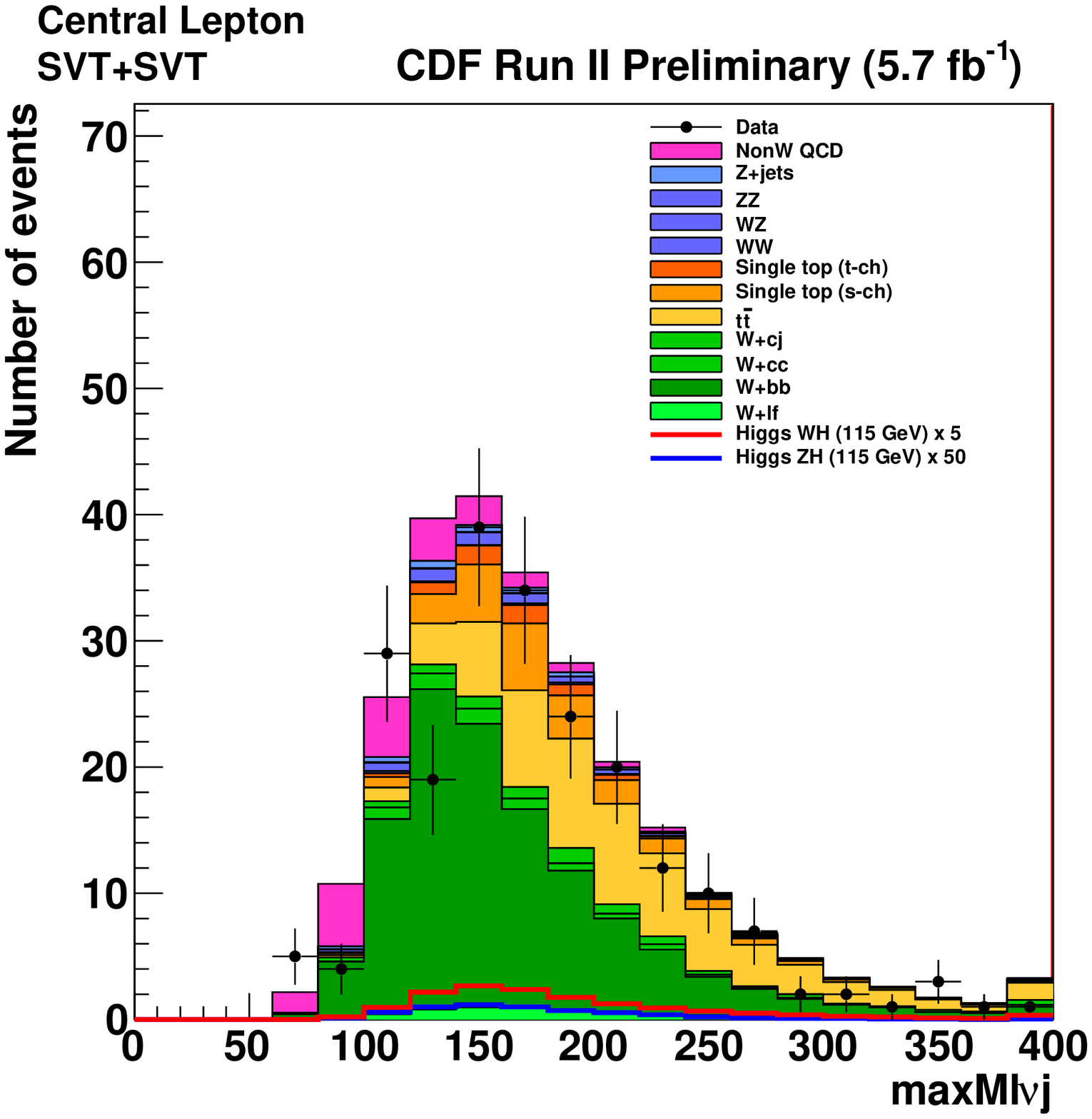}
    \caption [Control plots for TIGHT SVTSVT BNN Input and Output 1/1]{First half of the control plots for TIGHT charged lepton SVTSVT BNN Input Variables.}
  \end {center}
\end {figure}

\begin{figure}[ht]
  \begin{center}
    \includegraphics[width=6.7cm]{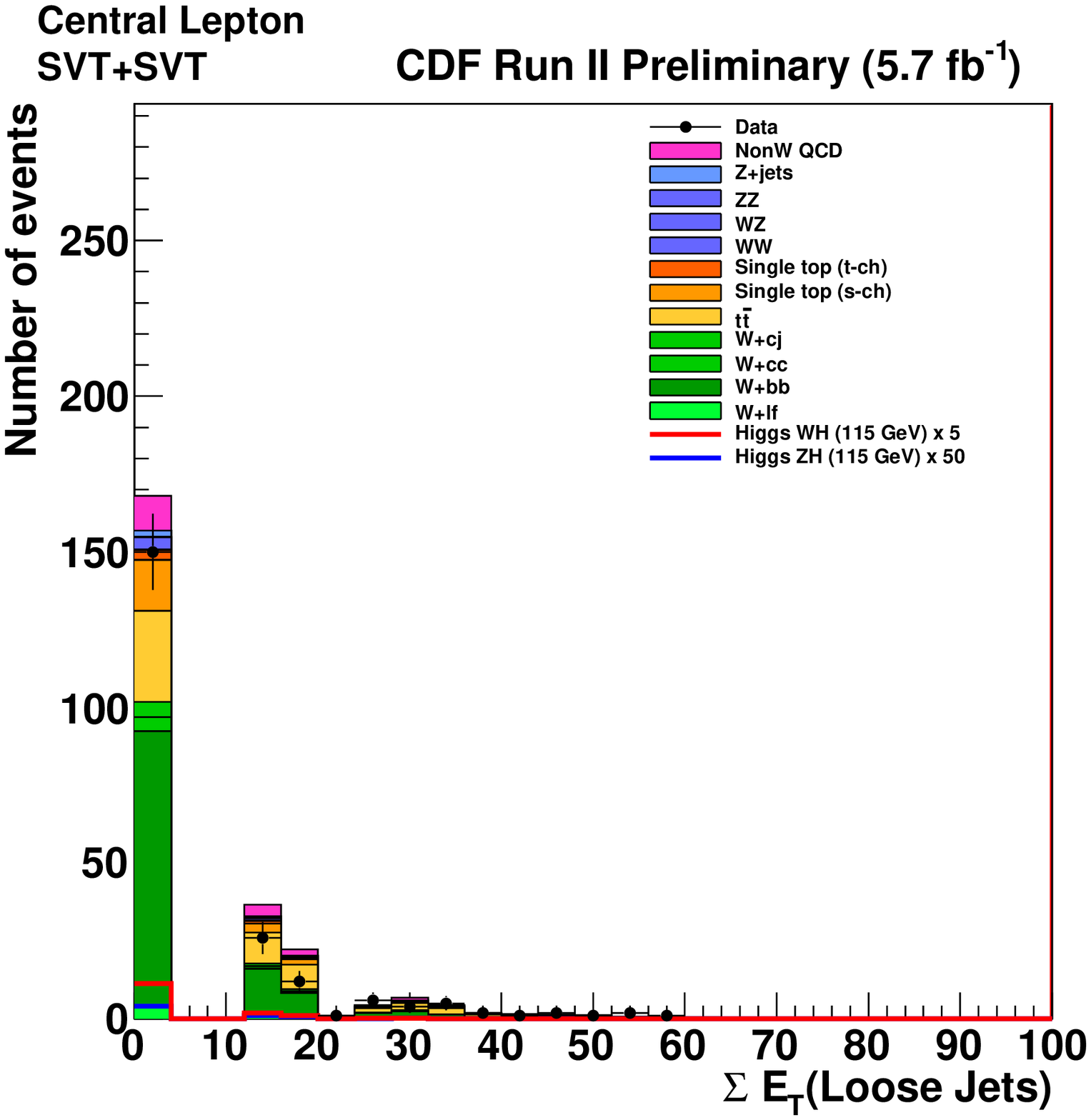}
    \includegraphics[width=6.7cm]{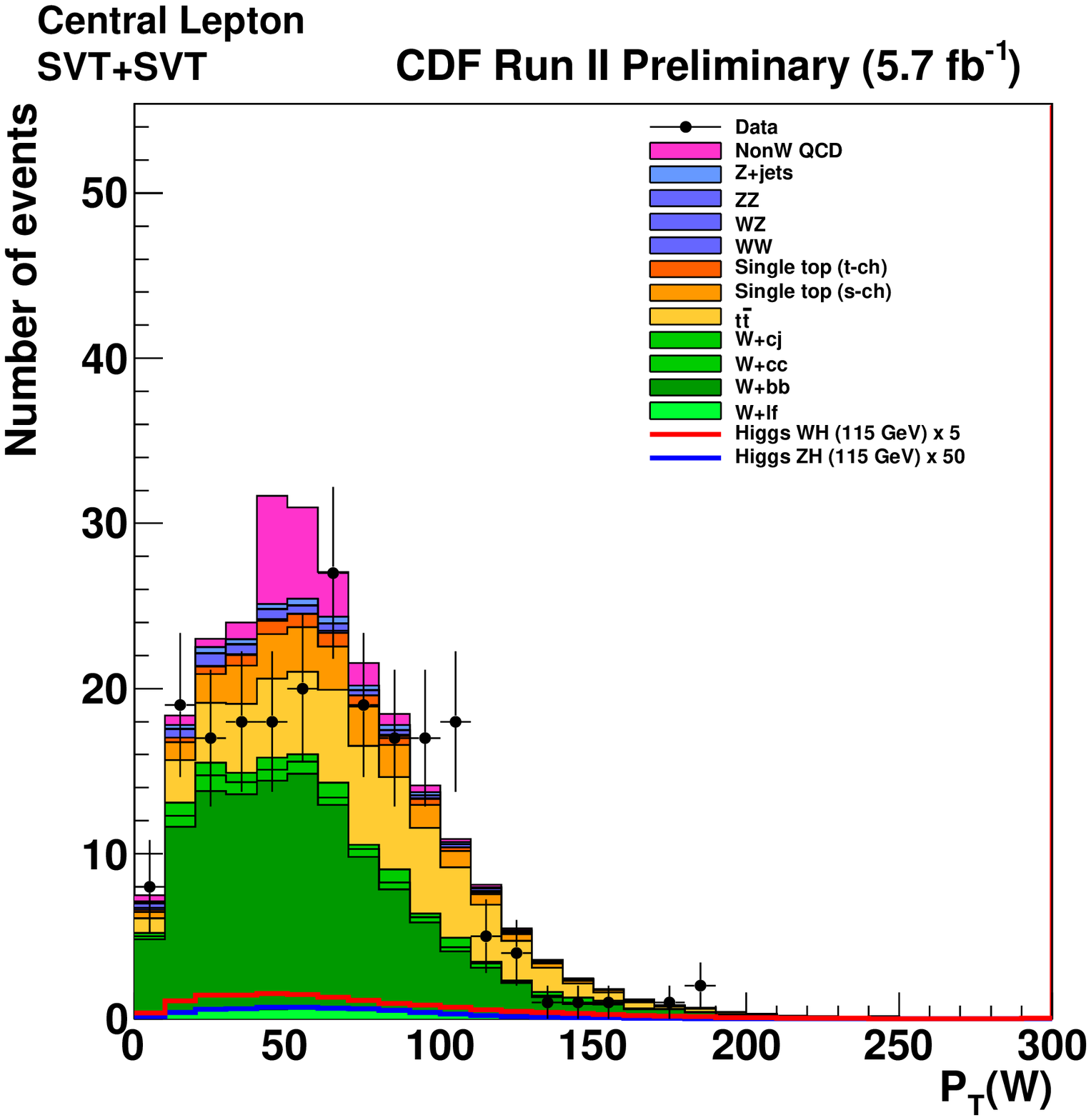}
    \includegraphics[width=6.7cm]{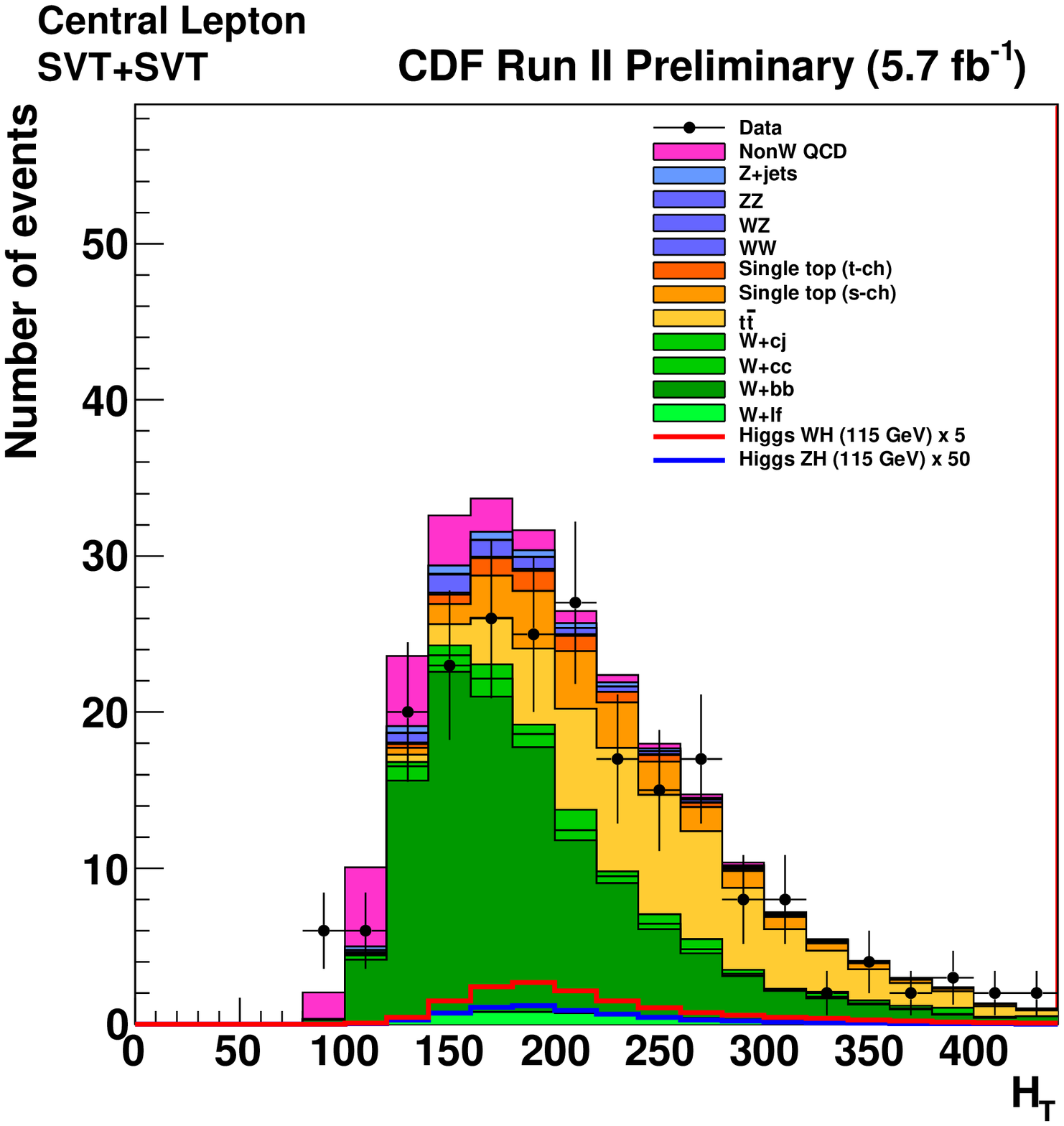}
    \includegraphics[width=6.7cm]{./appendix/ControlPlots/Kinematics/WH_SVTSVT_CEMCMUPCMX_met.eps}
    \includegraphics[width=6.7cm]{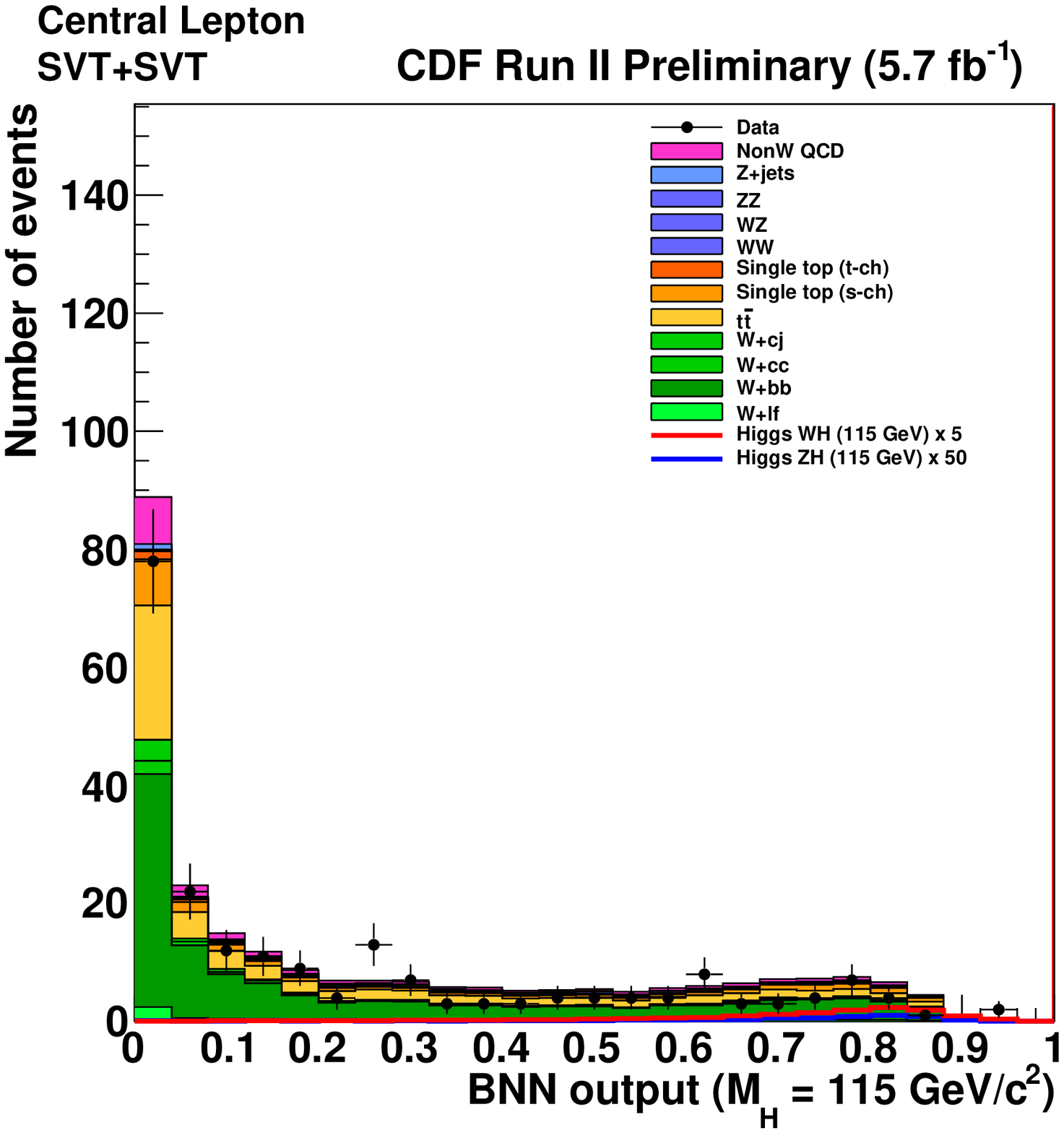}
    \includegraphics[width=6.7cm]{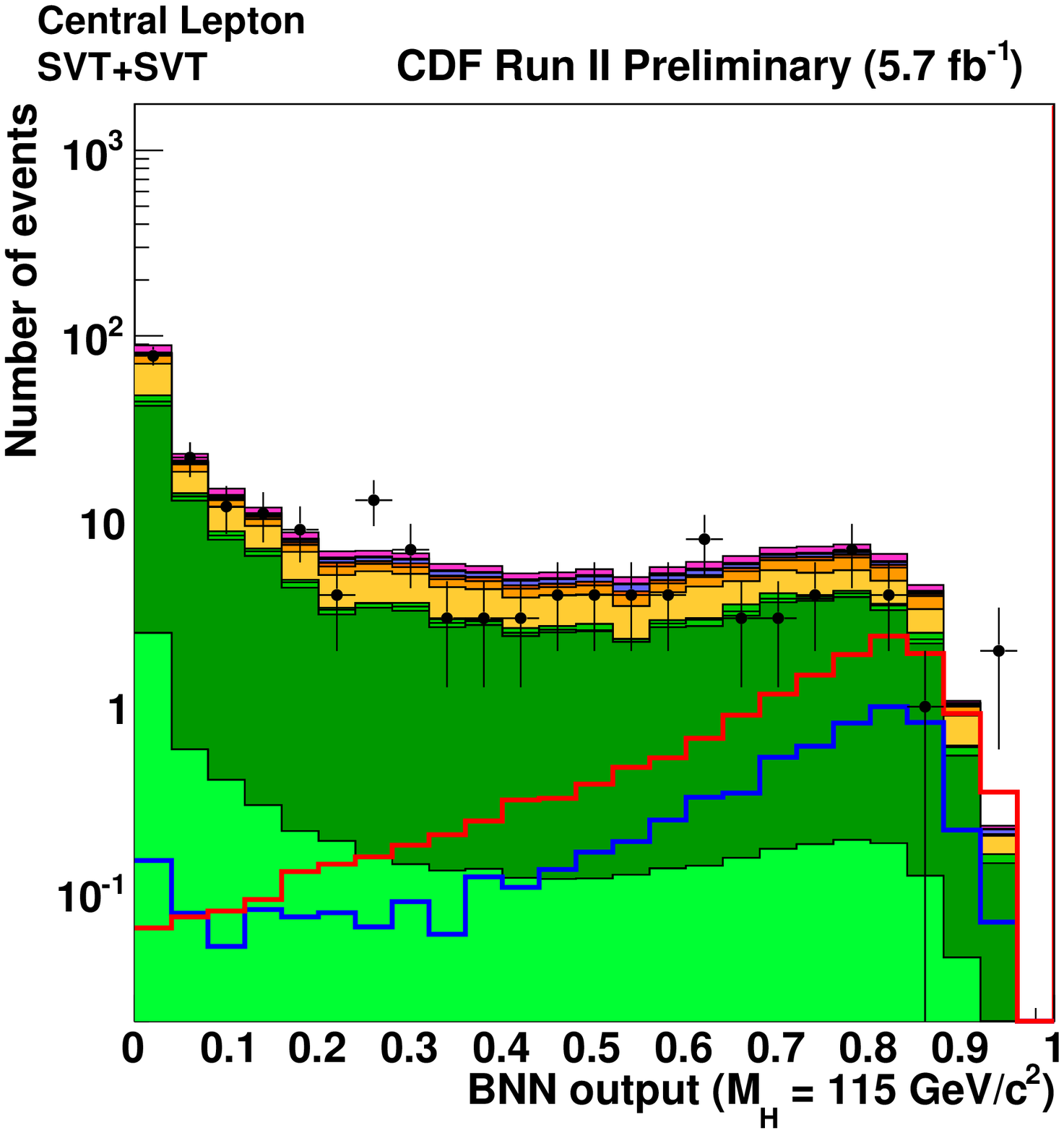}
    \caption [Control plots for TIGHT SVTSVT BNN Input and Output 1/2]{Second half of the control plots for TIGHT charged lepton SVTSVT BNN Input and the BNN Output Variable for the Higgs mass of 115 $\gevcc$.}
  \end {center}
\end {figure}

\clearpage

\clearpage{\pagestyle{empty}\cleardoublepage}
\chapter{Glossary\label{chapter:Glossary}}

\begin{itemize}

\item {\bf{$b$-tagging}} -	The process of identifying if a jet originates in a bottom quark
\item {\bf{ABCDF}} -	The software package I wrote for trigger combination
\item {\bf{ALPGEN}} -	Monte Carlo event generator
\item {\bf{ANN}} -	Artificial Neural Network
\item {\bf{BNN}} -	The final analysis discriminant, the output of an artificial neural network
\item {\bf{BSM}} -	Theories beyond the Standard Model theory
\item {\bf{CDF}} -	Collider Detector at Fermilab
\item {\bf{CEM}} -	Central Electromagnetic Calorimeter
\item {\bf{CES}} -	Central Electromagnetic Shower Maximum Detector
\item {\bf{CHA}} -	Central Hadronic Calorimeter
\item {\bf{CL}} -	Confidence Level (in the frequentist approach) and Credibility Level (in the Bayesian approach)
\item {\bf{CLC}} -	Cherenkov Luminosity Counter
\item {\bf{CMP}} -	Central Muon uPgrade Detector
\item {\bf{CMU}} -	Central Muon Detector
\item {\bf{CMUP}} - 	Muon candidates with hits both in the CMU and CMP detectors
\item {\bf{CMX}} -	Central Muon eXtension Detector
\item {\bf{COT}} - 	Central Outer Tracker, the drift chamber used for tracking
\item {\bf{CSL}} -	Consumer Server/Logger
\item {\bf{DiTop}} -	The background sample of top quark pair production
\item {\bf{FNAL}} -	Fermilab National Accelerator Laboratory
\item {\bf{FSR}} -	Final State Radiation
\item {\bf{ISL}} -	Intermediate Silicon Layers, the third subdetector of the silicon detector
\item {\bf{ISOTRK}} -	The central loose charged lepton category (mainly loose muon candidates)
\item {\bf{ISR}} -	Initial State Radiation
\item {\bf{JES}} -	Jet Energy Scale, one of the most important sources of systematic uncertainty for this analysis, as well as the only shape systematics
\item {\bf{JetProb}} -	Jet Probability  $b$-tagging algorithm
\item {\bf{L00}} -	Layer 00, the first subdetector of the silicon detector
\item {\bf{L1}} -	The first trigger level
\item {\bf{L2}} -	The second trigger level
\item {\bf{L3}} -	The third trigger level
\item {\bf{LHC}} -	The Large Hadron Collider
\item {\bf{MADEVENT}} -	Monte Carlo event generator
\item {\bf{MC}} -	Monte Carlo simulation
\item {\bf{MET}} -	Missing transverse energy
\item {\bf{MET2J}} -	The first of the MET-based triggers
\item {\bf{MET45}} -	The second of the MET-based triggers
\item {\bf{METDI}} -	The third of the MET-based triggers
\item {\bf{Obs}} -	The number of data events observed
\item {\bf{PDF}} -	Parton Distribution Function
\item {\bf{PEM}} -	Plug Electromagnetic Calorimeter
\item {\bf{PES}} -	Plug Electromagnetic Shower Maximum Detector
\item {\bf{PHA}} -	Plug Hadronic Calorimeter
\item {\bf{PMT}} -	Photomultiplier Tube
\item {\bf{Pretag}} -	The sample of events that pass all the event selection requirements, before any $b$-tagging requirement is applied
\item {\bf{PYTHIA}} -	Monte Carlo event generator and parton showering program
\item {\bf{QCD}} -	The background sample of pure QCD production faking a $W$ boson production
\item {\bf{SecVtx}} -	Secondary Vertex $b$-tagging algorithm
\item {\bf{SM}} -	The Standard Model of elementary particles and their interactions
\item {\bf{STopS}} -	The background sample of single top production in the s channel
\item {\bf{STopT}} -	The background sample of single top production in the t channel
\item {\bf{SUSY}} -	The principle of supersymmetry, which lead to several theories beyond the Standard Model
\item {\bf{SVT}} -	Secondary vertex reconstruction at the L2 trigger level
\item {\bf{SVTJP05}} -	One $b$-tagging category, where one jet is tagged by the Secondary Vertex algorithm and the other jet by the Jet Probability algorithm
\item {\bf{SVTnoJP05}} -	One $b$-tagging category, where one jet is tagged by the Secondary Vertex algorithm and the other jet is not tagged by the Jet Probability algorithm
\item {\bf{SVTSVT}} -	One $b$-tagging category, where both jets are tagged by the Secondary Vertex algorithm
\item {\bf{SVX-II}} -	Silicon Vertex Detector, the second subdetector of the silicon detector
\item {\bf{Technicolor}} -	A family of theories beyond the Standard Model
\item {\bf{TIGHT}} -	The central tight charged lepton categories (central electrons and central muons)
\item {\bf{TOF}} -	Time of Flight
\item {\bf{Wbb}} -	The background sample of $W$ boson + $b\bar{b}$
\item {\bf{Wcc}} -	The background sample of $W$ boson + $c\bar{c}$ and  $W$ boson + $cj$
\item {\bf{WH}} -	The main signal process for the Higgs boson search described in this thesis
\item {\bf{WH115}} -	The signal sample of $WH$ associated production when the Higgs boson has a mass of 115 $\gevcc$
\item {\bf{WHA}} -	Wall Hadronic Calorimeter
\item {\bf{WHAM}} -	WH Analysis Modules, the data analysis framework of which I am coauthor
\item {\bf{Wlf}} -	The background sample of $W$ boson + light flavour jets incorrectly tagged as heavy flavour (mistags)
\item {\bf{WW}} -	The background sample of $WW$
\item {\bf{WZ}} -	The background sample of $WZ$
\item {\bf{XFT}} -	eXtremely Fast Tracker - track reconstruction at L1 trigger level
\item {\bf{ZH}} -	The second signal process for the Higgs boson search described in this thesis
\item {\bf{ZH115}} -	The signal sample of $ZH$ associated production when the Higgs boson has a mass of 115 $\gevcc$
\item {\bf{Zjets}} -	The background sample of $Z$ boson + jets
\item {\bf{ZZ}} -	The background sample of $ZZ$

\end{itemize}

\clearpage{\pagestyle{empty}\cleardoublepage}

\newpage

\bibliography{reference}
\bibliographystyle{h-physrev5}

\end{document}